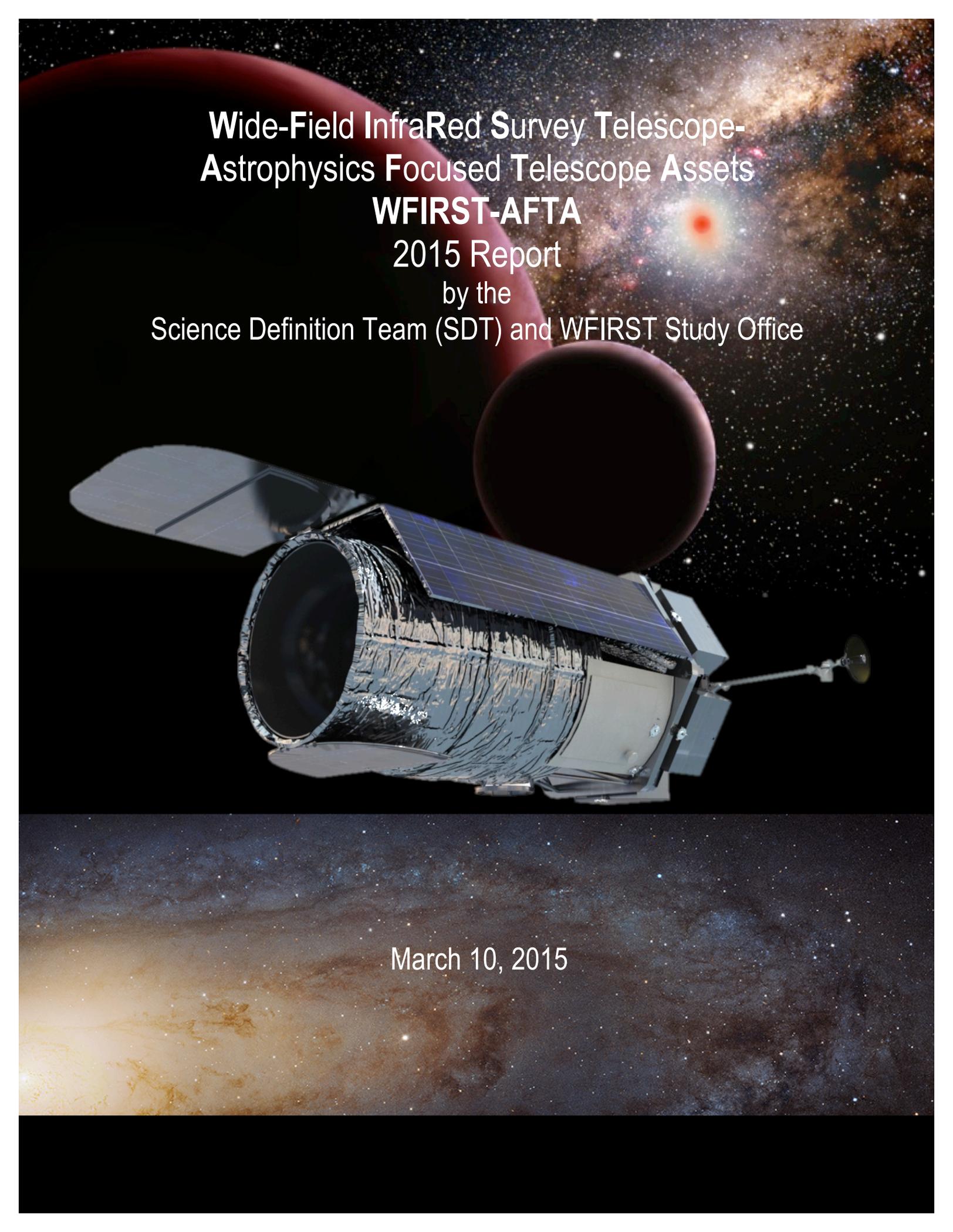

# **W**ide-**F**ield **I**nfra**R**ed **S**urvey **T**elescope-
# **A**strophysics **F**ocused **T**elescope **A**ssets
# **WFIRST-AFTA**
## 2015 Report
### by the
### Science Definition Team (SDT) and WFIRST Study Office

March 10, 2015

# **W**ide-**F**ield **I**nfra**R**ed **S**urvey **T**elescope-**A**strophysics **F**ocused **T**elescope **A**ssets
# WFIRST-AFTA
## 2015 Report

## Science Definition Team (SDT) and WFIRST Study Office

D. Spergel[1], N. Gehrels[2]
C. Baltay[3], D. Bennett[4], J. Breckinridge[5], M. Donahue[6], A. Dressler[7], B. S. Gaudi[8], T. Greene[9], O. Guyon[10]
C. Hirata[8], J. Kalirai[11], N. J. Kasdin[1], B. Macintosh[12], W. Moos[13], S. Perlmutter[14], M. Postman[11], B. Rauscher[2]
J. Rhodes[15], Y. Wang[16,17], D. Weinberg[8],

Ex Officio
D. Benford[18], M. Hudson[19], W. -S. Jeong[20], Y. Mellier[21], W. Traub[15], T. Yamada[22]

Consultants
P. Capak[17], J. Colbert[17], D. Masters[17], M. Penny[6], D. Savransky[23], D. Stern[15], N. Zimmerman[1]

Study Team
R. Barry[2], L. Bartusek[2], K. Carpenter[2], E. Cheng[24], D. Content[2], F. Dekens[15], R. Demers[15], K. Grady[2], C. Jackson[25]
G. Kuan[15], J. Kruk[2], M. Melton[2], B. Nemati[15], B. Parvin[15], I. Poberezhskiy[15], C. Peddie[2], J. Ruffa[2]
J.K. Wallace[15], A. Whipple[24], E. Wollack[2], F. Zhao[15]

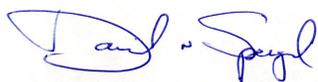

_______________________   3/10/15

David Spergel, SDT-Co-Chair        Date

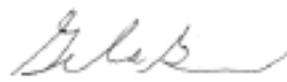

_______________________   3/10/15

Neil Gehrels, SDT Co-Chair        Date

1  Princeton University
2  NASA/Goddard Space Flight Center
3  Yale University
4  University of Notre Dame
5  California Institute of Technology
6  Michigan State University
7  Carnegie Institution for Science
8  Ohio State University
9  NASA/Ames Research Center
10 University of Arizona
11 Space Telescope Science Institute
12 Stanford University
13 Johns Hopkins University
14 University of California Berkeley/Lawrence Berkeley National Laboratory
15 Jet Propulsion Laboratory/California Institute of Technology
16 University of Oklahoma
17 Infrared Processing and Analysis Center/California Institute of Technology
18 NASA Headquarters
19 University of Waterloo (Canadian Space Agency representative)
20 Korea Astronomy and Space Science Institute
21 Institut d'Astrophysique de Paris (European Space Agency representative)
22 Tohoku University Astronomical Institute (Japan Aerospace Exploration Agency representative)
23 Cornell University
24 Conceptual Analytics
25 Stinger Ghaffarian Technologies

## EXECUTIVE SUMMARY

The transfer of the Astrophysics Focused Telescope Assets (AFTA) to NASA enables a truly compelling version of the Wide Field Infrared Survey Telescope (WFIRST), the top priority space mission identified in the 2010 National Academy of Sciences' (NAS) decadal survey, New Worlds New Horizons (NWNH). This report summarizes the science program of WFIRST-AFTA, presents a design reference mission, overviews the mission concept, describes recent progress in advancing the technical readiness of the key components, and shows how WFIRST-AFTA complements other planned ground-based and space-based telescopes to create an observing systems for the 2020s that will transform our understanding of the universe.

The mission envisioned in this report uses the re-purposed 2.4-m telescope equipped with a wide field instrument (WFI), comprising a wide field camera with a field of view (FoV) 200 times larger than HST's powerful WFC3 and an integral field unit (IFU) that will characterize supernovae to trace the evolution of the universe, and a coronagraphic instrument (CGI), which will be the first instrument able to characterize the atmospheres of super-Earth planets and Neptune-like planets around nearby Sun-like stars. This mission addresses all three of the questions identified for astrophysics in the NASA 2014 Science Plan[1]:

- *How does the universe work?*
- *How did we get here?*
- *Are we alone?*

The WFIRST-AFTA design reference mission (DRM) is a four part observing program comprising (1) a high-latitude survey optimized to study dark energy but enabling an enormous variety of other investigations, (2) a galactic bulge survey that will use micro-lensing observations to complete the planetary census begun by Kepler, (3) coronagraphic observations of nearby planets and proto-planetary systems, and (4) a Guest Observer program that will utilize the power of the WFI and the CGI to address a wide-ranging set of open problems in astrophysics. This planned program will make major contributions towards all three of the goals identified for astrophysics in the NASA Science Plan:

- *Probe the origin and destiny of our universe, including the nature of black holes, dark energy, dark matter and gravity.*
- *Explore the origin and evolution of the galaxies, stars and planets that make up our universe*
- *Discover and study planets around other stars, and explore whether they could harbor life*

Together with JWST, WFIRST is identified as a bedrock near-term mission for all three of these science themes in the recent roadmap report Enduring Quests, Daring Visions: NASA Astrophysics in the Next Three Decades (Kouveliotou et al. 2014).

NWNH emphasized that a wide field infrared survey telescope could produce compelling science in areas ranging from the study of dark energy to the search for exoplanets. The 2013 Science Definition Team report concludes: "If used for the WFIRST mission, the 2.4-meter telescope would be significantly more capable than the smaller versions of WFIRST studied in previous SDTs." The National Academy of Sciences' "Evaluation of the Implementation of WFIRST-AFTA in the Context of New Worlds, New Horizons in Astronomy and Astrophysics (2014)"[2] chaired by Fiona Harrison reinforced this view noting:

> "…a configuration with a 2.4-m telescope utilizing hardware made available to NASA by the National Reconnaissance Office, would enable a mission of the scientific scope envisioned for WFIRST. The larger aperture of WFIRST-AFTA offers the potential of substantially greater scientific reach than is possible with WFIRST/IDRM. The large aperture also makes inclusion of a coronagraph attractive. This addition was not envisioned by NWNH as part of WFIRST, but it has the potential to advance NWNH objectives aimed at the eventual realization of a future Earth-like planet imaging mission that was a high priority of the NWNH survey."

The 2.4-m aperture of WFIRST-AFTA collects almost three times as much light as the originally envisioned version of the mission and offers a factor of 1.9 improvement in spatial resolution (point spread function

---

[1] http://science.nasa.gov/media/medialibrary/2014/05/02/2014_Science_Plan-0501_tagged.pdf

[2] http://www.nap.edu/openbook.php?record_id=18712

effective area -- PSF), which itself provides another factor-of-two improvement in accomplishing many science programs. In particular, having the greater speed of the

*Probe the origin and destiny of our universe, including the nature of black holes, dark energy, dark*

---

**Box 1**

## MAPPING ANDROMEDA
(see http://hubblesite.org/newscenter/archive/releases/2015/02/image/a/format/small_web/)

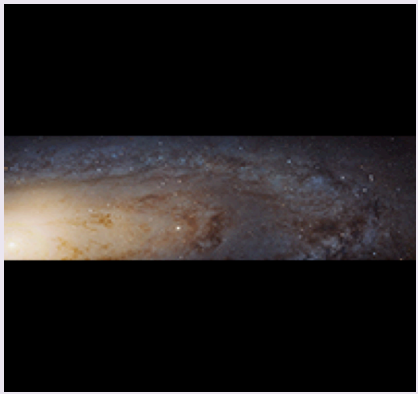

At the January 2015 AAS meeting, astronomers presented the largest NASA Hubble Space Telescope image ever assembled, this sweeping bird's-eye view of a portion of the Andromeda galaxy (M31) [http://hubblesite.org/newscenter/archive/releases/2015/02/image/a/]. The Hubble telescope is powerful enough to resolve individual stars in a 61,000-light-year-long stretch of the galaxy's pancake-shaped disk. It's like photographing a beach and resolving individual grains of sand. And, there are lots of stars in this sweeping view — over 100 million, with some of them in thousands of star clusters seen embedded in the disk.

The Panchromatic Hubble Andromeda Treasury (PHAT) assembled together into a mosaic image using 7,398 exposures taken over 411 individual pointings. The panorama is the product of the program. Images were obtained from viewing the galaxy in near-ultraviolet, visible, and near-infrared wavelengths, using the Advanced Camera for Surveys and the Wide Field Camera 3 (WFC3) about Hubble. **With 200 times the field of view of WFC3, WFIRST-AFTA could replicate the WFC3/IR observations with only 2 pointings.**

---

larger aperture and the sharper PSF allows WFIRST-AFTA to reach a factor-of-two deeper per unit time over an unprecedentedly large field for a large space telescope, ~90 times bigger than the HST–ACS FoV, and ~200 times bigger than the IR channel of WFC3 that has been a tremendously successful scientific tool. Our international partners (Canada, Japan, South Korea, UK, France, Germany and Spain) have expressed significant interest in partnering in developing this tremendously powerful telescope. Their potential contributions are described in Appendix I of the report.

Section 2 describes the WFIRST-AFTA science program. Section 3 describes the DRM that achieves this science program. Section 4 suggests policies for maximizing WFIRST-AFTA science through developing well thought out surveys, making its remarkable images available as quickly as possible and selecting Guest Observer programs through a competitive process. Sections 5 and 6 outline the path forward and conclude.

This science program addresses all three of NASA's goals for astrophysics in multiple ways:

*matter and gravity.*

By making precision measurements of both the growth rate of structure and the expansion history of the universe, WFIRST-AFTA will test general relativity on the largest scales and could reveal the nature of dark energy, the driver of cosmic acceleration, which is one of the great mysteries of modern science. In 2011, the Nobel Prize committee awarded its physics prize to three American scientists who used observations of supernovae to provide the first convincing evidence for cosmic acceleration. The use of a 2.4-meter telescope will enhance our ability to use the three basic techniques that were the focus of previous WFIRST mission studies (baryon acoustic oscillations, supernovae, and weak gravitational lensing) and will also significantly improve the use of two other important techniques: galaxy cluster counts and redshift space distortions. Specific strengths of WFIRST-AFTA for dark energy studies include:

- In addition to using lines of hydrogen atoms in galaxies to trace the large-scale distribution of matter at redshifts between 1 and 2, WFIRST-AFTA's larger collecting area will enable it to also use lines of ionized oxygen in more distant galaxies. This added capability enables measurements of large-scale structure out to nearly redshift 3, just two billion years after the big bang.
- WFIRST-AFTA's added sensitivity will enable the detection of a 60% higher density of galaxies in 10% less time at z=1.5 in its spectroscopic survey vs. WFIRST-DRM1 (Green et al. 2012). With this higher density of galaxies, the redshift space distortion measurements (measuring the effects of cosmic acceleration on structure growth) are also improved.
- WFIRST-AFTA will conduct a three tiered deep supernova survey, a shallow survey over 27.44 deg$^2$ for SNe at z <0.4, a medium survey over 8.96 deg$^2$ for SNe at z <0.8, and a deep survey over 5.04 deg$^2$ for SNe out to z =1.7 and then will use an Integral Field Unit (IFU) to take spectra of ~2700 distant supernovae discovered in its synoptic imaging survey. This enables more precise distance measurements and reduces uncertainties associated with cosmic evolution.
- WFIRST-AFTA will measure the shapes of 380 million galaxies over 2200 square degrees. By making measurements at multiple bands in an observing strategy designed to control for systematics, the DRM is designed so that a potential discovery of deviations from general relativity will not be limited by systematic error. The DRM's deeper observations will complement Euclid's and LSST's wider and shallower observations and will enable precision measurements of non-Gaussian features in the lensing maps, thus, significantly improving the dark energy measurements. These deeper lensing observations will also enable accurate determination of the masses of galaxy clusters whose abundance then provides a complementary method to determine the effects of dark energy on the growth rate of large structure.

The report reinforces the NAS Harrison report's findings that "WFIRST-AFTA observations will provide a very strong complement to the Euclid and LSST data sets" and "For each of the cosmological probes described in NWNH, WFIRST-AFTA exceeds the goals set out in NWNH".

***Explore the origin and evolution of the galaxies, stars and planets that make up our universe***

With the spatial resolution of HST's powerful WFC3/IR camera and more than 200 times its field of view, WFIRST-AFTA will impact a broad range of astrophysics.

- WFIRST-AFTA will make an accurate three-dimensional map of the distribution of dark matter across 2200 square degrees of sky. This map will reveal new insights into the relation between baryonic structures and dark matter. When combined with WFIRST-AFTA tracing galaxies across cosmic time through its high-resolution images, this dark matter map will be a Rosetta stone for understanding the process of galaxy formation and evolution.
- WFIRST-AFTA will be able to make "Degree-Deep-Fields" almost 100 times larger than the famed Hubble deep fields. These samples of millions of galaxies over billions of years will make fundamental contributions to our understanding of galaxy evolution. As discussed in Appendix H, a strategy of combining these deep fields with LSST's deep drilling field will significantly enhance the science return from both programs. Appendix H describes the many synergies between LSST, Euclid and WFIRST-AFTA: a synthetic observing program and data analysis system will produce even richer science from the sum of the whole system.
- WFIRST-AFTA will determine the positions and motions of more than 200 million stars in our Galaxy to unprecedented precision. With its large collecting area, it will be able to study stars up to 500 times fainter than the European Gaia telescope, a dedicated astrometric mission. Its larger aperture will enable WFIRST-AFTA to achieve a given astrometric (positional) precision for a faint star nine times faster than a 1.3-meter unobscured version of WFIRST. These measurements should be able to trace the dark matter distribution not only in our Galaxy, but also in our poorly understood dwarf companions, an important la-boratory for determining the mass and interaction properties of dark matter.

- WFIRST-AFTA's wide field and superb image quality will make it the ideal telescope to find counterparts to the gravitational wave merger events that will likely be detected in large numbers by the upgraded LIGO experiment in Washington and Louisiana. More generally, WFIRST-AFTA will provide a newer, deeper view of the transient universe.

- WFIRST-AFTA will enable detailed studies of the properties of stars not only in our own Milky Way, but also over the full extent of all neighboring galaxies. These comparative studies will greatly inform our understanding of both stellar evolution and galaxy formation.

The Hubble telescope has been making astounding contributions across a wide range of astrophysics. WFIRST-AFTA's Guest Observer program will extend this process of discovery. Appendix D contains a rich set of ~40 potential Guest Investigator (GI) and Guest Observer (GO) science programs that are uniquely enabled by WFIRST-AFTA and address many of the key questions identified in the decadal survey. Appendix E describes some of the conclusions from the Wide-Field InfraRed Surveys: Science and Techniques conference held in Pasadena, CA on November 17-20, 2014, where over 200 scientists gathered to discuss the wide range of potential uses of such a powerful survey telescope. As has proven extremely successful on other NASA missions, we expect that the selection of GO science programs will be made by a peer-reviewed Time Allocation Committee process and GI programs will be enabled through archival analysis of WFIRST-AFTA's survey data sets and the examples in Appendix D just give a sense of possibilities. Historical experience with HST and other great observatories suggests that our current list and discussions do not include many of the most important results that will be achieved by WFIRST-AFTA, as many of the visionaries who will use the telescope for novel applications are still in high school!

If the two missions are operating simultaneously, WFIRST-AFTA and the James Webb Space Telescope (JWST) are highly complementary and will be mutually enhancing. Appendix B describes some of the synergies between the two missions. By surveying thousands of square degrees of the sky, WFIRST-AFTA should be able to detect galaxies forming in the first few hundred million years after the big bang and could potentially detect supernova explosions occurring in the early universe. JWST's powerful spectroscopic capabilities will enable the detailed study of these rare objects. WFIRST-AFTA will discover myriads of strongly lensed quasars and galaxies that will be exciting objects for JWST follow-up observations. This powerful combination of WFIRST-AFTA wide field survey finding rare, but important, objects and JWST performing detailed follow-up study should also yield important results in galactic and extragalactic astronomy.

### Discover and study planets around other stars, and explore whether they could harbor life

Over the past five years, the Kepler mission has revolutionized our understanding of the properties of extrasolar planets. However, Kepler tells only part of the story: the transit method used by Kepler is not sensitive to planets in orbits significantly larger than that of the Earth, including analogs to all of the outer giant planets in the Solar System. Kepler is only sensitive to "hot" to "warm" exoplanets. WFIRST-AFTA's gravitational microlensing survey of the Galactic Bulge will complete Kepler's program through a technique most sensitive to "cold exoplanets". It will systematically survey the cold, outer regions of planetary systems throughout the Galaxy, detecting ~2600 total bound exoplanets in the range of 0.1-10,000 Earth masses, including ~1000 "Super-Earths" (roughly 10 times the mass of Earth), ~370 Earth-mass planets, and ~50 Mars-mass planets. This will enable the measurement of the mass function of "cold" exoplanets better than ~10% per decade in mass for masses >0.3 $M_{Earth}$, and an estimate of the frequency of Mars-mass embryos accurate to ~15%. When combined with Kepler, WFIRST-AFTA will provide statistical constraints on the characteristic density of rocky planets in the outer habitable zones of Sun-like stars.

WFIRST-AFTA will measure the frequency of free-floating planetary-mass objects in the Galaxy over nearly six orders of magnitude in mass, and will detect ~30 free-floating Earth-mass planets if there is one per star in the Galaxy.

WFIRST-AFTA will determine the basic properties of the microlensing-detected planetary systems, the planetary masses and the separation between star and planets, to better than ~10% for most systems. In addition, spectroscopic follow-up with JWST will allow for the measurement of host star temperature and metallicities for a subset of the most massive hosts in the disk and bulge.

WFIRST-AFTA will be sensitive to the transits of giant planets around 200 times more stars than Kepler, and 40 times more stars than TESS. WFIRST-AFTA will discover about 20,000 transiting planets with orbits less than a few tenths of an AU, with sensitivity down to Neptune-radius objects. By combining the transit and microlensing discoveries, WFIRST-AFTA will determine the frequency of giant planets over all orbital separations (from 0 to infinity) from a single survey.

Appendix G describes the needed precursor observations to maximize the science return from the WFIRST-AFTA microlensing program

One of NASA's long-term goals is to image and characterize an Earth-like planet around a nearby star. WFIRST-AFTA's coronagraph will be a major step towards this goal and will achieve NWNH's top "medium scale" priority for NASA by advancing high-contrast imaging in space. The CGI directly responds to this recommendation, providing an unparalleled opportunity for space-flight experience with a very high-contrast coronagraph and wavefront control system. WFIRST-AFTA is by no means a coronagraph-optimized telescope – but it is likely that future coronagraphs will also not have the luxury of designing to a perfect unobscured telescope, since flagship missions will have to satisfy a range of science goals just as WFIRST does. Thus, the WFIRST-AFTA coronagraph will be the prototype for a future mission that characterizes Earth-like planets. In addition to advancing towards this goal, the coronagraph will enable an exciting science program:

- The CGI will be the first instrument to characterize the atmospheres of gaseous planets around nearby Sun-like stars using reflected visible-light images and low-resolution (R ~ 70) spectra. These data, along with masses from existing radial velocity (RV) measurements, will give us unique information on planetary compositions, temperatures, and cloud layers that are not available via transmission and emission spectroscopy of transiting planets that have been observed by HST and Spitzer. Currently, only a handful of planets have been directly imaged and spectrally characterized, all of which are large, young, self-luminous gas giants far from their host star. WFIRST-AFTA will characterize over a dozen currently known RV planets down to Jupiter size and as close as 1 AU to their parent star. Upcoming RV surveys will likely provide a dozen more candidate targets.

- WFIRST-AFTA can study "Super-Earths" and "mini-Neptunes", classes of planets that were unknown until we were able to see planets beyond our solar system. One of the most striking discoveries about extrasolar planets made in the past decade is the large population of objects with radius between 1 and 4 times that of the Earth. Based on Kepler observations, these planets – which have no analog in our solar system – may be the most common kind of planet in the universe. The exact nature of these planets is poorly known. They could be so-called "super-Earths" with rocky compositions, or "mini-Neptunes" with massive hydrogen envelopes, or even stranger objects, partially or mostly water and ice. In a blind search, WFIRST-AFTA could potentially discover up to 6 mini-Neptunes and potentially 4 Super-Earths. Although WFIRST-AFTA's ability to characterize the smallest planets will be limited, the detections themselves will yield new science as well as provide targets for future missions dedicated to terrestrial habitable-zone planets.

- The WFIRST-AFTA CGI will deliver uniquely sensitive images of the planetary disks around nearby stars, resolving much smaller amounts of dust at much closer distances than possible with HST or other observatories. The small inner working angle and high contrast of the WFIRST-AFTA CGI will allow unique study of several types of circumstellar disks enabling it to address many of the most important questions in planet formation: What are the levels of circumstellar dust in, and exterior to, the habitable zones of exoplanetary systems? Will dust in habitable zones interfere with future planet-finding space observatories? How do dust sub-structures seen in disks trace the presence of seen or unseen planets? What veneer is delivered to planetary atmospheres and surfaces by asteroids, comets, and other material? How do disks of protoplanetary materials evolve to make Solar System-like architectures? The CGI imaging data collected in the planet surveys described above will be sufficient for detecting disks around nearby stars down to a level of ~10 times that of our solar system's zodiacal dust. This combined planet and disk dataset will provide the first full pic-ture of planetary systems around tens of near-by stars.

Appendix F describes the basis for estimating the expected CGI science yield for three groups of targets, known RV planets, "new" as-yet unknown blind-search planets, and debris disks.

While adding a coronagraph to the WFIRST mission enables very compelling science and advances our planet-imaging capabilities, the NAS Harrison report notes that there are significant risks associated with developing new technologies. The NAS Harrison report recommends:

"NASA should move aggressively to mature the coronagraph design and develop a credible cost, schedule, performance, and observing program so that its impact on the WFIRST mission can be determined."

Enabled by the enhanced funding in the 2014 Federal budget, the WFIRST-AFTA team has made significant progress in realizing this recommendation. As described in §3.4, the coronagraph design is based on the highly successful High Contrast Imaging Testbed (HCIT), with modifications to accommodate the telescope design, serviceability, volume constraints, and the addition of an Integral Field Spectrograph. Soon after the WFIRST-AFTA CGI architecture downselect results were announced in December 2013, the coronagraph technology development plan (TDP) was drafted and approved in March 2014. The main objective of the TDP is to mature the WFIRST-AFTA coronagraph to Technology Readiness Level (TRL) 5 by the planned Project Phase A start at the beginning of FY 2017. §3.4.2 describes nine key milestones in FY 2014-2016 needed to achieve this goal. During calendar 2014, the WFIRST program achieved the first three milestones and is on schedule to ensuring that the coronagraph is ready for integration into the WFIRST-AFTA telescope.

The NAS Harrison report also identifies the concern that "The mission may have to compromise some science performance to ensure that issues associated with the low thermal margins do not lead to significant cost growth and schedule delay". The present study mitigates the risk of low thermal margins by baselining a telescope operating temperature of 282 K, within the currently qualified range of the telescope. Appendix A considers the impact of the 282 K baseline and concludes that the loss in sensitivity is small at Y, J and H bands and 0.28 mag at F184. The section concludes that this sensitivity loss does not have a significant impact on any of the key science programs and that at the new baseline operating temperature the mission achieves its scientific objectives and now has larger thermal margins.

To fully take advantage of this larger telescope, the wide-field instrument uses higher performance infrared detectors (H4RGs) that provide four times as many pixels in about the same size package as the current generation of detectors. With eighteen of these detectors in a single focal plane, the wide field imager covers 0.28 deg$^2$ of sky in a single image at a pixel scale of 0.11 arcsec/pix. §3.3.2 details the significant progress made in the past two years in advancing the H4RG-10 detectors, a critical technology for the mission.

In the mission presented in this report, WFIRST-AFTA is deployed in a 28.5 degree inclined geosynchronous orbit; however, this is not the only option for science operation. Appendix C discusses the relative merits of inclined geosynchronous and Sun-Earth L2 orbits.

WFIRST-AFTA will address many of the most profound questions in astrophysics and should be a key part of NASA's mission portfolio for launch by 2024. As described in Appendix J, it is an important part of the constellation of tools that astronomers will use to obtain novel insights in the coming decade. With the availability of the 2.4-meter telescope, it is an even more compelling mission than originally envisioned in the decadal survey.

# 1   INTRODUCTION

This report contains the findings of a NASA-appointed Science Definition Team (SDT) to study the Astrophysics Focused Telescope Assets (AFTA) implementation of the Wide-Field Infra-Red Survey Telescope (WFIRST) mission. It follows the study of other implementations by the previous SDT as described in the 2012 report (Green et al. 2012). In July 2013, the Director of the Astrophysics Division of NASA's Science Mission Directorate charged the WFIRST-AFTA SDT to work with the WFIRST Study Office to continue studying the design reference mission (DRM) for WFIRST, using an existing 2.4 meter telescope which was provided to NASA and baselined a coronagraph instrument for exoplanet imaging in addition to the wide-field infrared imager. The existing telescope significantly reduces the development risk of the WFIRST. This document fulfills that charge.

The NRC's 2010 decadal survey of astronomy and astrophysics, "New Worlds, New Horizons" (henceforth NWNH) gave WFIRST the highest priority for a large space mission. The NWNH science goals for WFIRST are quite broad. They include: tiered infrared sky surveys of unprecedented scope and depth; a census of exoplanets using microlensing; measurements of the history of cosmic acceleration using three distinct but interlocking methods (weak lensing, baryon acoustic oscillations, supernova standard candles); and a guest observer/investigator program. What brought these very different science goals together was the realization, across the astronomical community, that recent advances in infrared detector array technology have, for the first time, made it possible to launch a wide field infrared telescope with a very large number of diffraction limited "effective pixels" in the focal plane. The telescope prescribed for WFIRST in the NWNH report had a primary mirror diameter of 1.5 m.

In addition to the science and payload envisioned for WFIRST by NWNH, NASA charged the current SDT with implementing the addition of a coronagraph instrument to the payload for direct imaging of exoplanets. One rationale for the addition is that the increase in the primary mirror size from 1.5-m to 2.4-m results in a finer angular resolution (smaller point spread function) that enables meaningful coronagraphy.

In October 2011 ESA selected Euclid as one of two Cosmic Vision medium class missions. Euclid is also a wide field telescope (1.2-m primary mirror), with the majority of its detector arrays (36 of 52) working at optical rather than infrared wavelengths. While its optical pixels properly sample diffraction-limited images, its infrared pixels are eight times coarser in area and do not take advantage of the angular resolution that is gained by going into space. The Euclid mission prime science is solely in the area of cosmology with dark energy surveys using the weak lensing and baryon acoustic oscillations techniques. In January 2012 the NRC's *Committee on the Assessment of a Plan for US Participation in Euclid* recommended a modest contribution to Euclid, saying "This investment should be made in the context of a strong U.S. commitment to move forward with the full implementation of WFIRST in order to fully realize the decadal science priorities of the NWNH report."

In December 2013 the NRC performed a review of WFIRST as requested by NASA. The committee was chaired by Fiona Harrison and co-chaired by Marcia Rieke. The charge was to compare the NWNH WFIRST to WFIRST-AFTA, with and without the coronagraph, and, based on the comparison, assess responsiveness of WFIRST-AFTA to the science objectives of NWNH WFIRST and assess responsiveness of WFIRST-AFTA with the coronagraph to the science and technology objectives of the NWNH technology development program. The conclusion of the report was that WFIRST did fulfill the objectives of NWNH, and in some cases exceeded the requirements. The coronagraph was found to address some of the NWNH recommendations to develop exoplanet direct imaging technology. The report expressed concern about the added risk of developing the larger WFIRST-AFTA mission compared with the simpler version described in NWNH. It was a highly valuable report and we appreciated the committee's efforts. The Study Office and SDT have worked diligently to address the concerns in the report.

Science has become even more exciting since NWNH – a growing realization of the importance and mystery of dark energy and a startling abundance and diversity of exoplanet systems – and the 2.4-m telescope gives NASA and the community a cost effective way to exploit the resolution of a Hubble scale telescope which would otherwise have been prohibitively expensive if originally proposed back in 2010. In October 2011, the Nobel Prize in Physics was awarded to Saul Perlmutter (a member of the SDT), Adam Riess, and Brian Schmidt "for the discovery of the accelerating expansion of the Universe through observations of distant supernovae." The mechanism of that acceleration is unknown. WFIRST will measure the expansion history of the universe and growth of structure to better than 1% in narrow redshift bins using several independent methods, providing constraints that will greatly narrow the range of possible mechanisms.

For exoplanets, we are in the midst of a revolution in detections and understanding. Five different techniques have resulted in nearly 1500 confirmed detections and an even larger number of candidates to be confirmed. The microlensing technique has detected free-floating planets with an indication that there are as many unbound planets as there are orbiting stars. When combined with the results from Kepler, WFIRST will produce the first statistically complete census of exoplanets. The coronagraph will make the major first step in direct imaging of exoplanets and produce results from exoplanet images and spectra. In addition, although WFIRST-AFTA is by no means a coronagraph-optimized telescope, exoplanet imaging technology – coronagraphy, wavefront control, and post-processing – has advanced to the point where very high levels of performance appear practical with WFIRST-AFTA. This opens up enormous science possibilities, including photometrically and spectroscopically characterizing a large sample of giant planets, mapping of structure in extrasolar zodiacal disks, and potentially characterizing 'super-Earth' planets around the nearest stars.

The DRM studies by the SDT, called WFIRST-AFTA DRM has the following characteristics:

- 2.4-m primary mirror
- Eighteen HgCdTe IR detector arrays in the Wide Field Imager (WFI) instrument
- Six imaging filters in the Wide Field Instrument (WFI) from 0.76 to 2.0 microns
- Slitless spectroscopy with a grism
- Integral Field Unit spectrograph for slit spectroscopy of SNe and other sources
- Coronagraph instrument for direct imaging of exoplanets (along with an additional year of operations)
- Integral Field Spectrometer within the coronagraph instrument for exoplanet spectroscopy
- 28.5° inclined geosynchronous orbit

The choice of the more-advanced H4RG detector arrays for WFIRST-AFTA is based on the rapid state of advancement in detector array manufacturing. Ground instruments are being built with 4k x 4k detector arrays and we have judged that H4RGs will be the state-of-the-art at the time of the WFIRST development. A vigorous program is underway to space qualify the devices in time for WFIRST.

The detector array layout for WFIRST-AFTA is shown in Figure 1-1. The size of the detector arrays is shown as a footprint on the sky to illustrate the wide field of view (FoV). The FoV is a factor of ~200 larger than the instruments on HST (WFC3/IR) and JWST (NIRCam), enabling deep sky surveys of unprecedented size. This report gives the science return from WFIRST-AFTA and a comparison to other observatories in Section 2 and a detailed description of the mission in Section 3.

To present definite forecasts for the scientific performance of WFIRST-AFTA, we have adopted a notional observing program for the 6-year prime mission: 3 deg$^2$ in the Galactic bulge to discover exoplanets through microlensing, an imaging and spectroscopic survey of 2000 deg$^2$ at high Galactic and ecliptic latitude, a 3-tiered imaging and spectroscopic survey for supernovae, a year of direct exoplanet observations with the coronagraph and 1.5 years (i.e., 25% of the 6-year mission) devoted to a competed Guest Observer program. The actual allocation of observing time will be decided closer to the launch of WFIRST, based on scientific and technical developments between now and then. At the end of its prime mission, WFIRST-AFTA will remain a facility of exceptional scientific power.

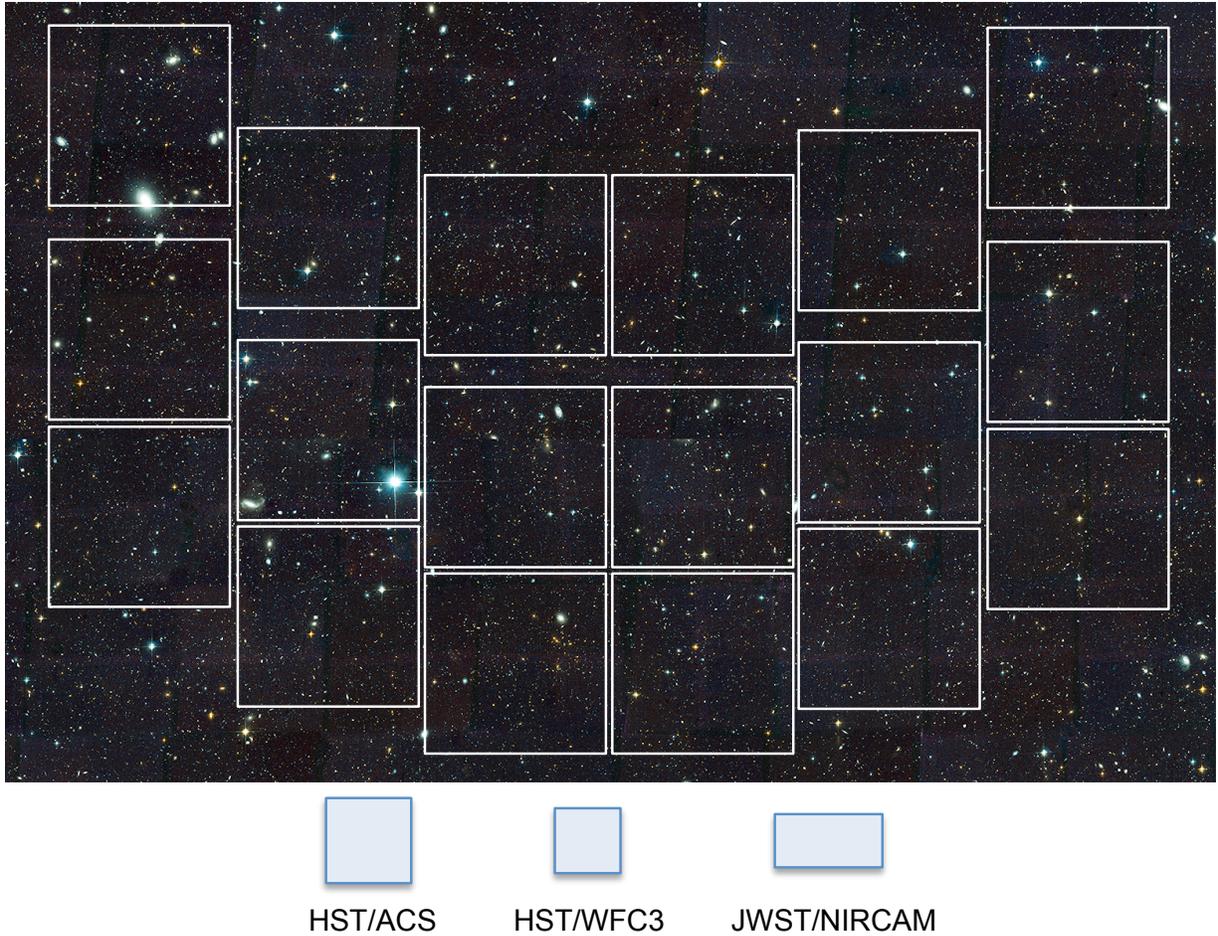

HST/ACS     HST/WFC3     JWST/NIRCAM

Figure 1-1: Field of view comparison, to scale, of the WFIRST-AFTA wide field instrument with wide field instruments on the Hubble and James Webb Space Telescopes. Each square is a 4k x 4k HgCdTe sensor array. The field of view is 0.28 degrees$^2$. The pixels are mapped to 0.11 arcseconds on the sky.

## 2  WFIRST-AFTA SCIENCE

### 2.1  Science Overview

The 2010 Decadal Survey, New Worlds, New Horizons (NWNH), recommended that two major programs be accomplished with one telescope: the search to understand the acceleration of the universe and the search for planets around other stars. The Decadal Survey also made clear that the same mission would provide opportunities for advancing knowledge in many other areas of astronomical science, both through a robust Guest Observer Program for new observations and a strong Guest Investigator Program for utilization of the data sets from the two core programs. It became clear to the SDT very early in the study of WFIRST-AFTA that even in a time of constrained resources, the 2.4-m diameter telescope made possible the rich programs envisioned by the Decadal Survey. This is discussed in detail in our 2013 Final Report (Spergel et al. 2013).

It also became clear to the SDT that the superb imaging capability of this telescope made a coronagraph on WFIRST-AFTA an exciting extension of its capabilities that not only would characterize giant planets and debris disks about nearby stars, but would also be an important step towards detecting habitable Earth-like exoplanets. NASA has decided to incorporate a coronagraph and selected a design concept. Thus, in this report we are able to discuss the potential scientific advances in more detail.

We emphasize the power and flexibility of the WFIRST-AFTA mission for scientific studies, see e.g., §2.2.1. This mission is exploring frontier areas and it is quite possible that there will be scientific surprises requiring adjustments to the core programs. Also, there may be changes in the scientific landscape by the time the mission commences requiring changes in the observing program. The SDT expects that the WFIRST-AFTA science team will evaluate the data available early in the mission and, if necessary, modify the strategy to take advantage of their findings. Also because of the mission's power and flexibility, we expect, just as in the case of HST, the Guest Observer and Guest Investigator communities will find ways to exploit the mission capabilities in novel and unforeseen ways.

In §2.2, we discuss in detail the dark energy program with a three pronged campaign of baryon acoustic oscillations, weak lensing of galaxies, and Type Ia supernovae to measure the evolution of the universe and possible departures from the model of gravitation as detailed in the general theory of relativity.

This program will produce a survey map of 2000 $deg^2$ with superb imaging in the field of the LSST. In addition to outstanding imaging, the WFIRST-AFTA survey will provide both infrared colors and redshifts accurate to better than 1 part in 1000. Such a map will be invaluable for a wide range of astrophysical studies. Both these astrophysical studies and the LSST data will significantly improve the analysis of the WFIRST-AFTA data, reducing systematic errors, both those we know about or suspect as well as the surprises.

§2.3 describes the potential impact on a wide range of astronomical science by the WFIRST-AFTA High-Latitude Surveys, both the wide field map discussed above as well as deep drilling surveys based on the supernovae fields. These surveys will provide exciting new information on issues ranging from the first billion years of cosmic history to stellar streams in the Milky Way.

In §2.4 we show how WFIRST-AFTA will advance our understanding of exoplanets through two complementary studies: the statistical approach of determining the demographics of exoplanetary systems and the detailed approach of characterizing the properties of a handful of nearby exoplanets. In §2.4.2 we set out the gains to be had with WFIRST-AFTA for the second core program, the microlensing search for thousands of planets down to below Mars mass, in orbit around F, G, K, and M stars "beyond the snow line." This survey is an essential complement to the Kepler mission. WFIRST-AFTA will provide significant advantages over previous WFIRST designs in determining the absolute, not just the relative, masses of these planets.

§2.4.3 presents how the WFIRST-AFTA mission will push ahead one of the Decadal Survey's highest priorities: not just finding, but measuring the properties of exoplanets, with an extremely high-contrast coronagraph. The large aperture WFIRST-AFTA telescope can provide the first platform to test this new and rapidly improving technology in a way that no other available telescope can do and take the first scientific steps in detailing the characteristics of non-transiting extrasolar planets, something that the last two Decadal Surveys have identified as one of the most important fields in astrophysics in 2020 and beyond. Directly imaging planets around nearby stars is a step that is in keeping with the spirit of the mission as described in NWNH. The Coronagraph Instrument (CGI) on WFIRST-AFTA will directly image exoplanets and disks around nearby stars. For disks, the CGI will produce images in blue and red filters, at two polarizations, measuring the amount of dust, and its interaction with planets. For ex-

oplanets, the targets include about a dozen known radial-velocity (RV) planets. A significant number of currently-unknown planets will be within our reach as well. One of the most striking discoveries from the Kepler mission is the large number of planets with radii 2-4 times that of Earth – planets with no analog in our solar system but detectable by WFIRST-AFTA. The CGI will image these planets in multiple filters and two polarizations, and obtain spectra of many of these at red wavelengths with a resolving power of 70, using water, methane, and ammonia features to characterize their atmospheric composition and properties. Combined with systematic Doppler measurements, WFIRST will extend our understanding of planetary properties around solar-type and earlier stars, complementing the late-type focus of TESS..

§2.5 points out that the wealth of data on the galactic bulge field obtained by the microlensing program will enable a wide range of studies of stars, time dependent objects and the structure of the Milky Way.

A robust Guest Observer Program is a requirement for the WFIRST program. We present some of the opportunities for guest observations to accomplish a broad range of astronomical objectives in §2.6. As emphasized earlier in this subsection, WFIRST-AFTA is a powerful and flexible observatory. We fully expect, as in the case of HST, the guest observer community will find ways to use the WFIRST-AFTA observatory for objectives that, at this time, no one has even considered.

The 2020 and beyond time frame will provide remarkable opportunities to synergize and relate the WFIRST-AFTA data sets with those from other major facilities such as JWST and LSST. We discuss some of the possibilities in §2.7, but it is likely that combinations of these data sets will provide new, unexpected insights not even contemplated here.

We have not presented a final subsection listing the WFIRST-AFTA Science Requirements similar to the one in the 2013 Final Report. During its 2014 activities, the SDT has worked with the WFIRST-AFTA Study Office scientists and engineers towards the preparation of detailed documents describing the Science Requirements and the Mission Requirements derived from the Science Requirements. This work is ongoing. High-level requirements have not changed substantially from those in the 2013 Final Report, and we take these as the basis for the Design Reference Mission presented in the following sections.

## 2.2 Dark Energy & Fundamental Cosmology

The accelerating expansion of the universe is the most surprising cosmological discovery in many decades, with profound consequences for our understanding of fundamental physics and the mechanisms that govern the evolution of the cosmos. The two top-level questions of the field are:

1. Is cosmic acceleration caused by a new energy component or by the breakdown of General Relativity (GR) on cosmological scales?
2. If the cause is a new energy component, is its energy density constant in space and time, or has it evolved over the history of the universe?

A constant energy component, a.k.a. a "cosmological constant," could arise from the gravitational effects of the quantum vacuum. An evolving energy component would imply a new type of dynamical field. Gravitational explanations could come from changing the action in Einstein's GR equation, or from still more radical modifications such as extra spatial dimensions. Observationally, the route to addressing these questions is to measure the histories of cosmic expansion and growth of structure with the greatest achievable precision over the widest accessible redshift range.

As defined by NWNH, one of WFIRST's primary mission goals is to "settle fundamental questions about the nature of dark energy, the discovery of which was one of the greatest achievements of U.S. telescopes in recent years." (Following common practice, we will use "dark energy" as a generic term that is intended to encompass modified gravity explanations of cosmic acceleration as well as new energy components.) It will do so using three distinct surveys that enable complementary measurements of the expansion history and structure growth. The Supernova Survey will combine wide-field imaging and Integral Field Unit (IFU) spectroscopy to measure precise distances to thousands of Type Ia supernovae (SNe). The High-Latitude Imaging Survey (HLS Imaging) will enable weak lensing shape measurements of hundreds of millions of galaxies, which will in turn yield precise measurements of distances and matter clustering through measurements of cosmic shear, galaxy-galaxy lensing, and the abundance and mass profiles of galaxy clusters. The High-Latitude Spectroscopic Survey (HLS Spectroscopy) will measure redshifts of tens of millions of galaxies via slitless (grism) spectroscopy. Measurements of baryon acoustic oscillations (BAO) in these enormous 3-dimensional maps will pin down the cosmic distance scale and ex-pansion rate at lookback times of 8 – 11 billion years (redshift z = 1 – 3), while measurements of redshift-space distortions (RSD) will determine the growth rate of matter clustering over the same period. Background on these measurement techniques and the forecasting methods used below to predict their performance can be found in the Green et al. (2012) report on WFIRST-DRM1, in papers by Wang et al. (2010, 2012, 2013) on cosmological constraints from galaxy redshift surveys, and in the comprehensive review article of Weinberg et al. (2013).

Roughly speaking, current cosmological observations measure distances and expansion history with uncertainties of 1 – 3% and matter clustering with uncertainties of 5 – 10% (see, e.g., Aubourg et al. 2014 for a recent analysis that combines many state-of-the-art measurements). WFIRST-AFTA, like other dark energy experiments of the 2020s, seeks to improve the precision of these measurements to the 0.1 – 0.5% level for *both* expansion history and structure growth observables, while simultaneously extending them to previously unexplored redshift regimes. This ten-to-hundred-fold improvement in multiple measurement techniques has two key implications. First, the newly opened discovery space is large: many theories of cosmic acceleration that are consistent with current data will be easily distinguished and stringently tested by WFIRST-AFTA. Second, the great improvement in statistical precision demands equal improvements in the control of systematic uncertainties associated with the measurements themselves and with their theoretical interpretation.

The WFIRST-AFTA dark energy program has been designed with control of systematics as a paramount consideration. For the supernova survey, IFU observations enable powerful approaches to photometric calibration, exact matching of rest-frame wavelength intervals for comparisons across redshifts, and spectroscopic diagnostics to control evolutionary effects in the supernova population. For weak lensing observations, the use of 4 – 5 dither positions at each of two telescope roll angles enables excellent reconstruction of image shapes, and shape measurements in three bands enable cross-correlation analyses that can test for and diagnose systematic effects while simultaneously adding statistical power. For galaxy clustering analyses, the high sampling density of the galaxy redshift survey enables studies that divide the galaxies into populations with different clustering properties, allowing consistency checks across these populations and multi-tracer approaches that amplify statistical precision.

Figure 2-1 presents an overview of the baseline dark energy program described in §2.2.2-2.2.4. The methodology of performance forecasts is described in §2.3 and Appendix C of the 2013 report (Spergel et al. 2013). We have updated these forecasts based on the expectations for the data sets described in §2.2.2-2.2.4 below. The supernova survey will measure "standard candle" distances with an aggregate precision of 0.20% at z < 1 (error-weighted z = 0.50) and 0.34% at z > 1 (z = 1.32). Galaxy clustering provides "standard rulers" for distance measurement in the form of the baryon acoustic oscillation (BAO) feature and the turnover scale of the galaxy power spectrum. The galaxy redshift survey (GRS) enables measurements of the angular diameter distance $D_A(z)$ and the expansion rate $H(z)$ using H$\alpha$ emission line galaxies at 1 < z < 2 and [OIII] emission line galaxies at 2 < z < 3. The aggregate precision of these measurements ranges from 0.49% to 2.05% (see Figure 2-1). The imaging survey will enable measurements of dark matter clustering via cosmic shear and via the abundance of galaxy clusters with mean mass profiles calibrated by weak lensing; we expect 40,000 M ≥ $10^{14}M_{sun}$ clusters in the 2200 deg$^2$ area of the high-latitude survey. These data constrain the amplitude of matter fluctuations at redshifts 0 < z < 2, and they provide additional leverage on the distance-redshift relation. Treating the fluctuation amplitude $\sigma_m(z)$ as a single-parameter change, we forecast aggregate precision of 0.24% (z < 1) and 0.88% (z>1) from clusters and 0.21% (z<1) and 0.78% (z>1) from cosmic shear. In the GRS, the distortion of structure in redshift space induced by galaxy peculiar velocities provides an entirely independent approach to measuring the growth of structure, with forecast aggregate precision of 1.1% at z = 1 – 2 and 3.5% at z = 2-3.

These high-precision measurements over a wide range of redshifts in turn lead to powerful constraints on theories of cosmic acceleration. If the cause of acceleration is a new energy component, then the key physical characteristic is the history w(z) of the equation-of-state parameter w = P/ε, the ratio of pressure to energy density. A cosmological constant has w = -1 at all times, while dynamical dark energy models have w ≠ -1 and an evolutionary history that depends on the underlying physics of the dark energy field. If the cause of acceleration is a breakdown of GR on cosmological scales, then it may be detected in a deviation between the measured growth history G(z) and the growth predicted by GR given the measured expansion history. Alternatively, some modified gravity theories predict a mis-

# WFIRST-AFTA Dark Energy Roadmap

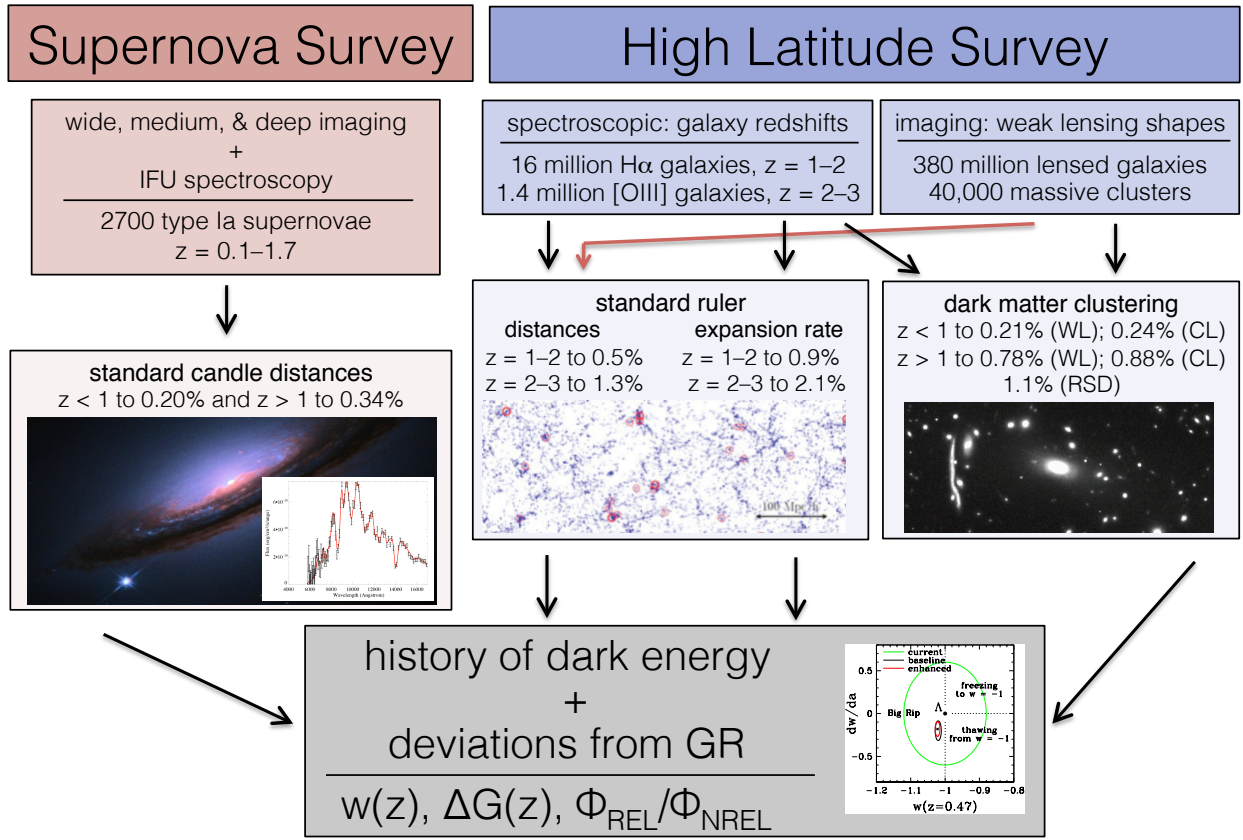

**Figure 2-1: A high-level view of the WFIRST-AFTA dark energy program.** The supernova (SN) survey will measure the cosmic expansion history through precise spectrophotometric measurements of more than 2700 supernovae out to redshift z = 1.7. The high-latitude survey (HLS) will measure redshifts of 18 million emission-line galaxies and shapes (in multiple filters) of 380 million galaxies. The former allow measurements of "standard ruler" distances through characteristic scales imprinted in the galaxy clustering pattern, while the latter allow measurements of matter clustering through the "cosmic shear" produced by weak gravitational lensing and through the abundance of galaxy clusters with masses calibrated by weak lensing. As indicated by crossing arrows, weak lensing measurements also constrain distances, while the galaxy redshift survey provides an alternative measure of structure growth through the distortion of redshift-space clustering induced by galaxy motions. Boxes in the middle layer list the forecast aggregate precision of these measurements in different ranges of redshift. These high-precision measurements of multiple cosmological observables spanning most of the history of the universe lead to stringent tests of theories for the origin of cosmic acceleration, through constraints on the dark energy equation-of-state parameter w(z), on deviations $\Delta G(z)$ from the growth of structure predicted by General Relativity, or on deviations between the gravitational potentials that govern relativistic particles (and thus weak lensing) and non-relativistic tracers (and thus galaxy motions).

match between the gravitational potential inferred from weak lensing (in Figure 2-1, cosmic shear and clusters) and the gravitational potential that affects motions of non-relativistic tracers, which governs redshift-space distortions (RSD) in the galaxy redshift survey.

This flow diagram necessarily simplifies some key points, most notably (a) that the SN distance scale is calibrated in the local Hubble flow while the BAO distance scale is calibrated in absolute ($H_0$-independent) units, so they provide complementary information even when measured at the same redshift, and (b) that cosmic shear, cluster abundances, and RSD probe the expansion history as well as the growth history. The combination of WFIRST-AFTA dark energy probes is far more powerful than any one probe would be in isolation, allowing both cross-checks for unrecognized systematics and rich diagnostics for the origin of cosmic acceleration. We discuss the anticipated constraints on theoretical models below, in §2.2.5.

As described in §2.1 and subsequent sections of this report, the data sets obtained for the dark energy programs will be an extraordinary resource for a vast range of astrophysical studies through Guest Investigator programs, while the WFIRST-AFTA facility enables revolutionary approaches to a still broader range of astronomical problems through Guest Observer programs.

### 2.2.1 The Power of WFIRST-AFTA for Dark Energy Investigations

The strength of WFIRST-AFTA as a tool for dark energy investigations derives from its wide FoV coupled to an unprecedented near-IR pixel count (~300Mpix), its 2.4-m aperture that affords both high sensitivity and exquisite angular resolution, and a design of both hardware and operations that enables excellent control of systematic uncertainties, including, for example, IFU spectrophotometry of supernovae and fully sampled weak lensing shape measurements in multiple bands. It is these elements that allow WFIRST-AFTA to pursue extremely powerful programs with three complementary methods: Type Ia supernovae, weak lensing and clusters, and redshift-space galaxy clustering.

The baseline design of the supernova and high-latitude surveys, and forecasts of their expected performance for dark energy studies, are described in §2.2.2-2.2.4 below. However, an essential strength of WFIRST-AFTA is the *great flexibility* it offers in designing a dark energy program that responds to the changing landscape of the field. Between now and the launch of WFIRST-AFTA we will learn more about the power of and critical challenges for individual measurement methods; we will develop better understanding of theoretical modeling uncertainties and strategies to mitigate them; we will develop a clear picture of the capabilities of complementary ground-based and space-based dark energy experiments; and ongoing observations may highlight particular redshift ranges or classes of measurements as the most important for probing tensions in the data and decoding their implications for cosmic acceleration. The WFIRST-AFTA dark energy program can respond flexibly to these developments by varying the balance among surveys and making trades among area, depth, and wavelength coverage within individual surveys. It can also respond to lessons learned from early observations with WFIRST-AFTA itself.

Figure 2-2 and Figure 2-3 illustrate the trades between area and depth for the imaging and spectroscopic components of the HLS. The imaging exposure time of the baseline survey is 174 sec per filter, for each of the 4-5 dithers at each of two roll angles, which with overheads yields a time of about 0.055 days per filter to cover 1 deg$^2$, at the depths marked by the gray squares in Figure 2-2. With 1.33 years allocated to the HLS imaging survey, the total area covered is 2227 deg$^2$. An H-band only survey could cover about four times the area in the same

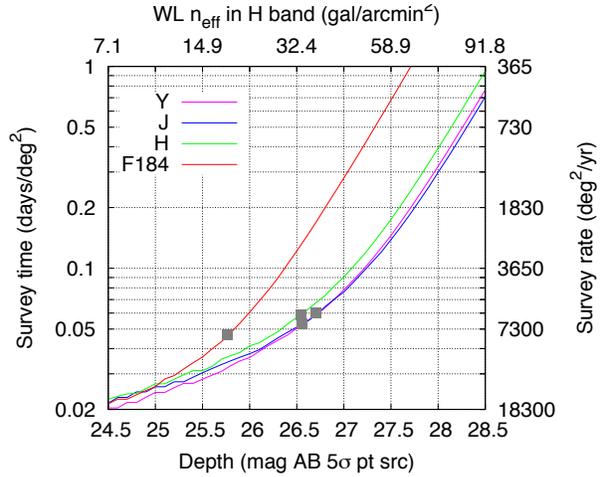

**Figure 2-2: The survey time per unit area per filter, including overheads, for imaging surveys. The gray squares indicate the baseline HLS survey depth. The right hand axis marked in deg$^2$ covered per filter per year of observing time. The upper axis shows the effective weak lensing source density in H-band as function of depth.**

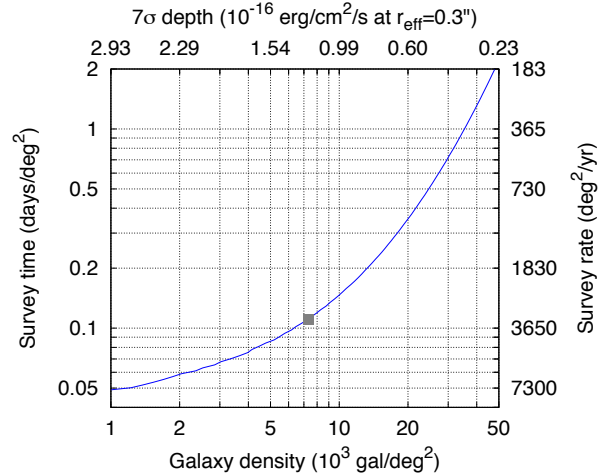

**Figure 2-3: The survey time per unit area for the grism spectroscopy required to observe a certain number density of Hα emitters at 7σ, at redshift z = 1.5. The corresponding flux limit for an $r_{eff}$ = 0.3 arcsec source is marked on the upper axis. The gray square indicates the baseline strategy. The right axis is marked in number of deg$^2$ covered per year of observing time.**

amount of time and thus provide shape measurements for many more galaxies, at the price of losing color information for photometric redshifts and cross-correlations of shape measurements in different filters. Alternatively, a four-filter survey could cover about twice as much area if the limiting depth were reduced by 1.2 magnitudes, or it could go 0.7 magnitudes deeper and cover half the area. The asymmetry of this trade reflects the transition from being limited by read noise and overheads to being limited by sky noise. The baseline exposure time is at the "knee" between these limits, a choice that yields (approximately) the largest number of source galaxies in a given observing time. There are of course many variations within this trade space, including hybrid strategies that obtain deep four-band imaging over a smaller area and shallower imaging in one or two bands over a larger area.

For spectroscopy, the baseline strategy covers the same 2227 deg$^2$ area as the imaging survey to a limiting 7$\sigma$ line flux of 1.2 x 10$^{-16}$ erg/cm$^2$/s (for an $r_{eff}$ = 0.3 arcsec extended source at z = 1.5), in a total of 0.68 years. Figure 2-3 shows the relation between survey speed and the density of galaxies in the redshift survey at z = 1.5. Raising the flux limit to 2.4 × 10$^{-16}$ erg/cm$^2$/s would make it possible to cover twice as much area in the same amount of time. Because the H$\alpha$ luminosity function is so steep, the total number of galaxies would be smaller by a factor of 2.1, although because of the larger volume surveyed the BAO constraints would only weaken by 13%. As with the imaging survey, the baseline exposure time is close to the value that yields the largest number of source galaxies in a given observing time.

To maximize the discovery potential of the WFIRST-AFTA dark energy program, it will be essential to review the survey strategies and balance among methods periodically, both before and after launch. Early observations will provide testing and validation of the assumptions about instrument performance that underlie these strategies, which can be adjusted accordingly.

## 2.2.2 Supernovae

For the baseline WFIRST-AFTA supernova survey, we adopt a total observing time of just over six months, spread over a two-year interval. The survey described here is an existence proof of a possible WFIRST-AFTA observing sequence. Its purpose is to demonstrate that an observing plan meeting WFIRST-AFTA science objectives is possible with the reference hardware and orbit. This survey design is based on a first optimization of the dark energy Figure of Merit (FoM) (defined in

§2.2.4) from SNe for a fixed observing time, given starting assumptions about statistical and systematic errors described below.

For this survey, supernova observations take place with a ~5-day cadence (with possibly a somewhat shorter cadence for the lowest-redshift tier), with each set of observations taking a total of 30 hours of imaging and spectroscopy. There are three nested tiers to the survey: a shallow survey over 27.44 deg$^2$ for SNe at z < 0.4, a medium survey over 8.96 deg$^2$ for SNe at z < 0.8, and a deep survey over 5.04 deg$^2$ for SNe out to z = 1.7. These survey regions will be observed in advance with multi-filter imaging both from the ground and with the WFIRST Wide Field Imager (and, ideally, with the spectra from the Wide Field Grism, as well) before the cadenced supernova observations begin. These observations will provide photometric redshifts (and, where possible, spectroscopic redshifts) for most of the galaxies that will later host supernovae.

### 2.2.2.1 Supernova Discovery

The survey will use the wide-field imager to discover supernovae in two filter bands: Y and J for the shallow, low redshift tier and J and H for the medium and deep tiers. The exposure times are designed to be sensitive to supernovae 12 rest-frame-days before peak brightness in each redshift tier. Supernova candidates are selected that show up with S/N greater than 4 in two filters, that were not detected on previous visits. Supernovae will thus be discovered no later than between 7 and 12 rest-frame-days (depending on redshift) before peak brightness, and this two-band ~5-day observing cadence will continue through the entire lightcurve, at least 40 rest-frame days past maximum light for any supernova that will be followed (as shown schematically in the upper panel of Figure 2-4). With two-band detections and an estimate of redshift from the pre-cadence observations, early screening and triggering of supernova candidates will be possible before the first IFU spectroscopy begins.

### 2.2.2.2 Supernova ID and Typing

To get the detailed light curves started as early as possible, the photometry-and-redshift-screened candidates will be scheduled for an IFU spectrum in the next supernova visit (IFU Visit 1, as shown schematically in the bottom panel of Figure 2-4). This first IFU spectrum while the supernova is still just brightening will only be used later to build the lightcurve, if the candidate ends up being followed, and is not intended to have sufficient signal-to-noise for supernova typing. During that visit there will also be another observation of the candidate xwith the imager, as part of the ongoing regular search. Candidates whose brightnesses have then increased since their discovery 5 days before, and whose color differences are consistent with a Type Ia supernova at the host-galaxy redshift, will be scheduled for an SN-typing spectrum during the next visit (IFU Visit 2), when the supernova is yet brighter. This spectrum and the two-band photometry will be used to screen out most of the remaining non-Type Ia supernovae and to sharpen the photo-z (or grism) redshift estimate; the observing time budget allows for twice as many of these screening spectra as there are Type Ia supernovae that will then continue to be followed. The selected supernovae will almost all be Type Ia, and these will then be scheduled for a deep IFU spectrum (IFU Visit 3). Both the typing spectrum and the deep spectrum will be near peak SN brightness, and together they will (1) confirm the Type Ia classification and remove any possible small remaining contamination of non-Type Ia supernovae, (2) provide spectral diagnostics that can help distinguish intrinsic color variations from the effects of dust extinction (Chotard et al. 2011, Mandel et al. 2014), and (3) match high and low-redshift SNe with similar properties to suppress evolutionary effects in the mix of supernovae.

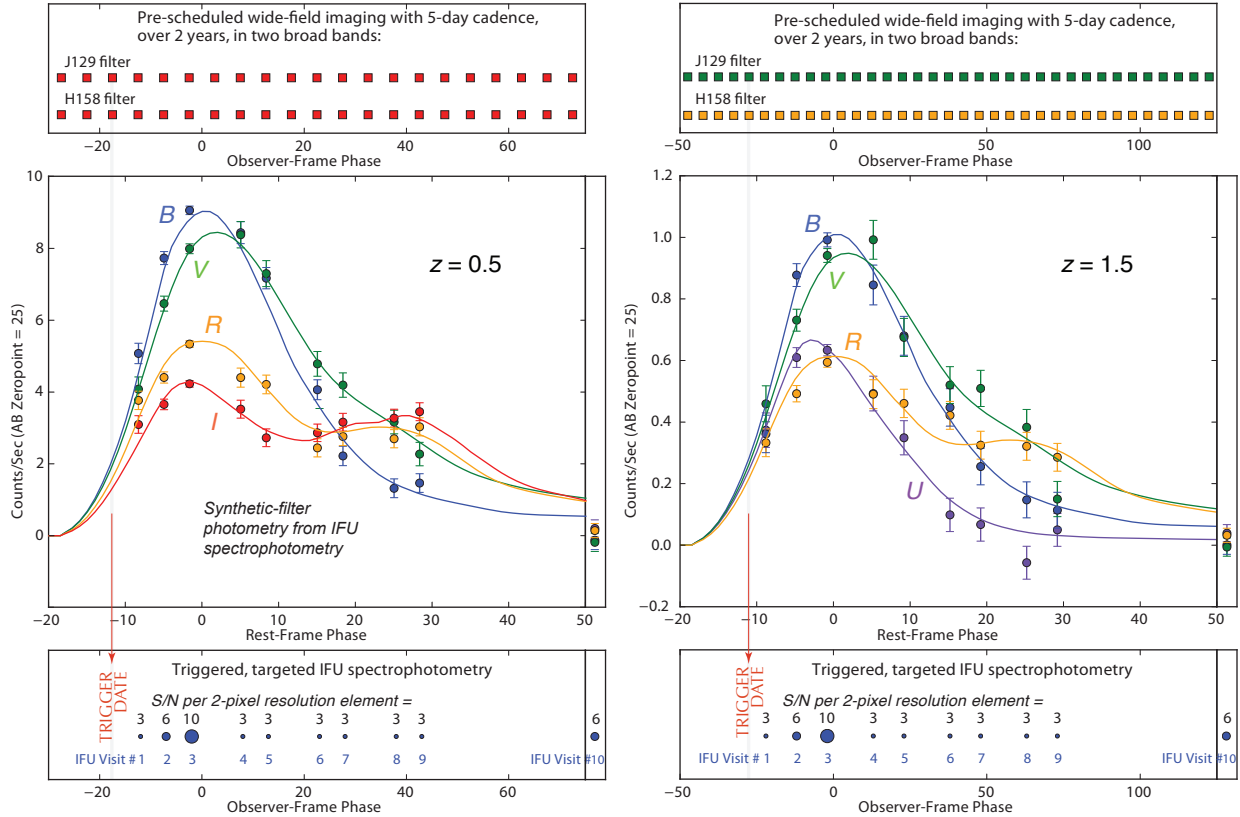

**Figure 2-4: An example WFIRST-AFTA observing sequence for discovering and studying SNe Ia across the range of redshifts between z = 0.2 and 1.7, demonstrating an observing plan that meets WFIRST-AFTA science objectives. The left panels show an example sequence for SNe discovered at z ~ 0.5, and the right panels are for SNe discovered at z ~ 1.5. The upper panel shows the cadence of the broad-band filter observations that will, over 2 years, find the supernovae and provide continuous broad-band photometry with two observer-frame filters. At the specific redshifts of the supernovae, these observer-frame filters will be sampling points on the SN lightcurve with a "blue-shifted" restframe filter (indicated schematically by the color of the scheduled points). The supernova discoveries will be "triggered" before peak brightness (as shown by the grey line and red-arrow connecting to the bottom panel) and then the IFU follow-up spectrophotometry Visits will begin on one of the next 5-day-observer-frame cadences. These Visits are shown schematically in the bottom panel, along with their visit number and S/N per 2-pixel resolution element. Photometry points calculated from the IFU spectrophotometry integrated over synthetic-filter bandpasses are shown in the middle panel, plotted as a function of the *rest*-frame days for this panel (to be compared with the *observer*-frame days of the top and bottom panels). The same restframe B, V, and R filters are used for both the z=0.5 and 1.5 example supernovae to show the similar observations available at the full range of redshifts.**

### 2.2.2.3 Spectra for Obtaining Light Curves

The spectrophotometric measurements used for the light curves for the cosmological analysis will also be derived from observations with the IFU spectrometer. Thus, for each supernova, six further IFU spectra (IFU Visits 4 through 9) will continue to be obtained along the light curve with a roughly 5-day *rest*-frame cadence (but, for efficiency, observed during the nearest 5-day-*observer*-frame cadence visit), until ~30 rest-frame days past peak. IFU spectrophotometry reduces systematic uncertainties by (a) removing filter-photometry's need for K-corrections (Saunders et al. 2014), as the same rest-frame wavelength range can be chosen for SNe at different redshifts and (b) allowing better separation of SN and background galaxy light as a function of wavelength than is possible with a grism. Moreover, (c), flux calibration is a limiting systematic uncertainty in current ground-based SN surveys (e.g., Sullivan et al. 2011) and a critical consideration for a space-based survey, and highly accurate flux calibration is much more straightforward for an IFU system than for a wide-field filter photometry system. For the IFU, the effective wavelength of a spectral resolution element is independent of the spectral distribution of a SN, but that is not the case for a broadband filter because SN features vary strongly relative to the filter resolution. Moreover *in situ* calibration of the spectral response is trivial for an IFU, but is not generally feasible for a wide-field imager.

Even though IFU observations are one object at a time, they prove more efficient than slitless spectroscopy over the 0.28 deg$^2$ FoV because the exposure time can be chosen for each supernova individually instead of being driven by the faintest object in the field, and because IFU observations have dramatically lower sky noise per pixel, allowing much better isolation of the faint SN signal.

### 2.2.2.4 Signal-to-Noise for the IFU Observations

The IFU measurements used for most of the lightcurve photometry (IFU Visits 1, and 4 through 9) are intended only to provide S/N = 15 when integrated over broad synthetic filter bands, so these observations have relatively short exposure times calculated[3] to give

S/N ~ 3 per 2-pixel resolution element. (The S/N per 2-pixel resolution element is defined here as the *average* signal-to-noise per 2-pixel resolution element in the observer-frame 1.02 to 1.28 micron band. Thus the S/N, as well as the actual resolution in the IFU design, will vary with wavelength.) The SN-typing spectrum (IFU Visit 2) is deeper and observed near peak SN brightness, with exposure times calculated to yield S/N = 6 per 2-pixel resolution element, which is equivalent to S/N=24 per synthetic filter band, when used as part of the lightcurve photometry set. The deepest spectrum (IFU Visit 3) is also observed near peak SN brightness, with exposure times calculated to yield S/N = 10 per 2-pixel resolution element, which is equivalent to S/N=40 per synthetic filter band. (The S/N for each IFU Visit is listed in the lower panel of Figure 2-4, and shown in the corresponding error bars of the synthetic-filter-photometry points in the middle panel.)

Finally, one galaxy-only reference spectrum (IFU Visit 10) will be taken about a year later after the supernova has faded, for galaxy subtraction, with S/N = 6 per 2-pixel resolution element. Since the galaxy reference spectra must be taken well after the SN has faded, they will not be observed during the first year of the survey. The time saved will be used in the third year to take the galaxy spectra for the supernova discovered in the second year. On average, in each 30 hour visit of observations roughly 8 hours will be spent on imaging discovery, 5 hours for typing spectra, 9 hours on spectra for light curves, 5 hours on the deep spectra, and 3 hours for the galaxy reference spectra.

Figure 2-5 shows simulated spectra with the S/N levels of the SN-typing spectrum (IFU Visit 2) and the deepest spectrum (IFU Visit 3) at a range of redshifts. The comparison of the two solid lines shows that the template of a Type Ia SN spectrum (dark line) can be distinguished from the (light gray line) example of a Type Ib/c SN spectrum. The Type Ibc have the most similar spectra to the Type Ia, but are clearly spectroscopically distinguishable with S/N = 6 per 2-pixel resolution element, by, e.g., the region around the Type Ia "sulfur W" feature, visible around 1.08 microns in the z

---

[3] The exposure times to obtain a given S/N were here calculated for an IFU design with one spatial pixel per 0.15" image-slice, while the design outlined in §3.3.1.2 uses 2 spatial pixels per 0.15" slice. More detailed studies of the optimal slice scale, and the optimal sampling of these slices, will drive the later iterations of this design, and hence the actual exposure times for a given S/N. These calculations will also be affected by the final choice of detector chip characteristics specified, particularly the asymptotic read-out noise and dark current. Finally, the optimizations will also have to account for the actual distribution of host-galaxy-to-SN light ratio in a pixel.

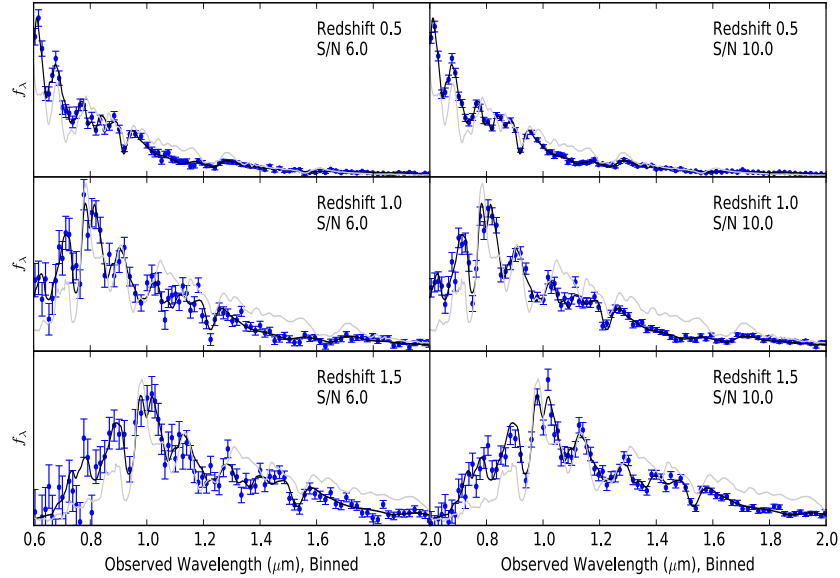

**Figure 2-5: Simulations of Type Ia SNe observations with the IFU, showing S/N = 6 and 10 (left and right columns) per 2-pixel resolution element, for redshifts z = 0.5, 1.0, and 1.5 (top, middle, and bottom rows). (The S/N per 2-pixel resolution element is defined here as the *average* signal to noise per 2-pixel resolution element in the observer frame 1.02 to 1.28 micron band.) The dark solid line shows the template Type Ia SN spectrum, while the light gray line shows an example of a Type Ibc SN spectrum. The Type Ibc have the most similar spectra to the Type Ia, but are clearly spectroscopically distinguishable with S/N = 6 per (2-pixel) resolution element, by, e.g., the region around the Type Ia "sulfur W" feature, visible around 1.08 microns in the z = 1 plots.**

= 1 plots. It can also be seen how the photometric accuracy of IFU spectroscopy allows broad-wavelength color differences between the SN types to be used in the type identification. (The SN typing is also much more accurate because of the host-galaxy redshift knowledge available.) As a final cross-check, the two-band imaging lightcurves, which extend at least 40 rest-frame days past maximum light, can also provide SN identification information up to $z \sim 1.3$, since they will show the characteristic second lightcurve peak in the redder restframe wavelengths.

### 2.2.2.5 *The Data Sample*

Figure 2-6 shows the number of SNe expected in each $\Delta z = 0.1$ bin of redshift. The drops at the redshift boundaries of the shallow and medium tiers are evident. In the deep tier we choose a fixed number of N=136 in each redshift bin for IFU follow-up, from the larger number discovered in the imaging survey, to keep the time for IFU spectra within the time allocated to the supernova survey. The total number of predicted SNe is 2725, with a median redshift z = 0.7.

To guide the design and forecast the performance of the supernova survey, we have adopted the following error model. The photometric measurement error per supernova is $\sigma_{meas} = 0.08$ magnitudes based on the seven IFU spectra with S/N=15 per synthetic filter band, the one typing spectrum, the one deep IFU spectrum, and the one galaxy reference spectrum. We assume an intrinsic dispersion in Type Ia luminosities of $\sigma_{int} = 0.09$ magnitudes (after correction/matching for light curve shape and spectral properties). This is based on stud-ies of nearby supernovae that use full spectral time series (which the IFU spectroscopy will provide). The other contribution to statistical errors is gravitational lensing magnification, which we model as $\sigma_{lens}$ = 0.07 × z mags. The overall statistical error in a $\Delta z$ = 0.1 redshift bin is then

$$\sigma_{stat} = [(\sigma_{meas})^2 + (\sigma_{int})^2 + (\sigma_{lens})^2]^{1/2} / \sqrt{N_{SN}} ,$$

where $N_{SN}$ is the number of SNe in the bin. For the current estimates, we assume a systematic error per bin of

$$\sigma_{sys} = 0.01 \ (1+z) \ / \ 1.8 \ \text{mag},$$

with no correlation of errors between redshift bins. (Development of a more realistic, fully-correlated systematics error-matrix study has also been initiated, and the preliminary results yield somewhat better FoM results than those used in this document; this work will continue.)

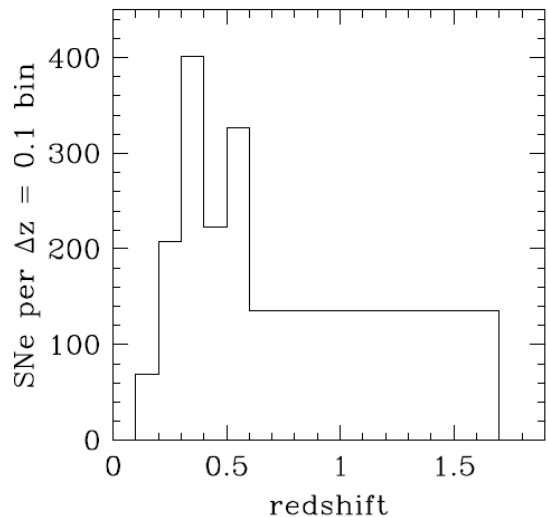

**Figure 2-6: Expected number of Type Ia SNe to be followed in each $\Delta z = 0.1$ redshift bin. For z > 0.6 there are, by design, 136 SNe followed up with spectroscopic observations in each bin (from a larger number detected). The total number of SNe is 2725.**

The total error for a bin is just the quadrature sum of the statistical and systematic errors. Diamonds in Figure 2-7 show the total distance error (which is half the flux error) based on the WFIRST-AFTA design and the error model above; these are the same errors shown for luminosity distance in Figure 2-1. By design, the statistical and systematic errors are comparable, except in the lowest redshift bin where the volume is small. In general, one gains by improving statistics in low redshift bins (where the exposure times are shorter) until one hits the systematics limit, and then one gains by going to higher redshifts, which continue to provide new information. (A similar result is seen in the preliminary analysis from the fully-correlated systematics error matrix study.) The total error in distance is about 0.5% per bin near z = 0.5, then climbs to 1% at the highest redshifts because of the increased contribution of lensing to the statistical error and the assumed redshift behavior of the systematic error.

Our error model is based on current empirical understanding of the supernova population, which is improving rapidly thanks to extensive surveys of nearby and moderate redshift SNe. The design and analysis of the WFIRST-AFTA SN survey, including the optimal division of observing time among redshift tiers, will ultimately depend on lessons learned from ground-based surveys and HST/JWST observations between now and the launch of WFIRST-AFTA. It is clear, however, that the near-IR coverage, sharp imaging, and large collecting area of WFIRST-AFTA will allow a supernova cosmology survey of extraordinary quality, far surpassing

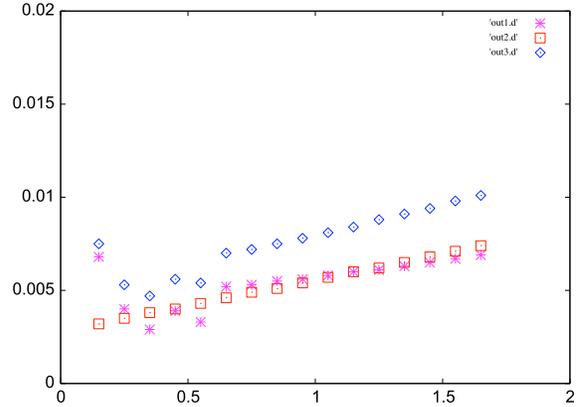

**Figure 2-7: Fractional errors in distance per $\Delta z$ = 0.1 bin. Squares and crosses show the statistical and systematic contributions, respectively, and diamonds show the total error.**

that of any other planned facility on the ground or in space.

### 2.2.2.6 *Complementary Ground-Based SN Observations*

A large sample of well studied low redshift (0.03 < z < 0.1) supernovae is crucial for the WFIRST-AFTA supernova survey: to anchor the Hubble Diagram, to identify subclasses of the Type Ia supernovae which can be paired with similar subclasses at high redshift to reduce the intrinsic spread of the supernova peak luminosities, and generally to sharpen our understanding of these objects as cosmological distance indicators. Because of the volume effect, the low redshift supernovae are much rarer per area on the sky, so while the supernova at z > 0.2 can be found with WFIRST-AFTA by searching tens of square degrees, the low redshift supernova search has to extend over thousands of square degrees for many years. The low redshift (z < 0.1) supernovae are, however, much brighter, and their light is in the optical, so this part of the supernova work is best done in a ground-based survey.

In the WFIRST-AFTA supernova survey discussed in this report, a sample of 800 low-redshift (z < 0.1) supernovae was assumed, each one measured with a spectrophotometric time series comparable to that obtained by WFIRST-AFTA for the high-redshift SNe (though with better S/N and cadence). Simply from the improved anchoring of the Hubble diagram, the availability of such a sample reduces the errors on the cosmological parameters obtained by the supernova survey by a factor of two, as shown in Figure 2-8, and the resulting FoM of the survey is increased by a factor of close to three as shown in Figure 2-9. However, even these large improvement factors underestimate the importance of the low-redshift spectrophotometric sample for WFIRST-AFTA, because of their key role in identifying and matching the subclasses of Type Ia supernovae that would otherwise introduce significant systematic errors after accounting for population drift over time and redshift.

Currently, the Nearby Supernova Factory is completing the spectrophotometric time series, comparable to those planned for WFIRST-AFTA, for approximately 300 low-redshift Type Ia supernovae. To build the larger sample of ~800 low-redshift supernovae with spectrophotometric time series, there are many projects (such as the iPTF, ZTF, SkyMapper, La Silla/QUEST, and eventually LSST) that can find the low-redshift supernovae, but few that can perform the spectrophotometric follow up. It will be important to develop the capability to follow these discoveries with precision spectrophotometry. Fortunately, this work requires relatively small telescope(s) and instruments.

### 2.2.3 Wide-Area Imaging and Weak Lensing

#### 2.2.3.1 High-Latitude Imaging Survey: Baseline Design

The WFIRST-AFTA high-latitude survey (HLS) will image 2227 deg$^2$ in 4 NIR bands (Y, J, H, and F184) spanning the range from 0.92—2.00 $\mu$m to magnitudes 25.8—26.7 AB (depending on band), and reaching the diffraction limit in J and redder bands. Figure 2-10 shows this imaging depth in comparison to LSST (after 10 years of operation) and Euclid. In an AB-magnitude sense, WFIRST-AFTA imaging is well matched to the i-band depth of LSST. The AB magnitude limits for LSST in g and r are fainter, but galaxies are red – for example, the median WFIRST-AFTA weak lensing source galaxy has i-H=0.8, r-H=1.2, and g-H=1.7, so even here the WFIRST-AFTA and LSST imaging depth remains well matched. The Euclid NIR imaging covers a wider area than WFIRST but is 2.6 magnitudes shallower. It is under-sampled with 0.3" pixels, and not used for galaxy shape measurements. The Euclid visible imaging depth is comparable to LSST, but in a single very wide filter rather than the ~5 optical bands necessary for photometric redshift determination. WFIRST-AFTA provides multicolor space-resolution imaging (PSF half-light radius 0.12—0.14"). The combination of Euclid/VIS and the WFIRST-AFTA imaging will provide 5-band photometry over the area of overlap. They will provide approximately the same physical resolution and sensitivity and cover a similar rest-frame wavelength range at z=1

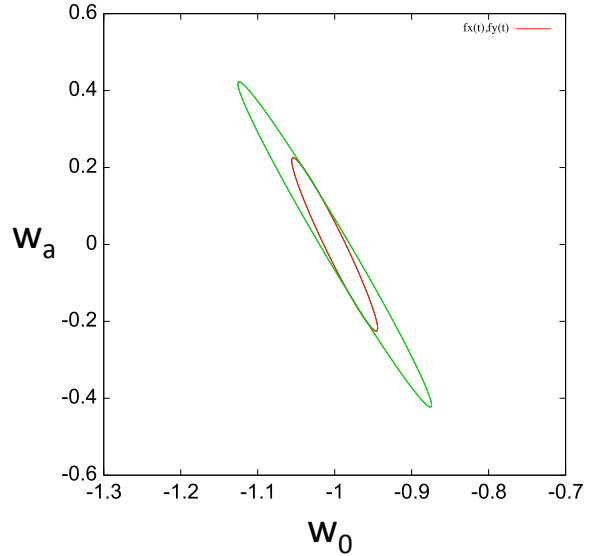

**Figure 2-8: Error ellipse in the $w_a$ vs. $w_0$ plane without a low redshift sample, the large ellipse, and with 800 ground based low redshift supernova, the small ellipse. This forecast assumes a flat universe and a dark energy equation of state w(a) = $w_0$ + $w_a$(1-a) with a = 1/(1+z). It incorporates anticipated constraints from Planck CMB measurements and WFIRST-AFTA supernova measurements, and the low redshift SN data set, but no other data.**

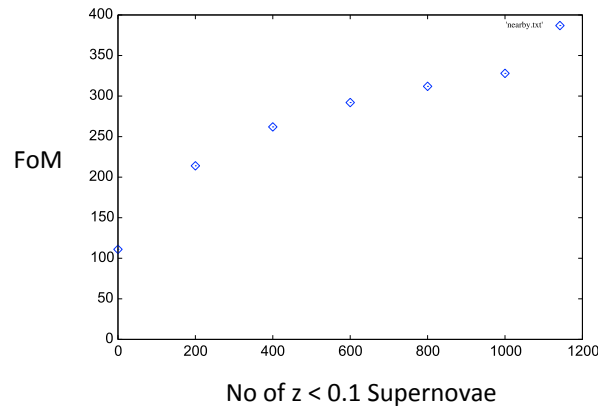

**Figure 2-9: The DETF Figure of Merit (see definition in §2.2.5) of the WFIRST-AFTA supernova survey as a function of the number of ground based low redshift supernovae, taking into account just the improvement in the anchoring of the Hubble diagram provided by larger numbers of low-redshift SNe. (Improvements from the better constraints on SN-Ia subclass-population drift are not shown here.)**

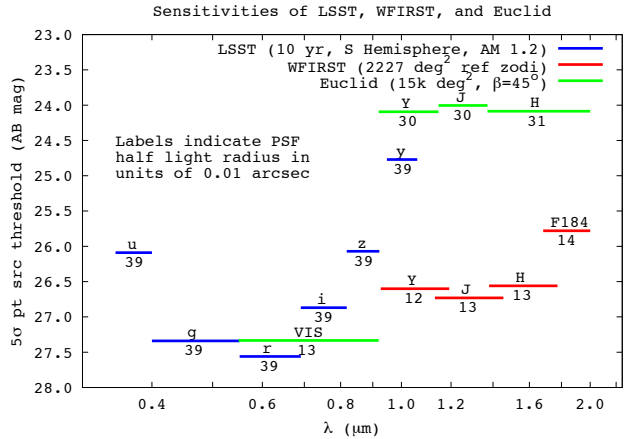

Figure 2-10: Depth in AB magnitudes of the WFIRST-AFTA (red), Euclid (green), and LSST (blue) imaging surveys. Labels below each bar indicate the size of the PSF (specifically, the 50% encircled energy radius) in units of 0.01 arcsec. The near-IR depth of the HLS imaging survey is well matched to the optical depth of LSST (10-year co-add).

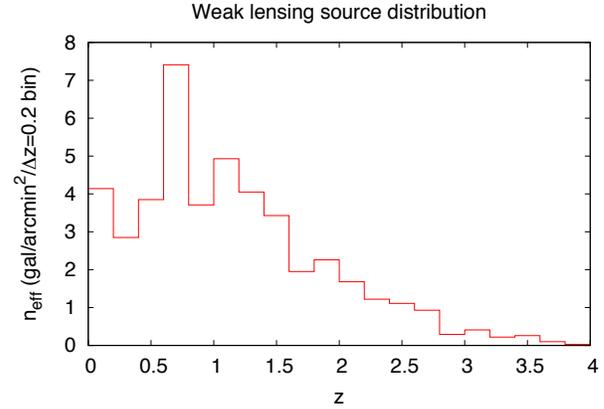

Figure 2-11: The distribution of weak lensing source redshifts for WFIRST-AFTA, as forecast using data from the CANDELS GOODS-S field (Guo et al. 2013, Hsu et al. 2014). The total effective number density for the J+H band combined sample is 45 arcmin⁻². The median redshift is 1.0 and the mean is 1.1.

as the Sloan Digital Sky Survey (SDSS) has done with its 5-band imaging at z=0.1.[4]

#### 2.2.3.2 Weak Lensing

The HLS imaging survey will make weak lensing (WL) shape measurements of 380 million galaxies over an area of 2200 deg². Weak lensing directly probes the clustering of matter between the observer and the lensed source galaxies, so it measures the growth of structure without uncertainties associated with galaxy bias. The weak lensing signal also depends on distances to the sources and the lensing matter distribution, so it provides expansion history constraints that are competitive with those from other methods. However, away from massive clusters the typical distortion of source galaxy shapes is only about 1%. Measuring the lensing signal with high precision, in the face of intrinsic ellipticity variations that are ~ 0.4 rms, requires enormous galaxy samples *and* exquisite control of systematic errors. Space-based measurements offer potentially enormous advantages for weak lensing because of high angular resolution and stability of the observing platform, allowing accurate characterization of the instrumental point spread function (PSF). The large aperture of WFIRST-

AFTA yields a high effective surface density[5] of lensed source galaxies, $n_{eff}$ = 45 arcmin⁻² in the HLS and potentially 200-300 arcmin⁻² in longer, targeted observations, much higher than any other ground-based or space-based facilities equipped for large area surveys.

The HLS follows a dithering strategy (see Figure 3-54 below) so that images in J, H, and F184 are fully sampled for shape measurements, even when some exposures of a galaxy are lost to chip gaps, cosmic rays, or detector array defects. We have chosen the exposure time of 174 secs for the HLS so that read noise and sky noise make roughly equal contributions in individual exposures. The survey area then follows from the total observing time devoted to the HLS imaging survey, which is here taken to be 1.33 years. Basic characteristics of the wide-angle HLS imaging survey are shown in Table 2-1, and the anticipated redshift distribution of weak lensing sources is shown in Figure 2-11. These yield forecasts are lower than those in the 2013 WFIRST-AFTA report primarily because we have updated the COSMOS Mock Catalog used for these forecasts to incorporate an empirical NIR color-magnitude diagram for faint galaxies based on recent observations from the CANDELS survey. The increase of the adopted telescope operating temperature to 282K has resulted in a separate reduction of $n_{eff}$ by 1%, 6%, and 18% in the J, H, and F184 bands, respectively.

---

[4] The EE50 SDSS resolution at z=0.1 is 1.5 kpc, versus 1.1 kpc for Euclid/VIS + WFIRST-AFTA at z=1. The aggregate SDSS sensitivity for a νL_ν = constant source is 4.1x10⁴¹ erg/s at z=0.1, versus 4.6x10⁴¹ erg/s at z=1 for Euclid/VIS + WFIRST.

[5] The effective surface density is the surface density of measured objects, reduced by a down-weighting factor due to the finite measurement noise in the galaxy ellipticities. For noiseless measurements, $n_{eff}$ is equal to n, the number of galaxies per arcmin² with measured shapes in the catalog.

| Characteristic | Y | J | H | F184 |
|---|---|---|---|---|
| $\lambda_{min}$ ($\mu$m) | 0.927 | 1.131 | 1.380 | 1.683 |
| $\lambda_{max}$ ($\mu$m) | 1.192 | 1.454 | 1.774 | 2.000 |
| PSF ½ light radius | 0.12" | 0.13" | 0.13" | 0.14" |
| Exposure time (s) | 5x174 | 6x174 | 5x174 | 5x174 |
| 5$\sigma$ depth AB pt src | 26.56 | 26.70 | 26.54 | 25.76 |
| 5$\sigma$ depth AB (exp. prof. $r_{1/2}$=0.3") | 25.39 | 25.56 | 25.44 | 24.71 |
| WL $n_{eff}$ (gal/am²) (combined filters) | N/A | 32.8 | 35.2 | 19.0 |
| | | 44.8 | | |

**Table 2-1: Properties of the HLS imaging survey in each filter. Weak lensing yields are averaged over the distribution of depths of the tiling, taking into account losses due to cosmic rays. Weak lensing cuts are based on detection S/N>18, ratio of half light radii of galaxy to PSF > 0.82, and ellipticity error $\sigma_e$<0.2.**

Figure 2-12 shows the predicted cosmic shear angular power spectrum for 17 tomographic bins of source galaxy photometric redshift together with the projected statistical errors for the WFIRST-AFTA WL survey. For multipole L less than a few hundred, the statistical errors are dominated by sample variance in the lensing mass distribution within the survey volume, while for high L or the extreme redshift bins the errors are dominated by shape noise, *i.e.*, the random orientations of the source galaxies that are available to measure the shear. For a given photo-z bin the statistical errors at different L are uncorrelated. Across photo-z bins, the errors at lower L are partly correlated because the same foreground structure contributes to the lensing of all galaxies at higher redshift; the errors decorrelate at high L when they become dominated by shape noise, which is independent for each set of sources. While Figure 2-12 displays the auto-spectra for the 17 photo-z bins, there are also 17×16/2 = 136 cross-spectra that provide additional information.

In addition to statistical measurements, the HLS imaging survey can produce maps of the projected or 3-dimensional dark matter distribution. The dark matter maps from the 2 deg² COSMOS survey (Massey et al. 2007) have been among the most popular cosmological results from HST, yielding tests of theoretical models, as well as an accessible illustration for public outreach (see Figure 2-13).

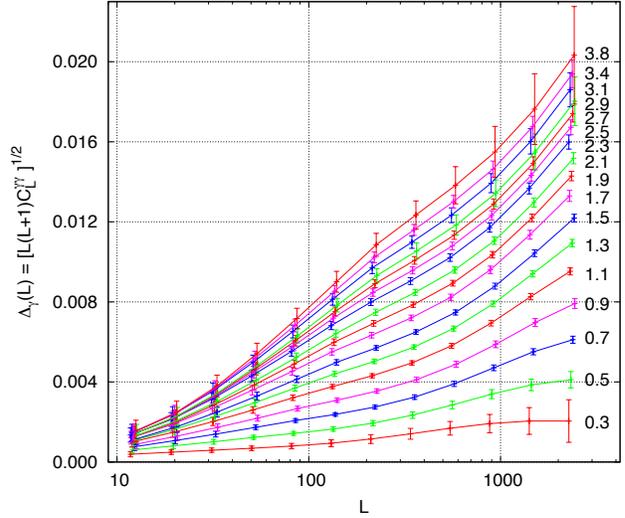

**Figure 2-12: The predicted cosmic shear angular power spectrum and associated statistical errors for the WL survey, for 17 tomographic bins of source photometric redshift as labeled. L is the angular multipole, and $L(L+1)C_L^{\gamma\gamma}$ is the contribution to the variance of cosmic shear per logarithmic interval $\Delta \ln L$.**

The HLS weak lensing survey provides the raw statistical precision to measure the amplitude of matter fluctuations $\sigma_8$ to 0.12% with other parameters fixed. Measuring the weak lensing signal (itself only a 1% distortion of source galaxies that are typically <1 arcsec across) to an accuracy commensurate with this statistical precision makes enormous demands on the control of systematic errors, including the optical PSF, understanding of the detector arrays, and calibration of photometric redshifts (discussed below). The required knowledge of PSF ellipticity ($4.7\times10^{-4}$ per component) will be obtained via observations of stars through the same optical path and on the same detector arrays as used for the galaxy measurements. Wavefront drifts must be kept to <0.7 nm rms per exposure so that the PSF can be measured faster than it changes, and short-timescale wavefront disturbances from the reaction wheels and high-gain antenna must be kept to a similar level. Detailed integrated modeling shows that WFIRST-AFTA meets these requirements with margin. (see §3.8) Line of sight motion will be tracked using the FGS, and high-frequency jitter will be fit to the stellar images in each observation. Finally, and most importantly, WFIRST-AFTA will carry out multiple shape measurements of each galaxy, using 3 different filters, 2 roll orientations per filter, and 4-5 dither positions per orientation (with at least 2 remaining after accounting for chip gaps and cosmic ray losses). This enables a robust program of internal cross-checks and null tests in the weak lensing program. Cross-correlations between observations taken many months apart can be compared with auto-correlations. Features in the effective PSF fixed to the focal plane, their dependence on wavelength, and dependences on source and sky background brightness will be measured. Finally, the multiple passes per filter support internal self-calibration as has been performed in SDSS (e.g. Padmanabhan et al. 2008). Exquisite relative photometric calibration, as required for e.g. photometric redshifts, will be enabled by combining these techniques with the inherent stability of a temperature-controlled telescope in space.

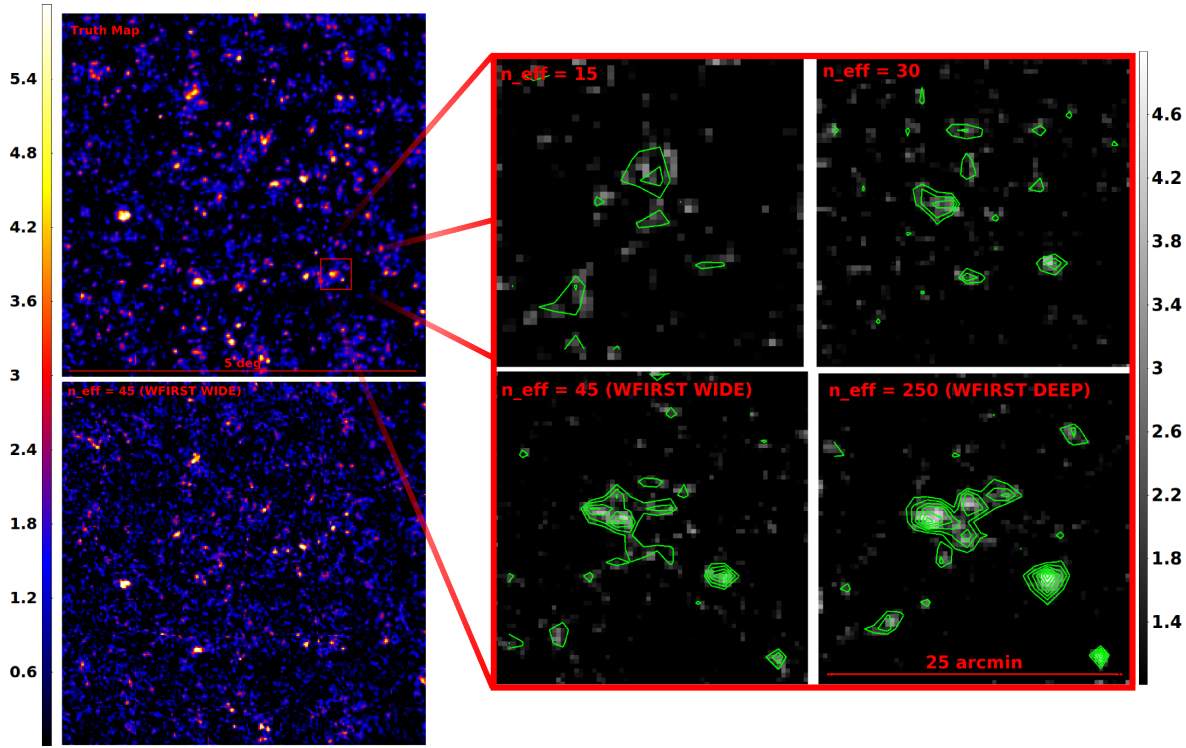

**Figure 2-13:** Maps of the projected surface density of dark matter of a 5 deg² field of view (left panels) and an expanded region (25 arcmin²) on a side (right panels). The underlying mass distributions are taken from a large cosmological N-body simulation. The bottom left panel shows a map with the weak lensing shape noise expected from the WFIRST-AFTA HLS, at a source density n_eff = 45 arcmin⁻². The lowest levels shown (blue) are 1σ above the background. This map accurately recovers the large scale features of the true mass distribution (top left panel). The zoomed panels (with contours spaced at 0.5σ starting at 1.5σ) compare the fidelity of maps at the HLS depth to maps with the 15-30 arcmin⁻² source density characteristic of ground-based or Euclid observations, which miss much of the fine detail at small scale. Over selected areas of tens of deg², WFIRST-AFTA could carry out deeper imaging to achieve source densities of 200-300 arcmin⁻² (bottom right panel), yielding projected mass maps that are virtually indistinguishable from noiseless maps (not shown). Figure credit Julian Merten/Alina Kiessling (Caltech/JPL).

WFIRST-AFTA is the only platform that will provide *both* the high resolution and stability offered by space and the internal redundancy required of a precision measurement. However, the WFIRST-AFTA weak lensing measurements are strongly synergistic with those of other large surveys. The cosmological interpretation of the signal requires optical data such as that provided by LSST (see below), and the combination with LSST and Euclid weak lensing maps provides additional opportunities for consistency checks. The synergy with Euclid and LSST is discussed extensively in Appendix H.

### 2.2.3.3 Clusters of Galaxies

The abundance of rich galaxy clusters as a function of mass and redshift offers an alternative route to measuring the growth of structure. The key uncertainty in this approach is accurate calibration of the cluster mass scale --- the average virial mass of clusters at redshift z as a function of a mass-correlated observable such as galaxy richness or X-ray luminosity --- which must be known to sub-percent accuracy to exploit the statistical potential of cluster surveys. The HLS imaging survey is an ideal tool for carrying out this calibration through measurements of the average weak lensing profiles of large cluster samples. The clusters themselves can be identified in WFIRST-AFTA imaging, together with optical imaging from LSST, or from X-ray surveys (the eROSITA mission in particular) or radio surveys that utilize the Sunyaev-Zel'dovich (SZ) effect.

The strategy of stacked weak lensing mass calibration of clusters for cosmological constraints has been successfully implemented by Rozo et al. (2010) for optically selected clusters and Mantz et al. (2014) for X-ray selected clusters, yielding some of the tightest current constraints on the amplitude of matter clustering. Forecasts for surveys like WFIRST and LSST by Oguri & Takada (2011) and Weinberg et al. (2013) indicate that constraints from cluster weak lensing should be competitive with and largely independent of those derived from cosmic shear analyses of the same weak lensing surveys. These forecasts may underestimate the power of this approach because they do not consider the additional constraints from cluster weak lensing at large separations, well beyond the virial radius. Furthermore, comparison of weak lensing mass profiles with galaxy velocity distributions for matched cluster samples can directly test modified gravity models that predict differences in the gravitational potentials that affect relativistic and non-relativistic particles.

Cluster weak lensing is analogous in many ways to galaxy-galaxy lensing (GGL): both of them involve correlating the galaxy shear map with a foreground tracer population. These forms of lensing analysis bring in the additional information provided by the cluster or galaxy maps, which mark the expected locations (including photometric and in some cases spectroscopic redshifts) of dark matter overdensities. The measurement systematics of cluster-galaxy and galaxy-galaxy lensing are also likely to be smaller than those affecting cosmic shear. GGL provides insight into the dark matter halo masses and extended environments of different classes of galaxies (e.g., Mandelbaum et al. 2006). In combination with galaxy clustering measurements, it can also be used to derive constraints on dark matter clustering. We have not included cosmological forecasts for GGL because they depend on uncertain assumptions about how well one will be able to model the non-linear bias of galaxies with respect to dark matter. However, plausible forecasts suggest that constraints on structure growth from this approach will be competitive with and perhaps even more powerful than those from cosmic shear alone (Yoo & Seljak 2012). We have also not included constraints from higher order lensing statistics (e.g., the shear bispectrum), which could have substantial power at the high $n_{eff}$ of the WFIRST-AFTA WL maps.

### 2.2.3.4  *Photometric Redshifts*

Accurate photometric redshifts are required for the WFIRST-AFTA weak lensing program. The predicted cosmic shear signal for a sample of source galaxies depends on the redshift distribution of the sources; a 1% change in the mean redshift for typical source galaxies produces a 1-2% change in the predicted shear power spectrum, an observable that WFIRST-AFTA will measure with an aggregate precision better than 0.2%. Dividing the sources into tomographic bins of different redshift is necessary to break degeneracies in cosmological parameter constraints. Photometric redshifts are also essential for modeling and removing the effects of intrinsic galaxy alignments, which can be distinguished from true weak lensing signals by cross-correlating shear maps derived from source samples with distinct redshift separations. These considerations lead to two different challenges for WFIRST-AFTA photometric redshifts (as for any other weak lensing survey). The first is to obtain precise photometric redshifts with small catastrophic outlier fractions for all sources, to enable tomography and intrinsic alignment analysis. The second is to calibrate the error distribution of the photometric redshifts – mean offset, rms width, skewness, outlier distribution – to high accuracy, so that uncertainties in this distribution do not limit the cosmological constraints.

The photometric redshifts will rely on a combination of WFIRST-AFTA data and optical data from the ground. The baseline HLS survey embeds the WFIRST-AFTA footprint in the sky area to be observed by LSST, although Northern Hemisphere fields could also be considered if appropriate optical data were obtained. We have simulated the photometric redshift performance of the combination of WFIRST-AFTA (NIR) + LSST (optical) photometry. These simulations started with the WFIRST-AFTA lensing cuts and a mock catalog based on the COSMOS intermediate band, UltraVista (H=24.5), Spitzer SPLASH (reaching a 3.6 $\mu$m depth of 25.5 AB), and HST/ACS data. The mock catalog is generated using the spectral energy distribution fits to the data catalog following the method outlined in Ilbert et al. (2013) to obtain photometric redshifts. This simulation produces a catalog of galaxies matching the redshift and color distribution of actual galaxies on the sky. The WFIRST-AFTA depths are as reported here; the LSST depths assumed are those at the end of the planned 10-year LSST survey.

The simulated performance is displayed in Figure 2-14, where it is seen that the combination of WFIRST+LSST provides excellent performance out to

z>3. The simulated sample has $\sigma_z/(1+z)=0.043$, and the fraction of objects with fractional redshift uncertainty <0.04 is 95%.

However, this simulation is still somewhat idealized. The actual redshifts of some galaxies (~3-5%, typically AGN and z>3 galaxies) are mis-assigned in the simulation, meaning the simulated outlier fraction is likely a lower bound. Furthermore, the simulation does not account for blending with nearby objects, calibration error and other issues with photometry. Based on experience with COSMOS and GOODS, adding these effects to simulations will increase the width and outlier fraction of the photo-z error distribution. When we correct based on scaling from such experiments, our estimated rms uncertainty increases to $\sigma_z/(1+z)=0.061$, and our estimated fraction of objects with fractional redshift error < 0.04 decreases to 78%. While there is more work to be done, both developing more complete simulations and optimizing photo-z algorithms, these results are an encouraging indication that photometric redshifts based on WFIRST-AFTA + LSST can meet the needs of the WFIRST-AFTA weak lensing program. Of course, photometric redshifts for a sample of hundreds of millions of galaxies out to z=3 will also be a huge asset to studies of the evolution and clustering of galaxies and AGN.

Calibrating photometric redshift performance to extremely high accuracy is a separate challenge that will likely require a multi-faceted approach. The most straightforward approach to direct calibration requires a sample of ~30,000 – 100,000 spectroscopic redshifts that spans the full flux and color range of the weak lensing source galaxies (Abdalla et al. 2008, Newman et al. 2015). These samples must have reliable spectroscopic redshifts for a very high fraction of the galaxies. However, existing large spectroscopic surveys from 8-m telescopes have flux limits 1-2 magnitudes brighter than the HLS photometric limit and achieve high reliability redshifts for only 60-75% of sample galaxies.

There are several strategies for overcoming this problem, discussed at length in the community white paper by Newman et al. (2015). The first is to improve the direct spectroscopic calibration samples using new wide-field spectroscopic instruments and large observing programs designed with photo-z calibration in mind. The advent of 30-meter class telescopes in the 2020s will make it feasible to obtain spectroscopic redshifts down to the WFIRST-AFTA photometric limit, though doing so for samples of tens of thousands of galaxies may be prohibitively expensive. Spectroscopic redshifts obtained with the WFIRST-AFTA IFU will play a crucial role in creating large samples to the HLS flux limit. Our preliminary estimates indicate that the IFU will yield ~30,000 spectroscopic redshifts for galaxies that serendipitously fall within the IFU FoV during HLS observations, to the flux limit of the WL measurements.

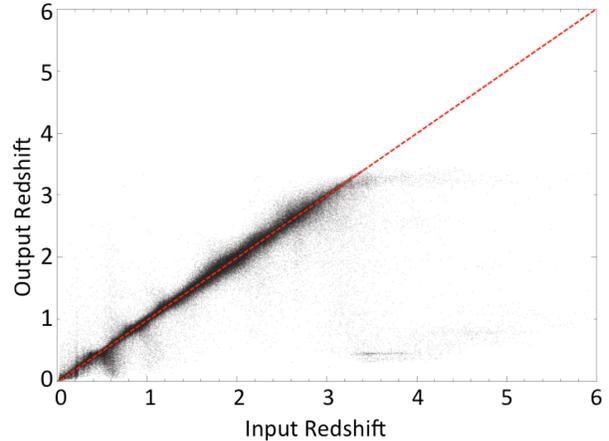

**Figure 2-14: The simulated photometric redshift performance for combined LSST (10-year) and WFIRST-AFTA data, for the WFIRST-AFTA lensing sample. The red dashed line indicates "perfect" performance, $z_{out}=z_{in}$.**

A second strategy is to focus on calibrating the conditional color-redshift distribution (Bordoloi et al. 2012) rather than directly calibrating photo-z error distributions. With optimal choices, this approach probably allows significantly smaller spectroscopic samples, making it more feasible to pursue high completeness with large telescopes. One can also decide to focus the cosmological weak lensing analysis on source galaxies in bins of color-flux space where the spectroscopic calibration is most reliable, though this comes at a statistical price as the source galaxy population is reduced in size.

A third strategy (Newman 2008) is to calibrate the redshift distribution of faint sources through their angular cross-correlation with spectroscopic samples of bright sources, from the WFIRST-AFTA GRS and ground-based surveys like the SDSS, DESI, and Subaru PFS. The amplitude of the cross-correlation is proportional to the fraction of the photo-z sample that is at the same redshift as the spectroscopic galaxies. This method is a subject of active investigation with simulations and observational data, examining performance, optimal methods of incorporating non-linear information and modeling the evolution of galaxy bias, and requirements for the spectroscopic clustering samples.

The final analysis of WFIRST-AFTA weak lensing data will likely involve describing photo-z error distributions with "nuisance" parameters that are constrained through combinations of direct spectroscopic calibration from IFU observations and ground-based data, cross-correlation with large spectroscopic samples, and the redshift dependence of the cosmic shear, cluster lensing, and galaxy-galaxy lensing measurements themselves. Fully defining and testing this strategy is an important objective for continuing work.

### 2.2.4 Wide-Area Spectroscopy and Redshift-Space Galaxy Clustering

#### 2.2.4.1 High-Latitude Spectroscopic Survey: Baseline Design

Over the same 2200 deg$^2$ area as the HLS imaging survey, the baseline WFIRST-AFTA program incorporates a grism slitless spectroscopic survey to map the distribution of emission line galaxies. The primary target is H$\alpha$ (0.6563 μm) at 1.05<z<1.88, but it is also possible to extend out to z=2.77 using galaxies with strong [O III] emission (0.5007 μm). The large aperture of WFIRST-AFTA allows a survey significantly deeper than the Euclid galaxy redshift survey, thus providing much more complete sampling of structure in the high-redshift universe. Figure 2-15 shows the limiting flux of the HLS spectroscopic survey for a point source and for an extended source with $r_{eff}$ = 0.3 arcsec and an exponential profile. Table 2-2 and Table 2-3 present forecasts of the total number and redshift distribution of galaxies detected in the HLS spectroscopic survey.

Our forecasts for H$\alpha$ emitters are based on a compilation of luminosity function estimates in the literature (Pozzetti et al. in prep, average of Models 1—3). There are fewer studies on the [O III] luminosity function in blind surveys; we have based our forecasts on WFC3 grism data (Colbert et al. 2013). For WFIRST-AFTA sensitivity and 70% completeness, we expect to measure 16.4 million H$\alpha$ emitting galaxies and 1.4 million [O III] emitters at 1.88<z<2.77. Forecasts are based on ≥7σ detections to provide margin.

The predicted number of H$\alpha$ emitting galaxies was estimated at 20 million in the 2013 SDT report. The changes are due to improvements in the grism efficiency and wavefront budget and operational efficiency (+18%), which are, unfortunately, offset by a downward revision of the luminosity function (-21%). The change of the assumed telescope operating temperature to 282 K makes a further reduction (-12%), for an overall change of -18%.

A simulation effort is underway to refine the completeness estimates for WFIRST-AFTA and optimize grism parameters and observing strategies. An exam-ple image from the first phase of this effort is shown in Figure 2-16.

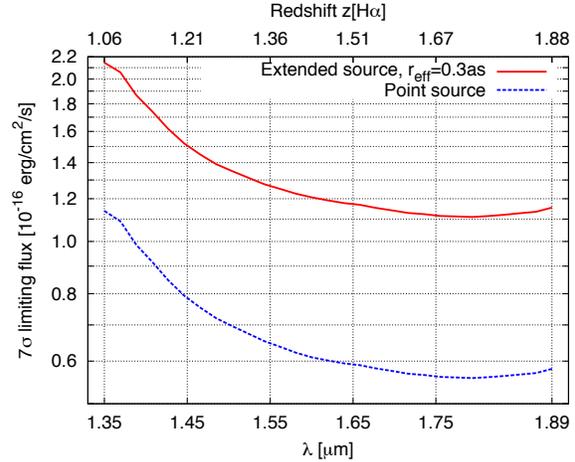

**Figure 2-15:** The emission line sensitivity for the WFIRST-AFTA HLS spectroscopy survey. The blue curve shows 7σ point source sensitivities (for six exposures), and the red curve shows extended source ($r_{eff}$ = 0.3 arcsec, exponential profile) sensitivities. The depth is observed-frame (not corrected for Galactic extinction). This depth is computed for the mean sky brightness in the GRS.

#### 2.2.4.2 Baryon Acoustic Oscillations and Redshift-Space Distortions

The galaxy redshift survey (GRS) allows measurement of the cosmic expansion history through the use of baryon acoustic oscillations (BAO). Sound waves that travel in the photon-baryon fluid of the pre-recombination universe imprint a characteristic scale on the clustering of matter, which is subsequently imprinted on the clustering of galaxies and intergalactic gas. The BAO method uses this scale as a standard ruler to measure the angular diameter distance $D_A(z)$ and the

| Characteristic | Hα | [O III] | Total |
|---|---|---|---|
| Number of galaxies (millions) | 16.4 | 1.4 | 17.8 |
| Redshift range | 1.06—1.88 | 1.88—2.77 | 1.06—2.77 |
| Statistical BAO errors | | | |
| Radial (1σ) | 0.90% | 2.05% | 0.82% |
| Tangential (1σ) | 0.49% | 1.31% | 0.46% |
| Overall scale (1σ) | 0.34% | 0.86% | 0.31% |

**Table 2-2:** Properties of the galaxy redshift survey, including forecast BAO errors in the tangential and radial directions and the overall error on the scale of the universe. Systematic errors are projected to be sub-dominant, at 0.1% or lower (Weinberg et al. 2013).

| z | n (Mpc⁻³) | dN/dz/dA (deg⁻²) |
|---|---|---|
| 1.10 | 6.62E-4 | 5986 |
| 1.15 | 7.77E-4 | 7296 |
| 1.20 | 8.85E-4 | 8598 |
| 1.25 | 9.44E-4 | 9457 |
| 1.30 | 9.67E-4 | 9971 |
| 1.35 | 9.70E-4 | 10262 |
| 1.40 | 9.54E-4 | 10329 |
| 1.45 | 9.29E-4 | 10269 |
| 1.50 | 8.96E-4 | 10099 |
| 1.55 | 8.62E-4 | 9886 |
| 1.60 | 8.31E-4 | 9674 |
| 1.65 | 7.93E-4 | 9367 |
| 1.70 | 7.53E-4 | 9005 |
| 1.75 | 7.05E-4 | 8527 |
| 1.80 | 6.50E-4 | 7948 |
| 1.85 | 5.96E-4 | 7360 |
| 1.90 | 8.03E-5 | 999 |
| 2.00 | 7.70E-5 | 973 |
| 2.10 | 7.06E-5 | 901 |
| 2.20 | 6.29E-5 | 809 |
| 2.30 | 5.42E-5 | 702 |
| 2.40 | 4.71E-5 | 611 |
| 2.50 | 4.02E-5 | 523 |
| 2.60 | 3.33E-5 | 433 |
| 2.70 | 2.65E-5 | 344 |

**Table 2-3: The comoving number density and surface density of emission line galaxies that WFIRST-AFTA will detect. We include Hα when it is in the grism wavelength range, and [O III] at higher redshift.**

Hubble parameter H(z), from clustering in the transverse and line-of-sight directions, respectively. The BAO method complements the Type Ia supernova method in several respects: it measures distances against an absolute scale calibrated via CMB observations (instead of relative to the low-redshift Hubble flow), its precision increases towards *high* redshifts because of the larger comoving volume available for clustering measurements, and it can directly measure H(z), which is tied to the dark energy density through the Friedmann equation. Our current understanding of the BAO method suggests that it will be statistics-limited even for the largest foreseeable surveys; small corrections are required for the effects of non-linear gravitational evolution and non-linear galaxy bias, but these can be computed at the required level of accuracy using analytic and numerical techniques. Forecasts for the precision of BAO distance and expansion rate measurements are included in Table 2-2.

The broad-band shape of the galaxy power spectrum P(k) provides a second "standard ruler" for geometrical measurements via the turnover scale imprinted by the transition from radiation to matter domination in the early universe, as well as a diagnostic for neutrino masses, extra radiation components, and the physics of inflation.

The galaxy redshift survey conducted for BAO also offers a probe of the growth of structure via redshift-space distortions (RSD), the apparent anisotropy of structure induced by galaxy peculiar velocities. RSD measurements (in the large scale regime described by linear perturbation theory) constrain the product of the matter clustering amplitude and the clustering growth rate (specifically, the logarithmic derivative of clustering amplitude with expansion factor), complementing weak lensing observations that measure the matter clustering amplitude in isolation. Comparison of weak lensing and RSD growth measurements can also test modified gravity theories in which the potential that governs space curvature can depart from the potential that governs "Newtonian" acceleration of non-relativistic tracers.

For measurements of BAO distances and expansion rates, the most important metric of a redshift survey is its total comoving volume and the product $nP_{BAO}$ of the mean galaxy space density n and the amplitude of the galaxy power spectrum P(k) at the BAO scale, approximately given by P(k = 0.2h Mpc⁻¹). For $nP_{BAO}$ > 2, the BAO measurement error is dominated by sample variance of the structure within the finite survey volume and would not drop much with a higher tracer density. For $nP_{BAO}$ < 1, the measurement error is dominated by shot noise in the galaxy distribution. For any given $nP_{BAO}$, the error on $D_A(z)$ and H(z) scales with comoving volume as $V^{-1/2}$. To predict $nP_{BAO}$ for the space densities in Table 2-3, we adopt the prescription of Orsi et al. (2010) for the bias factor between galaxy and matter clustering, b=1.5+0.4(z-1.5), which is based on a combination of semi-analytic model predictions and observational constraints. The clustering measurements of Geach et al. (2012) suggest a somewhat higher bias for Hα emitters, which would lead to more optimistic forecasts.

Figure 2-17 plots $nP_{BAO}$ vs. z. The BAO scale is fully sampled ($nP_{BAO}$ > 1) over the whole range 1.05 < z < 1.88 probed by Hα emitters, with $nP_{BAO}$ > 2 at z < 1.8. The strong decline at z > 1.5 arises because we assume that the Hα luminosity function does not evolve beyond the maximum redshift probed by the Colbert et

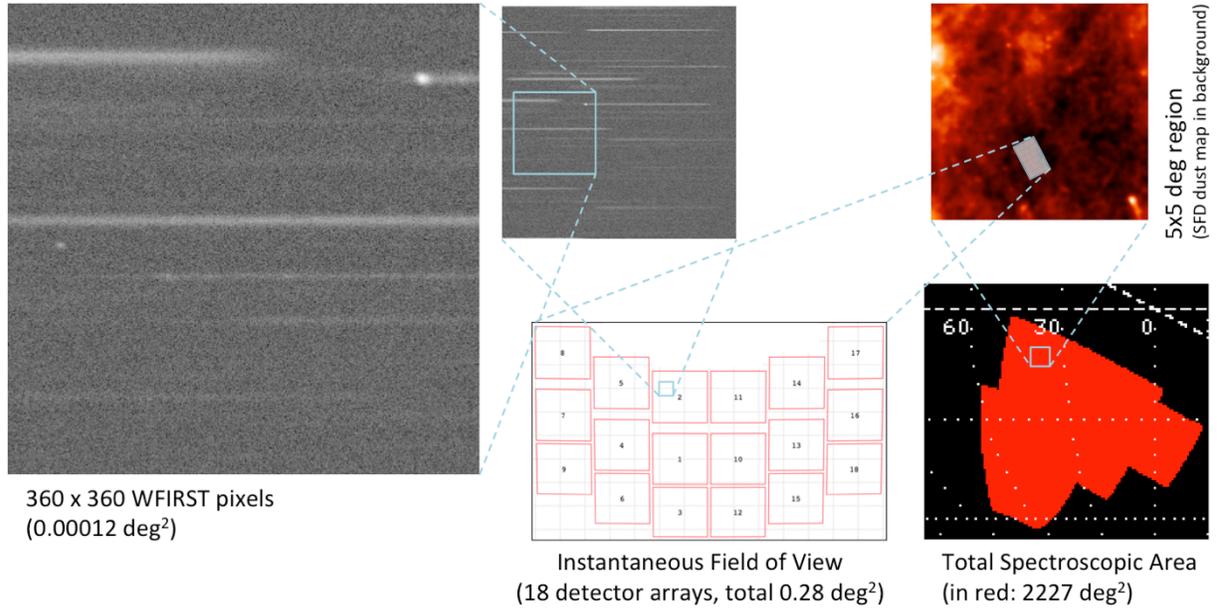

360 x 360 WFIRST pixels
(0.00012 deg²)

Instantaneous Field of View
(18 detector arrays, total 0.28 deg²)

Total Spectroscopic Area
(in red: 2227 deg²)

**Figure 2-16: A simulated 1ˢᵗ order grism image at WFIRST-AFTA resolution and sensitivity, and at 347 s exposure time as planned for the HLS. Each field in the HLS will be covered with a median of 7 grism exposures at 4 roll angles. This image covers 40x40 arcsec; the instantaneous field of view of WFIRST-AFTA is 2300 times larger.**

al. (2013) data; this assumption could prove pessimistic, though extrapolation of a fixed luminosity function to z = 2.2 gives reasonable agreement with the measurements of Sobral et al. (2013) at this redshift. While $nP_{BAO} \approx 3$ is "overkill" for BAO measurement at this scale, a high density survey allows better measurements of structure at smaller scales, better measurements of higher order clustering statistics, better characterization of galaxy environments, and more complete sampling of the population of star-forming galaxies, all beneficial for studies of galaxy formation and galaxy evolution. [OIII] emitters provide a sparse sampling of structure at z > 2; because of the large comoving volume, this sample of ~ 2 million galaxies yields useful cosmological constraints despite its relatively high shot noise.

Ground-based surveys like BOSS and the proposed MS-DESI experiment are likely to achieve $nP_{BAO}$ > 1 out to z ~ 1.1, but reaching full sampling with ground-based observations becomes very difficult at higher z. Estimated space densities from the Euclid Red Book (Laureijs et al., 2011), corrected for a known factor of ln10 error that arose in luminosity function conversion, imply space densities roughly 3.0, 8.6, and 19 times lower than those of WFIRST-AFTA at z = 1.1, 1.5, and 1.85, respectively, so Euclid BAO errors will be dominated by galaxy shot noise. Figure 2-18 presents a visual comparison of structure sampled at WFIRST-AFTA density and Euclid density, based on slices from

the Millennium simulation (Springel et al. 2005) populated with semi-analytic galaxy formation modeling

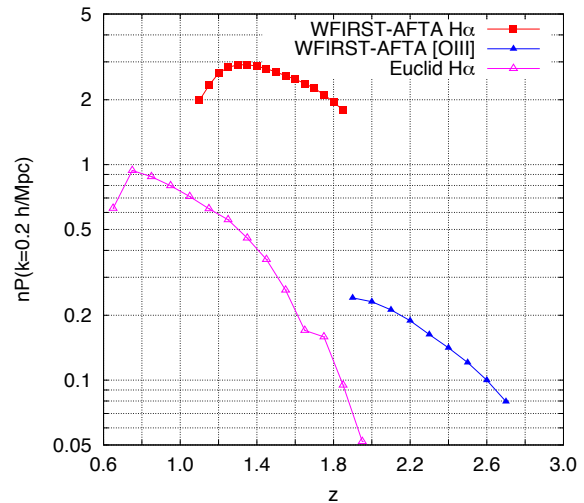

**Figure 2-17: Product $nP_{BAO}$ of the mean galaxy space density and the amplitude of the galaxy power spectrum at the BAO scale as a function of redshift for the WFIRST-AFTA GRS, based on the luminosity function of Hα emitters (squares) and [OIII] emitters (filled triangles). Open triangles show our estimate of the Euclid sampling density based on values from the Euclid Red Book. For $nP_{BAO}$ > 2, the statistical errors of BAO measurements are dominated by the sample variance of structure within the survey volume, while for $nP_{BAO}$ < 1 they are dominated by shot noise in the galaxy distribution.**

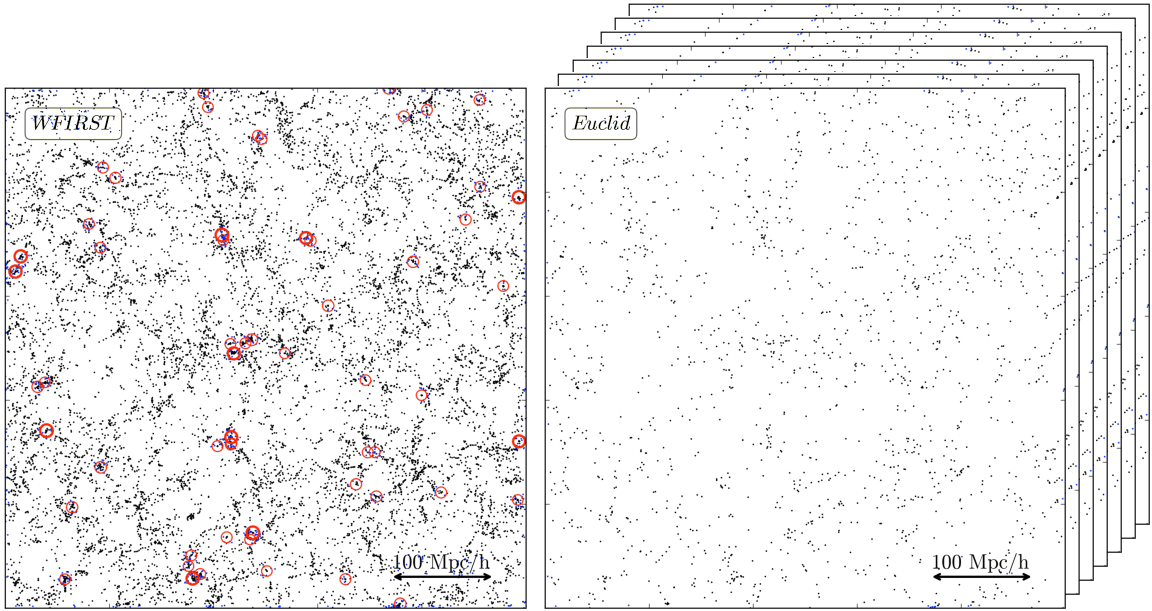

**Figure 2-18: Tracing large scale structure at the sampling density of WFIRST-AFTA (left) and Euclid (right). Each panel shows a slice 500 $h^{-1}$ Mpc on a side and 30 $h^{-1}$ Mpc thick from the Millennium simulation at z = 1.5. Points in the left panel show semi-analytic galaxies (Bower et al. 2006) selected at a luminosity threshold that yields our predicted space density for WFIRST-AFTA. Thin and thick red circles mark clusters with virial mass exceeding 5 × $10^{13}$ $M_{Sun}$ and $10^{14}$ $M_{Sun}$, respectively. The right panel shows the galaxy distribution with a higher luminosity threshold that yields the space density predicted for the Euclid GRS at this redshift, which is lower by a factor of 8.6. Conversely, the Euclid area is about seven times larger, so the two surveys have complementary properties. At z = 1.5 this slice would subtend a solid angle (9.2 × 9.2) deg² with redshift depth Δz = 0.022, so it represents a minuscule fraction (~ $10^{-3}$) of the GRS survey volume. Figure courtesy of Ying Zu.**

(Bower et al. 2006). A survey at the WFIRST-AFTA depth recovers much of the fine detail present in the full dark matter distribution, which is lost at the much sparser sampling of the Euclid survey. Conversely, Euclid covers a volume several times larger. We note that WFIRST-AFTA could carry out a shallow-wide GRS of 10,000 deg² in about one year of observing time, but this would be largely redundant with Euclid, while the deeper survey adopted in the baseline HLS design is complementary.[6]

### 2.2.5 Tests of Cosmic Acceleration Models

The WFIRST-AFTA supernova, imaging, and spectroscopic surveys will enable multiple independent measurements of cosmic expansion history and structure growth over the redshift range z = 0-3, each with aggregate precision at the ~0.1 – 0.5% level. This extremely high statistical precision demands that system-atic biases be very tightly controlled to avoid compromising the measurements. The WFIRST-AFTA mission is designed with control of systematics foremost in mind, so that it can in fact realize the promise of its powerful statistics. For SNe, the use of a space-based observatory and near-IR observations already mitigates key systematics affecting ground-based surveys, and the use of an IFU on WFIRST-AFTA reduces systematics associated with photometric calibration and k-corrections and provides spectroscopic indicators that can be used to mitigate evolutionary effects. For WL, unique aspects of WFIRST-AFTA are the high surface density of sources and the control of systematics enabled by eliminating the atmosphere and having highly redundant multicolor data, with an observing strategy that provides good sampling even when some of a galaxy's exposures are lost to cosmic ray hits or detector array defects. For the GRS, the high space density of galaxies in the WFIRST-AFTA survey will make it possible to measure higher order clustering statistics and split the data into subsamples to constrain models of galaxy bias, which are the primary source of uncertainty in deriving cosmological constraints from RSD and intermediate-scale P(k) measurements. Cross-correlation with the WL shear maps (which have significant statistical power even out to z=2) will provide further tests of galaxy bias models and improved cosmological constraints. In all of these aspects of systematics control, the large aperture and sharp PSF of WFIRST-AFTA play a crucial role.

---

[6] For example, a WFIRST-AFTA survey with 2×268s exposures could cover the 1.1 < z < 2 range to ~ 1.4× Euclid depth over 12,000 deg² in 1.0 years, but it would be observing largely the same galaxies as Euclid.

To characterize the ability of these measurements to test theories for the origin of cosmic acceleration --- i.e., to address the two questions posed at the start of §2.2, we adopt the framework put forward by the Dark Energy Task Force (DETF) (Albrecht et al. 2006). We assume a parameterized model in which the dark energy equation-of-state parameter evolves with redshift as $w(a) = w_0 + w_a(1-a) = w_p + w_a(a_p-a)$, where $a = (1+z)^{-1}$ is the expansion factor and $z_p = a_p^{-1}-1 \approx 0.5$ is the "pivot" redshift at which the errors on $w_p$ and $w_a$ are uncorrelated and the value of $w(a)$ is best determined. Additional parameters that must be constrained by the measurements are the baryon and matter densities $\Omega_b h^2$ and $\Omega_m h^2$, the Hubble parameter h, the curvature $\Omega_k$, and the amplitude and spectral index of the inflationary fluctuation spectrum. For some forecasts we also allow deviations from GR-predicted structure growth parameterized by a free value of the index γ in the growth rate evolution equation $d\ln G/d\ln a \approx [\Omega_m(a)]^\gamma$, where $\gamma \approx 0.55$ for GR-based dark energy models. Our analysis takes into account the degeneracies among these parameters and the ability of complementary cosmological observables to break these degeneracies.

With the assumptions described in the preceding sections, we have computed Fisher matrices for the WFIRST-AFTA SN, GRS, and WL surveys, which we combine with the Fisher matrix for Planck CMB measurements. In addition to WFIRST-AFTA and Planck, our forecasts assume a local SN calibration sample with 800 SNe, and they include the anticipated BAO measurements at z < 0.7 from the SDSS-III BOSS survey (Anderson et al. 2014). Our WL Fisher matrix is based on cosmic shear only and does not include more uncertain contributions from higher order statistics or galaxy-galaxy lensing, which might significantly strengthen the anticipated constraints. We do include the expected constraints from clusters calibrated by weak lensing, using a Fisher matrix computed for us by Michael Mortonson and Eduardo Rozo following the methodology described by Weinberg et al. (2013). Our overall forecasting methodology is similar to that described in the 2013 report (Spergel et al. 2013). For BAO/RSD we use the approach described in detail by Wang et al. (2013), assuming no non-linear correction ($p_{NL}$=1.0) and maximum fitting wavenumber $k_{max}$ = 0.25 h/Mpc (comoving).

For clusters, we have made the approximation of scaling error variances and covariances computed for the 2013 report by 480/380 to account for the changed estimate of the number of WL sources in the HLS.

For the fiducial combination of measurements, we forecast 1σ errors of 0.0089 on $w_p$ and 0.114 on $w_a$, after marginalizing over all other cosmological parameters in the model. The solid black ellipse in Figure 2-19 illustrates these constraints in the form of a $\Delta\chi^2$=1 error ellipse. This represents a dramatic improvement over the current state of the art, represented in Figure 2-19 by the green ellipse, with errors $\Delta w_p$ = 0.12 and $\Delta w_a$ = 0.6 taken from Aubourg et al.'s (2014) combination of the main current BAO, SN, and CMB data constraints. While current cosmological data are consistent with a cosmological constant, we obviously do not know what WFIRST-AFTA will find. For the purpose of Figure 2-19, we have imagined that the true model (marked by the x) has $w_p$ = -1.022 and $w_a$ = -0.18. A value of w < -1 is exotic but possible in some theories, and it leads eventually to an accelerating rate of acceleration that causes the universe to end in a "Big Rip" at a finite time in the future. The baseline WFIRST-AFTA would discriminate this model from a cosmological constant at > 2σ significance ($\Delta\chi^2$= 8.6). The red ellipse, smaller in area by a factor of 1.7, represents the more stringent constraints that could be obtained with a factor of 1.5 improvement over our forecast errors. Such improvement could plausibly be achieved by a combination of improved control of supernova measurement systematics, improved theoretical modeling that extracts non-linear information from galaxy-galaxy lensing and redshift-space clustering in the HLS, and some level of additional observing in an extended mission. These measurements would distinguish the adopted model from a cosmological constant with $\Delta\chi^2$= 14.2.

As a FoM for dark energy experiments, the DETF proposed the inverse of the constrained area in the $w_p$-$w_a$ plane. We find FoM = $[\sigma(w_p)\sigma(w_a)]^{-1}$ = 987 assuming GR (and 922 when we allow a free growth index γ). This value is represented by the shaded "Baseline" block in the upper panel of Figure 2-20. As a representation of likely pre-WFIRST-AFTA constraints, the dashed blue block in this panel marks the forecast FoM = 131 for a combination of Planck and "Stage III" dark energy experiments such as BOSS and the Dark Energy Survey, taken from Table 8 of Weinberg et al.[7] The program is $\sigma_\gamma = 0.015$, compared to the Stage III forecast of 0.15.

---

[7] The model space used by Weinberg et al. (2013) is similar but not identical, as it also allows for an over-

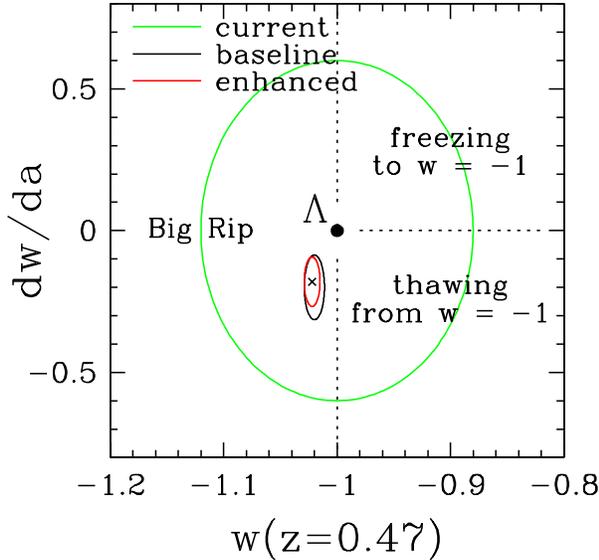

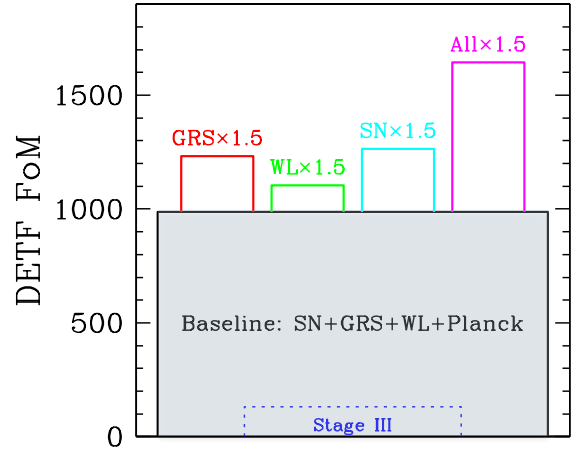

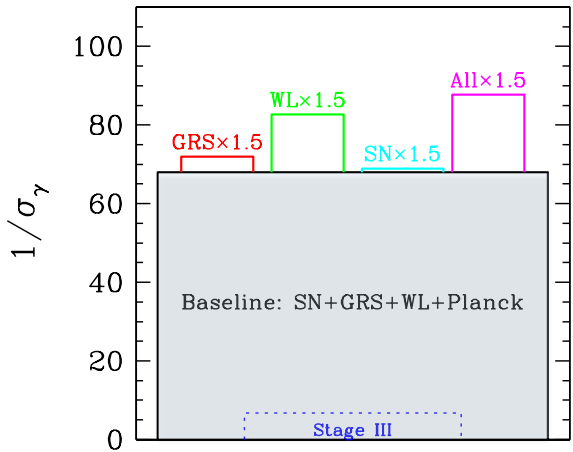

**Figure 2-19:** $\Delta\chi^2 = 1$ error ellipses on the value of the dark energy equation-of-state parameter w at redshift z = 0.47 (the redshift at which it is best determined by WFIRST-AFTA) and its derivative with respect to expansion factor dw/da. The green ellipse, centered here on the cosmological constant model (w = -1, dw/da = 0), represents current state-of-the-art parameter uncertainty from a combination of CMB, SN, and BAO data (Aubourg et al. 2014). For this figure, we have imagined that the true cosmology is w(z=0.47) = -1.022 and dw/da = -0.18, well within current observational constraints. The black ellipse shows the error forecast for the baseline WFIRST-AFTA SN, GRS, and WL surveys, combined with CMB data from Planck, a local supernova calibrator sample, and BOSS BAO and RSD measurements at z < 0.7. The red ellipse shows the "enhanced" case in which the precision of the WFIRST-AFTA measurements (but not the Planck, local SN, or BOSS measurements) is increased by a factor of 1.5, as a result of better control of systematic uncertainties, improved modeling of non-linear clustering, and possible additional observations in an extended mission. Legends indicate physically distinct regions of the parameter space: a cosmological constant ($\Lambda$), scalar field models that are "freezing" towards or "thawing" from w = -1, and models with w < -1 (often referred to as "phantom energy") in which increasing acceleration leads to a "big rip" at a finite time in the future. lower panel represents the growth index $\gamma$ --- as with the FoM, we plot $1/\sigma_\gamma$ so that higher values correspond to better constraints. Our forecast error for the baseline

**Figure 2-20:** Top: Figure of Merit FoM = $[\sigma(w_p)\sigma(w_a)]^{-1}$ for various assumptions. The shaded block shows the baseline case of FoM = 987 corresponding to the solid black contour of Figure 2-19. The blue dashed block shows the forecast FoM = 131 from Stage III experiments from Weinberg et al. (2013). Red, green, and cyan bars show the effect of increasing the measurement precision from the GRS, WL, or SN survey by a factor of 1.5, while the purple bar shows the effect of increasing the precision of all three sets of measurements by a factor of 1.5, as described in the text. Errors for Planck, local SNe, and BOSS are held fixed throughout. Bottom: Same as top, but for the (inverse) $1\sigma$ error on the growth index $\gamma$.

The measurement forecasts that are the basis of our FoM calculations are necessarily uncertain, and we think they are more likely to prove conservative than to prove overoptimistic. In the case of weak lensing, for example, we are not including the potential gains from galaxy-galaxy lensing. For the GRS, we have not included gains in full P(k) and redshift-space distortion analysis that could be enabled by theoretical modeling all offset in the amplitude of matter fluctuations in addition to free $\gamma$. This makes little difference to the FoM, but it makes the error on $\gamma$ larger.

improvements over the next decade. More generally, as emphasized in §2.2.1, there is considerable flexibility to optimize the WFIRST-AFTA dark energy program based on new knowledge about techniques and their sources of uncertainty. This optimization will likely improve performance relative to our baseline strategy. The colored bars in Figure 2-20 show the impact of reducing the error bars of the GRS, WL, or SN measurements by a constant factor of 1.5 (multiplying the Fisher matrix by 2.25), improvements that could plausibly be achieved without increasing the observing time allocated to the dark energy program. These bars illustrate the complementarity of the three prongs of the program: a 1.5× improvement of the SN measurements has the largest single impact on the DETF FoM, a 1.5× improvement of the WL measurements has the largest single impact on $\sigma_\gamma$, and a 1.5× improvement in the GRS measurements has a noticeable impact on both constraints. If all three sets of measurements are improved by 1.5×, yielding the purple bar at right, then the DETF FoM increases from 987 to 1645 and $1/\sigma_\gamma$ increases from 68 to 88.

For Figure 2-19 and Figure 2-20, we have included only minimal external data that are strictly independent of WFIRST-AFTA: Planck CMB, local SN calibrators, the BOSS BAO and RSD measurements at z < 0.7, and (implicitly) ground-based data for the photometric redshifts of the WFIRST-AFTA lensing source galaxies. These are important complementary data sets that are impossible (or do not make sense) to acquire from WFIRST-AFTA itself. The inclusion of other pre-WFIRST-AFTA data sets would, of course, increase the cumulative FoM. By the time of WFIRST-AFTA, there is a good chance that ground-based redshift surveys such as eBOSS, the DESI experiment, and the SuMIRe project with the Subaru PFS, will have measured BAO and RSD out to z = 1 over $10^4$ deg$^2$ or more. At z > 1 they may no longer be strictly independent of WFIRST-AFTA, but because they will cover large sky areas at low sampling density ($nP_{BAO}$ < 1) instead of 2200 deg$^2$ at high sampling density, the measurements can probably be combined as though they are effectively independent. The ability to add pre-WFIRST-AFTA SN and WL measurements to the cosmological constraints depends on whether they are limited by statistics or systematics and whether the systematics are correlated with any systematics in the WFIRST-AFTA analyses.

If WFIRST-AFTA development follows the schedule proposed in this report, then the LSST, Euclid, and WFIRST-AFTA dark energy experiments will be essentially contemporaneous, with all three of them reporting major results in the mid-to-late 2020s. As with ground-based surveys, it should be straightforward to combine WFIRST-AFTA GRS results with Euclid GRS results because of the complementarity between their narrow-deep and wide-shallow data sets. For SNe, we expect that the quality of IFU observations from WFIRST-AFTA will make its (spectro)photometry far better than anything achievable from the ground, but there may be advantageous ways to coordinate WFIRST-AFTA and LSST campaigns to get smaller statistical or systematic errors through additional light curve monitoring, better data on host galaxies, or the use of galaxy counts and shear maps to reduce the impact of lensing noise at high redshift (where it dominates the statistical error budget). The inclusion of SNe is a critical difference between WFIRST-AFTA and Euclid, so it is worth noting that our flat-universe forecast for *just* WFIRST-AFTA SNe + local SN calibrators + Planck CMB (and including Stage III priors, unlike the forecasts in Figure 2-9) yields an FoM of 582, more than a threefold improvement over the Stage III forecast.

For WL, the critical first task will be to cross-check LSST, Euclid, and WFIRST-AFTA measurements to test for systematics. Direct cross-correlation of shear maps from different observatories is a powerful test for shape measurement systematics, as these should be quite different among the three facilities. Photometric redshift uncertainties can be well constrained by combining WFIRST-AFTA and LSST data, to get deep 10-band photometry and good morphological information for all galaxies in the joint sample, to calibrate photo-z distributions via clustering cross-correlation with the imaging catalog (Newman 2008), and perhaps to obtain a direct spectroscopic calibration sample at faint magnitudes with the WFIRST-AFTA IFU. If these cross-checks and joint calibrations demonstrate that all three experiments are statistics-limited rather than systematics-limited, then their measurements can be combined to yield results much stronger than those from any one in isolation.

## 2.3 High-Latitude Surveys: General Astrophysics

WFIRST-AFTA will produce large data sets with homogenous observing conditions for each set. These data sets will be a treasure trove for Guest Investigators. The HLS will map ~2,200 deg$^2$ of sky in four broad NIR passbands down to a 5-sigma limiting AB magnitude of J=26.7 and will include a slitless spectroscopic survey component that will obtain R=~600 spectra with a 7-sigma line flux sensitivity of $10^{-16}$ erg cm$^{-2}$ sec$^{-1}$ over the same region of sky (see Figure 2-15). While the survey is designed to obtain constraints on the dark energy equation of state from weak lensing and BAO measurements, a remarkable range of new astrophysical investigations will be enabled by the HLS data. We focus below on a few of the science cases that are especially well enabled by the access to a 2.4-meter wide-field survey telescope in space and that require no additional time other than that already allocated to the HLS.

### 2.3.1 The First Billion Years of Cosmic History

Mapping the formation of cosmic structure in the first 1 billion years after the Big Bang is essential for achieving a comprehensive understanding of star and galaxy formation. Little is known about this critical epoch when the first galaxies formed. Determining the initial condition and abundances of the primordial interstellar gas as well as the dynamical properties of the first galaxies is critical in order to understand (1) how the first galaxies were formed through a hierarchical merging process; (2) how the chemical elements were generated and redistributed through the galaxies; (3) how the central black holes exerted influence over the galaxy formation and the overall star formation process; and (4) how these objects contributed to the end of the "dark ages."

Finding candidate z > 7 galaxies has been pursued in two ways - from moderate area but very deep imaging surveys (e.g., UDF, Subaru Deep Field, CANDELS) and from images of strongly lensing clusters of galaxies (e.g., Bradley et al. 2008, Postman et al. 2012). Recent observations have identified at least 3 candidate objects at z > 9 (Bouwens et al. 2011, Zheng et al. 2012, Coe et al. 2013). Two of these candidates were found in a survey of massive galaxy clusters. One of the lensed candidates is shown in Figure 2-21. While unlensed sources at these redshifts are expected to be extremely faint, with total AB magnitudes greater than 28, lensed sources can be 10 – 30 times brighter. This enhancement is particularly important because it puts some of these sources within the reach of the spectrographs on the James Webb Space Telescope and from very large ground-based telescopes. Luminous z > 7 galaxies are extremely valuable as their spectra can be used to determine the amount of Lyα photons produced by stars and galaxies escaping from neutral IGM. These objects will also enable the determination of the time when the epoch of reionization occurred up to its completion. Since only a tiny fraction of neutral hydrogen is needed to produce the high opacity of Lyα observed at z ~ 6, the damped Lyα absorption profile that results from even a partially neutral IGM (Miralda-Escude 1998) can be measured at low-spectral resolution. Furthermore, the early star formation rate (via Lyα and Hα emission measurements; Iye et al. 2006) can be estimated from the spectra of bright high-z galaxies.

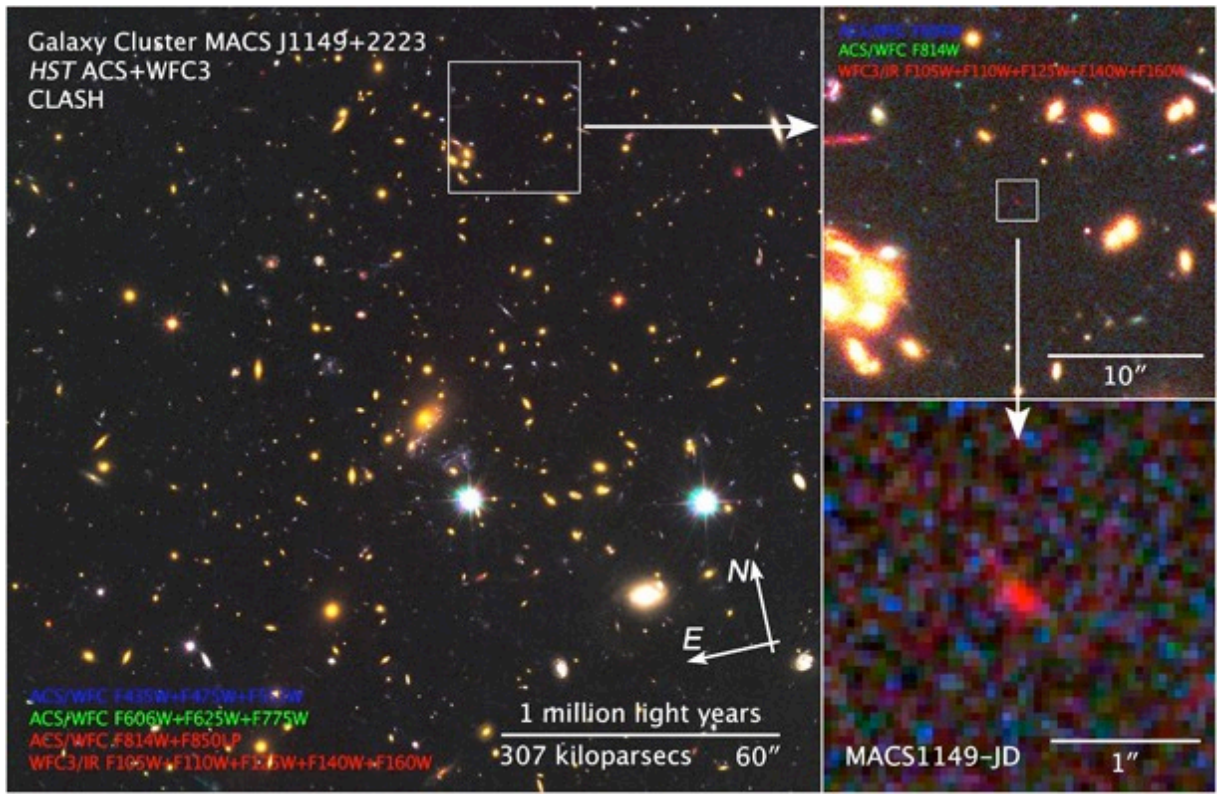

**Figure 2-21: A candidate for a galaxy at z=9.6, magnified by a factor of ~15 by the foreground cluster MACS J1149+2223 (z = 0.54). The object was found in an HST survey using the WFC3IR camera (Zheng et al. 2012). This young object is seen when the universe is only about 500 million years old.**

The HLS performed by the WFIRST-AFTA telescope as part of its weak lensing program will be superb for finding and studying objects in the early universe. A key advantage that the 2.4-meter aperture provides is that, in the same amount of time, the WFIRST-AFTA HLS will reach ~0.9 magnitudes fainter than the corresponding DRM1 HLS. The final 5-sigma limiting depth of the WFIRST-AFTA HLS is estimated to be J=26.7 AB mag. Figure 2-22 shows the predicted cumulative number of objects that would be found at or higher than a given redshift for both gravitationally-lensed and unlensed regions in a 2,000 square degree survey. The predicted counts (Bradley et al. 2012) for z~8 are extrapolated to higher redshifts using the few known z>9 candidates and bounded by assuming both a pessimistic (dM*/dz = 1.06) and optimistic (dM*/dz = 0.36) evolution of the characteristic magnitude of galaxies. The lensed source count predictions assume ~1 strongly lensing cluster per square degree. The mass models were adopted from the CLASH Multi-cycle treasury program (Postman et al. 2012). The larger collecting area of WFIRST-AFTA relative to DRM1 will yield, in some redshift ranges, as much as 20 times as many z > 8 galaxies (Figure 2-23).

Having 10,000 or more luminous z > 8 galaxies will allow very stringent constraints to be placed on the early star formation rate density, on the amount of ionizing radiation per unit volume, and on the physical properties of early galactic structures. With its ~0.1 – 0.2 arcsecond resolution, combined with lens magnifications in the 10 – 30 range, the HLS images will allow us to measure structures on scales of 20 to 50 parsecs, thanks to the boost in spatial resolution provided by the cluster lenses. The HLS will certainly herald a remarkable era in probing the first 1 billion years of cosmic history.

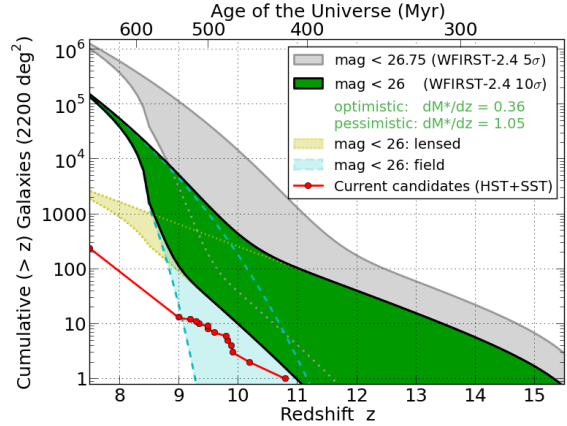

**Figure 2-22: Cumulative number of high-z galaxies expected in the HLS. JWST will be able to follow-up on these high z galaxies and make detailed observations of their properties. For understanding the earliest galaxies, the synergy of a wide-field telescope that can discover luminous or highly magnified systems and a large aperture telescope that can characterize them is essential; WFIRST-AFTA and JWST are much more powerful than either one alone.**

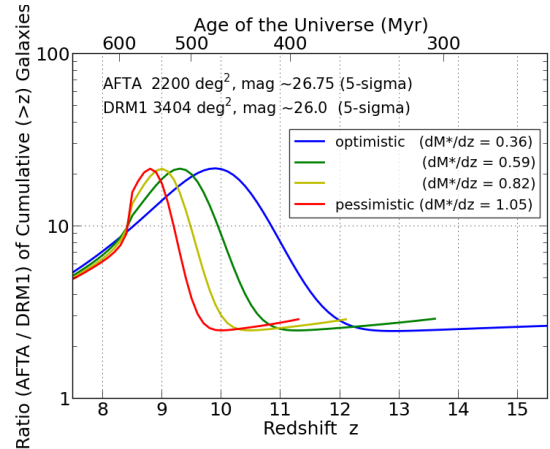

**Figure 2-23: The ratio of the cumulative number of high redshift galaxies detected with WFIRST-AFTA to the number detected with a smaller DRM1 version of WFIRST. The 2.4 m aperture yields up to 20 times more high-z galaxies.**

### 2.3.2 Mapping Dark Matter on Intermediate and Large Scales

Clusters of galaxies are important tracers of cosmic structure formation. All of their mass components including dark matter, ionized gas and stars are directly or indirectly observable. By its design, the HLS will be superbly suited to mapping weak lensing signatures. The signal-to-noise ratio of the shear signal is proportional to the square root of the surface density on the sky of galaxies that can be used to map the lensing. The WFIRST-AFTA HLS will produce catalogs with surface densities of 60 – 70 galaxies per square arcminute in a single passband and potentially as high as 80 – 100 galaxies per square arcminute in co-added multi-band images. Achieving such densities for weak lensing measurements has already been demonstrated with the WFC3/IR camera on the Hubble Space Telescope. The WFIRST-AFTA HLS will thus enable WL maps that are a factor 3 – 5 higher number density of galaxies than any maps produced from the ground (even with 8 to 10 meter telescopes) as illustrated in Figure 2-24. The larger telescope and multiple bands will produce much more robust and higher quality WL maps than Euclid (see Figure 2-13). On cluster scales (200 kpc – 2 Mpc), this higher galaxy density will allow the dark matter to be mapped to a spatial resolution of ~40 – 50 kpc. When combined with strong lensing interior to clusto-centric radii of ~200 kpc, the central dark matter distribution can be mapped down to a resolution of 10 – 25 kpc. Comparing such maps around the hundreds of intermediate redshift clusters expected in the HLS to those from numerically simulated clusters would give us unprecedented insight into the main mechanisms of structure formation.

An especially interesting class of clusters are those in the process of merging / colliding (see Figure 2-25) where all mass components are interacting directly during the creation of cosmic structure. Multiple such mergers have been observed, with the Bullet Cluster being the most prominent example (Clowe et al. 2006). Paired with follow-up numerical simulations, such systems gave important insight into the behavior of the baryonic component (Springel & Farrar 2007) and set upper limits on the dark matter self-interacting cross section (Randall et al. 2008), which is of great importance in the search for the nature of dark matter. Recently, more complicated merging systems have been identified (e.g. Merten et al. 2011, Clowe et al. 2012, Dawson et al. 2012) offering a great opportunity to better characterize the nature of cosmic dark matter. The HLS will be well suited to producing highly accurate strong and weak lensing mass maps in the environs of these information-rich merging systems.

On the largest scales (>10 Mpc), the distribution of matter can best be measured via weak gravitational lensing, owing to the fact that the mass is predominantly dark matter and thus can only be detected gravita-tionally. The WFIRST-AFTA HLS's unique combination of spatial resolution, infrared observations and wide area will provide mass maps that have extremely high scientific impact. Mapping the large-scale distribution of dark matter can help identify and eliminate systematic effects: unexpected spatial variations in the mass distribution can hint at exciting new physics but could also indicate observational systematics not discoverable via a power spectrum measurement like the one that will be used to constrain dark energy via the HLS.

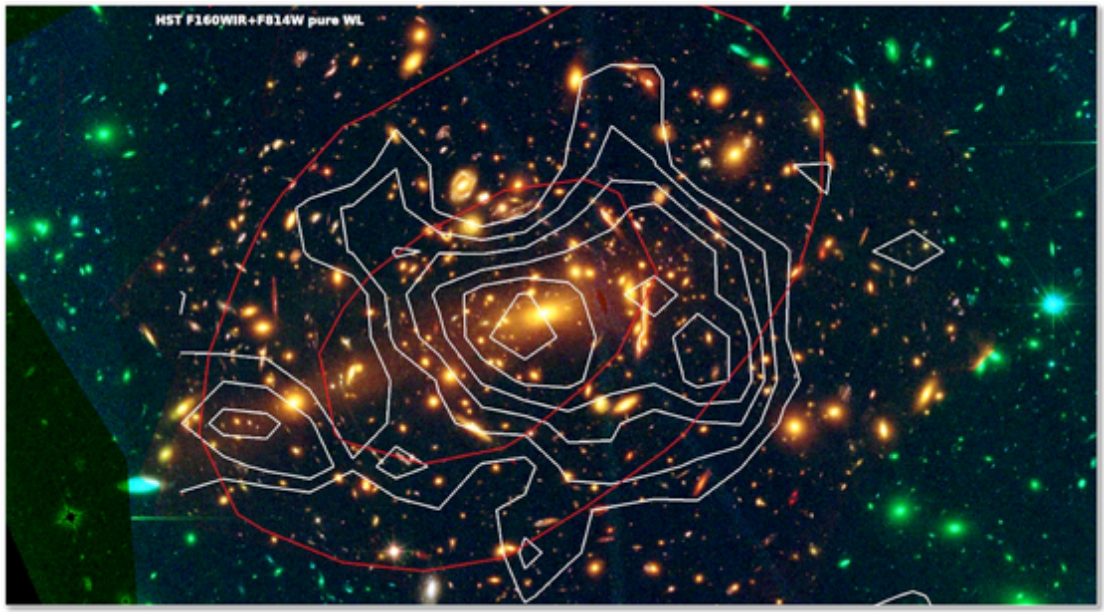

**Figure 2-24: Mass density contours around the cluster MACS J1206.2-0848 derived from a ground-based weak lensing survey with Subaru (red) vs. a weak lensing study with HST/ACS+WFC3 (white). The 10x higher surface density of lensed galaxies achieved from space yields ~3x higher spatial resolution maps. The HST data shown here is representative of the WFIRST-AFTA HLS. WFIRST-AFTA will make a map of this quality over 2,000 square degrees as part of its high latitude survey, a thousand-fold increase over the HST COSMOS mass map.**

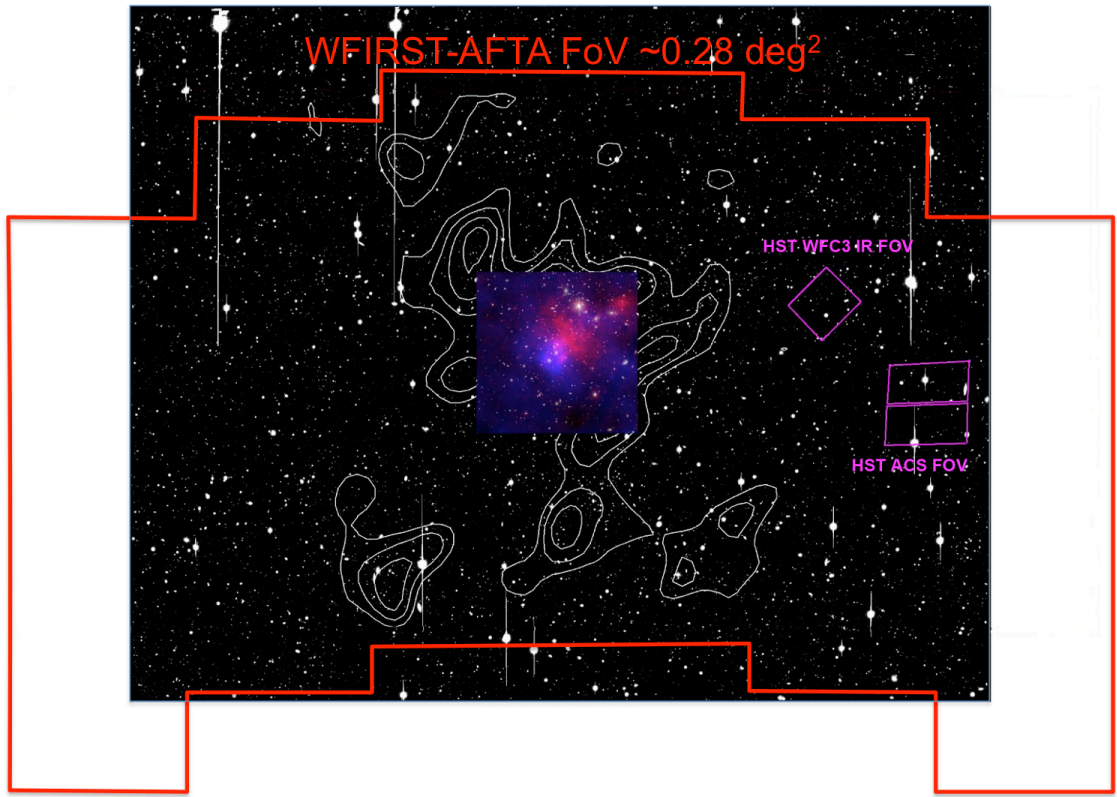

**Figure 2-25: Dark matter and cluster gas distribution around the merging cluster Abell 2744. Such maps require wide-field imaging. To map the entire area above with HST WFC3 or ACS would require 100 - 200 separate pointings, due to their small field of view. The same field is mapped with a single WFIRST-AFTA pointing (red outline.**

### 2.3.3 Kinematics of Stellar Streams in our Local Group of Galaxies

The HLS will sample each patch of sky twice for each filter, with the visits distributed over the duration of the full survey. This will enable absolute proper motions to be derived for a very large number of field stars. From the ground, only bright galaxies and quasars can be used as high-precision absolute astrometric reference points. Unfortunately, their low density requires either very wide-field transformations or a bootstrapping approach that measures a target star relative to its local field population and larger field population relative to fixed reference points. WFIRST-AFTA will provide access to an enormous number of slightly resolved medium-brightness galaxies, and absolute motions can be measured directly within the field of each detector array.

This local approach to measuring absolute motions has recently been used in the optical with HST to measure the absolute motions of the globular clusters, LMC, SMC, dwarf spheroidals and even M31, in addition to individual hyper-velocity and field stars. The strategy involves constructing a template for each galaxy so that a consistent position can be measured for it in exposures taken at different epochs. This template can be convolved with the PSF to account for any variations in focus or changes of the PSF with location on the detector array. The same approach that worked with HST in the optical should work with WFIRST-AFTA in the IR. The WFIRST-AFTA detector array pixel scale and sensitivity should be very similar to that of HST's WFC3/IR detector array pixel scale. HST images of the Ultra Deep Field through the F160W (1.6 micron) pass-band show that there should be about 30 galaxies per square arcminute for which WFIRST-AFTA could measure a position to better than 5 mas in a 360 second exposure (see Figure 2-26). This gives us about 500 reference objects in each WFIRST-AFTA detector array, enabling us to tie down the absolute frame to better than 0.5 mas in each exposure. The HLS 2-year baseline would allow absolute motions to be derived with systematic accuracies of about 125 μas/year. Folllow-on GO programs could extend this baseline to 5-years, enabling accuracies of about 50 μas/year. There are about 10 stars in the ~5 square arc-minute FoV of the UDF that can be measured with this accuracy, implying about 18 million halo stars in the high-latitude imaging survey.

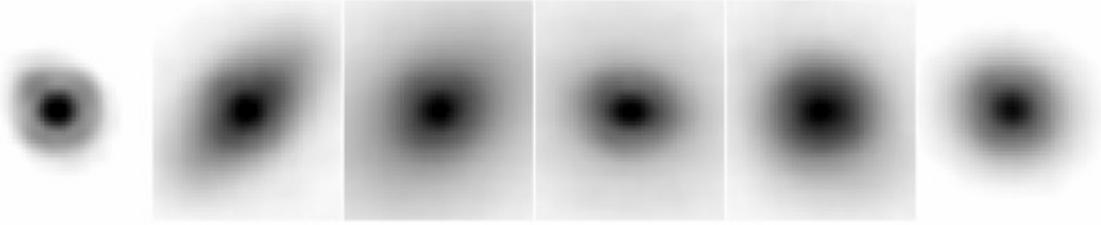

Figure 2-26: A stellar PSF (left most image) compared to 5 galaxy images from the Ultra Deep Field as seen with HST's WFC3/IR detector - an imager with similar resolution to the WFIRST-AFTA imager. Each of these objects enables a position uncertainty good to 0.02 pixel (2 mas) in a *single* exposure.

Gaia will overcome the sparse-galaxy issue by measuring positions across the entire sky in a single global solution, but it will not be nearly as sensitive as WFIRST-AFTA. Gaia will achieve the above precision for G stars down to about V = 17, which allows the plentiful turnoff-star population to be probed out to about 2.5 kpc. WFIRST-AFTA will achieve the above precision for G stars down to V=20 and will allow turnoff stars to be probed out to about 10 kpc (see Figure 2-27). There are about five times more stars within one magnitude of the turnoff than there are on the entire giant branch, and ten times more within two magnitudes of the turnoff, so WFIRST-AFTA's depth will allow us to probe well beyond the thin disk with very good statistics.

The typical dispersion of stars in the halo is ~150 km/s, but since stream stars were gently stripped from objects with low dispersion, they typically have motions that are coherent to better than 10 km/s. Such 225:1 concentrations in phase space are easy to detect. Current ground-based stream studies are focused on HB/RGB/SGB stars, but access to the much more plentiful stars below the turnoff will increase by more than an order of magnitude the number of stars we have access to and the distance out to which we can study streams. By comparing the physical locations of streams and the motions within them, we can tease out the structure of the Galactic potential.

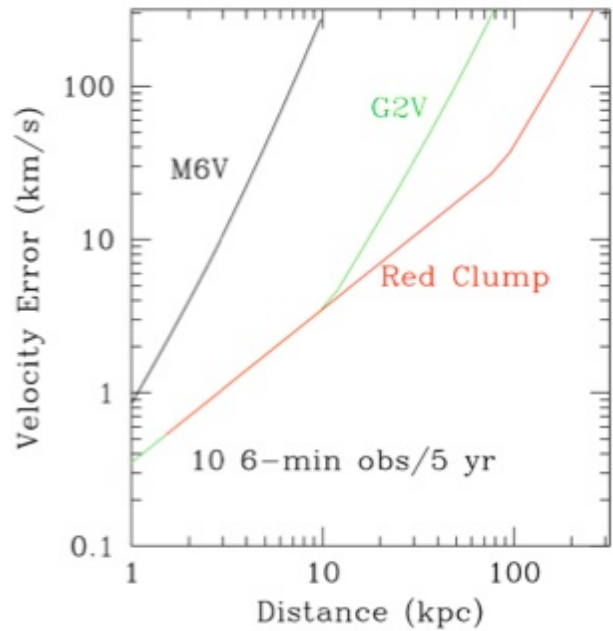

Figure 2-27: Velocity uncertainty of different components of the Galaxy's stellar population as a function of distance from the Sun. These errors are typical of what WFIRST-AFTA can achieve over a 5-year period. As part of its high latitude survey, WFIRST-AFTA will be able to measure parallax distances to over 200 million stars and will be able to track the motions of stars much fainter than those observed by the Gaia satellite.

These stellar streams are powerful probes of the distribution of dark matter in our Galaxy and can determine whether our Galaxy has the large number of million solar mass subhalos predicted in the cold dark matter model. If the dark matter is not cold, but warm as suggested in some theories, these massive subhalos will not exist, and the stellar streams will be smooth and unperturbed (Johnston et al. 2002).

### 2.3.4 *Discovering the Most Extreme Star Forming Galaxies and Quasars*

The supermassive black holes that sit in the centers of galaxies power active galactic nuclei (AGNs). There has been growing evidence that these supermassive black holes not only power these AGNs and quasars, but also play an essential role in the evolution of galaxies: (1) the stunning correlation between the masses of supermassive black holes, the AGN, with the masses of their host galaxies over a wide range of galaxy masses and galaxy types (Gebhardt et al. 2000, Ferrarese & Merritt 2000) and (2) numerical simulations show that star formation in the most massive galaxies must be suppressed by processes more powerful than supernovae and stellar winds in order to simulate galaxies resembling those we actually observe (e.g. Borgani et al. 2004). The most massive galaxies in the universe are the most affected by AGN feedback – they host the most massive black holes, and their star formation was truncated drastically about 10 billion years ago. The more massive the galaxy, the earlier this truncation happened, a trend called "downsizing" (Heavens et al. 2004, Thomas et al. 2005, Neistein et al. 2006). With the HLS spectroscopic survey, we will be able to probe the epoch of downsizing of the most massive galaxies in the universe: z~2, which is the peak of the star formation density and near the peak of the quasar activity.

The HLS spectroscopic survey will enable the largest census yet performed of powerful emission-line galaxies and quasars up to (and possibly beyond) lookback times of ~90% of the current age of the universe. It will enable us to better understand the relationship between the accretion of matter by active galactic nuclei and the star formation occurring in the most massive galaxies in the universe. Several key star formation indicators in the spectral range of the grism will be usable to track star formation rates (SFR) in galaxies over the range $1 < z < 4.2$. The H-alpha line will probe SFR that are at least $10 - 20$ solar masses per year over the range $1 < z < 2$, a cosmic epoch that is particularly challenging to study from the ground. The estimated surface density of such galaxies is about $10^4$ per square degree, implying the total survey will produce H-alpha measurements for over nearly 20 million galaxies. The OII[3727] line can be detected from star-forming galaxies lying between z = 2.6 and z = 4.2, but only for those systems with SFR in excess of ~200 solar masses per year. While such systems are less common, the WFIRST-AFTA HLS spectroscopic survey will allow us to accurately determine their space density over the key redshift range where the cosmic star formation rate has reached its highest value. At cosmic times from 200 – 600 million years since the Big Bang ($8 < z < 15$), Lyman-alpha emitting galaxies may be detectable if they have "attenuated" SFRs of at least $100 - 200$ solar masses per year.

The HLS grism survey will discover ~2600 z > 7 quasars with H magnitude < 26.5, with an estimated 20% of those quasars being at z > 8. These are the luminous quasars whose existence tracks the assembly of billion solar mass black holes a mere few hundred million years after the Big Bang, and whose light can potentially illuminate the transition of intergalactic gas in the universe from neutral to an ionized phase (Gunn-Peterson effect Gunn & Peterson 1965, Becker et al. 2001, Djorgovski et al. 2001, Fan et al. 2003). Such quasars must be found before their visible red light can be studied at high spectral resolution with JWST and ground-based 10-30-m class telescopes. The fainter quasars are particularly important to find because their region of local influence is smaller (e.g. Goto et al. 2011). These are the same quasars that would be used to identify CIV in the IGM allowing the study of chemical enrichment as a function of cosmic time. The HLS grism survey will also be a rich source of backlight targets for higher resolution spectroscopy with other facilities.

### 2.3.5 *HLS-Embedded Deep Fields Option*

The notional design for the HLS calls for a uniform depth survey of about 2200 $deg^2$ in four passbands. The survey depth and area are dictated by the requirement to ensure an ample yield of usable galaxies for the weak lensing component of the dark energy survey. However, there are advantages to consider a version of the HLS that, in addition to a wide area survey of at least 2000 $deg^2$, includes one or more regions of significantly greater depth (1.5 to 2.5 magnitudes fainter) but covering more modest areas (e.g., 20 to 50 $deg^2$). At depths of +1.5 mag fainter than the nominal HLS wide field survey, a WFIRST deep field, performed as part of a larger HLS, would provide NIR depths that are comparable to the *ugriz* limits of ~28 AB mag expected for the LSST survey. This deeper imaging would provide a superb training set for assessing the reliability of the photometry and shapes of galaxies near the limits of the wide-field shallower HLS and would also improve the robustness of such measurements for the full LSST survey. These HLS-embedded deep fields could cover 4 to 10 times more sky area than the WFIRST SN deep fields and in more passbands. An extra 1.5 mag of depth would require exposure times that are ~10x long-er than the nominal HLS exposures. A deep tier to the HLS of 20 to 50 deg$^2$ could thus be completed in about 10% – 20% of the time required for the full HLS. The exciting science enabled by including selectively target-ed deep fields within the HLS includes (1) Measure-ments of the large-scale distribution and clustering of z > 7 galaxies and of high-redshift AGN, (2) Characteriz-ing the properties of z > 2 galaxy clusters, (3) Ultra-faint surveys of the Milky Way Halo, and (4) Tracking AGN and QSO variability.

## 2.4    Exoplanet Science

### 2.4.1    *Extrasolar Planet Science Landscape*

The first discovery of planetary companions to Sun-like stars was, along with the discovery of dark energy, one of the greatest breakthroughs in modern astronomy (Latham et al. 1989, Mayor & Queloz 1995, Marcy & Butler 1996). These discoveries have excited the astronomical community and the broader public as well. Since then, the pace of exoplanet discovery has increased each year. There are now nearly 1500 confirmed exoplanets and Kepler has identified another 3300 candidates that await confirmation (Burke et al. 2014).

Nature has surprised astronomers with the enormous and unexpected diversity of exoplanetary systems, containing planets with physical properties and orbital architectures that are radically different from our own Solar System. Since the very first discoveries, we have struggled to understand this diversity of exoplanets, and in particular how our solar system fits into this menagerie.

WFIRST-AFTA will advance our understanding of exoplanets along two complementary fronts: the statistical approach of determining the demographics of exoplanetary systems over broad regions of parameter space and the in-depth approach of characterizing the properties of a handful of nearby exoplanets.

First, through its comprehensive statistical census of the outer regions of planetary systems using microlensing, including planets with separations spanning from the outer habitable zone to free floating planets, and analogs of all of the planets in our Solar System with the mass of Mars or greater, WFIRST-AFTA will complete the statistical census of planetary systems begun by Kepler. This science is described in §2.4.2.

Second, using a coronagraphic instrument, WFIRST-AFTA will be capable, for the first time in human history, of directly imaging planets similar to those in our Solar System. It will make detailed studies of the properties of giant planets and debris disks around nearby stars. It will also measure photometric properties of the 'mini-Neptune' or 'super-Earth' planets - objects that Kepler has shown to be the most common planets in our galaxy, but with no analog in our own solar system. It will also be the testbed for future coronagraphs capable of detecting signs of life in the atmospheres of Earth-like exoplanets. This science is explored in §2.4.3.

With these two complementary surveys, WFIRST-AFTA will provide the most comprehensive view of the formation, evolution, and physical properties of planetary systems. In addition, information and experience gained from both surveys will lay the foundation for, and take the first steps toward, the discovery and characterization of a "pale blue dot" — a habitable Earth-like planet orbiting a nearby star.

### 2.4.2    *Completing the Census of Planetary Systems*

After nearly two decades of exoplanet discovery, our statistical census of exoplanets, as well as our understanding of planet formation and evolution, remain incomplete. Ground-based surveys for exoplanets, initially using the radial velocity method, and now including the transit, direct imaging, and microlensing methods, are generally only sensitive to exoplanets occupying relatively narrow ranges of parameter space. Furthermore, this parameter space is largely disjoint from both the planets in our solar system, and from the three largest reservoirs of planets that are predicted by many planet formation theories (e.g. Ida & Lin 2004, Mordasini et al. 2009).

Just over five years ago, *Kepler* ushered in the next phase in the study of exoplanets by beginning the process of creating a broad and complete statistical census of exoplanetary systems. *Kepler* has been enormously successful, discovering thousands of candidate exoplanets in a bewildering array of architectures. However, Kepler can only tell part of the story: the transit method used by Kepler is not sensitive to planets in orbits significantly larger than that of the Earth, including analogs to all of the outer giant planets in the Solar System. Thus Kepler is only sensitive to one of the primary reservoirs of planets predicted by planet formation theories: "hot" to "warm" planets (Howard et al. 2012, Ida & Lin 2004, Mordasini et al. 2009).

The crucial next step is to assay the population of planets in the cold, outer regions of planetary systems, and to determine the frequency of free-floating planets. These latter two populations of planets are invisible to Kepler, yet constitute the two other main reservoirs of planets predicted by many formation theories. Fortunately, the gravitational microlensing method provides the perfect complement to the transit method, and so can be used to complete the census of exoplanetary systems begun by Kepler, and provide a nearly complete picture of exoplanet demographics.

While providing a complete census of exoplanets is, by itself, sufficient justification for a space-based gravitational microlensing survey, there are also specific and compelling science questions that can be ad-dressed with such a survey. Below we outline two of these, namely understanding the origin and evolution of planetary systems, and informing the frequency of potentially habitable worlds.

### 2.4.2.1 Understanding the Origins and Evolution of Planetary Systems

Canonical theories of planet formation and evolution originally developed to explain our Solar System (e.g., Lissauer 1987) did not anticipate the incredible diversity of planetary systems that have been observed. They have since been expanded and altered to better describe the wide range of planetary systems that we see. For example, the discovery of gas giant planets orbiting at periods of only a few days, as well as evidence for the migration of the giant planets in our own Solar System, have highlighted the fact that these theories must also account for the possibility of large-scale rearrangement of planet positions during and after the epoch of planet formation (Lin et al. 1996, Rasio & Ford 1996). Many of these theories also predict a substantial population of "free-floating" planets that have been ejected from their planetary systems through interactions with other planets (Juric & Tremaine 2008, Chatterjee et al. 2008).

In the most general terms, these formation theories should describe all of the relevant physical processes by which micron-sized grains grow through 13-14 orders of magnitude in size and 38-41 orders of magnitude in mass to become the terrestrial and gas-giant planets we see today. These physical processes are ultimately imprinted on the architectures of exoplanetary systems, specifically, the frequency distributions of planet masses and orbits (e.g. Ida & Lin 2004, Mordasini et al. 2009). Measuring these distributions, i.e., determining the demographics of large samples of exoplanets, is our best opportunity to gain insight into the physical processes that drive planet formation.

WFIRST-AFTA will be uniquely sensitive to three general classes of planets, which will provide crucial constraints on planet formation theories: cold planets, free-floating planets, and sub-Mars-mass planets.

**Cold planets:** Kepler's intrinsic ability to detect planets declines rapidly for small planets with large separations. As a result, Kepler has been unable to find small planets with periods longer than about one year. Simply put: Kepler is sensitive to "hot" and "warm" planets, but not to the "cold" planets in the outer regions of planetary systems, including analogs of all of the planets in our Solar System from Mars outward. In contrast, the ex-oplanet survey on WFIRST-AFTA is sensitive to planets from roughly the outer habitable zone outwards, including rocky planets with the mass of Earth up to the largest gas giant planets, and analogs of the ice giant planets in our Solar System.

**Sub-Mars-mass embryos:** WFIRST-AFTA is uniquely capable of detecting planets with mass as small as the mass of Ganymede (twice the mass of the moon) in significant numbers. Since Mars-mass bodies (about 0.1 $M_{Earth}$) are thought to be the upper limit to the rapid growth of planetary "embryos", determining the planetary mass function down to a few hundredths the mass of the Earth uniquely addresses a pressing problem in understanding the formation of terrestrial-type planets.

**Free-Floating Planets:** WFIRST-AFTA's microlensing survey can detect old, "free-floating" planets in numbers sufficient to test planet-formation theories. It will extend the search for free-floating planets down to the mass of Earth and below, a task not possible with other techniques and not possible from the ground. This will allow it to address the question of whether ejection of planets from young systems is a phenomenon associated only with giant planet formation or also involves terrestrial planets and planetary embryos.

### 2.4.2.2 Informing the Frequency and Habitability of Potentially Habitable Worlds

Obtaining a census of planetary systems is also an important first step in the paramount goal of determining how common life is in the universe. Particularly essential is a measurement of the frequency of potentially habitable worlds, commonly denoted $\eta_\oplus$. An accurate measurement of $\eta_\oplus$ provides a crucial piece of information that informs the design of direct imaging missions intended to search for biomarkers around nearby potentially habitable planets. Indeed, the primary goal of Kepler is to provide a robust measurement of $\eta_\oplus$. However, this number is so important that a confirmation of the value determined by Kepler is highly desirable, which can be provided by WFIRST-AFTA.

WFIRST-AFTA will improve the robustness of our estimate of $\eta_\oplus$ by measuring the frequency of terrestrial planets just outside the habitable zone, as well as the frequency of Super-Earths in the habitable zone. With this information, WFIRST-AFTA will be able to estimate $\eta_\oplus$ with only modest extrapolation. The combination of WFIRST-AFTA and Kepler data will make it possible to robustly interpolate into the habitable zone from regions just outside of it, even if the frequency of habitable planets turns out to be small. In addition, Kepler's measurement of radius combined with WFIRST-AFTA's measurement of mass will allow for statistical constraints on the densities and heavy atmospheres of small planets in the habitable zone.

Furthermore, a measurement of $\eta_\oplus$ will only provide our first estimate of the frequency of *potentially* habitable planets. The factors that determine what makes a potentially habitable planet actually habitable are not entirely understood, but it seems likely that the amount of water is a contributing factor. However, the origin of the water in habitable planets is uncertain. One possible source is delivery from beyond the snow line, the part of the protoplanetary disk beyond which it is cold enough for water ice to exist in a vacuum. As a result, the number, masses, and orbits of planets beyond the snow line likely have a dramatic effect on the water content of planets in the habitable zone (e.g., Raymond et al. 2004, Lissauer 2007). Therefore, an understanding of the frequency of habitable planets requires the survey of planets beyond the snow line enabled by WFIRST-AFTA.

### 2.4.2.3 *Expecting the Unexpected*

As compelling as these questions are, it is perhaps the unexpected and unpredictable returns of an exoplanet microlensing survey with WFIRST-AFTA that will prove to be the most enlightening for our understanding of the formation, evolution, and habitability of exoplanets. One of the most important lessons from the past two decades of exoplanet research is that there exists an enormous diversity of exoplanetary systems, generally far exceeding theoretical expectations. Indeed, one hallmark of the field is the fact that, whenever new regions of parameter space are explored, the subsequent discoveries necessitate revisions of our planet formation theories. *Kepler* is currently revolutionizing our understanding of the demographics of small, short-period planets. Because it will open up a similarly broad expanse of parameter space with a similarly large yield of planets (thus implying good sensitivity to rare systems), WFIRST-AFTA is essentially guaranteed to do the same. Furthermore, it will do so in a region of parameter space that is almost certainly critical for our understanding of both planet formation and habitability.

### 2.4.2.4 *Summary of Science Motivation*

In summary, a microlensing survey from space with WFIRST-AFTA, when combined with Kepler, will yield a composite census of planets with the mass of Earth and greater on both sides of the habitable zone, overlapping

---

**Box 2**

**WFIRST-AFTA Exoplanet Discoveries**

WFIRST-AFTA will be sensitive to exoplanets with mass greater than Ganymede, or roughly twice the mass of the moon. It will detect **2600** total bound exoplanets in the range of 0.03-1,000 Earth masses, including **1030** "Super-Earths" (roughly 10 times the mass of Earth), **370** Earth-mass planets, and **50** Mars-mass planets. This would enable the measurement of the mass function of cold exoplanets to better than ~7% per decade in mass for masses >0.3 $M_{Earth}$, and an estimate of the frequency of Mars-mass embryos accurate to ~15%.

WFIRST-AFTA will measure the frequency of free-floating planetary-mass objects in the Galaxy over nearly six orders of magnitude in mass, and will detect **30** free-floating Earth-mass planets, if there is one per star in the Galaxy.

These estimates are likely conservative. With more observing time, better optimization of the target fields, and more accurate event rates in the target fields, the exoplanet yield of WFIRST may be up to a factor of two higher.

**Characterization of Systems Discovered by WFIRST-AFTA**

Data from WFIRST-AFTA will likely allow estimates of the mass and distance to the host stars and planets to better than ~10% for the majority of the detected systems. In addition, spectroscopic follow-up with JWST will allow for the measurement of host star temperature and metallicities for a subset of the most massive hosts in the disk and bulge.

**Detection of transiting giant planets**

WFIRST-AFTA will be sensitive to the transits of giant planets around 200 times more stars than Kepler, and 40 times more stars than TESS. WFIRST-AFTA will discover about 20,000 transiting planets with orbits less than a few tenths of an AU, with sensitivity down to Neptune-radius objects. By combining the transit and microlensing discoveries, WFIRST-AFTA will determine the frequency of giant planets over all orbital separations (from 0 to infinity) from a single survey.

---

at almost precisely that zone, thus allowing it to achieve its primary goal to "*Complete the statistical census of planetary systems in the Galaxy, from the outer habitable zone to free floating planets, including analogs of all of the planets in our Solar System with the mass of Mars or greater.*" With such a census, WFIRST-AFTA will be able to address the three fundamental and inter-related questions: "*How do planetary systems form and evolve?*", "*What determines the habitability of Earth-like planets?*", and **"*What kinds of unexpected systems inhabit the cold, outer regions of planetary systems?*"** Without WFIRST-AFTA's survey, we will have essentially no knowledge of the frequency of solar systems like our own, no knowledge of the demographics of exoplanets in vast regions of parameter space that include analogs of Mars, Saturn, Neptune, and Uranus. Without WFIRST-AFTA, basic questions of the formation, evolution, and habitability of exoplanets will remain unanswerable.

#### 2.4.2.5 *The WFIRST-AFTA Microlensing Survey*

The microlensing technique for discovering exoplanets, illustrated in Figure 2-28, is well developed and mature (Gaudi 2012). Its capabilities have been amply demonstrated from the ground through several discoveries that have provided important new insights into the nature and diversity of planetary systems. These include the discovery of a substantial population of cold "Super-Earths" (Beaulieu et al. 2006, Bennett et al. 2008, Muraki et al. 2011), the first discovery of a planetary system with an analog to our own Jupiter and Saturn (Gaudi et al. 2008, Bennett et al. 2010), and the detection of a new population of Jupiter-mass planets loosely bound or unbound to any host star (Sumi et al. 2011). Importantly, ground-based microlensing surveys have determined that cold and free-floating giant planets are ubiquitous: on average, every star in the Galaxy hosts a cold planet (Cassan et al. 2012) and "free-floating" giant planets may outnumber the stars in our galaxy by two to one.

While exoplanet microlensing from the ground has had well-documented successes, realizing the true potential of the microlensing method, and achieving the primary science goals outlined above, is only possible from space with an instrument like WFIRST-AFTA (Bennett & Rhie 2002). Microlensing requires monitoring a very large number of stars in very crowded, heavily reddened fields toward the Galactic bulge continuously for months at a time. Therefore, reaching the needed sensitivity and number of planet detections requires steady viewing in the near-infrared without in-

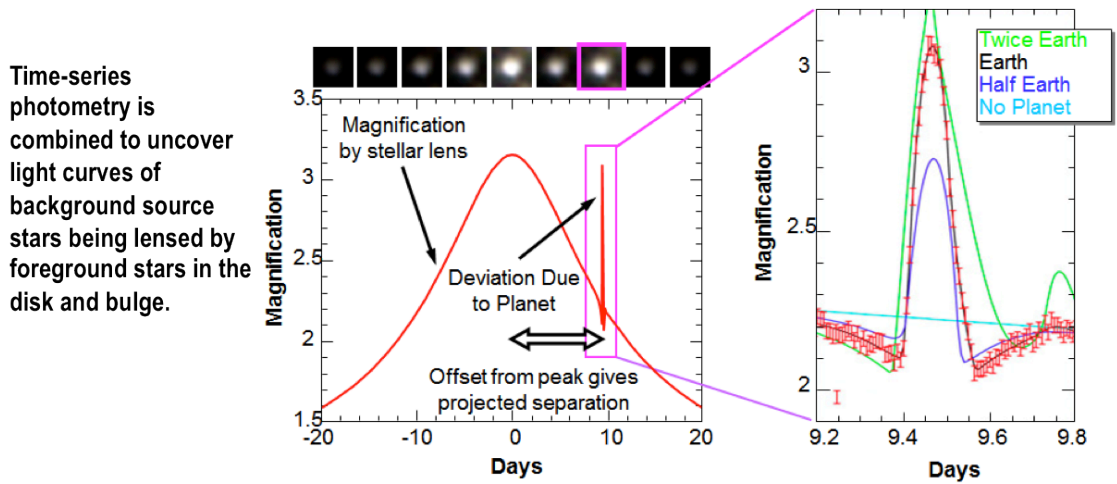

Time-series photometry is combined to uncover light curves of background source stars being lensed by foreground stars in the disk and bulge.

Planets are revealed as short-duration deviations from the smooth, symmetric magnification of the source due to the primary star.

Detailed fitting to the photometry yields the parameters of the detected planets.

**Figure 2-28: Schematic illustration of how WFIRST-AFTA discovers signals caused by planetary companions in primary microlensing events, and how planet parameters can be extracted from these signals. The left panel shows a simulated primary microlensing event, containing a planetary deviation from an Earth-mass companion to the primary lens. The offset of the deviation from the peak of the primary event, when combined with the primary event parameters, is related to the projected separation of the planet. The right panel shows an enlargement of the planetary perturbation. The width and precise shape of the planetary deviation yield the mass of the companion relative to that of the primary host lens.**

terruptions for weeks at a time, as well as high angular resolution, stable images over very wide-area fields. In other words, it requires a wide-field infrared survey telescope in space. Figure 2-34 shows a comparison of a field toward the Galactic center observed from the ground with a simulated field with similar crowding as would be observed with WFIRST. Without such a mission, we will not be sensitive to Mars-mass embryos, we will not determine the frequency of low-mass free-floating planets, and we will not complete the census of planets begun by Kepler.

#### 2.4.2.6 *Methodology for the Estimate of WFIRST-AFTA Yields*

We estimate the exoplanet yields for WFIRST-AFTA using the simulation methodology described in a series of papers by Penny et al. (2013, 2015a, b, c) and Penny (2014). This simulation methodology was also used to estimate the yields of the WFIRST-DRM1 and WFIRST-DRM2 designs, as well as Euclid. We expect the relative yields between different survey designs (given our assumptions) to be more robust than the absolute rates, for reasons described in Appendix K. To convert sensitivities to expected yields, we have adopted our best estimate of the planet distribution function from Cassan et al. (2012), which has the form

$$dN/dlogM_p dloga = f (M_p/95M_\oplus)^\alpha$$

where $f = 0.24/dex^2$ and $\alpha = -0.73$. We assume that this power-law distribution saturates at a constant value of $2/dex^2$ for $M_p < 5.2$ $M_\oplus$. This is likely conservative, as Kepler has found several multi-planet systems that have more than 2 planets per decade in semi-major axis (e.g., Lissauer et al. 2011, Gautier et al. 2012). Note that the exponent of this power-law distribution with planet mass is slightly steeper but of opposite sign to the exponent of the approximately power-law dependence of the detection efficiency on planet mass, and thus the number of expected planets is a relatively slowly rising function of decreasing mass in this model in the regime where we assume the exponent $\alpha$ holds. Note also that the cold planet frequency below ~$10M_{Earth}$ is very poorly constrained. This ignorance is, of course, one of the primary justifications for an exoplanet survey with WFIRST-AFTA.

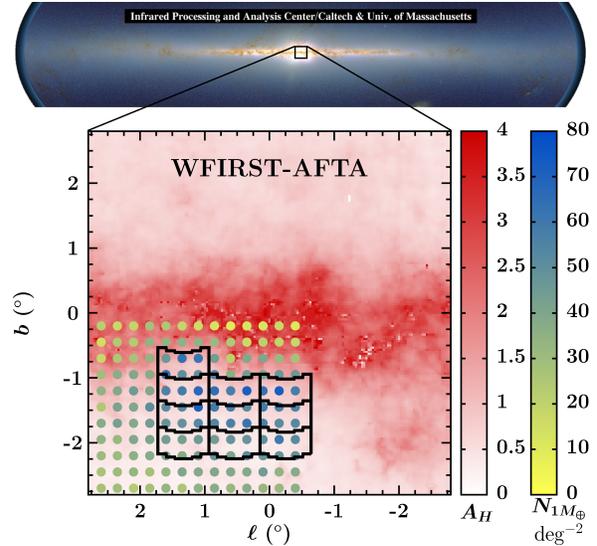

**Figure 2-29: The upper figure shows an image of the Galactic plane from 2MASS, whereas the bottom panel zooms in on the ~36 square degrees around the Galactic center. In the bottom panel, the solid black lines show the footprints of the 10 baseline WFIRST-AFTA fields in Galactic coordinates, with a total active area of 2.81 deg². These are compared to rates of bound Earth-mass planet detections as a function of location shown as the solid circles, and a map of the extinction in H-band from Gonzalez et al. (2012). The WFIRST-AFTA fields are centered on locations of relatively low extinction and relatively high planet detection rates.**

We chose 10 target fields toward the Galactic center[8], with locations indicated in Figure 2-29. The total survey area is 2.81 square degrees. Note that we did not fully optimize the survey relative to the number, location, and cadence of the target fields. We therefore expect our results to be conservative. We also note that our choice for the number of fields in the baseline survey design (10) is better optimized for high-mass planets than low-mass planets. Therefore, choosing a smaller number of fields could further increase the yields of low-mass planets, although of course this would come at the expense of the number of high-mass planet detections. It may also be possible to increase the sensitivity to other classes of planets that are at the edges of the sensitivity of WFIRST-AFTA, e.g., very small separation planets and massive habitable planets. As outlined in Appendix K, optimizing the survey strategy with respect to the survey parameters and target planet population is an important future activity.

---

[8] Although the primary microlensing survey will focus on fields located toward the Galactic center, a unique capability of microlensing is its sensitivity to very distant planetary systems, including planetary systems in other galaxies. Relatively modest surveys of serveral weeks towards M87 or the bulge of M31 would have significant sensitivity to Jupiter-mass planets (Baltz & Gondolo 2001) and be able to probe the frequency of planets in environments with stellar populations and star formation histories that differ substantially from those of the Milky Way disk and bulge.

As outlined in the operations concept (§3.10), we assume a total of 6 microlensing seasons of 72 days each. Because of the proximity of the Galactic center to the ecliptic, the moon moves through our target fields two to three times per season, resulting in losses of 4 to 5 days per lunation due to the moon avoidance angle. Thus for our nominal 6 seasons observing campaign, there are a total of 357 days of observations. We also consider the yields for an extended campaign of 7 seasons, which yields a total of 417 days of observations.

Each field will be observed every 15 minutes in the wide W149 filter, and once every 12 hours in the Z087 filter. The exposure times in these filters will be 52 secs (W149) and 290 secs (Z087). Due to the layout of the detectors in the focal plane, ~85% of the survey footprint is observed for the full time while the remaining 15% is observed only in either the Spring or Fall seasons.

Figure 2-30 illustrates an example simulated microlensing event, which includes the signal of a Ganymede-mass planet, with roughly twice the mass of the moon. The clear, unmistakable, and unambiguous planet signature highlights two of the intrinsic strengths of the microlensing method, namely that the majority of planetary signals are detected at high confidence (in this case with a total signal-to-noise ratio of roughly 27), and the signals are not prone to astrophysical false positives.

### 2.4.2.7    *Expected Results*

#### 2.4.2.7.1    *Bound Planets*

The basic results are summarized in Box 2, Table 2-4, and illustrated in Figure 2-31. In order to highlight the intrinsic sensitivity of the survey, Figure 2-31 shows the region of parameter space where WFIRST-AFTA is expected to detect at least 3 planets, assuming that every star has one planet at the given mass and semimajor axis. For example, if every star has a planet with a mass of 0.1 $M_\oplus$ and a semimajor axis of 1 AU, WFIRST-AFTA will detect roughly three such planets. Table 2-4 then reports the predicted yields for planets of a range of masses assuming the Cassan et al. mass function, where we have averaged the yields for a given mass over a range of semi-major axes between 0.03-30 AU.

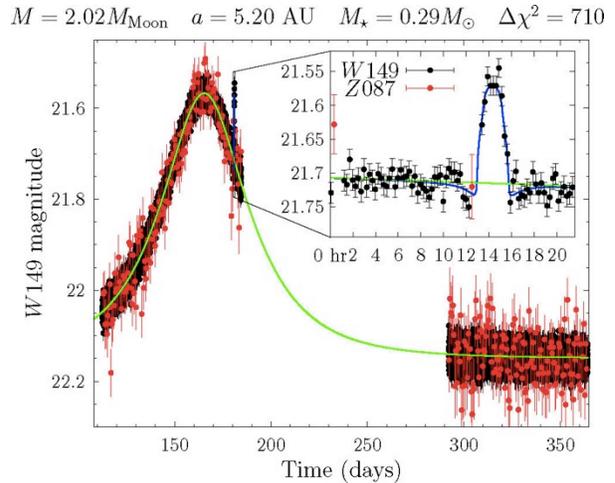

Figure 2-30: Example of a microlensing event light curve from simulations of a WFIRST-AFTA exoplanet survey, illustrating the signal of a planet with a mass of roughly twice the mass of the Moon (roughly the mass of Jupiter's giant moon Ganymede) orbiting a 0.29 solar mass star with a semi-major axis of 5.2 AU. The detection is highly significant, with a $\Delta\chi^2$ =710, roughly corresponding to a 27 σ detection.

We find that for the nominal six-season observing campaign, ***WFIRST-AFTA will detect roughly 2600 bound planets*** in the range of 0.1-1,000 $M_\oplus$ and semi-major axes in the range of 0.03-30 AU. This is compa-

| M/M$_{Earth}$ | WFIRST-IDRM (432 days) | WFIRST-DRM1 (432 days) | WFIRST-DRM2 (266 days) | WFIRST-AFTA (357 days) | WFIRST-AFTA (417 days) |
|---|---|---|---|---|---|
| 0.1 | 22 | 30 | 18 | 50 | 58 |
| 1 | 208 | 233 | 173 | 367 | 429 |
| 10 | 575 | 793 | 551 | 1030 | 1203 |
| 100 | 470 | 629 | 439 | 726 | 849 |
| 1000 | 298 | 367 | 261 | 426 | 497 |
| Total | 1701 | 2052 | 1442 | 2599 | 3036 |

Table 2-4: Predicted yields for bound planets for various mission designs. The yields adopted the planet distribution function for cold exoplanets as measured from ground-based microlensing surveys by Cassan et al., and normalized to the most recent microlensing event rates (Sumi et al. 2013) measured in fields that overlap a subset of the WFIRST-AFTA target fields.

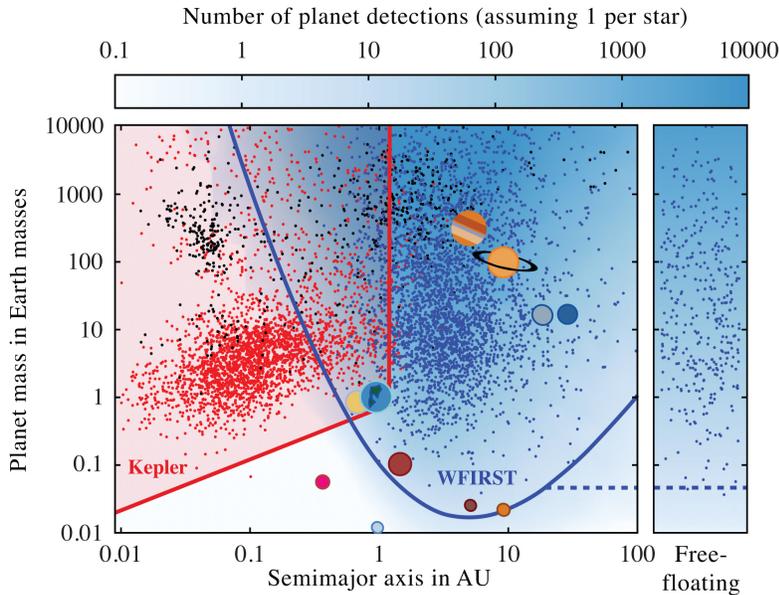

Number of planet detections (assuming 1 per star)

**Figure 2-31: The colored shaded regions show approximate regions of sensitivity for Kepler (red) and WFIRST-AFTA (blue). The solar system planets are also shown, as well as the Moon, Ganymede, and Titan. Kepler is sensitive to the abundant, hot and warm terrestrial planets with separations less than about 1.5 AU. On the other hand, WFIRST-AFTA is sensitive to Earth-mass planets with separations greater than 1 AU, as well as planets down to roughly twice the mass of the moon at slightly larger separations. WFIRST-AFTA is also sensitive to unbound planets with masses as low as Mars. The small black points show planets detected by radial velocity, ground-based transits, direct imaging, and microlensing surveys (i.e., all known exoplanets not found by Kepler). The small red points show candidate planets from Kepler, whereas the small blue points show simulated detections by WFIRST-AFTA; the number of such discoveries will be large, with roughly 2600 bound and hundreds of free-floating planet discoveries. *Thus, WFIRST-AFTA and Kepler complement each other, and together they cover the entire planet discovery space in mass and orbital separation, providing the comprehensive understanding of exoplanet demographics necessary to fully understand the formation and evolution of planetary systems. Furthermore, the large area of WFIRST-AFTA discovery space combined with the large number of detections essentially guarantees a number of unexpected and surprising discoveries.***

rable to (about 60% of) the yield of planet candidates from Kepler's prime mission, although of course WFIRST-AFTA's planets will be located beyond the "snow line", a region that is largely disjoint from that accessible to Kepler. With an additional seventh season, WFIRST-AFTA will detect another 440 planets, for a total of roughly 3040 bound planets. As illustrated in Figure 2-31, while massive planets can be detected over nearly the entire range of separations between 0.03-30 AU, lower-mass planets can only be detected over a narrower range of separations of 0.1-10 AU. At least 15% of the detected bound planets will have mass less than three times the mass of the Earth and WFIRST-AFTA will have significant sensitivity down to planets of roughly twice the mass of the Moon. WFIRST-AFTA will measure the mass function of cold planets to ~3%-15% in bins of one decade in planet mass down to the mass of Mars.

#### 2.4.2.7.2 Estimating the Frequency of Habitable Planets

Microlensing's sensitivity to planets falls rapidly for projected separations smaller than the Einstein radius. At projected separations of the order of 1/3 of the Einstein ring radius or smaller we expect the size of planetary deviations of the order of 1% or smaller. For typical combinations of the lens and source distances, a solar-mass star has an Einstein radius of ~4 AU, compared to a habitable zone that stretches from ~1-1.7 AU (Kopparapu et al. 2013). Therefore, the signals of rocky habitable zone planets will be short and of small amplitude (~1% or less), so careful control of photometry systematics to a millimagnitude or less on bright stars will be essential for reliable detections. Figure 2-32 shows an example WFIRST-AFTA simulated light curve of the perturbation due to an Earth-mass habitable-zone planet, demonstrating the low amplitudes and short durations expected for such detections. Furthermore, the detection probabilities for low-mass planets at projected separations much less than the Einstein ring are also small, implying that the signals of Earth-mass habitable-zone planets will be rare, even if the intrinsic frequency of such planets is high. Finally, because the ratio of the habitable zone distance to the Einstein ring radius scales as the host mass $M^{1.5}$, WFIRST-AFTA will only be able to detect habitable zone planets around stars close to or above the mass of the sun. This is in contrast to Kepler, which is most sensitive to (and has found) planets in the habitable zones of M- and K-dwarf stars, but is less sensitive to such planets orbiting solar-mass stars. Note however, that the old age of the bulge means that stars more massive than the Sun have already evolved off the main sequence.

Table 2-5 lists the expected number of habitable zone planets that WFIRST-AFTA will find. Assuming that there is one Earth-mass planet in the habitable

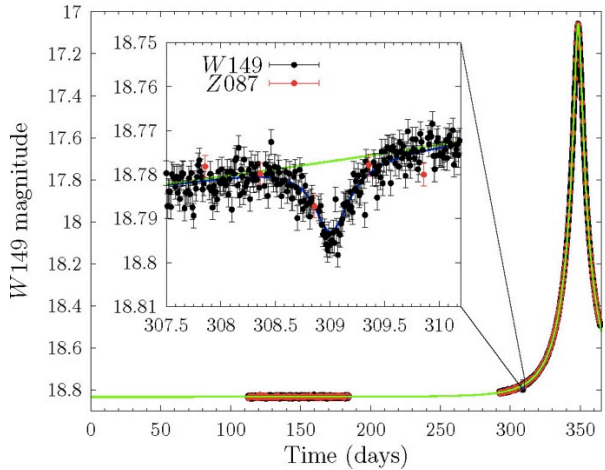

$M = 0.94 M_\oplus$   $a = 1.46$ AU   $M_\star = 0.95 M_\odot$   $\Delta\chi^2 = 939$

**Figure 2-32: Example of a WFIRST-AFTA detection of a habitable-zone planet. Such a detection of an Earth-mass habitable-zone planet is unlikely, but WFIRST-AFTA will be able to detect only slightly more massive planets in the habitable zone and Earth-mass planets just outside the habitable zone. The HZ planets that WFIRST-AFTA is sensitive to will orbit F-, G- or upper K-dwarf stars, complementing the planets Kepler and TESS can detect in the habitable zones of lower G-, K- and M- dwarf stars. The signals of rocky habitable zone planets will be short and of small amplitude (~1% or less), so careful control of photometry systematics to a millimagnitude or less on bright stars will be essential for habitable zone planet detection.**

| M/M$_{Earth}$ | HZ x 1.0 0.99–1.68 AU | HZ x 1.5 1.49–2.52 AU | HZ x 2.0 1.98–3.35 AU |
|---|---|---|---|
| 10 | 8 | 25 | 53 |
| 3 | 3 | 11 | 24 |
| 1 | 1 | 5 | 10 |

**Table 2-5: The number of WFIRST-AFTA planet detections expected in and near the habitable zone, assuming a planet abundance of one planet of a given mass per habitable zone (or 4.35 per decade of semimajor axis) per star.**

| M/M$_{Earth}$ | One per star | Cassan et al. |
|---|---|---|
| 0.001 | 0.002 | 0.004 |
| 0.01 | 0.54 | 1.1 |
| 0.1 | 5.7 | 11 |
| 1 | 28 | 57 |
| 10 | 108 | 134 |
| 100 | 356 | 82 |
| 1000 | 1161 | 50 |
| Total | 1659 | 364 |

**Table 2-6: Predicted yields for free-floating planets, assuming one free-floating planet per star in the Galaxy, and assuming the Cassan et al. (2012) mass function.**

zone of each main-sequence star, WFIRST-AFTA would detect one planet. However, for slightly larger-mass planets or at slightly larger distances, WFIRST-AFTA can expect to detect many more planets. So, while it is unlikely that WFIRST-AFTA will measure the abundance of Earth-mass habitable-zone planets, it can measure the abundance of planets just outside the habitable zone and the abundance of more-massive habitable-zone planets. This situation is very similar to that of Kepler, which, for solar-type stars, can measure the abundance of planets with radii similar to that of the Earth just *inside* the habitable zone, or the abundance of larger planets in the habitable zones. Combined, the results of both Kepler and WFIRST-AFTA can be used to effectively and robustly interpolate into the habitable zone of Sun-like stars.

### 2.4.2.7.3   Free-floating Planets

Free-floating giant planets cause a simple, one-time, symmetric, and achromatic point-mass microlensing event with a duration that is proportional to the square root of the planet mass (but degenerate with distance and velocity). For the overwhelming majority of sources, i.e., main sequence stars in the Galactic bulge, the angular source radius θ· is larger than the angular Einstein ring radius θ$_E$ for free-floating planets with mass less than about 0.1 Earth masses. In this case, the source is effectively 'resolved' by the planetary lens, and so microlensing events due to such low-mass free-floating planets have a morphology similar to a one-time inverse transit, with a duration that is proportional to twice the source crossing time, and an amplitude of θ$_E$/(2θ·). Therefore, in this regime, the durations of the events are independent of the free-floating planet mass, but the amplitude decreases with decreasing mass. Figure 2-33 shows an example WFIRST-AFTA simulated light curve due to a free-floating Mars-mass planet, which illustrates these features.

We show the expected yields for free-floating planets in Table 2-6 under two assumptions: one free-floating planet with a given mass per star in the Galaxy, and assuming the same Cassan mass function as we assumed for the bound planets. In both cases, the yields are given for a 357-day survey. For the augmented campaign of 417 days, the yields are ~17% higher. The lowest mass free-floating planet that WFIRST-AFTA is sensitive to is set by the source-diameter crossing time that is critically sampled by WFIRST-AFTA's 15-minute cadence, while at the same time providing enough photons for a significant detection of magnification at each epoch. These criteria result in an optimum free-floating planet mass of the order of 0.1 $M_{Earth}$ (roughly the mass of Mars), and a relatively sharp cutoff below about $10^{-3}$ $M_{Earth}$ (roughly half the mass of Pluto).

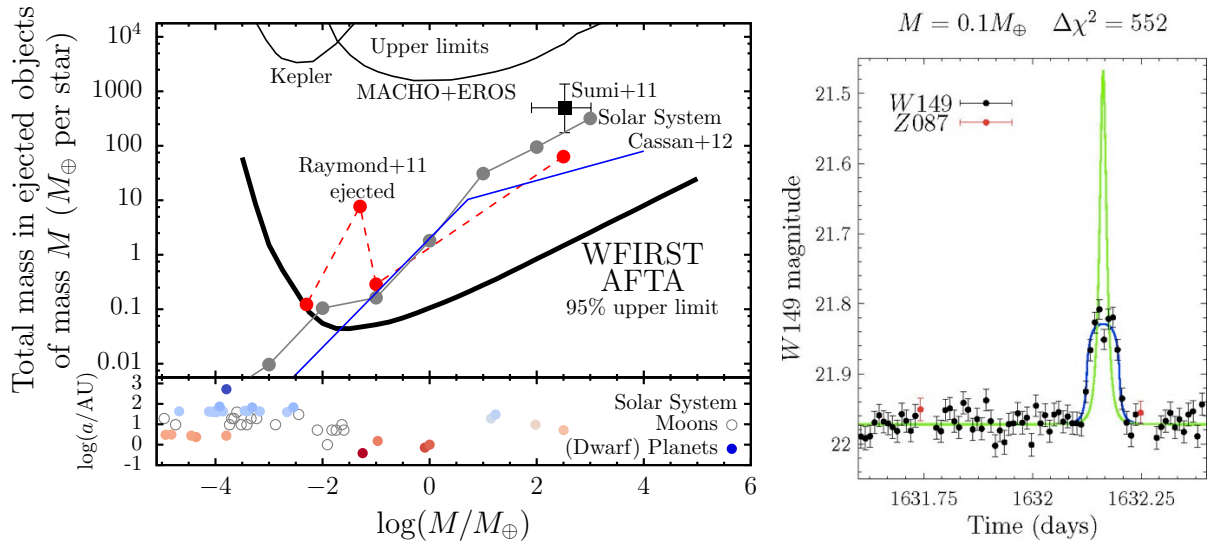

**Figure 2-33: WFIRST-AFTA will be uniquely sensitive to free-floating planets down to the mass of Pluto.** On the left is shown the 95% confidence upper limit on the abundance of free-floating objects that WFIRST-AFTA would infer if it did not detect any free-floating planets. The abundance is plotted as the total mass of ejected objects per star, rather than the number of ejected objects. For comparison, we show (a) the Jupiter-mass free-floating planet abundance that was measured by Sumi et al. (2011), (b) the mass distribution of solar system objects, (c) the Cassan et al. (2012) mass function that was used for the free-floating planet realization in Figure 2-31 and yields in Table 2-4, (d) predictions of the number of planetesimals and planets ejected in dynamical simulations of newly-formed systems with gas giant planets, assuming that only 20% of stars form giant planets (Raymond et al. 2010, 2011) and (e) the upper limit on the abundance of compact objects as measured by the MACHO and EROS surveys (Alcock et al. 1998) and from 2 years of Kepler data (Griest et al. 2014). Also shown in the lower panel are the masses of Solar System objects (planets, dwarf planets and moons), with the planets and dwarf planets color-coded by their semimajor axis - objects like those with blue hues would stand a reasonable (Neptune, Uranus) or high chance of being identified as free-floating planets by WFIRST-AFTA due to their large distance from the sun. On the right is an example of the detection by WFIRST-AFTA of a free-floating Mars-mass planet. From Penny et al. 2015b.

One complication with identifying free-floating planets is that they can be mistaken for planets on very wide orbits. Indeed, a planet at a large separation (greater than a few tens of AU, Han et al. 2005) from its host can produce an identical light-curve to that produced by a free-floating planet, with no detectable magnification due to the host star, unless the source trajectory has a special geometry aligned with the host-planet separation axis. It is therefore impossible to determine whether such short timescale light curves are due to bona fide free-floating planets, or instead due to a bound planet in a wide orbit around a faint star. How-ever, by analyzing blended light (or the lack thereof), together with analysis to rule out subtle binary-lens magnification features in the light curve, it will be possible to put strong limits on the brightness of any distant host. Follow-up AO imaging with JWST or ELTs may be able to further constrain such scenarios (see §2.7).

Rather than thinking in terms of a bound/free-floating dichotomy, it may be more productive to think in terms of the total mass budget of loosely bound objects with binding energies lower than a certain threshold. As shown in Figure 2-33, WFIRST-AFTA will be able to measure the mass budget of loosely- or un-bound objects down to a total mass of one Earth masses per star, over nearly 6 orders of magnitude in mass, from objects the mass of Pluto ($10^{-3}$ $M_{Earth}$) to the mass of Jupiter (300 $M_{Earth}$). At its most sensitive point, it will measure the mass down to less than 0.1 Earth masses per star for objects the mass of Mars (i.e., down to a limit of one such object per star). The loosely- or un-bound mass function that WFIRST-AFTA measures will be a boon to planet formation studies, potentially enabling an era of `precision planet formation`. This is be-cause in comparison to the multidimensional problem of bound planets (mass, semimajor axis, eccentricity, multiplicity, etc.) in one dimension (mass) it will be relatively simple to both empirically measure WFIRST-AFTA's sensitivity to free-floating or loosely bound objects (allowing conversion of the number of detections to a mass function) and produce predictions of the mass function from dynamical planet formation simulations. This is illustrated in Figure 2-33, which compares the predictions of the total mass of ejected planets and embryos from an illustrative planet formation simulation to the expected limits that can be achieved with WFIRST-AFTA.

### 2.4.2.8 *Physical Parameters of the Detected Systems*

In addition to a large yield of planets over a broad region of parameter space, the other dramatic improvement of a space-based microlensing mission over ground-based surveys is the ability to automatically and systematically estimate the masses of the majority of the planetary host stars and thus planetary companions. This qualitatively new capability is enabled by the small and stable point-spread functions (PSFs) in space, which allow for many different types of measurements that are difficult or impossible from the ground.

- **Measurement of the Host Starlight:** The WFIRST-AFTA PSF with angular size of ~0.14" in the W149 filter essentially resolves out the majority of the background stars even in the crowded fields monitored by the microlensing survey. In particular, we find that the probability of any given source being blended with an interloper star with $H_{AB} < 26$ is <15%. Given that the overwhelming majority (>90%) of the main-sequence lenses are brighter than this, interloper stars are likely to contribute negligibly to the total flux of the target. Therefore, during or soon after most microlensing events, any photometered target will contain only flux from the lensed source and the lens star itself. Therefore, the flux of the lens can be measured by subtracting the time-variable flux of the microlensed source from the target. One complication is that the target may also contain light from unresolved stellar companions to the lens or source, resulting in an overestimate of the lens flux (and thus lens mass) if this light is falsely attributed to the lens. In fact, based on simulations of binary companions to the lenses and sources (Henderson 2015), we find that such companions will result in catastrophic misestimates (>50%) of the lens mass for less than 10% of events. Nevertheless, even these rare situations may be resolved via direct measurements of the lens-source proper motion, as described below. A measurement of the lens flux can be combined with a measurement of the lens-source proper motion, inferred from the light curves or directly from images taken well after the microlensing event, to infer the lens (and thus planet) mass.

- **Direct Measurement of the Lens-Source Proper Motion:** For luminous lenses, the relative lens-source proper motion can be measured directly from images taken over a very wide time baseline before or after the microlensing event (Bennett et al. 2007). While the time baseline of the nominal WFIRST-AFTA six-year mission is sufficiently short that essentially none of the lens-source pairs will be completely resolved, it will be possible to measure the lens-source proper motion for marginally resolved pairs by measuring either the elongation of the PSF, or the differential centroid shift of the PSF in two different pass bands if the lens and source have substantially different colors. Importantly, these measurements also yield not only the magnitude, but also the direction of the lens-source proper motion. This vector proper motion can then be combined with one-dimensional "microlens parallax", which will be routinely measured for roughly half of WFIRST-AFTA microlensing light curves with detectable Earth-mass planets, to directly measure the lens mass (Gould 2014a). In many cases, it will be possible to measure the lens-source proper motion from the light curve as well as the images. This redundancy will allow one to check for contamination from light from companions to the lens or source.

- **Measurement of Higher Order Effects in the Microlens Lightcurves**: The small and stable PSFs delivered by WFIRST-AFTA will result in precise and accurate photometry during the microlensing events, with substantially higher precision and smaller systematics than is possible from the ground. This will enable the measurement of subtle but information-rich "parallax" distortions in the microlens light curves due to the orbit of WFIRST-AFTA about the Sun (Gould 1992) and in the case of a geosynchronous orbit, the orbit of WFIRST-AFTA about the Earth (Gould 2013). In the case of an L2 orbit, simultaneous observations from the ground will enable parallax measurements for high-magnification events (Yee 2013). Similarly, for a geosynchronous orbit, simultaneous observations from a satellite at L2 (e.g., JWST) would enable parallax measurements for high-magnification events. One-dimensional parallax measurements will be routinely measured for WFIRST-AFTA survey data alone (Gould et al. 2003), which can then be combined with the vector lens-source proper motions, inferred directly from images or from astrometric microlensing effects, to infer the lens mass and distance (Gould 2013, Gould & Yee 2014). For a subset of detections, it will be possible to measure the full two-dimensional microlens parallax; when combined with the relative lens-source proper motion that is routinely measured in planetary microlens lightcurves (Gould 1994), this will also allow one to measure the mass and distance to the host. Subtle distortions in the microlens light curves due to the orbital motion of the planet about the host will also enable constraints on the period, inclination, and eccentricity of the orbit in a subset of cases.

- **Astrometric Shifts due to Microlens Image Motions:** During a microlensing event, multiple images of the source are created whose positions and brightnesses vary as a function of time. While these images are typically separated by of order the Einstein ring radius $\theta_E \sim 1$ mas and thus cannot be resolved, the changing fluxes and positions lead to a time-varying centroid shift of the unresolved source image, whose magnitude is proportional to $\theta_E$ (Walker 1995, Dominik & Sahu 2000). Ensemble (mission-integrated) astrometric precisions of order 100 µas (roughly $10^{-3}$ of a pixel) or better should be easily achievable with WFIRST-AFTA for a large fraction of microlensing events, enabling the robust measurement of $\theta_E$ using astrometric centroid shifts for more massive and/or nearby lenses. When combined with the measurement of the host light for luminous lenses, this will allow for a measurement of the distance and mass of the host. When combined with full two-dimensional microlensing parallax measurements, this effect will allow for the detection and mass measurement of a significant sample of isolated neutron stars and black holes, thereby allowing for a determination of the mass function of isolated high-mass remnants, providing crucial constraints on supernovae explo-sion mechanisms (Fryer et al. 2012). These measurements cannot be obtained in any other way. If systematics can be controlled at the level of 10µas (roughly $10^{-4}$ of a pixel), then astrometric centroid shift measurements will be possible for substantially lower-mass lenses, including the majority of main-sequence hosts (Gould & Yee 2014). When combined with the routinely-measured one-dimensional microlens parallax, it will be possible to measure the masses and distances to the host stars (and thus planets) of a large fraction of the planetary systems detected by WFIRST-AFTA, thus enabling a relatively uniform and unbiased determination of the cold planet mass function and Galactic distribution of planets.

A quantitative assessment of the potential of WFIRST-AFTA to measure the above effects and thus infer host star masses and distances generally requires detailed simulations, as well as an understanding of systematics at a level that is one to three orders of magnitude smaller than has been explored to date with existing analogous observations. While there have been some preliminary explorations of these effects (Gould 2013, Yee 2013, Gould & Yee 2014), substantially more work needs to be done. Nevertheless, using conservative assumptions, Bennett et al. (2007) demonstrated that a significantly less capable space-based microlensing mission would enable measurement of the host star and planet masses to an accuracy of ~10% for the majority of the detected systems. We expect WFIRST-AFTA to measure the masses to a similar accuracy for an even larger fraction of systems.

### 2.4.2.9 *Detection of Giant Exoplanets via Transits*

WFIRST-AFTA will measure time-series photometry at precisions 1% or better for around 20 million dwarf stars, sufficient to detect transits of giant planets. It will thus survey 200 times more stars for giant planets than monitored by Kepler, and a factor of 40 more than will be monitored by TESS. Detailed simulations of the transiting planet yield of WFIRST-AFTA have not been performed, but McDonald et al. (2014) estimated that a Euclid optical microlensing survey would find about 5000 transiting planets, most of them giants but with some limited sensitivity down to Neptune-radius objects. Because WFIRST-AFTA's infrared microlensing survey will collect nearly six times as many epochs as McDonald et al. assumed for Euclid, and will monitor nearly twice the area of sky, we conservatively estimate twenty thousand transiting planet detections by WFIRST-AFTA. These planets will have separations out to a few tenths of an AU. Since this region of sensitivity to transits overlaps significantly with that of the microlensing survey, comparison of the results from the two techniques allows for a further statistical check on the planet candidates, and ultimately the determination of the frequency of giant planets over all orbital separations (from 0 to infinity) from a single survey.

Contamination by false positives is a significant concern, but the rate of this contamination can be quantified statistically using population synthesis analyses, and verified through follow-up observations in favorable cases (Sahu et al. 2006). That said, confirmation of WFIRST-AFTA's transiting planet candidates will not be trivial, and there will be far too many for individual follow-up efforts. However, a combination of proper motions and in/out-of-transit centroid shifts can be used to rule out background eclipsing binaries and the ~800 color epochs can be used to reject a range of hierarchical triples. Many of the planets will have more than just planet-star radius ratio measurements, with a measurement of the 1-2 micron secondary eclipse depth possible for hundreds of the planets. It should be possible to obtain metallicity and kinematic data for thousands of hosts using upcoming multi-object infrared spectrographs on 8-m telescopes, and many more using instruments on JWST and ELTs. These measurements would be particularly interesting for measuring the Galactic distribution of hot Jupiters, as the planet hosts will belong to both the disk and bulge populations.

### 2.4.2.10 *Improvements Over Previous Designs*

WFIRST-AFTA is intrinsically more capable than all previous WFIRST designs. For comparison, Table 2-4 shows the yields of the previous WFIRST designs (WFIRST-IDRM, WFIRST-DRM1, and WFIRST-DRM2), along with the yields of WFIRST-AFTA for the two observing scenarios discussed above. All of these yields assume the same Cassan et al. mass function. Note that the yields assume campaigns of duration 432 days (IDRM and DRM1) and 266 days (DRM2), as compared to 357 days for our nominal WFIRST-AFTA campaign, or 417 days for the extended campaign. Under these assumptions, WFIRST-AFTA will be substantially more capable than any of these missions, resulting in overall yields (averaged over planet mass) that are larger by factors of ~1.3 to 1.8. The improvements in the yields of WFIRST-AFTA range between factors of ~1.2 to 1.6 times larger for giant planets, to factors of 1.7 to 2.8 for Mars-mass planets. The larger relative increase in the yields of the lowest-mass planets simply reflects the fact that these planets are near the edge of the mission sensitivity for IDRM.

In addition to the improvement in the planet yields, we expect the WFIRST-AFTA to have substantially improved ability to infer the physical parameters of the detected planetary systems. This enhanced capability follows directly from the smaller PSF of the WFIRST-AFTA design (see Figure 2-34) due to the larger aperture. The better image quality results in more precise photometry and astrometry, reduced blending due to chance alignment of unrelated stars, and reduced time baselines required to partially or completely resolve the lenses and sources after the microlensing events. The magnitude of this improvement has not been quantified, but is expected to be substantial.

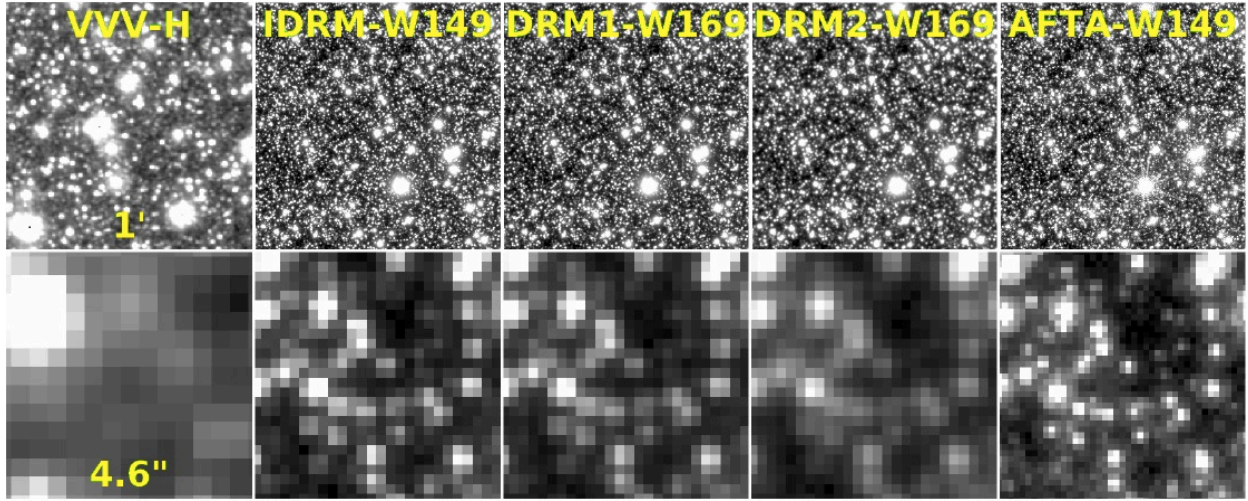

**Figure 2-34:** Left column: Section of a VVV H band image (Saito et al. 2012) from near (l, b) = (1.1°, −1.2°), which lies close to the center of the assumed WFIRST-AFTA fields (see Figure 2-29). Right four columns: Simulated images in the primary wide band of each WFIRST-AFTA design of the same mock field drawn from the Besancon model sightline at (l, b) = (1.1°, −1.2°). The top panels show a 1 x 1 arcminute region and the bottom panels show a 4.6 x 4.6 arcseconds (~13x) zoom-in. The pixel sizes are 0.339, 0.18, 0.18, 0.18 and 0.11 arcseconds from left to right respectively. The Vista Variables in the Via Lactea (VVV) image is based on data products from observations made with ESO Telescopes at the La Silla or Paranal Observatories under ESO programme ID 179.B-2002. From Penny et al. 2015a.

### 2.4.3 Discovering and Characterizing Nearby Exoplanets with a Coronagraph

Our understanding of the internal structure, atmospheres, and evolution of planets was originally developed through models that were tuned to explain the detailed observations of the planets in our own solar system. Our surveys of exoplanetary systems have led to the realization that there exists a diversity of worlds with very different properties and environments than those in our solar system, including gas giants under strong stellar irradiation, gas giants with massive heavy cores, water worlds, and "super-Earths" with masses intermediate to the rocky planets and ice giants in our solar system. Subsequently, these models have had to be expanded and generalized to explain the properties of these new worlds but our understanding is still rudimentary.

The best hope of understanding the physical properties of this diversity of worlds is through comparative planetology: detailed measurements of, and comparisons among, the properties of individual planets and their atmospheres. These measurements provide the primary empirical constraints on our models.

Understanding the structure, atmospheres, and evolution of a diverse set of exoplanets is also an important step in the larger goal of assessing the habitability of Earth-like planets to be discovered in the habitable zones of nearby stars. It is unlikely that any such planets will have exactly the same size, mass, or atmosphere as our own Earth. A large sample of characterized systems with a range of properties will be necessary to understand which properties permit habitability and to properly interpret these discoveries. A key uncertainty is the nature of the so-called "Super-Earths" – are they large rocky worlds, potentially habitable, or "mini-Neptunes" with icy cores and hydrogen atmospheres?

Currently, detailed characterization is only possible for relatively rare transiting systems. Unfortunately, transiting planets are a relatively small subset of systems. Those that are bright enough for significant follow-up tend to be discovered from the ground. Ground-based transit discoveries are strongly biased to large planets with short orbital periods that are subject to strong stellar irradiation. Of course, Kepler has detected many smaller and longer period transiting planets, but nearly all of these systems are far too faint for atmospheric characterization of the planets. Furthermore, atmospheric studies of transiting planets are only sensitive to very specific planet geometries and/or atmospheric pressures, therefore providing an incomplete view of the physical processes at work in the atmospheres of these systems. TESS and JWST will extend this sample, providing near-IR characterization of planets orbiting bright nearby stars, though still emphasizing short periods and lower-mass stars (see Figure 2-36).

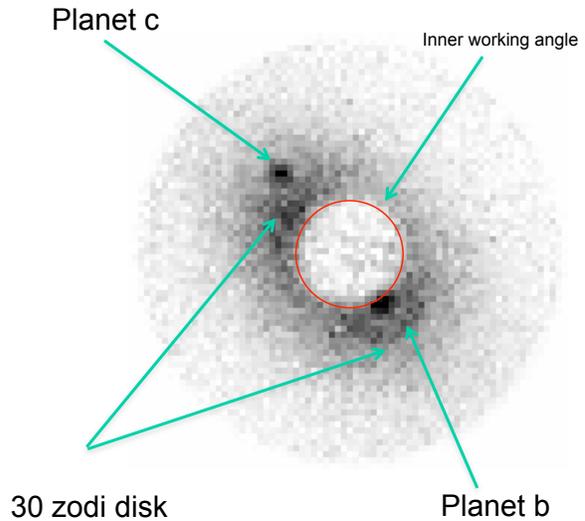

**Figure 2-35: Simulated WFIRST-AFTA coronagraph image of the star 47 Ursa Majoris, showing two directly detected planets. Simulation parameters: 10 hr exposure time, 525-580 nm, Hybrid Lyot Coronagraph. Field of view of 0.518 arcsec radius. A PSF reference has been subtracted, improving the raw contrast by a factor of 10.**

Direct imaging surveys of planets orbiting nearby stellar systems offer a complementary and critical approach to studying the detailed properties of exoplanets. First, planets detected by direct imaging tend to be at longer orbital periods and are less irradiated than those found by transits. Second, spectra of directly imaged planets provide powerful diagnostic information about the structure and composition of the atmospheres. Finally, these planets can be found around the closest stars, which tend to be the best characterized. While ground-based direct imaging surveys have made enormous strides in recent years, with new, much more capable surveys coming on line soon, such surveys are ultimately limited by the contrast achieved in the kinds of planets they can study, being sensitive only to warm and massive young planets. WFIRST-AFTA will detect many analogs to our cold Jupiter for stars in the solar neighborhood, and reach into the regime of the mini-Neptunes or Super-Earths.

Photometry or spectra will be obtained from more than a dozen of the currently known radial velocity (RV) planets and from new, lower-mass planets discovered through future RV surveys or by WFIRST-AFTA itself. Combined with demographics from Kepler and micro-lensing, astrometric information from RV and Gaia, and TESS/JWST transits, the atmospheric characterization achieved by WFIRST-AFTA with a coronagraph will provide comprehensive comparative planetology for planets outside the solar system. This new information will not only revolutionize exoplanet science, but it will also serve as an essential first practical demonstration of high-contrast imaging technology. This lays the groundwork for a future mission that will achieve the ultimate goal of directly detecting a "pale blue dot" around a nearby star and searching for signatures of life on that planet.

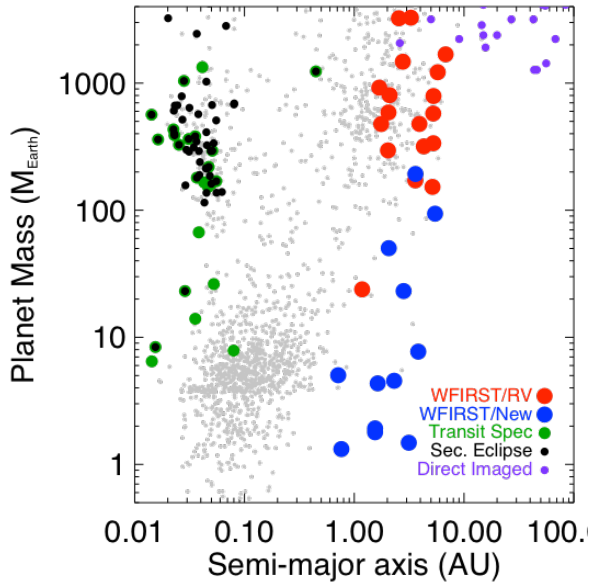

**Figure 2-36: Known exoplanets as of 1/2015. Although more than a thousand extrasolar planets are now known, only a much smaller subset have had their atmospheric composition characterized through transit spectroscopy, secondary eclipse studies, or direct imaging; these are highlighted. Also shown are the currently-known planets (detected through RV techniques) that are within the WFIRST-AFTA coronagraph's reach, and planets drawn from a Monte Carlo simulation of a WFIRST-AFTA search for new planets.**

The WFIRST-AFTA Coronagraphic Instrument (CGI) is designed to provide this characterization. It has a flexible coronagraph architecture (§3.4), using two deformable mirrors and selectable masks to block diffracted light and control residual speckles. In the baseline design, the CGI has two primary modes – a Hybrid Lyot Coronagraph (HLC) using a focal-plane phase-plus-amplitude mask, allowing a small inner working angle that will be used primarily for exoplanet discovery and photometry, and a Shaped Pupil Coronagraph (SPC), relying primarily on binary apodization of the input pupil, giving broader-band performance that is ro-bust against telescope aberrations or jitter; it is used primarily for spectroscopy. These coronagraphs feed either a high-spatial-resolution imager or an integral field spectrograph (§3.4).

In the remainder of this section we describe the exoplanet science that can be expected from a WFIRST-AFTA coronagraph with our current best estimate of performance. The key direct imaging targets fall into two categories: planets (giant planets and sub-Neptunes), and disks (zodiacal disk and Kuiper belt analogs). We will show that with reasonable assumptions on coronagraph parameters, the advances in our knowledge of exoplanet atmospheres and disk structure as well as the increase in our understanding of high-contrast imaging technology will be great. Furthermore, the importance and value of such an instrument as a technology advancement cannot be overstated. The number one priority at medium scale for NASA in the 2010 Decadal Survey was technology advancement of high-contrast imaging in space. A coronagraph on WFIRST-AFTA directly responds to this recommendation, providing an unparalleled opportunity for spaceflight experience with a very high-contrast coronagraph and wavefront control system. WFIRST-AFTA is by no means a coronagraph-optimized telescope–but it is likely that future coronagraphs will also not have the luxury of a perfect unobscured telescope, since flagship missions will have to satisfy a range of science goals just as WFIRST does. The gap between the current coronagraph state of the art and what will someday be needed to characterize a truly Earth-like planet is large (Figure 2-44). Ultimately, the best way to address that gap is to fly an intermediate-performance coronagraph instrument such as the one on WFIRST-AFTA.

### 2.4.3.1 Giant Planet Atmospheres

The Coronagraph Instrument (CGI) on WFIRST-AFTA will be the first instrument to characterize the atmospheres of gaseous planets around nearby Sun-like stars using reflected visible-light images (430-970 nm) and low resolution (R ~ 70) spectra in three bands from 600 to 970 nm. This characterization technique has been refined and calibrated over the past 100 years on our Solar System planets Jupiter, Saturn, Uranus, and Neptune. These data will give us unique information on planetary compositions, temperatures, and cloud layers that are not available via transmission and emission spectroscopy of transiting planets that have been observed by HST and Spitzer.

Reflected light spectra probe deep into the atmospheres of giant planets, up to ~10 bar depths (e.g.,

Marley et al. 2014). Near-IR spectra of transiting planets taken with HST WFC3 (and soon JWST) probe only much higher altitude regions, only to ~1 mbar (e.g., Kreidberg et al. 2014). Therefore WFIRST-AFTA data will probe much greater volumes of exoplanet atmospheres than do transit spectra, and this is the only technique that can be used to do so for cool planets.

We already have a sample of dozens of RV-discovered planets that the CGI will be able to detect via imaging. Knowing their masses (with m*sin(i) from RV and inclination (i) from the CGI) will constrain their gravities. There will also be adequate mission time to characterize a subset further via spectroscopy, and knowing their gravities will aid the interpretation of their spectra.

Most planets observable by the WFIRST-AFTA CGI have equilibrium effective temperatures of ~150 – 350 K, and they are expected to show numerous bands of $CH_4$ (methane) and a moderately strong band of $H_2O$ (water) in their spectra over the $\lambda$ = 600 - 970 nm region, the sensitivity range of the CGI IFS (see 3.4.1.2). The strengths of these bands and the shape of the continuum at shorter wavelengths ($\lambda$ =430 – 600 nm) are diagnostic of the atmospheric temperature, the mixing ratios of C and O, and the amount and location of clouds in the planet's atmosphere. Figure 2-37 shows that signal-to-noise (SNR) = 5 - 20 WFIRST-AFTA spectra of Jupiter-like planets would show multiple spectral features of $CH_4$ (to the right of the green line). The blue continuum in Figure 2-37 is a combination of Rayleigh scattering from $H_2$ molecules (not modeled here) plus cloud reflectivity and will be measurable with the 465 nm and 565 nm filters.

The detection and measurement of the 600-970 nm absorption features constrain the methane mixing ratio and cloud-top pressures of giant planet atmospheres (Hu 2014, Marley et al. 2014, Burrows 2014). Warmer planets will also show $H_2O$ absorption, allowing constraints on water abundance. Combining methane and water abundances will constrain the C/O ratio, a value that is diagnostic of determining the location of planet formation in circumstellar disks (Bond et al. 2010, Helling et al. 2014). The Hu (2014) and Marley et al. (2014) studies show that WFIRST-AFTA CGI spectra will provide good constraints (to factors of ~ 3) on these abundance values and planet temperature-pressure profiles. These studies and Burrows (2014) show that that the expected WFIRST-AFTA CGI planets (from known RV) are expected to have diverse photometric and spectroscopic signatures, so WFIRST-

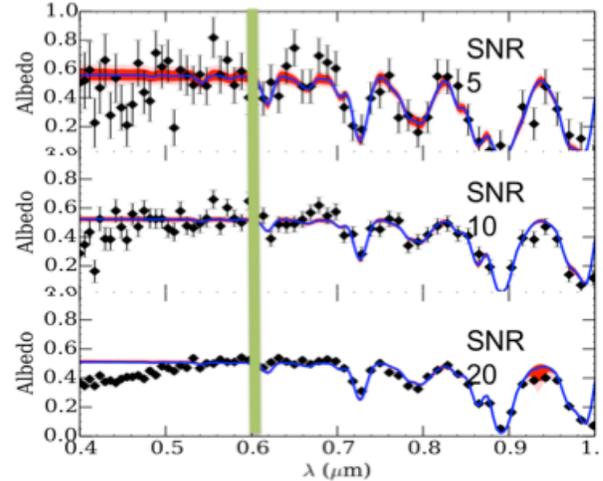

**Figure 2-37: Simulated WFIRST-AFTA CGI spectra of a Jupiter-like planet at SNR = 5 – 20. The green line indicates the short wavelength limit of the CGI IFS. Multiple methane absorption features are detectable over 600 – 970 nm, and the blue continuum slope is measurable (Marley et al. 2014).**

AFTA CGI observations will reveal their physical and chemical differences and diversities.

Measuring atomic abundance ratios (e.g., C/H and O/H) is also important for understanding how planet compositions are related to those of their host stars and how and where planets formed in circumstellar disks (Figure 2-38). The Marley et al. (2014) retrieval molecular technique also holds promise for determining or constraining atomic abundances from WFIRST-AFTA CGI spectra, and it is important to pursue future investigations to evaluate how well the WFIRST-AFTA CGI will be able to constrain planet atmosphere atomic compositions.

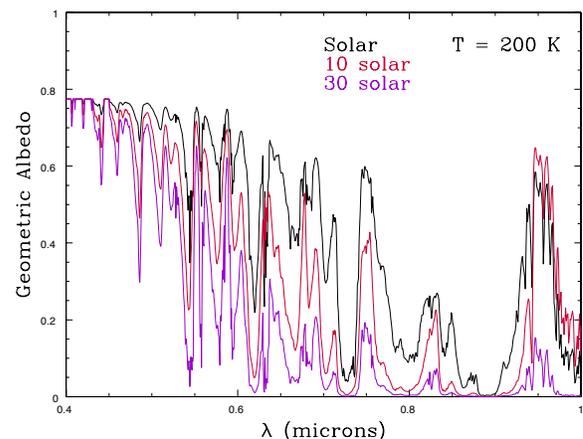

**Figure 2-38: Giant planet albedo spectra for homogenous model atmospheres at T=200K as a function of atmospheric metallicity. These models are cloud-free. From Burrows (2014).**

Photometric linear polarization measurements may also be helpful in characterizing exoplanets. Madhusudhan & Burrows (2012) show that Rayleigh scattering in giant planet atmospheres causes a very large variation (0 to ~50% or more) of linear polarization with planet phase. This effect may be very useful in using RV measurements to determine unambiguous orbit parameters and planetary masses (see also Fluri & Berdyugina 2010). Successful application will require planets with strong Rayleigh scattering with high signal-to-noise polarized light detections at blue (Rayleigh) wavelengths. Polarization could also be useful at longer visible wavelengths for discriminating between giant planet atmospheres that are clear, have tropospheric clouds, or stratospheric hazes (Stam et al. 2004).

In short, high contrast sensitivity and these photometric and spectroscopic capabilities make the WFIRST-AFTA CGI uniquely suited for probing the atmospheres of cool giant planets around nearby Sun-like stars. Existing techniques of transiting planet transmission and emission spectroscopy have given us the first glimpses into the atmospheres of hot giant planets strongly irradiated by their host stars, but these methods are not applicable to cool, non-transiting planets and are often less able to probe the bulk atmospheres of planets. The WFIRST-AFTA CGI will give us a clearer and deeper look.

### 2.4.3.2 Super-Earths and mini-Neptunes

With the performance of the baseline coronagraph, WFIRST-AFTA can only see the smallest planets (<1.5 RE) when they are physically close to their parent stars, and hence only around the very nearest stars. It is unlikely (but not impossible) that WFIRST-AFTA will be able to see a truly Earth-like planet (<1.5 RE, located in the classical habitable zone.) Still, WFIRST-AFTA has the potential for major discoveries in the study of subgiant planets.

One of the most striking discoveries about extrasolar planets made in the past decade is the large population of objects with radius between 1 and 4 times that of the Earth. Based on Kepler observations (Fressin et al. 2013, Marcy et al. 2014) these planets – which have no analog in our solar system - may be the most common kind of planet in the universe (see Figure 2-39). The exact nature of these planets is poorly known. They could be so-called "super-Earths" with rocky compositions, or "mini-Neptunes" with massive hydrogen envelopes, or even stranger objects, partially or mostly water and ice. Only a small number, mostly in tight orbits,

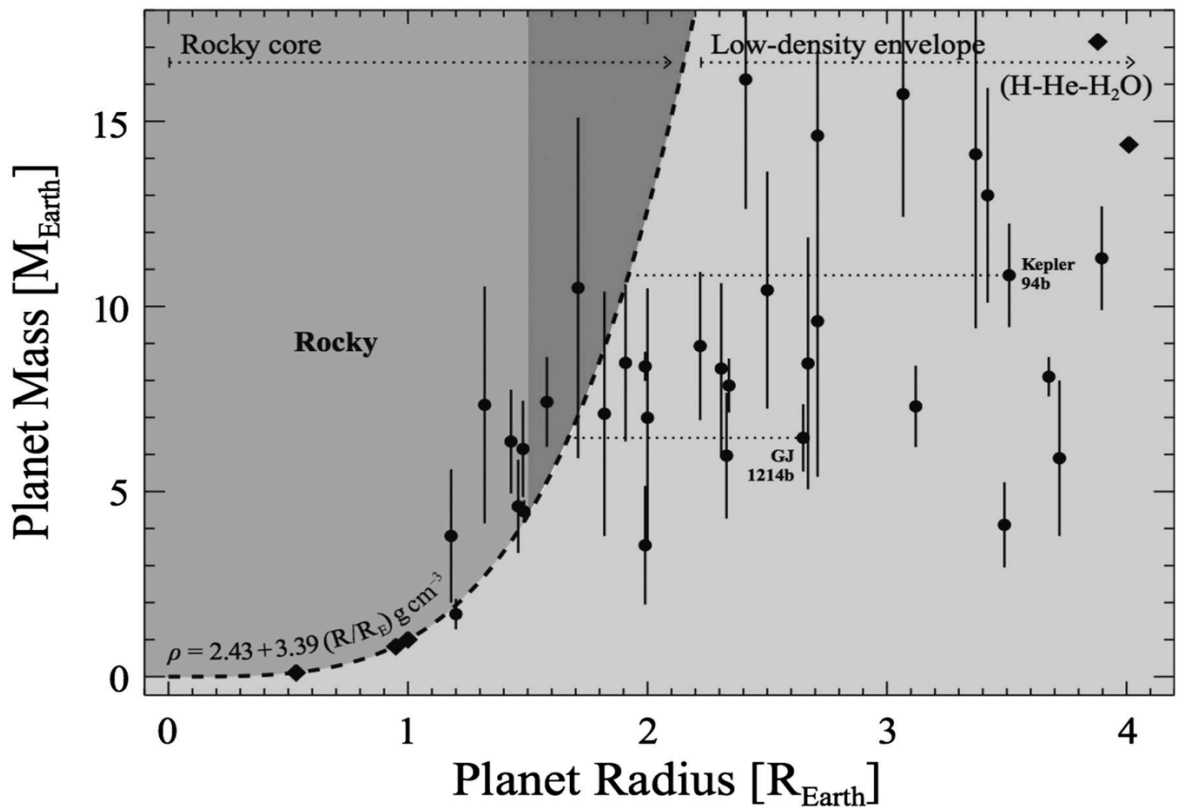

**Figure 2-39: Mass vs. radius of known exoplanets, with the regions of rocky composition and low-density (hydrogen/helium/water) indicated. From Marcy et al. 2014**

have measured densities or even density upper limits. The smallest planets seem to generally (but not always) have high densities consistent with rocky composition (Marcy et al. 2014) while planets >1.6 RE seem to have lower densities consistent with large hydrogen envelopes and/or water compositions (Rogers 2014). The atmospheres of these planets could span a huge range from gas-giant hydrogen to water vapor to $CO_2$ to oxygen or hydrocarbons (Hu et al. 2012), with or without clouds. Even fewer planets in this size range have had their atmospheric properties measured through transit and secondary-eclipse observations (Morley et al. 2013, Caceres et al. 2014) and those spectra are relatively featureless, indicating that high-altitude clouds are veiling the lower atmosphere.

The K2 and TESS missions, combined with JWST, will help illuminate this population through transit and secondary eclipse observations of planets orbiting brighter nearby stars for planets with short <60 day orbits. Direct imaging with the WFIRST-AFTA coronagraph is a powerful complement. Even though these planets are challenging targets for direct imaging (contrast ~$10^{-9}$), because they are so common, there will likely be multiple sub-Neptune planets within the coronagraph's reach. Compared to transits, WFIRST-AFTA will primarily be sensitive to planets in wider orbits, 0.5 - 5 AU (Figure 2-50). Transit observations are most sensitive to the higher layers of a planet's atmosphere (P < 1 mbar), particularly for cloudy or hazy atmospheres. Imaging measures the planet's reflectance, which probes deep into the atmosphere (potentially all the way to the surface, if the atmosphere is sufficiently thin). Imaging photometry, or spectroscopy of the brightest targets, can constrain atmospheric parameters including clouds even at relatively low SNR (Figure 2-40).

### 2.4.3.3 *Circumstellar Disk Science*

The WFIRST-AFTA CGI will deliver uniquely sensitive images of the planetary disks around nearby stars, resolving much smaller amounts of dust at much closer distances than possible with HST or other observatories. The small inner working angle (IWA) and high contrast of the WFIRST-AFTA CGI will allow unique study of several types of circumstellar disks:

1. Low density, zodiacal-like disks down to the HZs of nearby Sun-like stars
2. Inner regions (HZ to ~10 AU) of known massive, extended debris disks
3. Inner regions (HZ to ~10 AU) of warm disks discovered but not resolved at IR wavelengths

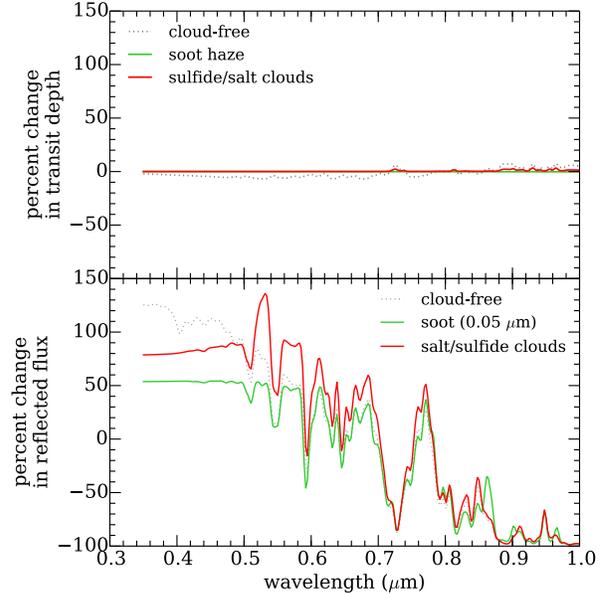

**Figure 2-40: Comparison of transit (top) and direct-imaging (bottom) spectra of a GJ1214b analog (a hot super-Earth) for three different cloud compositions. Figure from Caroline Morley. SNR of 5-10 broadband photometry would distinguish between these three cases.**

Observing a significant sample of systems spanning these areas will significantly address a number of currently outstanding questions about circumstellar disks and their role in planet formation and planetary systems evolution (c.f. Schneider 2014):

1. What are the levels of circumstellar dust in, and exterior to, the habitable zones of exoplanetary systems?
2. Will dust in habitable zones interfere with future planet-finding space observatories?
3. How do dust sub-structures seen in disks trace the presence of seen or unseen planets?
4. What veneer is delivered to planetary atmospheres and surfaces by asteroids, comets, and other material?
5. How do disks of protoplanetary materials evolve to make Solar System-like architectures?

Studying zodiacal-like disks may be the most exciting circumstellar disk frontier that WFIRST-AFTA CGI will open. Disks with dust masses on the order of ~10 times that of the solar system's zodiacal disk can be detected in multiple resolution elements in under 1 day of integration time for many Sun-like stars within ~15 – 20 pc. Figure 2-41 shows a simulated CGI image of a 30 zodi disk around the nearby star 47 UMa (G2V, 14 pc distance), easily detected at SNR ~ 10 in multiple resolution elements (Schneider 2014). The disk position angle and radial profile parameters are recovered from the simulated data with only small biases and small uncertainties.

The stars selected for both blind and known RV planet searches will make an ideal target sample for zodiacal disk searches. In fact, the planet search data alone can be used for these disk searches given that both investigations require similar imaging integration times (~10 hours). The presence or absence of disks and planets together will address some specific aspects of these nearest solar systems:

- Planets without disks indicate little recent collisional activity, like our solar system
- Disks without planets indicate significant interaction of small bodies only
- Disks and planets together will reveal the dynamics of their interactions via disk non-uniformities
- Neither disks nor planets will indicate efficient clearing of the pre-planetary disk

These will likely be the highest priority WFIRST-AFTA CGI disk observations since they probe planetary systems uniquely down to the habitable zones of nearby stars with good sensitivities. The ground-based LBT-I instrument will also search for HZ zodiacal dust in a number of the same stars, but in the thermal IR range with less spatial resolution. WFIRST-AFTA probes the same wavelength that future space coronagraphs will use. Combined visible WFIRST and mid-IR LBT-I data will provide the albedos of the grains, constraining their compositions.

WFIRST-AFTA CGI images will also extend our knowledge of known, more massive debris disk systems. Their similar resolution, high contrast, and small IWA will perfectly complement the much larger area of HST visible and near-IR scattered-light images of massive extended debris disks around relatively young stars (~ 50 Myr) in nearby (~20 – 50+ pc) moving groups. Approximately two dozen such disks have been studied by HST (e.g., Schneider et al. 2014), but we know very little about their inner regions due to the poor contrast and IWA of existing observations. WFIRST-AFTA CGI images will reveal whether the inner regions (~few – 20+ AU) are clear or dusty and will show the dynamical impact of any discovered planets.

Debris disks detected but poorly resolved by warm thermal IR emission will also be good targets for WFIRST-AFTA CGI. Spitzer detected λ = 24 μm (T ~ 150 K) disks around several nearby main sequence stars (e.g., Beichman et al. 2006), indicative of recent collisions that likely occur in the asteroid belt regions of these systems. More recently, WISE and Herschel have detected dozens of such disks (e.g., Padgett et al. 2012), and WFIRST-AFTA CGI images should be sensitive enough to detect and resolve these in scattered light. This combination of IR and visible-light images will reveal the albedo and hence constrain the compositions of the grains in these disks.

Determining the compositions and sizes of the dust in debris disks is important for understanding the formation and collisions of planetesimals in other solar systems. Debris disks have been found to be highly linearly polarized, varying from ~0 to ~50% as a function of angle from the central star. Studying this variation of linear polarization with scattering angle constrains the size distributions and compositions of disk materials (e.g., Perrin et al. 2009). Measuring the degree of polarization in two well-separated filters (e.g. ~465 nm and ~885 nm wavelength) in multiple resolution elements significantly improves the separate determination of dust compositions and sizes, allowing unique determinations in some cases (see §8.1 of Schneider 2014).

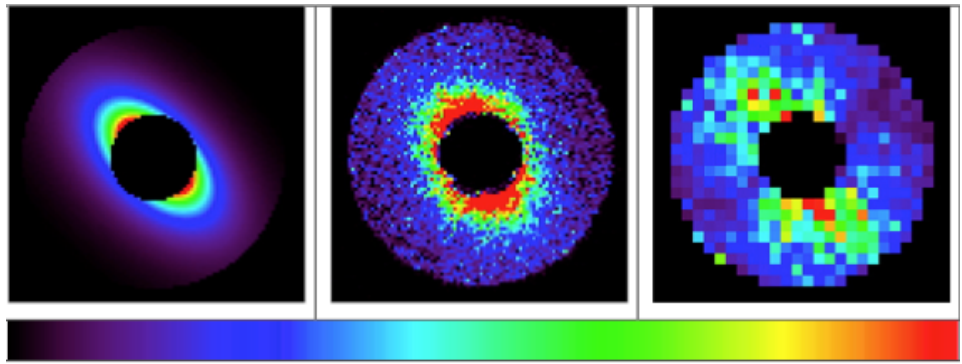

**Figure 2-41: Simulated WFIRST-AFTA CGI images of a 30 zodi disk around 47 UMa. The noiseless zodipic model is shown at left, a 10 hour simulated shaped pupil coronagraph image is in the middle, and the significance of the detection is shown on the right. The right panel shows signal-to-noise per 1.2λ/D resolution element (about 10 pixels), scaled from SNR = 1 – 15 (from Schneider 2014). The system's 2 known RV planets are detectable with WFIRST-AFTA but are not included in the simulations.**

The imaging capabilities of the WFIRST-AFTA CGI instrument are sufficient for making progress in all of these circumstellar disk science areas. Its images will measure the amount, location, size, and composition of dust in nearby solar systems, indicate the presence of both seen and unseen planets, and trace the history, frequency, and severity of planetary collisions. These observations are essential for understanding the formation and evolution of the nearest planetary systems.

### 2.4.3.4 *Principals and Concepts of Coronagraphy*

This section provides a narrative overview of the principles of coronagraphs in the context of WFIRST-AFTA. For a quantitative overview, see the chapter "Direct Imaging of Exoplanets" by Traub and Oppenheimer, in the book "Exoplanets", edited by S. Seager (2010).

Direct imaging is challenging because exoplanets are orders of magnitude fainter than their parent stars. To image these planets thus requires a system to create high contrast at small angular separations, by which we mean the suppression of the diffraction pattern of the host star, allowing the faint planet to be detected. Such a system of internal optics is called a *coronagraph*, named for the first such device invented by Bernard Lyot in the early '20s and used to image the Sun's corona. Nevertheless, astronomers could make little use of it for imaging other stars to find planets due to the distorting effects of the atmosphere. The advent of space-based observatories, however, removed this impediment and motivated the first discussions on how to isolate the light of a planet outside the Solar System from its host star (Spitzer 1962, Kenknight 1977, Davies 1980). In fact, several coronagraphs were included onboard HST. However, various aberration and alignment problems limited their efficacy (Krist et al. 2003).

Adaptive optics (AO) technology eventually advanced to the point of making exoplanet imaging feasible from the ground (Angel 1994). In the mid 1990s, coupled with the atmospheric correction provided by new AO systems, the Lyot coronagraph was revised to cancel starlight instead of sunlight, thereby enabling searches for faint companions around nearby stars (Lyot 1939, Golimowski et al. 1992, Nakajima 1994). The illustration in Figure 2-42 shows how on-axis starlight is rejected by a classical Lyot coronagraph, when there are no aberrations present. In the first focal plane, an opaque occulting spot blocks the core of the star's point spread function (PSF). Light from the fainter diffraction rings surrounding the PSF continues to propagate, and is collimated to a reimaged pupil plane. Fraunhofer diffraction theory dictates that the field distribution in this reimaged pupil is equal to the difference between the entrance pupil (a bright annulus in this case), and the convolution of a circular tophat function of the same width as the occulting spot with the entrance pupil. The latter subtrahend term is equivalent to a low-pass filtered copy of the entrance pupil. The resulting energy distribution is concentrated near the edges of the open annulus. Therefore, by adding a mask of similar shape as the original telescope pupil, but with the width of the open annulus slightly reduced, much of the remaining starlight is blocked. Because of this *Lyot stop*, the diffraction rings around the occulting spot in the final focal plane are strongly suppressed with respect to the original telescope PSF. For a succinct mathematical description of this concept, see Sivaramakrishnan et al. (2001).

Pupil apodization is another tool for suppressing starlight. As a natural consequence of Fraunhofer diffraction theory, altering the shape or transmission profile of the entrance pupil redistributes the on-axis starlight in the image plane. As shown in Figure 2-43 this relationship can be used to form a PSF with a zone of very high-contrast near the star, without any additional coronagraphic masks. A particular subset of apodized pupils is the shaped pupil, where the transmission is forced to be binary, that is, fully transmissive or fully opaque at any given point across the pupil. Various classes of shaped pupils have been developed using optimization tools (Spergel 2001, Kasdin et al. 2003, Vanderbei et al. 2003). Another class of apodized pupil coronagraphs is the Phase Induced Amplitude Apodization where the pupil apodization is achieved by aspheric mirrors remapping the distribution of light. These have the advantage of much higher throughput and smaller inner working angle at the cost of complexity. The Apodized Pupil Lyot Coronagraph (APLC) combines both an apodized entrance pupil with a focal plane mask and Lyot stop to improve on the starlight rejection of the classical design and better allow for central obstructions (Aime et al. 2001, Soummer et al. 2003).

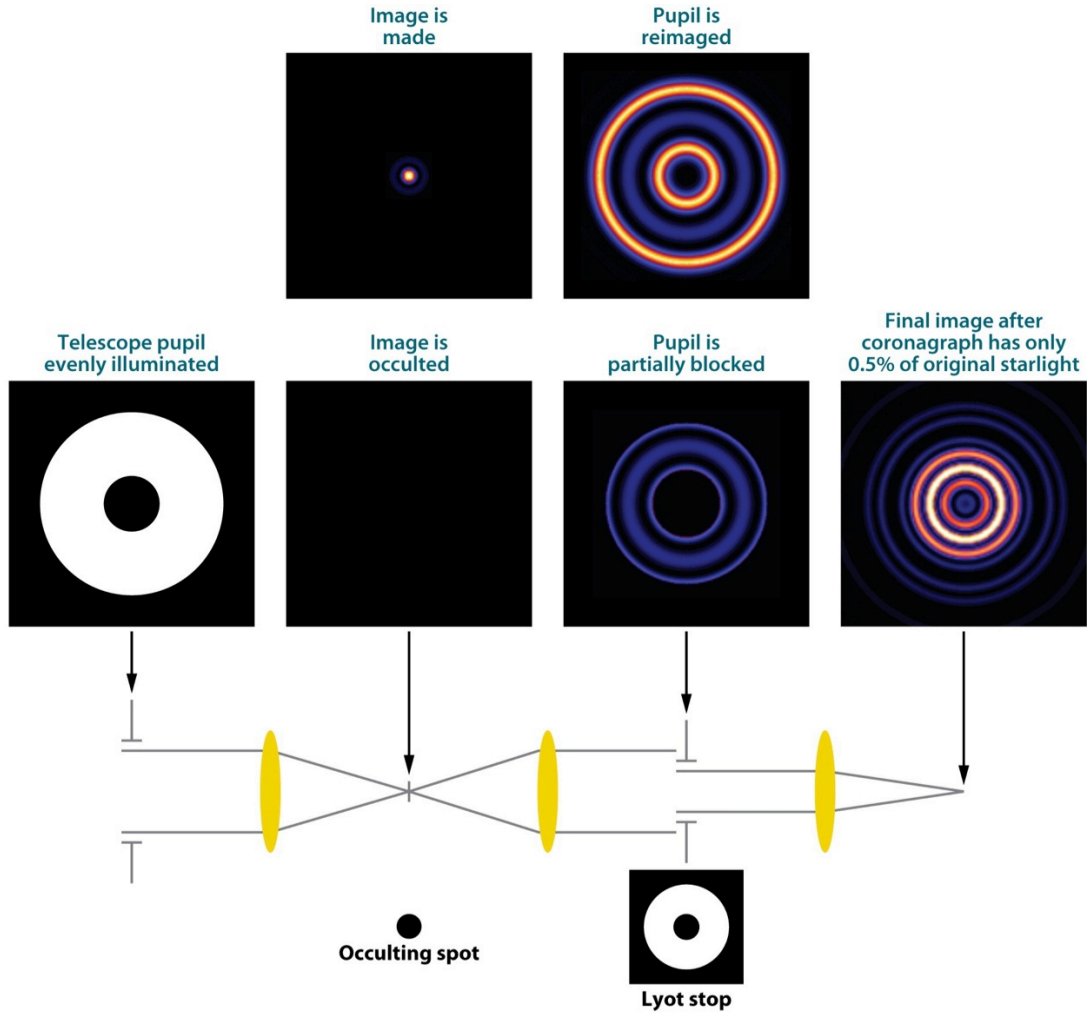

**Figure 2-42: Diagram of the propagation of on-axis starlight through a classical Lyot coronagraph, for the case of a perfect, monochromatic wavefront. In each of the two critical intermediate planes, the intensity is displayed immediately before and after passing through the mask. This diagram, based on simulations by A. Sivaramakrishnan, appears in a review article by R. Oppenheimer & S. Hinkley (2009).**

#### 2.4.3.5 *Ground-Based Coronagraphs*

Ground-based coronagraphs typically operate in the near-IR, where AO correction is better and surveys targeting young stars can take advantage of the self-luminous characteristic of gas giants and brown dwarfs, which radiate in the infrared for tens of millions of years after their formation (Burrows et al. 1995). They have produced a dozen notable successes (Kalas et al. 2008, Marois et al. 2008, Lagrange et al. 2010, Rameau et al. 2013, Kuzuhara et al. 2013, Currie et al. 2014, Kraus et al. 2014) but the systems discovered, with supermassive (3-10 MJ) young (<100 Myr) planets in wide (20-200 AU) orbits are very different than our own. Taking the Gemini/NICI campaign as a representative example of a large completed survey, statistical analysis shows that for a typical target star, the planet mass corresponding to 95% detection likelihood at 0.5 arcsec separation (~30 AU) was about 10 $M_J$ (Biller et al. 2013).

Even with high-order AO correction, coronagraph detection limits are severely limited by uncorrected quasi-static aberrations within the instrument, and their resultant speckle noise in the image (Racine et al. 1999, Marois et al. 2005, Hinkley et al. 2007). Speckles can be partially subtracted if certain diversity properties of the data are exploited — in practice either FoV rotation (angular differential imaging) or spectral differencing (Marois et al. 2006, Sparks & Ford 2002). The latter approach, relying on the chromatic scaling of the speckle halo, motivated the next generation of ground-based instruments to rely on integral field spectrographs (IFS) to improve the dynamic range. As an added benefit, the spectroscopic data provided by the IFS reveal precious insights about the atmosphere of the detected planet. Exoplanet imaging surveys combining AO, coronagraphy, and integral field spectroscopy are underway at Palomar Observatory, Gemini Observatory, and the ESO Very Large Telescope; soon another will begin at Subaru Telescope (Hinkley et al. 2011, Macintosh et al. 2014, Langlois et al. 2014, Groff et al. 2014). As illustrated in the angular separation-contrast parameter space of Figure 2-44, only these newest instruments (GPI and VLT-SPHERE) can reach true, young Jovian analogs (planets with mass 1 $M_J$ and semi-major axis 5 AU), and only around a small number of stars. No ground-based coronagraph can detect the reflected light from mature planets.

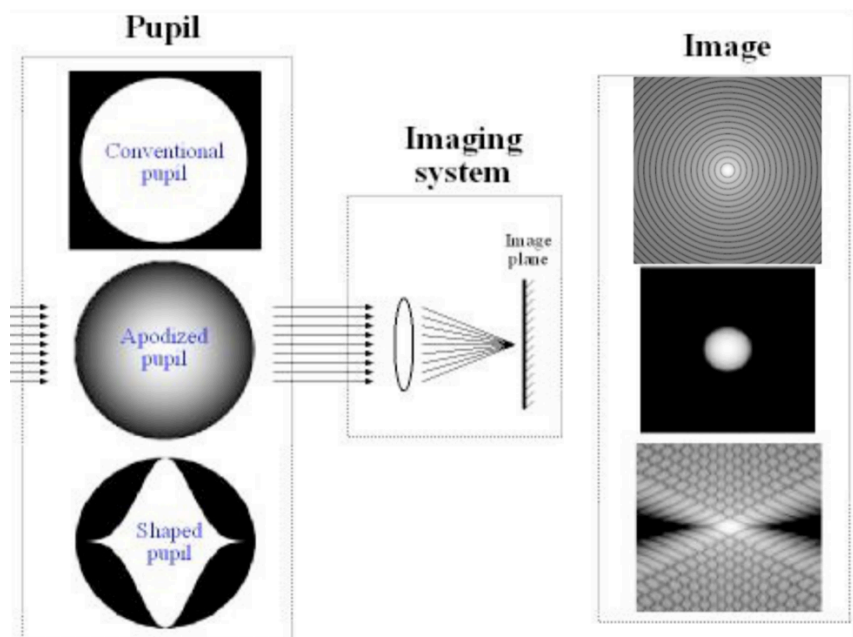

**Figure 2-43: Diagram showing how shaped pupil coronagraphs use apodization to create regions of high contrast in the image plane. For reference, the upper pair of graphics show the Airy function PSF created by a clear, unapodized circular aperture. When a smooth apodization is applied—in this case a circular prolate spheroidal profile—the concentration of energy within some radius is maximized, and the outer diffraction rings are suppressed by many orders of magnitude. Purely binary apodizations, as shown in the bottom pair, can achieve the same effect, although generally at some cost in the size of the search area.**

Future 20-40 m ground-based telescopes may also incorporate advanced adaptive optics coronagraphs (though likely not until ~2030.) Such large aperture telescopes will dramatically reduce the accessible angular separation, simply due to the scaled-down diffraction width. However, even under the most optimistic wavefront calibration (speckle control) assumptions, the contrast floor will be set by the AO performance.

ELTs observations will complement WFIRST-AFTA exoplanet imaging in wavelength coverage, as ground-based adaptive optics is most effective in near-IR. While WFIRST-AFTA will provide reflected light images of mature planets in visible light, ELTs will be probing planet formation processes and image a large number of self-luminous giant planets around young stars. The physical and chemical evolution of giant planets will be probed from formation / young age phases dominated by internal heat (observed by ELTs and JWST) to mature phases during which stellar radiation is a significant/dominant source of energy (observed by WFIRST). At intermediate ages and/or massive (~1 MJ and more) planets, both thermal emission in the near-IR with ELTs and reflected light in visible light with WFIRST can be measured, allowing the energy input (stellar radiation, internal heat) and output (emission in near-IR) to be measured. Together with mass measurement from RV, the main physical characteristics of the planet (radius, mass, effective temperature, internal heat output, surface gravity) can be derived, allowing detailed and robust interpretation of WFIRST-AFTA exoplanet spectra. While the overlap between ELT-observable and the WFIRST-imaged exoplanets may be limited to a few giant planets, it will provide representative templates for interpretation of additional WFIRST exoplanet spectra.

ELTs will also observe reflected light exoplanets in near-IR. While ELTs will not be able to reach the deep and stable contrast levels accessible from space with WFIRST, they offer IWA about 5x smaller in near-IR than accessible with WFIRST in visible light. The same type of planet (radius, albedo), if moved 5x closer to its host star, becomes 25x brighter in reflected light: ELTs will compensate for lack of deep contrast by smaller IWA. ELTs will therefore observe planets that are statistically closer in to their host stars, and more strongly illuminated, than WFIRST-AFTA. Direct comparison between ELTs and WFIRST spectra will help understand how exoplanets physical and chemical character-istics transition from cold Jupiter-like planets to hot Jupiters observable by transit spectroscopy.

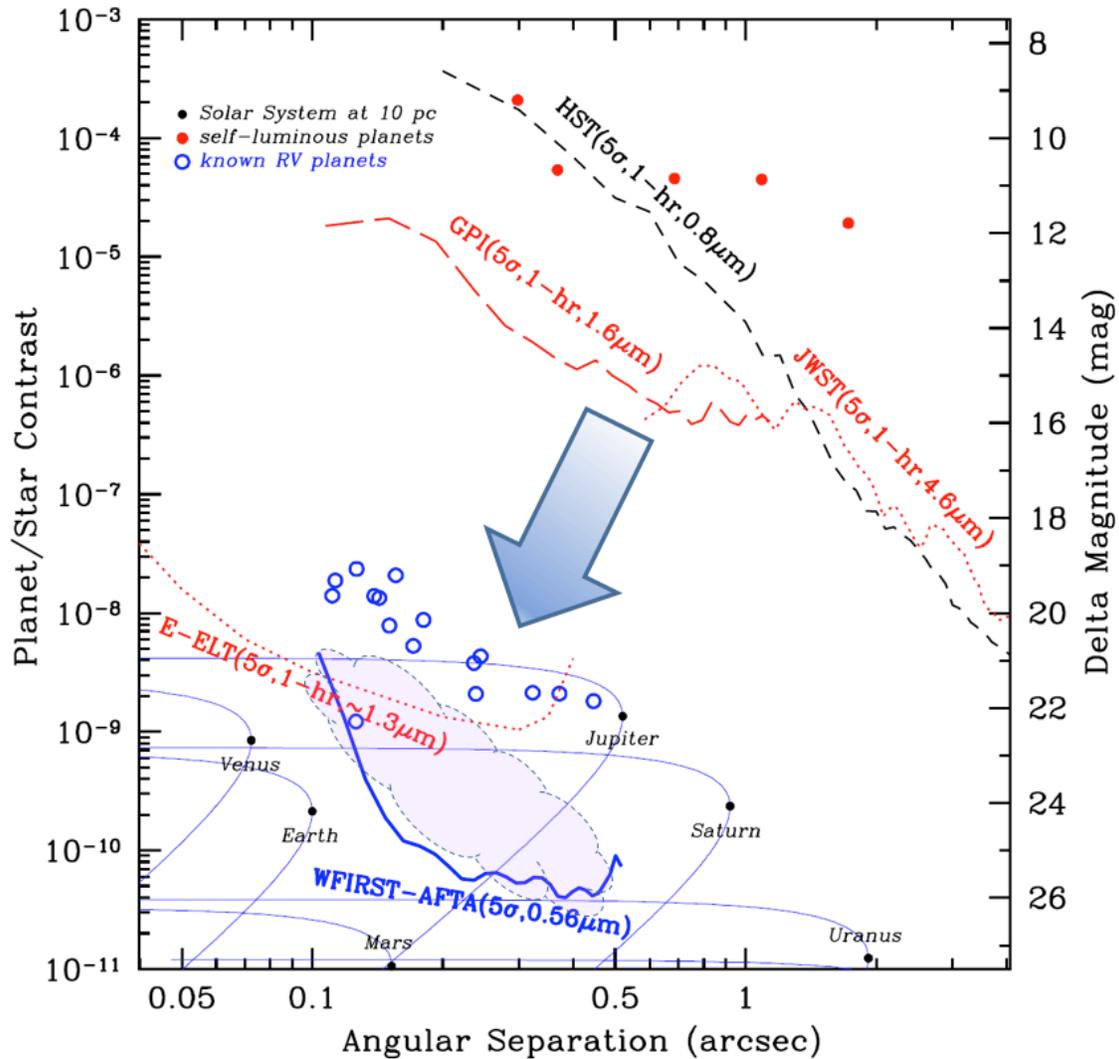

**Figure 2-44: Exoplanet detection limit of the HLC coronagraph on WFIRST-AFTA, compared to other high-contrast systems. The contrast values of RV planets detectable by HLC on WFIRST-AFTA are shown as open blue circles, along with the detection floor set by residual speckle noise (solid blue line, from Figure 2-48); note that this does not include photon noise. Other high contrast systems (Hubble Space Telescope, James Webb Space Telescope, Gemini Planet Imager, and the European Extremely Large Telescope) are shown for 1-hour exposures on fiducial targets, including photon noise. There are two important complementary areas here: (1) WFIRST-AFTA vs. GPI and JWST, and (2) WFIRST-AFTA vs. E-ELT and TMT. Regarding (1) for GPI and JWST, the limiting sensitivities are much poorer in absolute terms (see the labeled curves), but the wavelength range of operation is the near-infrared, where hot, young planets are bright, so the type of planet probed is completely different than for WFIRST-AFTA, which will observe the much more numerous mature, cool planets. Regarding (2) for the E-ELT and TMT, which have roughly similar sensitivities, the complementarity is that the ELTs will be able to observe planets closer to their stars than WFIRST-AFTA (owing to their 12 to 17 times larger diameters), and collect more photons per planet (allowing a poorer raw contrast but enabling a greater post-processing factor), so the ELTs will be best at observing habitable zones of nearby late-type stars, whereas WFIRST-AFTA will be best at nearby solar-type stars. The shaded blue cloud indicates the range of expected WFIRST-AFTA discoveries of new nearby Neptunes and Super-Earths.**

WFIRST will provide deep, well-calibrated contrast levels not accessible from the ground, and will thus allow direct imaging of sub-Jupiter mass exoplanets in/near habitable zones of Sun-like stars. Such planets will be very challenging for ELTs to observe due to the extreme contrast and ELTs detection limits are currently poorly understood in this regime. Ultimately, WFIRST-AFTA will push the low-mass detection limit for direct imaging further than possible with any other observatory (Figure 2-44).

### 2.4.3.6 *WFIRST Coronagraph Design and Performance*

As described above, a coronagraph is a set of optical elements that modifies the PSF of the telescope to create a region around the stellar image where a dim companion can be extracted. Due to optical errors in any system, all coronagraphs must be designed together with wavefront control via one or more deformable mirrors (DMs) to correct wavefront errors. The combined coronagraph and wavefront control system is characterized by the contrast, inner working angle, and stability achieved. Contrast is the degree to which the instrument can suppress scattered and diffracted starlight in order to reveal a faint companion. Inner Working Angle (IWA) is the smallest angle on the sky at which it can reach its designed contrast. This angle is typically only a few times larger than the theoretical diffraction limit of the telescope. The resulting residual stellar halo must also be stable over the time scale of an observation, so that the halo can be subtracted to reveal an exoplanet or disk. In principle, with a well characterized and stable PSF, various subtraction methods that have been developed and used on both ground and space images can be employed to average the background photon noise and extract faint planets that are below the raw contrast level. The limit of an observable planet is then determined only by the photon noise and the available integration time. In practice, there is a systematic limit to the contrast of a recoverable planet given by the stability of the telescope (how rapidly the background speckle pattern changes) and the structure of the exozodiacal light (a confusion limit set by clumping in a disk).

The performance of a coronagraph instrument on WFIRST will also be partially determined by the orbital environment and resulting stability of the telescope. Additionally, the recent history of planet imaging shows that recovering planets with up to factors of 10 fainter contrast than the background is regularly accomplished, both on large ground telescopes and from the Hubble Space Telescope. In some cases, factors of 100 have been achieved. We thus characterize the coronagraph instrument by its detection limit, that is, the limiting magnitude of a recoverable planet relative to the star from the combination of the coronagraph and wavefront control and data processing.

Over the last decade and a half there has been an astounding growth in research on coronagraphy with an outpouring of new coronagraph concepts, differing in architecture (pupil plane masks vs. image plane masks, phase vs. amplitude), performance (contrast, inner working angle, and throughput), and robustness (sensitivity to wavefront aberrations, thermal distortions, and jitter). Until recently, almost all approaches assumed off-axis telescopes, eliminating the need to cope with diffraction induced by central obstructions and spiders (a severe challenge for most coronagraph approaches). It has only been in the last several years that breakthroughs in the design of certain coronagraphs have allowed high contrast with obstructed telescopes or segmented mirrors (albeit with reduced performance). In the Fall of 2013 a study was performed (Traub et al. 2014) to examine the expected performance and robustness of 5 such coronagraphs if combined with the extremely complex pupil of the WFIRST-AFTA telescope: shaped pupil (SPC), hybrid Lyot (HLC), vector vortex (VVC), phase induced amplitude apodization/complex mask (PIAA/CMC) and visible nulling coronagraph (VNC). At the end of the study it was clear that the VVC could not be designed for the AFTA pupil, while VNC designs and components were not sufficiently mature to be developed in the time frame consistent with the anticipated WFIRST-AFTA mission schedule. While the PIAA / CMC had the potential for the best performance in inner working angle and throughput, designs with sufficiently high contrast were hard to achieve and it was by far the most sensitive to jitter and aberrations. The result was the selection of the SPC and HLC as the two baseline coronagraphs with the PIAA/CMC as backup pending further theoretical and laboratory work. The SPC had the advantage of being the simplest to implement with the least sensitivity to spacecraft jitter, and potentially broader bandwidth (well-matched to spectroscopy). HLC had potentially much higher contrast at a smaller inner working but with somewhat more manufacturing and performance risk and a higher sensitivity to jitter.

The HLC and SPC architectures are described in §3.4. The HLC (Figure 3-27) uses a combination of a focal plane mask and a Lyot stop to cancel on-axis starlight. However, unlike the classical Lyot coronagraph shown earlier, the focal plane mask has a continuous, non-monotonic, radial transmission profile. This transmission profile is band-limited in spatial frequency, so that field cancellation in selective regions of the Lyot plane is much deeper than the classical case, as well as achieving a smaller inner working angle (Kuchner & Traub 2002). The HLC focal plane mask is composed of layers of metal and dielectric deposits on glass. The pattern of layers makes use of the wavelength-dependent, phase-shifting film transmission function to achieve the desired reimaged pupil over a broad band-pass (Trauger et al. 2011).

The SPC (Figure 3-28) uses a binary-valued apodization of the telescope pupil to create a region of suppressed starlight in the image plane (Spergel 2001, Kasdin et al. 2003). The pattern of dark islands that define the shaped pupil are the outcome of a linear optimization program (Vanderbei et al. 2003). The program determines what parts of the pupil must be blocked, at minimum, and in addition to any existing obscurations, in order to create destructive interference over some range in separation and position angle (Carlotti et al. 2012, 2013). The SPC focal plane mask is a diaphragm matched to this dark region. This mask tends to diffract a large fraction of residual on-axis energy to the edge of the reimaged pupil and beyond it. Therefore, a Lyot stop serves to further reject starlight before it is brought to a final focus (Riggs et al. 2014).

The WFIRST pupil geometry — with its large central obstruction and six support struts — is challenging for any coronagraph design. The HLC and SPC work around these obscurations in different ways. Upstream of the HLC, the two DMs apply large static figures to remap the obscurations, and make use of the band-limited occulting mask (Trauger et al. 2013). The peak-to-valley DM stroke used in the latest HLC design is less than 300 nm. The SPC, on the other hand, uses the apodization pattern of the shaped pupil to confine diffraction from the obstructions to the masked-off area of the focal plane. As a consequence, more throughput is sacrificed than in a coronagraph with a purely phase-based approach. However, by not relying on the DMs to remap the discontinuities, the SPC operates in a regime of lower technological risk than the HLC.

Teams at JPL and Princeton are leading the HLC and SPC designs, respectively. Both teams have worked through several design iterations to maximize performance within the constraints of WFIRST. These efforts take into account the pupil geometry, the expected level of spacecraft jitter, low-order aberrations, and the expected distribution of planet signals in separation-contrast space. These designs are then used in a complete simulation of the entire optical system based on the PROPER propagation code developed by John Krist at JPL. PROPER accounts for all optical surfaces in the instrument and performs a diffraction simulation, including closed loop wavefront control, polarization, and jitter to reach a prediction for the final stellar image. More details on the modeling approach and simulation can be found in Appendix F. The resulting contrast prediction is then used in the science yield model (de-scribed in Appendix F) to estimate the number and type of detectable planets. For example, Figure 2-46 is the result of a polychromatic simulation of the shaped pupil shown in Figure 2-45. Figure 2-47 is the result of the Milestone 1 tests at the JPL High Contrast Imaging Testbed (HCIT) showing a contrast better than the $1\times10^{-8}$ that the shaped pupil was designed for, both monochromatically and in a 10% band. Detailed descriptions of the CGI instrument, the imaging camera, and the integral field spectrograph can be found in §3.4.

### 2.4.3.7 *Exoplanet Science Yields*

The SDT and collaborators have carried out detailed modeling of WFIRST-AFTA's ability to detect and characterize planets. See Appendix F for detailed discussion. We take the known RV planets in the solar neighborhood and compute their detectability (Table 2-7) based on their radius and separation from the parent star. Coronagraph models predict the residual speckle halo. Final sensitivity is set by an absolute floor from residual speckles after PSF subtraction, as well as the random noise contributions from foreground and background zodiacal light. The residual speckle floor is usually expressed as a fraction of the input speckle noise. Preliminary simulations show that even relatively simple algorithms can achieve a factor of ten attenuation of PSF halo noise, with a factor of thirty possible, and we evaluate science yield for both. We also evaluate science yield as a function of residual image motion (after correction by the coronagraph fast steering mirror) from 1.8 mas to 0.4 mas. For each target planet we estimate the exposure time required to obtain SNR=5 photometric measurements in a standard WFIRST-AFTA band (Figure 2-48, Table 2-7).

Although Doppler surveys of particular low-mass stars can be sensitive to near-Earth-mass planets in short-period orbits, Doppler surveys of the best coronagraph targets – higher-mass stars in the solar neighborhood - are marginally complete only to Saturn or Jupiter-mass planets in the range of orbits WFIRST-AFTA can access. (Figure 2-49, Howard & Fulton 2014), and of course are also incomplete for early-type stars and young stars. WFIRST-AFTA can potentially characterize much lower-mass planets (§2.4.3.2 and Appendix F), particularly if they are observed at optimal phase angle, but the individual planets that will be seen have not yet been discovered. Evaluating the ability of WFIRST-AFTA itself to discover these planets requires a complete DRM-like simulation. We can, however, evaluate the completeness of WFIRST-AFTA observations of any given star as a function of mass and semi-major axis (see e.g., Brown 2005, Savransky et al. 2010). Averaging these completeness values over a likely sample of targets gives a good idea of the range of planets WFIRST-AFTA is sensitive to (Figure 2-50). Also, in Appendix F, we use a random distribution of planets consistent with the measured Kepler populations to evaluate potential detectability of low-mass planets. We evaluate the number of such planets that could be discovered in a 50-day and 180-day survey for various target priorities. Table 2-8 summarizes the complete exoplanet yield of various categories.

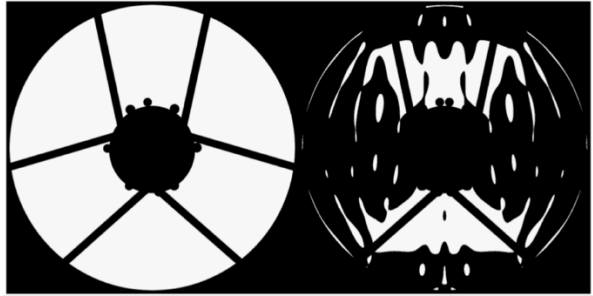

**Figure 2-45: Shaped pupil coronagraph characterization mask , shown to the right of the WFIRST-AFTA telescope pupil. This mask has 30% transmission with respect to the available open area, and is designed for a bowtie-shaped focal plane mask open between 2.5 lambda/D and 9 lambda/D, with a 65 degree opening.**

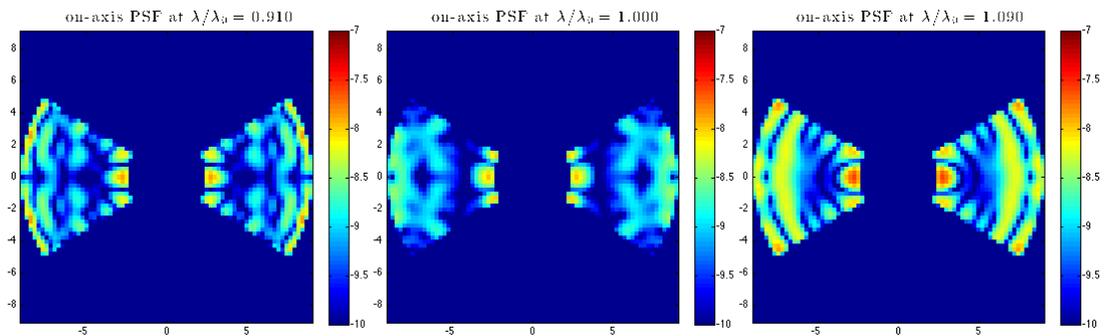

**Figure 2-46: Polychromatic simulation of the PSF corresponding to the mask shown in Figure 2-45, shown here at three wavelengths: 728 nm, 800 nm, and 872 nm (spanning an 18% bandwidth). This simulation includes a 2-DM wavefront control, but with no aberrations.**

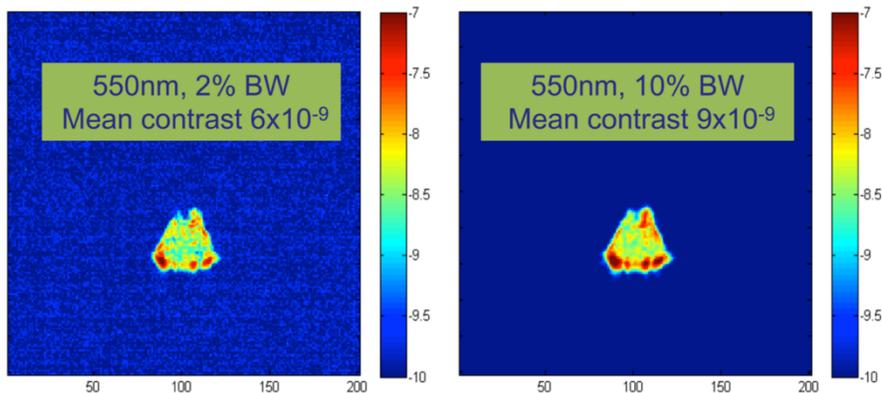

**Figure 2-47: Testbed measurements of the first-generation WFIRST-AFTA SPC design, demonstrating the achievement of two milestones: raw contrast below 10^-8 in a narrow band, and raw contrast below 10^-8 in a 10% band.**

| Planet name | Planet SMA (AU) | Planet mass (MJ) | Separation (arcsec) | Contrast (rel to star) | Integration time (days) |
|---|---|---|---|---|---|
| HD62509b | 1.69 | 2.9 | 0.1558 | 2.08E-08 | 0.0007 |
| HR8974b | 2.05 | 1.85 | 0.1389 | 1.39E-08 | 0.0128 |
| Ups And d | 2.55 | 10.19 | 0.1805 | 8.74E-09 | 0.0298 |
| 47 Uma b | 2.1 | 2.53 | 0.1427 | 1.34E-08 | 0.1093 |
| Ups And e | 5.25 | 1.059 | 0.3717 | 2.09E-09 | 0.1995 |
| HD192310c | 1.18 | 0.075 | 0.1265 | 2.35E-08 | 0.2092 |
| 47 Uma c | 3.6 | 0.54 | 0.2446 | 4.32E-09 | 0.2353 |
| HD176051b | 1.76 | 1.5 | 0.113 | 1.88E-08 | 0.4871 |
| HD114613b | 5.16 | 0.48 | 0.2384 | 2.08E-09 | 0.6197 |
| HD39091b | 3.28 | 10.3 | 0.171 | 5.28E-09 | 0.6321 |
| HD190360b | 3.92 | 1.502 | 0.2361 | 3.79E-09 | 0.8587 |
| HD160691e | 5.24 | 1.814 | 0.3227 | 2.13E-09 | 0.9988 |
| HD163917c | 6.1 | 27 | 0.1261 | 1.22E-09 | 1.4104 |
| HD10647b | 2.03 | 0.93 | 0.1112 | 1.40E-08 | 1.4249 |
| 14 Her b | 2.77 | 4.64 | 0.1506 | 7.83E-09 | 2.6133 |
| 55 Cnc d | 5.76 | 3.835 | 0.4458 | 1.81E-09 | 5.4068 |
| HD217107c | 5.27 | 2.49 | 0.2534 | 2.12E-09 | 5.6191 |
| HD142c | 6.8 | 5.3 | 0.2526 | 1.29E-09 | 6.2619 |
| HD154345b | 4.3 | 1 | 0.2209 | 3.12E-09 | 7.5659 |

**Table 2-7: WFIRST-AFTA detectable RV planets. Separation and contrast are given for a 60 degree phase angle. Integration times shown are to obtain SNR=5 photometry at 565 nm with 10% bandpass using Hybrid Lyot coronagraph, assuming albedos of 0.4 and 0.4 mas of post-control image jitter. See Appendix F for details.**

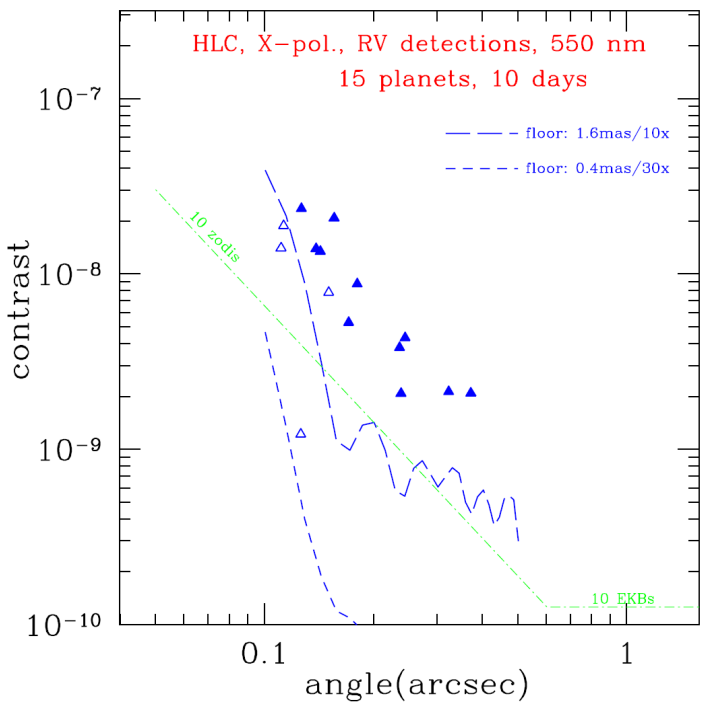

**Figure 2-48:** Contrast vs. radius plot showing known RV planets detectable with the WFIRST-AFTA Hybrid Lyot Coronagraph (HLC). Upper contrast curve is for worst-case performance modeling (1.8 mas residual image motion, x10 speckle attenuation from PSF subtraction) and lower curve is for the best-case (0.4 mas motion, x30 attenuation). Solid triangles represent detections of planets in less than one-day for the worst-case performance, open triangles are additional detections in the optimistic case. The dashed-green line is the estimated contrast of a target zodiacal dust cloud ten times that of the solar system, for a target at 10 pc.

WFIRST-AFTA could potentially discover these targets through blind observations of nearby stars, but such a survey will require significant amounts of telescope time. A sustained high-precision Doppler survey of nearby stars would both speed identification of potential targets, and provide mass constraints that will be crucial to interpreting WFIRST-AFTA spectra.

### 2.4.3.8  *Conclusions*

Although WFIRST-AFTA is by no means a coronagraphically optimized telescope, exoplanet imaging technology – coronagraphy, wavefront control, and post-processing – has advanced to the point where very high levels of performance appear practical with general-purpose telescopes. The baseline coronagraphic instrument will achieve detectable planet/star contrast ratios of at least $10^{-9}$ (after post-processing) at separations >0.2 arcseconds. This opens up enormous science possibilities, including photometric and spectroscopic characterization of a large sample of giant planets, mapping of structure in extrasolar zodiacal

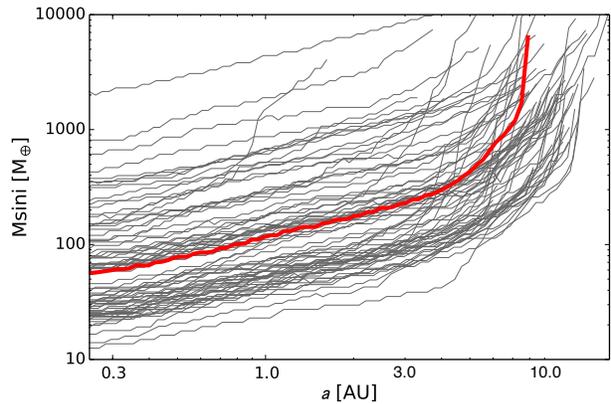

Completeness vs. Semi-major axis

**Figure 2-49: Completeness of Doppler surveys of 76 potential WFIRST-AFTA target stars, from Howard & Fulton (2014). The lines show the 50% completeness contour for each individual star, with the median completeness indicated by the thick line.**

disks, and potentially characterization of 'super-Earth' planets around the nearest stars.

| Science Program | Number of giant planets (4-15 $R_E$) with photometry | Number of sub-Neptune planets (2-4 $R_E$) with photometry | Number of super-Earth planets (1-2 $R_E$) with photometry | Number of planets (1-15 $R_E$) with R=70 spectral characterization |
|---|---|---|---|---|
| Known RV studies | 16 | 0 | 0 | 7 |
| 180-day new planet search | 2 | 6 | 4 | 9 |
| Total | 18 | 6 | 4 | 16 |

**Table 2-8: Summary of expected coronagraph science yields - number of RV planets that are potentially detectable and photometrically or spectrally characterizable and mean number of planets discovered in simulated searches for new planets. Not listed are the additional planets from future expanded RV searches. The numbers are uncertain, but may be comparable to the current known RV yield (see Figure F-6). The yield is evaluated for 0.4 mas residual image motion and x30 speckle post-processing attenuation. See Appendix F for details.**

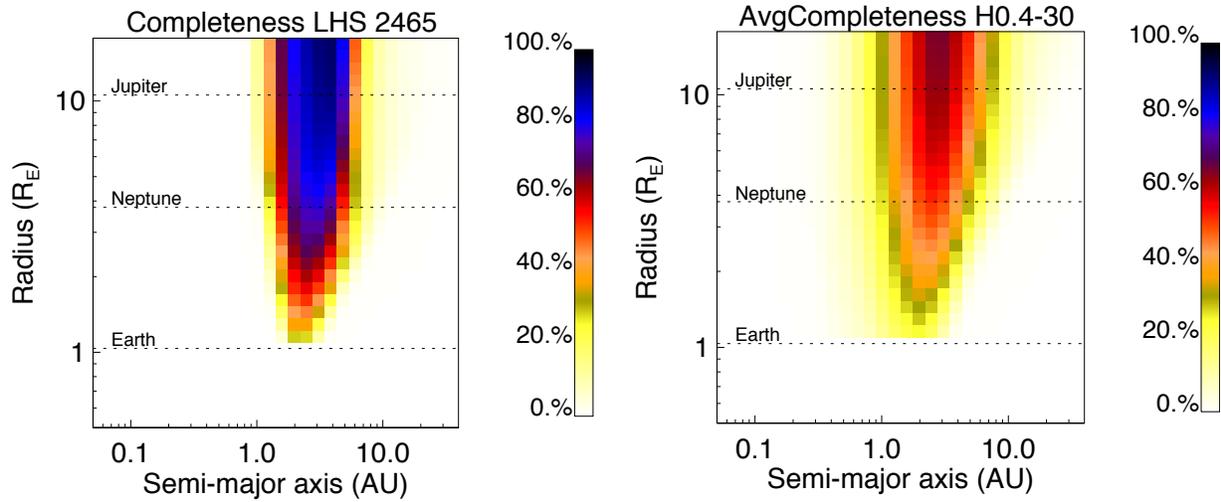

Figure 2-50: Single-visit completeness (probability of a planet being detected, if one is present, as a function of radius and semi-major-axis) for the WFIRST-AFTA HLC assuming 0.4 arcseconds of post-correction jitter and x30 speckle removal. Left: completeness for a single visit to a median star. Right: average completeness of a 6-month 46-star survey.

## 2.5 Galactic Bulge Field: General Astrophysics

The WFIRST-AFTA microlensing survey toward the Galactic bulge will potentially enable a very broad range of astrophysics, well beyond the primary goal of completing the census of cold exoplanetary systems. This enormous additional potential of the survey follows from the large number of stars being surveyed, the large number of measurements that will be made for each of these stars, and the quality of the photometric and astrometric precision that will be achieved for these stars, which will range from good to exquisite, depending on the brightness of the star and the level to which systematics can be controlled.

In this section, we provide a sampling of the range of the science that may be possible, including auxiliary science from the microlensing events themselves, stellar astrophysics, Galactic structure, and solar system science. In §2.4.2.9 we discussed the additional exoplanet science that can be obtained from the survey from the detection of tens of thousands of transiting giant planets.

There are a few things to note. First, this overview contains an admixture of potential applications, some of which have been worked out in some detail, and some of which have merely been suggested, with no real effort made to demonstrate their viability. These latter topics provide an important line of future work to vet their applicability. Second, we are primarily concentrating on auxiliary science that can potentially be extracted without changing the microlensing survey specifications. However, some of these applications will require new data reduction algorithms or precursor or follow-up observations in order to extract their full science potential. Finally, we note that several of these applications require excellent-to-exquisite control of systematics in the photometric and astrometric measurements. The extent to which these systematics can be controlled at the levels required is unclear. However, we note that it is likely that the microlensing survey itself will provide the best opportunity to characterize and remove these systematics by its very nature of requiring many repeated observations of a large number of point sources. We discuss this in a bit more detail below.

### 2.5.1 Properties of the Survey

We begin by summarizing the properties of the microlensing survey. As currently envisioned, the survey will consist of ten contiguous fields (see Figure 2-29) between Galactic longitudes of roughly -0.5 and 1.8, and Galactic latitudes of -1 and -2.2, for a total area of roughly 2.81 square degrees. The fields will be imaged in six 72-day campaigns spread over the nominal six-year mission lifetime, but due to losses from the moon moving through our field, the total survey time will be 357 days. During each campaign, each field will be observed every 15 minutes in a wide (W149) filter that spans 0.9-2 microns, and once every 12 hours in a bluer filter (Z087) that spans 0.77-0.97 microns. The exposure time per epoch in these filters with be 52 secs for W149 and 290 secs for Z087. Because of the layout of the detectors in the focal plane, ~85% of the survey footprint is observed for the full time while the remaining 15% is observed only in either the Spring or Fall seasons. For stars in the primary area, there will be a total of ~33,000 epochs in W149 and ~700 in Z087. The single-measurement photon-noise photometric and astrometric precisions will be ~0.06% and 0.05 mas for the one million stars with $H_{AB} < 14.0$, ~0.4% and 0.7 mas for the 12 million stars with $H_{AB} < 19.6$, and ~1.2% and 1.7 mas for the 56 million stars with $H_{AB} < 21.6$. WFIRST-AFTA will collect ~2.5 billion photons for a star with $H_{AB} = 19.6$ in the wide filter over the course of the mission.

### 2.5.2 Microlensing Auxiliary Science

The microlensing survey will be sensitive to microlensing due to isolated compact objects with masses from that of Mars to roughly 30 times the mass of the Sun. We expect 37,000 microlensing events due to known populations alone (stars and stellar remnants) (Penny et al. 2015a). Thus it will be able to measure or constrain the mass function of compact objects over roughly 8 orders of magnitude in mass. This includes free-floating planets, brown dwarfs, stars, and stellar remnants. For the upper end of this range, it will be possible to measure the masses of the objects directly from higher-order microlensing effects from the survey data itself (Gould & Yee 2014), whereas for lower-mass objects, either additional measurements may be required, or the mass function will only be determined statistically. It will also be possible to probe the binarity of these compact objects, including the companion frequency and mass ratio and separation distribution, for binaries with separations within roughly a decade of the Einstein ring radius of the primary. It will also be possible to detect tertiaries in some cases.

With the large number of microlensing events, it will be possible to measure the microlensing optical depth and event rate over the ~3 square degrees of the target fields. The optical depth, in particular, is proportional to the integral of the total mass density of compact objects along the line of sight, and as such is a very sensitive probe of the mass distribution of the Galactic disk and the Galactic bulge (Griest et al. 1991, Kiraga & Paczynski 1994). In particular, the optical depth can be used to probe the total mass, shape, and orientation of the Galactic bar (Paczynski et al. 1994, Zhao et al. 1995). The event rate is additionally sensitive to the velocity distribution and mass function of the compact objects in the stellar populations.

Roughly 5% of all microlensing events exhibit clear signatures of caustic crossings due to the lens being composed of a nearly equal mass binary with a projected separation near the Einstein ring radius (Alcock et al. 2000). With the near-continuous sampling and 15 minute cadence, nearly all of the caustic crossings of giant sources (which last ~10 hours) will be well-resolved, thus allowing one to measure the limb-darkening for these stars in the wide filter (Albrow et al. 1999).

### 2.5.3 Stellar Astrophysics

With a typical total exposure time of ~20 days per field in the wide (W149) filter and ~2.3 days per field in the blue (Z087), it will be possible to construct a very deep color-magnitude diagram over the ~3 square degrees targeted by the microlensing survey. The limiting magnitude in both filters is uncertain, but will likely be set by confusion, rather than photon noise. Nevertheless, it should be possible to measure the luminosity function of stars in the bulge down to nearly the bottom of the main sequence, as well as identify unusual stellar populations down to quite faint magnitudes, including potentially young stars, blue stragglers, stellar clusters, cataclysmic variables, and white dwarfs. Such studies will be enhanced by the fact that it will be possible to measure the proper motion and potentially parallax for a significant subset of the sources in the bulge field, thus allowing for discrimination between bulge and disk populations. Such science would be significantly enhanced by observations in one or more bluer filters at similar angular resolution, either with WFIRST-AFTA itself, or with Hubble Space Telescope precursor observations.

Of course, it will be possible to identify many variable sources in the fields, over a broad range of amplitudes and time scales. Variables can be identified with amplitudes greater than a few millimagnitudes, with time scales from tens of minutes up to the ~6-year duration of the mission. This will allow for the identification of a large number of stellar flares, eruptive variables, and pulsating variables, thereby enabling a wide variety of stellar astrophysics.

Perhaps most exciting, however, is the potential for WFIRST-AFTA to do asteroseismology on, and thus determine the mass and radii of, a significant number of stars in the survey area. Gould et al. (2015) estimate that WFIRST-AFTA will obtain asteroseismic information on the roughly one million giant stars with $H_{AB} <$ 14 in the target fields (see Figure 2-51). They also estimate that WFIRST-AFTA will obtain precise (~0.3%) distances to these stars via parallax. This is the only way to measure masses and radii of a large number of bulge giants, which are postulated to have abundances that are substantially different than local populations.

### 2.5.4 Galactic Structure

WFIRST-AFTA will obtain photon-noise limited parallaxes of <10% and proper motion measurements of <0.3% (0.01 mas/yr) for the ~56 million bulge and disk stars with $H_{AB}$<21.6 in its field of view. When com-bined with multicolor photometry in at least one bluer filter (to complement the measurements in W149 and Z087), it will be possible to estimate the effective temperature, metallicity age, luminosity, and foreground extinction for all of these stars. This is a larger number of stars with such measurements than will be achievable with Gaia, although of course the population of stars will be very different and complementary to that obtained by Gaia. From this dataset, it will be possible to extract an enormous amount of Galactic structure science, including a determination of the bulge mass and velocity distribution (including bar structure), the stellar density and velocity distribution of the Galactic disk, the metallicity and age distribution of the disk and bulge, and the three-dimensional distribution of dust along the line of sight toward the bulge fields.

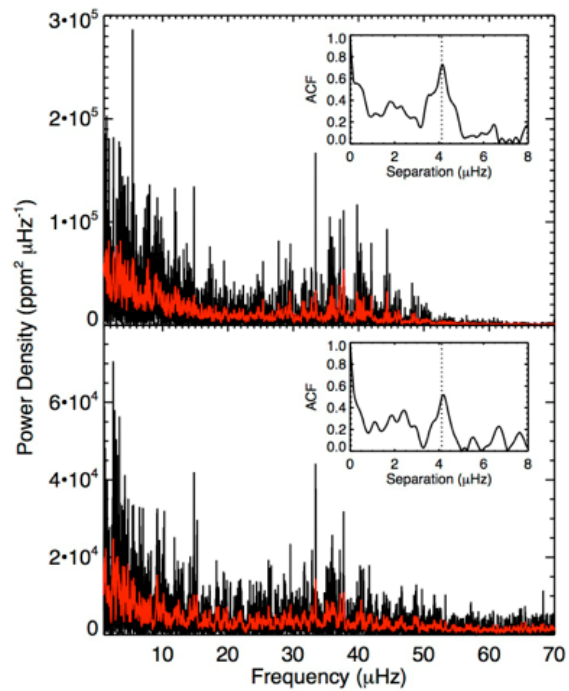

**Figure 2-51:** Asteroseimologic spectrum for KIC2836038 as observed by Kepler (top panel) and as simulated for WFIRST-AFTA (bottom panel). Red lines are the power spectra smoothed with a boxcar width of 1μHz. The insets show the autocorrelation function of the smoothed power spectrum between 20 − 60μHz after correcting the granulation background. The dotted line marks the published value for the large frequency separation Δυ. WFIRST-AFTA will be able to measure Δυ for the roughly one million giant stars with $H_{AB}$<14 in the microlensing survey fields. From Gould et al. (2015).

### 2.5.5 Solar System Science

Gould (2014b) estimates that WFIRST-AFTA will detect ~5000 Trans-Neptuninan objects (TNOs) down to absolute magnitudes of $H_{AB}$~29.6 (corresponding to diameters of D~10 km) over ~17 square degrees, enabling a precise determination of the size distribution of TNOs down to, and substantially below, the collisional break at ~100 km (see Figure 2-52). Furthermore, the orbital elements of these objects will be measured to fractional precisions of a few percent, allowing for dynamical classification into the canonical classical, resonant, and scattered populations, as well as identification of objects with new or unusual orbits. Binary companions to these objects will be discovered down to fainter magnitudes of $H_{AB}$~30.4, corresponding to diameters of 7 km, thus allowing for the determination of the binary size and separation distribution as a function of dynamical class. Finally, there will be ~1000 occultations of stars due to these TNOs during the bulge survey (Figure 2-52), allowing for a statistical estimate of the size and albedo distributions of TNOs as a function of magnitude and dynamical class. The several dozen TNOs discovered with $H_{AB}$<25 (D>100 km) will occult multiple stars during the survey, allowing for crude constraints on the shapes of these large TNOs.

### 2.5.6 Related Projects and Precursor Observations

The Galactic bulge survey of WFIRST-AFTA is unique in terms of its resolution, area, exposure time, total number of observations, and total number of sources. Therefore, it is essentially guaranteed to produce revolutionary science in all of the areas listed above. Nevertheless, there are a number of existing, planned, or proposed surveys with similar science goals, which will motivate, enhance, and/or complement the science potential of the WFIRST-AFTA bulge survey. These include the ground-based near-infrared: Vista Variables in the Via Lactea (VVV) Survey (Minniti et al. 2010), and optical: Blanco DECam Bulge Survey (BDBS), the Hubble Space Telescope Galactic Bulge Treasury Program (Brown et al. 2009) and Sagittarius Window Eclipsing Extrasolar Planet Search (SWEEPS) (Sahu et al. 2006), and the proposed Japan Astrometry Satellite Mission for Infrared Exploration (JASMINE) (Gouda et al. 2005). The science goals of the JASMINE mission concepts, in particular, have significant overlap with the auxiliary science topics listed above[9].

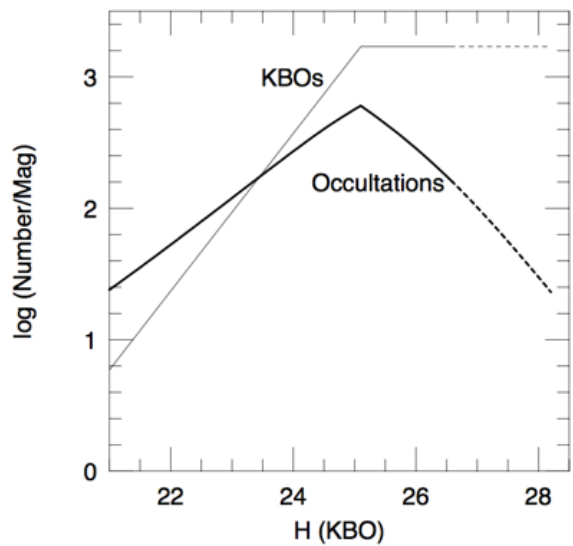

**Figure 2-52: The expected number of Trans-Neptunian Object (or KBO) detections (light solid line) and TNO occultations (heavy solid line) per magnitude that can be detected from the WFIRST-AFTA microlensing survey. Magnitudes here are in the Vega system, which is related to the AB system used in the text by $H_{Vega} = H_{AB} - 1.39$. WFIRST-AFTA will detect TNOs significantly fainter the current survey limits of $H_{Vega} = 26$. KBOs with $H_{Vega} < 23.5$ will occult more than one star during the survey, potentially allowing for crude constraints on their shape. From Gould (2014b).**

As discussed in Appendix G and advocated in the recent NASA ExoPlanet Analysis Group (ExoPAG) Study Analysis Group 11 report, precursor observations with HST of some or all of the area targeted by the WFIRST-AFTA bulge survey would significantly enhance many of the auxiliary science goals discussed above. HST imaging is well matched to that of WFIRST-AFTA in terms of resolution and sampling, while providing both bluer filters and the possibility of obtaining a longer time baseline of observations than is possible with WFIRST-AFTA alone. Precursor observations with HST could then be combined with the WFIRST-AFTA data to provide color-magnitude diagrams of the survey area in several colors, confirmation of low signal-to-noise ratio astrometric and proper motion measurements made internal to the WFIRST-AFTA bulge survey, and to provide locations and colors of all the point sources with higher resolution and fidelity than is possible with WFIRST-AFTA alone.

---

[9] http://www.scholarpedia.org/article/JASMINE

### 2.5.7  *Detector Characterization*

With its ~33,000 dithered images of ~$10^8$ point sources with known colors and nearly constant fluxes, the bulge survey will likely provide one of the best datasets with which to characterize WFIRST-AFTA's wide-field imager. In particular, we imagine that the bulge survey pipeline will measure the fluxes and positions of all of the stars in the field of view, while simultaneously measuring the response function of each of the WFI pixels, and how this response function varies with time. This time-variable response function will then be used to calibrate and remove detector artifacts such as persistence, inter-pixel capacitance, reciprocity failure, non-linearity, and intrapixel response variations.

## 2.6 Opportunities for the Guest Observer Program

The guest observer (GO) science program of WFIRST-AFTA will have high impact over a broad range of modern astrophysics. Taking Hubble as a guide, the GO program of WFIRST-AFTA will be at least as important to the astronomical community and the interested public as the dark energy, microlensing planet, and high contrast imaging components. The very wide field, large aperture, deep near-IR reach, and excellent spatial resolution of the observatory will ensure that a multitude of GOs pursue numerous programs requiring degree-sized fields in the statistical realm for the first time. Examples include studies of stellar populations in the Milky Way and its neighboring galaxies and mapping large-scale structure through cosmic time. Many of these will be logical outgrowths of single object and small sample studies with HST, JWST, and other observatories, but they are likely to have outsized impacts just as SDSS, Kepler, and other wide field surveys already have. GO science programs can also make full use of grism spectroscopy, IFU spectroscopy, and coronagraphic imaging on WFIRST-AFTA. For examples, the COPAG recently compiled a set of exciting cosmic origin science programs with the WFIRST-AFTA coronagraph.
(see http://cor.gsfc.nasa.gov/copag)

Some capabilities of WFIRST-AFTA are unrivaled. For the foreseeable future, atmospheric seeing ensures that no ground-based facility will produce comparably sharp ultra-wide field images that compare to WFIRST-AFTA. Likewise, the earth's atmosphere is opaque between the near-IR atmospheric J, H, and F184 bands, and even where it is not opaque, OH airglow ensures that the background will be large and highly time-variable compared to space. Compared to other space-based facilities, WFIRST-AFTA has 288x as many pixels as Hubble/WFC-3, and 210x JWST/NIRCam's FoV. Compared to Euclid, WFIRST-AFTA has approximately 4x the collecting area and about 2.5x the near-IR angular resolution (>6 improvement in PSF area). Euclid's dark energy surveys will certainly be useful for numerous other astrophysics investigations, but at this point Euclid is not planning to have a dedicated program for conducting GO pointed observations (Laureijs et al. 2011).

WFIRST-AFTA will be complementary to these other observatories, and their combined data will be invaluable for attacking numerous problems in modern astrophysics. In particular, the wide field and moderately deep reach of WFIRST-AFTA will be highly complementary to the narrow field, higher resolution, and deeper reach of JWST in their overlapping near-IR wavelength regions (see §2.7.1 and Appendix B). Likewise, WFIRST-AFTA will provide complementary wavelength coverage with much better spatial resolution and depth to LSST while still being able to sample wide areas of sky (see Appendix H). For many scientific investigations, WFIRST-AFTA's impressive imaging capabilities also provide natural synergies with powerful spectrographs from the next generation 30-meter telescopes (examples of TMT and GMT synergies are included in Appendix J)

Compared to previous, smaller-aperture concepts of WFIRST, the significantly increased sensitivity, and higher spatial resolution over a similar field of WFIRST-AFTA will substantially increase the potential and value of its GO program. This is true for both dedicated GO observations and for guest investigator (GI) use of data sets produced in the high-latitude, supernovae, and microlensing surveys. As for nearly all other missions, the astronomical community will be quite creative and will use WFIRST-AFTA for GO programs that the mission SDTs and SWGs will not conceive. If history is a guide, then these may prove to be among the highest impact programs executed with the telescope, rivaling or surpassing the results of the mission's primary dark energy and microlensing surveys. To illustrate this, the greater community has contributed ~40 GO ideas on a multitude of science themes. These 1 page descriptions span a very broad range of topics including the solar system, exoplanets, stellar astrophysics, nearby galaxies, extragalactic astrophysics, and the complementarity / synergy of WFIRST-AFTA and other missions and surveys. These papers clearly illustrate that WFIRST-AFTA will truly be a 'Great Observatory' for the Twenty-first Century. These papers are included as an appendix to this report.

We now highlight some illustrative GO observations that would take advantage of WFIRST-AFTA's strengths. These have been culled from the much more detailed GO examples presented in the earlier WFIRST SDT report (Spergel et al. 2013) and the Princeton Workshop white paper (Dressler et al. 2012), and many of these themes and programs are related to the community 1 page GO submissions. Because of the enhanced capability for a broad class of GO programs due to the larger aperture and the loss of capability for long wavelength (> 2.0 micron) studies, we have reallocate the galactic plane survey time to the GO program for the WFIRST-AFTA DRM.

Statistical studies of the stellar populations of the Milky Way and nearby galaxies may be one of the most obvious areas of WFIRST-AFTA GO science that will have outsized impacts. However, the higher resolution and speed of WFIRST-AFTA suggests a huge opportunity in deep studies of specific regions of the Milky Way disk. A prime example is the nature and origin of the initial mass function (IMF), particularly its extension to very high and very low masses. Surveys of the numerous low- to intermediate mass star forming regions within 1 kpc are still woefully incomplete. Ground-based near-IR surveys are either spatially complete with completely inadequate sensitivity (2MASS), or else they provide somewhat better but still inadequate sensitivity and resolution (e.g., UKIDSS now and VISTA in the future) over inadequate fields. WFIRST-AFTA surveys of the ~10 best studied young embedded clusters within 1 kpc to ~25 mag AB would reveal young brown dwarfs as well as some of the youngest Class 0 protostars in these nurseries. Combining WFIRST-AFTA and Spitzer mid-IR survey data will leverage these identifications and reveal much about the accretion properties of these stars early in their lives. Surveying more distant (D ~ 2–5 kpc) massive star forming regions will constrain high-mass star formation (numbers, luminosities) with only a single or a few deep WFIRST-AFTA fields in each near-IR broadband filter.

Closest to home, the wide field and excellent near-IR sensitivity will allow for sensitive surveys of both nearby field and open cluster brown dwarfs. A puzzling discovery from WISE surveys is a severe deficiency of brown dwarfs within the local volume, D < 8 pc: 33 brown dwarfs versus 211 normal stars is far less than typically predicted ratio of ~ 1:1 (Kirkpatrick et al. 2012). Cataloging the substellar populations of nearby young clusters with WFIRST-AFTA will reveal the brown dwarf fraction with age and location, giving insight to their dispersal into the field. Broadband 1 – 2 micron photometry with WFIRST-AFTA will locate brown dwarfs in color – color and color-magnitude diagrams (CMDs); these diagrams will also shed light on the mystery of why the near-IR colors and spectra of field brown dwarfs and extrasolar planets differ. Field brown dwarfs have colors and spectra consistent with a clearing of clouds as effective temperatures decrease to ~900 K (the L-T spectral type transition), but cool giant exoplanets, like those around HR 8799, seem to keep their clouds at these low temperatures, perhaps due to their lower surface gravities (e.g., Marley et al. 2012). Surveying the brown dwarfs in a number of different young open clusters will revel the precise age-dependence of this effect. Pro-gress in understanding the relations between brown dwarfs and giant planets will be crucial to interpreting WFIRST-AFTA's own coronagraphic exoplanet observations.

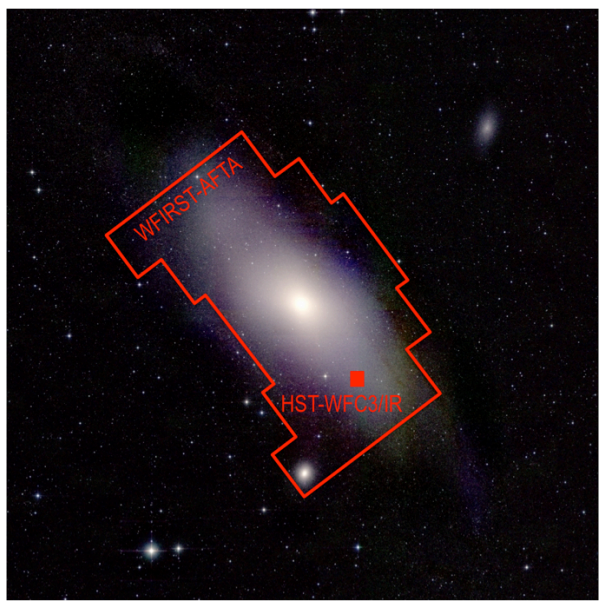

**Figure 2-53: 2MASS near-IR image of the Andromeda Galaxy M31. The dust lanes disappear at these wavelengths, revealing embedded and obscured stellar populations. The full field of the image is 1.4 x 1.4 degrees. The WFIRST-AFTA FoV is indicated by the red outline, and the field of the HST WF3 IR instrument is shown by the filled red rectangle. Both instruments have similar spatial resolutions and sensitivities, but WFIRST provides ~200 times the field of view of HST WFC3/IR.**

Continuing out from the Galactic disk to its stellar halo, deep maps of the Milky Way halo will reveal the distribution and ages of the oldest stars in the Galaxy as well as its dwarf satellites (see Kalirai 2012, Kalirai et al. 2012). WFIRST-AFTA will also be able to resolve individual stars in nearby galaxies and better separate stars from distant nearly point-like galaxies, providing accurate studies of their stellar populations and star formation histories for the first time. Near-IR performance is an important feature for these studies; the ability of near-IR light to penetrate dust lanes (Figure 2-53 of M31) and unique age-sensitive features in near-IR CMDs are key for these investigations. More details appear in §5.4.1 of the Dressler et al. (2012) Princeton whitepaper, Kalirai et al. (2012a, 2012b), and the Belsa & van der Marel, Geha, and van der Marel & Kalirai 1 page GO papers.

Many of the ~450 galaxies within 12 Mpc (Karachentsev et al. 2004) have been studied with Spitzer, GALEX, and other missions / surveys, but few have been observed at high spatial resolution.

WFIRST-AFTA images of these galaxies would provide CMDs that allow discerning their stellar populations and determining their ages. This would greatly leverage the existing data as well, providing a good look at the makeup of our local volume. Similarly, WFIRST-AFTA imaging data would probe the population and dynamical history of the core of the Virgo. These and related ideas appear in §5.4.2 of Dressler et al. (2012) Princeton whitepaper as well as in the Dalcanton & Laine 1 page GO contribution.

The HLS will discover numerous galaxy clusters that can be followed up by WFIRST-AFTA itself with deeper fields. Likewise, medium – deep WFIRST-AFTA observations of LSST deep drilling fields will probe structure formation and lensing discoveries will test $\Lambda$CDM models. For example, WFIRST-AFTA could carry out a deep galaxy census covering 100 times the area of the famous Hubble Ultra Deep fields. Numerous AGN will be detected in these and other surveys, and these data will allow for some study of their co-evolution with their host galaxies. Coronagraphic WFIRST-AFTA observations will allow more detailed study of individual AGN host galaxy morphologies and halos. These studies are presented in more detail in §2.6.3 and 2.6.5 of the Green et al. (2012) SDT report, §5.5 of the Dressler et al. (2012) Princeton whitepaper, and the Conselice, Merton & Rhodes, and Donahue 1 page GO papers.

WFIRST-AFTA data could also address the Cosmic Dawn theme of the NWNH Decadal Survey by characterizing the epoch, speed, and shape of cosmological reionization and perhaps characterizing reionization sources. This would be done with the luminosity function of high redshift quasars (mostly from the HLS), the high redshift galaxy luminosity function (Lyman-break galaxies), and Ly $\alpha$ emitters. The details of this exploration are presented in §5.6 of the Dressler et al. (2012) Princeton whitepaper, and related ideas appear in the Fan and Teplitz 1 page GO contributions.

WFIRST-AFTA will complement JWST in its quest for "first light" objects by searching for primordial stars too faint to be recognized as individuals by any telescope. The majority of the first light quasars and stars will not be seen individually, but their combined radiation may be detectable in the form of the cosmic infrared background (CIB). The source-subtracted CIB Spitzer/IRAC maps show no correlations with the visible HST/ACS sources down to AB mag >~28 and the signal has now been measured to ~ 1 degree scale (Kashlinsky et al. 2012). The amplitude and shape of this rise is a direct measurement of the clustering properties of the sources responsible for the fluctuations, thus a pri-mary key to understanding the nature of these sources. WFIRST-AFTA HLS data will reveal the power spectrum of the CIB at sub-degree scales and help determine whether it is dominated by radiation from the first stars (Kashlinsky et al. 2007), lower redshift stripped halo stars (Cooray et al. 2012), or other sources.

We list a number of these and other potential GO programs in Table 2-9. We intend this list to be a representative but by no means complete list of GO program areas that will likely take good advantage of the observatory's capabilities and have incredible scientific impact in the 2020 decade. WFIRST-AFTA will also be capable of conducting a reasonable number of these programs given that it will have ~1.5 years dedicated to its GO program.

| Program | Each pointing (sq deg) | # Targets | Comment |
|---|---|---|---|
| Embedded star formation < 1 kpc (Tau, Oph, Ser, Per, Ori) | 1 – 30 | ~10 | IMF: substellar to intermediate masses; calibrating models |
| 10-100 Myr open clusters (Pleiades, NGC clusters) | ~1 | ~10 | Mass Function to substellar masses, evolution, dynamics, brown dwarfs |
| Dense embedded star formation > 1 kpc (NGC 7538, W3, Gal Ctr) | ~ 0.25 (1 field) | ~15 | IMF: high mass end |
| Globular Clusters | 0.25 | 12 | Calibrate IR CMD with metallicity |
| Stellar populations of Galactic bulge, halo, satellites | HLS microlensing | HLS microlensing | Substructure, tidal streams, AGB ages from HLS and microlensing data |
| Ultra-faint MW dwarf galaxies including S/LMC | 0.25 – 4 | >10 | Star formation histories, metal-poor IMF, kinematics of LSST discoveries |
| Map local group galaxies (M31, M33, dwarf ellipticals) | ~2 - 20 | ~6 | Resolve disk structure, extinction maps, cluster dissolution to field |
| Mapping nearby galaxy thick disks and halos beyond local group | ~1 | > 24 | Halo substructure, population gradients, test dark matter, kinematics |
| Mapping the core of Virgo cluster | ~100 | 1 | Morphologies, luminosity function, superstar clusters, intracluster objects |
| AGN host galaxies (coronagraph) | 1E-6 | > 10 | Stellar populations, comparison to normal galaxies |
| Galaxy cluster (HLS) followup | 0.25 | > 24 | High z lensed objects, structure growth, interactions & dark matter |
| LSST Deep Drilling Fields | 0.25 | 1-4 | Galaxy Luminosity Function out to reionization epoch (z from dropouts) |
| QSOs as probes of cosmic dawn | HLS | HLS | Epoch, speed, patchiness of reionization using QSO spectra |

**Table 2-9: Examples of GO programs.**

## 2.7    Relation to Other Observatories

WFIRST-AFTA observations will be synergistic with other major astronomical facilities planned for the next decade. The WFIRST-AFTA science program has many unique aspects that will enhance the science return from upcoming space missions including JWST and Euclid, wide imaging and spectroscopic surveys being undertaken by Subaru, and ambitious upcoming ground based facilities including LSST and multiple 30-meter class telescopes.

In this section, we first briefly describe some of the general synergies between WFIRST-AFTA and JWST (§2.7.1). This includes science investigations that both take advantage of the WFIRST-AFTA HLS data and ones that would involve possible GO programs with various JWST and WFIRST-AFTA instruments. We also refer the reader to Appendix B of the WFIRST-2.4 2013 report for a broader discussion of WFIRST-JWST synergies. In the following subsections, §2.7.2 to 2.7.5, we briefly describe how LSST (§2.7.2), Euclid (§2.7.3), the giant segmented mirror telescopes (§2.7.4), and wide-field imaging and spectroscopy with Subaru (§2.7.5) all help aid in the primary dark energy science goals of WFIRST. These missions also provide tremendous synergies with WFIRST-AFTA in other areas of broad astrophysics, a few of which are captured in other areas of this report (e.g., Appendix H for LSST and Euclid; §2.3.5 for LSST and Deep Drilling Fields; Appendix J for TMT, GMT, and Subaru).

### 2.7.1    JWST

WFIRST-AFTA has its foremost synergy with the James Webb Space Telescope, the descendent of the Hubble Space Telescope and the premier space science facility for the 2020's. This synergy is a traditional one, pioneered by the construction seventy years ago at Palomar Observatory of the 48-inch wide-field *Schmidt Telescope,* designed to assist, complement and augment the science potential of the giant *Hale 200-inch Telescope*. A more recent example is the 2.5-m Sloan Digital Sky Survey (SDSS), which has surveyed vast areas of the sky, finding special and rare targets for today's largest and most powerful observatories including the Keck 10-m, Magellan, and VLT telescopes. These facilities provide the high spectral and spatial resolution, coupled with extreme sensitivity, that is essential for detailed study of faint and often complex sources. However, telescopes with these critical capabilities are necessarily limited to relatively small fields-of-view. For this reason, powerful large-aperture telescopes have been paired with smaller, wide-field telescopes that can survey very large areas of the sky to find the high value and/or rare targets for the larger telescope.

Just as importantly, wide-field surveys open new science frontiers. For example, the serious study of clusters of galaxies, which led to the foundational field of galaxy evolution, began with the Palomar Sky Survey by the Schmidt Telescope: such programs nourished the Hale telescope for a half-century. SDSS's measurements of large-scale structure in maps of galaxies and quasars have played a central enabling role in establishing today's consensus cosmological model. This dual capability --- providing the most important targets in a well-established class, and surveying for new phenomena --- enhances the value of the larger telescope and helps both kinds of facilities realize their full scientific potential.

We see both synergies in the pairing of WFIRST-AFTA with JWST. Appendix B, which was largely compiled from community input by interested astronomers, describes several exemplary studies that depend upon the science capabilities of WFIRST-AFTA and JWST. Perhaps the foremost example in JWST's prime mission is to study "first light," the appearance of the first stars, galaxies, and quasars in a previously dark universe. While the sky may be crowded with such inhabitants of the first-billion-years of cosmic history, comparatively rare are those bright enough to study, and to use as probes of the intergalactic medium – the womb of this cosmic birth. First-generation galaxies are the likely drivers of the reionization of the universe at redshifts $6 < z < 12$. Detailed study of the interactions between primordial gas and galaxies will be a major goal for JWST. Assembling sufficiently large samples of these early galaxies using JWST's own comparatively small-field cameras would be next-to-impossible. On the other hand, WFIRST-AFTA -- with a field-of-view 100 times greater -- will provide plentiful samples of embryonic galaxies and supermassive black holes that JWST can dissect through deeper imaging and pan-chromatic spectroscopy at near-to-mid-infrared wavelengths. A particularly powerful approach to studying the more numerous but fainter objects would be to use clusters of galaxies discovered in the HLS as gravitational telescopes that amplify the light (and magnify the size) of very distant galaxies.

JWST has powerful capabilities for studying high-redshift supernovae, possibly even those of *Population III* stars (the true `first stars,' lacking heavy elements). Again, the field-of-view of JWST is far too small to survey for such rare objects efficiently, so they must be discovered with other facilities. WFIRST-AFTA is unquestionably the best telescope to find such high redshift (z > 2) supernovae, because of its ability to conduct deep searches in the near-IR over very wide fields. JWST's superior IR spectrographs can make detailed characterizations of these supernovae, both for studying the first generations of stars in the early universe, and for cosmological studies to improve our understanding of supernovae as 'standard candles' at redshifts z < 1.5 and to perhaps extend such measurements to higher redshift.

Star formation in the Milky Way and its nearest neighbors, including the Magellanic Clouds, our Galaxy's dwarf companions, M33, and M31 (Andromeda), will be prime targets for joint study with WFIRST-AFTA and JWST. WFIRST-AFTA will be the first facility capable of deep near-IR observations, with high spatial resolution, over fields of many square degrees. Star birth in places like the Taurus Star Forming Cloud ranges across a wide area, covering a vast collection of young stellar objects in different phases of formation. The energy release from these young stars is powerful enough to drive winds through the birth cloud, stripping some young stars to their demise, and at the same time probably inducing the birth of other new stars. WFIRST-AFTA will provide a roadmap of the most interesting destinations in many star-forming clouds and young star clusters, in the Milky Way and its neighbors. However, understanding the dynamic processes underway will require detailed spectroscopy from the near-to-mid infrared, a prime capability of JWST.

WFIRST-AFTA panoramas of the youngest star forming regions will pinpoint the locations of young stars with accretion or debris disks. These will provide JWST with an abundance of targets for mid-IR spectroscopy that will inform our knowledge of how exoplanets are born. Many of these systems will be at least partly obscured by dense clouds of gas and dust. But, the unprecedented sensitivity and resolution of JWST's near-IR and mid-IR instrument will easily slice through the gas and dust with angular resolution comparable to WFIRST-AFTA's. WFIRST-AFTA and JWST will provide far more comprehensive insight into star and planet formation processes than either observatory alone, by collecting panchromatic images and spectra across a range of masses, ages, and environments.

Working together, WFIRST-AFTA and JWST will provide a tremendous opportunity for exploring how galaxies formed in the early universe by studying the stellar populations of nearby, *present-day* galaxies. Deep, wide-field imaging will uncover the archaeological history of how these galaxies were built in previous generations of star formation. For example, streams of stars have been found in the stellar halo of our Milky Way and its neighbor Andromeda that record the accretion and tidal disruption of satellite galaxies in the distant past, a primary process in the building of galaxy (See Appendix B "Galactic streams in nearby galaxies"). This is an excellent example of synergy among facilities: wide-field maps from SDSS have been used to find such structures, color-magnitude diagrams with HST have been used to obtain the histories of star formation, and large aperture telescopes such as Keck, Magellan, and VLT have made spectroscopic measurements for the kinematics and abundances of the stellar populations.

There are dozens of other more-distant neighbor galaxies, each with a unique history, which could be studied in this way. However, only WFIRST-AFTA has the sensitivity and wide field-of-view to make such studies practical. Following up with JWST for deeper imaging and spectroscopy of selected regions in the disks and halos of these galaxies will paint a more complete picture of their construction history, including the history of the buildup of heavy chemical elements over cosmic time. Likewise, globular clusters – richly populated star clusters that record a unique episode of star formation with a certain abundance – are valuable tools for deciphering the path of galaxy evolution (see Appendix B "Globular clusters in nearby galaxies"). WFIRST-AFTA's wide field will provide huge samples of these objects in many nearby galaxies, providing targets for deeper imaging with JWST to determine the age and heavy element abundance of each star cluster. Deep imaging of the dwarf galaxy companions of large neighbor galaxies will also help us understand how galaxies of very different masses and environments evolved over cosmic time.

Perhaps the richest form of synergy is when many facilities, each using its unique capability, combine to open and explore a new field of astrophysics. There is no better example than the search for and study of exoplanets. Planetary transit detections from Kepler, and in the future made by TESS, will provide a wealth of targets for JWST spectroscopy. This is an example of a synergy between Kepler, JWST, and TESS. With regard to WFIRST-AFTA, ground-based Doppler searches have provided a sample of nearby stars for *direct* imaging and spectroscopy with WFIRST-AFTA's coronagraph. Taken collectively, studies like these will greatly increase our knowledge of the orbits, mass, density, and atmospheres of both terrestrial and giant planets.

In addition, the demographic study of planets within ~1 AU of their host stars by Kepler, when combined with the WFIRST-AFTA microlensing survey that is sensitive to planets more than 1 AU from their stars, will lead to great progress in the theory of planet formation. Size and mass distributions for planets, when considered together with locations and orbits in stellar systems, will for the first time allow tests of theories of planet formation and migration as planetary systems evolve. WFIRST-AFTA's coronagraph, JWST's spectrographs, and the imaging cameras of both, building on the data from Kepler and the future TESS mission, will provide a multi-facility synergy that will establish exoplanet research as a leading field of 21st century astronomy.

### 2.7.2 Euclid

While WFIRST-AFTA produces deep images of the sky at infrared wavelengths, Euclid will produce deep images of the sky at optical wavelengths (see Figure 2-10). Euclid performs optical imaging to a magnitude of ~24.5 over 15,000 square degrees in a single very wide optical filter. This is complemented by NIR imaging in 3 filters to the same depth. Thus, Euclid will perform a survey that is on par with LSST in area but at significantly higher angular resolution and with the stable PSF afforded by space. WFIRST-AFTA's larger mirror just about balances the longer wavelength (IR vs. optical) to give the two observatories comparable angular resolution so their combined galaxy images will provide a more complete picture of galaxy properties. There are two primary areas of overlapping science between Euclid and WFIRST, but different strategies to accomplish those goals. Both Euclid and WFIRST aim to provide new insights into the nature of the dark energy via imaging and spectroscopy. Euclid uses a combination of NIR spectroscopic galaxy clustering (BAO and RSD) measurements and gravitational lensing shape measurements in a single optical band to study dark energy. WFIRST also uses these techniques. Euclid performs its wide survey in a 'single string' fashion, covering as much survey area as possible in a given amount of time. Euclid's survey is shallower and wider than that of WFIRST; hence the missions are highly complementary. While WFIRST covers less area than Euclid, its spectroscopic redshift survey has an order of magnitude higher density of galaxies. Its weak lensing survey measures a 2.5x higher density of galaxies at higher redshifts, and will observe in multiple filters in multiples passes over the sky to maximize the robustness of its results. WFIRST's philosophy differs from Euclid in that WFIRST seeks to make multiple redundant measurements to allow a variety of self-consistency checks and cross checks in both the galaxy clustering and weak lensing measurements. WFIRST will also use supernovae as distance indicators. This technique complements BAO in multiple ways, including unequaled precision in the late-time universe when dark energy becomes dominant. If launched early in the next decade, the 6-year prime mission of WFIRST will be contemporaneous with the 6.25-year Euclid prime mission, so the two projects will carry out their major dark energy analyses largely independently (at first) and be able to cross-check findings in detail. Later, more robust dark energy measurements can be attained by simultaneous processing of WFIRST, Euclid, and LSST data (see Appendix H). Per unit time, WFIRST is a very powerful "dark energy machine". Thus, if important discoveries about dark energy or modified gravity emerge from Euclid or other WFIRST precursors, they can be well characterized in WFIRST extended-mission observations beyond the 6-year primary mission.

### 2.7.3 LSST

The Large Synoptic Survey Telescope (LSST) will perform a 10-year survey over a significant fraction of the sky at optical wavelengths at depths comparable to WFIRST (see Figure 2-10). The combination of the two telescopes will measure the photometry of stars, galaxies, and quasars in nine colors spanning nearly a decade in wavelength. The WFIRST images will be much sharper, with a PSF that is 12 times smaller in area. However, LSST optical data will provide an important complement that will enable astronomers to identify a host of interesting astronomical objects in this 9-color space. LSST optical photometry (at or near the full depth of the LSST 10-year survey) is required to calculate photometric redshifts for the WFIRST weak lensing experiment (see the LSST Cadence paper in Appendix J). LSST will survey a larger area (and observe it many times over), but the ~2200 deg$^2$ overlap with the WFIRST HLS will allow better exploitation of the full LSST area. Benefits to LSST include characterizing correlations of optical colors with the near-IR spectral energy distributions (SED), high resolution morphology, and providing robust photometric redshifts in the overlap area to calibrate the wide LSST weak lensing survey. LSST will also require spectroscopic calibration of photometric redshifts (as does WFIRST) and will benefit from the WFIRST IFU spectra (see §2.2.5). The tightest constraints on dark energy will come from a combined analysis and processing of LSST and WFIRST (and Euclid) data. The benefits of jointly processing the LSST and WFIRST data include a more thorough understanding of SEDs and the effects of color gradients for weak lensing as well as better star/galaxy and blended object separation afforded in LSST data by using WFIRST-generated isophotes.

### 2.7.4  *Giant Segmented Mirror Telescopes*

Several thirty-meter class telescopes are being developed for first light in the 2020s (ELT, GMT, TMT). These telescopes, with their huge mirrors but limited (compared to WFIRST) fields of view, will be well placed to follow up on unusual and rare objects found in the groundbreaking deep 2,200 square degree WFIRST HLS. These telescopes will be geographically dispersed to provide coverage of nearly the whole sky, forming a powerful complement to WFIRST. Furthermore, the depth of the WFIRST HLS creates a challenge in calibrating photometric redshifts). Thirty-meter class telescopes play an important role in acquiring the complete sample of spectra to the full depth of the HLS imaging survey. Some of the spectra necessary for calibration will only be accessible from these telescopes and they thus will help in enabling WFIRST's weak lensing objectives. These facilities have not only larger aperture mirrors, but also near-infrared optimized instruments in combination with a new generation of adaptive optics. These instruments will allow single-object spectroscopy of z>7 candidates galaxies (TMT/IRIS) as well as multi-object capabilities which give such telescopes the possibility to cover a few arc minutes squared field at any given time (e.g. 2 arcmin squared for TMT/IRMS). The spectral resolution, provided by such instruments (R=2000-10000) enables the identification of emission lines (H$\alpha$, Oxygen nebular lines) needed to characterize physical properties of primordial galaxies (star-formation, metallicity, and dust content). Also, giant segmented mirror telescopes will help calibrate WFIRST redshifts via the determination of absorption line features red ward of the Ly$\alpha$ line usually due to metals (Iron, Magnesium, and Calcium) which are likely produced by supernovae explosions and AGB stars.

### 2.7.5  *Subaru*

The 8-m Subaru telescope in Hawaii will be the best survey telescope in the Northern Hemisphere when WFIRST flies. With powerful imaging (Hyper Suprime-Cam/HSC) and spectroscopic (Prime Focus Spectrograph/PFS) capabilities, it will be able to follow up on WFIRST surveys that are inaccessible to LSST. While the HLS will be, at least, mostly covered by LSST, there are benefits to spreading WFIRST deep and medium depth survey fields across the sky. These fields will be visited multiple times in SN searches, and obtaining optical photometry will be key to those science goals. These dispersed deep fields will also provide targets for the full range of the 30-meter telescopes described above.

## 3   WFIRST-AFTA DESIGN REFERENCE MISSION

### 3.1   Overview

The WFIRST-AFTA Design Reference Mission (DRM) defines the payload, spacecraft, and ground system architectures and hardware required to meet the WFIRST-AFTA science requirements. A wide field instrument provides the wide-field imaging and slitless spectroscopy capability required to perform the dark energy, exoplanet microlensing, and NIR surveys and a coronagraph instrument provides high contrast imaging and spectroscopy to perform the exoplanet coronagraphy science (see Figure 3-1 for a fields of view layout).

The payload features a 2.4-meter partially obscured aperture telescope, which feeds two different instrument volumes containing the wide field instrument and the coronagraph instrument (see Figure 3-2). The telescope hardware was built by Exelis under contract to another agency and was provided to NASA. This existing hardware significantly reduces the development risk of the WFIRST-AFTA payload (see §3.2.1).

The wide-field instrument provides two channels, a wide-field channel and an integral field unit (IFU) spectrograph channel. The wide-field channel includes three mirrors (two folds and a tertiary) and an element (filter/grism) wheel to provide an imaging mode covering 0.76 – 2.0 μm and a spectroscopy mode covering 1.35 – 1.89 μm. The wide-field focal plane uses 2.5 μm long-wavelength cutoff 4k x 4k HgCdTe detector arrays. The HgCdTe detector arrays are arranged in a 6x3 array, providing an active area of 0.281 deg$^2$. The IFU channel uses an image slicer and spectrograph to provide individual spectra of each slice covering the 0.6 – 2.0 μm spectral range over two subfields: a 3.00 x 3.15 arcsec field with 0.15" slices and a 6.00 x 6.30 arcsec field with 0.30" slices. Both subfields are imaged onto a singe 2k x 2k HgCdTe detector array. The instrument provides a sharp PSF, precision photometry, and stable observations for implementing the WFIRST-AFTA science.

The coronagraph instrument provides high-contrast imaging and spectroscopy. A flexible front-end provides low-order wavefront sensing and control, a fast steering mirror, and two coronagraphic modes selectable via interchangeable masks to produce a contrast on the order of 10$^{-9}$ (after post-processing). Direct imaging is provided over a bandpass of 430 – 970 nm, and spatially-resolved spectroscopy is provided by means of an integral field

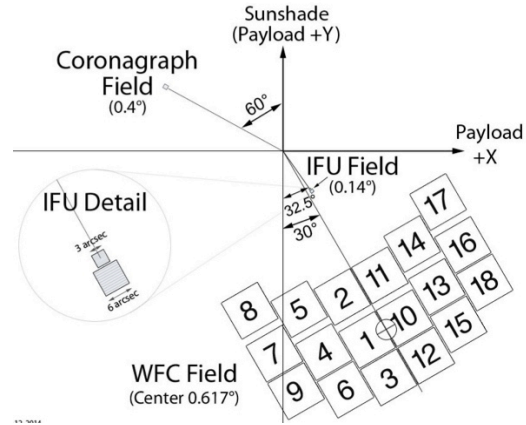

**Figure 3-1: The WFIRST-AFTA fields of view layout as projected on the sky showing the wide-field instrument fields (wide-field and IFU channels) and the coronagraph field (shared between the imager and the IFS).**

spectrograph over the spectral range of 0.6 – 0.97 μm with a spectral resolution of ~70.

Several different orbit choices were considered for WFIRST-AFTA including geosynchronous Earth orbit, Sun-Earth L2, and highly elliptical Earth orbits, with the best two options being Sun-Earth L2 and geosynchronous orbit. For this study, we have chosen to baseline a 28.5° inclined geosynchronous orbit with a mean longitude of 105°W and an initial right ascension of the ascending node (RAAN) of 228°. The primary factor that drove the selection of this orbit is the ability to continuously downlink data to a dedicated ground station enabling a much higher science data downlink rate. These benefits were weighed against the higher radiation environment and slightly less stable thermal environment versus the Sun-Earth L2 orbit chosen by the previous WFIRST SDT. Preliminary radiation analysis was performed during this study to assess the impact of the electron flux environment on the HgCdTe detector arrays and to size the radiation shield around the focal plane. A future analysis will optimize the shielding design by accounting for partial shielding provided by the remaining elements of the Observatory. As is typical for all spacecraft in GEO, additional shielding is required on both the spacecraft and payload electronics for the higher total dose radiation environment. Structural/Thermal/Optical (STOP) analysis was performed to assess the impact of the thermal environment in the geosynchronous orbit on the instruments performance. As detailed in §3.8, this slowly varying thermal environment is easily managed by the instruments with sufficient margin over the entire orbit. A more detailed assessment of the orbit trade study can be found in Appendix C. The Study

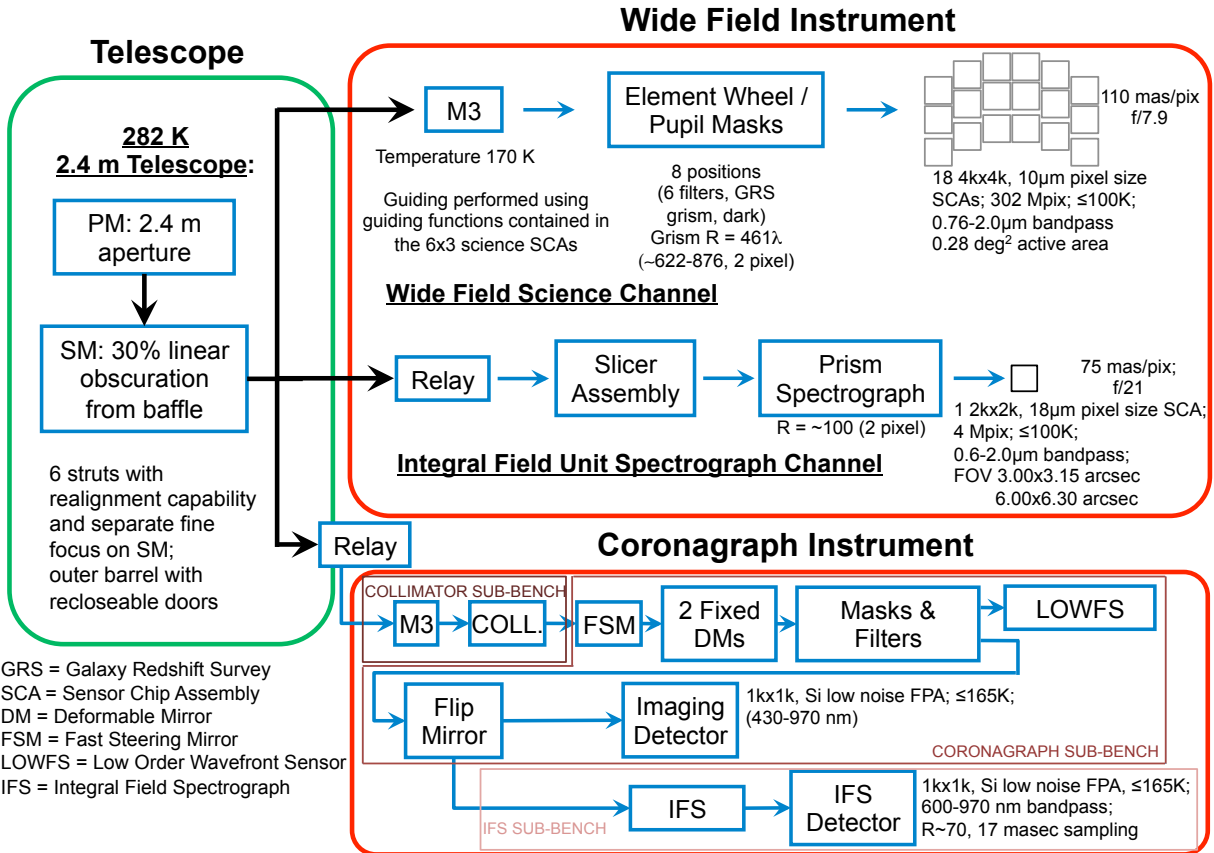

**Figure 3-2: WFIRST-AFTA payload optical block diagram.**

Office and SDT will continue to evaluate the orbit trade study to better quantify the assessment in Appendix C.

A Delta IV Heavy launch vehicle is baselined to directly inject the observatory into the baseline 28.5° inclined, geosynchronous orbit, eliminating the need for the large bi-propellant system carried in the 2013 SDT Report. The mission life is 6 years with consumables sized to allow an extension for a total of 10 years. The spacecraft uses mature technology and redundant hardware to protect against any one failure prematurely ending the mission. It provides a fixed solar array/sunshield that allows continuous operations over the full field of regard (see §3.6).

The SDT charter specified that the WFIRST-AFTA observatory be serviceable. For this study, it was decided to provide servicing at a module level, i.e. an entire instrument or a spacecraft module containing multiple electronics boxes. The modularity will also be a benefit during integration and test of the observatory (see Figure 3-3).

The WFIRST-AFTA design reference mission has been specifically tailored to meet the unique requirements of each of the WFIRST science programs.

The coronagraph, as a technology demonstration instrument, is not allowed to drive mission requirements beyond what is required to meet the science programs specified for WFIRST by NWNH (e.g. pointing requirements). For example, high-accuracy pointing, knowledge, and stability are all required to resolve galaxy shapes and precisely revisit both the microlensing and SN fields and these requirements support coronagraphy as well. Additionally, the design reference mission has been tailored to accommodate the coronagraph and analysis demonstrates that the observatory design provides an excellent platform enabling the coronagraph science program.

The GRS measurement requires NIR spectroscopy to centroid Hα emission lines and NIR imaging to locate the position of the galaxy image relative to the dispersion window. Dispersion at R=461λ (2 pixels) enables centroiding the Hα or other emission lines to a precision consistent with the redshift accuracy requirement. To address completeness and confusion issues, grism observations over at least 3 roll angles, two of which are approximately opposed, are obtained over ~90% of the mapped sky. The band-pass range of 1.35 – 1.89 μm provides the required redshift range for Hα emitters.

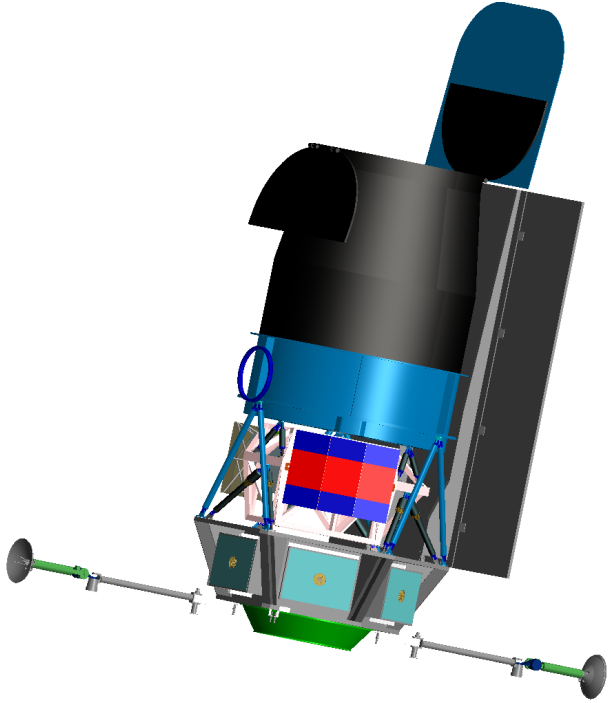

**Figure 3-3: WFIRST-AFTA Observatory configuration featuring the 2.4-m telescope, two modular instruments and a modular spacecraft.**

The SN survey measurement requires large light gathering power (collecting effective area times FoV) to perform the visible and NIR deep imaging and spectroscopy needed to discover, classify, and determine the luminosity and redshift of large numbers of Type Ia SNe. Precise sampling of the light curve every five days meets the photometric accuracy requirement, and the use of three NIR imaging bands in conjunction with IFU spectroscopy spanning 0.6 – 2.0 μm allows measurements of SNe in the range of 0.2 < z < 1.7, providing better control of systematic errors at low z than can be achieved from the ground and extending the measurements beyond the z ~ 0.8 ground limit.

The WL measurement requires an imaging and photometric redshift (photo-z) survey of galaxies to mag M ~24.6. A pixel scale of 0.11 arcseconds balances the need for a large FoV with the sampling needed to resolve galaxy shapes. Observations in three NIR filters, with ≥5 random dithers in each filter, are made to perform the required shape measurements to determine the shear due to lensing, while observations in an additional NIR filter are combined with color data from the shape bands and the ground to provide the required photo-z determinations. The GRS grism and the IFU, along with overlapping ground observations, are used to perform the photo-z calibration survey (PZCS) needed to meet the WL redshift accuracy requirement.

The microlensing survey measurement requires precise photometric observations of the Galactic Bulge and a fine PSF to detect star + planet microlensing events. Multiple fields are observed repeatedly to monitor lightcurves of the relatively frequent stellar microlensing events and the much rarer events that involve lensing by both a star and a planet. In the latter case the planetary signal is briefly superposed on the stellar signal. Microlensing monitoring observations are performed in a wide filter spanning 0.93 – 2.0 μm, interspersed ~twice/day with brief observations in a narrower filter for stellar type identification. Pointing to between 54˚ and 126˚ off the Sun enables the observation of microlensing fields for up to 72 continuous days during each of the twice yearly Galactic Bulge viewing seasons.

The coronagraphy measurements require high-contrast imaging and spectroscopy to detect and characterize exoplanets. Coronagraphy requires high thermal stability and low jitter of both the telescope and the coronagraph during observations. Although the coronagraph is not allowed to drive these observatory requirements, high-accuracy pointing, knowledge, and stability are all required to resolve galaxy shapes and precisely revisit both the microlensing and SN fields and these requirements support coronagraphy as well. Two deformable mirrors (DM) form a sequential wavefront control system (WFCS) that compensates for both phase and amplitude errors in the telescope and coronagraph optics. A Low Order Wavefront Sensor (LOWFS) provides the aberration measurements for the wavefront control system while a Fast Steering Mirror (FSM) is used in an internal pointing control loop to meet the coronagraph pointing requirements. The relatively short initial observations focus on discovery of planets near the target star, while longer observations are required for planet spectroscopic characterization.

### 3.2 Telescope

#### 3.2.1 Hardware Description

The WFIRST-AFTA telescope is existing hardware provided to NASA. It was built by Exelis and has been structurally qualified and partially assembled. The application for WFIRST-AFTA changes the ther-mal interfaces and orbit from that which the hardware was originally designed. These differences in environment and operation have different impacts on the telescope, from slight changes in the optical design, to a more robust thermal control system. However, all are within the design and qualification of the hardware. In particular, the nominal operating temperature of 282K leaves the telescope within its qualified range of performance.

The telescope is separated into two main assemblies, the forward optics assembly (FOA) and the Outer Barrel Assembly (OBA). The FOA includes the primary and secondary mirrors and their baffles, the secondary mirror support tubes and strut actuators (integral to the forward metering structure), the secondary mirror focus actuators located at the secondary mirror, and the aft metering structure (AMS). The aft metering structure is the primary interface to the rest of the observatory. The OBA acts as a thermal shroud for the FOA and AMS and includes recloseable doors. An outer barrel extension (OBE) will be added to lengthen the outer barrel for additional stray light control. The new suite of instruments (coronagraph and wide-field) will interface with the observatory through the instrument carrier rather than the AMS. This is done to provide load margins well within the original design and qualification of the AMS.

In order to restore the telescope to its operational state, some components will need to be rebuilt from existing designs as they were not provided with the telescope. These include the actuators for the secondary mirror struts, and the telescope control electronics. The actuators will be rebuilt to the design of the original hardware. The telescope control electronics design leverages existing electronics boards from Exelis' inventory of recent commercial flight hardware. Many of these boards are build-to-print, while a few of the boards will have minor modifications, e.g. adding a heritage circuit to a design or updating an FPGA.

The design and test documents for the heritage hardware and the flight hardware itself have been handled and maintained to retain full flight traceability. All of the heritage program documentation can be shared with no barriers to restrict use of information. Support hardware is also required to bring the reconfigured telescope to operational readiness, including test chambers, disassembly hardware, coating chambers and reassembly and integration hardware. Much of this hardware already exists at Exelis. Support hardware subsumed by other projects can be rebuilt to print.

The key components of the telescope are illustrated in Figure 3-4. The existing hardware is shown in magenta and includes the substantive elements of the assembly. The elements shown in green represent new designs, and include ancillary hardware such as baffles and struts. The elements shown in blue will be remade from existing designs (e.g., only four of the required six alignment drive actuators are available and two must be rebuilt). The elements in red represent existing hardware requiring slight modifications. This includes the Primary and Secondary Mirrors, which will be slightly refigured. The elements in yellow represent existing flight qualified designs requiring slight modifications, such as electronics boards.

Although the 2.4-meter telescope has been largely qualified and is flight ready, adaptation to the WFIRST-AFTA mission ten years after development brings its own unique set of potential risks. These include, but are not limited to, operation in a different environment for which it was designed, different loads from an all new suite of instruments, hardware degradation due to age, and the consistency of hardware heritage and pedigree with NASA standards.

The baseline WFIRST-AFTA design reference mission contributes to significantly reducing a number of these concerns. First, operation of the Telescope within its original operating temperature range (282K for the FOA, 232K for the OBA) mitigates any concern the telescope structure will not perform as designed due to thermo elastic deformations. Operating the telescope within its qualified temperature range also reduces the need to re-qualify the hardware, which minimizes the cost and schedule of telescope integration and test activities. Second, mounting the FOA as well as the instruments to a central Instrument Carrier structure, as opposed to mounting the instruments directly to the existing FOA instrument interface, mitigates the risk of the instrument mass growing beyond the designed load limits of the original telescope structure and actually reduces the load supported by the telescope.

Since the telescope was built and qualified over 10 years ago, aging of the structure and materials must be evaluated and this activity is already well underway. Specifically, the mechanical properties of the composite materials that comprise the bulk of the precision telescope structure must be verified. To address this concern, the coefficient of thermal expansion (CTE) of existing, aged coupons was remeasured and compared to original measurements. In the case of composite materials, CTE is a proxy for strength, stress, strain, and other properties. As stated in the Exelis report documenting the results of this testing "All samples were tested at room temperature to form a baseline to compare against historical measurements. The measured CTE for all laminates was the same as the measured CTE when originally fabricated within the test uncertainty."

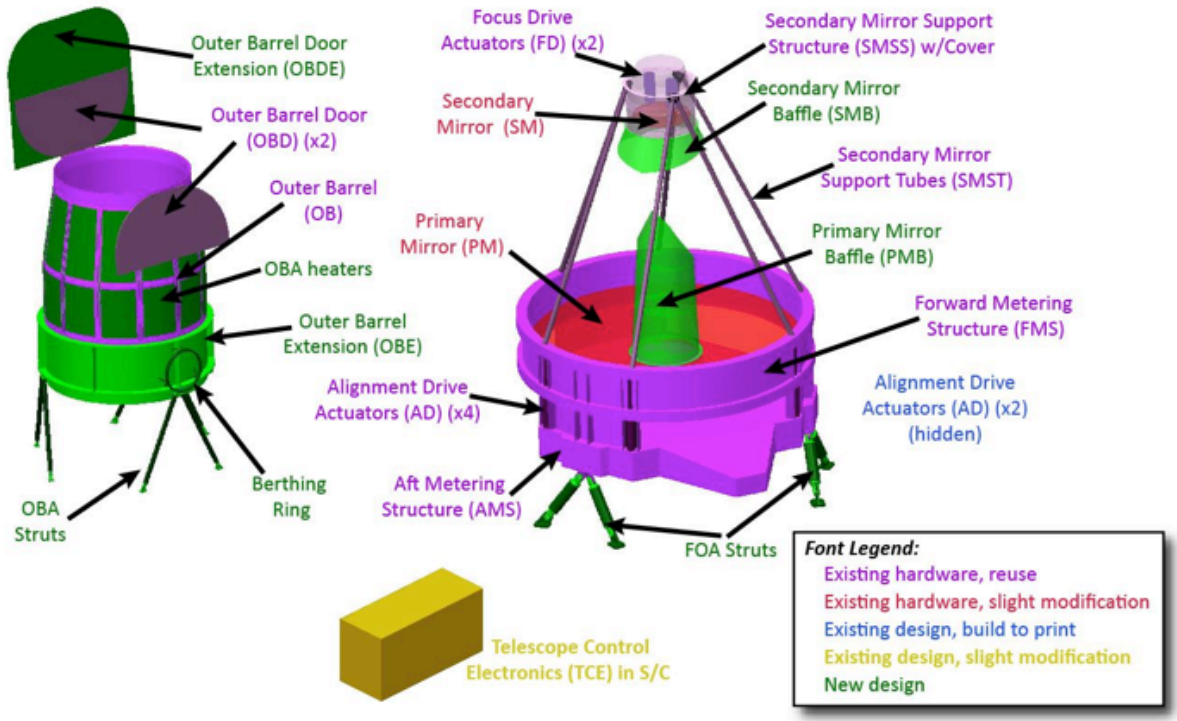

**Figure 3-4: The telescope key components identifying which components will be reused, remade, or will be built to a new design.**

A comprehensive list of all materials in the telescope sub-structures has been assembled with each material evaluated for aging and life expectancy. A test program is planned for all materials with unresolved life expectancy. Materials identified to be at risk of degeneration will be replaced.

Exelis, JPL, and GSFC mission assurance experts are developing a plan to evaluate the original build books, hardware pedigree and heritage documentation and reconcile those with NASA guidelines based on a Class B mission.

### 3.2.2  Integration & Test

A collaborative effort between Exelis, JPL and GSFC has defined an integration and test approach that produces a fully aligned and tested FOA and Tel-escope. The two highest-level planning flows of activity for the FOA and Telescope are shown in Figure 3-5 and Figure 3-6. It should be noted that the optical testing planned at the FOA and telescope levels intentionally provide traceability to mitigate risk and to streamline payload and observatory level optical tests.

The secondary mirror (SM) and primary mirror (PM) will be refigured and recoated at Exelis to meet mission requirements. These processes are routinely performed by Exelis so there is very low risk to this rework activity. In order to rework these mirrors, they must be removed from the FOA. Following the mirror rework, the FOA will be reassembled per flight certified processes with full flight documentation that meets mission assurance requirements. Component level tests will be conducted as required prior to FOA installation.

The first major optical test will occur at the complete FOA level, where the use of ground support equipment to off load the gravity effect on the PM can be incorporated. This approach has a long heritage on Exelis flight hardware and will use the gravity off loader support equipment from the original program.

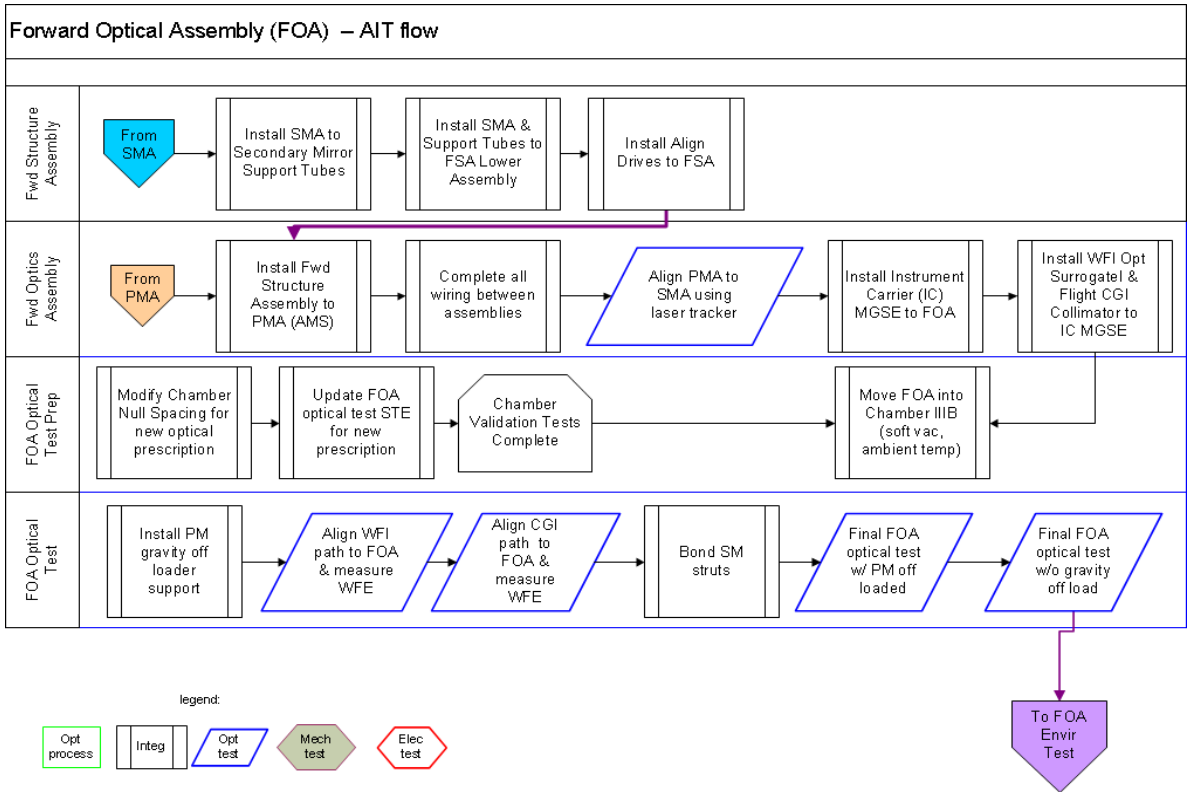

**Figure 3-5: An FOA high level assembly, integration and test (AIT) flow has been established and is analogous to the original program flow.**

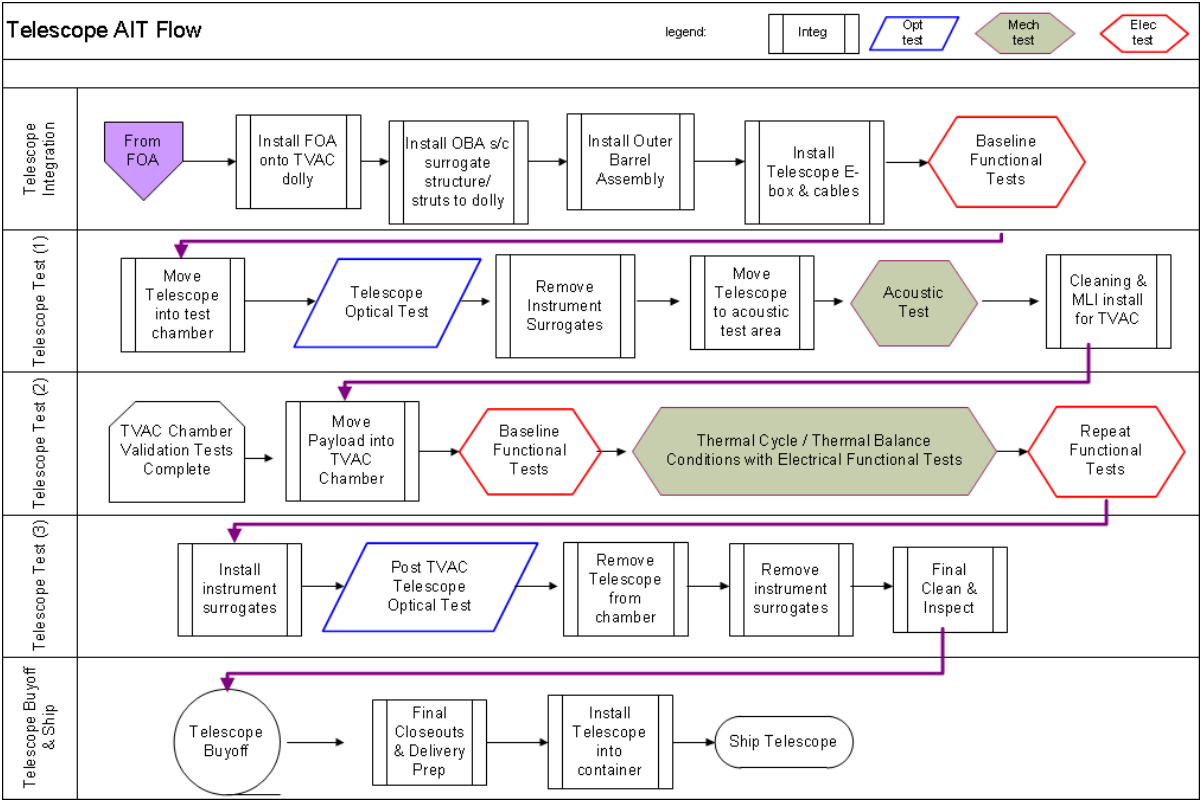

**Figure 3-6: Telescope high level AIT flow has been established using mission requirements and past experience.**

To provide a higher fidelity optical test with the PM off loaded, a simulator or Mechanical Ground Support Equipment (MGSE) for the flight Instrument Carrier (IC) will be used. This IC simulator will represent the critical mechanical and optical interfaces to the FOA. In addition, the use of the IC simulator allows a Wide Field optical verification unit (OVU) and the front-end optics of the flight Coronagraph to be installed as part of the FOA optical test. The FOA optical test will align the SM-PM system, as measured at the instruments optical interfaces. After the initial alignment, the tests will be repeated with and without the PM gravity off loader support equipment in place. This test will provide a baseline for subsequent telescope and payload tests where the configuration prevents use of the PM gravity off loader. Additionally, the integrated modeling team will develop a gravity-correlated model of the FOA after this testing.

The FOA will undergo baseline electrical functional tests prior to modal survey and vibration tests, which will be followed by a repeat of the optical test. The IC simulator and the Wide Field OVU will remain in place.

The telescope build will continue with installation of the FOA on the IC simulator on a piece of GSE representing the spacecraft interface. The OBA will then be integrated to this spacecraft GSE. The telescope electronics box and remaining cables will then be installed, which will be followed by a baseline electrical functional test of the telescope. A telescope level optical test will be conducted, which will essentially be a repeat of the FOA test without the PM gravity off loader. The telescope will undergo an acoustic test and a thermal vacuum (TVAC) test, which will consist of thermal cycles plus a thermal balance phase. A final post-environmental optical test will be conducted prior to final cleaning to verify that optical performance continues to meet all requirements after the environmental testing is completed.

### 3.2.3  *Development Schedule*

The overall schedule for the telescope delivery meets the WFIRST-AFTA program needs. Beginning with the start of long lead procurements, the telescope delivery can easily be completed in 40 months. This duration includes margin at key phases of the hardware build, per Exelis' standard practice (and consistent with industry standard practices) based on flight hardware heritage.

The schedule for telescope delivery includes design & fabrication of new baffles for the SM and PM.

Also, changes to the OBA door and the OBE will be designed, fabricated and tested in parallel with other telescope activities.

For the rebuild and retest of the FOA and telescope, Exelis defined a detailed integration and test sequence that reflects their long heritage of testing optical systems. In addition, Exelis evaluated similar integration and test tasks across a number of recent telescope and payload programs to validate the WFIRST-AFTA development schedule. Example tasks included installation of thermal control hardware, SM assembly test program, PM alignment and system level acoustic and TVAC tests.

For the telescope electronics, Exelis is building upon current flight hardware supply lines, so they have an established timeline for parts procurement and box integration and test. Thus, Exelis has high confidence in the integrity of the telescope schedule.

## 3.3 Wide-Field Instrument

### 3.3.1 *Hardware Description*

The Wide-Field Instrument (WFI) is divided into two modules, a cold instrument module (referred to as the instrument module below), containing the optics and the focal plane assembly, and the warm electronics module housed on the spacecraft. Optically, the instrument is divided into a wide-field channel, with both imaging and spectroscopy modes, and an integral field unit (IFU) channel. The key instrument parameters are shown in Table 3-1.

The instrument module is kinematically mounted to and thermally isolated from the instrument carrier, which provides the optical metered load path between the telescope and the instrument module structure. The instrument module design is similar to the HST/WFC3 design (see Figure 3-7) and reuses the design of the WFC3 latches, which can be robotically engaged and disengaged, to kinematically mount the wide-field instrument to the instrument carrier. The instrument module includes a grapple fixture to enable its removal by a robotic servicing vehicle. Guide rails on the instrument carrier and the instrument module align the instrument module latches with the instrument carrier latches. Connectors on the instrument module mate to a bulkhead connector panel on the spacecraft top deck. This panel provides access to enable robotic connector mate/demate to the harness that runs to the warm instrument electronics module on the spacecraft.

The instrument module consists of an outer enclosure structure, which is flexured off of the latches and supports the instrument radiators and MLI blankets, and an inner structure, the optical bench, which supports the instrument cold electro-optical components (see Figure 3-8). The mechanical latches interface directly to the optical bench, via thermally isolat-

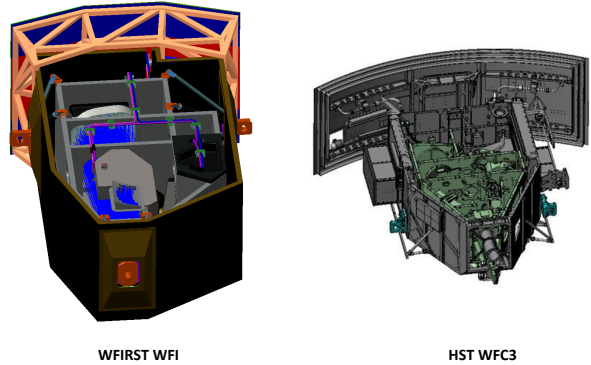

WFIRST WFI          HST WFC3

**Figure 3-7: Instrument comparison to HST/WFC3. WFIRST-AFTA wide-field is designed to be similar to the WFC3. Both use composite optical benches, radial latches, passive radiators, and heat pipes for thermal control.**

ing struts, providing a direct load path between the instrument carrier and the alignment-critical optical bench.

### 3.3.1.1 *Wide-Field Channel*

The wide-field channel optical train consists of a pair of fold flats, a tertiary mirror (M3), an element wheel (EW), and the HgCdTe Focal Plane Assembly (FPA), see Figure 3-2. The FPA uses 18 4k x 4k 10 µm pixel (H4RG) HgCdTe Sensor Chip Assemblies (SCAs), with a detector array wavelength cutoff of 2.5 µm (optical cutoff is limited to 2.0 µm via bandpass filters) and proximate Sensor Cold Electronics (SCE) units. The SCAs are arranged in a 6x3 layout with a pixel scale of 0.11 arcseconds/pixel (Figure 3-1). An 8-position EW provides 6 filters, a dark position (for calibration), a grism assembly for the GRS, and pupil masks to block in-band telescope radiation for the grism and the filters with bandpasses extending beyond 1.5 µm. The imaging mode is designed to be diffraction limited at 1.2 µm.

The filters are fused silica substrates with ion-assisted, highly stable bandpass filter coatings with heritage to recent HST instruments. The Study Office is procuring coupons of the most difficult to fabricate coatings from coatings vendors to validate that they can be manufactured to WFIRST-AFTA specifications. These will be fully characterized for spectral transmission, uniformity, and effects on wavefront error for orders zero and one.

| Mode | Wavelength Range (µm) | Sky Coverage (active area) | Pixel Scale (arcsec/pix) | Dispersion | FPA Temperature (K) |
|------|------|------|------|------|------|
| Imaging | 0.76 – 2.0 | 0.281 deg² | 0.11 | N/A | ≤100 |
| GRS Spect. | 1.35 – 1.89 | 0.281 deg² | 0.11 | R=461λ (2-pixel, grism in element wheel) | ≤100 |
| SN Ia Spect. / Photo-z Calib. | 0.6 – 2.0 | 3.00 x 3.15 arcsec 6.00 x 6.30 arcsec | 0.075 | R~100 (2-pixel; IFU spectrograph, 1 slice maps to 2 pixels) | ≤100 |
| Fine Guiding | 0.76 – 2.0 | 0.281 deg² | 0.11 | Guide off wide-field focal plane using windowing function of H4RG | ≤100 |

**Table 3-1: Key Instrument Parameters**

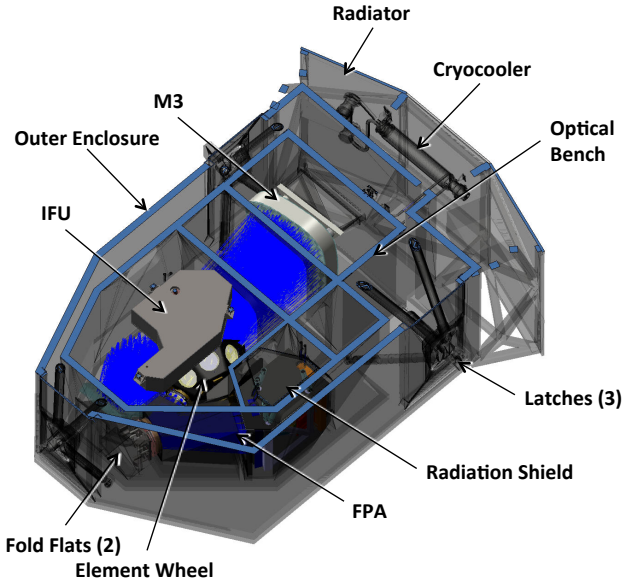

**Figure 3-8: Cut-away view of the Wide-Field Instrument showing the major elements. The instrument electronics are located in a serviceable spacecraft module. The optical bench and outer enclosure are Al honeycomb panels with composite facesheets. Harnessing is not shown.**

The 3-element GRS grism consists of 3 fused silica elements with diffractive surfaces on two of the elements. The grism has a spectral range of 1.35-1.89 $\mu$m, with a spectral resolving power (per 2 pixels) of R=461$\lambda$. The equivalent dispersion, $D_w = \lambda/(\delta\lambda/\delta\Theta)$, is obtained by multiplying by the an-gular size of a 2-pixel resolution element. Thus, for a value of $\lambda$=1.5, the corresponding value $D_w$ is 461*1.5*2*0.11= 152 arcsec. The grism wavefront error is diffraction limited across the bandpass. The grism design residuals are smaller than the diffraction limit and the design's chromatic error is improved over the design described in the 2013 SDT report.

The main challenge with the grism is the optical design due to its wide FoV, large dispersion, and relatively small F/#. The second challenge is to make high efficiency diffractive surfaces. The main challenge is solved using an innovative optical design with two diffractive surfaces. The second challenge is solved by early interaction with vendors. The Study Office is currently building an Engineering Development Unit to demonstrate that the grism assembly can be designed and fabricated with acceptable margins. This engineering activity will help establish the wide field spectrometer requirements. Multiple potential vendors have been engaged to study, design, and fabricate optical elements to WFIRST-AFTA specifications and the wide-field optical team will test the performance of the fully assembled grism at the nominal operating temperature.

A description of the EW complement is shown in Table 3-2. The H158, F184 and W149 filters and the grism are mounted with proximate cold pupil masks.

An Optical Test Demonstrator of the WFI is planned to mitigate schedule risks associated with alignment in a confined instrument structure, alignment requirements for large mirrors and FoV, and alignment procedure and analysis development. This room temperature test system will enable validation of the alignment portion of the wavefront error budget assumptions and support the development of metrology and wavefront sensing alignment procedures and aids. First use is targeted for October 2015.

| Band | Element Name | Min ($\mu$m) | Max ($\mu$m) | Center ($\mu$m) | Width ($\mu$m) | R |
|------|-------------|----------|----------|-------------|-------------|---|
| Element Wheel | | | | | | |
| Z | Z087 | 0.760 | 0.977 | 0.869 | 0.217 | 4 |
| Y | Y106 | 0.927 | 1.192 | 1.060 | 0.265 | 4 |
| J | J129 | 1.131 | 1.454 | 1.293 | 0.323 | 4 |
| H | H158 | 1.380 | 1.774 | 1.577 | 0.394 | 4 |
| | F184 | 1.683 | 2.000 | 1.842 | 0.317 | 5.81 |
| Wide | W149 | 0.927 | 2.000 | 1.485 | 1.030 | 1.44 |
| GRS | GRS Grism | 1.35 | 1.89 | 1.62 | 0.54 | 461$\lambda$ (2 pix) |

**Table 3-2: Filter and disperser descriptions in the wide-field channel spectral selection element wheel. GRS grism dispersion is wavelength dependent and the resolving power is over 2 pixels.**

### 3.3.1.1.1 Mechanical

The instrument structure, both the outer enclosure and the optical bench, will be made from aluminum honeycomb panels with composite facesheets of cyanate siloxane resin in a carbon fiber matrix. This mature, strong, light, low-moisture absorption composite material is a good low-thermal expansion match to the ULE® mirrors used in the instrument. The mirrors are stiff lightweight ULE® sandwich designs with heritage from Kepler, GeoEye and other programs. The optical bench structure has top and bottom panels with structural bulkheads between them, and is kinematically supported by and precisely aligned to the instrument carrier at three latch locations via thermally isolating struts. Mirror and filter mounts are made from Ti and other materials, as needed, to appropriately athermalize the design. Thermally isolating flexures on the outer enclosure tie it to the instrument side of the latches in three locations without distorting or impacting the alignment of the optical bench. The outer enclosure supports inner and outer MLI blankets that thermally isolate it from the optical bench, a two-zone radiator that cools the optical bench and provides the heat sink for a reverse Brayton Cycle cryocooler, and the instrument servicing grapple.

A radiation shield, consisting of a three-layer sandwich of graphite epoxy and lead, surrounds the focal plane, except for the light path leading to the focal plane, reducing the event rate in the detector arrays, caused by trapped electrons, to a level below that of Galactic cosmic rays. A thick outer layer of graphite epoxy stops the electrons, the lead layer stops bremsstrahlung photons produced by the electrons as they are stopped, and the inner graphite epoxy layer stops tertiary electrons produced by the photons. When sizing the individual layers, the radiation analysis accounted for the shielding provided by the rest of the wide-field instrument structure.

### 3.3.1.1.2 Mechanisms

The element wheel is the only mechanism routinely used in science operations. The EW's canted design allows for precise placement of any one of the elements in the space-constrained volume of the instrument. The EW assembly includes cold pupil masks for three filters and the grism, which blocks the image of the highly emissive struts and obscurations in the telescope pupil so as to limit parasitic thermal input into the focal plane. A DC brushless motor with redundant windings drives the EW mechanism. Motor control is closed loop using a rotary transducer for position feedback. A detent system acts on the wheel via roller bearings to provide repeatable positioning for each element. The detent system also serves as a restraint during launch. No power is required to hold the wheel's position during science operations. All drive electronics for the wheel are fully redundant with completely independent wiring.

A second mechanism is attached to the F2 fold flat and allows for tip/tilt and piston degrees of freedom. This mechanism provides compensation for optical alignment changes that occur both shortly following launch (e.g. gravity release and cool down corrections during initial commissioning) as well as longer term. Operation of this mechanism after commissioning is expected to be on an as needed basis to correct for misalignments due to long term changes such as moisture outgassing from the composite structural components in the telescope or instrument. The mechanism uses stepper motor and lead screw actuators that do not require power to hold position. The resolution of the actuator is on the order of 2 μm with a total travel of +/- 3 millimeters. All drive electronics for the actuators are fully redundant with completely independent wiring.

Early funding has allowed the Study Office to begin prototyping the Element Wheel and F2 Mirror Assemblies to mitigate the risk that design or materials deficiencies will be discovered late in the devel-opment flow. Engineering Development Units of both of these mechanisms are currently being built to verify their cold operational performance before and after vibration testing. Figure 3-9 shows the element wheel prototype.

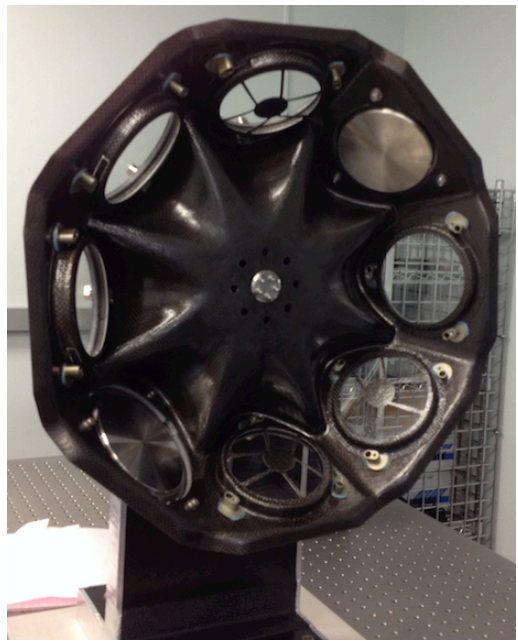

**Figure 3-9: An early prototype of the element wheel. Pupil masks are shown on representative filters and the grism is represented by a mass model.**

### 3.3.1.1.3 *Thermal*

The wide-field channel meets thermal requirements in the GEO orbit by combining a passive, cold-biased thermal design with precise thermal control of the critical focal plane hardware by a reverse Brayton Cycle cryocooler. Active cryocooler control is used to meet SCA and SCE thermal stability requirements (±0.3K over a day, and ±10mK over any HLS imaging observation, currently 173.5 secs) in the presence of environmental changes. Thermal gradients across the large focal plane array are on the order of 1 K (see Figure 3-10 and Figure 3-11).

The cryocooler circulates gaseous neon as a refrigerant to heat exchangers thermally coupled to the wide-field channel and IFU SCA base plates. The warm cryocooler compressor, along with the cryocooler electronics box, are mounted and heat sunk to a warm passive radiator. Constant conductance heat pipes are used to transport optical bench dissipated and parasitic heat loads to a separate, cold, passive radiator. This radiator includes integrated spreader heat pipes to cool the optical bench to ≤170 K. The cryocooler cools the SCA mosaic plate to ≤100 K. Both the top and bottom panel of the optical bench have embedded ethane heat pipes that transport parasitic loads from the latches, optical ap-erture, and outer enclosure to the bench radiator. Optical bench thermal stability is maintained using PID controllers that drive heaters attached to the optical bench heat pipes.

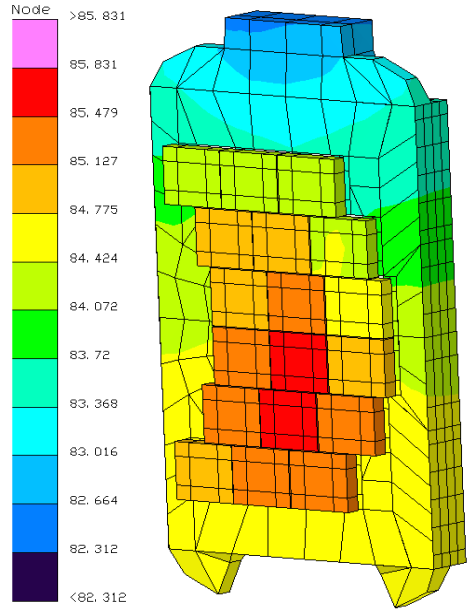

**Figure 3-10: Thermal map of the wide-field channel focal plane assembly, showing small (<~1.5K) gradients. This thermal stability is independent of observing season or Observatory pointing due to active thermal control.**

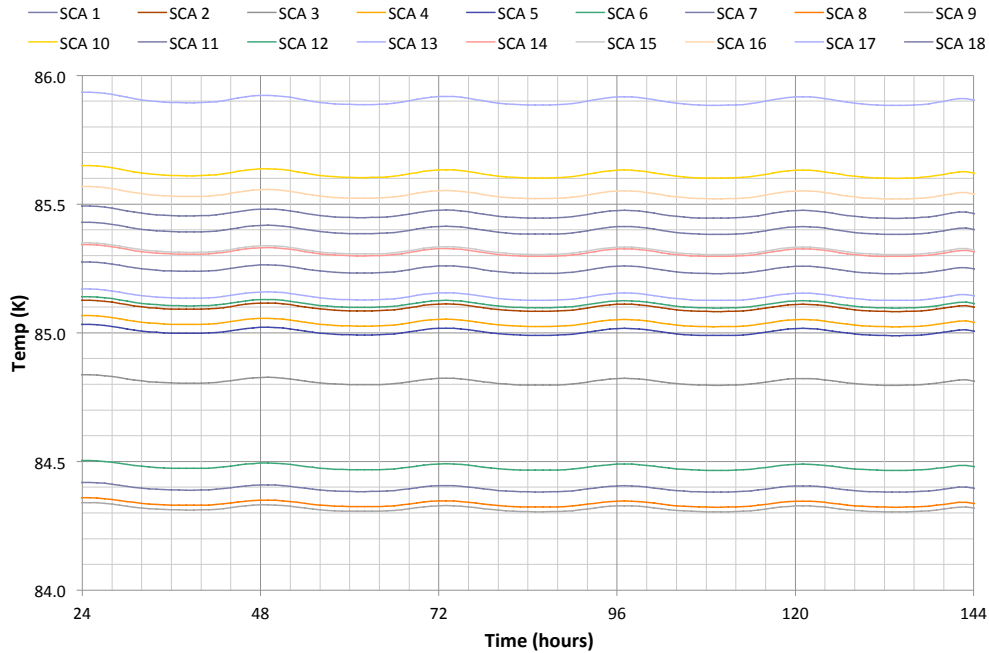

**Figure 3-11: Thermal stability of each of the SCAs over several days in an inertially fixed attitude. Short and long term stability requirements are met using a cryocooler. The power allocated to the cryocooler can be traded against the detector temperature.**

### 3.3.1.1.4  *Focal Plane Assembly*

The FPA uses 18 H4RG SCAs mounted in a 6x3 pattern in a SiC mosaic plate. Each SCA has 4088 x 4088 active pixels, 10 μm in size, plus 4 extra rows or columns of reference pixels on each edge. The f/7.9 optical system maps each pixel to 0.11 arcsec square on the sky, providing an FPA active FoV of 0.281 deg². Readout wiring and stray light considerations limit the extent to which the SCAs can be packed on the mosaic baseplate, with the active pixel gap in the x (6 SCA) direction being 2.5 mm and the 2 gaps in the y (3 SCA) direction being 8.564 and 2.5 mm (see Figure 3-1 and Figure 3-12). The FPA includes a light shield to control stray light, limiting direct illumination to active pixels. Each SCA is constructed by hybridizing an HgCdTe sensor to a Si Read Out Integrated Circuit (ROIC) and then mounting that unit to its own SiC base. Flex cabling connects each SCA to its own SCA Control Electronics (SCE), with all 18 SCEs being mounted to an aluminum mounting plate located <30 cm from the SCA mosaic plate. Each SCA is (simultaneously, in sync with all others) read out non-destructively by its SCE in ~5.5 s via 32 parallel outputs operating at 100 KHz. One 16 x 16 pixel guide window is also read out in sync at 5.8 Hz from each SCA (position in each SCA variable). All SCA-harness-SCE units are identical, simplifying production and sparing. Each SCE mounting plate is supported off the FPA's low thermal conductance com-posite truss structure using thermally isolating Ti flexures. The entire assembly is surrounded, except for the incoming optical beam, by a combined radiation and stray light shield enclosure. The shield extends around the second fold flat (F2) and is primarily designed to minimize the exposure of the SCAs and SCEs to the trapped electron environment in the geosynchronous orbit.

The H4RG 4k x 4k HgCdTe sensor itself is the only technology driver for the instrument. A discussion of the technology development requirements and plan is given below in §3.3.2.

### 3.3.1.1.5  *Wide-Field Electronics and Wide-Field Channel Signal Flow*

The wide-field instrument electronics (handling both the wide-field and IFU channels) are summarized (including redundancy) in the Figure 3-13 electronics block diagram. Three warm electronics boxes (Instrument Command and Data Handling, or ICDH, Mechanisms Control Electronics, or MCE, and Focal Plane Electronics, or FPE) are mounted to the serviceable instrument warm electronics module provided by the spacecraft. Fixed harness for the cold instrument module on the instrument carrier and the warm electronics module on the spacecraft connect at serviceable connector bulkheads on the spacecraft top deck. The FPE formats the raw image data from the SCEs, performs multi-accum processing and lossless compression before sending the data to the ICDH. A factor of 2 has been used in the data volume estimates while the Wide-Field team studies compression methods to select the best method for the data content and sampling strategy. ICDH functions are noted in detail in the block diagram, with key areas being multiplexing the 19 compressed data streams for delivery to the spacecraft and processing the 18 SCA guide windows to deliver guide star centroid motion updates to the spacecraft Attitude Control System. The MCE controls the EW to select filter/grism positions on a routine basis, and the F2 fold flat optical alignment mechanism to optimize the wide-field channel WFE performance during commissioning. The signal flow off the focal plane is controlled by the cold SCEs mounted in the FPA and the warm SCE Control Unit (SCU) boards mounted in the FPE box on the spacecraft (one dedicated SCE/SCU chain for every SCA, all processing data in parallel). During an integration, image frames interleaved with guide window data are delivered to an SCU for processing every ~5.5 secs. The guide window data is stripped out and sent to the ICDH for centroid processing.

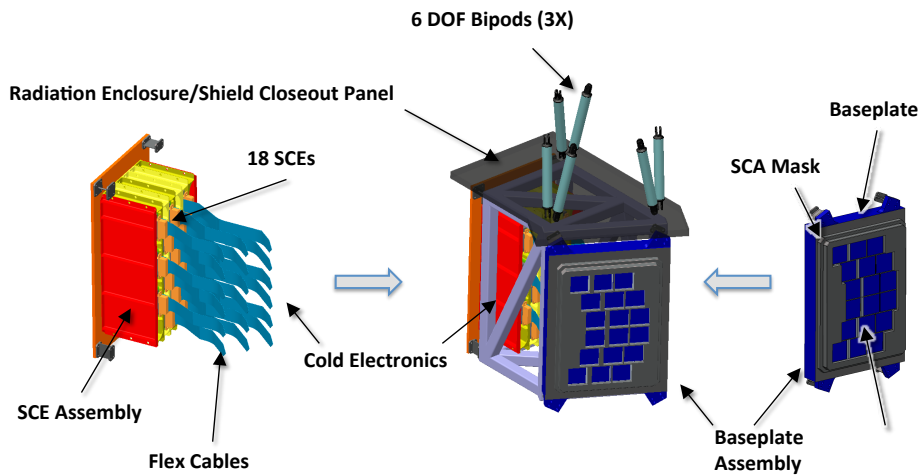

**Figure 3-12: Focal plane assembly (center) with SCE sub-assembly (left) and Baseplate Assembly containing the 18 SCAs (right). The remaining radiation shielding is not shown for clarity.**

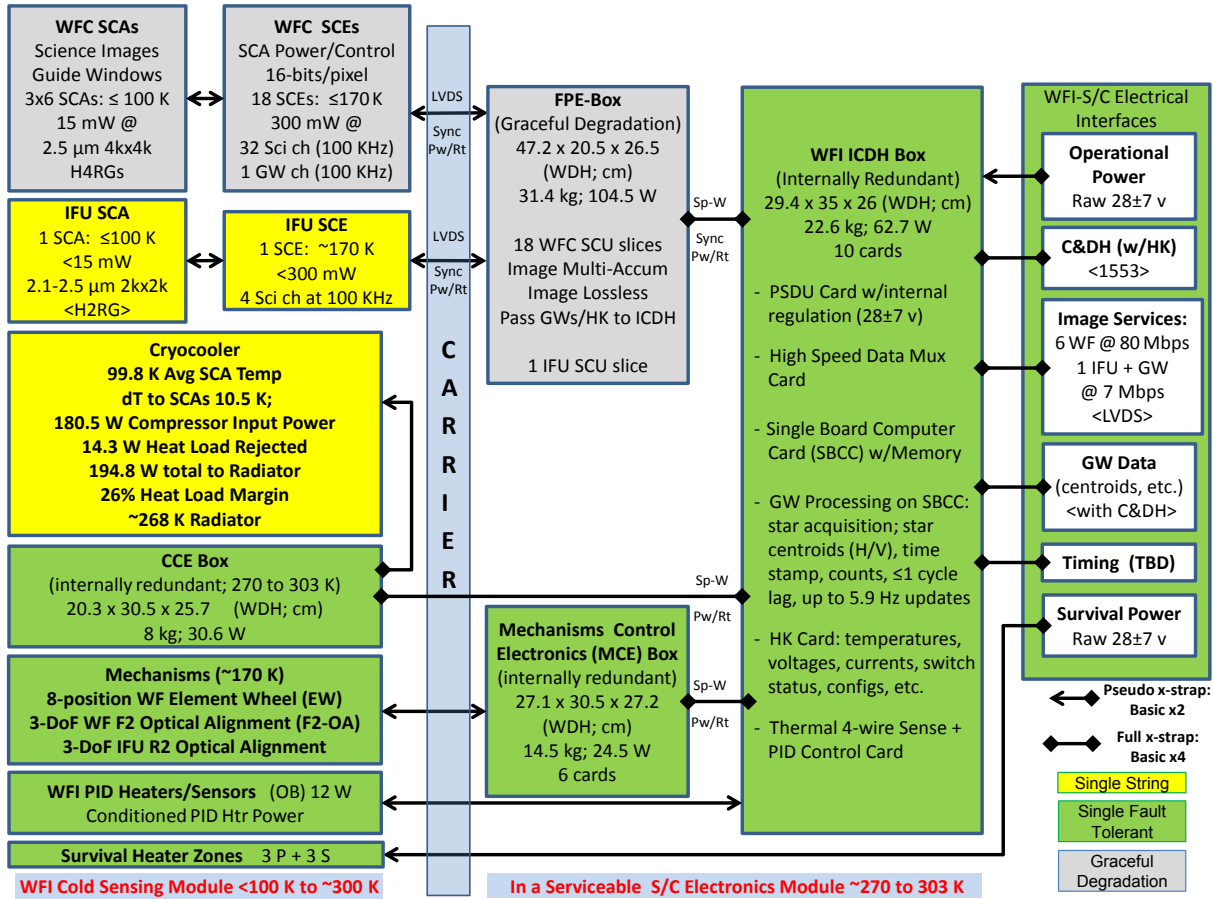

**Figure 3-13: Electronics block diagram for the wide-field instrument.**

The WFI outputs ~0.9 Gbps of science and guide-window data from 19 detector arrays to one Ka-band downlink so effective data handling and interface management between the data handling elec-tronics is essential. Therefore, the Study Office is focusing on early instrument science data stream requirements and architecture definition to identify drivers to instrument design and evaluate competing architectures. An early multi-detector test bed of the WF Channel focal plane and FPE concept (including guide mode readouts) is being developed to demonstrate and evaluate design details. This test bed will be used to demonstrate science data flow from the SCAs to the output of the transmitter. It is targeted for completion at the end of calendar year 2015.

### 3.3.1.2 Integral Field Unit Channel

The integral field unit is a separate channel contained within the wide-field instrument. It uses two small FoV apertures (3.00 x 3.15 arcsec and 6.00 x 6.30 arcsec) to limit the sky background entering the instrument channel (see Figure 3-1). The two fields serve different science needs. The smaller field is used for typing and spectroscopic follow-up of Type Ia SNe. When operated in parallel with the HLS, the larger field enables a spectroscopic redshift survey of a subset of galaxies observed in the HLS. This data will be used to calibrate the photometric redshifts determined in the HLS. In addition, in a GO program, the IFU channel could be used to observe objects of interest, as the data product is a data cube (imagex, imagey, spectral position) with good stability and high S/N.

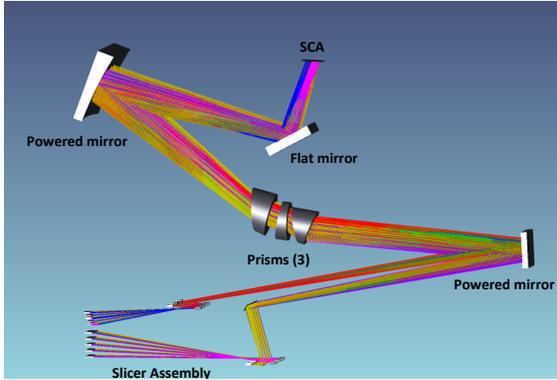

**Figure 3-14: Optical layout of the IFU. The relay is not shown.**

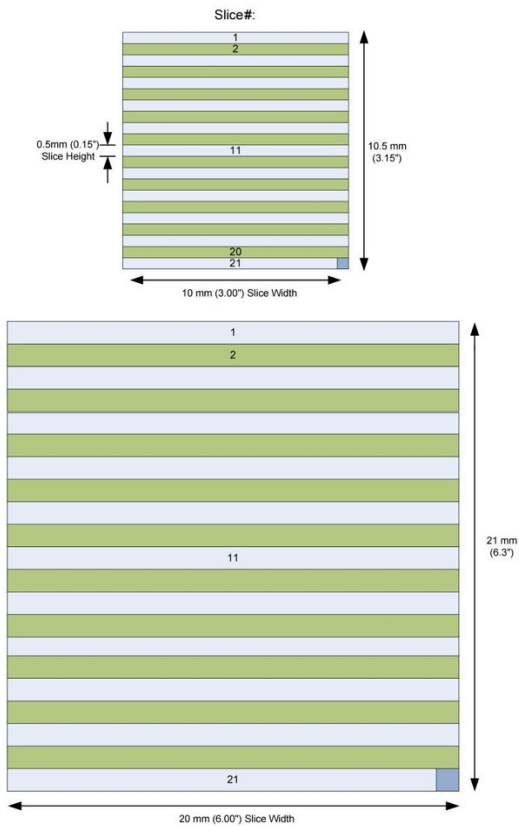

**Figure 3-15: The image slicer has 2 sets of 21 mirrors, 0.5 mm and 1.0 mm wide, yielding field slices 0.15 and 0.30 arcsec wide.**

An optical relay reimages the two small fields onto an image slicer and spectrograph covering the 0.6-2.0 μm spectral range (see Figure 3-14), resulting in a detector array format wherein 2 sets of 21 slices, 3.0 x 0.15 arcsec and 6.0 x 0.30 arcsec each (see Figure 3-15), are imaged onto separate pixel sections of a single H2RG detector. The pixel scale at the focal plane is 0.075 arcsec for the smaller field and 0.15 arcsec for the larger field. The resolving power is flattened by the use of a compound prism of materials SF6G05 and ZnSe to a typical value of 100 (2 pixels), see Figure 3-16a.

The IFU channel elements are packaged in a separate optical bench that is installed into the wide-field instrument housing. A tip/tilt/focus mechanism is included on the second relay mirror to allow removal of any residual pupil alignment or focus shifts due to gravity release, cool down, or composite desorption. The opto-mechanical assembly is designed to operate at the same temperature (170 K) as the wide-field instrument optical bench.

The slicer is based on a commercially available reflective image slicer, a version of which has been through space environmental testing. While there are a substantial number of elements, they are small (a maximum 5 cm) and work in a slow optical beam with relaxed stability tolerances.

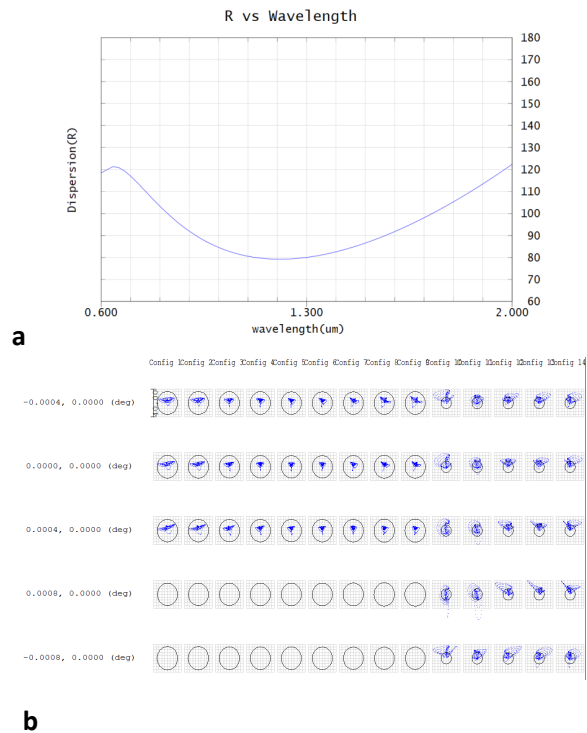

**Figure 3-16: The variation in resolving power of the IFU over the bandpass is shown in a. The diffraction limited imaging performance for the 3 arcsec aperture, across the instrument field of view is shown in b.**

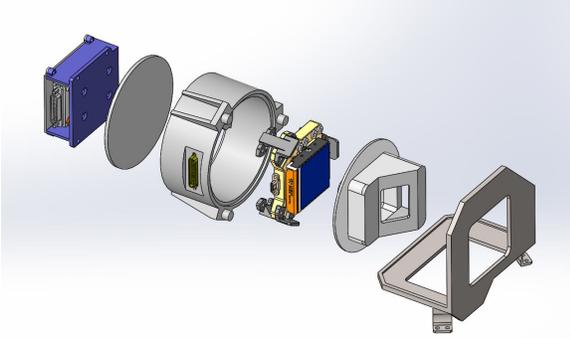

**Figure 3-17: Expanded view of the IFU channel focal plane assembly. This assembly shares design heritage with JWST.**

The IFU Focal Plane Assembly consists of a single SCA, an interconnect cable, cold electronics (SCE) and optical mask integrated into a small housing to form a single assembly. The 18 μm pixel H2RG SCA (2048 x 2048 pixels) is similar to those used on JWST and Euclid, but with a hybridized detector layer and ROIC mounted onto a silicon carbide base instead of a molybdenum base. The SCE is the same ASIC as used for the wide-field channel focal plane, mounted to the opposite side of the housing baseplate (see Figure 3-17).

The entire FPA structurally interfaces to the IFU metering structure through thermally isolating flexures, which include provisions for precise optical alignment. The focal plane thermally interfaces to the cryocooler through a simple heat exchanging mount on the housing. The cryocooler removes the SCE and SCA dissipated heat loads as well as parasitic heat loads to maintain the assembly (and specifically the SCA and to a lesser extent SCE) temperature of 100K for low dark noise and read noise. The housing components are silicon carbide to be low mass and match the detector base material. The entire assembly shares design heritage with JWST.

### 3.3.2  Enabling Technology

H4RG-10 near-IR array detectors are critical to enabling WFIRST-AFTA. The H4RG-10 is a 4096 x 4096 pixel HgCdTe detector array that has 10 μm pixels. While building on NASA's heritage developing 1k x 1k pixel H1R, 18 μm pixel detectors for Hubble/WFC3, and 2k x 2k pixel H2RG 18 μm pixel detectors for JWST, the H4RG-10 advances the state of the art by approximately quadrupling the number of pixels per unit focal plane area (see Figure 3-18). The higher pixel density simplifies mechanical and thermal system design, as well as integration and test. More-over, because the cost of making an astronomy-grade HgCdTe SCA scales roughly with the photo-sensitive area, the H4RG-10 has the potential to significantly reduce the cost per pixel.

Preliminary detector requirements have been developed as part of the detector development technology plan. These are mostly used to ensure that detector testing is sufficient to test (with margin) devices at this performance level. These requirements are listed in Table 3-3. Science simulations are in process to validate the detailed flowdown of these requirements.

The H4RG-10 ROIC was originally developed for use with visible wavelength Si PIN detectors. In this context, NASA worked closely with the U.S. Naval Observatory to mature the H4RG-10 ROIC to TRL-6 for the Navy's Joint Milli-Arcsecond Pathfinder Survey (JMAPS) satellite. The only new element for WFIRST was hybridizing HgCdTe detectors to the H4RG-10 ROIC instead of Si PINs.

As a proof of concept, in FY12 the WFIRST Study Office built and tested a small test lot of H4RG-10s. This was a success, both proving the concept and elevating the H4RG-10 to TRL-4. Out of a total of six devices constructed, five were functional, and four performed well enough to merit detailed characterization. Device testing is performed at GSFC in the Detector Characterization Laboratory, a state of the art facility that has been used to develop and test candidate detectors for many previous missions including HST and JWST. The performance was very promising, with the exception of about 20% higher readout noise than is desired. In FY13, a Process Optimization Lot was started to optimize the potential flight recipes. The growth and processing of the detector ma-terial was varied among different devices. "Banded" arrays with spatially dependent recipes for efficiently spanning parameters were built. In addition to pixel design parameters such as contact size, the passivation, doping, base material, and backfill epoxy were varied in a design of experiments across a total of 14 tested devices from three detector lots.

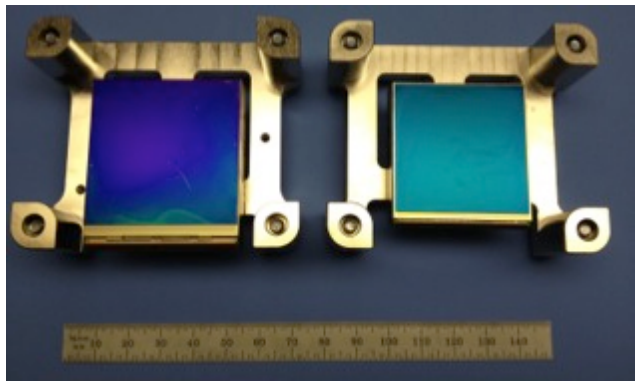

**Figure 3-18: Comparison of the size of an H4RG (4k x 4k pixels, 10 μm pixel size) on left to an H2RG (2k x 2k pixels, 18 μm pixel size) on the right.**

| Specification | Value | Units | PV1/2 (MS1) Value | PV3 (MS2) Value |
|---|---|---|---|---|
| Dark Current | < 0.1 | e-/sec/pixel | 0.0023 | 0.0066 |
| CDS Readout Noise | < 20 | e- rms | 12 | 17 |
| Quantum Efficiency | > 60% | Averaged over 0.79-2.0 μm band | 95% | >90% |
| Inter-Pixel Capacitance | ≤ 12% | Total in 8 adjacent pixels | 8-9% | 8-9% |

**Table 3-3: Detector technology development requirements. The right two columns summarize the results from the two lots of detectors based on the results below.**

The devices have all been tested and resulted in the downselect of a recipe to proceed to a full array fabrication. Some examples of the data are given below. Here we emphasize parameters tested for the first two technology milestones: interpixel capacitance, dark current, correlated double sample (CDS) noise, and quantum efficiency. Figure 3-19 shows the cross talk averaged per band for four devices. Figure 3-20 shows maps and band-averaged dark current of four devices from the initial lot. A faint glow is evident on some devices; while this glow is well below the technology development requirement, it will be addressed as part of an opportunity to improve performance of the initial H4RG readout integrated circuit (ROIC). Figure 3-21 shows the CDS (or short term) noise maps and band averages for the same example devices, showing margin against the technology development requirement.

Quantum efficiency was also measured over the 0.8-2.0 μm band for each device. Ongoing work includes improved reference photodiode calibration under the cold conditions and low ambient light levels necessary for these tests. The quantum efficiency is above the technology goal of 60% (averaged across this spectral bandpass), and detailed spectral efficiency will be available when the calibration activities

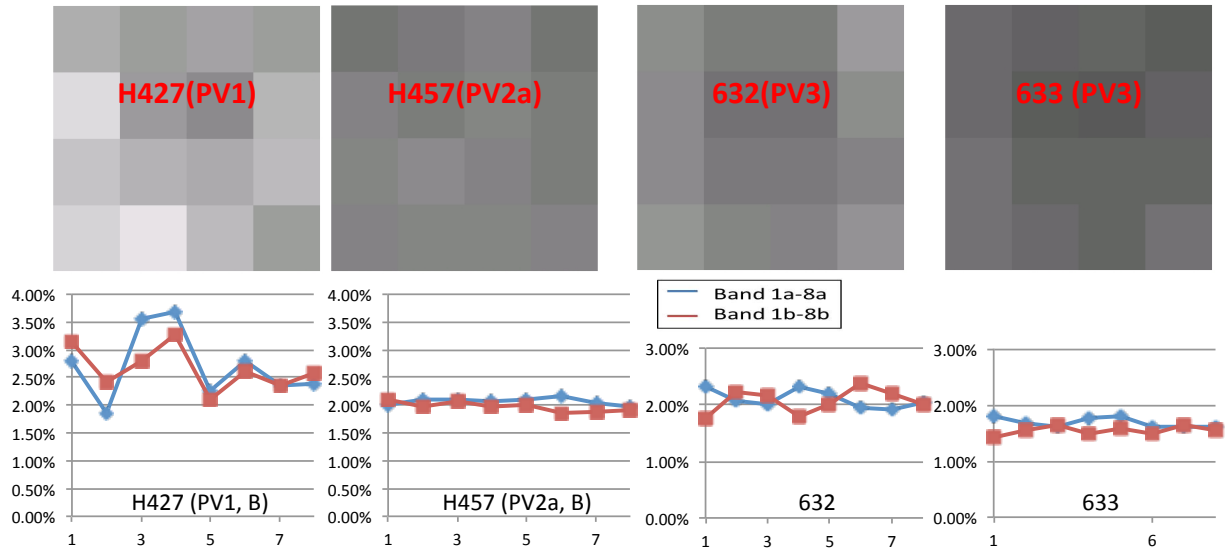

**Figure 3-19: Example cross talk for each of two repeats of 8 band designs over four example H4RG test devices. The x-axis in the bottom row is band #, with two sets of curves for the 1st and 2nd occurrence of each band design. The y-axes indicate the average crosstalk that is found by averaging the crosstalk in the 4 adjacent pixels. The total (sum of) crosstalk in the 8 adjacent pixels is approximately 4 times the average crosstalk per pixel (the bulk of the total crosstalk is within the 4 adjacent pixels). The selected band (Band 1) is the bottom left, with a repeat being the bottom left of the upper right quadrant, in each array. The upper row shows band-averaged cross talk with black to white range of 1-4%.**

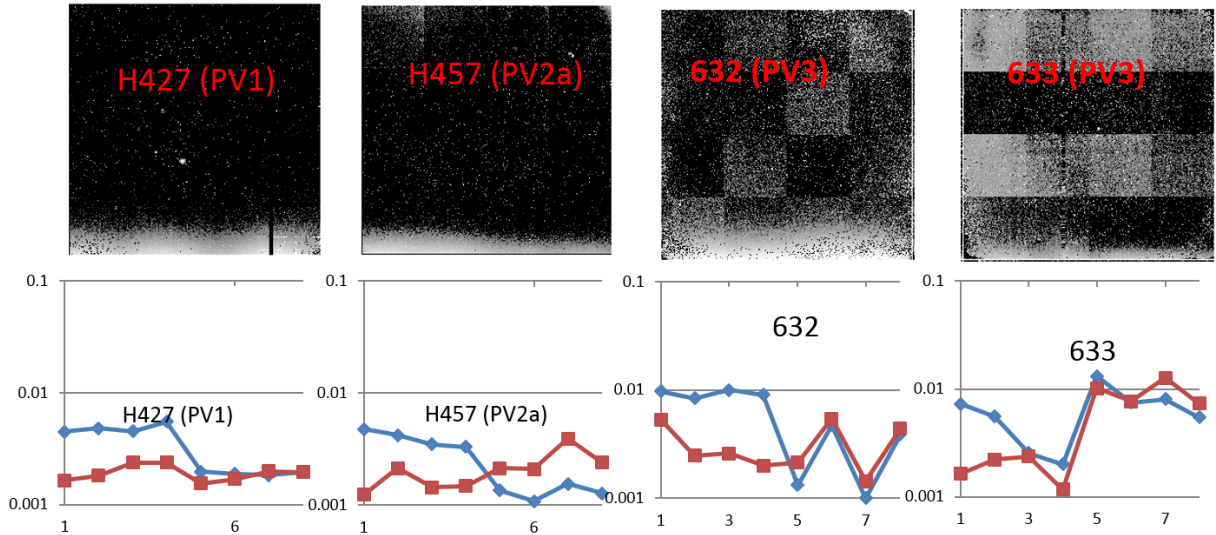

**Figure 3-20:** Dark Current maps of the same example devices shown in Figure 3-19. Uncertainty: 0.003 e-/s/pixel; i.e., any values below 0.001 e-/s or negative are set to be 0.001 in the plots. All plotted dark current numbers are median values. Images are in log scale [black = 0.005 e-/s, white = 0.1 e-/s]. While a low level glow along readout register (bottom of SCA) is evident, this is within the technology development requirement. However, this will be addressed in a redesign/remake of the H4RG readout integrated circuit (ROIC).

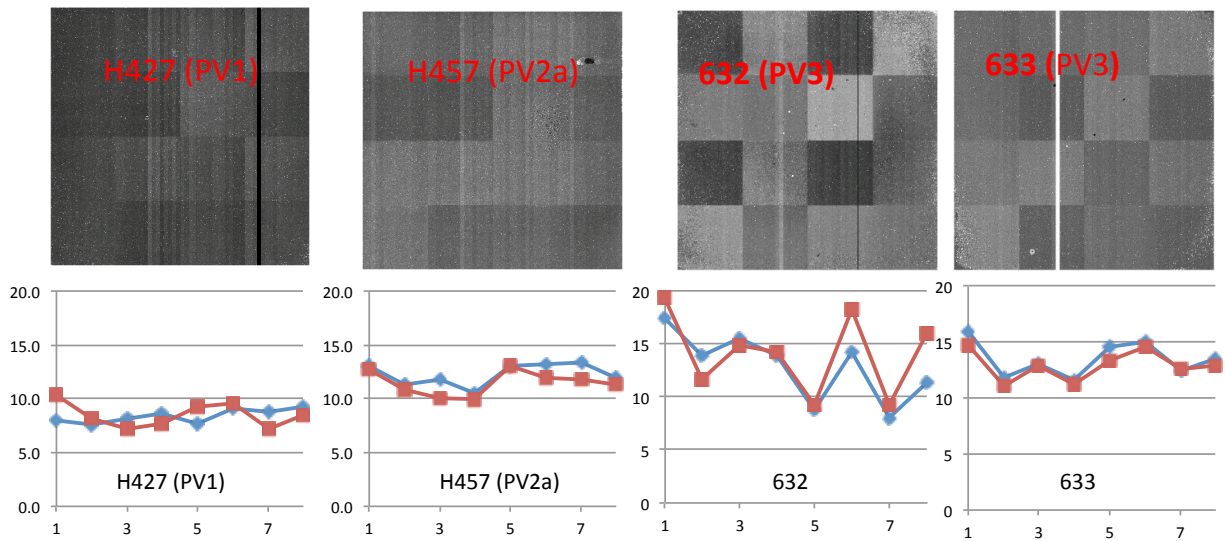

**Figure 3-21:** Correlated double sample or short term noise for the same devices in Figure 3-19 and Figure 3-20. All plotted CDS noise numbers are median values. Images are in linear scale: Black=0, White=30e-. Horizontal and vertical axes on the plots are band # and CDS rms noise averaged per band, respectively.

are completed.

The two rightmost columns in Table 3-3 above list the performance parameters in the selected band (Band 1) of the two lots of detectors, showing margin against the technology development requirements.

Full array lots of two passivations (PV2a, PV3) are now being fabricated based on the most promising pixel design (Band 1) based on performance and yield considerations. Testing of these arrays will be completed in 2015, and will result in a downselect of the passivation type for the flight design. This will be followed by the final pre-flight lot, a yield performance lot, by the end of FY16.

The NIR detector is the only technology development required for the WFI. A detector technology plan was approved in March 2013 formalizing the plan described above. Table 3-4 shows the technology milestones for the NIR detector technology development. Passing each milestone requires a formal review and concurrence by the inde-pendent Detector Technology Assessment Committee (DTAC). The first two milestones have been passed, with the 3rd milestone planned for the end of FY15.

| MS # | Milestone | Milestone Date |
|------|-----------|----------------|
| 1 | Produce, test, and analyze **2 candidate passivation techniques** (PV1 and PV2) in **banded arrays** to document baseline performance, inter-pixel capacitance, and shall meet the following derived requirements: dark current less than 0.1 e-/pixel/sec, CDS noise less than 20 e-, and QE greater than 60% (over the bandpass of the WFI channel) at nominal operating temperature. | 7/31/14 **Passed** |
| 2 | Produce, test, and analyze **1 additional candidate passivation techniques** (PV3) in **banded arrays** to document baseline performance, inter-pixel capacitance, and shall meet the following derived requirements: dark current less than 0.1 e-/pixel/sec, CDS noise less than 20 e-, and QE greater than 60% (over the bandpass of the WFI channel) at nominal operating temperature. | 12/30/14 **Passed** |
| 3 | Produce, test, and analyze **full arrays with operability > 95%** and shall meet the following derived requirements: dark current less than 0.1 e-/pixel/sec, CDS noise less than 20 e-, QE greater than 60% (over the bandpass of the WFI channel), inter-pixel capacitance ≤3% in nearest-neighbor pixels at nominal operating temperature. | 9/15/15 |
| 4 | Produce, test, and analyze final selected recipe in **full arrays demonstrating a yield > 20%** with operability > 95% and shall meet the following derived requirements: dark current less than 0.1 e-/pixel/sec, CDS noise less than 20 e-, QE greater than 60% (over the bandpass of the WFI channel), inter-pixel capacitance ≤3% in nearest-neighbor pixels, persistence less than 0.1% of full well illumination after 150 sec at nominal operating temperature. | 9/15/16 |
| 5 | Complete environmental testing (vibration, radiation, thermal cycling) of one SCA sample part, as per NASA test standards. | 12/1/16 |

**Table 3-4: The Technology milestones for the near infrared detector technology development follow a progression from pixel design through full array and yield demonstration. The first two milestones have been passed.**

### 3.3.3  Integration & Test

The wide-field instrument integration and test process provides early optical testing of small assemblies prior to full up instrument verification testing to minimize instrument I&T risk. All optical assemblies (optics plus mount) will be fully characterized prior to integration into the instrument. The two fold mirrors, the powered M3 mirror, and the Element Wheel will be installed and initial alignments performed using laser metrology. An engineering development FPA, with visible wavelength detector arrays, will be installed to enable optical alignment using wavefront sensing. The flight FPA will then be installed in the instrument and final optical alignment will be verified by wavefront sensing measurements at the operating temperature during instrument thermal-vacuum testing. Separately, the IFU will be integrated by building up the relay module, the slicer assembly, and the spectrograph. Again, these assemblies are verified separately prior to testing the full IFU channel and prior to its installation in the wide-field optical bench (see Figure 3-22).

The wide-field instrument verification effort is strengthened by the availability of the telescope and instrument carrier along with a full aperture autocollimation flat mirror. This allows the actual telescope to be used for performance of the instrument in a Test As You Fly configuration. During payload I&T, the wide-field instrument is tested with the telescope to provide the most realistic optical test.

### 3.3.4  Development Schedule

The WFIRST schedule includes 12 months to complete the conceptual design and 12 months to complete the preliminary design during formulation. 42.5 months are budgeted for the final design, fabrication, integration, and test of the wide-field instrument. These durations are standard for an instrument of this size and complexity. The only unusually long-lead subsystem is the detector subsystem with a budgeted development and test duration of 43 months from start of fabrication of the flight SCA lots, through characterization, build-up into the FPA, and through testing with the FPE. Fabrication of the flight SCA lots will start early in Phase B to meet the need date for instrument integration. The detector subsystem technology development plan is designed and funded to support this schedule (see §3.3.2).

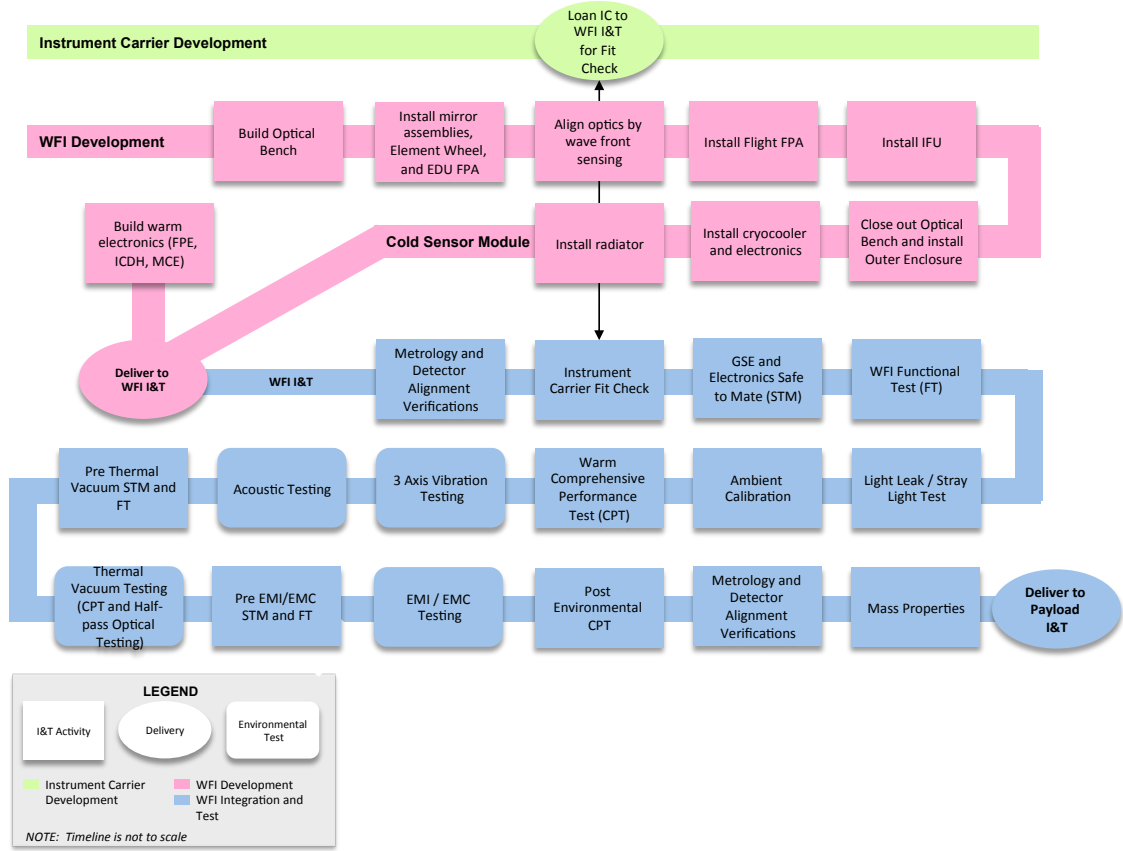

**Figure 3-22: I&T flow for the wide-field instrument. The instrument is built up and tested in small assemblies to minimize the risk of finding instrument optical anomalies late in the instrument development.**

### 3.4 Coronagraph Instrument

#### 3.4.1 *Hardware Description*

##### 3.4.1.1 *Coronagraph*

###### 3.4.1.1.1 *Coronagraph Optical Design and Performance*

The coronagraph instrument will take advantage of the 2.4-meter telescope's large aperture to provide novel exoplanet imaging science at approximately the same cost as an Explorer mission, including the additional operations time. The coronagraph is also directly responsive to the goals of the Astro2010 by maturing direct imaging technologies to TRL 9 for an Exo-Earth Imager in the next decade. The coronagraph design is based on the highly successful High Contrast Imaging Testbed (HCIT), with modifications to accommodate the telescope design, serviceability, volume constraints, and the addition of an Integral Field Spectrograph (IFS). The coronagraph is required to provide very high contrast between inner and outer working angles to maximize the science yield. Previous demonstrations in the HCIT have exceeded $10^{-9}$ contrast for on-axis unobscured pupil coronagraphs. More recent work in the HCIT with obscured pupils has made considerable progress. The Shaped Pupil Coronagraph, for example, has demonstrated better than $10^{-8}$ contrast at 10% bandwidth. A detailed summary of the recent success in the technology development is found in §3.4.2.

Figure 3-23 shows the coronagraph instrument layout, which comprises a single optical beam train that, using mask-changing mechanisms, can operate in either a Shaped Pupil Coronagraph (SPC) or Hybrid Lyot Coronagraph (HLC) mode. A single optical beam train can thus operate alternatively as two different coronagraphs.

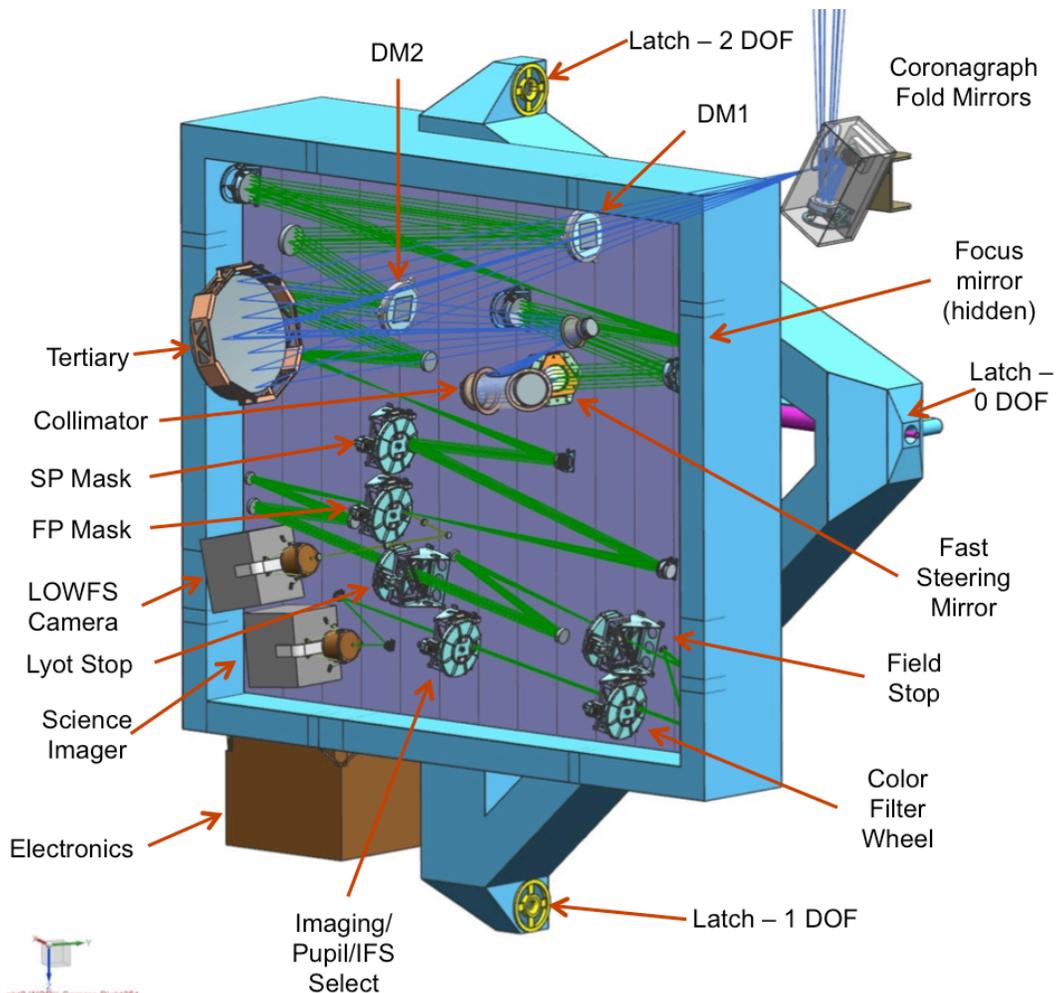

**Figure 3-23: Collimator assembly and coronagraph core optics bench. The IFS is on a separate optical bench, which is not shown for clarity.**

A pair of flat mirrors located at the output of the telescope serves to relay the off-axis output beam to the coronagraph input where a tertiary mirror (M3) and collimator mirror (M4)—both of which have low-order aspheric surfaces—significantly reduce the off-axis low-order wavefront error introduced by the aspheric primary mirror (PM) and secondary mirror (SM) (see Table 3-5). The output of this collimator assembly (M3+M4) feeds the coronagraph core with a collimated beam with extremely low residual wavefront error over the FoV of the instrument. Thus the residual wavefront error will be very insensitive to off-axis pointing changes resulting from the line-of-sight (LOS) jitter correction servo. The wavefront-corrected collimated beam forms a pupil image (primary mirror conjugate) on the two-axis Fast Steering Mirror (FSM) that corrects for LOS jitter. An off-axis parabola (OAP) relay following the FSM re-images the pupil conjugate onto the first of a pair of deformable mirrors (DM). Each DM is a monolithic module consisting of a 48×48 array of Lead Magnesium Niobate (PMN) electrostrictive force actuators. The DM face sheet, bonded to the actuators, has a static optical surface figure of 10 nm rms in the unpowered state and the surface figure can be actively controlled to better than 0.025 nm rms.

Just downstream from the DM pair, another OAP relay re-images the pupil onto a shaped pupil mask which controls diffraction before light reaches the image plane. At the image, an occulting mask rejects most of the starlight and passes the light originating from around the target star. The rejected starlight from the occulting mask is fed to the Low-Order Wavefront Sensor (LOWFS) that is used to provide real time control feedback to the FSM (for pointing correction) and to a passive mirror (for focus correction). This critical element offloads the thermal drifts of the telescope.

Following the image plane masks, an OAP collimates the beam and reimages the pupil onto a Lyot stop, a mask that strips away most of the remaining starlight while passing the planet light. The coronagraph core optical beamtrain between the FSM and the Lyot stop must be maintained at a cleanliness level of CL100 to ensure that instrument scatter is well below the exozodiacal background light level. After the Lyot plane, a filter wheel mechanism selects either a 10% or 18% wide band for wavefront sensing, planet detection and characterization or debris disk imaging. Following the color filter the light passes through a mechanical rotation mechanism, by which the output of the coronagraph core is directed to either an imaging camera or a lenslet-based Integral Field Spectrograph (IFS). The light transmitted to the imaging camera is split by a polarizing beamsplitter, allowing simultaneous detection of the two orthogonal polarization states. Since the SPC is highly insensitive to the polarization induced by the telescope, the IFS beamtrain does not require a polarizer. Both the imaging camera and the IFS utilize the same 1K×1K format electron multiplication CCD (EMCCD) with 13 micron pixel pitch. The same rotation mechanism can put in place a pupil imaging lens for focus diversity estimation.

| Design Specification | | Value | Units |
|---|---|---|---|
| Static WFE [rms] | After M2 | 13,570 | nm |
| | After M4 | 2.2 | |
| Dynamic WFE (14 mas LOS Jitter variation) [rms] | After M4 (all low order aberrations) | 100 | pm |
| | After M4 (Coma only) | 10 | |
| Throughput (excludes detector QE) | SPC (IFS) | 0.037 | N/A |
| | HLC (imaging) | 0.041 | |
| Imaging FoV [radius] | without masks | 1.62 | arcsec |
| Imaging pixel plate scale | | 0.01 | arcsec |

Table 3-5: Correction of wave front error by M3 + M4. The table shows the correction of the low-order aberrations introduced by the telescope primary and secondary mirrors. Static WFE is shown before and after correction by M3 + M4. The dynamic WFE residual after correction by M3 + M4 is shown for all low-order aberrations summed and for coma only.

The IFS has a 76×76 format lenslet array that samples the image plane with a 0.82x0.82 arcsec radius FoV for the entire array. A pinhole mask on the output of the lenslet array truncates the wings of the lenslet PSF's, thereby reducing spectral crosstalk of the IFS focal plane array (FPA). It has constant dispersion across the full 0.60 – 0.97 μm spectral range, with a free spectral range of 18%.

Some key coronagraphic optical design specifications such as static and dynamic wavefront error at the fast steering mirror, optical throughput and FoV are tabulated in Table 3-5.

The contrast for both SPC and HLC coronagraphs was simulated using an IDL based diffractive propagation code. The results of the simulations (see Figure 3-24, Figure 3-25, and Figure 3-26) show that the HLC produces better than $10^{-9}$ contrast in the presence of low LOS jitter, but is sensitive to increasing jitter, whereas the SPC produces better than $10^{-8}$ contrast and shows a lower sensitivity to increasing jitter.

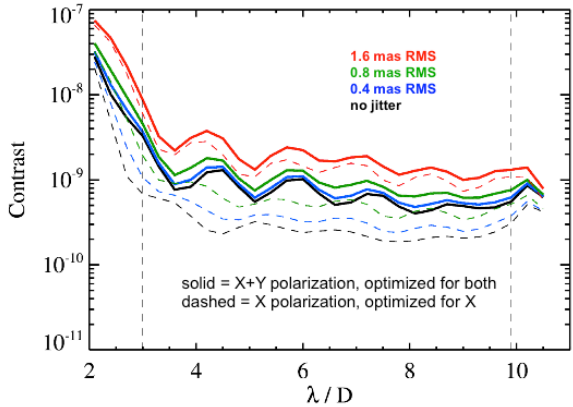

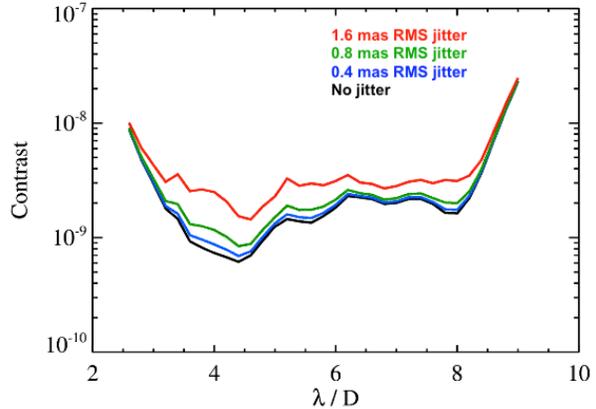

**Figure 3-24: Model-predicted contrast for the HLC in the presence of LOS jitter. Jitter values represent the variation after control by the LOWFS, which is expected to reduce rms jitter to a fraction of a milli-arcsecond.**

**Figure 3-25: Model-predicted contrast for the SPC in the presence of LOS jitter. The LOWFS is expected to reduce jitter seen by the coronagraph to a fraction of a milli-arcsecond.**

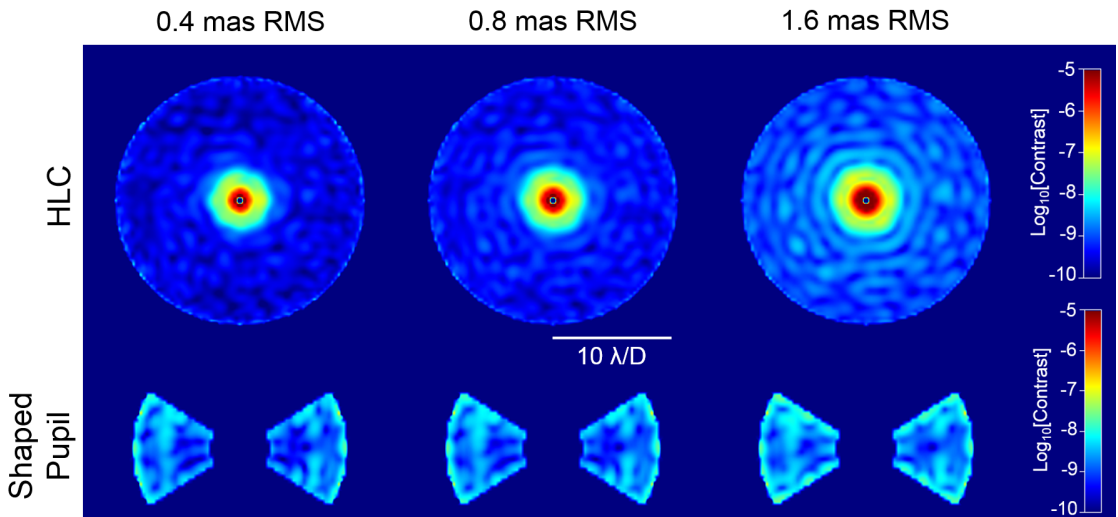

**Figure 3-26: Log scale contrast maps for the HLC and SPC, in the presence of different levels of post-LOWFS jitter. Both designs show good tolerance to jitter.**

### 3.4.1.1.2 Coronagraph Implementation

High-contrast coronagraphy requires extreme stability of the optical system over the observation time period. Coronagraph optics require ~0.25 μm relative stability during an observation. The coronagraph optical bench (shown in Figure 3-23) must be positioned with respect to the telescope with an accuracy of 1 mm. The thermal stability of this position must fall within 0.5 μm during an observation period. To achieve this positional accuracy and stability, all of the coronagraph core optics are mounted on a stiff, a-thermalized, optical bench. The optical bench - made of carbon fiber composite material - is actively controlled to <10 mK to achieve 0.25 μm internal stability. The coronagraph optical bench assembly will be maintained at 290 K to ensure that the DMs are flat when unpowered. The LOWFS detector will be operated at 250 K and both the coronagraph imaging and IFS detectors will be operated at 165 K to minimize dark current. All three detectors will be passively cooled with radiators. Blanketing will isolate the coronagraph radiatively from the ~220 K instrument carrier. The coronagraph temperature will be maintained

| Mask | SPC [Imaging] | SPC [IFS] | HLS [Imaging] |
|------|---------------|-----------|---------------|
| Pupil Mask | 1 | 3 | N/A |
| Focal Plane Mask | 2 | 9 | 2 |
| Lyot Mask | 1 | 1 | 1 |
| Field Stop Mask | N/A | N/A | 2 |
| Color Filters | 2 | 3 | 2 |

**Table 3-6: Coronagraphic mask/filter sets for each configuration. The table shows the number of masks or filters required in each configuration at each location in the optical beamtrain.**

| $\lambda_0$ (nm) | $\Delta\lambda$ FWHM/$\lambda_0$ (%) | Science Purpose | Polarization | Channel | Coron. |
|------------------|--------------------------------------|-----------------|--------------|---------|--------|
| 465 | 10.0 | continuum, Rayleigh | Pol. | Imager | HLC |
| 565 | 10.0 | continuum, Rayleigh | Pol. | Imager | HLC |
| 835 | 6.0 | $CH_4$ continuum | Pol. | Imager | SPC |
| 885 | 5.6 | $CH_4$ absorption | Pol. | Imager | SPC |
| 660 | 18.0 | $CH_4$ spectrum | Unpol. | IFS | SPC |
| 770 | 18.0 | $CH_4$ spectrum | Unpol. | IFS | SPC |
| 890 | 18.0 | $CH_4$ spectrum | Unpol. | IFS | SPC |

**Table 3-7: Bandpass filters for planet imaging, spectroscopy, and dust debris disk imaging.**

with its electronics power dissipation in addition to software-controlled heaters.

In order to optimally satisfy the three science objectives of planet imaging, planet spectral characterization and dust debris imaging, the coronagraph is designed to operate in two different modes. Using active mechanisms that change pupil masks, focal plane masks, Lyot masks, and band-pass filters, a single optical beamtrain can be operated as either a Hybrid Lyot Coronagraph or a Shaped Pupil Coronagraph. The Hybrid Lyot Coronagraph mode will be used for planet imaging, whereas the Shaped Pupil Coronagraph mode will be used for planet spectral characterization and dust debris imaging. This unique dual mode architecture not only allows the optimum coronagraph configuration to be used for the different types of science investigation, it also serves as redundancy to mitigate risk of an operational problem with one of the two modes. The two coronagraph modes are shown schematically in Figure 3-27 and Figure 3-28. There are eight electromechanical mechanisms in the coronagraph. Six of the eight use an identical design. The other two are the FSM and the focus adjust mechanism. The six identical wheel mechanisms are in sequence from entrance to exit: the pupil mask wheel, the focal plane mask wheel, the Lyot mask wheel, the field stop wheel, the waveband filter wheel, and the camera selector wheel. The coronagraphic mask sets and color filters are shown in Table 3-6 and Table 3-7, respectively.

### 3.4.1.1.2.1 Structural Design

The output of the telescope is relayed to the coronagraph via a pair of fold mirrors mounted on the instrument carrier. The coronagraph instrument (CGI) consists of a collimator assembly, the coronagraph core optics, an integral field spectrograph (IFS), optical benches, support structure, a thermal cover, cameras, and electronics. The optical beam train is laid out in two geometric planes, with the exception of the two fold mirrors at the telescope output. The collimator assembly and IFS optics exist in one plane and the coronagraph core optics including imaging camera and LOWFS fall in a single parallel plane. This is shown clearly in the right side image in Figure 3-29. The coronagraph instrument will be housed in a carbon fiber composite frame with a triangular support structure. The support structure will be attached to the instrument carrier (IC) via three latches of a design class that has been proven on HST (the 3-2-1 latch system). The coronagraph support structure is designed to enable robotic servicing of the entire instrument. The optical beamtrain consists of three separate opto-mechanical sub-systems: the collimator assembly, the coronagraph core optics and the IFS. Each sub-assembly consists of optics mounted to an optical bench. The optical benches, as well as the instrument supporting structure, are fabricated of carbon fiber composite material for its stiffness and athermal behavior. The structure and three optical benches are shown in Figure 3-29.

An integrated finite element model (FEM) of the coronagraph, instrument carrier and telescope was developed to evaluate the structural design. The mass for the coronagraph instrument including all structure is 120 kg. A modal analysis of the integrated hardware includes multi-layer insulation (MLI), heaters, radiators, thermal straps, temperature sensors, and heat pipes.

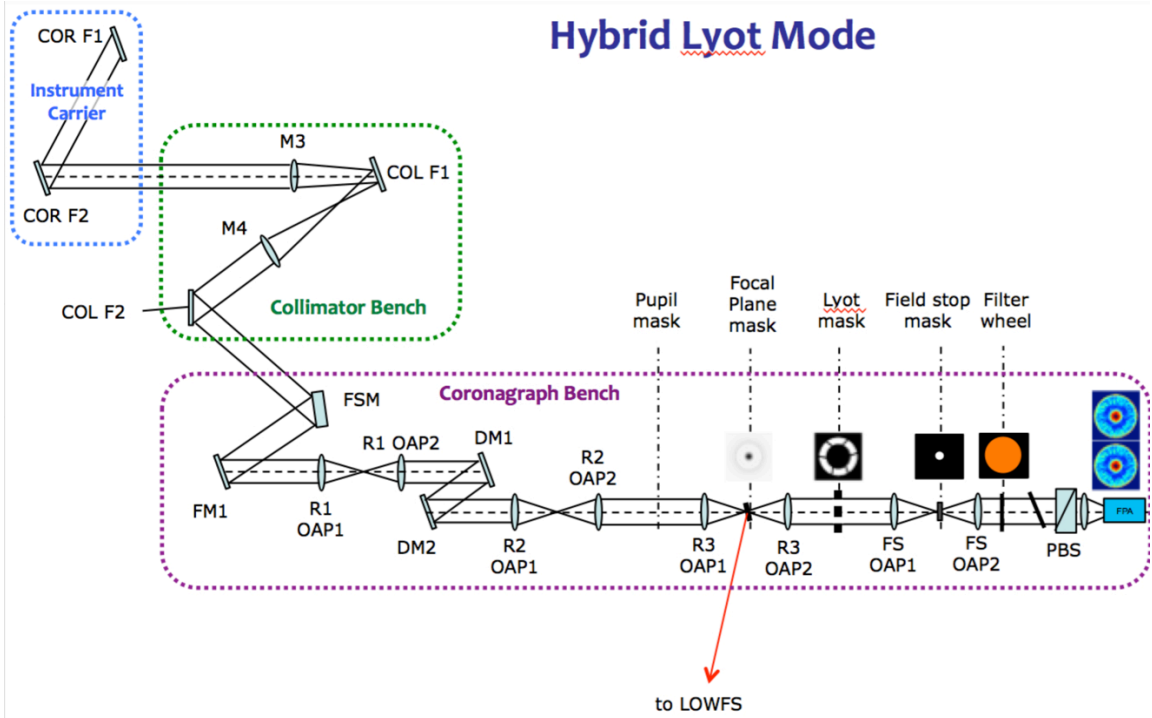

**Figure 3-27: The Hybrid Lyot Coronagraph mode is shown. The output is detected by the imaging camera (reimaging optics shown as lenses for simplicity).**

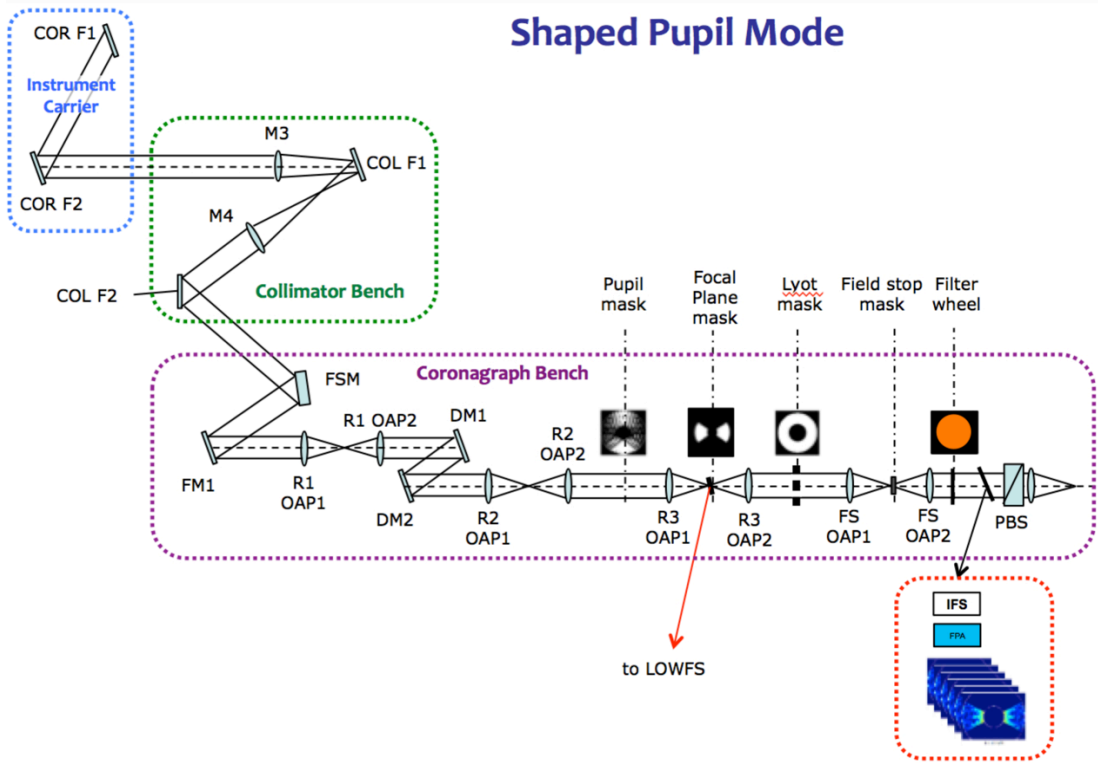

**Figure 3-28: The Shaped Pupil Coronagraph mode is used with the IFS for spectral characterization (reimaging optics shown as lenses for simplicity).**

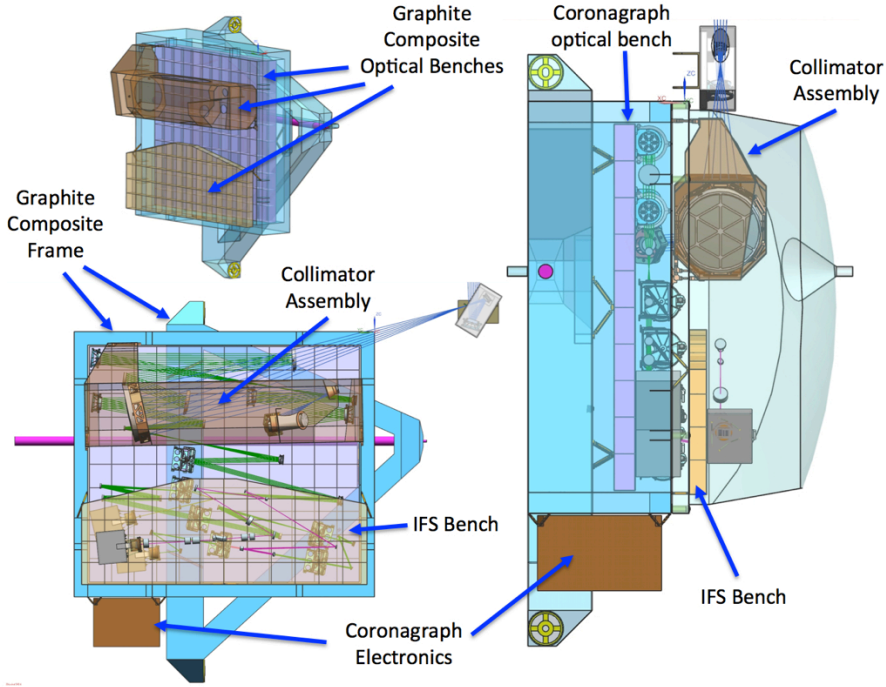

Figure 3-29: Coronagraph supporting structure and three opto-mechanical benches.

A thermal math model (TMM) of the Coronagraph was developed to support structural thermal optical (STOP) analysis and trade studies. The TMM simulates all the major components and structures of the Coronagraph. Figure 3-30 shows the TMM-generated temperature maps of the coronagraph optical benches. The TMM was used to evaluate the final design prior to proceeding to the STOP analysis of coronagraph performance.

FEM was carried out to ensure that the structure meets the resonance frequency requirements.

### 3.4.1.1.2.2    Thermal Design and Modeling

The thermal design of the Coronagraph utilizes both passive and active controls to achieve the desired operating temperature and thermal stability while minimizing the heater power required and the mass. The Coronagraph consists of the fold mirror assembly, optics, optical benches, support structure, thermal cover, cameras, and electronics. The support structure, thermal cover, cameras, and electronics are all individually wrapped with multi-layer insulation (MLI). The Coronagraph support structure is attached via kinematic mounts to the instrument carrier, on which the fold mirror assembly is mounted. The optical benches and thermal cover are isolated from the coronagraph support structure. There are three cameras that are supported off the electronics via bipods. The electronics are mounted to both the support structure and optical benches. The thermal

### 3.4.1.1.2.3    Electrical Design and Data Rates

The coronagraph instrument electronics (shown in Figure 3-31) has dual string redundancy in key electrical subsystems where significant risk can be retired. The spacecraft will supply redundant power at ≈29 V to redundant DC-DC converters in the instrument. A redundant 1553 interface will handle commands and housekeeping, while redundant Space-Wire interfaces will be used for science data. The spacecraft will read a limited number of internal coronagraph temperature sensors for thermal monitoring while the coronagraph is off. Command and data handling will be performed by a RAD750 flight computer and a field-programmable gate array (FPGA). The instrument computer is single string due to its proven reliability. The two DMs will be driven by separate power supplies and electronics. If one DM or driver completely fails, the second DM will still allow much of the planned science. The failure of discrete DM actuators or individual DM elements (or the drive electronics for individual DM elements) will result in a graceful degradation of DM performance. All actuators will have redundant windings, with redundant drive electronics (only one powered). Detector drive electronics will also be redundant.

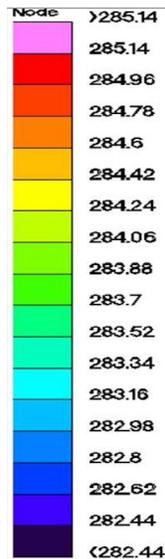
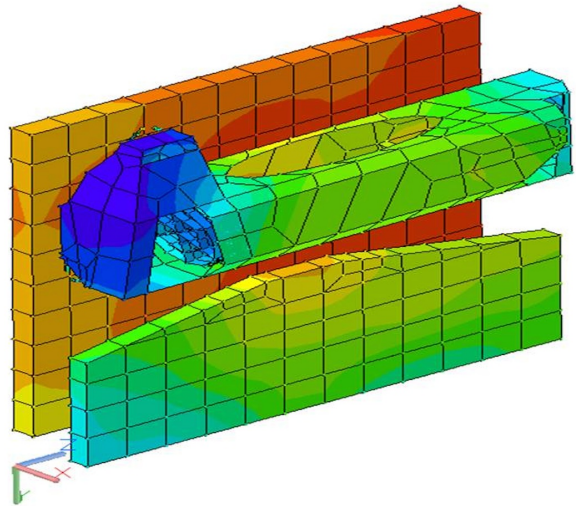

Figure 3-30: Coronagraph thermal math model simulation of the temperature profile on the three coronagraph optical sub-benches following a slew from a bright star to a science target star. This is representative of the temperature profiles that are used in the coronagraph STOP analysis. The gradients shown in the figure (~2.5 K) are from the preliminary thermal control design with further optimization possible.

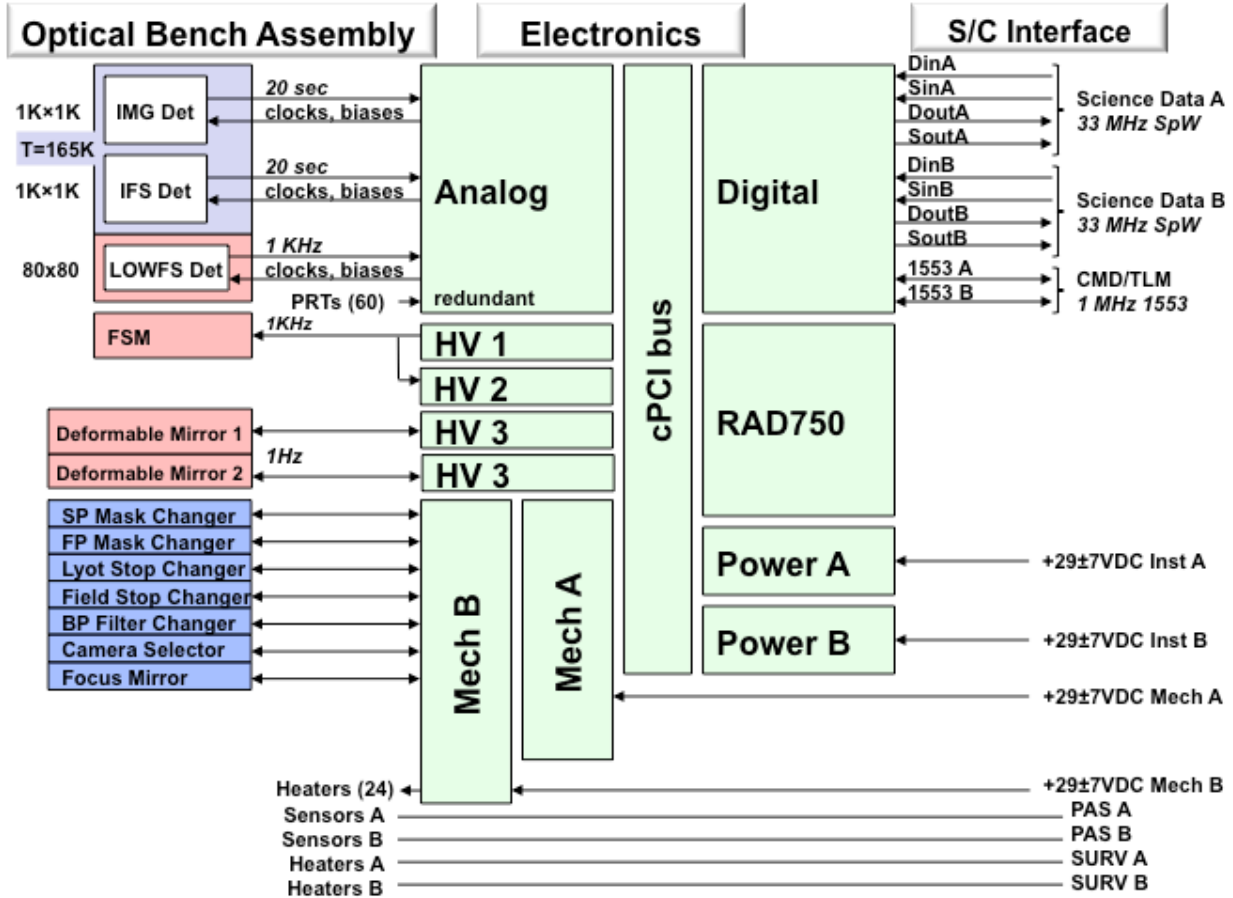

**Figure 3-31: Coronagraph electronics system block diagram.**

There are three cameras in the coronagraph, the imaging camera, the IFS camera and the LOWFS camera. The average raw, uncompressed data rate for the coronagraph instrument is estimated to be a very modest 235 kbps. The assumptions underlying this estimate are as follows.

Since the CGI will observe over a small field, the two coronagraph science cameras have only 1 million pixels each. Only one of the two CGI science cameras will be used during an observation or diagnostic run and frame times will be in the 100's of seconds for target stars and many 10's for bright stars. Moreover, only a fraction of the pixels from each imaging camera frame will be used. The nearly raw IFS frames will be sent to the ground where spectral data cubes will be produced as part of the data post-processing activity. Although the LOWFS camera will be operated nearly 100% of the observing time, only a very small subset of camera pixels will be used to capture the wavefront.

The coronagraph image plane and IFS detectors must have extremely low noise. Dark current noise must be no more than 1 e- over a thousand to several thousand seconds, while read noise must be no more than 1 e- for the sum of 100-1000 reads. Silicon detectors can provide the requisite low dark-current when cooled to ≈165 K. Very low read noise is more difficult to obtain. A trade study, carried out to identify the best possible detector, resulted in the selection of the e2v model CCD201 electron-multiplying charge-coupled device (EMCCD) as the baseline coronagraph detector for imaging and spectroscopy (see Figure 3-32 and Table 3-8). As part of the trade study, other devices considered were standard CCDs (like the type flown on Kepler), deep depletion CCDs (Gaia) and sCMOS sensors. Only EMCCDs have demonstrated the required low read noise, attained by means of the analog gain. The e2v model CCD201 has a format of 1K×1K pixels with 13 micron pitch and can be operated in three modes: i) standard CCD (no gain), ii) analog gain and iii) gain with photon counting. For most of the coronagraph operations the EMCCD detector will be operated in the latter two modes but the capability to operate as a standard CCD provides good risk mitigation. The CCD201 is a high TRL device that is offered by e2v as a standard product, used not only by astronomers but also by medical device suppliers. e2v is developing 4K×4K version of the CCD201 which offers performance advantages for the IFS. The WFIRST-AFTA Coronagraph team will monitor the progress of this larger ar-ea device and in the event that its TRL advances soon enough for the coronagraph, the baseline design could change to utilize this newer device.

| Specification | CCD201-20 Expected Value | Requirement | Unit | Notes |
|---|---|---|---|---|
| Active pixels | 1024 x 1024 | 1024 x 1024 | --- | --- |
| Pixel pitch | 13 x 13 | 13 x 13 | microns | Effective area: 177.2mm$^2$ |
| Effective read noise | 0.2 | 0.2 | e$^-$ | Achieved using EM gain |
| Saturation signal per pixel | 50863 | N/A | e$^-$ | EM amp full well @ 1MHz vertical frequency |
| Dark current | 9 x 10$^{-5}$ | 1 x 10$^{-4}$ | e$^-$/pix/sec | Temp 165K, IMO |
| Clock induced charge (CIC) @ 5$\sigma$ threshold | 0.0013 | 0.0018 | e$^-$/pix/frame | 10 MHz horizontal frequency; 1 MHz vertical frequency; EM gain=1000 |
| Quantum efficiency | 88 | 88 | % | Value at 660 nm, 165K |
| | 68 | 68 | % | Value at 770 nm, 165K |
| | 28 | 28 | % | Value at 890 nm, 165K |

**Table 3-8: Coronagraph detector requirements and e2v CCD201 expected performance. Expected quantum efficiency values correspond to the "Basic Multi2" AR coating for standard thickness silicon reported by e2v. Expected CIC value as reported by NüVü in the EMN2 camera with CCCP controller. IMO = inverted mode operation.**

In-orbit radiation damage to the detector can increase the read noise considerably. Work is currently underway to characterize the performance degradation of the CCD201 in the relevant environment over the mission lifetime. In addition, operating techniques such as pocket pumping will be tested to determine to what extent the radiation induced degradation can be mitigated in flight.

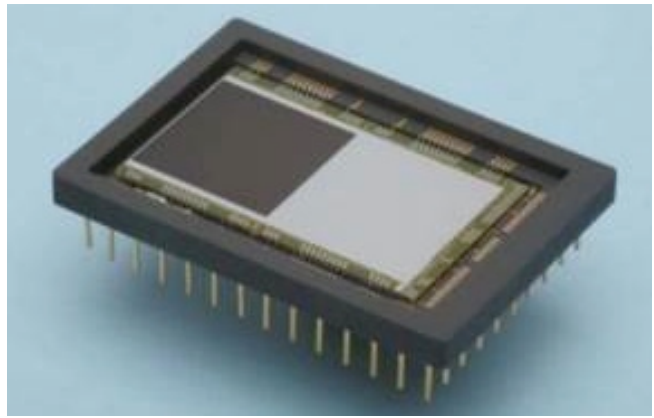

**Figure 3-32: e2v model CCD201 EMCCD is a frame transfer device. The dark region on the left is the light sensitive area whereas the frame transfer region on the right is coated with a blocking layer.**

| IFS Specifications | | | |
|---|---|---|---|
| Band center wavelength (nm) | 660 | 770 | 890 |
| λ.min (nm) | 600 | 700 | 810 |
| λ.max (nm) | 720 | 840 | 970 |
| # of dispersed pixels | 25.2 | | |
| Instantaneous bandpass | 18% | | |
| f/# | 870 | | |
| Lenslet pitch (μm) | 174 | | |
| Lenslet sampling at $\lambda_c$ [# lenslets/( $\lambda_c$ /D)] | 3.3 | 3.9 | 4.6 |
| FoV (# of  $\lambda_c$/D*) [radius] | 14.5 | 12.4 | 10.7 |
| FoV (arcsec) [radius] | 0.82 x 0.82 | | |
| Pinhole diameter [microns] | 25 – 30 | | |
| Lenslet array format | 76x76 | | |
| Magnification from lenslet to detector | 1:1 | | |
| Spectral resolution Δλ/λ [over 2 pixels] | 70 ± 5 | | |

**Table 3-9: Design specifications for the IFS**

### 3.4.1.2  *Integral Field Spectrograph*

The WFIRST-AFTA Coronagraph's integral field spectrograph (IFS) was designed to meet the baseline science requirements (see Table 3-9). The IFS is required to cover a spectral range from 600-970, corresponding to a full spectral bandpass of 47%. Due to chromatic effects, the wavefront control system is only able to correct at high contrast up to an 18% instantaneous bandwidth (see §3.4.1.3). To date, testbed demonstrations on unobscured pupils have achieved high contrast on up to 20% bandwidths. Simulations suggest similar bandwidths will be achievable with the selected coronagraphs, using the WFIRST-AFTA obscured pupil. Demonstration of high contrast with obscured pupils using the IFS will take place in the HCIT as part of the technology program. The 47% IFS bandwidth is separated into three different 18% spectral bands each covering diagnostic spectral features. These spectral bands have central wavelengths at 660nm, 770nm, and 890nm, where each band is independent and does not overlap with the next.

The IFS spectral resolution requirement of R~70 across the entire spectral bandpass requires a low-resolution dispersion element. R~70 means that each spectral channel must be 1.4% in width (B=1/R). At the shortest wavelength of 600nm, the spectral resolution element is then required to be no more than 8.6 nm in width. The line spread function, i.e., resolution element, must be critically sampled on the detector, leading to 4.4 nm/pixel. At the longest wavelength of 970nm, the resolution element is requested to be less than 14.2 nm with a pixel scale of 7.14 nm. The intermediate wavelengths follow the same scalings.

The IFS must achieve high astrometric and spectrophotometric precision. The spectrophotometric precision of the instrument should be 0.06 mag (5%) with respect to the host star. This level of precision enables a S/N~20 spectrum to not be limited by the IFS.

The spatial sampling is accomplished using a lenslet array placed at a focal plane within the instrument, where each lenslet focuses the incident light into a well-separated grid of compressed images. The process of spatial sampling (lenslet array), spatial filtering (pinhole mask) and spectral dispersion (prism assembly) is schematically represented in Figure 3-33. At the lenslet image focus, a small aperture

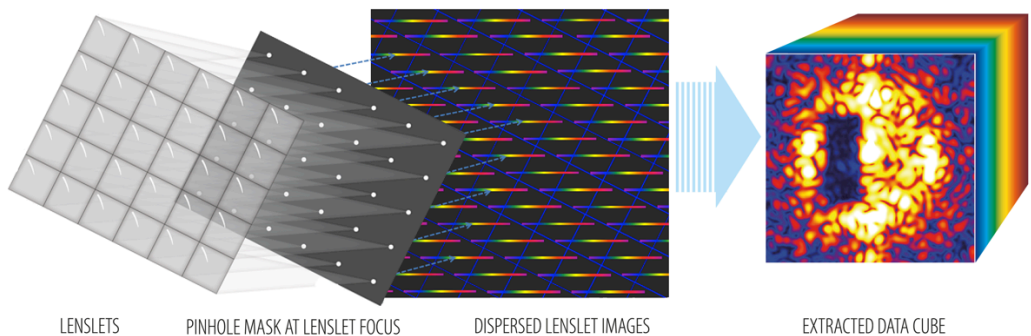

LENSLETS          PINHOLE MASK AT LENSLET FOCUS          DISPERSED LENSLET IMAGES          EXTRACTED DATA CUBE

**Figure 3-33: The coronagraph IFS uses a lenslet array for separating field points. The lenslets are clocked relative to the dispersion axis in the spectrograph so the spectra do not overlap on the focal plane.**

mask blocks the light diffracted out of the core. The compressed lenslet images are collimated (not shown), dispersed (not shown), and focused (not shown) onto an EMCCD detector to produce more than five thousand individual small spectra. These can then be extracted and assembled to form a data cube.

### 3.4.1.3  *Low-Order Wavefront Sensor*

To maintain the coronagraph contrast during the scientific observation, the coronagraph's wavefront needs to be stable at level of tens of picometers. Although the WFIRST-AFTA telescope is well-designed, it has LOS jitter originating from vibration sources such as the observatory attitude control reaction wheel and High Gain Antenna (HGA), as well as some low-order wavefront aberrations resulting from thermally induced optics surface distortions and optical misalignment. WFIRST-AFTA dynamic and thermal analysis shows that during a typical observation period the telescope suffers LOS jitter (fast) at the milli-arcsecond level, pointing drift (slow) of a few milli-arcseconds, and RMS wavefront variation dominated by focus and coma aberrations of approximately 100 picometers.

To achieve the required coronagraph performance in a realistic space environment WFIRST-AFTA incorporates a Low-Order Wavefront Sensing and Control (LOWFS/C) subsystem. LOWFS/C uses the rejected star light from the coronagraph to sense and suppress the telescope line-of-sight jitter, the pointing drift, and the low order wavefront errors. The measured wavefront information will also be recorded and used in ground operations for the coronagraph image post-processing needed to further remove the speckle field and enhance contrast.

The WFIRST-AFTA LOWFS (low-order wavefront sensor) uses a Zernike phase contrast wavefront sensor. The concept of phase contrast was first introduced in 1935 by Frits Zernike to enhance a microscope image by transforming an apparently transparent biological specimen fluid into a high-contrast object image outlining the specimen. The principle of the Zernike wavefront sensor works as follows: a small phase disk is placed at the center of the focused star image or point-spread-function (PSF). The phase disk introduces an optical path difference (OPD) of a quarter of a wavelength. The phase disk has a radius of about 1/2 that of the first Airy ring of the PSF. The PSF light outside the phase disk transmits the focal plane without any phase modulation. When light is observed at the reimaged pupil plane the star light forms an intensity variation across the pupil that results from interference between the light that experienced the phase delay induced by the phase disk and the light transmitted without modulation. The intensity pattern from the interference mimics the OPD map from the wavefront aberration, which is the phase variation across the pupil. In essence, the Zernike wavefront sensor's phase mask at the focal plane transforms the wavefront phase map into an intensity map at the pupil plane which can then be directly observed by a low-noise, fast-readout (~1000 frame-per-second) CCD camera. The LOWFS uses differential image based software to calculate the wavefront error from the intensity pattern. The measured LOS jitter and LOS drift, manifested as the wavefront slope across the pupil, is sent to the Fast Steering Mirror (FSM) to correct and stabilize the line-of-sight. The measured focus error, which is usually the dominant low-order wavefront error, is sent to a fast focusing mechanism for correction. The other measured low-order wavefront errors, such as comas and astigmatisms, are sent to the DM that is conjugated to the system pupil for correction. The measured wavefront error is recorded along with the coronagraph science observation and used in post processing to further reduce speckle and increase the image contrast.

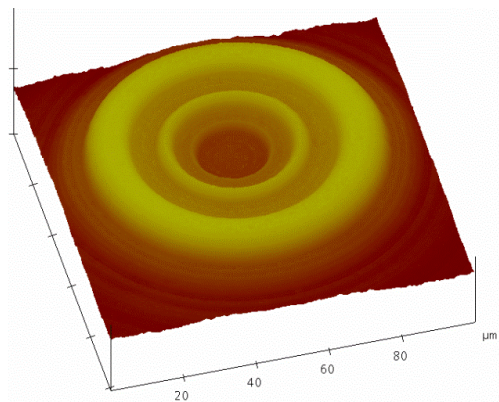

**Figure 3-34: A Hybrid Lyot Coronagraph (HLC) occulting (focal plane) mask dielectric profile map measured by an atomic force microscope. The area shown is about 100x100 microns. The recessed center area is the Zernike phase disk used by the LOWFS.**

To implement the Zernike wavefront sensor, the coronagraph's starlight rejecting mask and the LOWFS phase disk are consolidated into a single focal plane mask as shown in Figure 3-34. The combination of the coronagraph occulting mask and LOWFS phase mask into a single mask simplifies the design of the LOWFS optics and, more importantly, reduces the optical non-common-path error which is the difference between the LOWFS measured wavefront and wavefront error experienced by the coronagraph. The latter is what LOWFS/C is trying to sense and correct. Because the occulting mask reflects only starlight while transmitting the nearby planet light, the occulting mask acts as a low-pass spatial filter for LOWFS. The light fed into the LOWFS contains only low spatial frequency wavefront information and thus the LOWFS can only sense low-order wavefront error. Fortunately, the coronagraph wavefront error, resulting from vibrations and thermal drift, is primarily low order in nature. This ensures the LOWFS is suitable for the WFIRST-AFTA coronagraph.

Following the occulting mask, the rejected light is collected by the LOWFS imaging optics and forms a pupil image on the LOWFS sensor, a fast readout CCD camera. Since LOWFS senses only low-order wavefront error this pupil image is sampled by 8×8 pixels. Concentrating the LOWFS light in a relatively small number of pixels has the advantage of low photon statistical noise, which limits LOWFS performance, as well as rapid CCD frame read-out enabling fast jitter correction. A LOWFS test-bed is part of the Coronagraph technology development program. The goal is to advance the LOWFS technology to NASA TRL-5 in accordance with the Coronagraph Technology Development Plan (see §3.4.2.1).

As part of the technology development effort the HLC occulting mask with LOWFS phase disk was recently tested. Preliminary testing and calibration have shown that the Zernike sensor is quite sensitive to the low-order wavefront error and is capable of sensing an equivalent of sub-milli-arcsecond LOS tilt needed for the WFIRST-AFTA coronagraph. The sensor clearly measures the wavefront error caused by the atmospheric turbulence. When a wavefront tilt is introduced by a PZT-actuated tilt mirror, the Zernike sensor responds with corresponding sign and amplitude. An example of the Zernike sensor readout is shown in Figure 3-35. The test data agrees well with the model predictions.

### 3.4.1.4 Calibration Hardware

To simulate the WFIRST-AFTA telescope jitter and drift a representative Optical Telescope Assembly (OTA) Simulator is being designed and built. It will be used as the surrogate for the WFIRST-AFTA telescope. The OTA Simulator also serves as the calibration gauge for the LOWFS performance evaluation and testbed model validation.

The OTA Simulator, shown in Figure 3-36, consists of a pinhole star source, a small scale replica of the WFIRST-AFTA telescope with a primary mirror diameter of 2 inches, a pupil mask which contains the central obscuration and secondary mirror supporting struts, a jitter mirror, an actuated off-axis paraboloid (OAP), and pupil relay optics. The OTA telescope mirrors and an OAP are moved with PZT actuators and can be controlled precisely with displacements as small as a fraction of a nanometer. This displacement creates a precisely controlled low-order wavefront error on the order of picometers. The jitter mirror, actuated by PZT tilt stages, can introduce the high-temporal frequency LOS jitter, simulating the WFIRST-AFTA wavefront jitter from the vibration sources. The color-filtered light source is fed to the OTA Simulator via a single mode optical fiber. The spectral bandwidth and intensity is controlled in the source module located outside the OTA Simulator. The OTA Simulator optics sit on a small 1.5' x 2.5' optical bench. The light from the OTA Simulator is collimated and pupil imaged at the FSM and then transmitted into either the LOWFS/C optics for stand-alone testing or the dynamic High Contrast Imaging Testbed (HCIT). The entire LOWFS/C testbed will be operated in a vacuum chamber to avoid atmospheric turbulence. The OTA Simulator sub-bench is thermally insulated by MLI to enhance its stability.

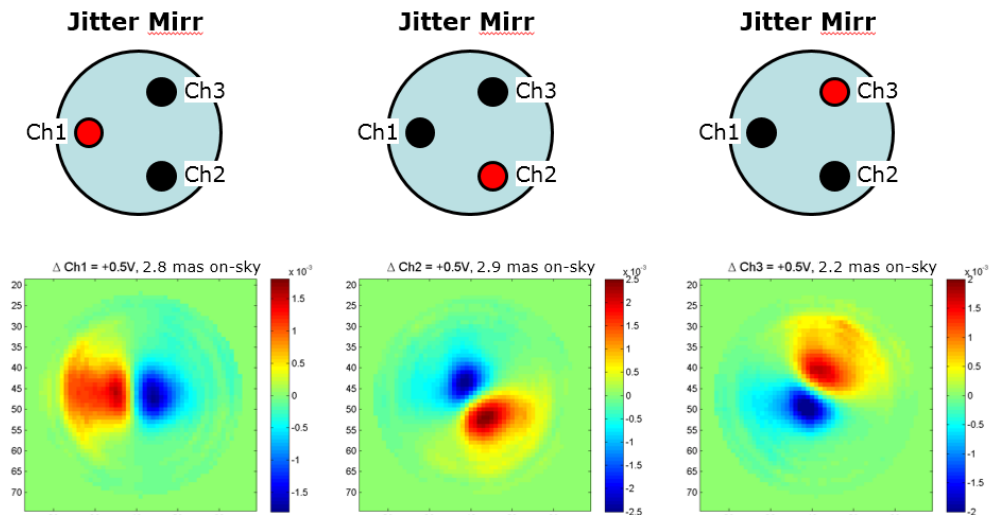

**Figure 3-35: LOWFS in-air experimental results. The jitter mirror has three PZT actuators, each was driven by +0.5V which introduces an equivalent on-sky line-of-sight tilt of ~3 milli-arcseconds. The driven actuator (channel) is indicated by the red dot in the upper row plots. The corresponding LOWFS signal, which is from the image differences between the tilted and reference images are shown in the plots in the lower row. The signature of wavefront tilt is shown as the positive and negative blobs of image intensity variation in the LOWFS signal. The signals morphology matches the modeled signal well.**

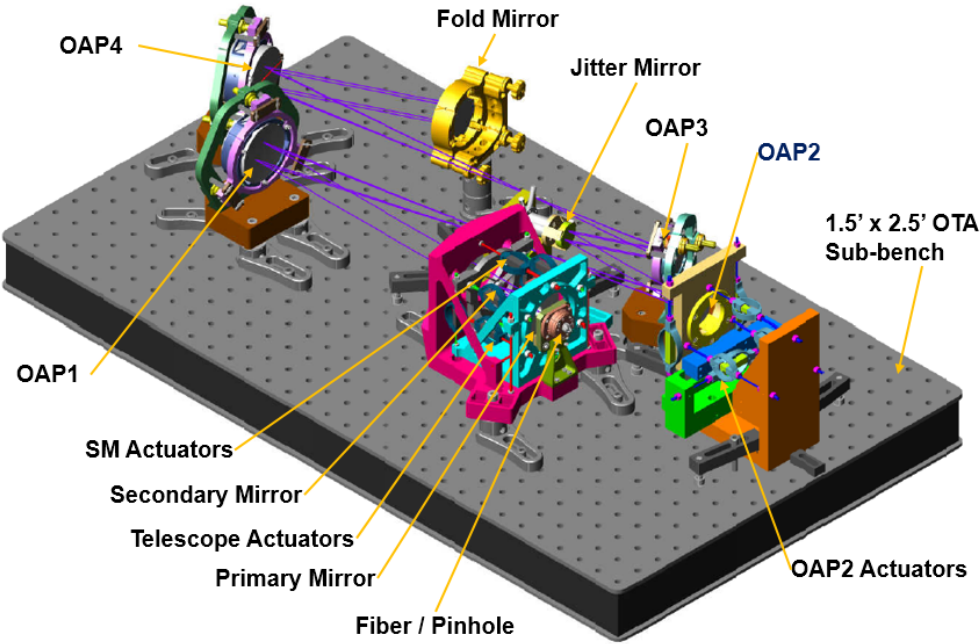

**Figure 3-36: OTA Simulator design.** The OTA Simulator uses precise motion of the optics generated by the PZT actuators to simulate the WFIRST-AFTA telescope jitter of order of milli-arcsecond and wavefront drift of 10s of picometers. The optics of the OTA Simulator sit on the 1.5' x 2.5' optical bench. The optical components are labeled in this CAD plot.

Since the OTA Simulator can produce small wavefront errors, it can be used to calibrate the LOWFS, as well as during the coronagraph testbed model verification. Later, it may also be used as the surrogate WFIRST-AFTA telescope during the Coronagraph Instrument integration and test.

### 3.4.2 *Enabling Technology*

#### 3.4.2.1 *Coronagraph Technology Development Plan and Milestones*

Soon after the WFIRST-AFTA CGI architecture downselect results were announced in December 2013, the coronagraph technology development plan (TDP) was drafted and approved in March 2014. The main objective of the TDP is to mature the WFIRST-AFTA coronagraph to Technology Readiness Level (TRL) 5 by the planned Project Phase A start at the beginning of FY 2017. To track the technology progress along the way, the TDP calls out 9 key milestones in FY 2014-2016, which are listed in Table 3-10. Passing each milestone requires a formal review and concurrence by the independent Technology Assessment Committee (TAC). The first 3 milestones scheduled for completion during calendar 2014 were successfully reviewed and approved by the TAC.

The coronagraph technology development is split into 2 phases: static and dynamic. During the static phase in FY 2014-2015, the experimental validation of both primary coronagraph models and the development of the low-order wavefront sensor are pursued in parallel on separate testbeds. This includes fabrication of pupil-plane and focal-plane masks de-signed for the WFIRST CGI, and demonstration of wavefront control using these masks to achieve the required starlight suppression with a static input wavefront. During the same time, the LOWFS/C demonstration task involves designing and fabricating the optical telescope assembly (OTA) simulator and using it to produce wavefront changes, which are then sensed and corrected by the LOWFS/C subsystem.

| M# | Milestone Title | Date |
|---|---|---|
| 1 | First-generation reflective Shaped Pupil apodizing mask fabricated with black silicon specular reflectivity of less than $10^{-4}$ and 20 μm pixel size. | 7/21/14 **Passed** |
| 2 | Shaped Pupil Coronagraph in the High Contrast Imaging Testbed demonstrates $10^{-8}$ raw contrast with narrowband light at 550 nm in a static environment. | 9/30/14 **Passed** |
| 3 | First-generation PIAACMC focal plane phase mask with at least 12 concentric rings fabricated and characterized; results are consistent with model predictions of $10^{-8}$ raw contrast with 10% broadband light centered at 550 nm. | 12/15/14 **Passed** |
| 4 | Hybrid Lyot Coronagraph in the High Contrast Imaging Testbed demonstrates $10^{-8}$ raw contrast with narrowband light at 550 nm in a static environment. | 2/28/15 |
| 5 | Occulting Mask Coronagraph in the High Contrast Imaging Testbed demonstrates $10^{-8}$ raw contrast with 10% broadband light centered at 550 nm in a static environment. | 9/15/15 |
| 6 | Low Order Wavefront Sensing and Control subsystem provides pointing jitter sensing better than 0.4 mas rms per axis and meets pointing and low order wavefront drift control requirements. | 9/30/15 |
| 7 | Spectrograph detector and read-out electronics are demonstrated to have dark current less than 0.001 e/pix/s and read noise less than 1 e/pix/frame. | 8/25/16 |
| 8 | PIAACMC coronagraph in the High Contrast Imaging Testbed demonstrates $10^{-8}$ raw contrast with 10% broadband light centered at 550 nm in a static environment; contrast sensitivity to pointing and focus is measured. | 9/30/16 |
| 9 | Occulting Mask Coronagraph in the High Contrast Imaging Testbed demonstrates $10^{-8}$ raw contrast with 10% broadband light centered at 550 nm in a simulated dynamic environment. | 9/30/16 |

**Table 3-10: Key coronagraph technology development milestones.**

In the dynamic testing phase during FY 2016, the components and subsystems developed and validated during the static phase will come together to demonstrate TRL 5 coronagraph performance. With the OTA simulator injecting wavefront disturbances expected on orbit – both the line-of-sight jitter and slow thermally driven wavefront changes – the Occulting Mask Coronagraph (OMC) testbed, convertible between HLC and SPC modes and integrated with LOWFS/C subsystem, will demonstrate starlight sup-pression in a representative environment, as required for meeting NASA's TRL 5. The PIAACMC backup coronagraph will demonstrate the required starlight suppression with a static input wavefront and go through an "open loop" testing of contrast sensitivities to wavefront tip, tilt, and defocus. The rest of this section will describe in more detail each coronagraph technology development area and the most significant results achieved during 2014.

### 3.4.2.2 Shaped Pupil Coronagraph Status

The shaped pupil coronagraph is one of the OMC instrument's two operating modes. With a design pioneered by the High Contrast Imaging Laboratory (HCIL) at Princeton University, the SPC has a relatively simple architecture and uses an optimized binary pupil apodizer to diffract on-axis starlight in a way that creates a dark hole in the PSF, as described in

§2.4.3.6. A field stop in the image plane blocks the bulk of the starlight outside the dark hole, while the slightly off-axis planet light is passed through the field stop and reimaged onto a detector. Thus, the starlight suppression is largely achieved by the pupil mask, with DM wavefront control improving the coronagraph performance, and the field stop reducing the signal dynamic range on the detector. The shaped pupil is relatively insensitive to observatory jitter, and the pupil mask design is achromatic.

Shaped pupil apodizing masks designed for the WFIRST-AFTA telescope pupil have small "island" features and are not suitable for the traditionally used transmissive mask fabrication approach (Balasubramanian et al. 2013). A new reflective mask fabrication approach, with high-reflectivity aluminum regions and absorptive black-silicon regions (Figure 3-37), was developed and validated at JPL and produced on a silicon wafer. A number of masks using this approach have been fabricated at JPL and Caltech in 2014, with significant quality improvements along the way. A mask designed to produce two 60° "bow-tie" shaped dark holes was delivered to the SPC static testbed in April of 2014. Prior to delivery, this mask underwent extensive characterization for a variety of imperfections and analysis/modeling of their expected impact on coronagraph contrast, with the results summarized in Table 3-11. These results showed that the reflective mask quality is consistent with achieving good SPC testbed (and on-orbit) performance, and were approved by the TAC as meeting Milestone 1 success criteria.

During the summer of 2014, high contrast (starlight suppression) was obtained with this mask in the static shaped pupil coronagraphic testbed, using both narrowband light (a 2% spectral band around 550 nm) and broadband light (a 10% spectral band around 550 nm), as shown in Figure 3-38. Since two deformable mirrors were not available at that time for the SPC testbed, these experiments were performed with 1

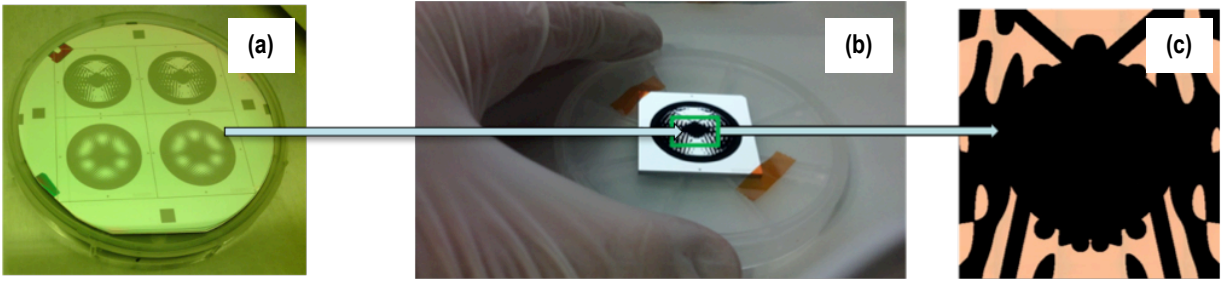

**Figure 3-37: Reflective shaped pupil: (a) a processed wafer with 2 discovery and 2 characterization masks, (b) a fabricated characterization mask, (c) design details showing absorbing black silicon and highly reflective aluminum. Colored tints on (a) and (c) are caused by microscope illumination.**

| Mask Imperfection Type | Measured Level | Predicted Change in Contrast after WFC | Comments |
|---|---|---|---|
| Black Si reflectivity, specular | $<7 \times 10^{-8}$ | $<2.1 \times 10^{-10}$ | Upper bound; limited by measurement setup |
| Black Si reflectivity, diffuse | $<0.6\%$ | $<10^{-11}$ | |
| Mask wavefront error | $\sim 0.036\lambda$ rms (above focus) | $7 \times 10^{-11}$ | Post WF control; better wafers received for next version |
| Isolated defects | Small pinholes and 2 scratches | $8 \times 10^{-12}$ | Post WF control |
| Al reflectivity variations | $\sim 0.5\%$ | Correctable with <1% of DM stroke | Post WF control; variations on SP similar to other coronagraph optics |
| **Total** | | **$<3 \times 10^{-10}$** | **Upper bound** |

**Table 3-11: Reflective Shaped Pupil Mask characterization results.**

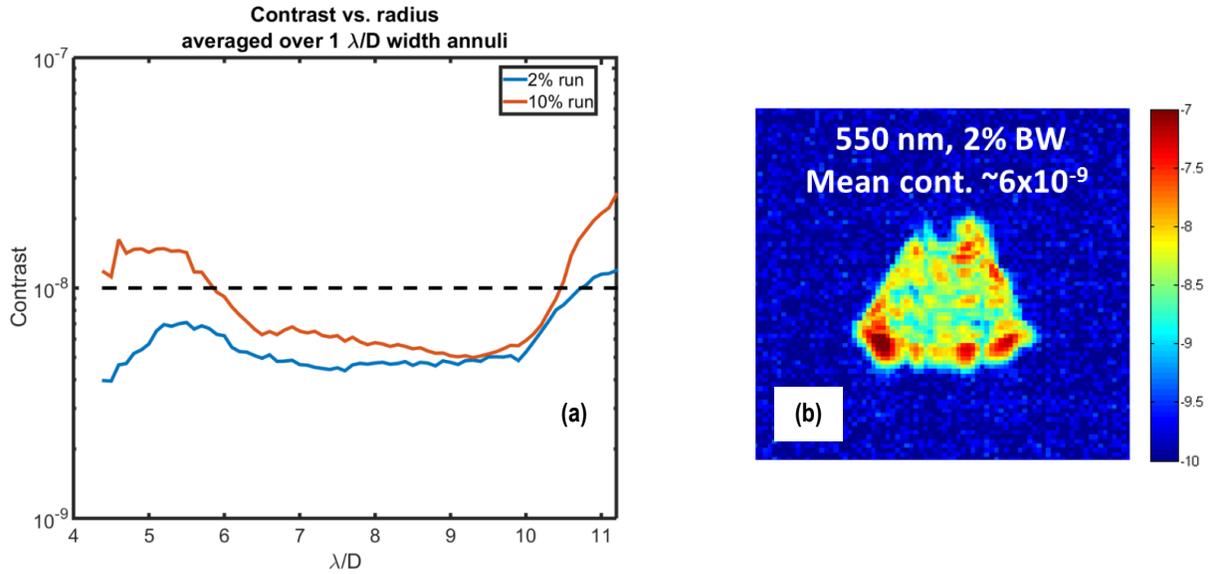

**Figure 3-38: Shaped pupil testbed results: (a) narrowband and broadband contrast as a function of working angle. (b) 2D image of the dark hole with 2% light. Light outside the dark hole is blocked by a field stop. A single-sided dark hole was created with 1 DM.**

DM, producing a one-sided dark hole. Testbed conversion to the 2 DM configuration was recently completed, which will enable demonstration of 2-sided dark holes in 2015. The results in Figure 3-38 were used to pass Milestone 2: narrowband shaped pupil contrast demonstration. In fact they are also consistent with the SPC portion of Milestone 5 that is due 9/30/2015: broadband OMC contrast demonstration.

While the shaped pupil design developed during the CGI architecture downselect was validated in terms of both mask fabrication and testbed performance, the HCIL group at Princeton University produced an improved "Shaped Pupil Lyot Coronagraph" (SPLC) design that incorporates a Lyot stop along with the shaped pupil mask. This design improves the inner working angle to ~2.8 λ/D thus enhancing the science performance. The masks for this SPLC design will be fabricated and demonstrated on the testbed in 2015. No fabrication process changes will be necessary to produce SPLC masks compared to the 1st generation WFIRST-AFTA CGI SPC masks.

### 3.4.2.3 Hybrid Lyot Coronagraph Status

The Hybrid Lyot Coronagraph is the other OMC operating mode. The main HLC starlight suppression component is a focal plane occulting mask with optimally shaped metal disc and dielectric layer profiles, as described in §2.4.3.6. The metal disc reflects the majority of the incoming on-axis starlight toward the LOWFS reimaging optics, while the transmitted starlight is diffracted to be blocked by a Lyot stop in a downstream pupil plane. At the same time, slightly off-axis planet light is passed through the Lyot stop and transmitted to the imaging detector.

A band-limited Hybrid Lyot coronagraph has previously demonstrated the best reported laboratory contrast results with an unobscured pupil (Lawson et al. 2013), exceeding $10^{-9}$ in a 10% band at ~3λ/D working angle. These coronagraph designs used linear occulting masks to create a one-dimensional (and one-sided due to the use of 1 DM) dark hole. The WFIRST-AFTA CGI HLC has a new design capable of achieving high broadband contrast with the WFIRST-AFTA telescope's obscured pupil in the presence of realistic observatory jitter. Since the initial WFIRST HLC design was presented during the downselect process in 2013, design revisions, consisting of optimizing both the occulting mask profile and initial deformable mirror settings, resulted in significant improvements to the predicted contrast performance at small inner working angles with realistic LOS jitter. At the same time, the profiles of metal and dielectric layers were simplified to make the mask fabrication less challenging. For the first time, circular HLC occulting masks were successfully fabricated at JPL's MicroDevices Lab and extensively characterized, as shown in Figure 3-39.

The HLC starlight suppression performance is currently being validated in a dedicated testbed, with work progressing toward Milestone 4 in 2015. The HLC testbed has included, from the start, two 48x48

AOX DMs baselined for the flight instrument. It is also important to point out that for the first time the HLC occulting mask has been designed to not only work as the coronagraph occulter in transmission, but also as the Zernike LOWFS mask in reflection.

#### 3.4.2.4 PIAACMC Status

The backup Phased-Induced Amplitude Apodization Complex Mask Coronagraph (PIAACMC) architecture is based on the concept of apodizing the telescope pupil using reflections from two carefully shaped aspheric mirrors (Guyon et al. 2005). The design variation proposed for the WFIRST-AFTA CGI combines PIAA mirrors with a milder profile and a focal plane phase mask. The PIAACMC concept promised the smallest inner working angle, and was selected as the backup technology to OMC. In July of 2014, the 2nd generation PIAACMC design was delivered featuring 1 deformable mirror, for a 1-sided dark hole, and a simplified optical layout different from the OMC. While the PIAA mirror design was more benign than those already demonstrated under TDEMs, a new and untested component was the focal plane mask, which in the updated PIAACMC design became reflective and divided into 35 rings with >1000 azimuthal zones of varying heights. Fabrication and characterization of this PIAACMC mask and demonstration through modeling that its fabrication quality is consistent with achieving the required contrast was required to pass milestone 3 in December of 2014. Using nanofabrication at JPL's MDL and an extensive suite of characterization equipment consisting of optical, atomic force, and scanning electron microscopes (Figure 3-40), it was shown that the PIAACMC mask quality is sufficient to meet milestone 3. A PIAA mirror design update and tolerancing, followed by the mirror procurement and testbed validation, are the next PIAACMC development steps for FY 2015-2016.

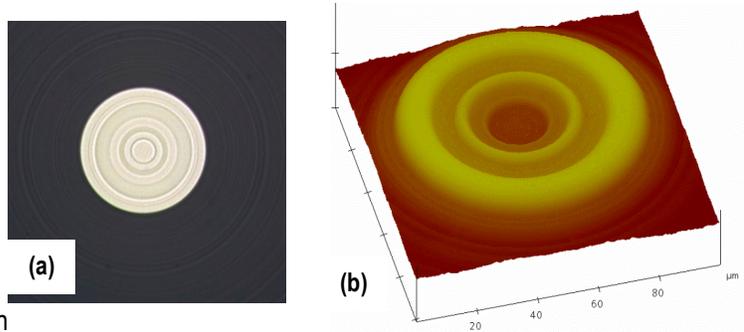

**Figure 3-39: The HLC occulting mask imaged with (a) a differential interference contrast microscope and (b) an atomic force microscope (the central portion of the mask).**

#### 3.4.2.5 Low-Order Wavefront Sensing and Control

To achieve the required coronagraph performance in a realistic space environment, a Low-Order Wavefront Sensing and Control (LOWFS/C) subsystem is necessary. This subsystem provides sensing and suppression of spacecraft LOS pointing drift and jitter as well as low-order wavefront errors driven by changes in thermal loading of the telescope.

The LOWFS/C performance will be initially validated in a dedicated LOWFS/C testbed, before its components are integrated into the dynamic OMC testbed. This demonstration will include the OTA simulator (a miniature version of the WFIRST-AFTA telescope) that generates realistic pointing jitter and low-order wavefront errors, as well as LOWFS/C components and algorithms that provide sensing and closed-loop suppression of these wavefront disturbances to the levels that enable each coronagraph mode to meet its science requirements. Sensing in-formation can also be used on the ground for PSF subtraction and data editing. Closed loop control will suppress the pointing jitter and drift using a Fast Steering Mirror (FSM) with ~50 Hz control bandwidth.

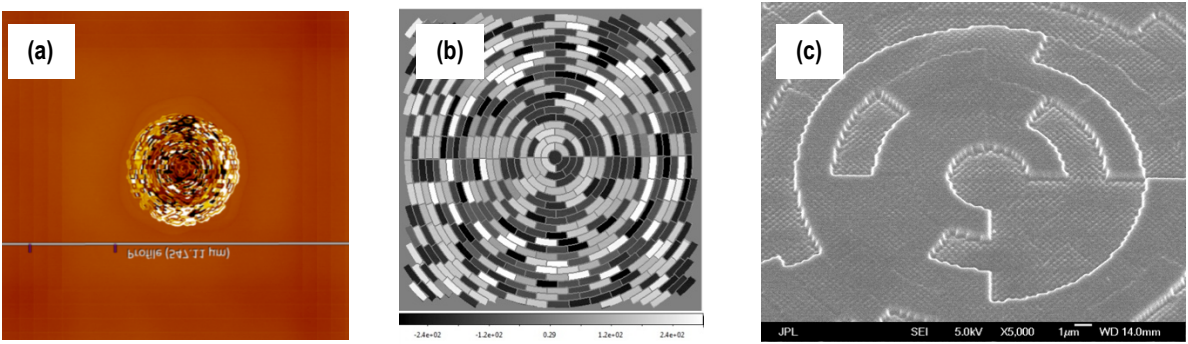

**Figure 3-40: A PIAACMC reflective focal plane mask imaged with (a) an interferometric optical microscope (for larger features), (b) an atomic force microscope image (for fine features), and (c) a scanning electron microscope (for zone transitions).**

The sensor uses starlight discarded by the coronagraph after it is reflected from a common location for both OMC technologies – a focal plane with HLC occulters and SPC field stops. The selected LOWFS architecture and recent progress are de-scribed in detail in §3.4.1.3.

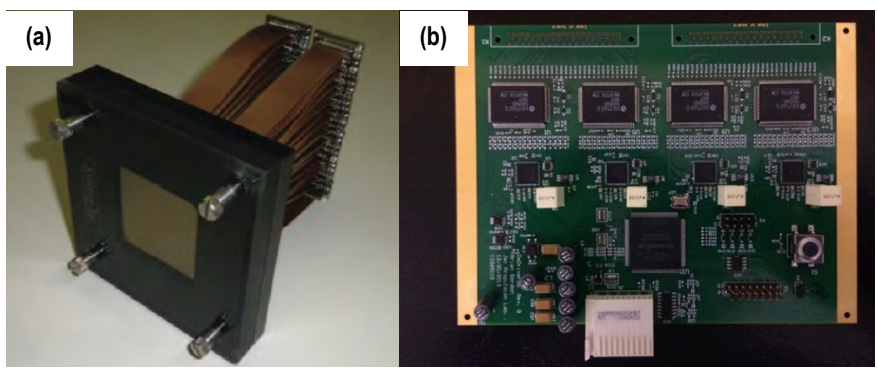

**Figure 3-41: (a) A 48x48 PMN deformable mirror with 24 flex ribbon and connectors, (b) a new generation 128 channel, non-multiplexed, DM driver board.**

The LOWFS/HLC mask shown in Figure 3-39 already underwent initial testing in the LOWFS lab. A calibrated PZT tilt mirror was used to test the sensitivity. These early lab tests have demonstrated that the LOWFS sensor has high sensitivity and is capable of measuring the sub milli-arcsecond tilt needed for meeting Coronagraph Milestone 6, as shown in Figure 3-35. The test image and the LOWFS signals closely matched the model predictions, limited by the laboratory air turbulence. The complete LOWFS/C validation testing will take place in 2015 in a vacuum chamber followed by the integration of LOWFS/C into the dynamic coronagraphic testbed.

### 3.4.2.6 *Deformable Mirror*

The WFIRST-AFTA coronagraph instrument has baselined the monolithic PMN (Lead Magnesium Niobate) Deformable Mirror (DM) technology developed by AOA Xinetics (AOX) (Wirth et al. 2013). JPL has a long history of working with AOX to develop large format, small actuator pitch PMN DMs for high-contrast imaging. Such deformable mirrors were used in the highest contrast testbed demonstrations of all Lyot-type coronagraphs: HLC, SPC, PIAA, and Vector Vortex (Lawson et al. 2013). Two 48x48, monolithic PMN DMs were manufactured by Xinetics for the first time and delivered to JPL in 2009 (Figure 3-41a) with the specifications listed in Table 3-12.

One of these two DMs passed a random vibration test with an overall level of 10.8 g rms in 2012. This test level was based on a survey of random vibration proto-flight test levels for 10 relevant JPL missions. Both DMs are currently being used in the static HLC testbed. New 48x48 DMs are currently being fabricated at AOX, with one of them slated to go through thermal cycling qualification testing.

Recently, a new generation of DM driving electronics for the HCIT has been designed and fabricated to move from multiplexed to non-multiplexed drivers (Figure 3-41b). This change is aimed at eliminating the stroke drift experienced by actuators with lower parallel resistance during multiplexing.

### 3.4.3 *Integration & Test*

The coronagraph instrument integration and test flow has been carefully developed by considering the following: (1) optical and mechanical interfaces for ease of verification; (2) parallel development and verification of key sub-assemblies; (3) key performance verification at the instrument level and minimizing payload level I&T efforts. Coronagraph optics can be divided into two categories: (1) passive optics that form a diffraction-limited beam with the primary, secondary, tertiary, and collimating mirrors; and (2) active optics that suppress starlight for planet imaging and spectroscopic characterizations. Figure 3-42 shows key optical and mechanical interfaces.

The coronagraph integration and test (I&T) flow is shown in Figure 3-43. First the passive optics, such as the fold mirror assembly and the tertiary/collimator sub-bench, are built and tested. They will be delivered to the telescope vendor early to be integrated and tested with the telescope primary and secondary mirrors. This allows early verification of the telescope and its wavefront quality to the coronagraph instrument.

In the meantime, the coronagraph sub-bench is integrated and tested. First the active components (DMs, detectors and actuators) are fully tested. Then the optical bench is assembled and aligned, with a possible seating shake and subsequent alignment check/adjustment. In parallel, the electronics are assembled and fully bench tested. Next the electronics are integrated with the optical bench, and full functional and performance testing is done in vacuum. It may be possible to do some testing in air, using short integration times to avoid detector saturation. Once functionality and performance have been verified, the full instrument assembly can be completed. EMI/EMC testing follows, with any necessary modifications to the electronics or shielding. Then the thermal blankets are installed, followed by mass properties and dynamics testing. After a functional test, the instrument is placed in a vacuum chamber for bake out, thermal cycling, and thermal balance tests. Finally, the full functional and performance testing is done in vacuum with cooled detectors.

| Parameter | Specifications | Notes |
|---|---|---|
| Array Size | 48 x 48 | |
| Actuator Spacing | 1 mm | |
| Stroke Range | ~0.5 $\mu$m | |
| Actuator Capacitance | 50 nF | Nominal |
| Facesheet material | 0 – 100V | Typical bias voltage 30V to 50V |
| Coating | Fused Silica | |
| Unpowered surface figure | Bare Aluminum | |
| Influence function at nearest neighbor | 10 nm rms | |
| Hysteresis | 1% | Hysteresis is temperature sensitive; listed specification is for room temperature operation |
| Electrical connection interface | 48 x 50 pin-grid array | Integrated at JPL |

**Table 3-12: The WFIRST-AFTA coronagraph baseline deformable mirror specifications.**

### 3.4.4  Development Schedule

The coronagraph instrument development schedule is divided into three key sections. The technology development section (TRL-5 and TRL-6) will demonstrate the system level at TRL-5 in the relevant environment, using the HCIT under vacuum and an OTA simulator to mimic the telescope jitter and thermal drift. Key assemblies such as the DM and ultra-low noise detector will also be developed to retire risk early during the technology phase. We plan to elevate the DM and detector TRL to 6 before the formulation phase.

The next section is the design reference mission development and risk reduction activities section. During the pre-formulation phase, significant resources are invested in the early design of a reference flight instrument. Multiple design cycles are performed to refine the design. During these design cycles, a full physics-based, end-to-end coronagraph model has been developed that can be used for instrument performance predictions and simulations. This physics-based model will help with design trades during the formulation phase. During this phase, a number of engineering risk reduction activities will be performed, e.g., prototyping the precision mask changer mechanism and the deformable mirror driver electronics. The focus will be on mechanism stability and repeatability for the high-precision mask alignment (shaped pupil and hybrid-Lyot), and flight parts, packaging and cabling for DM electronics. These efforts during the pre-formulation phase will help reduce technical and schedule risks during the implementation phase.

The final schedule section covers Phase A/B/C/D. The coronagraph instrument schedule includes 12 months to complete the conceptual design and 12 months to complete the preliminary design during formulation. 40.5 months are budgeted for the final design, fabrication, integration, and test of the coronagraph instrument. These durations are standard for an instrument of this size and complexity. After instrument delivery to the payload, additional time is also budgeted for abbreviated end-to-end performance tests at the payload level (see §3.7).

The HCIT will be converted into a coronagraph systems testbed after the technology demonstration phase. The coronagraph systems testbed will be upgraded when new hardware, such as prototypes, engineering development units, engineering models, etc., is available. It will be used for early flight software development during the formulation phase. The coronagraph systems testbed can be used during phase E to support science operations and possible new algorithm development.

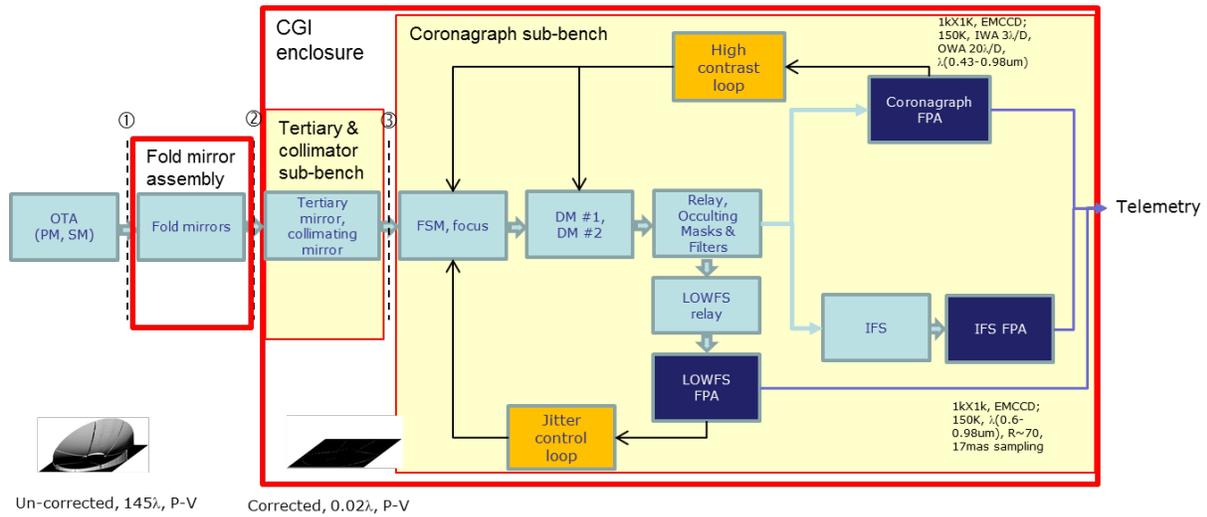

**Figure 3-42: The key coronagraph instrument sub-assemblies are shown. The fold mirror assembly in the left red box is separate from the instrument itself and is separately mounted to the instrument carrier. The Tertiary & Collimator sub-bench (highlighted in the left yellow box) will be tested with the telescope primary and secondary mirrors early. The two yellow areas are two separate sub-benches to be developed in parallel.**

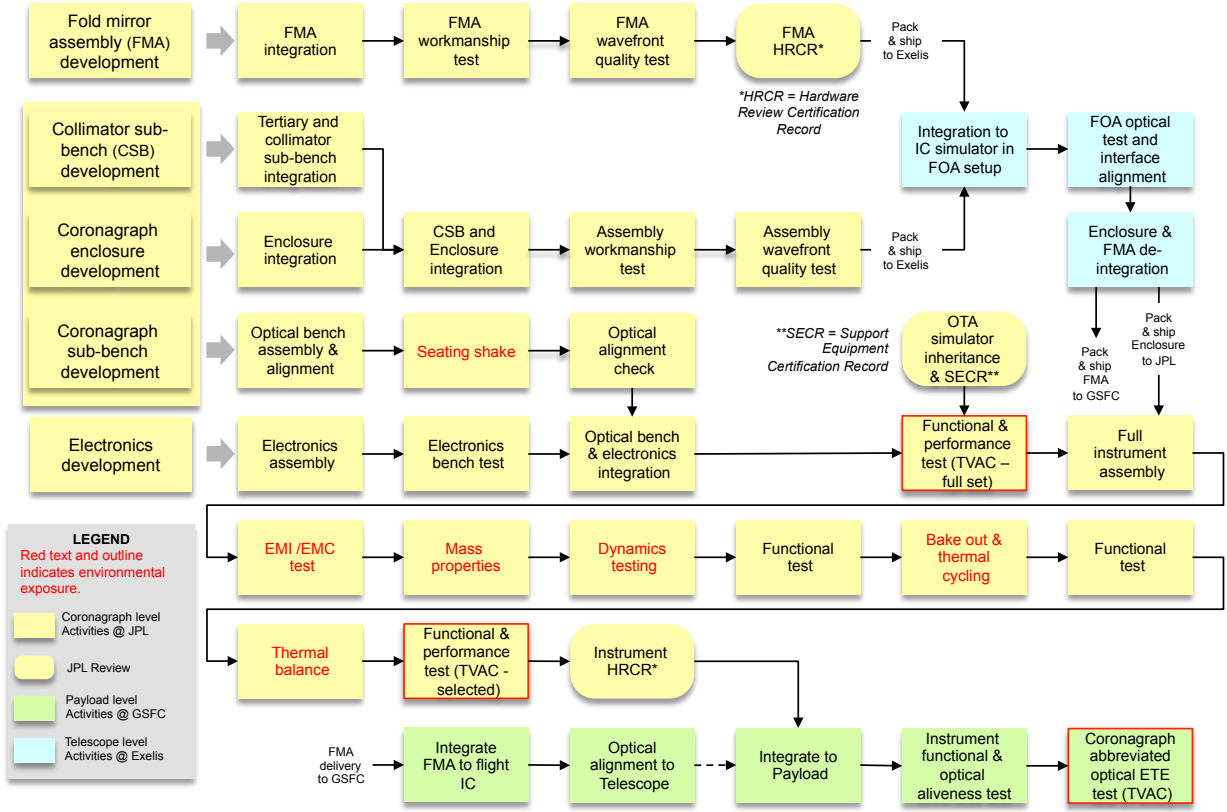

**Figure 3-43: The Coronagraph instrument Integration and Test flow chart.**

### 3.5    Fine Guidance System

The FGS function for WFIRST-AFTA is provided by tracking guide stars using the science focal plane of the wide-field channel in its imaging mode (the spectroscopy mode is discussed at the end of §3.5). One Guide Window (GW) containing a pre-selected guide star is tracked on each of the wide-field channel focal plane's 18 H4RG SCAs using its standard single GW output channel. The horizontal and vertical (h/v) motion of the centroid of each star's position is determined in the wide-field instrument's ICDH, and provided to the S/C ACS system with a lag time of no more than one GW update cycle. On each SCA, the readout of its GW (16 x 16 pixels in size while observing) is interleaved with the uninterrupted readout of 128 lines of science image data (Science Outputs and the GW Output cannot be active at the same time given the current H4RG design), resulting in a GW update rate of 5.86 Hz, and a Science Image frame time of 5.46 s. ACS simulations have indicated that a ≥4 Hz update rate is very comfortable for meeting the ACS slow drift pointing control requirement of 14 mas rms/axis. The interleaving of the GW readouts only increases Science Image frame time by ~3.5% (5.28 s to 5.46 s), with minimal read noise implications. Other choices are possible, but for this design reference mission concept the GW size and Science Image interleaving implementation provides a good balance between GW and Science Image update rates.

Operationally, all science survey pointing positions are uploaded from the ground, and each pointing includes a row/column position on each SCA identifying where the GW star ($12 < H\_AB < 17$ magnitudes acceptable) for that SCA will be found. The slew from one FGS controlled pointing position to the next is made under star tracker and gyro control, and a 64 x 64 pixel Acquisition GW size is temporarily used to ensure that the GW star selected on each SCA is within that initial GW given the star tracker/gyro pointing accuracy. Simulations indicate that guide star acquisitions will require only one GW update, and after the overall pointing error drops below 100 mas all GWs will return to their observing mode size of 16 x 16 pixels.

This FGS concept as implemented on WFIRST-AFTA is very robust in terms of both redundancy and scientific utility. As few as 4 of the 18 FGS GWs that are available are needed for the ACS to meet its pointing requirements, so there is substantial tolerance to any issue (e.g. the SCA reset at the start of an observation) that might interfere with the acquisi-tion or quality of a GW on multiple SCAs. From a science perspective, though non-common path pointing drifts will need to be considered for coronagraph and wide-field IFU pointings, wide-field channel observations will be immune to this concern since its focal plane provides the FGS star images, a risk reduction consideration for WL survey imaging. Even more critically, all GW star images are sent to the ground, providing PSF variation tracking at ~6 Hz across the wide-field channel focal plane, data valuable for validating the accuracy of WL galaxy shape measurements. Guiding during GRS observations is more challenging due to the use of dispersed star images, as h/v motion must be derived from the cross dispersion and bandpass cutoff edges of the spectral continua of bright stars. FGS performance is being evaluated relative to the mildly relaxed GRS pointing stability requirements (~20 mas rms/axis) for stars of magnitude $H_{AB} = 12.5$ and brighter (the probability of ≥3 such stars in the FoV at the North Galactic pole is >95%), for update rates of 2 Hz and 4 Hz.

### 3.6    Spacecraft

The WFIRST-AFTA spacecraft has been designed to provide all the resources necessary to support the payload in geosynchronous orbit using mature and proven technology. The design is based on the Solar Dynamics Observatory (SDO) spacecraft, which was designed, manufactured, integrated, tested, and qualified at GSFC and is currently operating in geosynchronous orbit, as well as other recent GSFC developed spacecraft like the Global Precipitation Measurement (GPM). The spacecraft bus design provides cross strapping and/or redundancy for a single-fault tolerant design.

***Structures and Thermal***: The spacecraft bus design features two decks stiffened by six gussets and faced with module support plates all composed of aluminum honeycomb panels with M55J composite facesheets and sized to meet minimum launch vehicle frequency requirements while minimizing the spacecraft structure mass. The spacecraft bus provides the interfaces to the payload and the launch vehicle, transferring payload loads through the instrument carrier struts, through the spacecraft gussets and into the payload attach fitting (PAF); see Table 3-13 for the Observatory mass breakdown. The spacecraft upper deck provides shear support of the bus, while also resisting bending. The propulsion subsystem is supported inside the bus on the lower deck, which also provides the interface from the gussets to the PAF.

The spacecraft structure includes an instrument carrier that is the primary metering structure for the payload. The instrument carrier is a composite truss structure that provides opto-mechanical support for both the telescope and instruments, provides mechanical latches, with design heritage from HST, to interface the instrument modules to the carrier, contains harness to route power and data between the spacecraft and the instrument modules, and provides thermal isolation between the two instrument volumes. The telescope mounts directly to the instrument carrier via bipods that attach to the existing telescope main mount interface. The telescope outer barrel assembly mounts directly to the spacecraft top deck via bipods, so that there is no direct load path from the outer barrel to the telescope or instrument optical systems.

The bus also supports a multi-panel solar array/sunshield to prevent the Sun from illuminating the instrument carrier, the telescope outer barrel past the secondary mirror, and the instrument radiators during science observations at roll angles up to 15° (see Figure 3-44). The side panels are stowed for launch and are released on-orbit to maximize the area available for generating electrical power.

The spacecraft design features 6 serviceable modules containing spacecraft, wide-field, or telescope electronics. These modules are designed to be robotically serviceable, enabling refurbishment or upgrades to the observatory by replacing individual modules. Each of the spacecraft modules includes a kinematic Module Restraint System (MRS) based on the Multimission Modular Spacecraft (MMS) design, which was proven during the on-orbit module change out of the Solar Maximum Repair Mission. This MRS includes a beam interfacing to kinematic constraints on the top of each module and an accompanying floating nut/ACME threaded fastener on the bottom of each module. These module interfaces can be robotically engaged and disengaged, providing a simple, repeatable mechanical interface to the spacecraft structure. The modules also include a grapple fixture to enable removal of the modules. The expected docking interface with the robotic servicer is a standard Marman band interface. Docking to the existing launch vehicle payload adapter fitting allows sufficient reach for the robotic servicer to remove and replace the spacecraft modules.

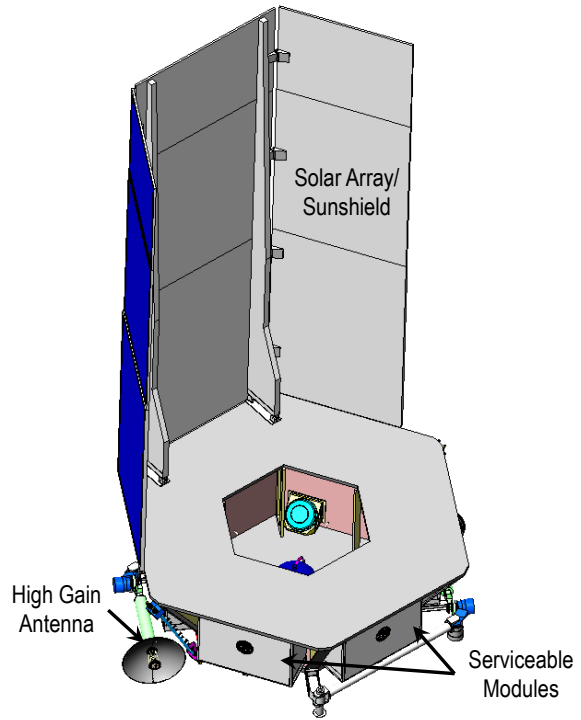

**Figure 3-44: The spacecraft bus with solar array/sunshield showing serviceable modules (both the sunshield side panels and the high gain antenna are shown in launch configuration). The serviceable modules house spacecraft, wide-field and telescope electronics.**

Each module contains a set of connectors with alignment features to allow "blind mating" of the module harness connectors to connectors on the harness internal to the spacecraft bus connecting between modules and to the payload. This electrical interface

| | Mass (kg) | Cont. (%) | Mass+Cont. (kg) |
|---|---|---|---|
| Wide-Field | 489 | 30 | 636 |
| Coronagraph | 120 | 30 | 156 |
| Telescope | 1582 | 11 | 1763 |
| Spacecraft (incl. IC) | 1868 | 30 | 2428 |
| Observatory (dry) | 4059 | 23 | 4983 |
| Propellant (3σ) | 107 | | 122 |
| Observatory (wet) | 4166 | | 5105 |

**Table 3-13: Observatory mass breakdown**

has been demonstrated on the HST radial instruments.

The electronics are mounted on the outboard panel of each of the modules with doublers used as needed for thermal conductivity. The spacecraft cold-biased thermal design maintains the spacecraft and payload electronics within their operational limits using surface coatings, heaters and radiators.

*Electrical Power*: The electrical power subsystem uses a direct energy transfer design, using heritage designs from the recent GPM mission, to provide regulated power to the Observatory. Three, body-mounted solar array panels provide the observatory power. The solar array is sized to provide full observatory power (3200 W) at end of life with 2 strings failed at the worst case observing angles of 36° pitch and 15° roll with 30% power margin. The solar array relies on multi-junction, gallium arsenide solar cells with at least 29.5% efficiency to provide Observatory power. To minimize thermal backload from the array to the instruments, the backside of the array is insulated and optical solar reflectors are interspersed between the solar cells to cool the array. An internally redundant power supply electronics box controls the distribution of power, providing unregulated 28 Vdc power to the spacecraft and payload, provides solar array power management, and controls battery charging. The baselined geosynchronous orbit has two eclipse seasons of ~23 days per year with a maximum eclipse period of 72 minutes. Three 80 A-hr batteries provide power to all Observatory loads during the maximum eclipse duration.

*Communications*: The communication subsystem design leverages the SDO design to provide high reliability, continuous data downlink from the geosynchronous orbit. S-band transponders receive ground commands and send real-time housekeeping telemetry to the ground via 2 omni-directional antennas. The high rate science downlink uses two dedicated Ka-band transmitters, each with a 0.75 m gimbaled antenna, to continuously downlink science data without interrupting science operations. The Ka-band modulator has been updated from the SDO design and is capable of downlinking at a data rate of 1.2 Gbps, providing 50% margin to the baseline science data rate of 600 Mbps. A prototype of the updated Ka-band modulator is currently being assembled and will be available for testing in early 2015.

*Command & Data Handling*: The internally redundant command and data handling (C&DH) subsystem hosts the spacecraft flight software, provides on-board real time and stored commanding, receives payload and spacecraft housekeeping data, and provides fault management for spacecraft health and safety as well as safing for the payload, when necessary. The observatory uses a MIL-STD-1553B command/telemetry bus and uses SpaceWire for the high rate science data. The C&DH provides the interface between the instruments high rate science data and the Ka-band system and formats and encodes the instrument science data for downlink. With both the wide-field and coronagraph instruments operating, the C&DH will interleave this data onto the Ka-band downlink. Due to the direct, continuous science data downlink, no science data recorder is required for WFIRST-AFTA. The C&DH also provides the interface to the S-band transponders, performing command decoding and distribution as well as encoding for S-band downlink.

*Propulsion*: The propulsion subsystem is a simple mono-propellant design in blow down mode and uses an SDO heritage design. A single fuel tank stores the high-purity hydrazine fuel and nitrogen pressurant and mounts to the spacecraft lower deck. Eight 22 N (5 lbf) attitude control subsystem (ACS) thrusters adjust for launch vehicle dispersions, provide east-west stationkeeping, provide momentum management and allow for disposal at the end of mission life.

*Attitude Control*: The attitude control subsystem provides three-axis control of the spacecraft and uses data from the inertial reference unit, star trackers, and the payload FGS to meet the pointing control and stability requirements. The internally redundant inertial reference unit provides precise rate measurements to support slew and settle operations. The star trackers (3 for 2 redundant) are mounted on the instrument carrier so they are directly related to the instrument pointing. The star trackers are used for coarsely pointing to within 3 arcsec RMS per axis of a target. After that, the FGS takes over to meet the fine pointing requirements for revisits and relative offsets (see §3.4.4). A set of four 75 N-m-s reaction wheels is used for slewing as well as momentum storage. The wheels are mounted to and passively isolated from the spacecraft lower deck to attenuate high-frequency wheel-induced disturbances. The ACS thrusters are used to desaturate the wheels to manage momentum buildup. The internally redundant Mechanism and Attitude Control Electronics (MACE) box controls the attitude control subsystem and provides an independent safe hold capability using coarse sun sensors and reaction wheels, which keeps the observatory thermally-safe, power-positive and protects the instruments from direct sunlight. The MACE controls the fine pointing of the observatory based on the guide star centroids provided by the wide-field instrument C&DH. The MACE also deploys the solar array side panels and the high gain antennas and controls the gimbals for the high gain antennas.

## 3.7 Payload and Observatory Integration & Test

Integration and test (I&T) activities for WFIRST-AFTA occur in two phases, Payload I&T and Observatory I&T. During Payload I&T, the Telescope, Wide-Field Instrument, Coronagraph Instrument and Instrument Carrier will be integrated together and the optical performance of the payload will be verified. For the wide-field instrument, the optical test is a high-fidelity end-to-end measurement of wavefront error and photometric throughput. For the coronagraph instrument, it is an interface test with a performance health check at a single wavelength. As shown in Figure 3-45, deliverables to the effort are:

- The Telescope, including the Outer Barrel Assembly (OBA), the Forward Optic Assembly (FOA), and the Telescope Electronics
- The Instrument Carrier
- The Coronagraph Instrument (CGI) Optics Package and Fold Mirror Assembly (FMA)
- The Wide-Field Instrument Cold Sensing Module and the Warm Electronics, including the Instrument Command and Data Handling Electronics (ICDH), Focal Plane Electronics (FPE) and the Mechanism Control Electronics (MCE)

At delivery to Payload I&T, these assemblies will have already been fully qualified against the required environmental tests, will have been verified to meet performance requirements and will have demonstrated compliance with internal optical alignment budgets.

Figure 3-46 shows the Payload I&T flow, includ-

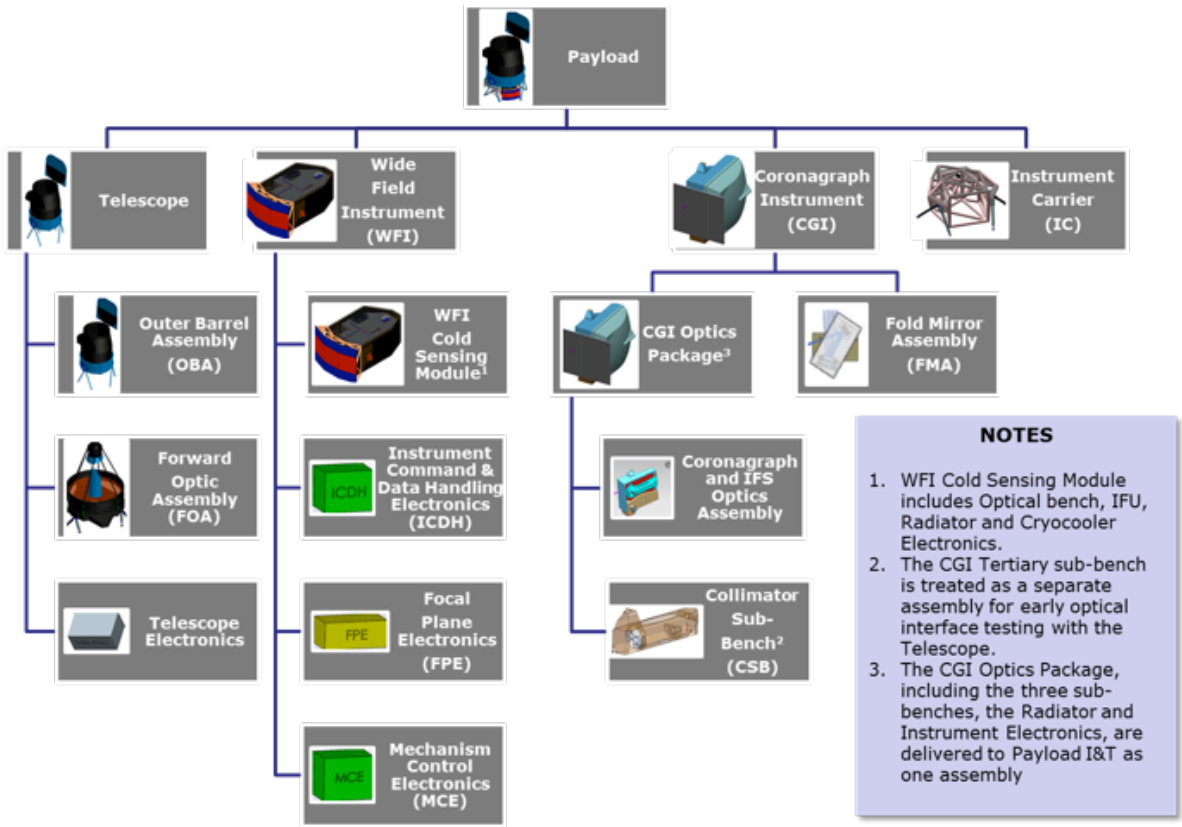

**Figure 3-45: Assemblies delivered to payload I&T.**

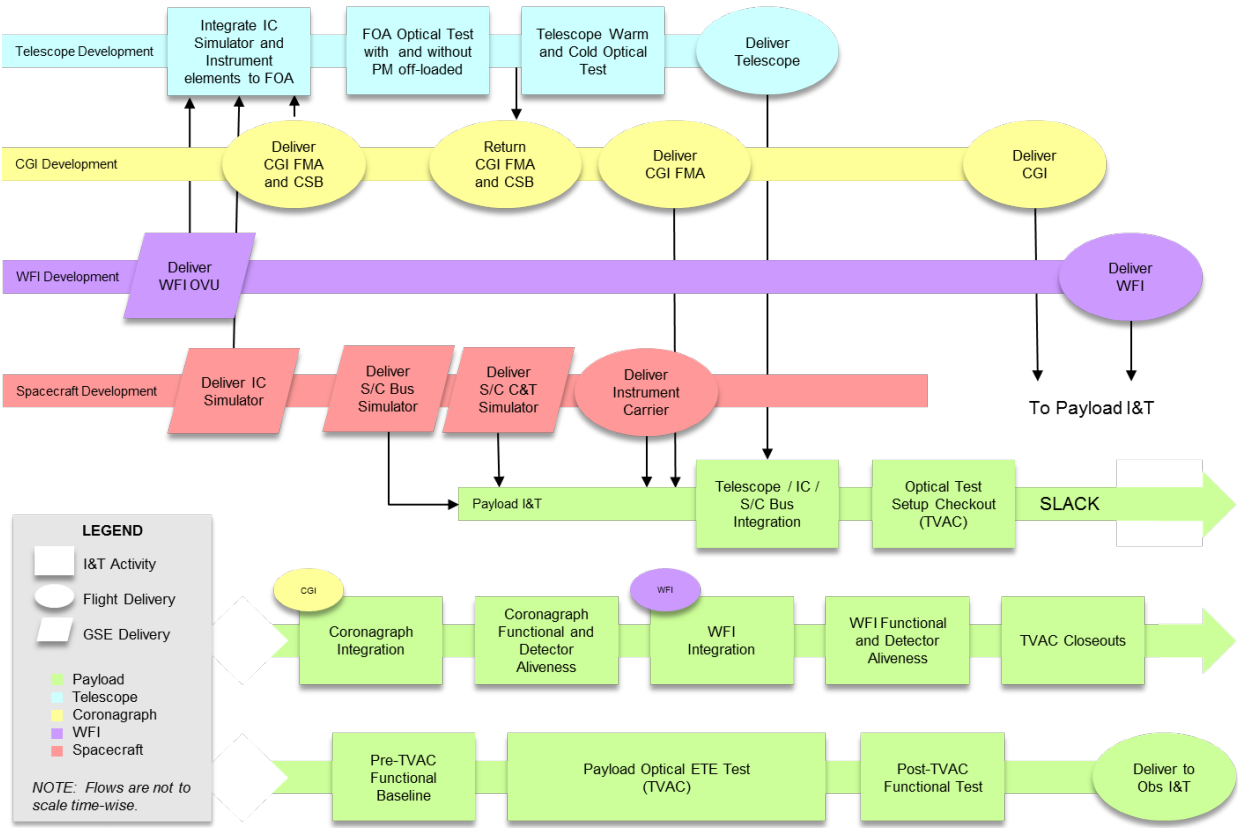

**Figure 3-46: Payload integration and test flow, including build-up of the payload and the key optical tests.**

ing the build-up of the Payload as well as the critical optical tests that occur between Payload elements. Prior to the start of Payload I&T, instrument surrogates are delivered to the Telescope development at the Exelis facility to enable optical interface verification and measurement of Telescope wavefront error (WFE). The WFI team provides a high-fidelity optical verification unit (OVU); the CGI team provides their flight FMA and flight Collimator Sub-Bench, which is temporarily separated from the CGI optics package; and the Spacecraft team provides an IC Simulator.

While the WFI and CGI continue their flight developments, the build-up of the Payload starts with the delivery of the Telescope and Telescope Electronics, the flight IC from the Spacecraft team and the flight CGI FMA to the Payload team in the Goddard Spacecraft Systems Development and Integration Facility (SSDIF). These assemblies are mechanically integrated together and aligned. After thermal closeouts are completed, the partial Payload assembly is put in the Space Environment Simulator (SES) chamber to check out the optical test setup needed for the later Payload end-to-end (ETE) test.

Payload integration resumes in the SSDIF with successive deliveries of the flight CGI and flight WFI. Each instrument is mechanically and electrically integrated to the Payload. Final ambient operation is confirmed with functional testing of the electronics and detector aliveness tests.

After thermal closeouts, the Payload is returned to the SES for Payload Optical ETE testing. This test consists of four optical test setups:

- <u>Primary Mirror (PM) Center of Curvature (CoC) test</u> to track PM sag due to gravity, the largest error source;

- <u>WFI Half Pass test</u> to replicate Instrument-level results and track WFI alignment separately;

- <u>Payload ETE test</u> to verify WFI+Telescope WFE under flight-like conditions; and

- <u>CGI ETE test</u> to check Coronagraph alignment to Telescope and verify CGI+Telescope WFE at one wavelength.

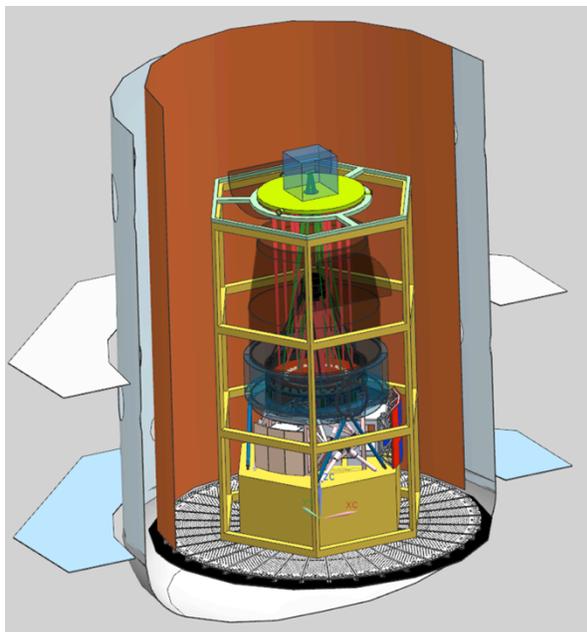

**Figure 3-47: Payload in the Space Environment Simulator chamber in the payload optical end to end test configuration. Thermal shrouds are not shown.**

Figure 3-47 shows the green ray trace of the PM CoC test (converging above the auto-collimating flat) and the overlapping red ray traces of the WFI half-pass and Payload ETE tests. The challenge of testing through gravity-induced aberration is solved by characterizing mirror sag in lower assembly-level tests and by using focus-diverse wavefront sensing with phase retrieval at the Payload level.

Observatory I&T begins after the completion of Payload Optical ETE testing when the integrated Payload assembly is delivered to the spacecraft and the two are mechanically and electrically integrated together. First motion tests of the solar array/sunshield and high gain antenna boom release mechanisms are performed and the baseline Comprehensive Performance Test (CPT) is run to verify full Observatory electrical performance across all interfaces. The CPT is performed before and after all major Observatory environmental activities. After the baseline Observatory performance is verified, the Observatory is subjected to a series of environmental tests to verify performance during and after exposure to the predicted launch and on-orbit environments. Electromagnetic Interference and Compatibility testing will verify that all Observatory functions are self compatible, i.e. they do not create an electromagnetic interference that disturbs another element. The Observatory is next subjected to vibration and acoustic testing to verify structural integrity against the simulated launch envi-ronment. A launch vehicle interface deployment test is also performed to demonstrate compatibility with the mechanical shock environment produced when the Observatory separates from the launch vehicle. Finally, the Observatory moves back to the SES chamber for Observatory thermal balance and vacuum testing. The thermal balance test is performed to allow correlation of the thermal models for final on-orbit temperature predictions while the thermal vacuum testing cycles the Observatory between its worst-case temperature extremes to verify performance across the operational temperature range. After thermal testing, the Observatory is shipped to the launch site for final preparations prior to launch.

## 3.8 Integrated Modeling/Mission Performance

Integrated modeling as realized on WFIRST combines structural, thermal, optical, and control systems models to provide Observatory level optical performance assessments of major design configurations. These assessments are typically WFE and/or LOS absolute/stability predictions at an interface defined by Science (e.g. WFI and CGI) or Spacecraft (e.g. star tracker platform) instrumentation teams.

These Observatory IM-provided input datasets enable detailed science performance simulations of key WFIRST science surveys (e.g. coronagraphy, galaxy redshifts, galaxy shapes, SNe luminosity, microlensing, etc.) to be performed, determining margins on key survey requirements (e.g. imaging mode PSF ellipticity stability for WFI, contrast and speckle noise for CGI) that when combined with sky model assumptions can be used to estimate science survey returns (see Figure 3-48).

### 3.8.1 Integrated Modeling Inputs

Structural, thermal, optical, controls, and systems leads from each of the major Observatory Flight Segments (S/C, Telescope, WFI, and CGI) join with the Study Office IM discipline and analysis leads to form the Observatory IM team. This team structure has matured dramatically over the last year, and is key to the high quality of the results being produced.

The major tools used are Thermal Desktop v5.6+, SINDA v5.6+, MSC/Nastran v2008, Zemax, SigFit, Code V, and the DOCS Toolbox. Linear Optical Models (LOMs) are generated internally by the IM team to the specifications of the Instrument teams to determine WFE/LOS sensitivities to both rigid body motions and figure changes of optical elements. The DOCS Toolbox is used to assess optical defor-mation/motion impacts using the LOM sensitivities. It is particularly valuable to have two independent optical performance assessment techniques available, SigFit and LOM/DOCS. They are used to cross-check each other's optical predicts, with SigFit also providing high spatial frequency information useful in the evaluation of optical mount designs.

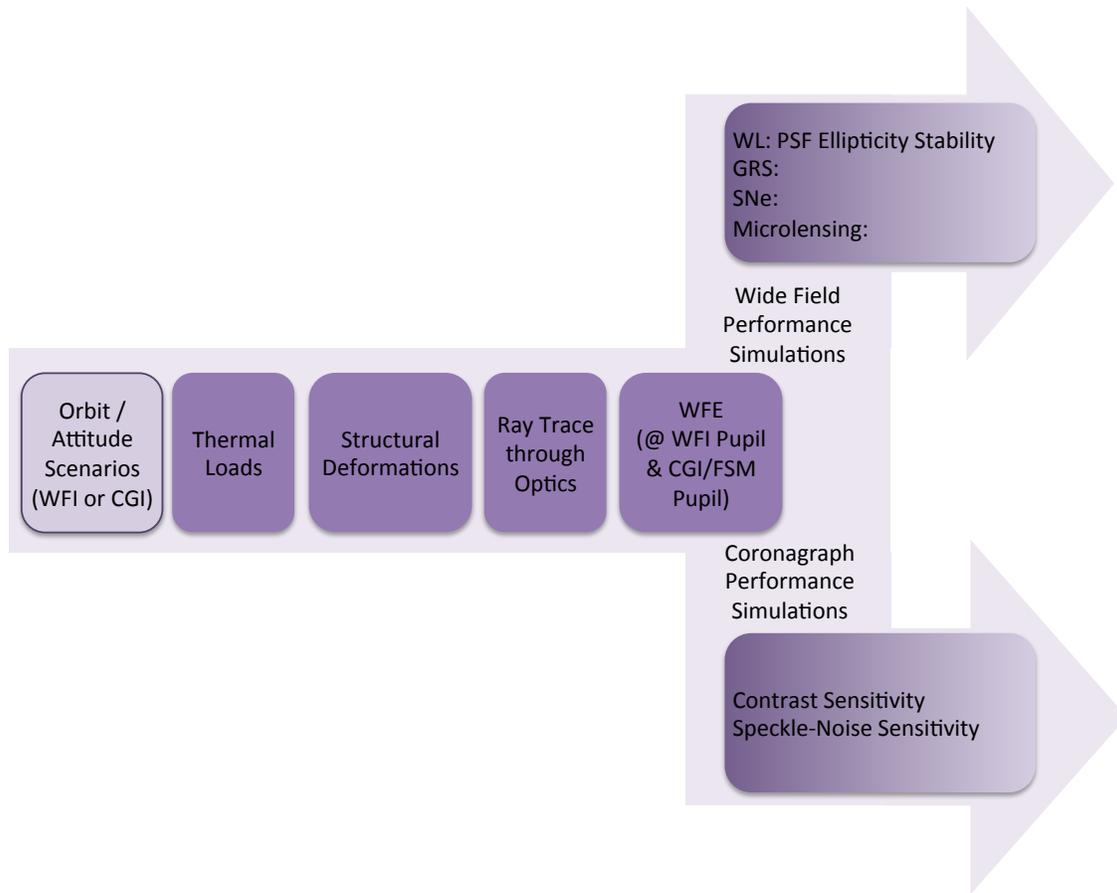

**Figure 3-48: The integrated modeling process assesses WFE and LOS stability at defined instrument interfaces to enable detailed science performance simulations for each science program.**

Each of the IM Flight Segment teams (S/C, Telescope, CGI, and WFI) provides the IM team with structural and thermal models meeting the Math Model Guidelines document requirements. These models are integrated by the IM team to form the Observatory Structural/Thermal/Optical (STOP) and Jitter models.

The STOP model assesses the optical response of the Observatory to environmental changes that result in thermal/gravitational/moisture disturbances (e.g. gravity release, temperature changes from ground to orbit, orbital/attitude-driven temperature fluctuations, and long term moisture desorption), and includes the effects of both rigid body motions and figure changes of optical elements.

The Jitter model is built in two steps. The Observatory Modal Structural Model is built first. It includes D-struts to isolate the Payload from S/C disturbances, includes an open OBA door and two deployed high gain antennas (HGAs), has passive vibration isolation for the reaction wheel assemblies (RWAs), and predicts modes out to 300 Hz. Prior to the final modal analysis run, an on-orbit thermal profile is applied to the structural model to set grid point temperatures so that temperature dependent material properties will reflect orbital conditions.

In the second step, the modal results are used to construct a MatLab Reduced Dynamics Response Model with 0.25% damping. Disturbance sources are applied to this model, and the rigid-body motions of the optical elements (figure distortions at high frequencies are not significant) are convolved with the LOM to estimate the resulting WFE/LOS instabilities.

As a reference point, the Observatory structural model element count is as high as the Global Precipitation Measurement (GPM) Observatory and Solar Dynamics Observatory (SDO) structural models were at CDR. Models of this detail are needed due to WFIRST-AFTA's large mirrors and tight WFE/LOS requirements, yet they are agile enough to allow for reasonably short analysis cycles once the Observatory model is created (~100 STOP solutions from 3 orbit/attitude cases in 1+ weeks).

Modeling Uncertainty Factors (MUF's) are multiplicative factors applied to the analysis results to account for model prediction uncertainties. A x2 factor is currently applied to the STOP optical element grid point motion predictions, a value which will decline as models are test-correlated and when statistical models are run. For Jitter predictions, the MUF value increases with frequency due to higher modal prediction uncertainties. For the current Observatory configuration, a MUF of x2.04 is used below 20 Hz, with this value increasing linearly to x4.13 at 40 Hz and remaining fixed at that value for higher frequencies. The values used reflect the use of some disturbance models that have been validated at the component level. Note that the Telescope model used in the current IM analysis has been modally correlated, but no credit has yet been taken for this in the Jitter MUF.

For STOP, two orbit/attitude slew cases were analyzed. A WFI Worst Slew case (Beta = 11° orbit) where the Observatory was stable in a Hot Case attitude (pitched 36° away from the Sun and rolled about the LOS 15°, with Azimuth angle chosen to place the normal to the WFI Optical Bench radiator in the orbit plane), and then the Observatory was instantaneously slewed to a Cold Case attitude (pitched 36° towards the Sun, rolled 15° about the LOS, with Azimuth angle chosen to place the radiator normal 48° out of the orbit plane). This is an extremely demanding case, selected from a triaged set of orbital cases from an earlier but similar configuration primarily based on selecting the attitudes that produced the hottest and coldest WFI optical bench temperatures.

A Coronagraph representative "Day in the Life" slew sequence designated "OS3" was defined by the CGI team and is a series of 4 pointings (at a calibration star spectrally similar to the eventual target star, then at a bright star to quickly create a dark hole, then at the target star (which has a planet), and then back to the original calibration star). A Beta = 42° orbit was chosen, with the time of year selected to maximize the absorbed solar flux delta resulting from the modest pitch variations (~8°) associated with this pointing sequence.

An additional thermal disturbance that is common to both slew cases is the initial cooldown from room temperature (293 K) to orbital temperatures. The first STOP solution time step of each slew case was arbitrarily used as the reference temperature at which the initial commissioning adjustments of the Telescope Secondary Mirror (SM) and any instrument-internal (i.e. the WFI F2 fold-flat mirror) optical alignment mechanisms are made. One SM setting is expected to apply to all four WFIRST Science optical paths, since it is intended to optimize Telescope focus, and this was confirmed in the current IM analyses with the CGI and WF Channel ideal SM refocus motions (for the same WFI worst slew case initial time step) differing only by 0.3 um in the despace (focus) direction. During flight, these initial settings (both SM focus and any instrument-internal focus settings) will be made prior to the initiation of Science Operations, and modifications are only expected to be necessary in response to long term temperature or moisture desorption drifts.

Pre-launch alignment offsets designed to have the Wide-Field Channel mirrors cool into alignment after launch are considered in the IM simulation of the WFI commissioning process, but CGI and IFU optical paths are assumed to be ideally aligned prior to launch and do not yet take advantage of pre-launch offsets.

For Jitter, RWA, HGA, and WFI cryocooler disturbance sources are all independently evaluated. Four standard (not fine-balanced) Honeywell HR14-75 wheels (LDCM heritage) are the basis for the RWA disturbances, including sub-super harmonics from a Honeywell-supplied bearing model. Four HGA stepper motors (SDO, GPM, and LDCM heritage) are the basis for the HGA pointing disturbances. The HGA actuator design includes microstepping and low detent features to reduce disturbances, and passive HGA boom damping. A linearized model of actuator stiffness is included in the structural model.

The reverse Brayton Cycle cryocooler (design down-scaled from HST/NICMOS) disturbances are so low that they cannot be measured during normal operation (pointing jitter due to the cryocooler was not detected during HST/NICMOS observations). The power spectral density disturbance input is scaled from vibrations that can be measured during initial cooldown, and a MUF of x10 is applied to these estimated values. To be conservative, the same cryocooler disturbances are input at its compressor and both the Wide-Field and IFU Channel focal planes.

### 3.8.2 Integrated Modeling Outputs

The integrated modeling process was first developed and run for the wide-field instrument after the 2013 WFIRST SDT report. With the decision to include the coronagraph as a baseline instrument on WFIRST-AFTA, the coronagraph instrument's structural, thermal, and optical models have been incorporated into this process. The integrated model results presented below focus on the wide-field instrument's critical imaging mode, but a full set of STOP and Jitter optical stability outputs (WFE and LOS) have been generated for the current design for the Wide-Field Imaging Mode (WIM), Wide-Field Spectroscopy Mode (WSM), IFU, and CGI optical paths, and are the basis for the Preliminary Performance Simulation Results presented in §3.8.3.

#### 3.8.2.1 STOP Optical Processing/Outputs for WFI and CGI

For WFI STOP, after an initial SigFit/LOM WFE cross-check for WIM and WSM using a 1 K bulk temperature change, the WFI Worst Slew case (see §3.8.1) was run in both WIM and WSM configurations (WIM with a filter in the WF Channel beam path, WSM with the grism disperser there, and with CGI and IFU optical paths always being available). The slew case was modeled for 32 time steps with 1 hour between each time step, covering 6 hours before the slew from the hot to the cold attitude, and 24 hours after, with the x2 MUF being applied to the optical element grid point motions predicted at each time step.

Telescope SM correction (tip/tilt/focus available, but only focus adjustment used) was applied to optimize WFE at the WF Channel WIM exit pupil, as previously discussed.

WF Channel F2 alignment correction (tip/tilt/focus all used) was then performed considering all optical motions/deformations along the WF Channel WIM optical path. The F2 focus adjust was an iterative procedure, with the optical motions resulting from the cooldown from the ground alignment temperature of 293 K to the initially analyzed time step being compensated, as planned, by ground alignment offsets to ensure that the cooldown optical error budget was met (this eventually required that 82% of the cooldown optical element rigid body motions be removed by the ground alignment offset, a reasonable value, with no figure compensation being required).

LOM sensitivities were then applied to optical grid point motions at each time step to estimate the effects on WFE and LOS stability. These WFE/LOS predictions were compared to specifications to assess preliminary margins (see Table 3-14 for the WIM margins summary for STOP/Jitter).

The CGI STOP processing sequence was very similar, the major differences being the use of the OS3 slew case (see §3.8.1) to define the analysis orbit/attitude scenario, the use of the WIM T2 position (no special T2 adjustment for the coronagraph) and the fact that no pre-launch alignment offsets were in-cluded. WFE (at the fast steering mirror) and LOS (at the first focus) were assessed following cooldown and at each of 42 output time steps (5000 seconds apart) that spanned the OS3 observing sequence. The LOM sensitivities were applied to the optical motions (and to the deformations of the PM and SM only) at each output time step to generate the WFE and LOS predicts as a function of time for delivery to the coronagraph team.

| Metric | Discipline | Source | | Predict w/MUF | Requirement | Margin |
|---|---|---|---|---|---|---|
| WFE Stability | STOP | WFI Worst Case Slew | | 0.04 nm RMS / 184 sec | 0.707 nm RMS / 184 sec | **17.5x** |
| | Jitter | RWA | 0.5 nm | 0.536 nm RMS (RSS) | 0.707 nm RMS | **1.3x** |
| | | HGA | 0.191 nm | | | |
| | | Cooler | 0.027 nm | | | |
| LOS Stability | STOP | WFI Worst Case Slew | | 1.8 marcsec radial (ACS Control) | 14 marcsec RMS/axis | **~10x** |
| | Jitter | RWA | 9.3 marcsec RMS/axis | 10.6 marcsec RMS/axis (RSS) | 14 marcsec RMS/axis | **1.3x** |
| | | HGA | 5.2 marcsec RMS/axis | | | |
| | | Cooler | 0.223 marcsec RMS/axis | | | |

**Table 3-14: Wide-field Imaging Mode integrated modeling results, with Modeling Uncertainty Factors included (x2 for STOP and x2-x4 for Jitter), show large STOP margins even for a demanding WFI worst case slew and good jitter margins for all major Observatory vibration sources. Note: 18 SCAs are included in the Wide-Field Channel STOP performance assessment; one central focal plane field point is used for the Wide-Field Channel Jitter assessment; HGA jitter responses are average values, with peaks to be assessed in the next design iteration.**

Detailed optical predicts based on cooldown and the above observing sequences have been generated for all optical paths and will be provided as inputs to all WFIRST science survey teams to aid in their detailed Performance Simulations. Two preliminary post-processing assessments based on the current STOP optical predicts are included in §3.8.3, the first addressing WIM PSF ellipticity stability during the WFI Worst Slew case (critical to weak lensing's galaxy shape survey), and the second addressing the contrast achieved by the coronagraph during the OS3 slew case (critical to its ability to detect planets).

### 3.8.2.2  *Jitter Optical Processing/Outputs for WFI and CGI*

The Jitter processing is more simple than STOP processing as the disturbance sources are applied to the Observatory structure and the resulting rigid body optical motions (no figure changes) are directly assessed using LOM optical sensitivities for each optical path. As previously noted, preliminary results/margins relative to the WFIRST 2013 SDT Report Jitter requirements are shown for the WIM configuration in Table 3-14. Detailed Jitter optical predicts have been generated for all optical paths and will be provided as inputs to all WFIRST science survey teams to aid in their detailed performance simulations (current results have been provided to the coronagraph team).

### 3.8.3  **Preliminary Performance Simulation Results**

#### 3.8.3.1  *Wide-Field Science Performance Simulations*

A key performance parameter for the weak lensing shape measurement is the knowledge of the ellipticity of the Point Spread Function (PSF) in the presence of thermal drifts when the Wide-Field Channel is in its WIM configuration and making WL imaging observations. The PSF is fit with a Gaussian:

$$W(x,y) = A_0 \exp^{\left(\frac{1}{2}\bar{x}^T M \bar{x}\right)}, \bar{x} = \begin{bmatrix} x - x_c \\ y - y_c \end{bmatrix}, M^{-1} = \begin{bmatrix} M_{xx} & M_{xy} \\ M_{xy} & M_{yy} \end{bmatrix}$$

and the ellipticities are calculated as:

$$e_1 = (M_{xx} - M_{yy})/(M_{xx} + M_{yy})$$
$$e_2 = 2M_{xy}/(M_{xx} + M_{yy})$$

The April 2013 WFIRST SDT report established a PSF ellipticity knowledge requirement of better than $4.7 \times 10^{-4}$. Demonstrating PSF ellipticity stability at that level over a WL imaging observation would greatly facilitate meeting this requirement. Initial analysis of the WFE results indicates that even during the worst case WFI slew, as previously described, with a x2 MUF applied, order of magnitude ellipticity stability margins are predicted over the full WF Channel focal plane (see Figure 3-49). Having large positive margins for this demanding slew case is extremely encouraging, as it suggests that we can design the weak lensing observing strategy (perhaps the most demanding WFI-based survey) without needing to make slew size a primary scheduling constraint.

#### 3.8.3.2  *Coronagraph Science Performance Simulations*

For the coronagraph, contrast and speckle-noise sensitivities to temperature and vibration disturbances are key parameters. The IM team has developed stability predictions for the wavefront error, due to both types of disturbances as evaluated at the fast steering mirror located at the exit pupil. The CGI team has generated preliminary science performance simulation predicts based on those predictions as described below. For science performance, different aspects of the contrast are of interest, with the two most important parameters being:

1. The mean contrast $\langle C \rangle$ over the dark hole (DH)
2. The (spatial) standard deviation of the post-processed image, $\sigma_{PP}$

Since the coronagraph alone can't produce a sufficient dark hole due to the wavefront error induced by imperfections in the optics, it must use actively controlled deformable mirrors to correct the wavefront errors to create the dark hole. The net contrast performance of the WFIRST-AFTA system is thus dependent on both the coronagraph optics and the wavefront control system, and it must be predicted using end-to-end modeling that incorporates both.

After science yield, the most important operational performance requirement is the efficiency of reaching a dark hole. Dark hole acquisition is an iterative process that takes place while the instrument ob-serves a bright star near the science star. The field at the focal plane is sensed, and a revised DM solution to eliminate the remaining light in the dark hole is computed and commanded. In successive iterations the light level drops and the time to estimate the field grows accordingly. The presence of incoherent light, such as from the zodi background, and detector noise can further lengthen the time to achieve the necessary signal-to-noise ratio (SNR).

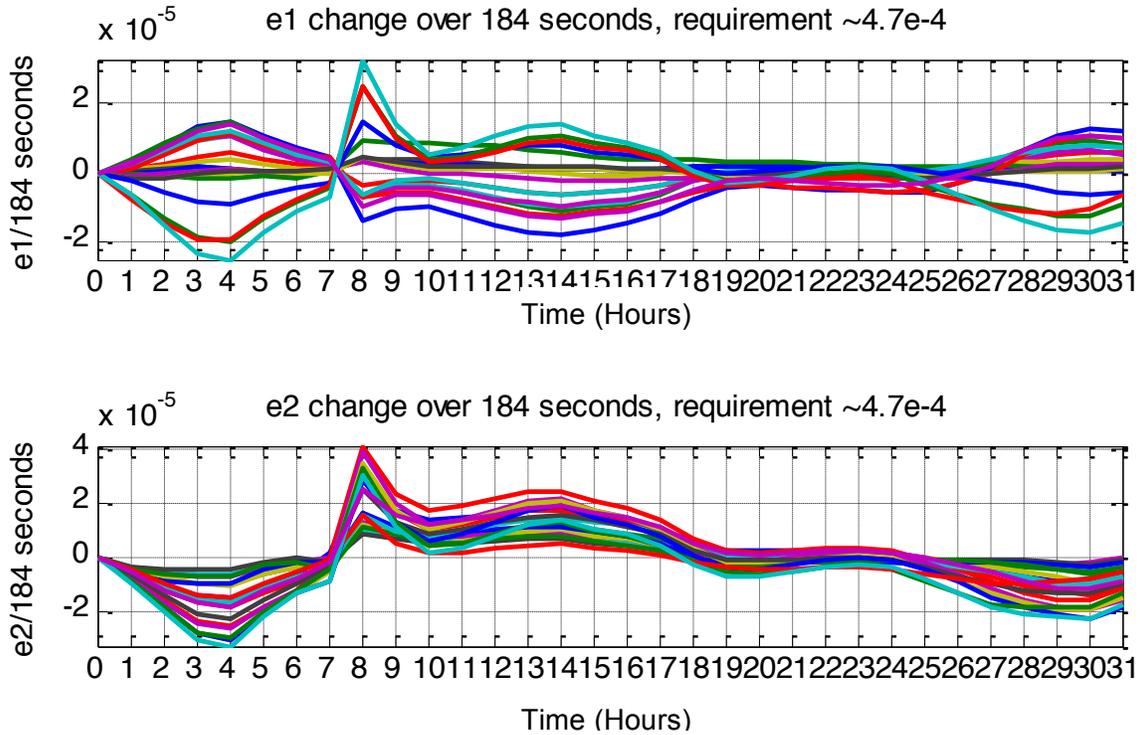

**Figure 3-49: Plot of delta ellipticity over a weak lensing observation with time steps of 3600 seconds after a slew at hour 7. $\Delta e1 \leq 3.3 \times 10^{-5}$ / 184 sec provides 14.2x margin on $4.7 \times 10^{-4}$ and $\Delta e2 \leq 4.1 \times 10^{-5}$ / 184 sec provides 11.5x margin on $4.7 \times 10^{-4}$. The lines represent the center field point of each of the 18 detectors and a point for the FoV center.**

The formation of speckles in the dark hole is modeled by a diffraction model. The performance of each WFIRST-AFTA coronagraph mode is derived from intensive numerical simulations of the combined telescope + coronagraph system. This involves calculating the propagation of a wavefront through the optics and accounting for diffraction from obscurations and scattering from hypothetical surface fabrication errors. The freely and publicly-available PROPER optical propagation library (Krist 2007) is used for this propagation. PROPER includes routines to perform near and far field propagation, create complex apertures, generate low and high order synthetic aberrations, and model a deformable mirror with measured actuator influence functions. PROPER is available for IDL (Interactive Data Language) and is being ported to Matlab. The PROPER library has been verified for coronagraphic modeling in NASA Technology Demonstrations for Exoplanet Missions (TDEM) studies (Krist et al. 2011, 2012a, 2012b) and validated against results from the High Contrast Imaging Testbed (HCIT) at JPL (Sidick et al. 2013). Some coronagraphs had particular components that could not be well represented directly with PROPER routines and thus required custom procedures (e.g., high sampling for the small focal plane mask used by PIAACMC).

PROPER is used to model a realistically aberrated WFIRST-AFTA system, including deformable mirrors. By itself, PROPER only represents the output of the system and does not perform any wavefront control. That is accomplished by using custom software that implemented the Electric Field Conjugation (EFC) wavefront control algorithm (Give'on et al. 2007). The aberrated wavefront for the system is computed using PROPER and then EFC is used to determine the deformable mirror settings to suppress the scattered light (this is an iterative process due to the nonlinear nature of the propagation). After the dark hole around the star is created, jitter is included by propagating multiple small telescope pointing offsets through the model. The end-to-end modeling procedure used is the same as was used in the WFIRST-AFTA coronagraph design downselect process described in Krist (2014).

The full performance model includes the STOP model that gives the wavefront errors at the pupil located at the CGI fast steering mirror along with the PROPER model that starts at this point and predicts speckles in the coronagraph image plane. As of this report we have begun using this model to assess the coronagraph performance in one key respect: the stability of the speckles in the presence of thermal variations resulting from normal operations on orbit.

STOP modeling of a realistic observation sequence was conducted to evaluate the wavefront stability and its effect on coronagraphic performance. A bright star (β Ursae Majoris) was imaged over 6.1 hours and then the telescope was slewed to the science target, a star with known radial-velocity-detected exoplanets (47 Ursae Majoris), for 22.3 hours of exposures. To optimize the thermal stability of the Payload during the scenario, a set of modest thermal accommodations (an MLI radiation shield to block the view of the IC to the MLI on the back of the Solar Array panels, MLI closure of the small gaps between the Solar Array panels, and fixed temperatures for the bipods running from the IC to the Telescope and the S/C) were added to the thermal analysis. The wavefront variations at each 5000 second time step due to the structural and optical deformations caused by the thermal changes were computed and input to the PROPER coronagraph model to generate the corresponding dark hole images, with planets added to 47 UMa. A realistic model of the low-order wavefront sensing and control system (LOWFS/C) was included to correct for wavefront variations.

As shown in Figure 3-50, in the *unprocessed* image (left), only the brightest planet can be seen with any confidence against the background scattered light speckles. Subtracting the β UMa image from the 47 UMa one significantly reduces the speckles and clearly reveals the three simulated planets with contrasts of ~1x10⁻⁹. This level of speckle suppression is only achievable because the wavefront remained stable (to within tens of picometers) during the observations of both stars, thanks to the thermal stability of the system and the use of LOWFS/C. Based on the encouraging performance shown in the contrast predictions, the next design iteration will incorporate design features to provide a more thermally stable instrument carrier.

### 3.8.4 Summary and Future Plans

The initial IM results based on the Observatory configuration described in this report indicate that the Observatory design is performing well and has positive margins relative to the WFI's WIM and WSM optical requirements as currently understood. WFI IFU channel WFE/LOS predicts have been completed and will be used to assess its preliminary performance margins and to refine requirements. The detailed evaluation of contrast and speckle noise performance by the CGI team using the IM-supplied STOP and Jitter data will continue, with the results presented in §3.8.3.2 being a major step forward in substantiating the feasibility of incorporating a coronagraph on WFIRST-AFTA. These data results inform the understanding of current margins and provide insight into future design trades.

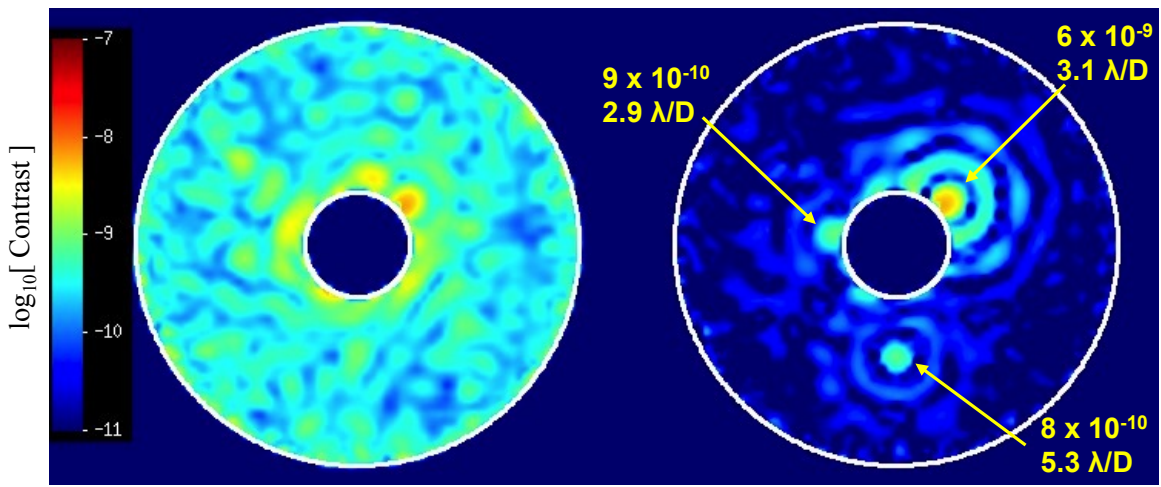

**Figure 3-50: Simulated coronagraphic images of 47 UMa using the Hybrid Lyot Coronagraph in broadband light (λ = 523 - 578 nm). On the left is the unprocessed image (sum of all images of the star taken during the STOP model time series), expressed in terms of equivalent contrast relative to the star. On the right is the absolute difference between the 47 UMa and β UMa images in which the background speckles are largely subtracted to reveal planets (the planet contrasts and distances from the field center are indicated).**

The current IM cycle has continued the maturation that has been seen in previous iterations. All four optical paths are now being modeled. Math Model Guidelines have been established to ease the integration of the S/C, Telescope, CGI, and WFI structural and thermal models to form the Observatory IM model. SigFit and LOM optical cross-checks have been implemented (both before and after the STOP runs). WFI cooldown offsets have been analyzed along with sequential Telescope and WF Channel refocusing. HGA design, modeling and analysis has been greatly refined. An interface for delivering WFE and Jitter optical results to the CGI team has been developed and will provide an excellent frame of reference as we engage with other science survey simulation teams to understand the optical data deliveries that will enable their work. There has not been space in this short summary to address the ACS pointing control work that is ongoing, but per the STOP LOS line item in Table 3-14 the pointing stability margins under ACS control are substantial, and future work will involve considering possible interfaces to the fast steering mirror function that is provided in the CGI.

Accurate, timely, team-oriented, and well-configured integrated modeling is essential to providing the optical performance feedback needed to support both trade evaluations/closures in the next design cycle, as well as, science survey simulations aimed at validating key requirements flowdowns. The progress made in the current IM iteration has been a big step towards realizing an IM process that responds to those challenges.

### 3.9 Ground System

The WFIRST-AFTA Ground System comprises three main elements: 1) the facilities used for space/ground communications and orbit determination, 2) the Mission Operations Center (MOC) and 3) the Science Operations Center (SOC), including the facilities for science observation planning, science and instrument operations, and ground data processing and archiving. For each element, existing facilities and infrastructure will be leveraged to provide the maximum possible cost savings and operational efficiencies. The functions to be performed by the ground system and the associated terminology are shown in Figure 3-51.

Two dedicated 18 m dual Ka and S-band antennas located in White Sands, NM are used for spacecraft tracking, commanding and data receipt. The antennas are separated by ~3 miles to minimize the chance of weather events interrupting the downlink. The two ground stations are within the beam width of the Ka-band antenna on the spacecraft so both receive the downlinked data simultaneously. Three 18 m antenna ground stations are currently in use at White Sands, one is dedicated to the Lunar Reconnaissance Orbiter (LRO) and two are dedicated to SDO. The two SDO antennas may be available for WFIRST-AFTA, depending on how long beyond 2015, the end of the baseline 5-year mission, SDO operates. The antenna dedicated to LRO is expected to be available for WFIRST-AFTA as LRO is currently in its extended mission phase. The WFIRST-AFTA development cost assumes that one new ground station will be built with one of the three existing ground stations available as the WFIRST backup. The White Sands ground stations interface with the MOC for all commanding and telemetry. Tracking data is sent to the GSFC Flight Dynamics Facility (FDF).

The MOC performs spacecraft, telescope and instrument health & safety monitoring, real-time and stored command load generation, command uplink and telemetry downlink, spacecraft subsystem trending & analysis, spacecraft anomaly resolution, safemode recovery, level 0 data processing, and transmission of science and engineering data to the science and instrument facilities. The MOC performs Mission-level Planning and Scheduling. With continuous downlink access, operations and communication scheduling complexity is reduced and DSN opera-tions are eliminated.

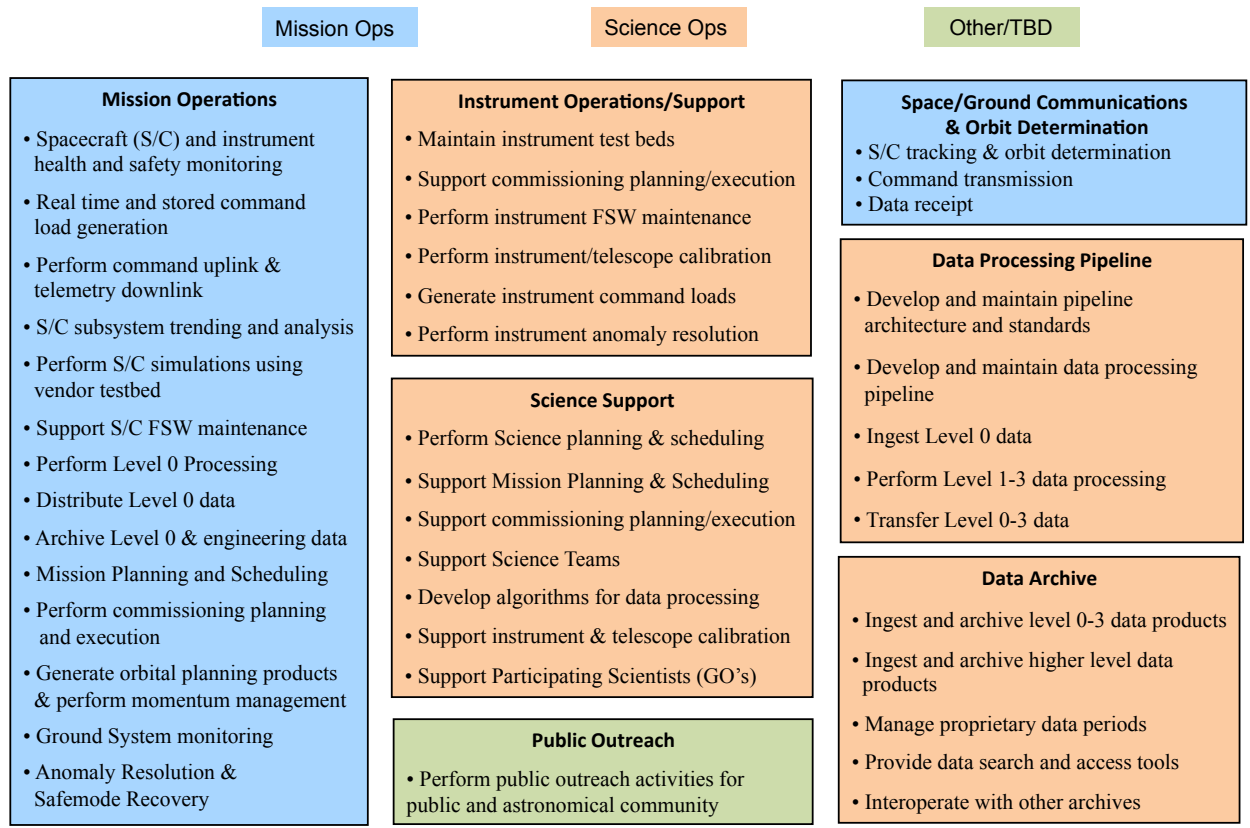

**Figure 3-51: WFIRST Ground System functions and associated terminology.**

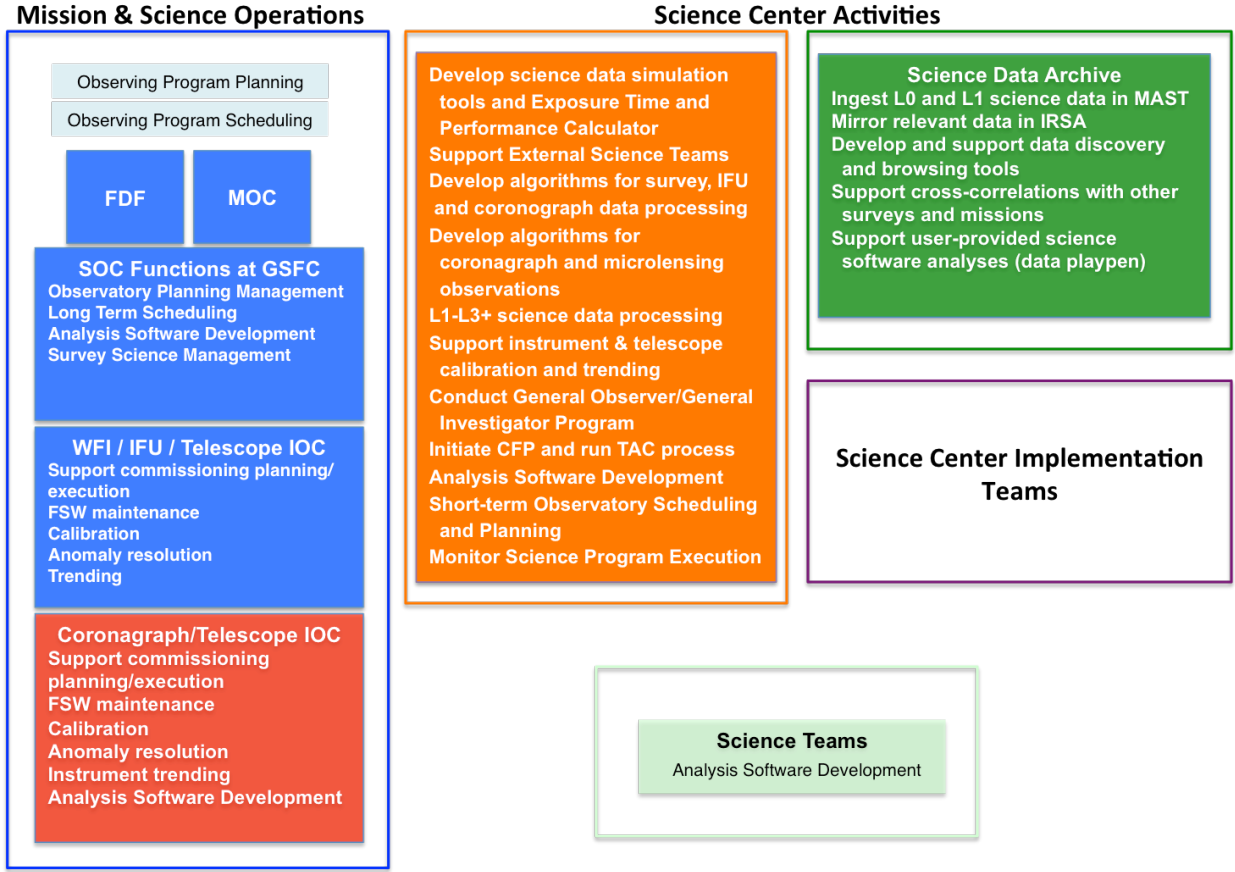

**Figure 3-52: Proposed assignment of ground system tasks to Mission, Science, and Instrument Operations Centers and the Science Teams.**

The SOC science and instrument divisions maintain instrument test beds, perform instrument & telescope calibrations, assist in the resolution of instrument anomalies, perform instrument flight software maintenance, and generate instrument command loads.

The SOC is also responsible for science planning & scheduling, supporting mission planning activities carried out by the MOC, running the Guest Investigator (GI) and Guest Observer (GO) Programs, providing Science Team, GI, and GO support, and performing public outreach activities for the public and the astronomical community. Data handling responsibilities of the SOC involve ingesting Level 0 science and engineering data from the MOC and performing Level 1-3 data processing for the Science Teams and GO community and transmitting these calibrated data to the archive, which will hold all science and engineering data from the mission. Data search and access tools are provided to the science community that enable efficient searches and delivery of archival data that ensure interoperability with NASA data archives and the Virtual Astronomical Observatory.

Several dedicated Science Teams will be funded over the prime phase of the mission to work with the WFIRST Project, Science Centers and broader astronomical community on the definition, execution, and analysis of the primary dark energy, exoplanet, and NIR survey observing programs. In this period, the GO/GI program provides funding for analysis of public data and for pointed observations selected competitively. Operations costs and grants for the GO/GI program in the primary mission are fully included in the lifecycle costs.

The proposed assignment of ground system tasks to the Mission, Science, and Instrument Operations Centers and to the Science Teams are shown in Figure 3-52.

### 3.10   Concept of Operations

We have constructed an existence proof of a possible WFIRST-AFTA observing sequence. **This is only an existence proof: the actual observing schedule will be determined at a time closer to launch. Its purpose is to demonstrate that an observing plan meeting WFIRST-AFTA science objectives is possible with the reference hardware and orbit.**

For the "existence proof" exercise, the following constraints were assumed:

- The total available observing time is 6 years, assuming a start of science operations on October 31, 2024.
- The orbit is a circular geosynchronous orbit with a mean longitude of 105 °W. It has initial inclination 28.5° and RAAN 228°. Over the course of the mission, Sun, Moon, and Earth asphericity perturbations cause the orbit to precess to inclination 26.4° and RAAN 187.6°. The initial inclination is driven by the latitude of the launch site, and the initial RAAN is chosen to provide optimal geometry for the science programs.
- The supernova program is carried out over a period of 2 years, with a 26% duty cycle. The SN observations require 0.53 years of live time plus 0.10 years of overhead. The SN field is located in a region near an ecliptic pole so that there are no annual cutouts due to Sun avoidance and no monthly cutouts due to Moon avoidance, and must have E(B−V)<0.02. The notional field chosen is centered at $b = −34°$, $l = 261°$.
- The microlensing program is carried out during six ~ 2.4 month "seasons" when the Galactic bulge is in the field of regard. In order to measure the stellar microlensing curve, we require observations over the full season, except for interruptions of ~5 days each month when the Moon passes near the Galactic bulge. To provide a time baseline for lens-source relative motion studies, 3 microlensing seasons are scheduled at the beginning of the mission and 3 are at the end.
- The HLS program requires imaging in multiple filters as well as counter-dispersed spectroscopy (dispersion in both the north and south directions). With prism dispersion in only one direction relative to the spacecraft and a constraint to point the solar array/sunshield toward the Sun, counter-dispersion requires two visits ~6 months apart (or N years + ~6 months), with the telescope pointed "forward" in the Earth's orbit in one case and "reverse" in the other. Other drivers on the HLS program are to: (i) reduce zodiacal background [all observations are at <2x the background at the ecliptic poles]; (ii) contain the HLS within the LSST footprint; and (iii) make as much of the footprint as possible visible from telescopes in both hemispheres. Since the zodiacal background is highest in the Ecliptic, these drivers compete. For this exercise, we have prioritized (i) and (ii), and done the best possible on (iii). Galactic dust columns are E(B−V)<0.02 (median) and E(B−V)<0.04 (95th percentile) according to the Schlegel et al. (1998) map. Of the HLS footprint, 72% is north of declination 40 °S and 44% is north of 30 °S.

- The coronagraph program is assigned 1 year of total observation time, notionally broken into 26 blocks of 2 weeks each. These blocks are required to avoid the GEO eclipse seasons.
- The GO program uses time not allocated to other programs, and has a requirement of ≥1.25 years. All celestial objects must be accessible at some point during the GO period.

The time not allocated to strategic programs is 1.33 years. All programs were required to operate within the Sun angle constraints (line-of-sight to Sun angle 54—126° and roll ±15°) to provide adequate power and stray light control, and stray light constraints were enforced for both the Earth and the Moon. The existence proof observing sequence is shown in Figure 3-53.

### High-Latitude Survey

The high-latitude survey is performed in 4 filters (Y, J, H, and F184) and the grism. For each filter, a sequence of 3 or 4 images is acquired via small-step dithers (roughly ¼ SCA horizontally and vertically simultaneously) to cover chip gaps (see Figure 3-54). The observatory then slews 1 field height and repeats the pattern. A sequence of strips along the sky is ac-quired in this way. A second pass over each field is acquired a half integer number of years later, to provide full sensitivity, Nyquist sampling (for J and redder bands), and to aid in the self-calibration solution. For calibration purposes, the second pass is rolled relative to the first pass to provide repeat observations of stars at all pairs of positions on the focal plane. The repeat passes are distributed throughout the mission so that long-term drifts in the calibration can be tracked and removed. The grism mode uses a similar pattern, except that 4 passes with 2 small-step dithered positions each are performed, so that WFIRST obtains grism spectra at 4 roll angles. Two of these are scheduled in the "leading" geometry relative to Earth's orbit and two in the "trailing" geometry, so that both northward and southward dispersion directions are obtained. This improves the wavelength determination of emission lines, since the offset of the emission line centroid relative to the continuum centroid (used as a reference in slitless spectroscopy) is suppressed as a source of systematic error.

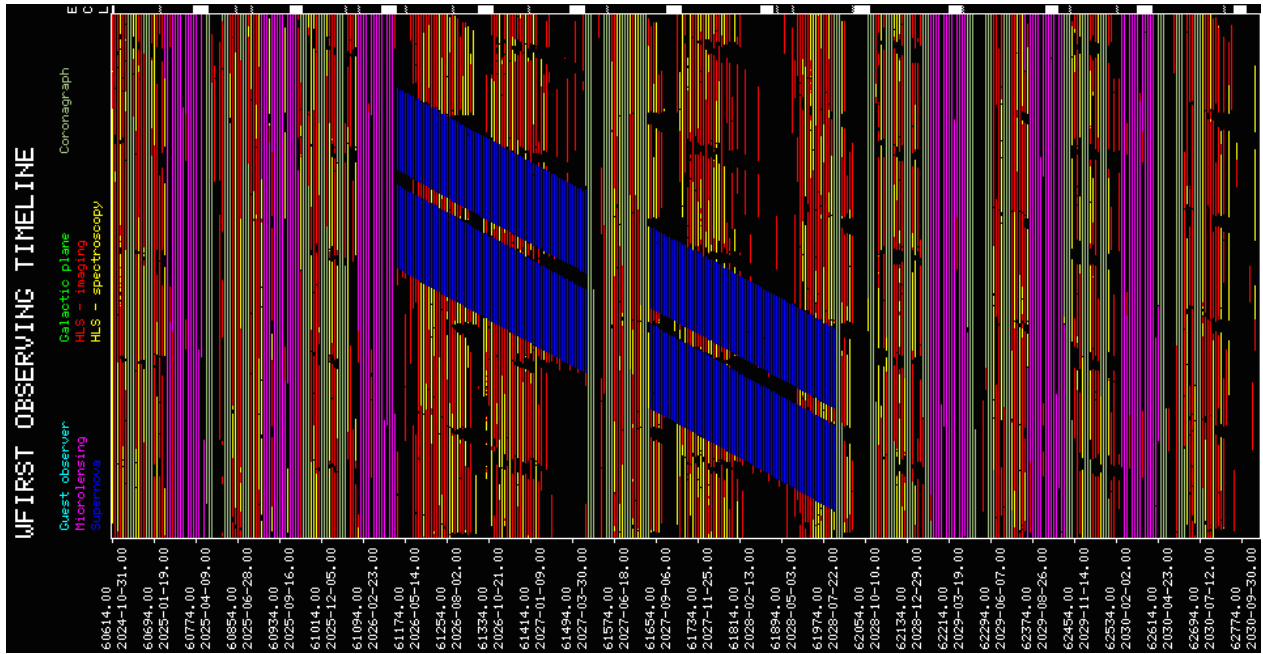

**Figure 3-53: The example WFIRST-AFTA observing sequence. Each vertical stripe denotes a 5-day period, and the 6-year mission proceeds from left to right. The microlensing seasons (magenta) and coronagraph blocks (gray) are indicated, as are eclipse seasons (white rectangles at top). The supernova survey (blue) uses portions of the telescope time for 2 years. High-latitude imaging (red) and spectroscopy (yellow) are interspersed.**

The HLS covers 2227 deg² (defined by at least 3 observations per filter, including the grism; the bounding box is 2279 deg²) and uses 2.01 years (including overheads). The example observing sequence is based not on analytic estimates of coverage rates, but on actual analysis of individual pointings, includ-ing the effects of field distortions on the tiling solution, the slew and settle times to the next field (including slews on all 3 axes, the reaction wheel torque and angular momentum allocations, and the observatory moment of inertia tensor). The HLS footprint is displayed in Figure 3-55. All HLS science forecasts assume that 3.3% of the observations are lost due to miscellaneous factors and are not recovered.

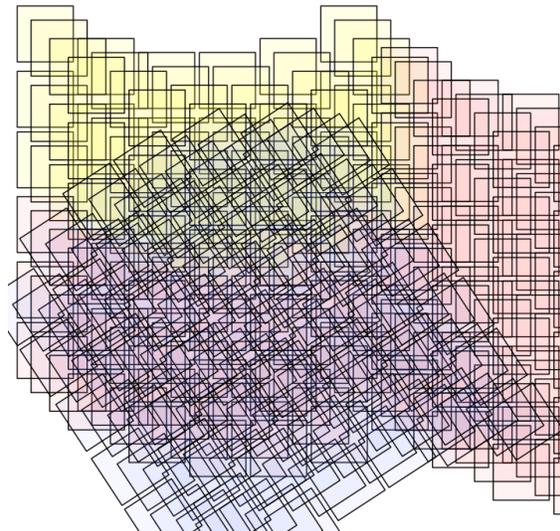

**Figure 3-54: Example of the sequence of observations for the high-latitude survey in H-band. The observatory first performs a sequence of 4 small-step dithers to cover the chip gaps (yellow). This pattern is repeated to cover the sky (red). A half-integer number of years later, the observatory returns to perform a second pass over the field (blue) at a general angle. The second pass enables internal relative calibration via repeat observations of stars at all pairs of positions on the focal plane, as well as monitoring of long-term drifts.**

The exposure times in the high-latitude survey are 32 frames (173.5 s) or 64 frames (347.1 s) for the imaging and spectroscopy exposures, respectively.

**Microlensing**

WFIRST carries out its microlensing program during 72-day "seasons" when the Galactic Bulge is between 54° and 126° from the Sun (hence accessible with WFIRST). Each season is interrupted only when the Moon passes near the Galactic Bulge, which occurs for ~5 days each month in geosynchronous orbit. The 6 microlensing seasons have a total length of 0.98 years (1.18 years of Sun angle accessibility for 6 seasons, minus a total of 0.20 years of lunar interruptions). The microlensing footprint is displayed in Figure 3-55 and in more detail in Figure 2-29.

The initial RAAN of the orbit is chosen to place the orbit normal within 48° of the Galactic Bulge, thereby guaranteeing 24-hour coverage of the Bulge region with no daily Earth viewing cutouts. The orbit remains favorably oriented for microlensing for 12 years following launch, and is thus acceptable for the 6-year primary mission.

The microlensing program observes 10 fields in the Galactic Bulge (although the final number and selection are still being optimized). Most observations are carried out in the wide (W149) filter, cycling through each field in sequence with 52 s observations. Every 12 hours, a cycle (one observation of each field) will be carried out with a bluer filter (Z087) and a longer exposure time (290 s) to provide color information on microlensed sources. Because of the layout of the detectors in the focal plane, ~85% of the survey footprint is observed for the full time while the remaining 15% is observed only in either the Spring or Fall seasons.

**Supernovae**

The time allocated to the supernova search in this "existence proof" operations plan is a total of approximately six months. The plan is to run the supernova survey for a 2-year period, using a 30-hour visit every 5 days for a total of 146 visits. The strategy is to use the Wide-Field Imager in two filter bands to discover the supernovae and to use the IFU spectrometer to take spectra to identify and type the supernova candidates, to obtain points on the light curves, take a deep spectrum near peak brightness for redshift measurements and to measure detailed spectral fea-tures, and to take galaxy reference spectra of the host galaxy after the supernovae have faded to a negligible level to allow galaxy background light subtraction from each light curve point.

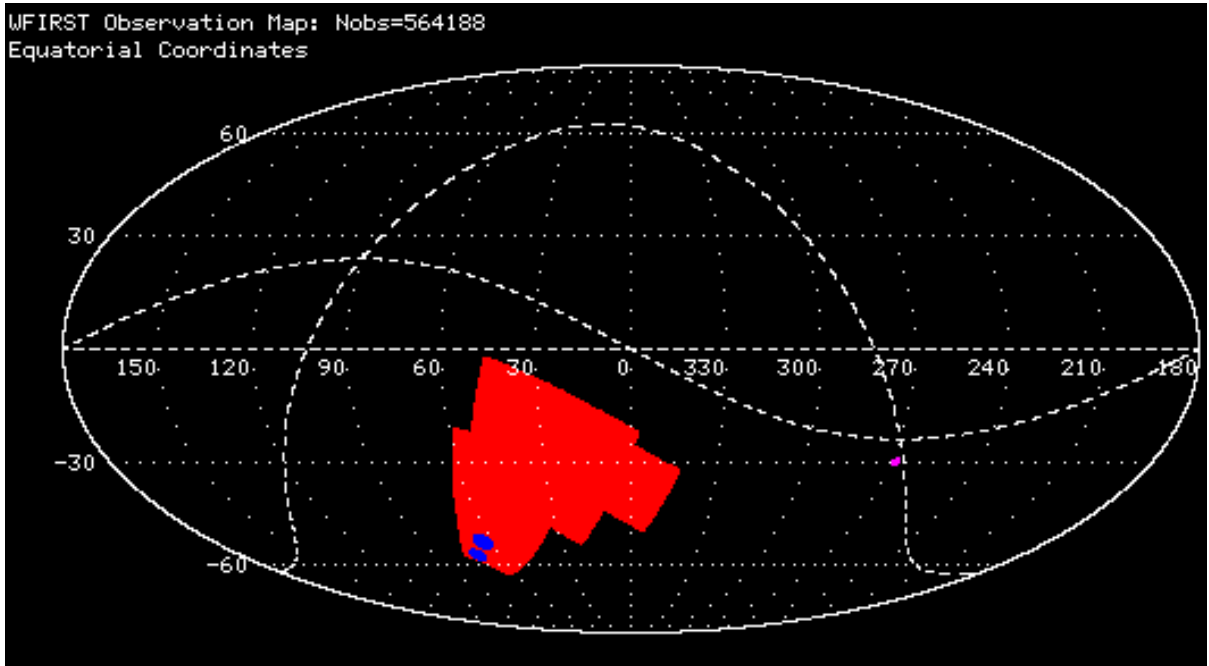

**Figure 3-55: The footprint observed during the WFIRST-AFTA DRM. The red area indicates the high-latitude survey. The magenta spot indicates the microlensing fields. The blue spots denote the proposed supernova survey fields.**

Since the number of supernovae per square degree increases significantly with redshift due to the volume effect, the plan is to carry out the survey in three tiers, searching larger areas at lower redshifts, all near the ecliptic pole to minimize interference from the Sun and the Moon:

- Low z    (z < 0.4) search    27.44 deg$^2$
- Mid z    (z < 0.8) search    8.96 deg$^2$
- High z   (z < 1.7) search    5.04 deg$^2$

With an imager field 6 H4RG detectors wide and 3 detectors high, the 5.04 deg$^2$ search area would be 3 fields wide and 6 fields high, making the area 18 detectors wide and 18 detectors high. Similarly the 8.96 deg$^2$ area would be 4 fields wide and 8 fields high for 24 by 24 detectors, and the 27.44 deg$^2$ area would be 7 fields wide and 14 fields high for 42 by 42 detectors. More elaborate cadences and tilings could be implemented with for example different cadences for the search for the different redshift ranges. The exposure times for the search are designed to be able to be sensitive to supernovae 12 days before peak brightness. The proposed supernova fields are displayed in Figure 3-55.

The spectroscopic observations in this plan will be designed to observe one supernova at a time, using the IFU. The exposure time is tailored to achieve the desired S/N ratio on the individual target supernovae, with the longest exposure times used for the z~1.7 SNe.

Each 30-hour visit will on average be used as follows:

- Imaging for discovery          8 hours
- Supernovae ID and typing    5 hours
- Lightcurves                         7 hours
- Deep Spectra                      7 hours
- Galaxy Reference                3 hours

There are some subtleties to the schedule that will be worked out in more detail when the final observing program is selected. A supernova will have to be followed for about 35 rest frame days past discovery to obtain its light curve. Thus out of the 146 visits over the two years the discovery search will be run only for the first 132 visits, leaving the last 70 days (depending on the redshift) for light curve follow up.

The galaxy reference spectrum will have to be taken approximately one year after the peak brightness of the supernova to make sure that the supernova light has faded to a negligible level. Thus for the first year of the survey galaxy reference spectra will not be taken, using only 27 hours per visit. During the second year galaxy reference spectra will be taken for the supernovae discovered during the first year, using all 30 hours per visit. Time for the galaxy reference spectra for the supernova discovered during the second year, roughly 3 hours every 5 days, will have to be allocated during the third year, using the time saved during the first year. This time does not have to be exactly 3 hours every 5 days but can be scheduled to minimize the impact on the third year program.

The example observing plan includes overheads and the brief (~3 hr) cutouts each day when the supernova fields are near the Earth. It currently schedules the reference spectra with the 5-day-cadence supernova visits, which is not realistic for supernovae observed in the second year; it is expected that incorporating this correctly will make the scheduling easier since reference spectra can be scheduled at times of lower demand by other programs.

**Coronagraph**

The initial observations will focus on discovery and characterization of planets near the target star. When a previously known or unknown planet is detected, additional observations will be made for longer time periods with full spectral resolution for planet characterization. A typical scenario for characterization of a planet will unfold as follows. For a given target star, the coronagraph will first acquire and observe a relatively bright nearby star in order to generate the dark hole. If the bright stars are taken from the roughly 100 brightest stars in the sky, then on average any given target star will have one such bright star within approximately 20 degrees of it. Once the dark hole is achieved, the telescope will slew to the science target. The deformable mirror configuration will be held constant to that generated from the bright star. The target stars have a median brightness of $V \sim 6$ mag. Full spectroscopic characterization of the planet will take place sequentially in the three, 18% wide IFS spectral bands. Finally, the telescope will slew to a calibration star for speckle characterization. The calibration star is chosen to be spectrally compa-rable to the science target. The required observation time for this observational scenario depends strongly on the coronagraph performance parameters and the actual planet contrast. Analysis was carried out to estimate the observation time. A $10^{-8}$ instrument contrast and a $10^{-9}$ planet contrast were assumed. The design has 10% overall throughput to the image plane and 6% throughput to the IFS detector (including reflection losses) in the entire annulus around the target star. Assumptions include the very low noise EMCCD detector.

Figure 3-56 shows a schematic of the coronagraph observation scenario. In a representative scenario, the bright, target and calibration stars would be beta Ursa Major, 47 Ursa Major and 61 Ursa Major, respectively. The slew, thermal settling, and acquisition of each of the three stars requires roughly twelve hours. The dark hole generation on the bright star takes place within 6 hours. Spectroscopy covering 600-970 nm in 3 sequential 18% bands takes approximately 20 hours. The total time required to complete the entire representative scenario is two days.

**Slew and Settle Operations**

As a survey mission, the slew and settle time of the observatory is a key performance characteristic. We have allocated a maximum of 40 N·m/s angular momentum for slews, with the remainder of the reaction wheel capacity reserved for angular momentum storage. This enables the most frequent steps (including all microlensing steps, as well as the small dithers and 1-field steps in the HLS) to be torque-limited.

Slew and settle performance has been estimated by time-domain simulation. The pointing profile was composed of 490 gap-filling slews (small dithers) of 0.025 deg in Y and Z, 84 short FoV (field steps) of 0.8 deg, and 13 tile-sized row slews of 5.5 deg. This proof-of-concept profile differs from the latest survey profile described in earlier paragraphs, but does provide a representative sample of slew lengths to use in the coverage rate estimates quoted above.

The slews followed a torque-limited profile, shaped at the ends to avoid excitation of structural oscillatory modes. Assuming the mass properties from an early observatory design iteration:

- Gap-filling slews require 11.25 sec for the slew, and <6.25 sec to settle.
- Short FoV slews require 50.25 sec for the slew, and <8.25 sec to settle (see Figure 3-57)
- Row slews require 132 sec for the slew, and <13.0 sec to settle.

All times have been re-scaled to the current moment of inertia tensor (plus margin) for the operations concept presented here.

Once the observatory has settled, a full frame of the H4RG-10s (~5.5 s) is acquired in reset/read mode, then sampling begins for the next exposure.

Overall observation efficiency is most sensitive to slew/settle performance of the most numerous slews: the gap-filling slews (HLS) and short FoV slews (microlensing). As the design matures, efforts to achieve high efficiency will focus on maintaining the level of slew/settle performance presented here, and in choosing efficient survey tiling strategies.

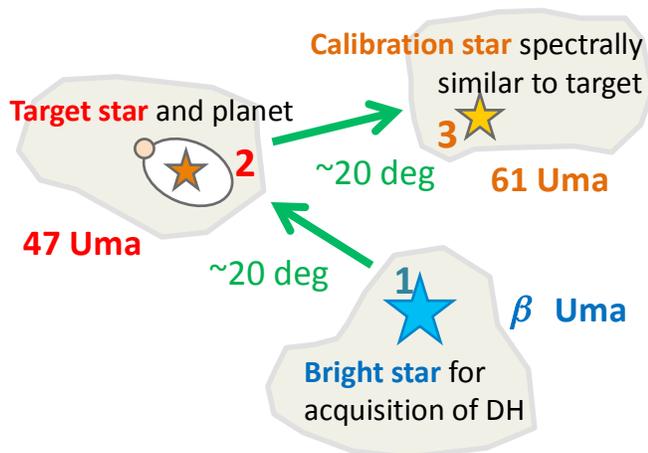

**Figure 3-56: Scenario for a typical known planet characterization observation.**

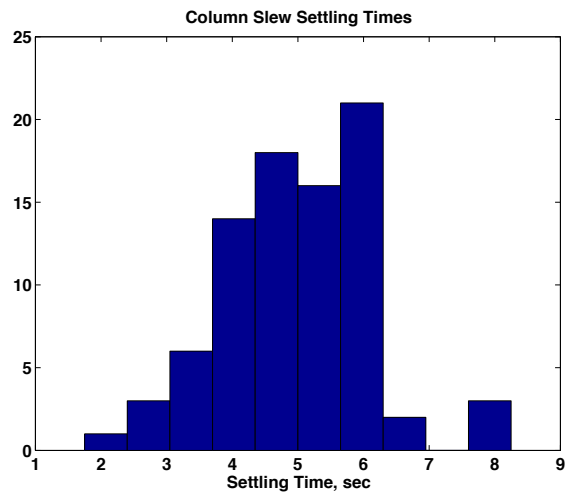

**Figure 3-57: Settling times for short FoV slews (0.8 deg) are typically between 4 and 6 sec, with a few outliers up to 8.25 sec.**

### 3.11  Cost & Schedule

The WFIRST-AFTA configuration in this study report builds on the past WFIRST configurations, as well as earlier JDEM mission design concepts. These encompass the JDEM-Omega configuration submitted to Astro2010, through the WFIRST design reference missions developed in collaboration with the WFIRST Science Definition Team (November 2010 thru July 2012), and more recently with the WFIRST Science Definition Team (October 2013 to January 2015). The current WFIRST-AFTA design reference mission incorporates the use of an existing 2.4-meter aperture telescope with a wide-field infrared instrument to meet the science objectives of the 2010 Decadal Survey. The most significant improvement of this payload combination over earlier smaller aperture WFIRST and JDEM designs, was the incorporation of considerably finer plate scale into the wide-field instrument, opening up game-changing Guest Observer and Guest Investigator science opportunities, in addition to the baseline dark energy and exoplanet science prescribed in the Decadal Report

In addition, a coronagraph has been added to this WFIRST payload baseline as a technology demonstration for high contrast exoplanet imaging. The baselined Occulting Mask Coronagraph (OMC) incorporates both Shaped-Pupil and Hybrid Lyot occulting techniques into the coronagraph design. These techniques were selected for their technical maturity and due to the fact that they place less stringent interface demands on the observatory.

The development phase for this WFIRST-AFTA baseline is 82 months, from preliminary design through launch (phase B/C/D). This development phase is preceded by a formulation phase (Phase A, nominally 12 months) and several years of pre-formulation studies, many of which have already taken place. The observing phase (Phase E) of the mission is baselined and costed for six years. The development schedule is shown in Figure 3-58. Slightly over seven months of funded schedule reserve is included in the development.

The WFIRST-AFTA design reference mission utilizes the 2.4-meter aperture telescope that was fabricated for another flight program, but subsequently not flown. The telescope optics and all of the supporting structure has been fabricated and integrated. The actuators will be rebuilt to the design of the original hardware and the telescope control electronics design leverages existing electronics boards from Exelis' inventory of commercial flight hardware to bring the telescope to full flight configuration. The availability of this telescope hardware eliminates this critical piece of the payload from getting on the mission critical path and simplifies the number of trades that need to be performed by establishing interfaces early. The early delivery of the telescope enables more extensive payload test opportunities (telescope + wide-field), including the certification of critical payload-level Ground Support Equipment. The WFIRST-AFTA wide-field instrument has incorporated some simplifications as a result of the larger aperture 2.4 m telescope. The wide-field instrument utilizes two optical channels, a simplification from earlier concepts. A single element wheel is accommodated by the use of a grism for the galaxy redshift survey. The wide-field focal plane has eighteen detectors, one-half the number of detectors of some earlier design reference missions. The wide-field instrument incorporates the addition of an Integral Field Unit; however, electrically this capability is essentially a nineteenth detector and electronics channel. The optics for the IFU are small and high TRL.

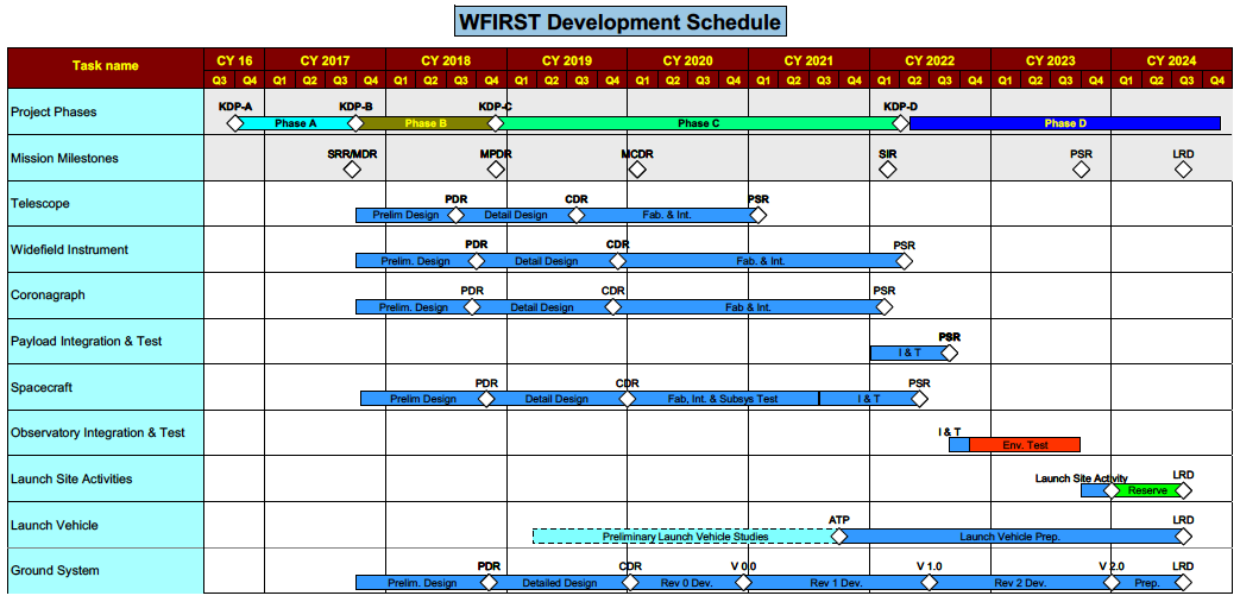

**Figure 3-58: The 82 month WFIRST-AFTA development schedule includes slightly over seven months of funded schedule reserve.**

The build-up and integration philosophy of the WFIRST-AFTA observatory is based on the well-established practice of building small assemblies of hardware, thoroughly testing them under appropriate flight environments, and then moving on to the next higher level of integration with those assemblies. The WFIRST-AFTA telescope and instrument will be developed and individually qualified to meet mission environments. The critical path of the mission is through the development of the wide-field instrument. The instrument is qualified prior to integration with the telescope. Following ambient checkout, the entire payload is tested at temperature and vacuum at GSFC to verify the end-to-end optical performance. The spacecraft is then integrated to the payload, and tested in ambient. Following a successful ambient checkout, a complete observatory environmental test phase is performed. Upon successful completion of the observatory environmental test program, the observatory is readied for shipment to the KSC, where the launch campaign is conducted.

The parallel build-up of all of the mission elements allows substantial integration activity to occur simultaneously, increasing the likelihood of schedule success. The early delivery of the telescope will allow the use of the telescope for early instrument optical checkout including ground support equipment, allowing the retirement of integration risk even earlier in the program. Because all of the major elements of the observatory (telescope, instrument, and spacecraft) are located at GSFC over 2 ½ years before the planned launch, there is considerable flexibility in optimizing the schedule to compensate for variation in flight element delivery dates, should workarounds be necessary. Over the 2 ½ year I&T period, the Project will have flexibility to reorder the I&T work flow to take advantage of earlier deliveries or to accommodate later ones. Should instrument or telescope schedule challenges arise there are options to mitigate the schedule impacts by reallocating the payload level environmental test period. Should instrument or spacecraft challenges arise, there are options to modify the workflow and pull other tasks forward to minimize risk and maintain schedule. The payload design includes access to the instrument volume when attached to the spacecraft, allowing late access to the instruments during observatory I&T. The WFIRST-AFTA observatory I&T flow is very achievable, given the planned schedule reserve and the opportunities available for workaround.

Fifty-five months are allocated to complete the wide-field instrument, from the start of preliminary design, through the delivery of the instrument, not including funded schedule reserve. The coronagraph delivers two months before the wide-field instrument. The overall 82-month observatory development schedule includes slightly over seven months of funded schedule reserve, further increasing the likelihood of executing the plan.

Early interface testing between the observatory and ground system is performed to verify performance and mitigate risks to launch readiness. Prior to payload integration, interface testing between the spacecraft and the ground system is performed. Immediately following payload integration to the spacecraft, end-to-end tests are performed, including the payload elements. These tests are performed numerous times prior to launch to ensure compatibility of all interfaces and readiness of the complete mission team.

Several NIR survey mission concepts have been costed by the Study Office over the past six years, and many of these have been validated with independent cost estimates requested by NASA HQ. These have tracked very closely, including the independent cost estimate performed as part of the Decadal Survey. The current design reference mission is uniquely different in that it also includes the coronagraph technology demonstration. HQ has directed that the coronagraph will impose no driving requirements on WFIRST, and during development the coronagraph implementation will be treated as a technology demonstration. These two points are critical, as they ensure that the coronagraph will not jeopardize the mission critical path and put at risk the science mandated by the Decadal Survey. However it does allow exoplanet high contrast imaging to take a huge leap forward in technological evolution with a modest additional budgetary investment, because of the fortu-itous timing of the WFIRST survey mission, along with the maturation of the high contrast imaging technologies. No other approach offers such a highly leveraged advancement in aerospace coronagraphic technology advancement.

The life cycle cost estimate for the current baseline (with coronagraph) is in process and is being developed using the same techniques (grassroots, parametric, and analogy) that have been applied to previous cost estimates. Project Management, Systems Engineering, Mission Assurance, and Integration & Test, are estimated using a grassroots approach and validated against analogous missions. Pre- and Post-launch Science will be projected based on the expected size of the WFIRST Science Announcement of Opportunity (AO). This estimate will include support for the AO-selected science teams in the mission development phase as well as the 6 years of operations. Additional funds for the Guest Observer and Guest Investigator Program are included in the Science estimate. The payload and spacecraft will be estimated primarily using parametric estimates, with the exception that the telescope estimates rely on historical estimates to cost the overall telescope effort. The parametric estimates will be constructed using master equipment lists (MELs) and historical cost databases and are adjusted for mass and complexity factors.

The WFIRST-AFTA design reference mission has an implementation strategy that is low risk, proven and amenable to workarounds. The early delivery of the telescope due to its unique nature, allows for a payload and observatory I&T program that will have the ability to retire risk early and affords the flexibility to adapt to changes should development problems arise. The current WFIRST design reference mission is executable and delivers the #1 ranked science defined in New Worlds New Horizons, while providing an unprecedented Guest Investigator/Guest Observer capability and an extraordinary opportunity in space-based high contrast exoplanet imaging.

## 4   SCIENCE POLICIES

### 4.1   Data Rights Considerations

The clear trend in the astronomical community during the past decade has been a move away from sequestered access to telescope resources in the form of guaranteed time and proprietary data rights. For facilities that are primarily intended for guest observer use, this approach held sway for NASA's Great Observatories. Hubble, Spitzer, and Chandra, for example, all had a one year default proprietary period for their General Observer data. However, these policies were not applied for some types of programs. In particular, data sets of general use to the broad community, such as the Hubble Deep Field, many Hubble Treasury projects, and Spitzer Legacy observations are released with no proprietary period. Large programs (e.g., Spitzer GO programs over 500 hours) often waived any proprietary period. During its extended mission, Spitzer projects for over 100 hours had a zero default proprietary period with an optional maximum of 90 days.

Interestingly, an inquiry into time-to-publication for various proprietary period models revealed there is no strong correlation between proprietary data periods and promptness of publication. This would seem to indicate that the cessation of proprietary rights does not provide a motivator to prompt publishing. Further, the willingness of teams to waive proprietary periods on large investigations points to an underlying situation wherein the value of a period of prior planning and preparation is more important than the value of a period of restricted data access.

More recently, the Fermi/LAT instrument, which provides a large area imaging capability for general astrophysics, used a data policy with a one year post-launch proprietary period for the instrument team, but then all data after the first anniversary are available immediately upon processing. The quantitative question to ask is then what the post-launch proprietary period is for, and therefore how long it should be. This can be answered when considering whom the proprietary period serves.

The WFIRST SDT subscribes to a broad open data / open skies policy, whereby all researchers across the world will have equal access to all data in the archive. As a general principle, this ensures maximal utilization of this publicly funded science mission. To quote from the *New Worlds, New Horizons* Decadal Survey, "the principle of open skies is compatible with the guiding principle of maximizing future scientific progress" (NWNH, p.85). It allows for competition within the community, which is a widely recognized means to ensuring the highest level of scientific merit. Open competition implies that pre-launch products such as science data simulations and analysis algorithms will be made available to the community to enable the broadest reach of scientists to use the data from WFIRST for a wide range of science investigations.

For WFIRST's major surveys (the HLS, supernova, and microlensing surveys), it is expected that Science Investigation Teams (SITs) will be funded to work with the WFIRST Project Office, Science Operations Center (SOC), and the broader astronomical community to develop the operational planning, science simulations, and analysis software development (see §3.10). These teams will be well prepared for analyzing survey data, and their algorithms and data products would be made public along with those provided by the WFIRST SOC.

Given that the instrument/science teams have significant knowledge and experience with WFIRST by the end of in-orbit checkout, it would not substantially diminish their competitive advantage to allow for a zero proprietary period for all WFIRST data in the archive. This is a policy consistent with that being implemented for LSST; the preface to the *LSST Science Book* states "the LSST project will deliver fully calibrated survey data to … the public with no proprietary period" (p. 9). The time to enter data into the archive needs to be prompt in order to facilitate rapid follow-up observations, but this time may be level-dependent. NASA categorizes the most basic data as Level 0: unprocessed instrument data, where only artifacts of communication have been removed. The first astronomically useful data are likely to be Level 1B: reconstructed, unprocessed instrument data with instrument and radiometric calibrations applied. The data manipulation required to ready data to this level is relatively modest, and is done with the best-available calibrations at the time of production. This level of processing and release is appropriate for WFIRST, enabling the SOC to produce transient alerts and to archive the raw images for the community to do near-real-time analysis. For the LSST, where transient alerts are also an important element of the science, the total time for this stage (i.e., from data storage to L1B availability) is 60 seconds (*LSST Science Book*, 2009, Section 2, p.39;

http://www.lsst.org/files/docs/sciencebook/SB_2.pdf).
Following this lead, prompt release in the archive of WFIRST L1B data products would maximize the output of time-domain astronomy from WFIRST, including the great number of supernova and microlensing events. Eventually, science data are processed to Level 3: im-aging data have been extensively calibrated and co-added to provide deep images and/or spectra, and other science data products (often catalogs) are produced. It is reasonable for this greater effort to have a longer timeline for release. Many of the high-level data products (e.g., basic source catalogs with source classifications) could and should be released to the public on rapid timescales (3 months or less). A longer time period can be adopted for releases of specialized high-level survey products that require more intensive processing and meticulous analyses.

WFIRST archival data are of substantial value, both to provide for General Investigator (GI) science and to enable competing teams to address the key science investigations of the mission. For example, a meritorious external team wishing to conduct microlensing investigations using the WFIRST data should be supported to do so, even though there would be a competitively selected SIT responsible for working with the WFIRST Project and SOC on the design and execution of a microlensing survey. This approach would yield the best science outcome through multiple independent and competing groups analyzing the same data. In a no-proprietary-period approach, both teams should be funded. The SDT does not support a policy wherein a single team selected early in the mission lifetime would be forever without viable competition for the analysis and publication of results with the WFIRST data. To this end, we suggest that the WFIRST project conduct a competition roughly two years prior to launch that would select teams to conduct compelling science investigations using the already-planned WFIRST survey data. The SDT also assumes there will be at least annual calls for Guest Observer (GO), GI and Theory proposals, to allow responsiveness to the latest scientific developments and insights. This approach should provide a level of support and sufficient timescale to permit the selected teams to be in a position to conduct these investigations as soon as WFIRST data become available, thereby reducing the discrepancy in preparation between the initially-selected WFIRST science investigation teams and the newer teams.

## 4.2    Science Team Selection Considerations

As no formal decision for the WFIRST acquisition strategy has been made, there remains the possibility of a few different approaches to selecting the SITs. Key in these options is whether there is a hardware contribution related to instrumentation that is a portion of the SIT selection. Note that in all options the role of the WFIRST SOC will need to be factored in when as-sessing the optimal strategy. While the SITs have the goal of achieving specific science outcomes, the WFIRST SOC and Project have as top goals optimizing the overall scientific productivity of the mission by the entire astronomical community. We consider three options here as representative of the range of paths that could be explored.

In the first option, a NASA Announcement of Opportunity (AO) is released to solicit proposals for providing a flight instrument along with scientists including one or more members of the SIT. For WFIRST, the two instruments would thus result in two groups from which members would be drawn to comprise the SIT, along with a number of Project Scientists, representatives of international participants, and others. Since WFIRST is a mission comprised predominantly of well-defined observations (e.g., Planck) rather than a general observer capability (e.g., Spitzer), the key focus of the two SITs would necessarily be on these high priority astronomical observations. To combine the selection of the instrument and its SIT would be to merge the instrument-building team with the astronomical observation team. This approach would not permit making a selection independently on the instrument-building and astronomical observation experience and techniques of competing groups.

A second option to consider would solicit members for a Science Working Group (SWG) with relevant and broad experience to provide oversight and counsel on the direction and progress of the WFIRST development. Prior to launch, a separate solicitation would result in selecting a number of science investigations to produce the required WFIRST science outcomes defined by the Decadal Survey, including measurements of dark energy and dark matter, the production of a large infrared survey for archival purposes, exoplanet microlensing study, and the demonstration of coronagraphy for observations of debris disks and exoplanets. The selected SWG would therefore resemble the present SDT, and would have no hardware deliverable and no defined science investigations to conduct. This approach could provide a prompt initiation of the SWG, and because of its advisory nature could remain dynamic, with members assigned for a term and being replaced by new members throughout the duration. However, it does divorce the SWG from the teams who must eventually conduct the Decadal-prioritized science; this will result in a set of instrumentation and an operational design that was not optimized for the selected science investigations. However, this approach does not reflect the role of the WFIRST SOC staff, who would already bring extensive experience in optimizing science from the WFIRST mission.

A third approach would be a scenario where NASA acquires the instruments separately from the science investigations. For the science investigations, multiple teams would be selected to address the science outcomes listed in the previous paragraph. These SITs could be selected by NASA Research Announcement (NRA) rather than by AO, and could be selected both sooner and more quickly as a result. The individual SITs would be tasked with conducting the Decadal-prioritized science, and would be responsible for addressing, in collaboration with the WFIRST Project and SOC, a significant amount of work during the course of WFIRST's mission lifecycle. Initially, during the formulation period, this would consist primarily of science requirements definition and flowdown, and could also involve early science observations using other facilities to promote the development of these requirements or other related future activities (cf. Appendix G). During mission development, each SIT would be responsible for the production and dissemination of high fidelity science data simulations, which serve both to inform the broader community of the output of WFIRST and to provide the WFIRST SOC with flight-like data products to run through the archive pipeline. The SITs would participate with the SOC in conducting data challenges to promote end-to-end evaluation of WFIRST operations and analysis. Nearing the time of WFIRST launch, the SIT activities would then trend toward participation in detailed observation planning and in-orbit checkout planning. During the initial verification and checkout period, the SIT would work with the SOC to process the early data and to freshen the survey execution plans. Once formal survey operations begin, the SITs would each then be tasked with conducting the necessary analysis and calibration to realize the final science outcome that was their responsibility.

To form the mission SWG, each SIT, instrument team, and international partner would contribute one member in addition to the project team and SOC members. A number of interdisciplinary scientists would also be appointed, in part to represent the interests of the broader scientific community, *à propos* the guest observer program. This SWG, numbering in the neighborhood of 15 members, would then feature representation for both the science outcomes and the instrument builds, but would not conflate the membership and selection of the two. It would provide representation for each of the science investigations, for the instrument teams, for the broader community, and for international partners.

In this third option, the hardware deliverables of the instruments would be acquired via independent means, and would be executed accordingly. There would be close connections between the SITs and the scientists and engineers working on the instruments and at the SOC: at first through requirements definition, then through data simulations, subsequently during integration and testing analysis, and finally through on-orbit calibration. In this fashion, the SITs would serve as a body of instrument scientists for the hardware without having been selected specifically for their knowledge of instrumentation; scientists with that expertise would be selected with the instrument build, and will also be contributed by the SOC.

One potential disadvantage with the above approaches is that SWG membership tends to be either static, where individuals selected near the beginning of the mission lifecycle remain in their designated role until the end of the mission, or tends to be dynamic, where individuals who help to conceive of the mission design are not necessarily those who eventually execute the mission observations and analysis. Both of these extremes have obvious undesirable downsides. To mitigate this, the third option would employ a mix of strategies. A SIT that is selected near the beginning of the mission for, e.g., a five-year term would then be reconsidered halfway through mission development for either re-competition or continuation. The interdisciplinary scientist positions could be provided a shorter term, as their deliverables are less concrete, and can be competed multiple times over the mission lifecycle. This approach should permit a consistent infusion of new people and new ideas into the project, while some of the people (in particular, the instrument teams) would be engaged for the longer term. Establishing an additional external Advisory or Users Committee with rotating membership might also be considered.

In all three of the scenarios described above, the SITs would work in close collaboration with the SOC to execute the primary science components of the WFIRST mission. In this scenario, the SIT members work with the WFIRST SOC to test and verify the performance of the two key science instruments prior to launch, although with some differences of responsibilities depending on the role of the SIT with regards to instrument production. After launch, the SITs and the SOC work in partnership to ensure the surveys are executed efficiently and that a vibrant GO program is enabled as a key component of the WFIRST mission. The SITs and SOC work to deliver promptly to the global astronomical community a broad range of high-level science products to maximize WFIRST science and to ensure the instruments are performing at their peak capabilities. The SITs would, as befits their charter, be amply funded to carry out the advanced analyses needed to extract the top-level scientific objectives of the primary surveys. The SITs would have the advantage of significant funding and an intimate knowledge of the instrument performance, gleaned during their participation in the preflight integration and testing of the science instruments, that will ensure completion of the necessary science outcomes. However, the community at large would be enabled, and ideally well funded, to perform similar studies using the reduced data in the public archive in order to provide essential cross-checks of results and to foster new ideas for extracting the key constraints on dark energy, dark matter, and the properties of exoplanets. This broad participation enabled by a well calibrated and well documented archive built by a collaborative effort of the SITs and SOC staff will ensure that the WFIRST mission is regarded by the entire astronomy community as a key asset for all of astrophysics and not an expensive experiment being run by a small fraction of the community.

Regardless of which approach is eventually taken, the principle guiding the selection of the Science Working Group should be one in which scientists are competitively selected for science investigations. The SDT suggests that a rigorous tasking for these scientists will ensure robust involvement of these scientists and will provide for a healthy and deep involvement in all phases of mission development. Their tasks would include collaboration with the WFIRST Project and SOC on science requirements flowdown, preparatory observations, data simulation, operations planning, data analysis, and publication. This is essentially the task in option three above, but could be employed with more or less emphasis in the other two options.

As recommended in §4.1, the SDT envisions a data rights model where a near-zero proprietary period is invoked for all WFIRST observations following in-orbit checkout. Scientists associated with the project would be provided incentive via direct support for their programmatic activities, not guaranteed time. In keeping with this, there would be no reserved observations catalog requiring that the GO program avoid observing in any of the fields selected for the surveys. The SDT did consider models of proprietary access for SWG members, but found no justification that satisfied the rigor of maximizing science results; instead, the justification for such proprietary access was typically to provide for incentives/rewards for the years of effort those individuals contribute to the success of the WFIRST mission.

Given the above, the SWG and the SITs would be entirely populated by members with significant funded work necessary for the success of the WFIRST mission in all phases. Specifically, the statement of work would include collaboration with the WFIRST Project and SOC on:

- Phase A/B: science requirements flowdown
- Phase C/D: simulated data flow; training sets; preparatory observations; algorithm preparation
- Phase D/E: operations planning; pipeline validation
- Phase E: data analysis; calibration/trending → science outcomes

A decision on the details underlying these broad tasks – which are best performed at the project level, which are best performed by the SOC, which are best performed by the science teams, and what the interfaces are – will depend on the nature of the selection of the SITs and the roles of the WFIRST Project and SOC. The SITs should be provided support, including funding for salary time for the PI and Co-Is, and additionally for such staff and postdocs as are necessary to conduct detailed modeling and analyses. The funding level would be modest early on, increasing pre-launch to a level sufficient to carry out the major science investigations incumbent upon WFIRST.

# 5  PATH FORWARD

This report marks the completion of the 2014 SDT and Project Office work on the WFIRST-AFTA mission. NASA will use this material and ongoing costing studies of the mission to decide on the path toward further development. There are a number of activities that are important for future science teams to undertake. The following tasks are recommended for study in the coming two years:

A.  Optimization of the Reference Design

• Study possible GO and GI program examples and how they fit in the observing program. The deep supernova fields, parallel coronagraph observations, and possible "deep drilling" GO programs will also be valuable as calibration data sets. Determine quantitatively which science programs and calibrations could be undertaken with the same data sets.

• Optimize the filter set and grism bandpass across the WFIRST science program. The optimization may be a function of the telescope temperature if at the time of the study a final temperature has not been selected. As part of this optimization, there should be a call for community input.

• Perform further investigations of the IFU instrumentation (e.g. a data cube simulator) and science (GO program capabilities and use for calibration of redshift distributions).

• Develop a more detailed operations concept for the supernova program, including the science return for alternate observing strategies.

• Explore non-exoplanet science cases for the coronagraph and its associated instruments (AGN, high spatial resolution visible imaging, evolved stars, etc.)

• Perform analysis to improve predictions of microlensing event rates, and conduct a detailed assessment of microlensing event parameter estimation (see Appendix K).

• Continue to track the science impact of engineering trades (in e.g. momentum management, communications, detector specifications...) as the need arises.

• Study the advantages and disadvantages of going to L2 (versus the inclined geosynchronous orbit baseline) for science yield and the cost, risk, and complexity of the mission (see Appendix C).

B.  Systematic Error Control

• Assess the relative photometric calibration of the wide-field instrument. Determine via simulation the extent to which HLS observations enable this calibration internally, and investigate the impact of color terms, nonlinearities, persistence, background/thermal variations, and secular degradation (e.g. coatings) on the calibration solution. Also determine the extent to which the microlensing data and deep WFI imaging in parallel with the coronagraph could be used to improve or cross-check this solution.

• Characterize anticipated astrometric performance of the wide-field channel. Estimate the extent to which the distortion map and plate scale can be measured internally to WFIRST observations, and assess the impact of nonlinearities, time dependence of the optical parameters, chromatic terms, and small-scale irregularities on the astrometric performance of the HLS and on pointed GO observations.

• Develop photometric and astrometric systematic error budgets (e.g. PSDs) for the microlensing program. Connect these requirements to the output of the integrated modeling and assess performance and margin.

• Build a calibration strategy and an absolute calibration error budget, including covariances between wavelengths, for the IFU and compare it to preliminary supernova program requirements. Study the ability to cross-check this solution for the IFU against observations with the wide-field instrument.

• As a result of the above studies determine what, if any, hardware modifications are required to meet the systematic error requirements of the dark energy and microlensing programs.

C.  Synergies with Other Observatories

• Survey the need for precursor observations for microlensing (see Appendix G), for low redshift supernovae, and for radial velocity studies of exoplanet coronagraph targets.

- Study opportunities for joint observations and requirements for joint analyses (including software requirements) with Euclid, LSST and other ground telescopes.

D.  Coronagraph

- Develop a more detailed coronagraph DRM model, including: the most recent estimates of performance and exposure times, a search strategy for finding new planets, statistical information about extrasolar planet populations, expected preparatory information from Doppler surveys, and post-processing capabilities.
- Perform a deeper investigation of coronagraph polarization, including polarization effects of the telescope, science gains from polarization measurements, and relative benefits of alternative polarization analyzer modules, and operational aspects.
- Assist with development of a wavefront control algorithm technology development plan including quantitative milestones on, e.g., convergence rate vs. star brightness, joint estimation algorithms that continuously update wavefront solutions to distinguish planets from speckles.
- Determine pre-flight and in-flight calibration requirements for the coronagraph.
- Study and analyze techniques for removing residual star light in coronagraph images and spectra in order to optimize sensitivity for detecting planets and circumstellar dust.

E.  Policy Issues

- Consider various topics related to foreign involvement in WFIRST through discussions with foreign members of the SDT.

## 6   CONCLUSION

The WFIRST-AFTA mission is highly compelling and is found in the current study to be feasible and affordable. It will revolutionize our understanding of the expansion of the universe, the birth and evolution of galaxies, and the formation and atmospheric composition of exoplanets. It has significantly enhanced performance compared to previous configurations and compared to the Astro2010 requirements, as concluded by the 2014 NRC review panel. The larger mirror gives finer imaging and improved sensitivity making the mission more powerful and more complementary to LSST and Euclid. In addition, JWST will greatly benefit from WFIRST's discovery of interesting objects in the wide-field survey. The coronagraph enables rich scientific return at much lower cost than a dedicated smaller coronagraphic telescope mission, enabling the first images and spectra of exoplanets like those in our solar system and of dust disks around nearby stars. The future conceptual design work described in §5 will further refine the observatory and its performance, enabling a faster and lower-risk development.

## Appendix A    Impact of Telescope Temperature

In order to reduce risk, the WFIRST-AFTA SDT has baselined a telescope operating temperature of 282 K, within the currently qualified range of the telescope. This section considers the impact of the 282 K baseline, relative to the 270 K operating temperature in the 2013 SDT report. Attention is focused on the wide-field instrument because the coronagraph operates out to only 1 µm and hence is not sensitive to the telescope thermal background.

An assessment of the wide-field channel sensitivity was carried out using the following assumptions:

- Telescope temperatures of 270 K and 282 K are assessed. The telescope was qualified to operate down to 277K.
- The standard detector read noise model is used: one sample every 5.423 s (100 kHz readout, using 32 channels of the H4RG-10 device, including the interleaved guide window). Read noise is assumed to be at the requirement level of 20 e⁻ per CDS, with a correlated noise floor of 5 e⁻ RMS. The super-Poisson variance in sample-up-the-ramp mode (1.2x in the limit of many samples) is included. Detector dark current (excluding telescope and instrument thermal backgrounds allocated separately) is 0.015 e⁻/pix/s.
- The equivalent emissivity per surface of the primary and secondary mirrors is 0.025. This includes the emissivity of the protected silver coating (0.017), a 2% contamination allowance distributed evenly among the optical surfaces, and 30% additional margin.
- For thermal calculations +2 K was applied to the telescope temperature, e.g. the "282 K" is actually evaluated at 284 K. This factor is intended to account for several miscellaneous effects that will increase the thermal backgrounds in WFIRST beyond the formal thermal emission in a rectangular filter. For example, a $\Delta\lambda/\lambda = 0.05$ width trapezoid-shaped cutoff is equivalent to an increase in temperature of +0.65 K, and a variation of $\sigma_T = 1$ K RMS results in an effective emission temperature increase of +0.037 K relative to the mean temperature of the mirror due to the nonlinearity of the intensity-temperature relation for an emissive surface.
- Because WFIRST-AFTA uses 2.5 µm cutoff detectors, a warm telescope creates an additional thermal background due to filter red-leak (or imperfect out-of-band rejection: a small fraction of the telescope thermal radiation near 2.5 µm penetrates the filter). An out-of-band transmission of ≤0.01% for all filters is assumed.
- In the F184 filter, microlensing-wide filter, and grism modes, a full pupil mask is included. A pupil mask was not included for the H band in the 270 K baseline, but is included at 282 K. In the other filters, the emissivity from the secondary mirror support tubes and central baffle are included.
- Zodiacal backgrounds are taken to be 1.40x, 1.60x, and 1.76x the annual mean at the ecliptic poles for the grism, Y/J/H filters, and F184 filter respectively. (For the grism, this is the mean sky brightness in the HLS; for the filters, this is the 75th percentile sky brightness, as described in the main report.)
- Field dependence of the filter bandpass is not yet included. It will result in higher telescope thermal emission in the center of the field of view and lower thermal emission near the edges. This variation in thermal background is estimated to be 20% peak-to-valley.
- The coronagraph has an internal operating temperature of 290 K, warmer than the telescope. Thus it emits thermal radiation that can reflect off the secondary mirror and enter the WFI. In the H, F184, and W149 filters it is blocked by the pupil mask. In the bluer filters with no pupil mask, some of this radiation can in principle reach the WFI detectors due to nonzero out-of-band filter transmission, but this is negligible compared to the thermal emission from the telescope central baffle (the small solid angle of the coronagraph more than compensates for the higher surface brightness at 290 K vs. 282 K). Thus the coronagraph thermal emission was not included.

### A.1.    HLS Spectroscopic Survey

The 2013 baseline grism survey wavelength range was 1.35—1.95 µm (270 K). Given the increased thermal backgrounds at 282 K, the wavelength range has been reduced to 1.35—1.89 µm. At 270 K and $\lambda_{max}$=1.95 µm, the thermal background was estimated at 0.17 e⁻/pix/s (versus 0.49 e⁻/pix/s for the sky), whereas at 282 K and $\lambda_{max}$=1.89 µm the backgrounds include 0.27 e⁻/pix/s (thermal) and 0.45 e⁻/pix/s (sky).

The yield of Hα detections at 7σ in the HLS spectroscopic survey is 7344 gal/deg² at 282 K, versus 8363 gal/deg² at 270 K. Figure A-1 shows the redshift

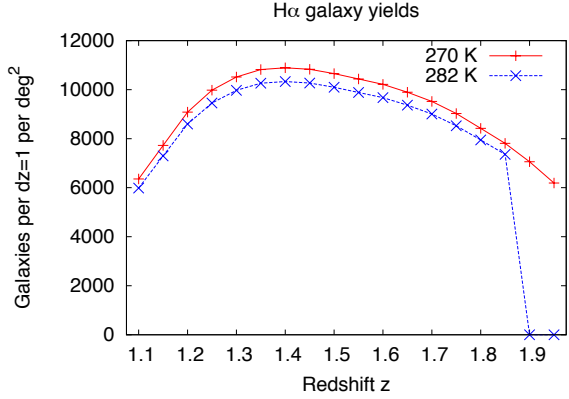

Figure A-1: The redshift distribution of the spectroscopic Hα galaxy sample at 270 K and at 282 K. Note the cutoff at z=1.88 due to the reduced wavelength range at 282 K.

distribution of the spectroscopic Hα galaxy sample at 270K and 282K.

## A.2. HLS Imaging Survey

The impact of the 282 K baseline temperature on the sensitivity of the HLS imaging survey is described in Table A-1. Both depths (point source sensitivity) and WL galaxy yields are shown. Overall the penalties associated with 282 K operation are mild: in H band we lose 0.07 mag of depth and 6% of the source galaxies. The F184 band (the reddest) suffers the greatest degradation, but the loss of depth is still only 0.28 mag, with an 18% reduction in the number of WL galaxies.

Extended source sensitivity is worse than point source sensitivity by the amount shown in Table A-1 (for an exponential profile source with a half light radius of 0.3 arcsec). This factor is independent of the telescope temperature.

The loss of only 0.28 mag of depth in F184 when going from 270 K to 282 K may seem counterintuitive since the telescope thermal emission (which is brighter than the zodiacal emission) increases by a factor of 3x. It is therefore instructive to review the intermediate results of the exposure time calculator (ETC) to understand why only a modest degradation is predicted. The most important contributions to the noise budget in the HLS are sky (zodiacal) background, thermal emission, and read noise. In a 32 frame exposure (173.5 s), the zodiacal background is 65 e⁻/pix, compared to thermal emission of 66 e⁻/pix (270 K) or 207 e⁻/pix (282 K). The read noise for 32 frames is 10 e⁻ RMS (variance of 100 sq. e⁻/pix), dominated by the CDS noise. These numbers alone would predict a degradation of a factor of 372/231 in variance when going from 270 K to 282 K, or 0.26 mag in depth. The more complete calculation in the ETC finds a degradation of 0.28 mag. Note that even at 282 K, thermal emission contributes only 58% of the total noise budget.

## A.3. Deeper Imaging Modes

Many GO programs will require deeper imaging than is planned for the HLS, and consequently the impact of telescope temperature is greater for these observations. This impact is shown in Figure A-2 for the H and F184 filters (the two most strongly impacted). The time penalty in F184 for the warmer telescope is 1.5x at 26 mag AB (0.2 mag deeper than the HLS), 2.0x at 27 mag AB, and 2.1x at 28 mag AB. Even at 28 mag AB depth, the time penalty in the bluer bands is small: 11% in H and 6% in J.

## A.4. Microlensing

The microlensing survey uses primarily the wide filter (W149), which due to the cutoff at 2 μm admits a significant thermal background. However, since microlensing observations are performed close to the ecliptic, the sky background is also higher. At the midpoint of a microlensing season, the zodiacal count rate is 3.28 e⁻/pix/s, in comparison to the thermal emission count rate of 1.21 e⁻/pix/s (at 282 K) or 0.39 e⁻/pix/s (at 270 K). The thermal rate, while significant, is not dominant even at the higher temperature. The distribution of starlight count rates is shown in Figure A-3, and it is seen that starlight is dominant over zodiacal and thermal emission in almost all pixels. The median starlight count rate is 12.06 e⁻/pix/s, which corresponds to a 2.6% degradation of the Poisson error bar in going from 270 K to 282 K. The degradation due to the higher thermal background is worse for fainter pixels, e.g. at the 10th percentile brightness the degradation worsens to 5.0%, and at the 1st percentile brightness it worsens to 6.6%. Given this very minor degradation in the photometric uncertainties in the microlensing fields, the higher operating temperature will have negligible impact on the WFIRST-AFTA microlensing program.

| Characteristic | Y | J | H | F184 |
|---|---|---|---|---|
| Exposure time (s) | 5x174 | 6x174 | 5x174 | 5x174 |
| 5σ depth AB (pt src, 270 K) | 26.58 | 26.72 | 26.61 | 26.04 |
| 5σ depth AB (pt src, 282 K) | 26.56 | 26.70 | 26.54 | 25.76 |
| WL n_eff (gal/am², 270K) | N/A | 33.7 | 35.1 | 24.5 |
| WL n_eff (gal/am², 282K) | N/A | 33.2 | 33.1 | 20.0 |
| Extended src depth penalty (mag @ r_{1/2}=0.3") | 1.17 | 1.14 | 1.10 | 1.05 |
| 270→282 K depth penalty (mag) | 0.02 | 0.02 | 0.07 | 0.28 |
| 270→282 K WL n_eff penalty | N/A | -1% | -6% | -18% |

Table A-1: Properties of the HLS imaging survey and their dependence on telescope temperature. The penalties for the 282 K telescope temperature for imaging depth and WL galaxy yield are shown in the last two rows. Calculations in this table are at the nominal number of exposures; HLS forecasts in the main text take into account the full distribution of the number of exposures.

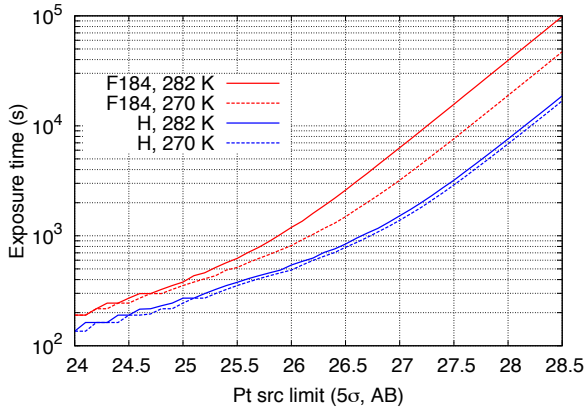

**Figure A-2:** The exposure time required to reach a given depth in the H and F184 filters. The zodiacal brightness is taken to be that at median conditions (30° ecliptic latitude and at quadrature). The total time is split into 5 dithered sub-exposures; depths beyond 28.5 mag AB would likely be split into more sub-exposures due to the cosmic ray rate in GEO.

## A.5. Integral Field Channel

The integral field channel produces independent spectra for each segment of the image slicer at the spectrograph entrance. The telescope thermal emission is therefore dispersed, and will increase the background only in the last few pixels of each spectrum. This will not have a significant impact on the supernova program.

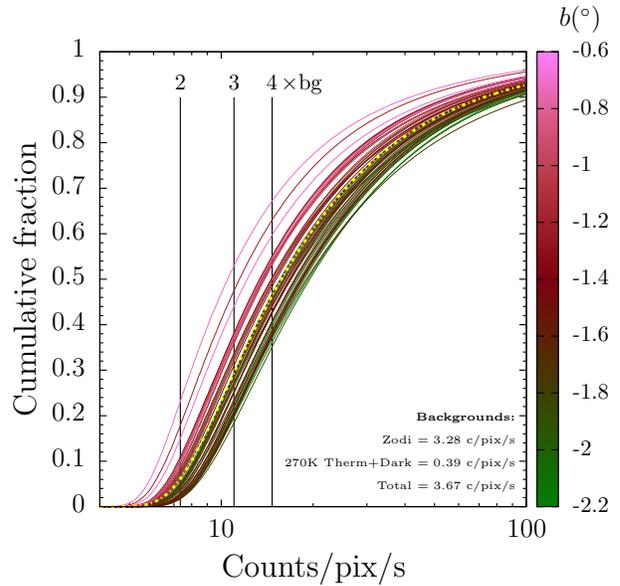

**Figure A-3:** The simulated distribution of count rates from starlight + zodiacal background in the microlensing fields. Colors indicate fields at different Galactic latitudes; the yellow curve is the mean over all fields simulated.

Appendix B     Scientific Synergies between WFIRST-AFTA and JWST

# Wide Field Infrared Survey Telescope

## *Synergistic Science Programs between WFIRST-AFTA and JWST*

## January 2015

## http://wfirst.gsfc.nasa.gov/

This appendix contains a rich set of 9 potential synergistic science programs that are uniquely enabled by WFIRST-AFTA observations coincident with JWST. The papers were put together mostly by STScI staff, with some contributions from other astronomers, and the SDT wishes to thank the community for this valuable contribution. The one pagers highlight the tremendous potential of WFIRST-AFTA to advance many of the key science questions formulated by the Decadal survey. These programs have not been vetted by the SDT, and we are not endorsing these specific studies. Additionally, the authors did not see the final version of the WFIRST-AFTA mission capabilities prior to their submissions, so inconsistencies may exist.

<u>Title</u>: Scientific synergies obtained by the simultaneous operation of WFIRST-AFTA and JWST

<u>Author</u>: John W. MacKenty, Space Telescope Science Institute

<u>Contributors</u>: Martha Boyer[1], Larry Bradley[2], Dan Coe[2], Harry Ferguson[2], Karl Gordon[2], Paul Goudfrooij[2], Jason Kalirai[2], Knox Long[2], Marshall Perrin[2], Massimo Stiavelli[2], and Tommaso Treu[3]
(1=GSFC, 2=STScI, 3=UCSB).

<u>Summary</u>: This White Paper considers a set of scientific investigations which require the availability of observations from WFIRST-AFTA coincident with JWST observations. These scientific synergies offer a substantial increase in the scientific return from both missions and will explore a range of pressing and compelling questions ranging from the origins of our Universe, the nature of dark matter, the assembly of galaxies, the fates of stars, and the properties of exoplanets.

<u>Background</u>: This White Paper assumes a WFIRST-AFTA mission built upon a NRO 2.4 meter telescope with a core scientific capability of near-infrared imaging comparable in sensitivity and resolution to HST/WFC3-IR over a field of view of order 0.25 square degrees. These investigations assumes wavelength coverage from 0.6 to ~2 microns and some investigations require or would benefit from extending coverage (but not spatial sampling) to 0.4 microns and/or from the inclusion of a richer set of visible and medium band filters.

<u>Science Investigations</u>:

1. Studies of the earliest galaxies (Dan Coe and Larry Bradley)
   - Large solid angle survey with WFIRST-AFTA to discover the most luminous sources at 10<z<20 using the Lyman break drop-out technique that is used by HST to discover z>11 galaxies.
   - JWST NIRSPEC and MIRI detailed spectroscopy enabled by the discovery with WFIRST-AFTA of sufficiently bright sources yields clues to ionization, metal growth, and possibly masses of these infant galaxies.

2. Finding the first stars via high-z pair instability supernovae (Massimo Stiavelli)
   - Discovery of PISN from zero metallicity first generation stars at z>10 to explore the first generation of stars, their initial mass function, and the initial enrichment of the ISM.
   - PISN have 1-2 year timescales and would be observable with WFIRST-AFTA to z~15-20. Their expected source density of 1-2 per square degree requires WFIRST-AFTA for discovery and JWST to determine the properties of their host galaxies. Working in the infrared, WFIRST-AFTA will (post-SWIFT), have an unparalleled capability to discover rare z>6 sources (e.g. GRBs and supernovae) due to the ultraviolet opacity of the IGM prior to reionization.

3. Probing dark matter via lensed QSOs (Tommaso Treu)
   - Measurement of dark matter sub-halos down to $10^6$ $M_{Sun}$ will test CDM versus self-interacting or warm dark matter models.
   - WFIRST-AFTA has the unique capability to discovery 1000s of quad lensed QSOs in a large solid angle sky survey which requires HST resolution. JWST/MIRI is then required to measure flux ratios of the lensed sources at mid infrared wavelengths to avoid stellar microlensing.

4. Progenitors of supernovae and other highly variable objects (Knox Long)
   - Within 50 Mpc, where significant resolution is possible to identify individual stars or their environments, we expect 50 core collapse SN and 15 SN1a per year.
   - JWST follow-up of these nearby supernovae will be significantly enhanced if pre-detonation images permit the identification of the type of progenitor star. WFIRST-AFTA surveys at HST resolution over a broad range of wavelengths provide pre-detonation images of nearby galaxies.

5. Globular clusters in nearby galaxies (Paul Goudfrooij)
   - Massive star clusters trace early star formation history and galaxy assembly.
   - Wide Field, high resolution and sensitive NIR photometry is required to select samples outside the local group. JWST/NIRSPEC MOS spectroscopy is required to obtain metallicities to derive star formation time-scales.

6. AGB stars and dust production in the local volume (Martha Boyer & Karl Gordon)
   - AGB stars are major producers of dust, especially carbon, in the ISM. Understanding both the mechanisms of dust production and its implications for the infrared properties of galaxies requires improved understanding of this short lived phase of stellar evolution.
   - Wide field near infrared observations with WFIRST-AFTA will establish a sample of AGB stars suitable for detailed follow-up with spectroscopy and longer wavelength observations using JWST.

7. Galactic streams in nearby galaxies (Harry Ferguson)
   - Tidal streams from the merging of dwarf galaxies into nearby large galaxies provide clues to the assembly history of these galaxies and a key test of CDM models.
   - WFIRST-AFTA has the unique capability to discover and map streams surrounding ~100 massive galaxies within 5 Mpc. JWST will then provide the deepest imaging and infrared spectroscopy to yield constraints on the ages, kinematics, and chemical abundances of the streams via observations of individual stars.

8. Simultaneous observations of exoplanets with WFIRST-AFTA and JWST (Marshall Perrin)
   - The inclusion of a white light coronagraphic instrument with an integral field spectrometer on WFIRST-AFTA opens the possibility of simultaneous observations with the JWST coronagraphs to achieve broader spectral coverage.
   - Studies of exoplanet atmospheres, indications of weather and seasonal variations, and (with the inclusion of a polarimeter capability within the WFIRST-AFTA coronagraph), detailed measurements of scattered light from circumstellar dust in exoplanetary systems will be possible.

9. Resolved Studies of Galaxy Formation (Jason Kalirai)
   - Measuring the surface brightness, abundance gradient, velocity dispersion, star formation history, and sub-structure in the halos of nearby Milky Way type galaxies provides a view into their formation history, and an input to test models of galaxy formation.
   - Observations with WFIRST-AFTA will reveal starcount maps for nearby galaxies, and can be used to locate a pristine population of halo stars. Follow up observations with JWST can be used to directly measure turnoff ages and ground based 30-meter telescopes to measure stellar velocities.

## 1. <u>Studies of the earliest galaxies</u> (Dan Coe and Larry Bradley)

The first galaxies likely formed ~ 100 – 400 Myr after the Big Bang (z ~ 11 – 30) contributing to reionization and the end of the "dark ages", a process likely complete by ~ 900 Myr (z ~ 6). The nature of these early galaxies and the degree to which they contributed to reionization are the current frontier outstanding questions of extragalactic astronomy. The Hubble and Spitzer Space Telescopes have yielded candidates with robust photometric redshifts as distant as z ~ 10.8 or 420 Myr (Coe13) and perhaps z ~ 11.9 (Ellis13, though the authors urge caution due to the single-band detection). The former is gravitationally lensed into multiple images, the brightest magnified by a factor of ~8 to AB mag 25.9, enabling more detailed follow-up study. Significant samples of high-redshift candidates have been discovered back to ~600 Myr after the Big Bang (~100 candidates at z ~ 8: Bradley12, Oesch12, Ellis13). Deep Hubble imaging has yielded the fainter, more numerous population, while wider, shallower Hubble surveys as well as very wide field (~square degree) ground-based surveys have discovered the rarer relatively brighter galaxies. No survey has yet achieved the area, depth, and wavelength coverage required to yield significant numbers of z > 9 galaxies within the universe's first 500 Myr.

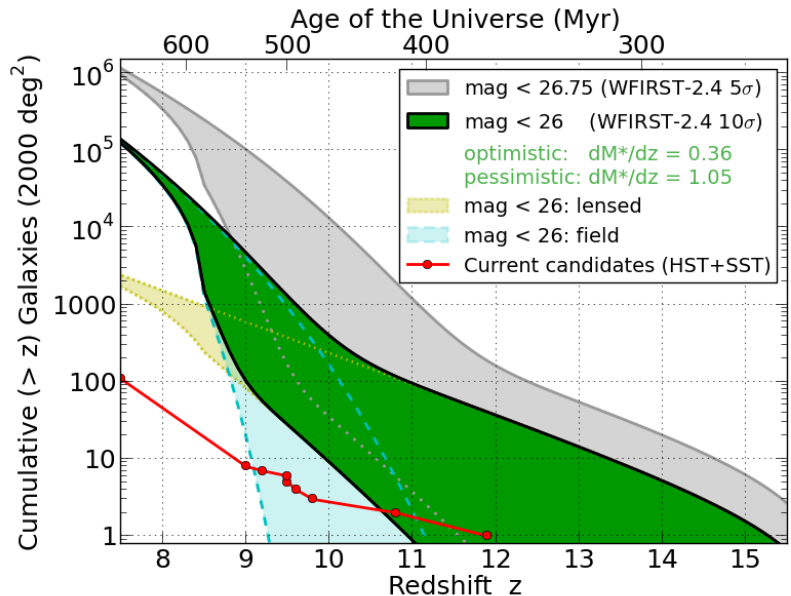

The wide and deep near-IR survey made possible by NRO WFIRST-AFTA will transform our understanding of the early universe by yielding ~100,000 high-redshift (7.5 < z < 15) candidates bright enough (AB mag < 26) for follow-up study, that could allow for detailed stellar population studies, metallicities, and kinematics with the James Webb Space Telescope and large ground-based telescopes. The figure at right shows a range of expectations for cumulative (in redshift) numbers of high-redshift galaxies in a 2,000 square degree survey. This is based on current measurements at z ~ 8 (Bradley12) extrapolated to higher redshifts both optimistically and pessimistically based on constraints from the handful of ~8 current z > 9 candidates (Bouwens11, Zheng12, Coe13, Bouwens13, Ellis13, Oesch13). Contributions from gravitationally lensed galaxies are estimated by roughly assuming one strongly lensing cluster per square degree and adopting an average lens model from the CLASH multi-cycle treasury program (Postman12). Most ten-sigma detections (green) will be properly identified, but pushing down to five-sigma detections will result in >50% incompleteness relative to the numbers shown in gray. Key questions to be addressed by these observations are: (1) how rapidly did early galaxies build up; (2) what was their contribution to reionization; (3) were early galaxies primarily composed of pristine Population III stars?

2. Finding the first stars via high-z pair instability supernovae (PISN) (Massimo Stiavelli)

A central goal of the JWST mission is the observation of the first stars and galaxies. Actually observing a single population III star may be possible when it dies as a pair-instability supernova. Such supernovae are very luminous but also very rare. WFIRST-AFTA provides the most viable means of discovering such objects as this requires deep large area near infrared surveys with high angular resolution and good photometric precision over long timescales. JWST would then provide detail studies of both the SN and its host galaxy.

Present day star formation mechanisms rely upon gas cooling via metals and dust to condense gas into stars. As these heavy elements originated in the first generation of stars (Population III) stars, those stars are expected to have formed via different and very uncertain mechanisms which likely resulted in a relatively higher proportion of very high mass stars. Such stars can end their lives as Pair Instability SuperNovae (PISN) with extremely high luminosities that visibly brighten their host galaxies. At z>10, time dilation will make these SN visible for periods of several years requiring long term (but infrequent) monitoring with deep images to discover PISN.

PISN have been discussed by Weinmann and Lilly 2005 (ApJ, 624, 526). They find SN rates in the range 0.2-4 sq deg per year at redshift 15-25. These rates are in some sense even an upper limit for the following reasons: (1) recent studies have found even lower masses for Pop III stars largely due to some evidence for fragmentation. If the mass function extends down to few tens of solar masses and at the same time the upper mass cutoff goes down to 300 $M_{Sun}$ or even much lower, the range of masses giving rise to PISN shrinks and they become even rarer. (2) The full impact of negative feedback (especially radiative but also chemical, i.e., if there is enrichment you don't form Pop IIIs) have been underestimated until the last 5 years or so (e.g. Trenti and Stiavelli 2009, ApJ, 694, 879). There is a lot of uncertainty on Pop III from $H_2$ and even higher uncertainty on those from atomic H. While 2 PISN per sq deg per year is a moderately optimistic but not unreasonable rate if fragmentation is rare, it would be hard to sustain that rate with the amount of fragmentation that some are finding now.

3. <u>Probing dark matter via lensed QSOs</u> (Tommaso Treu)

Measuring the mass function of subhalos down to small masses (e.g. $1 \times 10^6 M_{Sun}$) is a unique test of cold dark matter and of the nature of dark matter in particular. The CDM mass function is supposed to be a power law rising as $M^{-1.9}$ until Earth-like masses. A generic feature of self interacting dark matter and warm dark matter is to introduce a lower mass cutoff. Current limits set that cutoff at somewhere below $1 \times 10^9 M_{Sun}$. We know from MW problems that there is a shortage of luminous satellites at those masses, but that's never going to be conclusive because subhalos can only be detected with traditional methods if they host stars.

One powerful way to detect subhalos independent of their stellar content is via the study of large samples of strong gravitational lenses, in particular the so-called flux ratio anomalies (see Treu 2010, ARA&A, 48, 87 and references therein). Small masses located in projection near the four images of quadruply-lensed quasars cause a strong distortion of the magnification and therefore alter the flux ratio. However, anomalous flux ratio can also be caused by stellar microlensing if the source is too small. The only way to avoid microlensing is to go to wavelengths such as mid-IR where the lensed quasar emission is sufficiently extended. Large sample of thousands of lenses are needed to achieve sensitivity down to $1 \times 10^6$-$1 \times 10^7$ solar masses required to probe the nature of dark matter. Unfortunately, at the moment only 3 dozen quads are known, and only a handful of those are bright enough at mid-IR to be observable from the ground (e.g., Chiba et al. 2005, ApJ, 627, 53). JWST-MIRI is so powerful that it will be able to observe any quad discovered in the next decade with integration times of order seconds to minutes. However, JWST will be limited for this application by the number of known quads. Ground based surveys before JWST launches will probably discover of order 100 quads, still short of the number require to fully realize its potential as a dark matter experiment.

Discovering thousands of quads will require a close to full sky survey at HST-like resolution (Oguri & Marshall, 2010, MNRAS, 405, 2709), i.e. WFIRST-AFTA. The ability to resolve and therefore discover quads is a very strong function of resolution and therefore WFIRST-AFTA will be unmatched in its ability to find them. Of course if WFIRST-AFTA comes after JWST, those will not be available for mid-IR follow-up at JWST resolution.

4. The progenitors of SNe and other highly variable objects (Knox Long)

JWST will carry out detailed studies of SNe in nearby galaxies, particularly since its IR capability will allow it to probe the late phases of a SN explosion to determine the nature and distribution of the ejecta and to understand the conditions in SNe that lead to dust creation (see, e.g. Barlow 2009). However, a crucial aspect of understanding SNe is to connect a SN explosion to its progenitor star. The combination of WFIRST-AFTA and JWST are crucial to rapid progress in this area.

The importance of high quality imagery for the advancement of this subject has been made clear with HST. Indeed, some HST imagery exists for about 25% of the SNe (and luminous blue variable outburst) exploding in galaxies within 28 Mpc (Smart et al 2009). (Over about a decade, 50% of the SNe actually occurred in galaxies that were observed with HST, half occurred outside HST's small FOV) HST images have resulted in the identification of at least 4 progenitors, including discovery of the binary progenitor for SN1993J (Maund et al 2004) and enabled studies of the local stellar populations in the remainder. As a result, we know that the most common type of core collapse SNe (type II-p) arise from red supergiants and that the minimum and maximum masses for these SNe is about 8 Msol and 17 Msol (Smart et al. 2009).

But WFIRST-AFTA, with its large FOV, will do much better. For example, it would be quite straightforward to create a uniform treasure archive of nearby galaxies at distances within, say 50 Mpc, that extended out well beyond the D25 contours. About 50 core collapse SNe and about 15 Ia SNe are expected in this volume every year. It would also be possible with a little effort to survey larger galaxies, up to and including M31 and M33. Not only would this enable some of the science associated with the PHAT multi-cycle treasury program of M31 (which an allocation of 830 orbits with HST covers only a portion of the galaxy), but it would also ensure that when the next SN goes off in one of these galaxies, that pre-exposure images would be able to identify the progenitor down to (depending on crowding) a mass of about 3.5 $M_{Sun}$ (Dalcanton 2012). One could also imagine shallower surveys to map the entirety of the SMC or LMC down to considerably lower mass limits.

Furthermore, the needs for pre-explosion images are not just limited to the progenitors of SNe, but includes a wider range of other objects: the Eta Car-like explosions of luminous blue variables, supernova imposters, such as SN2008S and optical transients such as those seen in M85 and NGC300. All of these objects are likely candidates for JWST observations. JWST would be used not only for spectroscopic observations during the explosion that would characterize the SNR, but, if the WFIRST-AFTA mission were complete, would carry out the NIR imagery after the SN had faded to find out which object had been disrupted as a result of the SN explosion.

References: (1) Barlow, M. J. 2009, Prospects for Studies of Stellar Evolution and Stellar Death in the JWST Era, in Astrophysics in the Next Decade, Astrophysics and Space Science Proceedings. ISBN 978-1-4020-9456-9. Springer Netherlands, 2009, p. 247

5. <u>Near-field cosmology and galaxy evolution using globular clusters in nearby galaxies</u> (Paul Goudfrooij)

The increasing realization that the study of star clusters has direct relevance for the basic processes involved in how galaxies assemble and evolve over time has placed this field at the forefront of extragalactic research in recent years. A fundamental, yet unresolved, question in the formation of early-type galaxies is the assembly and chemical enrichment histories of their 'halo' and 'bulge' components which are represented by the metal-poor and metal-rich GC subpopulations, respectively. Working together, WFIRST-AFTA and JWST can make unprecedented progress in solving this question.

a) Globular Clusters as Fossil Records of the Formation History of Galaxies

Star formation plays a central role in the evolution of galaxies and of the Universe as a whole. Studies of star-forming regions in the local universe have shown that star formation typically occurs in a clustered fashion. Building a coherent picture of how star clusters form and evolve is therefore critical to our overall understanding of the star formation process. Most star clusters disrupt within a few Gyr after they form, building up the field star population in the process. However, the most massive and dense star clusters remain bound and survive for a Hubble time. These globular clusters (GCs) constitute luminous compact sources that can be observed out to distances of several tens of megaparsecs. Furthermore, star clusters represent the best known approximations of a "simple stellar population", i.e., a coeval population of stars with a single metallicity, whereas the field stars in galaxies typically constitute a mixture of populations. Thus, studies of GC systems can constrain the *distribution* of stellar ages and metallicities whereas measurements of the integrated light of galaxies can only provide luminosity-weighted averages of these key quantities. Consequently, GCs represent invaluable probes of the star formation rate and chemical enrichment occurring during the main star formation epochs within their host galaxy's assembly history (see, e.g., review of Brodie & Strader 2006).

The study of extragalactic GCs was revolutionized by the Hubble Space Telescope (HST). The main reason for this is that the size of GCs is well-matched to diffraction-limited optical imaging with a 2-m class telescope: a typical GC half-light radius of ~3 pc at a distance of 15 Mpc corresponds to 0.05 arcsec on the sky, which is roughly the diffraction limit (and detector pixel size) for HST at 600 nm. This yields very high quality photometry of GCs relative to ground-based imaging by beating down the high galaxy surface brightness in the central regions of galaxies. Furthermore, it also allows robust measurements of globular cluster radii, and hence of their dynamical status.

Notwithstanding the important progress that HST imaging has facilitated in this field, there is one critical property of GC systems that HST imaging *cannot* address well. GC systems around massive nearby early-type galaxies extend far into the galaxy halos, covering several tens of arcminutes on the sky (e.g., Goudfrooij et al. 2001; Zepf 2005), while the FOV of HST images is only 3.3 × 3.3 arcmin². This is illustrated in Figure 1. Obviously, wide fields of view (> 20' per axis) are required to accurately determine total properties of GC systems (e.g., total numbers of clusters per unit galaxy luminosity, color or metallicity distributions, trends with galactocentric distance). Furthermore, the faint outer halos of galaxies are thought to hold unique clues regarding the early assembly history of galaxies, and bright GCs constitute one of the very few probes that can be studied at high S/N in these environments.

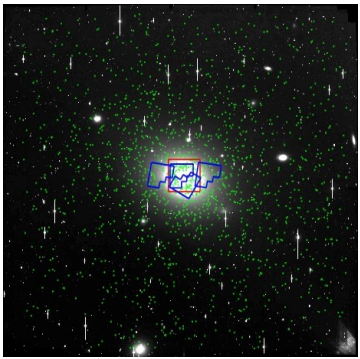

**Figure 1: R-band KPNO 4-m/MOSAIC image of the giant elliptical galaxy NGC 4472 in the Virgo cluster of galaxies, covering a 36' × 36' FOV. Footprints of available HST/ACS and HST/WFPC2 images are drawn in red and blue, respectively. Globular cluster candidates from Rhode & Zepf (2001) are indicated as green dots. Note the small fraction of globular cluster candidates covered by HST images, implying the need for large and uncertain extrapolations when trying to extend conclusions from the HST studies to the full systems of globular clusters. Figure taken from Zepf (2005).**

b) New Constraints on the History of Star Formation and Chemical Enrichment of Early-Type Galaxies

A key discovery of HST studies of GC systems of luminous galaxies was that their optical color distributions are typically bimodal (e.g., Kundu & Whitmore 2001; Peng et al. 2006). Figure 2 shows an example. Follow-up spectroscopy of bright GCs using 10-m-class telescopes indicated that both ``blue'' and ``red'' populations are typically old (age > 8 Gyr), implying that the color bimodality is mainly due to differences in metallicity (e.g., Puzia et al. 2005). In broad terms, the metal-rich GC population features colors, metallicities, radial distributions, and kinematics that are consistent with those of the spheroidal (`bulge') component of early-type galaxies. In contrast, the metal-poor GC population has a much more radially extended distribution, and is likely physically associated with metal-poor stellar halos such as those found around the Milky Way and M31 (e.g., Bassino et al. 2006; Goudfrooij et al. 2007; Peng et al. 2008).

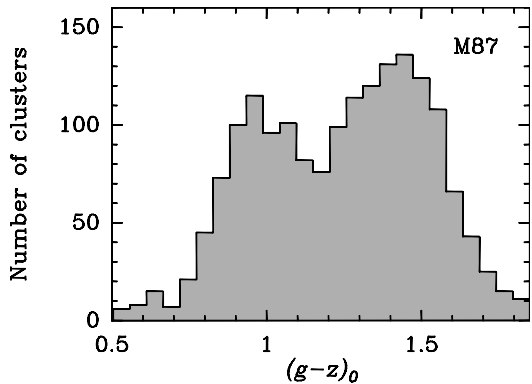

**Figure 2: *g–z* color distribution of globular clusters in the massive elliptical galaxy M87 from Peng et al. (2006). Note the obvious color bimodality, which has been confirmed to be mainly due to differences in metallicity, and which is common among massive early-type galaxies in the local universe.**

The bimodality in optical colors of GCs constitutes one of the clearest signs that star formation in luminous early-type galaxies must have been episodic. However, we emphasize that the optical color distributions do not significantly constrain *when* these events occurred, or in what order. This is because optical colors alone cannot generally distinguish between different combinations of age and metallicity (the ``age-metallicity degeneracy''). A general understanding of the age and metallicity distributions of GC systems requires braking this degeneracy. There are two primary and complementary ways to do this.

**(1) The combination of optical and near-IR photometry**. The main power of this method (using color-color diagrams) is the ability to identify age differences (of order 25% for high-quality data), due to the fact that near-IR colors are primarily sensitive to metallicity while optical colors (e.g., *r-z*) are sensitive to both age and metallicity. This approach resulted in the identification of substantial populations of intermediate-age metal-rich GCs in the inner regions of several early-type galaxies (Goudfrooij et al. 2001; Puzia et al. 2002; Hempel et al. 2007; Georgiev et al. 2012). The current limitation of this method is twofold. While HST has a powerful near-IR channel in its WFC3 instrument, its use is limited to the *innermost regions* of nearby galaxies due to its relatively small footprint of 2' × 2' (cf. Fig. 1 above). The NIRCam instrument to be installed on the 6.5-m James Webb Space Telescope (JWST) will reach 2 mag fainter than HST in a given integration time, but its footprint is similarly small. Conversely, while near-IR imaging instruments with reasonably large fields of view are starting to become available on large ground-based telescopes (e.g., 7.5' × 7.5' for HAWK-I on the VLT), contamination of GC candidate samples by compact background galaxies is a major concern for ground-based spatial resolution (see, e.g., Rhode & Zepf 2001). As demonstrated by HST, imaging at 0.1 arcsec resolution effectively eliminates this concern due to the marginally resolved nature of GCs (cf. Section 1). Thus, the study of galaxy formation and evolution by means of accurate GC photometry will benefit tremendously from space-based wide-field imaging in the 0.6 – 2 μm range. A relatively simple multi-chip Optical-NIR camera installed on one of the two 2.4-m NRO telescopes would be ideal for this (and many other) purpose(s). Their fast (f/1.2) primary mirror could easily yield a useful field of view of hundreds of square arcminutes per exposure at a resolution of 0.1 arcsec, providing accurate photometry of virtually *all* GCs associated with nearby galaxies with very little contamination. Such an instrument would not only place crucial constraints on the assembly history of massive early-type galaxies, it would also allow the selection of the best targets for follow-up multi-object spectroscopy to infer their chemical enrichment history (see below).

**(2) Follow-up Space-Based Multi-Object Spectroscopy**. The main strength of this technique lies in the presence of intrinsically strong absorption lines of several key elements in the 0.7 – 2.5 μm region (e.g., O, Mg, Al, Si, Ca, Fe), which facilitates accurate determinations of overall metallicities and element abundance ratios that can be used to infer typical timescales of star formation (e.g., Valenti et al. 2011). Currently, this technique is only available from the ground and is therefore significantly hampered by the high surface brightness of the diffuse light of the inner regions of the host galaxies. In practice, this limits the application of this technique currently to the outer regions of galaxies. This has caused a general lack of crucial spectroscopic information for the metal-rich GCs, which are located mainly in the inner regions. While future developments in the area of adaptive optics systems on large telescopes will enable high spatial resolution imaging and spectroscopy from the ground, they will do so only over a small (< 1') FOV which is not useful for spectroscopy of extragalactic GCs. This science can however be expected to advance dramatically with the advent of NIRSpec on the JWST with its multi-slit array and high-efficiency medium-resolution gratings.

6. AGB stars and dust production in the local volume (Martha Boyer & Karl Gordon)

When low- to intermediate-mass stars ($0.8 < M < 8 \ M_{Sun}$) begin to ascend the asymptotic giant branch (AGB), pulsations levitate material from the stellar surface and provide density enhancements and shocks, encouraging dust formation and re-processing (e.g., Bowen 1998; van Loon et al. 2008). This dust is subsequently released to the interstellar medium via a strong stellar wind driven mainly by radiation pressure on the grains. The composition of the dust depends on the atmospheric chemistry (the abundance of carbon relative to oxygen) which is altered by dredging up newly formed carbon to the surface of the star (the 3rd dredge-up; Iben & Renzini 1983). The efficiency of the 3rd dredge up (which depends on the stellar mass) and the initial metallicity of the star determine whether or not a star will ultimately become a carbon star, with metal-poor stars becoming carbon-rich more easily owing to a lack of oxygen available to tie up the newly dredged-up carbon into CO molecules.

While all stars between 0.8 and 8 solar masses will pass through the AGB phase, the phase itself is short, making AGB stars rare. Despite this, they may be among the most important contributors of dust in galaxies. The most significant known dust producers are AGB stars and supernovae (SNe) and it is unclear which dominate the dust production in the Universe since the dust-production rate from SNe is highly uncertain. However, even if assuming the most generous estimates for SNe dust production, the AGB stars come out as at least equally important over the lifetime of galaxies like the Magellanic Clouds (e.g., Matsuura et al. 2009; Boyer et al. 2012). AGB stars are certainly the most important contributors of carbon dust, as their total dust input is dominated by the carbon stars and SNe are thought to produce mainly silicates.

The infrared flux of galaxies is often used to infer the total masses and star formation rates of galaxies. Since AGB stars are among the brightest sources and radiate strongly from 1-20 microns, they can significantly affect these measurements. In the Magellanic Clouds, AGB stars account for up to 30% of the 3-4 micron flux, despite being < 5% of stellar population (Melbourne & Boyer 2013). Without accounting for the AGB stars, stellar masses measured in the near-IR could be biased too high, especially in galaxies with large intermediate-aged populations.

Despite their importance, very little is known about the details of the evolution and dust production in AGB stars. AGB stars in galaxies more distant than the Magellanic Clouds have been difficult to observe at IR wavelengths owing to limits in sensitivity and resolution, leaving us with observations of a handful of stars in a random smattering of nearby galaxies. In the Magellanic Clouds, the Spitzer SAGE programs (Gordon et al. 2011; Meixner et al. 2006) have provided an incredible wealth of new information about AGB stars with their fully complete photometric catalogs in the IR, but IRS spectra was limited to only the brightest AGB stars even in these nearby galaxies. AGB stars in the Milky Way have been difficult to identify, leaving us with samples biased towards only the most extreme examples. Optical surveys do not suffice for AGB observations, as up to 60% of a galaxy's AGB population can be obscured by circumstellar dust (Boyer et al. 2009).

The James Webb Space Telescope will go a long way towards understanding these important stars owing to its incredible sensitivity in the IR. For the first time, whole populations of AGB stars can be observed both photometrically and spectroscopically in the IR within the complete suite of galaxy environments available in the Local Volume, potentially shifting our understanding of AGB stars and AGB dust production in profound ways. Since the lifetime of JWST is limited, it is crucial that we identify AGB stars in the Local Volume before, or at least in tandem with the JWST mission so that IR spectra can be obtained for an unbiased sample of stars. The most efficient way to identify AGB stars is to use near-IR photometry since the spectral energy distributions of most AGB stars peak from 1-3 microns, and since circumstellar dust does not obscure the stars at these wavelengths in the vast majority of cases (95% of the AGB stars in the LMC are detected in the K-band with 2MASS data).

Near-IR photometry also has the potential to separate carbon-rich from oxygen-rich AGB stars, depending on the exact filters chosen. This is due to molecular absorption in AGB atmospheres (e.g., TiO, CN, $C_2$) in the 0.5-2 micron region. The J-K color successfully separates C from O stars in the Magellanic Clouds with 85% confidence (Riebel et al. 2012), allowing for the construction of a representative and un-biased sample of sources for follow-up spectrosco-py. The near-IR wide filters on HST/WFC3, on the other hand, do not separate C-rich from O-rich stars, making it difficult to calibrate stellar evolution models.

With filters similar to 2MASS J, H, and K, and with resolution and sensitivity similar to HST, WFIRST-AFTA could identify all AGB stars of interest within the Local Volume, allowing for the most efficient usage of JWST.

7. Galactic streams in nearby galaxies (Harry Ferguson)

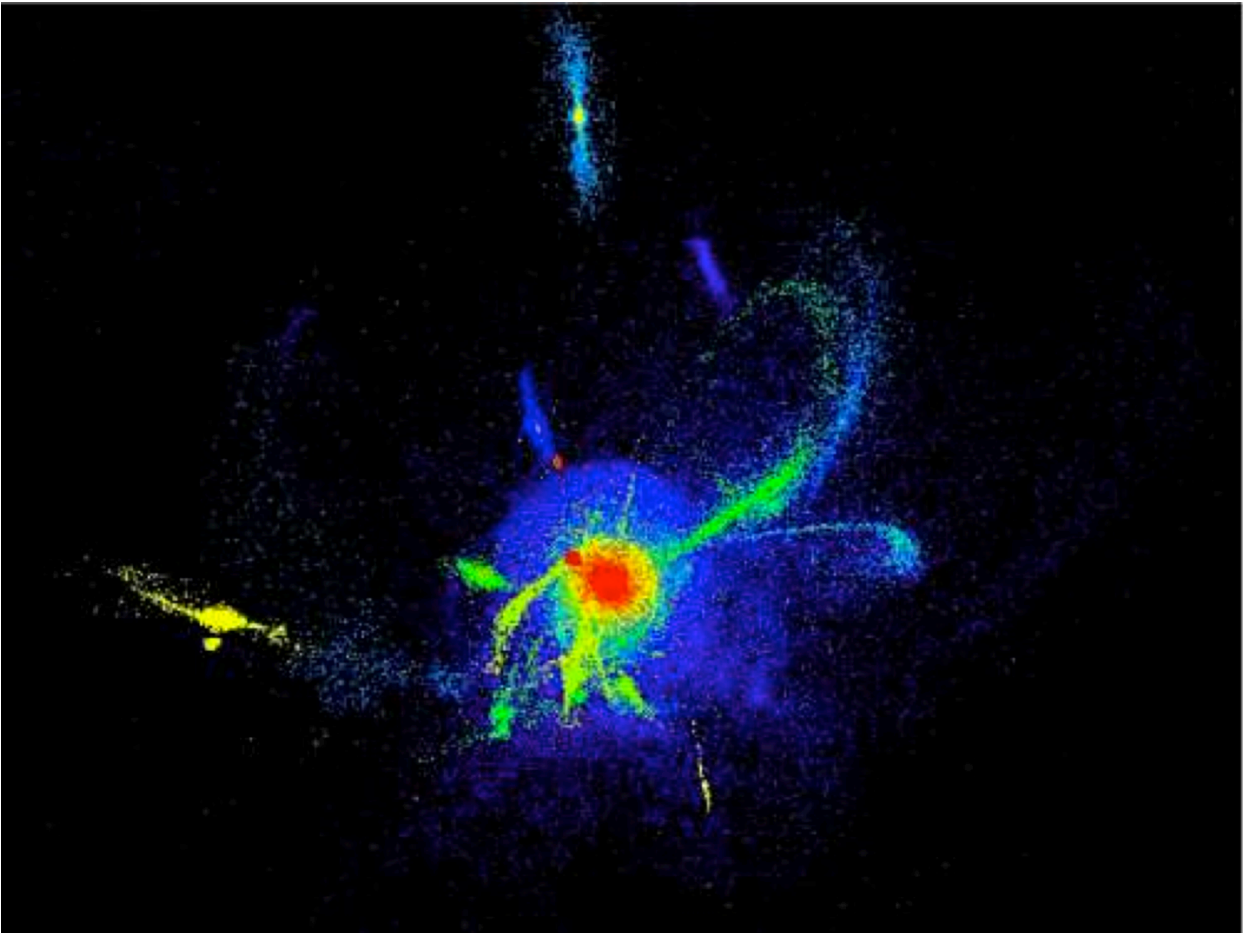

**Figure 1: Simulations of tidal streams around a z=0 elliptical galaxy; colors code stellar age (from Bournaud 2010 ASPC, 423, 177).**

WFIRST-AFTA's wide-field capabilities, low background, and high spatial resolution will revolutionize the study of streams of stars around nearby galaxies, formed by the tidal destruction of infalling dwarf galaxies. Outside of the Local Group, these streams are typically below the surface-brightness limits for detection from ground-based telescopes, but can be revealed with high-resolution observations, where individual RGB stars can be distinguished from background galaxies. Numerical simulations of galaxy formation suggest that there are streams around every L* galaxy.

The TRGB absolute magnitude is $M_J$ = -5.67-0.31[Fe/H] (Valenti 2004, MNRAS, 354, 815). With observing times of an hour or two per galaxy, WFIRST-AFTA can detect streams down to TRGB + 3 mag around all the 100 most massive galaxies within 5 Mpc. With observing times per galaxy of 10-100 hours, WFIRST-AFTA can detect streams in galaxies to the distance of the Virgo cluster. WFIRST-AFTA will provide breakthrough statistics on the number and morphology of the streams and the colors of their stars. The dissolution of dwarf galaxies in the halos of massive galaxies depends sensitively on (1) their dark-matter contents, (2) star-formation and feedback in the dwarf galaxies before they enter the massive halo and (3) interaction of the gas in the dwarf galaxies with the gas in the more massive halo. Item (1) is a fundamental assumption of LCDM models, but is currently hard to reconcile with all of the observed properties of satellite dwarf galaxies. Items (2) and (3) are very poorly constrained in galaxy-formation models; they provide the wiggle room for LCDM to survive the current discrepancies. So observations of streams will be very im-portant. The same observations, of course, will pick up low-surface-brightness dwarf galaxies, which will also be important.

JWST does not have the field of view to carryout such imaging studies, but has the spectroscopic sensitivity to follow up these observations. A rough rule of thumb is that JWST NIRSpec can obtain an R=100 spectrum of any star detected by WFIRST-AFTA, with the same exposure time that WFIRST-AFTA used to detect the star. This will improve the constraints on the metallicities of the streams, and, by doing so, help to constrain the ages of the streams. At higher resolution, R=2700, JWST, with observing times of 10-100 hours for galaxies within 5 Mpc, can obtain kinematics and more precise chemical abundance measures of individual elements. Phase-space densities of the streams will provide strong constraints on the mass and shape of the parent dark-matter halo (e.g. Penarrubia et al. 2006, ApJ, 245, 240), and may also allow tests of alternative models of gravity (e.g. Kesden 2009, Phys Rev. D. 80, 83530.)

The combination of JWST and WFIRST-AFTA greatly exceeds the individual contribution of each mission. WFIRST-AFTA is the finder scope. It obtains broad-brush statistics on the prevalence of streams. JWST probes the details that are necessary to really constrain the physics.

8. <u>Simultaneous observations of exoplanets with WFIRST-AFTA and JWST</u> (Marshall Perrin)

Motivated by recent dramatic growth in studies of exoplanets, there is significant community interest in support of including a very high contrast coronagraphic camera as a second science instrument alongside the wide-field instrument on WFIRST-AFTA . Such a coronagraph would operate at contrasts up to 1e9 in visible light (likely at least 0.6-1 micron total range with instantaneous spectral coverage ~ 10-20%, observed with an integral spectrograph with resolution R~50-100). This would enable low-resolution imaging spectrophotometry of Jupiter-mass planets around nearby stars, as well as extremely sensitive observations of scattered light from circumstellar dust. The inner working angle of this coronagraph would be well matched to JWST's coronagraphs, of order 0.2", with the smaller telescope diameter being offset by the shorter operating wavelengths.

We highlight here three specific areas of synergy between an WFIRST-AFTA coronagraph and the coronagraphs on JWST: (a) broad spectral coverage studies of atmospheric properties and chemistry in exoplanets, (b) the exciting possibility of time domain observations of weather or seasonal changes in exoplanetary atmospheres, which could benefit substantially from simultaneous operation of the two observatories, and (c) characterization of the complete range of bodies in nearby systems, including dusty debris belts and analogs to the still-poorly-understood Fomalhaut b.

a) Understanding Exoplanet Atmospheres.

Brown dwarfs and giant planets have atmospheric (photospheric) properties that change dramatically as they cool over time, due to complex processes such as cloud formation and on-equilibrium atmospheric chemistries that are not yet well understood (e.g. Fortney & Nettelmann 2010). While sample sizes remain small, initial indications from e.g. the HR 8799 planets and 2M1207 suggest that planetary-mass companions have colors distinctly different from brown dwarfs, hinting at significantly different physics. The complexity of atmospheric chemistries imprints itself in atmospheric spectra in a variety of ways, including for instance absorption and scattering features from clouds of $H_2O$ and $NH_3$, and also the effects of high altitude photochemical hazes (e.g. Cahoy et al. 2010). In the case of Jupiter, such photochemical hazes are one of the dominant contributors to atmospheric opacities, causing its overall reddish color (cf. Marley et al. 1999). Unlike transiting planets, directly imaged planets are (at least potentially) accessible to spectral characterization across a broad range of wavelengths spanning the bulk of their spectral energy distributions, offering a path to interrogate atmospheric properties through observations in both scattered light and thermal emission. Current spectral models for exoplanet atmospheres are subject to degeneracies when fitting only near-infrared data; optical spectrophotometry (0.6-1.0 microns) can break these degeneracies and aid in unambiguous retrieval of atmospheric parameters. For instance, a Cycle 19 HST program (PI: Barman) has recently obtained WFC3 0.9 $\mu$m observations of HR 8799 to allow robust tests of cloud models that are not possible on the basis of near-IR data alone. The combination of a high contrast optical coronagraph on WFIRST-AFTA with the near- to mid-infrared capabilities of JWST would enable extending such studies to the large sample sizes needed to robustly characterize extrasolar planet populations.

Advanced adaptive optics systems on ground-based 8-10 m (and potentially larger) telescopes will likely be competitive with JWST's contrast performance at near-IR wavelengths (1-2 microns), but atmospheric effects will in general prevent very high contrast (1e9 or better) observations from the ground, and will certainly preclude such observations at wavelengths substantially less than a micron even on timescales of 10-20 years from now. A very high contrast coronagraph in space is the only route to such data.

b) Weather and Seasons on Exoplanets

In addition to characterizing the bulk properties of exoplanets, we seek to understand them as dynamical systems, worlds in their own right. In several cases for close-in transiting planets, time-resolved observations have enabled a rough mapping of variations in brightness and temperature across their surfaces, providing insights into global circulation and weather patterns (e.g. Knutson et al. 2009, 2011; Cowan et al. 2009). Likewise, observations of brown dwarfs in the field indicate that substantial fraction (perhaps 10-20%) show variability on short time scales indicative of weather (Radigan et al. 2011); in some case the variability is as high as 25% in the J band. Initial models suggest that this variability can be explained by patchy clear regions in an otherwise cloudy atmosphere, possibly indicating holes or breaks in the cloud layer. (Radigan et al. 2012)

Recent modeling has shown that appears feasible to extend this technique of time-resolved mapping to direct imaging observations, and use the rotational modulation of exoplanets to recover their properties, including potentially surface properties, cloud cover variations and weather. See Kostov & Apai 2012 for a detailed discussion, and also Cowan and Strait 2013. These will be technically challenging observations but the potential payoff is large. This technique would strongly benefit from simultaneous studies with both JWST and WFIRST-AFTA to allow multi-wavelength characterization of any variations and better wavelength coverage to discern spectral signatures of spectral components. While additional study of optimal observing techniques is needed, initial simulations suggest greater modulation at shorter wavelengths (Kostov & Apai 2012), indicating that WFIRST-AFTA can be a key contributor here. Because of the time variability of these effects, simultaneous observations are required with both telescopes if broad spectral models are to be fit.

Depending on mission lifetimes, it may also be possible to observe seasonal changes. For instance, over the years of the Cassini mission to Saturn, we have witnessed the seasonal transformation of the northern hemisphere from pale blue to straw yellow (and the inverse in the southern hemisphere), believed to be driven by haze production due to increased ultraviolet illumination with the changing angle of incidence of sunlight. The potential extended mission lifetimes of JWST or WFIRST-AFTA are comparable to a significant fraction of an orbital period for Jovian planets on 3-5 AU orbits. Maximizing the temporal overlap between the two missions might improve our ability to detect such seasonal changes.

c) Studies of Planetary System Architectures as traced by Small Bodies

A coronagraphic instrument on WFIRST-AFTA could also potentially provide polarimetric measurements of light scattering from both planets and circumstellar dust - crucial because JWST lacks any polarimetry capabilities whatsoever. (In fact, several of the leading coronagraph concepts require polarization beam splitting for optimal operation, for instance the vector vortex coronagraph, so polarimetric capabilities may arise naturally as an inherent part of the starlight suppression methods.) Studies of solar system planets and theoretical modeling of exoplanets have demonstrated that polarimetry is a key diagnostic for assessing the presence of clouds or hazes in planetary atmospheres, so polarimetry could help with the above mentioned studies. But moreover, measurements of polarization can provide key constraints on dust grain properties in circumstellar disks (e.g. Graham et al. 2007, Perrin et al. 2009). When combined with observations of dust thermal emission in the mid-infrared with JWST and submillimeter with ALMA, scattered light imaging and polarimetry will enable comprehensive characterization of grain distributions, and therefore constraints on parent planetesimal populations in nearby debris disks.

Consider also the case of Fomalhaut b (Kalas et al. 2008), now confirmed by recent STIS coronagraphy to definitely be a physical companion on a disk-crossing-orbit, but whose physical nature remains very unclear. It has been suggested to possibly be a dust cloud from a recent collision of two KBO-type objects, or perhaps a giant planet surrounded by a swarm of moonlets in a collisional grinding process (See Currie et al. 2012 and Galicher et al. 2012). Having found one such object already with HST, it seems likely that further examples exist and may be found by JWST. Combined optical and infrared measurements, especially with optical polarimetry, would be a compelling tool to determine the physical nature of Fomalhaut b and similar objects. (Indeed, if the ACS HRC coronagraph was still operational, it would today be near mandatory to obtain ACS polarimetry of Fomalhaut b). Because JWST lacks polarimetric capabilities, a WFIRST-AFTA coronagraph including polarimetry would provide an extremely valuable adjunct capability that would enhance our ability to characterize small bodies and dust grains. When combined with infrared observations of giant planets and their atmospheres, JWST and WFIRST-AFTA together hold the potential to achieve comprehensive understanding of nearby planetary systems spanning from giant planets to asteroid and debris belts.

9. <u>Resolved studies of galaxy formation</u> (Jason Kalirai)

The detailed study of nearby galaxies can impact our fundamental knowledge of the processes that shape the formation and assembly of galaxies. The signatures of early accretion events can be directly seen in the halos of galaxies, where recent star formation is non-existent and stellar orbital time scales are of order the Hubble time. Perhaps the best example of a fully resolved spiral galaxy's halo is that of M31, as shown in the Figure below (McConnachie et al. 2009). From this star count map, we can measure the surface brightness profile, the level of substructure, and the abundance gradient of the stellar halo, and directly compare these signatures to those predicted in modern N-body simulations of galaxy formation (e.g., Bullock & Johnston 2005).

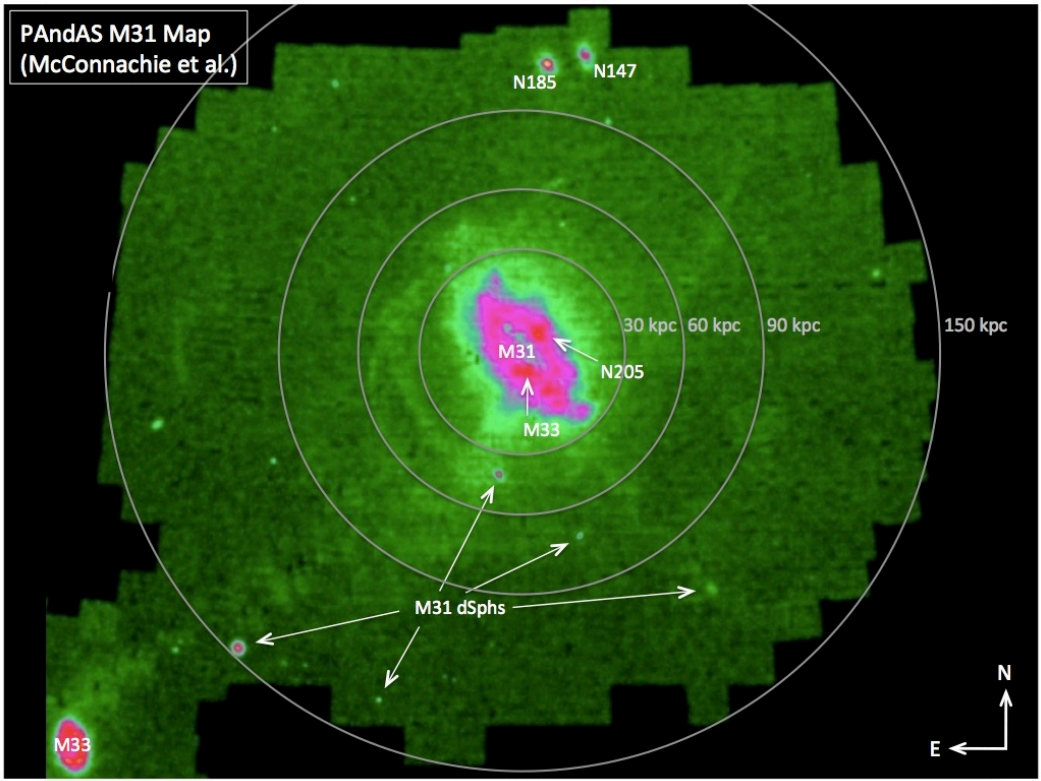

Unlike our map of M31, most nearby galaxies contain a patchwork of high precision data, often focused on the visible bulge and disk. Improving this situation is difficult since it requires a telescope with

1.) high throughput to measure the stellar tracers in old stellar population, red giants, out to several Mpc,
2.) high resolution to provide robust morphological separation between faint extragalactic contaminants and stars in the sparse halos,
3.) wide field of view to probe stellar distributions to out beyond 100 kpc in a limited number of pointings

WFIRST-AFTA satisfies all of these requirements. With a modest investment of telescope time of about 1 month, WFIRST-AFTA can survey several magnitudes of the red giant branch in 100 of the closest galaxies over their complete spatial extents. Such a survey would provide the means to study the diversity of stellar halos in detail, and to correlate their properties with galaxy size, luminosity, environment, and other factors.

Despite this wealth of new information, WFIRST-AFTA, like Hubble, will be limited to measuring direct ages for old stellar populations to galaxies within the Local Group. The main-sequence turnoff in a 13 Gyr population is too faint to measure with a 2.4 meter telescope beyond the distance of Andromeda. JWST's NIRCam and NIRISS instruments can easily do this, out to several Mpc. However, blind pencil beams will run the risk of encountering recent accretion events and therefore not revealing the age of the oldest component of the galaxy, and so the JWST observations need WFIRST-AFTA to establish the spatial maps beforehand. The JWST observations can be then fine tuned to explore quiescent areas of the stellar halos, with the appropriate density to reveal robust star formation histories.

Taken together, the combination of star formation histories (JWST), level of substructure (WFIRST-AFTA), abundance gradients (WFIRST-AFTA and JWST), surface brightness profiles (WFIRST-AFTA) and velocity dispersions (30-meter) could be established for a dozen galaxies, and would provide an unprecedented data set to test N-body models of galaxy formation and evolution.

## Appendix C    Orbit Discussion (L2 vs. GEO)

The baseline orbit for WFIRST-AFTA is a 28.5° circular inclined geosynchronous orbit with an initial RAAN of 228°.[1] However, all of the core WFIRST science programs are possible at either GEO or in a halo orbit around the Sun-Earth L2 Lagrange point; the SDT has deliberately avoided writing requirements that could not be met at either location. The Observatory design would also be similar at L2, but there would be changes in some systems as noted below.

While both orbits offer a viable mission concept, there are significant advantages and disadvantages to each choice. The Study Office and SDT intend to perform a more detailed assessment of the orbit trade in 2015. This will include quantitative assessment of the science performance in each area, cost differences, and the relative risk of the GEO and L2 options. A high-level discussion of the key factors at play in this trade is discussed below.

***Communications and Data Rate:*** The principal advantage of the baseline GEO orbit is that it enables WFIRST-AFTA to continuously downlink data to a dedicated ground station enabling a much higher science data downlink rate.

For an L2 mission, the Observatory communication and data system will need to move from this "bent-pipe" downlink concept at GEO to a store-and-downlink approach with multiple science downlinks per day to the Deep Space Network (DSN). This will require the addition of a science data recorder to store large amounts of science data between contacts. The expected number of contacts per day, the duration of each contact, and the requirement to be able to miss at least one contact (which will happen occasionally) will drive the size of the recorder. Although the WFIRST-AFTA communication system is capable of downlinking 300 Mbps from L2, the DSN is currently limited to data rates of 150 Mbps. These factors necessitate the science data volume be reduced by some combination of compression and less frequent sampling of the data. There is additional complexity in planning and scheduling the communications between the ground and the Observatory at L2 that does not exist in GEO with a continuous communications link. The impact of this to operations staffing will be assessed as part of the trade study.

From the science point of view, the higher data rate in GEO means that much more detailed information from the NIR detectors can be downlinked (every 1 or 2 samples of each pixel), versus L2 where only summary information is available (slopes in each exposure, plus quality flags and the full samples from a small sub-sample of the pixels to study the effects of the ramp-fitting algorithms). The availability of this information reduces risk in the dark energy analysis, since it enables additional internal cross-checks (e.g. first versus second half of each exposure). This information also provides additional diagnostics of subtle detector effects and the propagation of pointing drifts through the ramp-fitting algorithm.

The high data rate will also transmit information on microlensing sources that change brightness significantly during a 50-second integration (although we expect significant improvement for only a small fraction of the planetary events; note that the 15-minute observing cadence was chosen for measurement of the finite source size effects). For similar reasons, the handling of moving objects by the analysis pipelines is simpler with the data rate in GEO.

The continuous communications in GEO also allows immediate access to WFIRST-AFTA data, which would enable real-time alerts of events detected by WFIRST-AFTA to be sent to other telescopes. However, there are few anticipated planetary events for which observations from other telescopes will be useful for measuring microlensing parallax or possibly other science goals. Some of the brighter events can be observed from ground-based telescopes, but the optimal strategy may be to simply observe all the WFIRST-AFTA exoplanet survey fields from the ground. Real-time alerts might be useful for follow-up observations from larger, narrow-field telescopes, like JWST or 8m+ ground-based telescopes. JWST is unlikely to be able to respond more quickly than about 48 hours to alerts, so there will be little advantage for the GEO orbit, but there might be some advantage with the GEO orbit for follow-up observations from ground-based telescopes. However, the science return of such observations is likely to depend on developments in adaptive optics systems in the next decade, and no study of the science return of such a program has been performed.

***Observing Constraints:*** In addition to the Sun angle constraints, which are the same at GEO and at L2,

---

[1] Of all possible circular GEO orbits, the 28.5° inclination provides the maximum lift capability for a launch from Florida, and the RAAN was chosen to optimize visibility of both the Galactic Bulge and Southern Galactic Cap.

WFIRST-AFTA suffers from additional observing constraints in GEO due to the Earth and the Moon. The scheduling of observations is thus more complex and less efficient at GEO than at L2.

Due to Sun constraints, the microlensing fields are available during two ~72 day "seasons" each year, centered in March and September. At L2, these fields are available with no interruptions throughout the whole season. In GEO, there is a cutout of 4—5 days once each month as the Moon passes near the Galactic Bulge.[2] This means that to achieve the same total duration of the microlensing program, ~1 extra season must be allocated for a GEO mission versus an L2 mission. This allocation may impact some GO programs that need to observe near the Galactic Bulge.

The dark energy programs are less affected by these constraints. The supernova fields are located at high ecliptic latitude for year-round visibility (for either a GEO or L2 mission), and hence have no Moon cutouts; depending on final placement there may be a daily Earth cutout, but there is no requirement to schedule them at a particular time of day. The principal effect of GEO observing constraints on the dark energy program is that the telescope must move from field to field to avoid the Earth, making several large (tens of degrees) steps each day. These inefficiencies have been incorporated in our forecasts. However, the increased efficiency at L2 would be minor (of order several percent).

The coronagraph program is the most strongly impacted by GEO observing constraints. The Moon is not an issue, since coronagraph observations can be scheduled at a time of month when the Moon is not near the target. However, 66.5% of the sky (a "band" around the WFIRST-AFTA orbit) suffers a daily Earth viewing cutout of duration ranging from 0 to 5.6 hours. This band precesses, but still 59% of the sky is within the daily cutout band throughout the duration of the primary mission. For coronagraph targets within this band, WFIRST-AFTA would have to "nod" off the target and return after the Earth has moved out of the way. At L2, these complex and inefficient operations would not be necessary.

The scheduling software used for the GEO baseline in this report has also been run successfully for an L2 orbit. A detailed optimization of both mission profiles, incorporating ongoing improvements in the fidelity of the slew modeling, will provide a quantitative basis for this trade.

***Thermal and Mechanical Disturbances:*** The Observatory has been found to meet thermal and jitter stability requirements in GEO with large margins. However, the improved thermal stability of the L2 environment (without variable thermal loading from the Earth) and reduction of HGA stepping disturbances are expected to result in further increase in margins relative to the requirements of the weak lensing and coronagraph programs. These improvements in stability would simplify the science analysis and improve the science team's ability to simultaneously solve for any unanticipated sources of systematic error.

Due to the design of the science programs (particularly the pseudo-random dither strategy recommended for microlensing), the predictable daily effective plate scale changes in GEO due to relativistic aberration have no impact.

***Instrument Sensitivity:*** The principal backgrounds (zodiacal light, read noise, dark current, and telescope thermal emission) and hence instrument sensitivities are nearly identical at GEO or L2. Earthshine is a source of background in GEO but not at L2; however analysis has shown that WFIRST-AFTA already meets its Earthshine requirement of <10% of the zodiacal background in GEO with large margins, so the effect of eliminating it is small.

***Ionizing Radiation:*** Both GEO and L2 experience the flux of cosmic ray protons and heavy ions, but GEO is within the Earth's trapped electron belt. Thus in the baseline GEO orbit, a significant amount of radiation shielding is required around the wide-field focal planes to meet our preliminary charged particle flux requirement of $\leq 15/cm^2/s$ (versus $5/cm^2/s$ at L2). This shielding could be reduced due to the lower radiation environment at L2.

For the short HLS imaging exposures, pixels "recovered" by flagging the cosmic ray in the full ramp data are affected by increased read noise, and hence do not have the full sensitivity of an uncorrupted pixel. Thus the lower cosmic ray rate at L2 results in fewer lost pixels. Preliminary estimates suggest a minor (4%) increase in the number of weak lensing source galaxies at L2, relative to the numbers quoted in this report. The risk of degradation of the WFI science programs if the number of affected pixels per ionizing particle is greater than expected, or due to temporal variability in the electron flux, would be lower at L2.

---

[2] The duration of this cutout is determined by the range of angles over which the Moon violates stray light constraints. The variation in length is due to precession of the Moon's orbit, with the longest cutouts occurring when the Moon's apogee lies near the Galactic Bulge.

***Microlensing Parallax:*** Both the GEO and L2 orbits offer opportunities for microlensing parallax measurements. The most useful measurements are likely to ones that can give mass estimates for very short duration microlensing events, likely due to free-floating planets, because it is otherwise quite difficult to determine lens masses for these events. While it is possible to measure the microlensing parallax effect for some events from the GEO orbit (Gould 2013), a more promising method is probably to do simultaneous observations from the ground and WFIRST-AFTA (Gould et al. 2003). Only the brighter events can be observed from the ground, but high precision measurements are not needed to get interesting constraints on free-floating planet lens masses. While detailed state-of-the-art calculations of this question have not yet been performed, it seems likely that the L2 orbit is somewhat better for the estimation of free-floating planet masses with the microlensing parallax effect.

***Launch Vehicle and Propulsion:*** For this DRM, the Delta IV Heavy launch vehicle was baselined. This vehicle enables either a GEO or L2 mission with healthy mass margins.

The Delta IV Heavy has the capability to lift the Observatory directly to the circular inclined geosynchronous orbit, which allowed the simplification of the Observatory propulsion system from a large bi-propellant system (needed to circularize from GTO and baselined in 2013) to a much smaller mono-propellant system. A mono-propellant propulsion system is also required for an L2 configuration; however, a flight dynamics analysis will need to be performed to determine the $\Delta V$ required for any course corrections to L2, for orbit insertion, and station-keeping. It is expected that the fuel required to meet these $\Delta V$ requirements will be larger than for the GEO configuration.

Another factor in the trade study is the requirement to make the Observatory compatible with servicing by a robotic servicer. Early in the trade study, a location for servicing the Observatory will need to be determined, with options including L2 or at an intermediate location, such as an Earth-Moon Lagrange point. The trade study will assess implications to the Observatory design for servicing at either location.

***Structures:*** In general, the baseline Observatory mechanical and thermal design supports an L2 configuration, but some updates may be required for an L2 configuration.

***Cryocooler:*** A mechanical cryocooler is required for a GEO mission using the 2.5 μm-cutoff detectors (the most mature) since with the thermal loading from the Earth the ~90 K operating temperature is not achievable with a passive radiator. Studies have shown that the requirements of the WFIRST observing programs do not allow us to keep the radiator continuously pointed away from the Earth in GEO (the strategy usually employed by survey missions in Sun-synchronous orbit).

Passive cooling using large radiators was the baseline for past WFIRST DRMs at L2. However, a large radiator may have some implications for servicing the wide-field instrument. More study will be required to assess whether a cryocooler is required at L2.

Although the WFIRST-AFTA DRM was designed for a 28.5° inclined geosynchronous orbit, the design of the Observatory for an L2 mission would be very similar. A detailed trade study, considering the factors above, to assess the impacts to the WFIRST-AFTA DRM will be performed in 2015. The resulting modifications to the DRM will be costed by the Project to understand the total cost delta for an L2 configuration.

**Appendix D    GO and GI One Page Science Ideas**

# Wide Field Infrared Survey Telescope – Astrophysics Focused Telescope Assets (WFIRST-AFTA)

## *A Collection of Short Science Programs from the Astronomical Community*

## Jan 2015

## http://wfirst.gsfc.nasa.gov/

This appendix contains a rich set of ~40 potential Guest Investigator (GI) and Guest Observer (GO) science programs that are uniquely enabled by WFIRST-AFTA (many of which were submitted for the May 2013 report). The one page science programs ("one pagers") were put together by the broader astronomical community, and the SDT wishes to thank the community for this valuable contribution. In addition to this appendix, the recent November 2014 "Wide Field Infrared Surveys" meeting in Pasadena brought together ~200 scientists to share in exciting science opportunities with WFIRST-AFTA. An archive of the presentations given at this meeting is available at this website, http://conference.ipac.caltech.edu/wfirs2014/page/sched.

The one pagers below (as well as the presentations linked on the website above) highlight the tremendous potential of WFIRST-AFTA to advance many of the key questions formulated by the 2010 Decadal Survey, "New Worlds, New Horizons in Astronomy and Astrophysics". These programs have not been vetted by the SDT, nor are we endorsing these specific studies. We have removed specific contributions that require WFIRST-AFTA capabilities that are inconsistent with the current baseline (e.g., the science can not be done without imaging beyond 2 microns). As has proven extremely successful on other NASA missions, we expect that the selection of GO science programs will be made by a peer-reviewed Time Allocation Committee process and GI programs will be enabled through archival analysis of WFIRST-AFTA's survey data sets.

The table below highlights each of the one-page science contributions and an indication of whether the science program requires GI data from the High-Latitude Survey (HLS), Supernova Survey (SN), or Microlensing Survey (ML), or whether the program requires dedicated GO observations. The categorization of these papers into the bins below was done by the WFIRST SDT, and not by the authors of the papers. The table also includes a list of the science cases that highlight synergies between WFIRST-AFTA and JWST (Appendix B) and synergies with other future missions (Appendix J).

| Community 1 Page Science Programs | GI - HLS | GI - SN | GI - ML | GO |
|---|---|---|---|---|
| **Solar System / Exoplanets** | | | | |
| Norwood - Solar System Science | | | | X |
| Schlichting - Survey of KBOs | | | | X |
| Ardila – Free Floating Planets | | | | X |
| Holman - Transit Timing Variations | | | | X |
| Grillmair – Exoplanet Spectroscopy | | | | X |
| Tanner – Exoplanet Transits | | | X | |
| Tanner – Exoplanet Astrometry | | | X | |
| **Stellar Astrophysics** | | | | |
| Tanner – Coldest Brown Dwarfs | | | | X |
| Kalirai – Stellar Fossils in the Milky Way | X | X | X | X |
| Kalirai - IR Color-Magnitude Relation | | | | X |
| Ardila – Closest Young Stars | | | | X |
| Martinache - Low Mass Stars | | | X | X |
| Sahu – Neutron Stars and Black Holes | | | X | |
| Gaudi – Bulge Parallaxes | | | X | X |
| **Galactic Astrophysics / Local Volume** | | | | |
| Stern – LMC Proper Motion | | | X | X |
| Besla – Counterpart of the Magellanic Stream | | | | X |
| Geha – Faintest Milky Way Satellites | X | | | X |
| Deason – Mass of the Milky Way | X | | | |
| Strigari – Cold vs. Warm Dark Matter | | | | X |
| Johnston – Finding or Losing Missing Satellites | X | | | X |
| Johnston – Potential of the Milky Way | X | | | X |
| van der Marel – Nearby Galaxy Halos | | | | X |
| Guhathakurta – Extragalactic Halo Ages | | | | X |
| Laine – Substructure in Nearby Galaxies | | | | X |
| Dalcanton – Resolved Stellar Populations | | | | X |
| Abraham – Resolving ICL in Virgo | | | | X |
| **Extragalactic Astrophysics** | | | | |
| Mihos - Surface Photometry of Galaxies | | | | X |
| Conselice – Galaxy Morphologies | X | X | | |
| Stern – Strong Lensing | X | X | | |
| Appleton – Shocked Galaxies | X | X | | |
| Merten – Distribution of Dark Matter in Clusters | | | | X |
| Merten – Merging Clusters | | | | X |
| Natarajan – Group-Scale Lenses | X | | | |
| Donahue – Weighing Clusters | X | | | |
| Donahue – Cluster Evolution & Red Sequence | | | | X |
| Stern – Obscured QSOs | X | X | | |
| Stern – Faint High-z QSOs | X | X | | |
| Fan – Strongly Lensed QSOs | X | | | |
| Fan - High z Quasars and Reionization | X | | | |
| Teplitz - Reionization Sources | X | | | |
| Scarlata – Resolved z=2 SF Galaxies | X | | | X |
| Kasliwal – Counterparts of NS Mergers | | | | X |
| Gladders – Strong Lensing | X | | | X |
| **WFIRST and JWST Synergies (Appendix B)** | | | | |
| Coe – Earliest Galaxies | X | | | |
| Stiavelli - Pair Instability Supernovae | X | X | | X |
| Treu – Lensed QSOs as Probes of Sub-Halos | X | | | |
| Goodfrooij – Globular Clusters | | | | X |
| Boyer – AGB Stars | | | | X |
| Ferguson – Tidal Streams | | | | X |
| Perrin – Exoplanet Atmospheres | | | | X |
| Kalirai – Galaxy Structure | | | | X |
| **Other WFIRST-AFTA and Future Mission Synergies (Appendix J)** | | | | |
| Rhodes – LSST Cadence | | | | |
| Dell'Antonio - WFIRST and TMT - Cluster, Galaxy, Supernova, & Near Field Cosmology | | | | |
| Tanaka - WFIRST and TMT - GW Sources | | | | |
| Tanaka - WFIRST and TMT - First Supernovae | | | | |
| Jiang - WFIRST and TMT - High-z QSOs #1 | | | | |
| Shen - WFIRST and TMT - High-z QSOs #2 | | | | |
| Shen - WFIRST and TMT - BH Scaling Relation | | | | |
| McCarthy - WFIRST and GMT - Galaxy Evolution | | | | |
| Finkelstein - WFIRST and GMT - Reionization | | | | |
| Green - WFIRST and the Subaru PFS | | | | |
| Strauss - LSST and WFIRST | | | | |
| Capak - Synergies between Euclid & WFIRST | | | | |

James Norwood (New Mexico State University), Kunio M. Sayanagi (Hampton University)
Michael H. Wong (University of California Berkeley); kunio.sayanagi@hamptonu.edu

## WFIRST's Potential Contributions to Solar System Science

Abstract

Among the many disciplines of astronomy expected to benefit from the upcoming Wide-Field InfraRed Space Telescope (WFIRST) mission will be Solar System science. This white paper explores potential WFIRST observations of Solar System targets, focusing on giant planet investigations as an example, and discusses special software and operations implementations that would be of particular use to the Solar System community.

WFIRST: Continuation of the Hubble Space Telescope Legacy

With anticipated spatial and spectral resolution comparable to Hubble Space Telescope (HST)'s Wide Field Camera 3, WFIRST will be seen, in multiple ways, as a continuation of HST-quality observations in the visible and near-infrared parts of the spectrum. The contributions of WFIRST to Solar System science will be will be significant, as proven by HST's legacy. In the broader picture, WFIRST's capabilities will allow it to serve a role similar to that of HST in publicity, education, and public outreach, providing high-quality imagery of various objects in the visible, portraying them as the human eyes would see them. WFIRST can also capture stunning, newsworthy phenomena such as erupting storms, impact events, and photogenic satellite transits. However, there are myriad ways in which WFIRST will uniquely contribute to scientific investigations as well.

WFIRST's Potential Contributions to Giant-Planet Science

The visible and near-infrared spectra of the four giant planets - Jupiter, Saturn, Uranus, and Neptune - are dominated by the signature of methane, which varies in absorption strength by several orders of magnitude. Because of this, different parts of this spectral range probe a wide variety of depths in the atmospheres of these planets, especially when combined with center-to-limb comparisons. Observations at different wavelengths will reveal the altitudes and optical properties of the upper tropospheric cloud layers on these planets, and how they vary with location. These conclusions will in turn have significant implications for our understanding of chemistry, thermodynamics, and circulation on these planets.

Recent findings on Uranus and Neptune have shown that the upper tropospheric methane abundance varies with latitude (unlike on Jupiter and Saturn) (Sromovsky et al. 2014, *Icarus* 238, 137-155; Karkoschka and Tomasko 2010, *Icarus* 205, 2, 674-694). Comparison with nearby hydrogen lines will enable determination of the methane abundance across the disks of these planets for accurate interpretation of the methane opacity, as well as clarifying the extent of this abundance variation. In addition to determining the abundance variation to greater precision, further investigation will reveal the depth to which these variations persist, and whether they evolve over time - all three items of great interest to modelers of the Uranian and Neptunian atmospheres.

Importance of Time-Domain Science

While observations with the Hubble Space Telescope have already enabled much of the above analysis, it must be remembered that many of these discoveries have been made through observing transient events that evolve in timescales of weeks to months, and capturing future events of different types should bring further progress. We emphasize that the planets are not static objects; these time-domain studies are critically important because they are one of the main drivers of new discoveries in solar system science today. With Jupiter and Saturn, we have seen dramatic atmospheric changes such as the disappearance of Jupiter's south equatorial belt; the ongoing shrinking of Jupiter's Great Red Spot; the development, evolution, and dissipation of immense Saturnian storms in 1990-1991 and 2010-2011; and intermittent and yet intense storm activities on Uranus. The orbital periods of Uranus and Neptune are long enough that seasonal change on these planets is not well understood, and ongoing observations will be critical as the seasons proceed, both to discover new development and to maintain a suitable temporal baseline of observations.

WFIRST Sensitivity: Critical to Small Body Observations
The Hubble Space Telescope has been employed in successful campaigns to discover new satellites around the giant planets as well as Kuiper Belt objects (e.g., Showalter et al. 2013, DPS meeting #45, #206.01; Showalter et al. 2012, IAU Circ., 9253, 1). As the retrieval techniques continue to improve, WFIRST will likely be able to discover further new satellites and KBOs. In addition to new discoveries, WFIRST will enable further investigations of known small objects whose orbital elements are poorly constrained.

WFIRST: Complementing JWST Capabilities
WFIRST will also be valuable when used to complement other datasets. For example, the upcoming James Webb Space Telescope (JWST) will set the new standard for near- and mid-infrared observations; however, the JWST's missions exclude many solar system science topics. For example, Mars, Jupiter, and Saturn will not be observable over JWST's full spectral range because those targets are too bright and saturate JWST pixels even with the shortest possible exposure. Furthermore, JWST may observe Solar System targets only at solar elongation angles between 85° and 135°, typically granting two 50-day windows per year. Should a time-sensitive event such as an impact or sudden storm occur when the body in question is outside of JWST's field of regard, other facilities such as WFIRST will be able to step forward.

Solar System Observation Requirements
There are two special requirements of WFIRST implementation that would be of unique importance to Solar System observations: tracking and targeting of moving objects, time-critical observation scheduling.

*Moving Object Targeting and Tracking*
Because targets in the Solar System are constantly moving with respect to the background sky, the ability observe moving targets will be essential. The maximum movement rates of several Solar System targets are given in the table below; if such rates are unachievable with WFIRST, the target movement rates may be lowered at the expense of freedom in scheduling. It must also be noted that more swiftly moving targets may also be desired, such as comets and near-Earth asteroids. Additionally, since Solar System targets do not have fixed locations, observers would appreciate a system pipeline that could recognize the names of Solar System targets and determine their ephemerides, rather than requiring target positions to be supplied by the user.

Table 1: Maximum Tracking Rate Necessary for Observing Solar System Targets

| Object | Max rate (mas/sec) |
|--------|--------------------|
| Mars | 31.5 |
| Ceres | 18.6 |
| Jupiter | 10.1 |
| Saturn | 5.1 |
| Uranus | 2.4 |
| Neptune | 1.6 |
| Pluto | 1.4 |
| Eris | 0.45 |

*Time-Critical Observation Scheduling*

Lastly, it must be mentioned that time-sensitive events are not uncommon in Solar System astronomy. Some events may be anticipated, such as close approaches by asteroids, optimal observing geometry for the giant planets' ring systems, and mutual events among satellites. Observing plans must be able to request such time constraints. Other time-sensitive events have little advance warning, such as sudden changes in comet outgassing, meteorological events in planetary atmospheres, and witnessing the aftermath of impact events. The ability to conduct target-of-opportunity considerations will be of immense value in such scenarios.

Hilke E. Schlichting (UCLA), hilke@ucla.edu

**A Full Portrait of the Kuiper Belt, including Size Distributions, Colors and Binarity**

Background

The Kuiper belt is a remnant of the primordial Solar system. It consists of a disk of icy bodies located at the outskirts of our planetary system, just beyond the orbit of Neptune, and is the likely source of short period comets. More than 1200 Kuiper Belt Objects (KBOs) have been detected since its discovery in 1992 (Jewitt & Luu 1993). In the Kuiper Belt, planet formation never reached completion and as a result it contains some of the least processed bodies in our Solar system. Its dynamical and physical properties illuminate the conditions during the early stages of planet formation and have already led to major advances in the understanding of the history of our planetary system. However, despite all these successes, many open questions remain: is the Kuiper Belt undergoing collisional evolution, grinding small KBOs to dust and therefore a true analogue to the dust producing debris disk around other stars? How did Neptune's outward migration proceed (Malhotra 1993, Tsiganis et al. 2005)? And what do the colors of KBOs imply about their formation and consequent evolution? Only a deep, uniform wide-field survey will be able to definitively answer these and other questions.

WFIRST

WFIRST will enable a uniform wide-field survey with unprecedented sensitivity of the Kuiper Belt. Since there are estimated to be more than $4 \times 10^4$ KBOs with diameters greater than 100km (R ~ 24), WFIRST will increase the current number of known KBOs by two orders of magnitude. WFIRST has the potential to provide us with an almost complete census of KBOs with magnitudes of R< 26.7 ($m_{F087}$ ~ 27.4). This will yield the best measurement of the KBO size distribution below the observed break at R ~ 24, which will provide important constraints on the material properties of KBOs and their collisional evolution. In addition, WFIRST will enable us to make, for the first time, detailed comparisons between the size distributions of KBOs in different dynamical classes, shedding light onto the origin of the break in the KBO size distribution and the planet formation process itself. Furthermore, WFIRST will provide a detailed census of the resonant population in the Kuiper Belt and should discover 100s – 1000s of binaries, which together provide important constraints on Neptune's migration history. Finally, it will provide a uniform survey of KBO colors over a wide range of sizes. Comparison between the colors of small KBOs whose sizes are below the break radius with that of larger KBOs will, for example, show if some of the color diversity in the Kuiper belt can be attributed to collisional resurfacing.

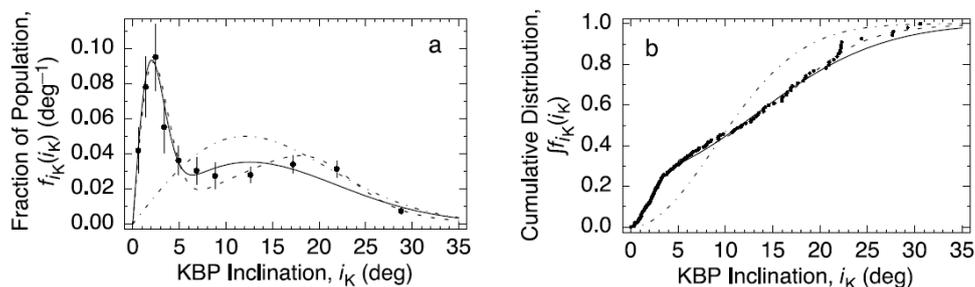

Caption: Unbiased inclination distribution of KBOs from Elliot et al. (2005).

Key Requirements

Coverage – Ideally ~14,000 deg$^2$ (+/- 20$^o$ of the ecliptic), but ~7,000 deg$^2$ suffices to discover the majority of KBOs (assuming distribution of fainter (smaller) KBOs is similar to brighter counterparts)

Cadence – 15 minute exposures, revisit each field three times separated by ~30-60 minutes

Wavelength Coverage – Need two NIR filters to get colors

David R. Ardila (NHSC/IPAC), ardila@ipac.caltech.edu

# Free-floating Planets in the Solar Neighborhood

## Background

As part of its primary mission, WFIRST is likely to refine the estimates for the population of very faint isolated objects, all the way down to planetary masses. Indeed, using microlensing observations, Sumi et al. (2011) conclude that isolated, free-floating Jupiter-mass planets are twice as common as main-sequence stars, and at least as common as planets around stars. Delorme et al. (2012), using data from the 335 deg$^2$ Canada-France Brown Dwarf Survey InfraRed project, report a free-floating 4-7 $M_{Jup}$ candidate which may belong to the AB Doradus moving group (50-100 Myrs).

The possibility of free-floating planets in the solar neighborhood is an exciting one: as with the study of low-mass brown dwarfs, isolated planets provide excellent laboratories to understand the evolutionary processes at play in planets around other stars. They also provide constraints on the efficiency of molecular cloud collapse mechanisms of planet formation versus expulsion from protoplanetay disks due to scattering with other planets (e.g. Veras et al. 2009). Ultimately, the discovery of a nearby free-floating planet will have a strong and lasting public impact. It is not impossible that the closest body to the solar system could be a Jupiter mass planet expelled from a young planetary system!

## The Investigation

We propose to search for young (<100 Myrs) Jupiter-mass planets within 5 pc of the Sun. According to Spiegel and Burrows (2012) a 100 Myr, 1 $M_{Jup}$ has $M_J$=24.7 mag ($m_J$ AB=24.1 mag at 5 pc), $M_H$=25.9 mag ($m_H$ AB = 25.7 mag at 5 pc), and $M_K$=27.4 ($m_K$ AB=27.7 mag at 5 pc). In J and H these magnitudes are well within the capabilities of the Galactic plane WFIRST survey, if performed with a 2.4 m telescope. We limit the estimate to t<100 Myrs, d<5 pc as beyond those limits the luminosity function falls down very quickly, detection efficiency becomes poor, and ground-based follow-up is difficult. The survey area required to find a significant number of objects is uncertain, but this investigation could be carried out as ancillary to the 1240 deg$^2$ Galactic Plane survey. Multi-epoch observations are crucial to determine proper motions and rule out distant objects. Ground-based spectroscopic follow-up is required to confirm the surface gravities.

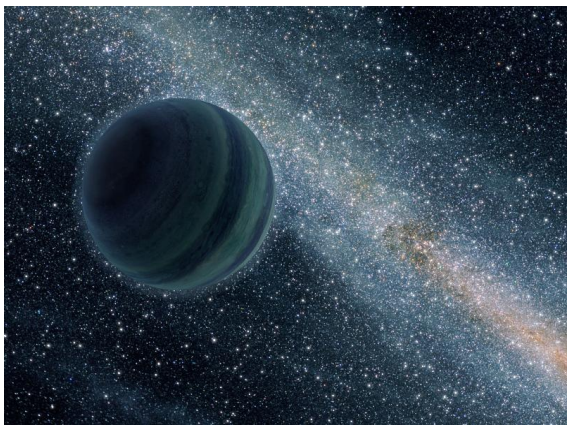 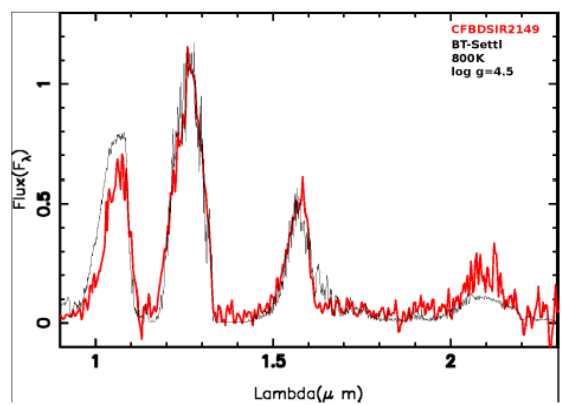

Fig. 1. Left: Artist conception of a free-floating planet in the solar neighborhood. Right: Spectrum of CFBDSIR2149-0403, a candidate free-floating Jupiter-mass object (Delorme et al. 2012).

## Key Requirements

Depth – 24 AB mags in J, 26 AB mags in H; Field of View – Galactic plane survey; Cadence – Three epochs to obtain proper motion; Wavelength Coverage – YJHK

Matthew J. Holman, Charles Alcock, Joshua A. Carter (Harvard-Smithsonian Center for Astrophysics), Eric Agol (University of Washington), Michael Werner (JPL)
mholman@cfa.harvard.edu

# Measuring Planet Masses with Transit Timing Variations

### Background

The *Kepler* mission has demonstrated that systems with multiple Earth-size to Neptune-size planets are common. However, these planets' masses are inaccessible to current and future radial velocity (RV) or astrometric measurements; hence, we are blind to their bulk densities, which have important implications for planet composition, habitability, and formation.

Gravitational interactions among transiting planets in such systems can lead to observable deviations of the transit times from strict periodicity (Agol et al. 2005, Holman et al. 2005). In fact, roughly sixty of these systems show transit timing variations (TTVs). Analysis of these TTV signals has become a key tool in confirming *Kepler* planets and measuring their masses (Holman et al. 2010; Lissauer et al. 2011; Ford et al. 2012; Fabrycky et al. 2012; Steffen et al. 2010; Carter et al 2012).

The mass sensitivity of the TTV method is governed by the orbital elements of the planets, the precision with which individual transit times can be measured, the total number of observations, and the total time span of those observations. Transit times measured by WFIRST would be ~2.4 times more precise than those measured by *Kepler*, the ratio of aperture diameters. By obtaining more precise transit light curves of selected *Kepler* target stars, and by significantly extending the time baseline of *Kepler*, WFIRST could measure the masses of Earth-size planets near the habitable zones of their stars. As mentioned, this is well beyond the limits of RV observations and will not be feasible for *Kepler* stars even with the TMT.

### WFIRST

The idea would be to use either WFIRST's imager or an auxiliary camera optimized for highly precise photometry to observe transits of selected *Kepler* target stars. These stars would host multiple transiting planets, one (or more) of which would be near its habitable zone. Furthermore, we would select for systems that already show TTVs in the *Kepler* data. The frequency of transits in any given system would range from monthly to yearly, depending upon the stellar properties. The duration of the observations would typically be tens of hours. Only a small number of the most promising targets would be selected to minimize the impact on WFIRST's surveys. Future targets might come from missions such as TESS. (If, as has been proposed by Perlmutter, Werner, et al., the WFIRST IFU is used for transit spectroscopy of exoplanets, the same data might be used for transit timing, depending upon the performance of the instrument.)

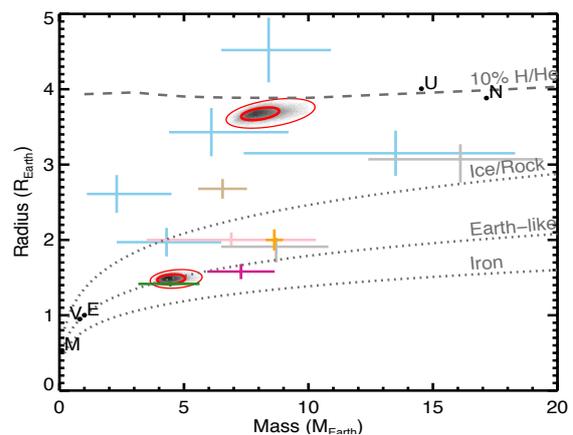

Estimates of masses and radii for a number of exoplanets, mostly from Kepler. The error ellipses refer to the well-studied Kepler 36 system. WFIRST will provide mass estimates for upward of 100 Kepler exoplanets with known radii, determining their densities, placing them on the theoretical curves, and revealing the frequency of planets of different type and composition (Carter et al., 2012).

### Key Requirements

Photometric precision – The precision with which an individual transit is measured is directly related to the depth of the transit, the S/N of the observations, and the duration of the ingress and egress of the transit. Ideally, as is the case with *Kepler*, these observations would be photon-limited.

Saturation – We would be observing bright stars. It would be essential to know that all of the photons are captured, even for heavily saturated targets.

Observing cadence – To resolve the ingress and egress of the transit of a small planet, exposure times less than a few minutes are required. This is easily satisfied by WFIRST.

Scheduling flexibility – The idea would be to use WFIRST to observe selected transits of *Kepler* targets. The approximate times of these transits would be known in advance, but it would be necessary to insert these periods of transit observation into the overall schedule.

Carl J. Grillmair (Caltech), carl@ipac.caltech.edu

## Exoplanet Spectroscopy with WFIRST

Provided a low-resolution spectroscopic capability is incorporated into the final design, NRO WFIRST could be used without serious modification for spectroscopy of transiting exoplanets. Exoplanet atmospheres could be studied both in transmission and in emission, and the effective wavelength range (.7 to 2 microns) of NRO WFIRST is well-matched to the most interesting molecular signatures we expect to see in hot-Jupiter-like atmospheres. Key diagnostic molecules would include water, methane, carbon dioxide, and carbon monoxide, which are particularly important for investigating atmospheric radiation balance, temperature structure, equilibrium/nonequilibrium chemistry, and photochemical processes.

Exoplanet spectroscopy can be carried out with either a slit or a slitless system. A slitless system necessarily entails some risk of spectral contamination by other sources in the field, but in most cases this could be overcome if the spacecraft has some degree of rotational freedom. For a traditional slit, to avoid jitter-induced, time-variable slit losses, exoplanet spectroscopy would benefit from using a slit many times wider than the FWHM of the spatial PSF. While the critical information is contained in the time domain rather than the spatial domain, an integral field unit would also be suited to exoplanet spectroscopy, provided that pointing-induced "slit losses" can be minimized. The optimal system would again use fibers, lenslets, or image slicers whose elements subtend an area several times larger than the spatial PSF.

Since the primary molecular bandheads are quite broad, high spectral resolution is not required, particularly for an initial survey. The low resolution, R = 75 grism considered in DRM1 for would be more than adequate. By obtaining spectra over multiple eclipses, WFIRST can build up signal-to-noise ratio for the faintest or most interesting sources, or search for variations in composition, temperature structure, or other signs of global climate change.

Systematics will be dominated by pointing performance and detector stability. With the large field of view available, pointing drifts can be very accurately determined in post-processing using the many thousands of other sources in the field. Without any design optimizations specific to exoplanet observations, HST and Spitzer have both demonstrated that, with suitable calibration and decorrelation techniques, systematics can be brought down to a level of 30-70 ppm (Demory et al., 2011, A&A, 544, 113, Todory et al. 2012, ApJ, 746, 111).

Some of the brightest exoplanet spectroscopy candidates will saturate NIRSpec and thus be essentially unobservable with JWST. If NRO WFIRST includes the higher resolution, R=600 prism for the galaxy redshift survey in DRM1, then WFIRST will in principle be capable of obtaining exoplanet spectra for targets ~1.6 magnitudes brighter than JWST. In the post-HST era, WFIRST may become our only means for obtaining high-precision, space-based spectra for such objects in this particularly interesting wavelength range.

Angelle Tanner (MSU), at876@msstate.edu
David Bennett (U Notre Dame)

**WFIRST: Additional Planet Finding Capabilities - Transits**

Background
Since the detection of the first transiting planet, HD209458, in 2000 over 200 planets have been detected with this method. The recently extended Kepler mission will eventually open the flood gates of such systems with over 2000 planetary candidates. The goal of the Kepler mission is to detect habitable earth-mass planets around solar-type stars and it is close to that goal. An advantage to the detection of a transiting planet is that we are able to place stringent limits on the mass and radius of the planet, thus determining its density. Finding transiting planets also allows us to study the atmospheric composition of hot Jupiters and super-Earths through transit spectroscopy and photometry.

WFIRST
The primary objective of the WFIRST exoplanet science program is to complete the statistical census of planetary systems in the Galaxy with a microlensing survey. This program will cover 500 days of observation time over five years and is expected to be sensitive to habitable Earth-mass planets, free floating planets and all solar system analog planets except Mercury. Using estimates of the anticipated signal-to-noise of the data as well as the observing cadence and number of stars in the microlensing study, It is expected that WFIRST will obtain light curves with this cadence for ~$3 \times 10^8$ stars and a photometric precision of 1%. With this sensitivity and observing cadence the WFIRST microlensing light curves can also be used to detect up to 50,000 Jupiter transits around main sequence stars and about 20 super-Earth transits around the brightest M dwarfs in the field of view. The statistics on the populations of hot Jupiters from the jovian transits will shed light on the mechanisms responsible for planetary migration when combined with information on the properties of the host star.

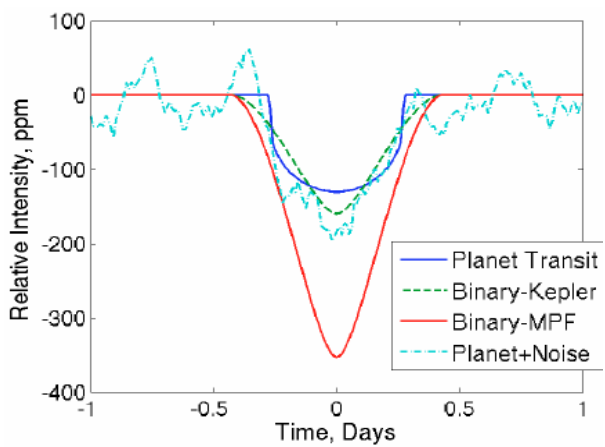

Caption: Simulated planetary transit light curve from the Microlensing Planet Finder mission (light blue, noise added) similar to WFIRST (Bennett et al. 2010) compared to Kepler (Borucki et al. 2010).

Key Requirements
Duration – 500 days (µL survey)
Precision – 1% photometry
Cadence – Every 15 minutes
Sensitivity – J < 19 at 5 sigma
S/N – ~100 for all 300 transits of a Jupiter mass planet in a 3 day orbit

Angelle Tanner (MSU), at876@msstate.edu

# WFIRST: Additional Planet Finding Capabilities - Astrometry

## Background

While the first exoplanet discovered solely through astrometric observations remains elusive, this method of exoplanet detection has been used to confirm their existence and determine the mass of the planet. While not all planets transit their star, they do all make them wobble. It is just a matter of getting a large number of data points with good astrometric precision. For instance, the $0.64 M_J \sin i$ Jupiter mass planet discovered around the M dwarf GJ 832 at an orbital separation of 3.4 AU, period of 9.35 years and distance of 5 parsecs should produce an astrometric wobble with about a 1 milli-arcsecond amplitude.

## WFIRST

The primary objective of the WFIRST exoplanet science program is to complete the statistical census of planetary systems in the Galaxy with a microlensing survey that will cover 500 days of observation time over five years. In order to maximize the number of microlensing events, WFIRST will observe a set of adjacent fields in the Galactic bulge. Observing in the near infrared significantly reduces the effects of extinction relative to visible wavelengths, increasing the number and apparent brightness of the background stars. A total of seven fields will be observed. The exposure time will be 88 seconds per field, with a slew and settle time of 38 seconds between fields. With this observing cadence and WFIRST's 0.15" spatial resolution, the positions of the $\sim 3 \times 10^8$ stars collected for the microlensing survey can also be used to detect hundreds of $> 10\ M_J$ planets in 100 day orbits around $0.05 - 0.3\ M_R$ M dwarfs. While the astrometric precision of a single observation may be on the order of 1 milli-arcseconds, WFIRST will collect a few thousand observations over the course of the microlensing program.

Caption: The astrometric signals of 10 $M_J$ (dashed) and 40 $M_J$ (solid) companions in a 100 day orbit around 10% of the 28000 M dwarfs in the FOV. With > 20000 observations, final astrometric precisions will be < 0.05 mas

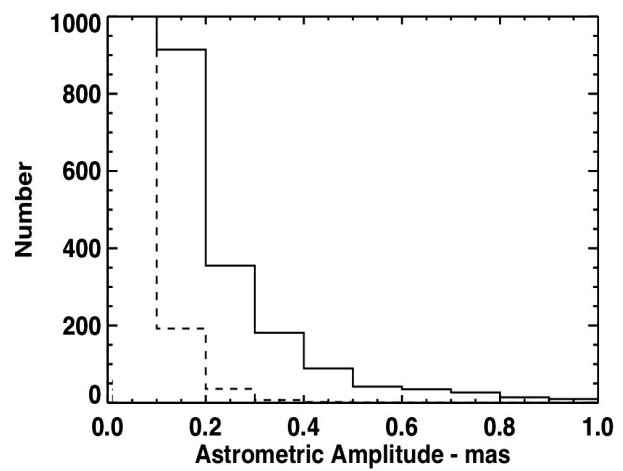

## Key Requirements

Duration – 500 days (µL survey)
Resolution – 0.15"
S/N – 100 for a J < 19 mag star
Precision – 1 mas single measurement

*Angelle Tanner (MSU), at876@msstate.edu,*
*Adam Burgasser (UC San Diego), aburgasser@ucsd.edu*

## Identifying the Coldest Brown Dwarfs

Background

One of the key scientific contributions of large infrared surveys such as 2MASS, SDSS, UKIDSS and WISE has been the discovery of brown dwarfs in the immediate vicinity of the Sun, sources with insufficient mass to sustain core hydrogen fusion. These include members of the newly-defined L, T and Y spectral classes whose atmospheres are distinctly planetary-like in composition ($H_2O$, $CH_4$ and $NH_3$ gases, mineral and ice clouds). While roughly 1000 brown dwarfs have been uncovered to date, some as cool as ≈300 K, the bulk of the Galactic brown dwarf population has likely cooled to temperatures that have only become recently accessible due to WISE. Because to brown dwarf thermal evolution, both the shape of the substellar mass function below 0.1 $M_\odot$ and the minimum "stellar" mass remain too poorly constrained (uncertain by factors of 3-5) to robustly constrain formation models. Current surveys of cool brown dwarfs are also highly incomplete to modest distances (e.g., up of 100 pc), inhibiting studies of Galactic spatial and velocity distributions, key statistics for dynamical formation models. A large-scale, volume-complete assessment of the space density of brown dwarfs over a broad temperature range (300—3000 K) would allow us to probe stellar formation processes as a function of both mass and time (given the inherent time-dependent cooling of brown dwarfs), providing simultaneous tests to brown dwarf formation and evolutionary models.

WFIRST

The combination of wide-field coverage, depth, time resolution and filter selection makes WFIRST a highly efficient survey machine for the coldest brown dwarfs. A focused infrared GO survey using at least three WFIRST filters has the potential to detect a few thousand L dwarfs and a few hundreds T dwarfs based on LSST expected performance and a smaller WFIRST field of view. The combination of LSST visual rizy and WFIRST NIR photometry will allow for a thorough census of brown dwarfs over the MLT spectral range, as LSST will be less sensitive to the cooler L dwarfs. A WFIRST brown dwarf photometric survey could be completed as a separate GO program as it would only require a couple deep (mag AB ~ 25) fields separated over a couple years for additional proper motion information independent from LSST. Some of the stars found with a LSST+WFIRST combined data set could have their parallaxes determined using the LSST data or with a focused infrared parallax programs. The brown dwarfs discovered from this survey could also be followed up with high-resolution infrared spectra from >10-m class telescopes to further constrain spectral types of the object and investigate multiplicity.

Caption: Near-IR CMD of nearby brown dwarfs utilizing data from the 2MASS and other near-infrared surveys that have wavelength coverage (JHKs) similar to WFIRST (Kirkpatrick et al. 2005). A targeted WFIRST GO brown dwarf survey will go much deeper than 2MASS and, thus, will be able to extend the bounds of a complete a census of nearby MLT brown dwarfs.

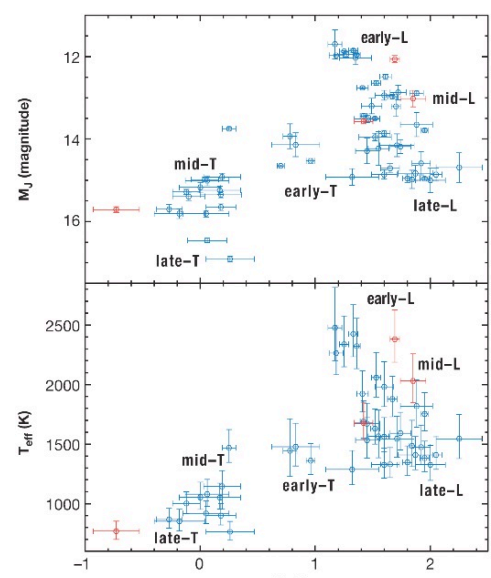

Key Requirements

Coverage – > 2500 deg$^2$ high galactic latitude, > 1500 deg$^2$ galactic plane (HL survey)
Bands – Three of the four bands - F087, F111, F141, or F178
Sensitivity – Mag AB ~ 25

Jason Kalirai (STScI), jkalirai@stsci.edu

# Stellar Fossils in the Milky Way

## Background
98% of all stars will end their lives quiescently and form white dwarfs. These remnants are remarkably simple as they contain no nuclear energy sources. Over time, white dwarfs will radiate away thermal energy and cool, thereby becoming dimmer and redder. In old stellar populations, such as the Galactic halo, a significant fraction of the mass is now tied up in white dwarfs and the properties of these stars hold clues to infer the nature (i.e., the age and mass function) of their progenitors.

## WFIRST
The largest sample of white dwarfs studied to date comes from the SDSS, which has increased the known population to over 20,000 remnants (Eisenstein et al. 2006). The brightest white dwarfs have $M_V$ = 10, therefore SDSS is mostly sensitive to the luminosity function out to less than a 1 kpc. WFIRST will discover and characterize the luminosity function of white dwarfs down to a much larger volume, across different sightlines. In the Galactic disk, the spatially-dependent structure of these luminosity functions correlates to peaks in the star formation history (see Figure below). The faintest stars, also in the Galactic halo, provide a robust estimate of the formation time of the first populations (Harris et al. 2006). Follow up spectroscopy of these stars can yield their fundamental properties (temperature, gravity, and mass), which can be connected to the progenitor masses through the well-measured initial-final mass relation (Kalirai et al. 2008). A wide field survey of the Milky Way disk and halo will provide a complete characterization of this remnant population in the Milky Way.

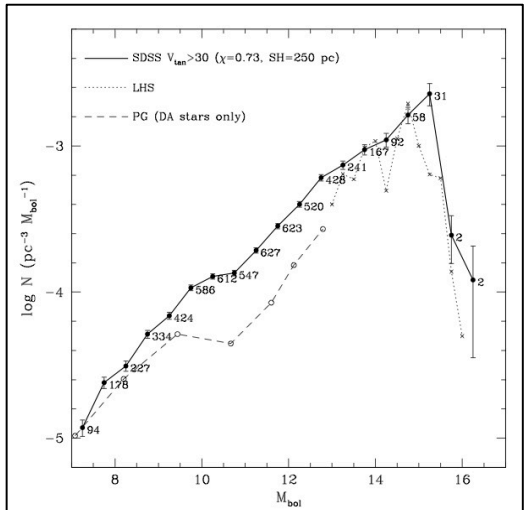
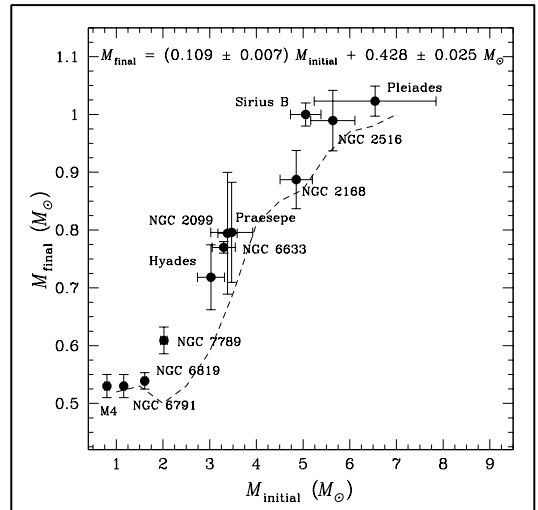

Caption: The luminosity function of white dwarfs in the Galactic disk (Harris et al. 2006) shows a turnover at the faint end (left), corresponding to the age of the oldest stars in the Galactic disk. The initial-final mass relation of stars is shown in the right panel (Kalirai et al. 2008).

## Key Requirements
Depth – Measure faintest white dwarfs in the local halo ($M_V$ = 17).
Field of View – Sample range of stellar environments in Galactic disk and different probes in halo.
Cadence – Second epoch for proper motions would enable kinematic separation of components and a much cleaner selection from contaminants.
Wavelength Coverage – Two NIR filters for color-magnitude diagram analysis

Jason Kalirai (STScI), jkalirai@stsci.edu

# The Infrared Color-Magnitude Relation

## Background
Star clusters in the Milky Way (MW) have served as the primary tools to measure the color-magnitude relation of stars, and to calibrate its dependency on stellar properties such as age and metallicity. This relation is a key input to test stellar evolution models, and in turn to carry out population synthesis studies that aim to interpret the integrated light of astrophysical sources across the Universe (e.g., Bruzual & Charlot 2003). For decades, this work has primarily focused on the interpretation of visible-light color-magnitude diagrams (CMDs).

## WFIRST
WFIRST will enable high-precision IR CMDs of stellar populations. The figure below illustrates the morphology of the IR CMD of the globular cluster 47 Tuc, from a 3 orbit (depth) exposure with the WFC3/IR camera on HST (Kalirai et al. 2012). The sharp "kink" on the lower main sequence is caused by collisionally induced absorption of $H_2$. Unlike the visible CMD, the inversion of the sequence below the kink is orthogonal to the effects of distance and reddening, and therefore degeneracies in fitting fundamental properties for the population are largely lifted. The location of the kink on the CMD is also not age-sensitive, and therefore can be used to efficiently flag 0.5 Msun dwarfs along any Galactic sightline with low extinction. A WFIRST two-stage survey will first establish the IR color-magnitude relation and the dependency of the "kink" on metallicity through high-resolution, deep imaging of Galactic star clusters. Second, this relation can be applied to field studies to characterize the stellar mass function along different sightlines, the dependency of the mass function on environment, and to push to near the hydrogen burning limit in stellar populations out to 10's of kpc.

### Key Requirements
Depth – Well dithered exposures extending down to the H burning limit in clusters with [Fe/H] = -2.2 to 0.0 (i.e., 10 kpc)
Field of View #1 – Single pointings for globular clusters covering appreciable spatial extent
Field of View #2 – Wide field survey of Galactic plane sampling over star forming regions and spiral arms
Cadence – One image per galaxy
Wavelength Coverage – Two NIR filters for color-magnitude diagram analysis

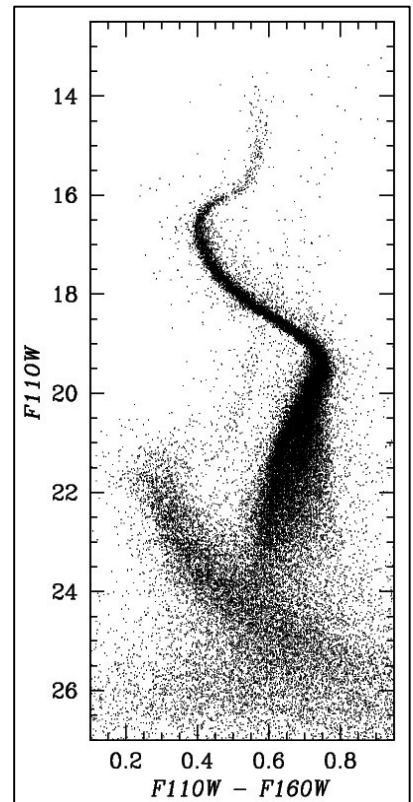

Caption: IR color-magnitude diagram for the nearby globular cluster 47 Tuc, constructed from a 3 orbit (depth) observation with HST/WFC3/IR (Kalirai et al. 2012). The kink in the lower main-sequence of the cluster is caused by $H_2$ opacity. The fainter main-sequence represents stars from the background SMC galaxy.

David R. Ardila (NHSC/IPAC), ardila@ipac.caltech.edu

# Finding the Closest Young Stars

### Background

Young, nearby stars provide us with unique laboratories to study stellar evolution and planet formation. Suspected planets in their surroundings will be hot and inflated and therefore easier to image that in older systems. Since the identification of TW Hya as a nearby (50 pc), young (10 Myrs) T Tauri star (Rucinski & Krautter 1983), about 100 stars within 100 pc have been classified as young (<100 Myrs). The co-moving, co-eval associations to which they belong are extended and sparse, which makes them difficult to identify. While determining the age of a single star is subject to large systematic uncertainties, placement of the group members on a color-magnitude diagram provides a robust way to determine their age.

### The Investigation

There are two components to this investigation: completing the low-mass population of young nearby associations and finding new associations. The first part involves imaging known nearby associations: the β Pictoris moving group, the ε and η Chamaeleontis clusters, the TW Hya association itself, the Tuc/Hor, Columba, AB Doradus, and Argus Associations. These all range in ages from 5 to 50 Myrs. For the second component, we will select known isolated young stars to search for low-mass stars with their know space motions. Determining membership to the association requires determination of the UVW velocity components and medium resolution spectroscopy to search for youth indicators. The 2.4m WFIRST will be able to detect all the Y0 ($M_H$=20 mag, Kirkpatrick et al. 2012) dwarfs within 100 pc ($m_H$(AB)≈26.3 mag). The role of WFIRST is to determine proper motions of very low-mass objects (best done in the NIR, where they are brightest) in an efficient way, given its large field of view. Repeated visits (at least three and preferably five) and ground-based spectroscopic follow-up will be necessary.

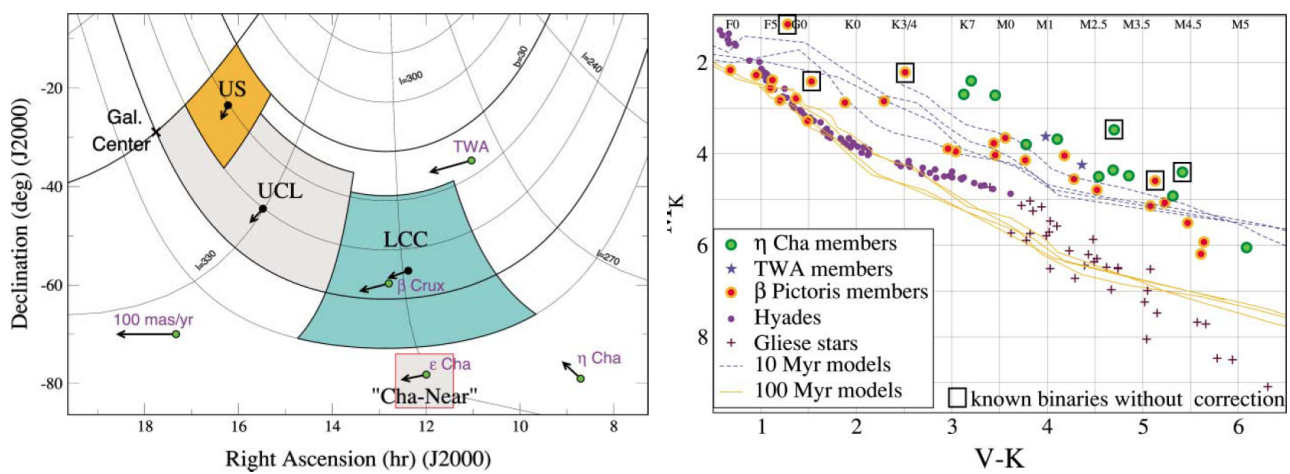

Fig. 1: Left: Nearby southern young associations. Right: CMD diagram for associations nearby the sun. From Zuckerman & Song (2004)

### Key Requirements

Depth – 26.3 AB mags in H to reach all Y0s within 100 pcs
Field of View – 0.375 square degrees (DRM1)
Cadence – At least three epochs, separated by one year
Wavelength Coverage – YJHK

Frantz Martinache (Subaru Telescope) frantz@naoj.org

# Super-resolution imaging of low-mass stars with Kernel-phase & precision wavefront calibration with Eigen-phase

## Background

In the high contrast regime or at very small angular separations, traditional image analysis techniques fail to interpret some of the content, as faint companions and structures are buried under diffraction features. In such situations, it is advantageous to adopt an interferometric point of view instead, and examine the Fourier properties of images. With an instrument providing Nyquist-sampled non-saturated images, and when wavefront errors become small (Strehl ~ 50 %), it is possible to extract self-calibrating observable quantities called kernel-phases (a generalization of the notion of closure-phase), which exhibit a compelling property: they do not depend on residual wavefront errors, therefore enabling the detection of features in the super-resolution regime. The technique was already successfully applied on HST/NICMOS archival data, and uncovered previously undetected companions around nearby M- and L-dwarfs in an unambiguous fashion (Martinache, 2010, ApJ, 724, 464; Pope et al, 2013, PASP, submitted). The same linear model that leads to kernel-phases also provides Eigen-phases, which in combination with an asymmetric pupil mask enable high precision wavefront characterization (Martinache, 2013, PASP, submitted), important for coronagraphs and general telescope performance.

## Kernel-phase on NRO telescope

A space-borne telescope is the ideal observatory for the study of cool dwarfs and their companions in the near-to-mid IR. Ground based observations of such objects indeed typically require Laser Guide Star Adaptive Optics (LGSAO) observations, which do not exhibit satisfactory performance yet; and the detection limits are strongly limited by the background, beyond the K-band.

The self-calibration properties of Kernel-phase enable detection of companions and high-precision relative astrometry in a regime of angular separation (< 100 mas in the near-IR) that currently eludes any other imaging technique. Repeated observations allow to fully characterize orbital parameters, which combined with RV or wide field astrometry, provide high precision dynamical masses of these otherwise elusive objects (e.g. Martinache et al, 2007, ApJ, 661, 496; Martinache et al, 2009, ApJ, 695, 1183).

The application of the technique is however not restricted to nearby low-mass stars. Any observation usually performed with high-performance AO, for instance, resolved observations of the Galactic center could benefit from Kernel-phase.

## Key requirements

Sampling: PSF should be Nyquist sampled at the shortest wavelength of interest

Wavefront quality: Strehl better than 50 % at the shortest wavelength of interest

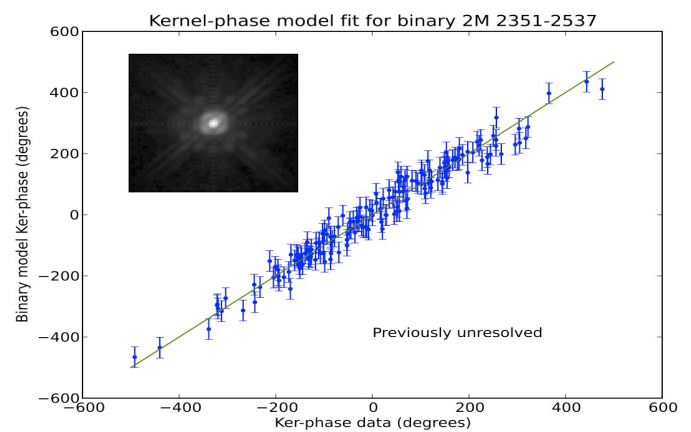

Example of NICMOS Ker-phase super-resolution (64 milli-arc second separation) discovery around the nearby L-Dwarf 2M 2351-2537 (Pope et al, 2013).

Kailash Sahu (STScI), ksahu@stsci.edu

**Detecting Neutron Stars and Black Holes through Astrometric Microlensing**

Stars with initial masses of ~8–20 M$\odot$ are expected to end their lives as NSs, and those with greater masses as BHs (Heger et al. 2003). As much as 10% of the total mass of our Galaxy should be in the form of NSs and BHs (Oslowski et al. 2008). However, a vast majority of stellar remnants are expected to be single, either primordially or due to disruption of binaries by supernova (SN) explosions (Agol & Kamionkowski 2002). Such isolated massive remnants are extremely difficult to detect directly, and in fact no isolated BH has ever been unambiguously found within our Galaxy. Mass measurements of NSs and BHs mostly come from observations of binary systems which show NS masses to be concentrated around 1.4M$\odot$, and BH masses to be a narrow Gaussian at 7.8 ± 1.2 M$\odot$ (Ozel et al. 2010). However, theoretical models (Fryer & Kalogera 2001) predict that the compact-remnant distribution should be a continuous distribution from 1.2 to 15M$\odot$. The discrepancy between the observed and predicted mass distribution of BHs is generally attributed to the fact that, BHs in binaries are a biased and minority sample. What is missing is an unbiased mass distribution for isolated stellar remnants.

**Astrometric Microlensing:** Isolated stellar remnants can be detected through microlensing. The microlensing survey programs such as OGLE and MOA have so far detected more than 8000 microlensing events. If stellar remnants constitute a few percent of the total mass, many of these observed microlensing events must be due to stellar remnants. However, microlensing light curves are degenerate with respect to the mass, velocity and distance of the lens. A route to resolving these degeneracies arises from the fact that microlensing, in addition to amplifying the brightness of the source, produces a small shift in its position (Dominik and Sahu, 2000). Thus if high-precision astrometry is added to the photometry, the deflection of the source image can be measured, and thus the mass of the lens determined unambiguously. The sizes of the astrometric shifts, however, are such that they require a 2m-class space based telescope.

**WFIRST/NRO:** Astrometric precisions of order 300 microarcsec have been achieved with HST/WFC3, so we expect that a similar astrometric precision is achievable with NRO. The maximum astrometric deflection caused by a 0.5 M$\odot$ lens at 2 kpc is 400 μas (see Figure 1). This implies that NRO can measure the astrometric deflection caused by almost all NSs and BHs.

In order to unambiguously measure the mass of the lens, the remaining quantity to be determined is the distance to the lens. Fortunately, the timescales of microlensing events caused by NSs and BHs are long ( >30 days), so photometric observations of these events with NRO will provide clear measurements of the parallax distances.

From our HST observations towards the galactic bulge, we estimate that NRO will enable monitoring of about 20 million stars towards the Galactic bulge in a single pointing, covering about 0.25 square degrees. By monitoring 10 fields in the Galactic bulge, 200 million stars can be easily monitored continuously. Since the optical depth towards the Galactic bulge is about 3 * 10(-6), this will lead to the detection of 600 microlensing events at any given time, and 5000 events per year. Taking the current statistics of microlensing events, about 10% of them (500 events) are expected to have T$_E$ > 30 days, and at least three dozen of them are expected to be due to NSs and BHs. Observations taken over 5 years will lead to the detection and mass measurement of well over 100 NSs and BHs. These measurements will provide (i) constraints on SN/GRB explosion mechanisms that produce NSs and BHs, and (ii) a quantitative estimate of the mass content in the form of stellar remnants.

Scott Gaudi (OSU), gaudi@astronomy.ohio-state.edu, Matthew Penny (OSU), Naoteru Gouda (NAOJ), Yoshiyuki Yamada (Kyoto U.)

# Proper Motions and Parallaxes of Disk and Bulge Stars

## Background

Measurements of the kinematics and three-dimensional structure of stars in the Galactic Bulge and inner disk allow for the determination of the dynamical mass in these populations, and provide important clues to their formation and evolutionary history. The dominant formation mechanism of bulges in the universe (i.e., secularly-grown pseudobulges versus merger-driven classical bulges) remains poorly understood (i.e., Kormendy & Kennicutt 2004), and our bulge provides the nearest and thus most accessible example with which to test the predictions of various formation models. Furthermore, the above measurements can provide useful information with which to clarify the evolutionary processes of the supermassive black hole at the Galactic center and its relation to the evolution of the Galactic bulge. Nevertheless, there are a number of challenges to obtaining such measurements for the bulge, in particular the small proper motions and parallaxes, and the large and variable extinction towards the bulge. To date, information on the kinematics of bulge stars has been limited to primarily radial velocities, luminous stars or stars with large proper motions, or a few deep but narrow pencil-beam surveys with HST. Direct geometrical distances to individual stars in the bulge have been essentially unavailable.

## WFIRST-AFTA

WFIRST-AFTA will enable high-precision proper motion measurements and parallaxes to essentially all the ~56 million bulge and foreground disk stars in the ~2.8 square degree microlensing survey field-of-view with magnitudes of $H_{AB}$<21.6. For the exoplanet survey, WFIRST-AFTA will achieve photometric precisions of ~1.2% per 52s observation for stars with $H_{AB}$<21.6. Assuming a resulting astrometric precision of a $\sigma_{AST}$ ~ 2 mas per observation (i.e., $\sigma_{AST}$ ~ FWHM/SNR~0.14"×0.012~1.7 mas), and N~33,000 observations, the final mission uncertainty on the measured proper motions over a T~5 year baseline will be $\sigma$ ~ $(12/N)^{1/2}(\sigma_{AST}/T)$ ~ 0.01 mas/yr. The typical proper motion of a star in the bulge is µ~100 km/s/(8000 pc) ~ 3 mas/year, and thus individual stellar proper motions will be measured to ~0.3%. Similarly, the fractional uncertainty on the parallax of a star in the bulge will be $\sigma_\Pi/\Pi$ ~ $(\sigma_{AST}/\Pi)(2/N)^{1/2}$ ~ 10%, where $\Pi$ ~ 1/8 mas is the typical parallax of a star in the bulge. Note that these estimates assume that systematic uncertainties can be controlled to better than ~0.01 mas, or 1/1000th of a pixel. Because these observations will be taken in the NIR, they will reach below the bulge MS turnoff and will be relatively unaffected by extinction. The occasional observations in bluer filters will help distinguish between bulge and disk populations. With these measurements, AFTA-WFIRST will provide unprecedented measurements of the kinematics and structure of the bulge.

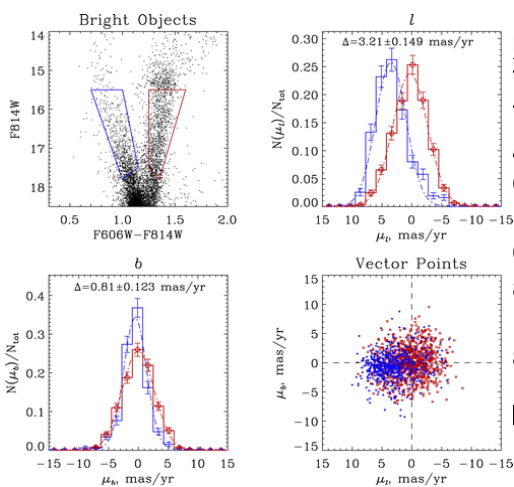

## Key Requirements

Total Span of Observations: T~ 5 years
Total number of epochs: ~33,000
Astrometric precision per epoch for $H_{AB}$<21.6 of ~2 mas
Control of systematic errors to better than ~0.01 mas

Caption: CMD and bulge, disk proper motion distributions from an HST study using ASC/WFC (Clarkson et al 2008, ApJ, 684, 1110). WFIRST-AFTA will provide individual proper motions that are roughly an order of magnitude more accurate than HST.

Daniel Stern (JPL), daniel.k.stern@jpl.nasa.gov

# Quasars as a Reference Frame for Proper Motion Studies

## Background

Luminous quasars behind the Magellanic Clouds or the Galactic bulge provide several unique and important scientific applications. First, they provide the best reference frame for proper motion studies (e.g., Kallivayalil et al. 2006a,b; Piatek et al. 2008). For example, recent improvements measuring the proper motion of the Magellanic Clouds all relied on Hubble observations of fields centered on quasars. The results were surprising, as the Clouds were found to be moving significantly faster than previous estimates (e.g., van der Marel et al. 2002). The tangential motion of the SMC was also found to differ from that of the LMC. This implies that the Clouds may not be bound to each other or the Galaxy, and may instead be on their first pericentric passage. Precisely measuring the proper motion of the Magellanic Clouds will also improve modeling of the Magellanic Stream, which is a sensitive probe of the Galactic potential. Finally, bright background quasars are useful background probes for absorption studies of the interstellar medium.

## WFIRST

The deep, wide-field, high-resolution imaging capabilities of WFIRST will provide a significant improvement for this science, vastly improving the statistics relative to previous Hubble studies. The precise astrometry afforded by space-based observations, combined with the large number of background AGN identified from their broad-band spectral energy distributions (SEDs), including mid-IR data (e.g., Kozlowski et al. 2011, 2012), will allow large numbers of both AGN and Galactic/Magellanic Cloud stars to be identified in each field.

## Key Requirements

Depth – To provide high S/N detections of large numbers of AGN

Morphology – To precisely measure quasar positions

Grism – For spectroscopic confirmation of quasar candidates

Field of View – Wide-area to improve statistics

Wavelength Coverage – single band sufficient

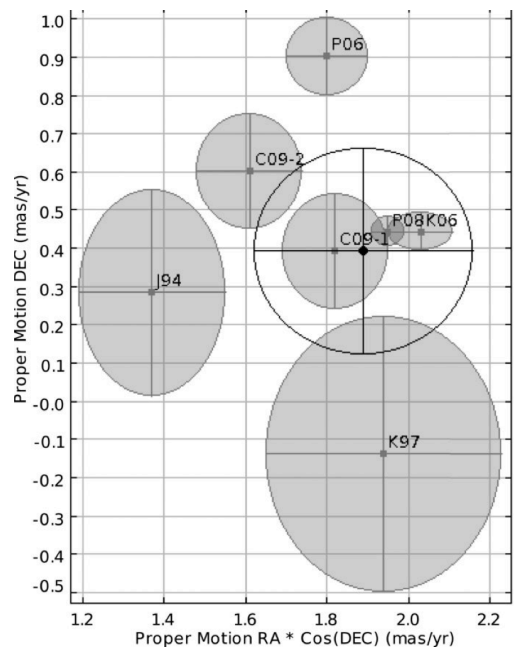

Caption: Recent measurements of the proper motion of the Large Magellanic Clouds, from Vieira et al. (2010).

Gurtina Besla (Columbia) gbesla@astro.columbia.edu; Roeland P. van der Marel (STScI) marel@stsci.edu

**The Detection of the Elusive Stellar Counterpart of the Magellanic Stream**

Background

As the most massive satellites of the Milky Way (MW), the Large and Small Magellanic Clouds (LMC and SMC) play an important role in our still developing picture of the buildup and evolution of the Local Group. In the ΛCDM paradigm, MW-type halos are expected to have built up the majority of their mass by the accretion of LMC-type subhalos; *the evolution and disruption of the Magellanic System is thus directly relevant to our understanding of how baryons are supplied to the MW.*

The Magellanic Clouds (MCs) are undergoing substantial gas loss, as is evident by the stream of H I trailing (the Magellanic Stream) and leading the MCs (the Leading Arm), in total stretching 200° across the sky (Mathewson et al. 1974; Nidever et al. 2010). Despite extensive modeling and multi-wavelength studies of the system, the dominant mechanism for the formation of this extended gas distribution is unknown. Leading theories are: tidal stripping of the SMC (by the MW, Gardiner & Noguchi 1996; or by the LMC alone, Besla, Hernquist & Loeb 2012; see Figure) or hydrodynamic processes (ram pressure stripping, Mastropietro et al. 2005; stellar outflows, Nidever et al. 2008). *Distinguishing between these formation scenarios is critical to the development of an accurate model for the orbital and interaction histories of the MCs with each other and with the MW.*

In the tidal models, stars are removed in addition to gas. In contrast, stars are not removed in any of the hydrodynamic models. *The detection of stellar debris in the Magellanic Stream would prove conclusively that the Stream is in fact a tidal feature, ruling out models based on hydrodynamic processes.* Although previous searches for stars in sections of the Stream have yielded null results (e.g., Guhathakurta & Reitzel 1998, Bruck & Hawkins 1983), the predicted optical emission from the stellar debris in e.g., the Besla, Hernquist & Loeb 2012 model ( >32 mag/arcsec[2]; V-band ), is well below the observational limits of these prior surveys.

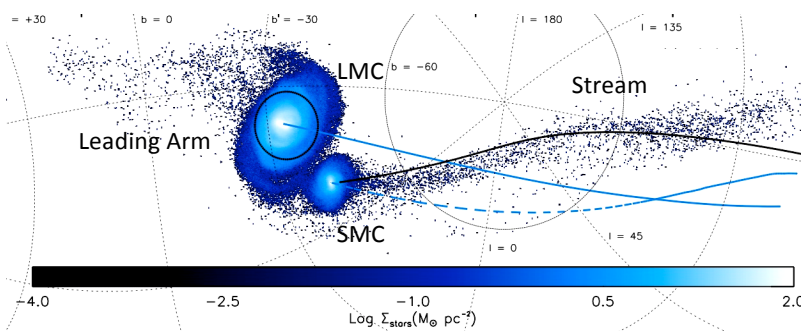

Caption: The stellar counterpart to the Magellanic Stream from Besla, Hernquist & Loeb (2012); color indicates stellar density. Galactic coordinates are overplotted. The past orbits of the MCs are indicated by the blue lines. The current location of the gaseous Stream is marked by the solid black line. Densities < 0.01 M⊙/pc[2] are typical.

WFIRST

WFIRST will enable a uniform, wide-field, deep survey across sections of the gaseous Magellanic Stream to ascertain the existence of a faint stellar counterpart. Using MC main sequence stars as tracers of stellar debris has been proven to reach equivalent population surface brightness of >33 mag arcsec[−2] based on CTIO Mosaic-2 imaging (Saha et al. 2010); with WFIRST's wider field of view, larger surface brightness limits can be reached. Furthermore, a wide area survey will uncover any spatial offsets between the gaseous and stellar stream, which may occur owing to hydrodynamic gas drag.

Key Requirements

Depth – NIR mag limit of >23.5 (to reach I-band surface brightness >33 mag/arcsec; cf. Saha et al. 2010)
Field of View – contiguous mapping across the Stream (~ 20 degrees)
Wavelength Coverage – Two NIR filters for color-magnitude diagram analysis

Marla Geha (Yale University), marla.geha@yale.edu

**Near-field Cosmology: Finding the Faintest Milky Way Satellites**

Background
Since 2005, fourteen dwarf galaxy satellites have been discovered around the Milky Way, more than doubling the satellite population. These 'ultra-faint' galaxies are the least luminous and most dark matter-dominated galaxies in the known Universe. However, the census of Milky Way satellites is far from complete. Accurately estimating the total number of luminous satellites is key to understanding the 'Missing Satellite Problem', and constraining galaxy formation models on the smallest scales.

All of the ultra-faint galaxies have so far been discovered in the SDSS as slightly statistical over-densities of resolve stars. Due to the SDSS magnitude limits, less than 1% of the total Milky Way volume has been searched for these faint galaxies. Various assumptions for incompleteness suggest that the Milky Way may contain anywhere from 60 to 600 luminous satellites out to 300 kpc (Tollerud et al. 2008). Given even optimistic predictions, we expect one Milky Way satellite per 100 square degrees.

Detecting ultra-faint galaxies with WFIRST
Detecting ultra-faint galaxies throughout the full Milky Way volume requires imaging down to the main sequence turnoff at 300 kpc (Walsh et al. 2008), corresponding to roughly r = 27 mag. High spatial resolution is critical, as background galaxies which are otherwise unresolved from the ground, vastly overwhelm the number of stars at these magnitudes. Star-galaxy separation will severely limit the number of distant Milky Way satellites found in ground-based surveys. A deep wide-field WFIRST survey would allow an unprecedented search for the faintest dwarfs to the outer limit of the Milky Way halo.

Characterizing ultra-faint galaxies with WFIRST
In addition to finding ultra-faint galaxies, WFIRST is well positioned to characterize these galaxies. High-precision photometry of the stellar main-sequence turnoff will yield accurate ages for these systems which appear to have exclusively old stars (e.g., Brown et al 2012). Furthermore 80 microarc/year proper motions will allow *internal* measurements of motion in these systems, allowing estimates of dark matter mass and density (i.e., <2 km/s for 50 stars @ 100 kpc, in 3 years).

Depth – Deep point sources imaging to the main sequence turnoff at 300 kpc ($m_r \sim 27$)

FOV – Wide field imaging (1000+ sq deg) to cover large volume of Milky Way halo

Cadence – Repeat observations for proper motions

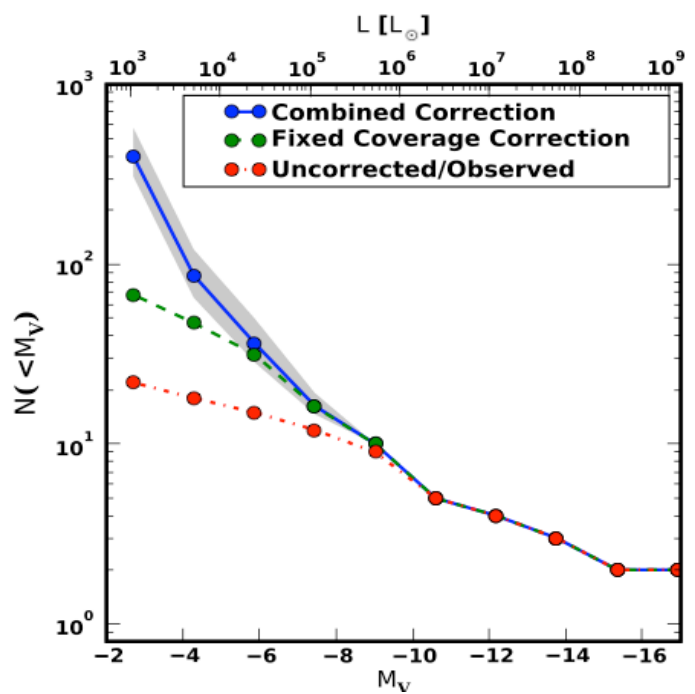

Caption: The luminosity function of known Milky Way satellites (red) as a function of luminosity. Green and blue curves are the predicted satellite numbers based on different assumptions for incompleteness (taken from Tollerud et al. 2008).

Appendix D: GO and GI One Page Science Ideas

Alis J. Deason (UCSC) alis@ucolick.org; Kathryn V. Johnston (Columbia);
David N. Spergel (Princeton University); Jay Anderson (STScI)

# The Mass of the Milky Way

## Background

The mass of our Galaxy is a fundamental --- yet poorly constrained --- astrophysical quantity. Several attempts have been made to measure the total mass of the Milky Way using kinematic tracers, the orbits of the Magellanic Clouds, the local escape speed and the timing argument. The results of this extensive list of work are distressingly inconclusive with total masses in the range 0.5-3 x $10^{12}$ Msol. This somewhat confused picture is partly caused by the difficulties of the task. Full kinematic analyses of tracer populations are hampered by small sample sizes and lack of complete information on the phase-space coordinates. However, the dearth of distant halo tracers is now beginning to be rectified. Deason et al. (2012) have compiled a sample of stellar halo stars, with measured line-of-sight velocities, out to r ~ 150 kpc. This project is ongoing, and it is anticipated that the number of distant halo stars tracers will substantially increase in the next few years. However, regardless of the number of halo tracers, the well-known *mass-anisotropy* degeneracy still plagues any mass estimate. At large distances in the halo, our measured line-of-sight velocity closely approximates the radial velocity component, so we have very little handle on the tangential motion of the halo stars.

## WFIRST

The WFIRST high latitude survey (2500 deg$^2$) will achieve an absolute proper motion accuracy of 80 mu as/yr for a carbon giant at r ~ 100 kpc (J ~ 16). This corresponds to an uncertainty of ~ 40 km/s in tangential velocity, which is below the expected velocity dispersion at these distances. This will allow us, for the first time, to break the mass-anisotropy degeneracy in the distant halo. Furthermore, with full 3D velocity information we will be able to confirm or rule out any associations between random halo tracers, which is a key assumption in any dynamical analysis.

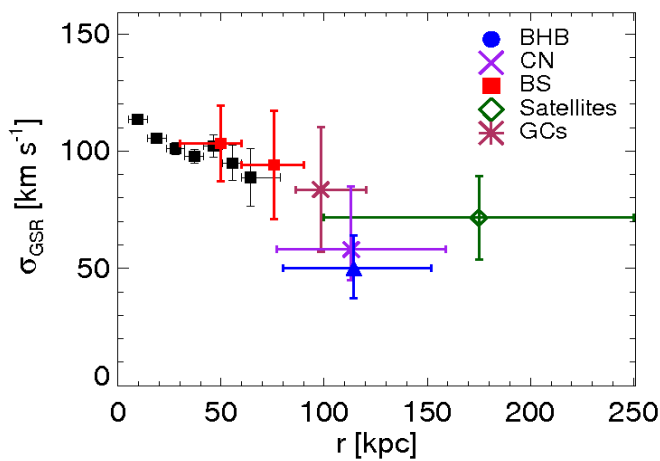

## Key Requirements

High latitude survey (2500 deg$^2$) needed for good statistics of distant tracers. The anticipated sample size (approx. several tens) will be comparable to current samples of distant halo tracers (r > 50 kpc), with radial velocity measurements.

Caption: Radial velocity dispersion of halo stars (adapted from Deason et al. 2012). At large distances the dispersion drops remarkably: the implications for the MW mass depend strongly on the (as yet unknown) velocity anisotropy.

Louis E. Strigari (Stanford University), strigari@stanford.edu
Kathryn V. Johnston (Columbia University), David N. Spergel (Princeton University),
Jay Anderson (STScI) jayander@stsci.edu

# Distinguishing between cold and warm dark matter with WFIRST

## Background

The theory of cold dark matter predicts that the central density profiles of dark matter halos are cusped, rising like 1/r towards the centers of galaxies. Warm dark matter theories, on the other hand, predict that halos are much less dense in their centers. In order to distinguish between these theories, and place strong constraints on the nature of particle dark matter, it is essential to measure the density profiles of the most dark matter-dominated galaxies in the universe.

The dwarf spheroidal (dSph) satellite galaxies of the Milky Way are particularly interesting systems in which it is possible to probe the nature of dark matter. They are near enough that it is possible to measure the velocities of individual stars, and from these measurements extract the dark matter mass profiles. In recent years, through high-resolution multi-object spectroscopy the line-of-sight velocities of hundreds of stars have been measured, providing important measurements of their integrated dark matter mass profiles within their half-light radius. Though there are some hints that their dark matter distributions may be less cuspy than is predicted by cold dark matter theory, precise measurements of their central dark matter mass profiles have remained elusive. Because line-of-sight velocities provide only one component of the motion of a star, there is a well-known degeneracy between the central slope of the dark matter mass profile and the velocity anisotropy of the stars. The only way to break this degeneracy is to measure stellar proper motions in dSphs.

## WFIRST

Because of its pointed nature and large field-of-view, WFIRST will play a key role in studying the dark matter distributions of dSphs. For example, Draco and Sculptor are at a Galactocentric distance of 80 kpc. At this distance, a tangential velocity of 10 km/s corresponds to a proper motion of 27 micro-arcseconds per year. The figure illustrates the projected uncertainty on the log-slope of the dark matter density profile of a dSph at a distance of 80 kpc, as a function of the number of stars with measured proper motions for different assumed errors for the transverse velocities. These simulations assume that the dSph has 1000 measured line-of-sight velocities, in addition to the indicated number of proper motions. The addition of proper motions clearly improves the constraint on the central dark matter density profile, and will ultimately provide a valuable method to rigorously distinguish between cold and warm dark matter candidates.

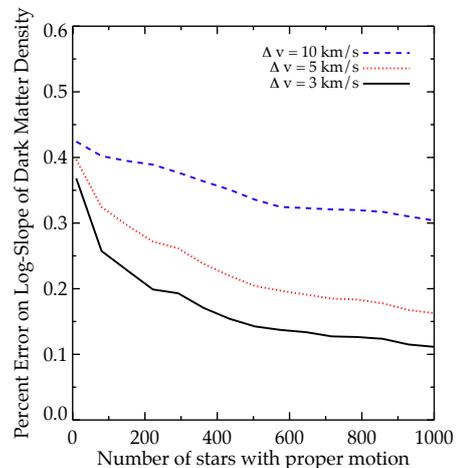

## Key Requirements

Astrometry figures suggest that 10 6-minute integrations over 5 years would achieve 90 micro-arcsecond/year accuracy for positions of the brightest stars in these objects. This survey would require additional visits to increase the accuracy to levels where the slope could be measured.

Kathryn V. Johnston (Columbia University) kvj@astro.columbia.edu
David N. Spergel (Princeton University); Jay Anderson (STScI)

**Finding (or Losing) those Missing Satellites**

Background

Within the standard cosmological context, the Milky Way's dark matter halo is thought to be embedded with thousands of dark matter subhalos with masses greater than $10^6 M_\odot$. If such a population exists, its influence is expected to be detectable in the form of gaps in tidal globular cluster tidal streams such as those associated with GD-1 (d=10kpc) and Pal 5 (d=20kpc). Indeed, current photometric studies suggest there may be gaps in Pal 5's streams. An alternative explanation for these features in streams is that they are a natural consequence of the clumping of debris along the orbit. These scenarios may be distinguished – and hence the presence of "missing satellites" either proved or ruled out – by mapping the debris in phase-space. Stars on either side of gaps caused by encounters with subhalos should exhibit sharp discontinuities in energy and angular momenta, and can potentially deviate beyond the spread expected for tidal stripping alone. In contrast, if the gaps are simply a natural clumping, the phase space structure should be more continuous and stay within the range expected for tidal stripping.

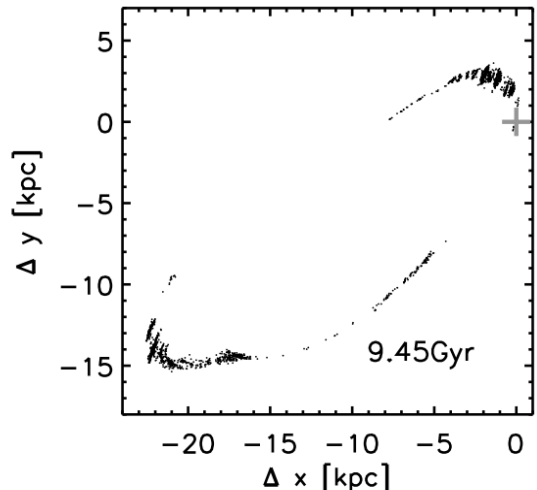

Caption: End point of a simulation of a Pal 5-like stream bombarded by the expected subhalo population. Cross indicates where the main body (not shown) would be.

WFIRST

WFIRST can contribute uniquely to this goal by virtue of its pointed nature and large field of view, which will allow proper motions of stars in selected fields in streams to be probed. With 10 6-minute integrations spread over 5 years, proper motions for dwarfs in globular cluster streams (such as those associated with GD1 and Pal 5) could be assessed with few km/s accuracies. This is only a little larger than the expected dispersion of stream stars, and comparable to the velocity offsets expected from the satellites original orbit. Hence any differences in orbital properties for stars around a gap beyond this range could be assessed by averaging properties on either side.

Kathryn V. Johnston (Columbia University) kvj@astro.columbia.edu
David N. Spergel (Princeton University); Jay Anderson (STScI)

**Mapping the Potential of the Milky Way with Tidal Debris**

Background

The disruption of satellite stellar systems around the Milky Way has left tidal debris in the form of streams of stars, spanning tens of degrees (in the case of globular clusters) or even entirely encircling the Galaxy (in the case of the Sagittarius dwarf galaxy). By virtue of their origin, we know more about samples from these structures than we do about random samples – we know that each sample consists of stars that were once apart of the same object. We can exploit this knowledge to use tidal debris as much more sensitive probes of the Galactic potential than possible with a purely random sample. Moreover, while debris structures have been found at small (~5kpc) and large (~100kpc) distances from the Sun, they are more apparent and longer-lived in the outer halo of the Galaxy – precisely the region where we know least about the potential. Using stellar streams, the Milky Way is the one galaxy in the Universe where we can hope to map the 3-D shape, orientation and mass of a dark matter halo at all radii within the virial radius.

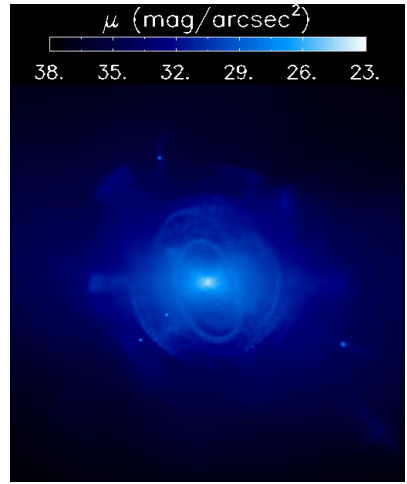

Caption: External view of simulated debris around a galaxy. Box is 300kpc on each side.

WFIRST

WFIRST can contribute uniquely to this goal by virtue of its pointed nature and large field of view, which will allow proper motions of stars in selected fields in streams to be probed even at large Galactocentric radii. With 10 6-minute integrations spread over 5 years, proper motions for dwarfs in the globular cluster streams (such as those associated with GD1 and Pal 5) in the inner halo (< 20kpc) and for giants in satellite debris (e.g. Sagittarius and the Orphan Stream) in the outer halo (> 50 kpc) could be estimated with few km/s and tens of km/s accuracies respectively. These numbers are only a little larger than the expected dispersion of stream stars, and comparable to the velocity offsets expected from the satellites original orbit. Hence, averaging over the debris stars in several fields along each stream would produce an accurate map of stream centroids in phase-space and at different orbital phases – something beyond the reach of current surveys and a key step forward for mapping the Galactic dark matter halo.

Roeland van der Marel and Jason Kalirai (STScI), marel@stsci.edu; jkalirai@stsci.edu

**Dissecting Nearby Galaxies**

Background
Recent wide-field imaging surveys of the Milky Way and M31 (Juric et al. 2008; McConnachie et al. 2009) have enabled a new landscape to test cosmological models of galaxy formation on small scales (Bullock & Johnson 2005). Direct imaging of stellar halos can provide detailed insights on the surface brightness profile, chemical abundance gradients, stellar ages, level of substructure, and quantity of dwarf satellites and star clusters. This input represents a key constraint to unravel the formation and assembly history of galaxies within the hierarchical paradigm (Font et al. 2011), but is limited to the detailed study of just two fully resolved spiral galaxy halos. Current surveys of galaxies outside the Local Group (Spitzer/SINGS, HST/ANGST) have primarily involved either deep pencil-beam probes of narrow fields of view, or wide-field coverage at shallow depth. Similarly, optical ground-based surveys such as SDSS and Pan-Starrs lack the depth to study the tracer population, red giant branch stars, outside the Local Group.

WFIRST
WFIRST will enable a uniform 1.) wide-field, 2.) deep, and 3.) high-resolution study of the full extent of over a hundred nearby galaxies in the Local Volume. Accurate photometric characterization of the top three magnitudes of the red giant branch in each galaxy can map the halo structure, surface brightness profile, substructure content, and metallicity gradient. Variations in these properties, and their connection with the host environment and galaxy luminosity will provide tight constraints to cold dark matter models of galaxy formation.

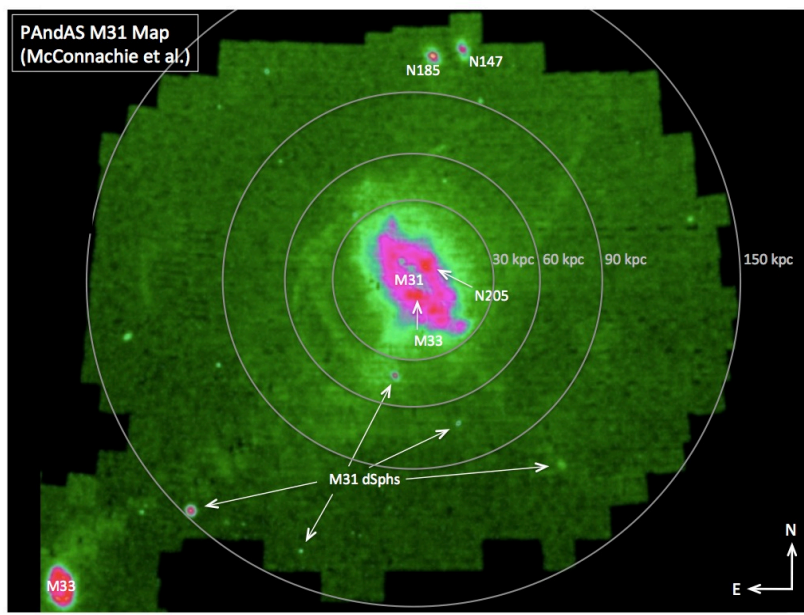

Caption: A (very) wide-field map of M31 from the PAndAS survey (McConnachie et al. 2009) reveals the clearest picture of substructure in a spiral galaxy's halo.

Key Requirements
Depth – Several magnitudes of the red giant branch in ~100 galaxies (i.e., out to d = 5 Mpc)
Field of View – Full halo extent out to 150 kpc (tiling in nearby galaxies)
Cadence – One image per galaxy
Wavelength Coverage – Two NIR filters for color-magnitude diagram analysis

P. Guhathakurta (University of California, Santa Cruz), raja@ucolick.org
R. Beaton (University of Virginia), rlb9n@virginia.edu

# Galaxy Evolution from Resolved Stellar Populations: Halo Age Distributions of the Local Volume

<u>Background</u>: Large galaxies like the Milky Way are formed hierarchically through dozens of satellite accretion events that not only build up the stellar halo, but also contribute to the evolution of the galaxy as a whole. Statistical studies comparing satellite and field dwarfs indicate that star formation is quenched early in the accretion process (e.g., Geha et al. 2012). Thus, the youngest stars in a satellite can be used to date the initial infall of that satellite into its parent halo. Asymptotic giant branch (AGB) stars are prime tracers of these accreted populations as they are present in stellar populations over a wide range of ages (4 Myr to 10 Gyr). As shown in Figure 1 for the LMC, the NIR magnitude and color of an AGB star is strongly dependent on its age. The NIR CMD is therefore an ideal tool for quantifying the stellar age distribution of nearby galaxy halos.

<u>WFIRST</u>: Observational programs of this nature are difficult to carry out from the ground and to date this technique has only been applied to the SMC, LMC and, as a pencil-beam survey in M31. With WFIRST, however, it is feasible to explore the halo age distribution in a representative sample of nearby galaxies to complement Hubble and Spitzer Legacy programs (LVLS, ANGST), that have observed recent and old star formation tracers in all galaxies within 3.5 Mpc and all S/Irr within 11 Mpc. Further, WFIRST will allow an expansion beyond existing HST NIR imaging to obtain full spatial coverage of the M31 disk, bulge and halo. Key questions addressed by these observations will be: (1) what is a 'typical' galaxy's accretion history? (2) over what physical scales does satellite accretion vs *in-situ* star formation dominate? and (3) is the age distribution smooth or discontinuous, and what does this imply about the typical progenitor over time?

<u>Key Requirements</u>
Depth – Reliable well below the TRGB for galaxies over 1-11 Mpc (Ks~22-30)
Spatial Resolution – Morphological distinction between stars and galaxies
Field of View – Wide-field for efficient mapping of nearby galaxy halos
Wavelength Coverage – J,H,Ks to separate AGB stars as in Figure 1

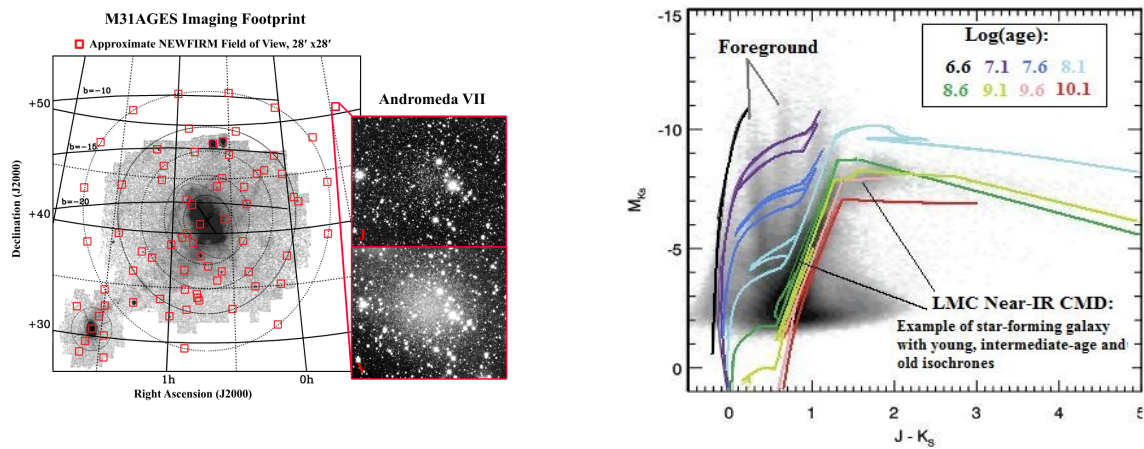

Figure 1.  (Left) An example CMD for the LMC emphasizing the clean separation of AGB stars by age (Martha Boyer, private communication). (Right) The M31 Asymptotic Giant Exploration Survey footprint is the first large scale age mapping project in a large galaxy. With WFIRST it is feasible to complete filled area surveys of this nature for a representative sample of galaxies.

Seppo Laine (SSC, Caltech), seppo@ipac.caltech.edu
David Martinez-Delgado (MPIA), Carl Grillmair (SSC, Caltech), Steven R. Majewski (UVa)

## Substructure around Galaxies within 50 Mpc

### Background

A lot of the attention on galaxy evolution has been focused on interactions between nearby roughly equal mass galaxies (e.g., Toomre & Toomre 1972). However, minor mergers and dwarf galaxy accretion events are far more common, and are critical in the evolution of larger galaxies by building their halos, bulges, bar and spiral and even globular cluster systems. These events may leave relics in the circumgalactic environment in the form of discrete tidal stellar streams and intracluster light. The investigation of such streams and their significance in hierarchical galaxy formation has just begun (e.g., Martinez–Delgado et al. 2010). The prerequisites include deep red visual and near-IR imaging of the peak of the stellar energy distributions of galaxies over large areas.

### WFIRST

WFIRST can probe the circumgalactic environments with sufficient depth and spatial resolution around galaxies up d = 50 Mpc, and with its large field of view it can perform these observations very efficiently. Combined with ground-based surveys in the visual wavelengths, the ages, metallicities and masses of the stellar populations of disrupted companions can be studied for a large sample of galaxies, providing much tighter constraints on the minor merging history in the CDM models of galaxy formation.

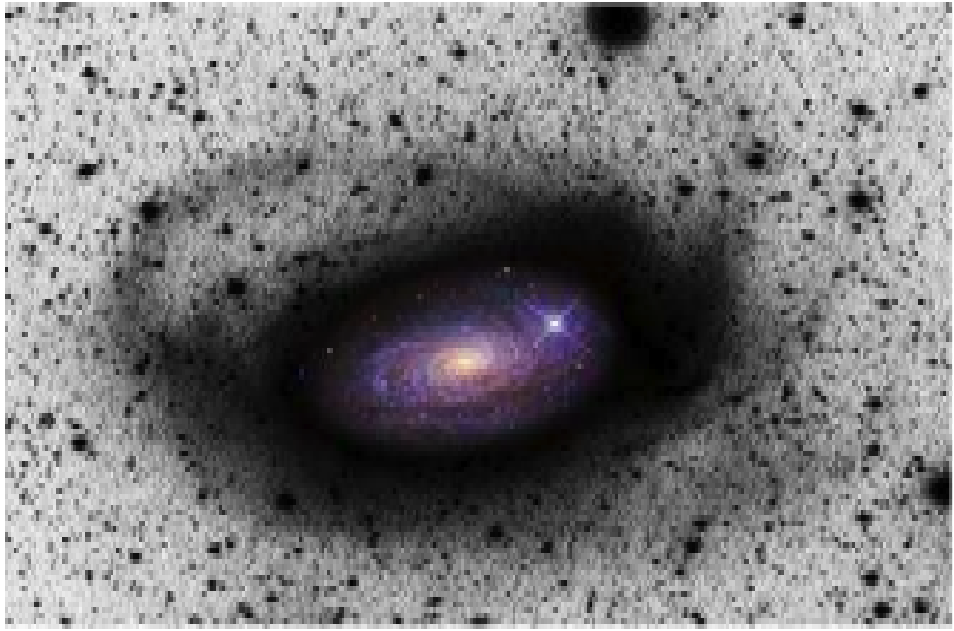

### Key Requirements

Field of View – Mapping the halos out to 100 kpc (35'–7' at 10–50 Mpc)
Depth – 27 mag arcsec$^{-2}$ (AB mag) at 1.2 $\mu$m
Resolution – Resolving the width of features down to 100 pc (2" – 0.4" at 10-50 Mpc)
Cadence – Only one epoch of imaging is required (assuming sufficient depth achieved)
Wavelength Coverage – ~0.7 $\mu$m to (>1.5 $\mu$m) is optimal for C–M diagnostics of stellar populations.

Julianne Dalcanton (University of Washington), jd@astro.washington.edu
**Resolved Stellar Populations in Nearby Galaxies**

Background
From the UV through the near-IR, galaxies' SEDs are dominated by stars. These stars encode a rich record of key astrophysical processes. From the distribution of stars in a CMD, one can map out the history of star formation, which in turn can be used to anchor models of the physics that controls galaxy evolution (i.e., what physical conditions lead to star formation? how does energy from evolving stars affect the ISM? how do evolving stars affect the global SED?), and models for interpreting the observations made of galaxies across cosmic time (i.e., what is the IMF? what timescales and star formation rates are associated with traditional star formation rate indicators? etc). CMDs also contain rich, redundant information about the evolution of the stars themselves, allowing one to constrain the physics of some of the most rapidly evolving, rare stars. Many of these classes of stars, such as AGB and red core Helium burning stars, are incredibly luminous, and can have a significant impact on the overall SED in the near-IR. Unfortunately, the internal evolution of these stars is sufficiently complex that it cannot be derived theoretically from first principles, and instead must be extensively calibrated for a wide range of metallicities and stellar masses. The net result is that stars in nearby galaxies lay the foundation for building an accurate interpretation of extragalactic observations.

WFIRST:
The main factor limiting stellar population studies is angular resolution. With ground-based telescopes, light from individual stars is blended together, prohibiting the study of stars beyond the Local Group. While adaptive optics can alleviate this problem, stellar population studies require a level of photometric accuracy and stability that cannot yet be achieved routinely for long integrations over large fields of view. In contrast, the superb resolution and stability available with a 2.4m space-based telescope allows individual red giant stars to be resolved at distances out to ~5 Mpc in the main body of galaxies, or out to even larger distances in low surface brightness regions (halos, intercluster light, dwarf galaxies, etc). High luminosity stars, which more easily rise above the crowding limit, can also be detected out to greater than ~10 Mpc, allowing the detection of massive main sequence stars, blue and red supergiants, and AGB stars. WFIRST's wavelength coverage is optimized for the cooler of these stars, whose bolometric luminosity peaks longward of 7000 Angstroms; however, many hotter stars can be sufficiently luminous to still be detectable at WFIRST's wavelengths. The near-IR coverage of WFIRST also minimizes the impact of dust on interpretations of CMDs. Existing studies in the Magellanic Clouds (Nikolaev & Weinberg 2000, Boyer et al 2011) have demonstrated that there would be significant added value if WFIRST detectors were sensitive out beyond 1.7 microns, even with the high background expected for a non-cryogenic mission.

Key Requirements
Depth – Less necessary than resolution, given that most stars are in regions that are limited by crowding/confusion rather than photon counting. In uncrowded regions, depth allows detection of fainter, more numerous stars, giving greater sensitivity to stellar halo substructure
Morphology – Resolution absolutely critical to maximizing depth and separating stars/galaxies
Grism – Spectral characterization of AGB subtypes and separation from core He-burning stars
Field of View – Wide-area surveys required to identify rare populations
Wavelength Coverage – minimum of 2 filters. Including an optical filter as far to the blue as possible maximizes sensitivity to stellar temperature, separating features on the CMD. Including a filter longward of H-band increases ability to separate AGB subtypes (O-rich, C-rich, etc).

Roberto Abraham (University of Toronto); Patrick Coté (NRC Herzberg)
roberto.abraham@utoronto.ca

**Resolving Intracluster Light in the Virgo Cluster**

Intracluster light (ICL) is an important tracer of the dynamical state, growth history and structural composition of galaxy clusters. Relevant observables include the fractional luminosity, radial light profile, and presence of substructure in the ICL, all of which are a strong function of cluster mass, substructure, and dark matter distribution (Krick, Bernstein & Pimbblet 2006, Krick & Bernstein 2007, Rudick et al. 2010, Montes & Trujillo 2014). Numerical simulations indicate that the ICL in clusters accounts for anywhere from 10% to 70% of the total cluster luminosity (Purcell et al. 2008, Rudik et al. 2010, Martel et al. 2012, Contini et al. 2014), though this remains largely untested.

In spite of its importance, very little is known about the ICL. It is notoriously difficult to study the ICL by direct imaging, since even the brightest components of the ICL have surface brightnesses around B~28 mag/arcsec$^2$ (Mihos et al. 2005) and the vast majority of the ICL is expected to be orders of magnitude fainter than this. The fundamental limits to low surface brightness imaging are set by control of systematics (scattering, detector stability, and flat fielding errors), so the limiting depth for ground-based low surface brightness imaging (B~29.5 mag/arcsec$^2$; see Slater et al 2009) has remained fairly near the detection threshold of the ICL for the last 40 years. New observational techniques for low surface brightness imaging from the ground (such as the Dragonfly Telephoto Array) hold some promise for allowing the ICL to be probed to around B ~ 32 mag/arcsec$^2$ from the ground (Abraham & van Dokkum 2014), but going much fainter than this will likely remain challenging even a decade from now. The most powerful technique for investigating the ICL in very nearby clusters will therefore be to forego direct imaging, and to instead exploit the resolution of space-based observatories by directly resolving the ICL into its constituent stars. Pioneering observations with HST (Williams et al. 2007) have set the stage for this type of work, but truly transformative observations require the coupling of high resolution with wide areal coverage. As a result, WFIRST seems set to make important contributions to this important area of astrophysics. We can best illustrate the potential of WFIRST by showing how it might be used study the ICL in the Virgo cluster.

To properly populate a color-magnitude diagram we seek to image stars two magnitudes below the tip of the red giant branch. We can estimate the number of such stars in the Virgo cluster in two ways:

1. Using NGC2419 (a very compact globular cluster which has a mass of about $10^6$ M$_\odot$ and about 5 million solar luminosities) as a reference, we count 556 stars within 2 mag of the RGB tip. From this we expect to find $5.56 \times 10^{-4}$ detectable stars per unit solar mass in a low metallicity old stellar population. Assuming Virgo has a total mass of $3 \times 10^{14}$ M$_\odot$ (Ferrarese et al. 2012), and a baryonic M/L of 100, we have a baryonic mass of $3 \times 10^{12}$ M$_\odot$. If 10% of this mass is in the ICL, we have $3 \times 10^{11}$ M$_\odot$ of baryons in the Virgo ICL. The corresponding number of detectable stars within two magnitudes of the tip of the RGB is $1.7 \times 10^8$ stars. This suggests something like 100 million detectable stars within 2 mag of the RGB tip in Virgo. Spreading these over 100 sq. deg. (the area covered by the Next Generation Virgo Survey) results in a space density of $1.7 \times 10^6$ intracluster stars deg$^{-2}$, or about 500,000 per WFIRST field.

2. An alternative approach is to scale from the work of Williams, Ciardullo et al. (2007), who report ~5000 intracluster stars in a deep ACS field (with a field of view of 11.39 arcmin$^2$). Based on their Figure 9, the tip of the RGB is at F814W = 27 mag. Their data goes down to about F814W = 29 mag so they probe about 2 mag below the RGB tip. The space density of intracluster stars in this paper is ~ 450/arcmin$^2$, corresponding to 1.6×10$^6$ intracluster stars deg$^{-2}$, in good agreement with our previous estimate of about half a million targets per WFIRST field within two magnitudes of the RGB tip.

The required integration time can be determined from knowledge of the absolute magnitude of the tip of the red giant branch. Since the ICL is thought to be comprised mainly of stripped stars, we can estimate the magnitude of the RGB as a function of abundance by using two globular clusters with different metallicities, namely ω Cen at [M/H] ≃ -1.5 and 47 Tuc at [M/H] ≃ -0.6. According to Bellazzini et al (2004), the H-band AB absolute magnitude of the RGB tip is -4.55 +/- 0.18 for ω Cen and -4.94 +/- 0.18 for 47 Tuc. At the distance modulus of Virgo (31.1) stars two magnitudes fainter than the tip of the red giant branch would be at $H_{AB}$=28.55 mag based on a ω Cen stellar population, and at $H_{AB}$=28.16 based on a 47 Tuc stellar population. Taking the fainter of these (to be conservative) the HST/WFC3 exposure time calculator indicates that S/N = 5 at $H_{AB}$=28.5 mag would be achieved in about 54,000s with WFIRST. A similar amount of time would be required in order to obtain a comparable depth in at least one other band, for a total of ~100ks. Covering Virgo in its entirety is out of the question (it would mean 360 WFIRST pointings!), but an ambitious program to radially sample the cluster ICL at several positions from the core to the virial radius would be feasible. This would be a very large program, but the legacy value would be enormous.

Chris Mihos (mihos@case.edu) and Paul Harding (paul.harding@case.edu)

**Deep Surface Photometry of Galaxies and Galaxy Clusters**

Background

The low surface brightness outskirts of galaxies hold a wealth of information about processes driving their evolution. Interactions and accretion events in galaxies leave behind faint, long-lived tidal tails and stellar streams which can be used to trace their interaction history (e.g., Martinez-Delgado et al 2010). In galaxy clusters, tidally stripped material mixes to form the diffuse intracluster light (ICL; e.g., Mihos et al 2005), whose structure and luminosity provides constraints on the dynamical history of the cluster (Rudick et al 2009). The photometric properties of the extended disks of spiral galaxies probe star formation in low-density environments (e.g., Bigiel et al 2010), as well as dynamical models for stellar migration in galaxy disks (Sellwood & Binney 2002; Roskar et al 2008). However, accessing this information is quite difficult, as optical surface photometry must reach down to $\mu_V > 27$, or < 1% of the ground-based night sky. Working to an equivalent depth in the near IR is impossible from the ground, since the IR sky is much brighter and significantly more variable than in the optical. Furthermore, this level of accuracy must be achieved over a wide field of view (>~0.5 degree to study nearby galaxies and clusters.

WFIRST

WFIRST can take advantage of the low IR background levels from space to deliver deep, wide-field surface photometry in the near IR. Compared to optical surface photometry, working in the IR has the advantage of maximizing the signal from the old stellar populations that comprise the ICL, providing a more direct tracer of stellar mass at low surface brightness, and minimizing the effects of scattered light from Milky Way galactic cirrus and internal extinction from the target galaxies. Proper baffling and PSF stability will be critical to reduce scattered light and allow for the subtraction of extended stellar wings in the imaging; high spatial resolution will help resolve and reject faint background sources from contaminating the deep photometry.

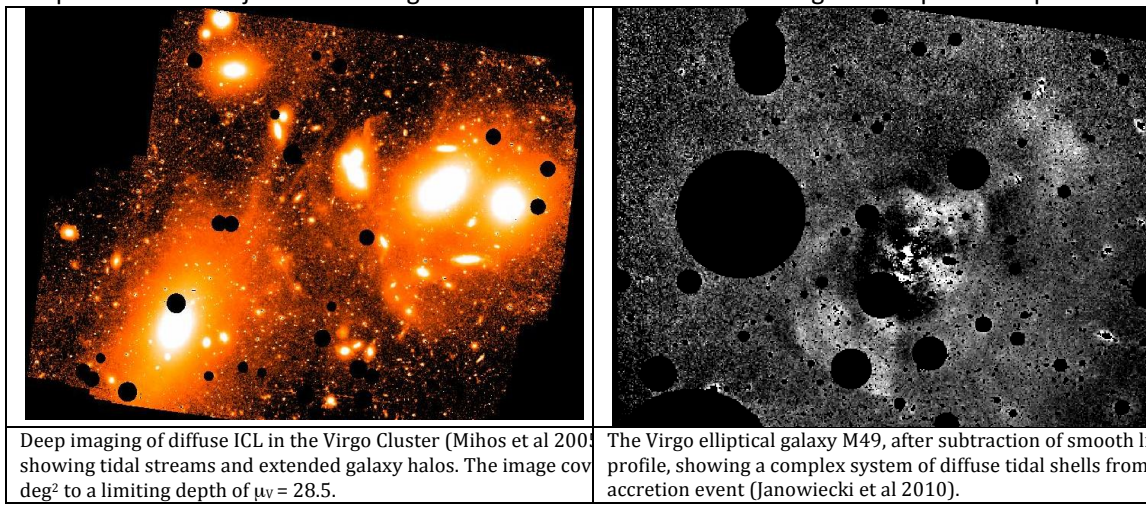

Deep imaging of diffuse ICL in the Virgo Cluster (Mihos et al 200[?] showing tidal streams and extended galaxy halos. The image cov[ers ?] deg² to a limiting depth of $\mu_V = 28.5$.

The Virgo elliptical galaxy M49, after subtraction of smooth lig[ht] profile, showing a complex system of diffuse tidal shells from a[n] accretion event (Janowiecki et al 2010).

Key Requirements

*Depth* – Limiting J-band surface brightness of ~ 27 mag/arcsec² (~ 0.001 MJy/sr)
*FOV* – Wide field necessary to cover nearby galaxies and galaxy clusters
*High spatial resolution* – To reduce photometric contamination from faint background objects.
*Baffling and PSF stability* – To limit scattered light and allow accurate subtraction of extended stellar wings
*Wavelength* – Infrared bands needed to sample peak of old population SED, better trace stellar mass distribution, and minimize contamination due to scattering and absorption from Galactic and target galaxy dust

Christopher J. Conselice (Nottingham), conselice@nottingham.ac.uk

## Galaxy Structure and Morphology

<u>Introduction</u>

There are several major issues within galaxy formation and evolution where WFIRST/NRO telescope will make a big impact.   One of these critical aspects of a WFIRST/NRO mission will be to study galaxy structure - i.e., sizes and morphologies of galaxies on a larger area than is possible with Hubble Space Telescope, or from the ground using adaptive optics.    Galaxy structure is critical for deciphering the processes which drive galaxy assembly and allow us to move beyond simply counting properties of galaxies to study directly their assembly processes through e.g., mergers, gas accretion, and star formation.   Yet this has proved difficult to do for galaxies at z > 1.  Properties such as the asymmetry of a galaxy's light, its concentration, and the clumpy nature of this light all reveal important clues to the galaxy formation process.

<u>WFIRST/NRO</u>

The Hubble Space Telescope has shown how powerful this resolved structural approach is for understanding galaxies. Only very small fields have been observed with Hubble, and most of those within the observed optical. To study the formation of galaxies will require high resolution imaging in the near-infrared, which WFIRST/NRO will provide, especially if a large 2.4m mirror is utilized.  This will bring the pixel size down to 0.13-0.11 arcsec, allowing us to perform WFC3 type surveys over at least a few degrees, which will be much larger than any near infrared survey of this type for distant galaxies yet performed.   This will allow us to directly measure in an empirical way how galaxies, and the stars within them, assembled over most of cosmic history to within a Billion years of the big bang.

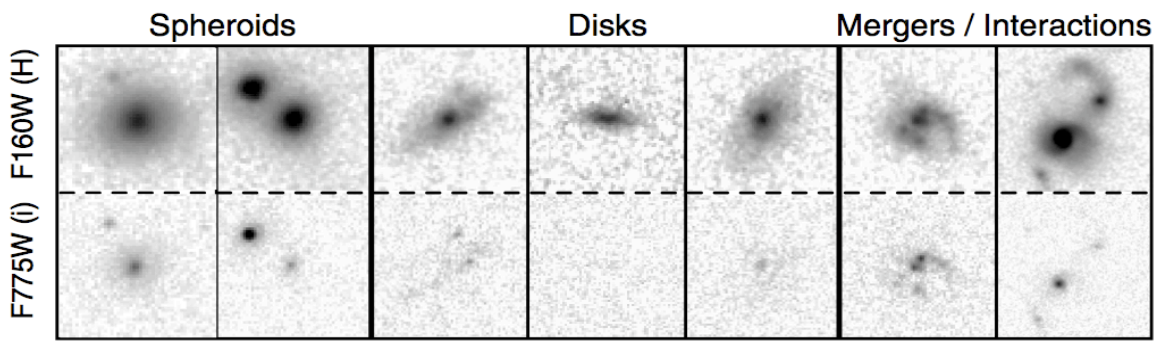

Caption: How infrared light (F160W) shows features of distant galaxies not seen in optical
light (F775W)  (from Kocevski et al. 2011)

<u>Key Requirements</u>

Depth – Deep enough to examine galaxy structure down to ⅓ L* up to z = 4
FoV – Large survey areas to obtain statistically representative galaxy populations
Wavelength – Need > 1.6 microns to examine the rest-frame optical structures of galaxies

Daniel Stern (JPL), daniel.k.stern@jpl.nasa.gov
**Strong Lensing**

Background
Strong gravitational lensing is a remarkable physical phenomenon. Massive objects distort space-time to the extent that light sources lying directly along the line of sight behind them can appear multiply-imaged (see Schneider 2006 for a review). When this rare alignment occurs, we are given: (1) the opportunity to accurately infer the mass and mass distribution of the lensing object; and (2) a magnified view of the lensed source, often probing luminosities and size scales that would not be accessible with current technology. While the approximately 200 galaxy-scale strong lenses we know of currently generally come from low- resolution, ground-based surveys, the bulk of the scientific potential of these rare systems requires (and has relied upon) high-resolution follow-up studies with the Hubble Space Telescope (Fig. 1; e.g., Browne et al. 2003, Bolton et al. 2006). The same has been true for the similar number of cluster-scale strong lens systems (e.g., Smith et al. 2005).

Simply making a precise measurement of lens statistics provides cosmographic information from the total lens counts, the so-called "lens redshift test" (e.g., Capelo & Natarajan 2007). If the lens mass distribution is well-constrained, as in the case of "compound lenses", where multiple sources line up behind the lensing galaxy or cluster, the lens geometry can be used to measure ratios of distances — compound lenses are standard rods for probing the Universe's expansion kinematics with high precision (Golse et al. 2002, Gavazzi et al. 2008). However galaxy-scale compound lenses are rare, typically just 1% of galaxy-scale lenses, and only by imaging a substantial fraction of the sky can we find such golden lenses.

WFIRST
Based on Hubble surveys of a few square degrees, we expect strong lenses to have an abundance of about 10 per sq. deg. (Faure et al. 2008), suggesting that the Hubble-era sample of lenses observed at high resolution will be expanded by several orders of magnitude by the WFIRST weak-lensing survey (e.g., Marshall et al. 2005), the majority of these will be galaxy-scale lenses. The number of observable galaxy-scale lenses is a very strong function of angular resolution: the factor of six degradation in resolution from a diffraction limited 1.5-meter space-based telescope relative to typical ground-based conditions incurs a two order of magnitude decrease in the number of lenses identifiable from the ground.

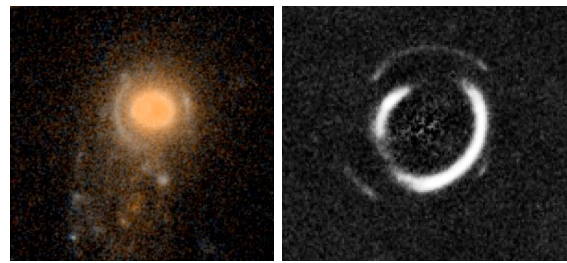

Caption: Strong gravitational lenses imaged with Hubble. WFIRST will increase the number of strong lenses known by a factor of ~100, thus also identifying rare configurations — such as lensed SNe, lenses with higher order catastrophes in their caustics, and compound (e.g., double lens-plane) lenses — which have even richer scientific potential. [From Moustakas et al. 2007 and Gavazzi et al. 2008.]

WFIRST, with its high-resolution and wide-field capabilities, will revolutionize the field of strong gravitational lensing, enabling fundamental, new cosmological and astrophysical probes.

Key Requirements
Depth – To provide high S/N detections
Morphology – To precisely measure positions
Field of View – Wide-area to detect rare events
Wavelength Coverage – Single band sufficient

P. Appleton, J. Rich , K. Alatalo, P. Ogle (Caltech), S. Cales (U. Concepcion) &  L. Kewley (ANU)
apple@ipac.caltech.edu

# Searching for Extreme Shock-dominated Galaxy Systems from 1 < z < 2

<u>Background</u>: The importance of turbulence and shocks at heating gas in galaxies is only just beginning to emerge through both visible IFU imaging and mid-IR molecular hydrogen observations of collisional galaxies, and galaxies with powerful AGN winds. Recent models of fast shocks predict optical emission-line ratios, which overlap the Low Ionization Nuclear Emission Line (LINER) region of the traditional BPT diagnostic diagram (Fig1a).  Amazingly (Fig1b & c) the distinct nature of shocked gas can be diagnosed even when Hα and [NII] are blended—as in the case of a low-res GRISM spectrometer. Although many LINERs are likely to be heated by sub-Eddington accretion onto a black hole, there are clear examples of LINER emission where galaxy-wide shocks strongly dominate the optical spectrum, e. g. NGC 1266 (Alatalo et al. 2012), or the giant shocked filament in Stephan's Quintet (Xu et al. 2003; Appleton et al 2006). The shocks may suppress star formation, as in the case of some radio galaxies, where the jet may drive the heating (Ogle et al 2010; Nesvadba et-al 2011). Even old radio galaxies may still show remnants of shocked gas (Buttiglione 2013). Discovering more "pure-shock" objects at higher z, will allow us to investigate places in the universe where strong turbulence is operating to quench star formation. Finding "nascent" pre-starforming disks in a highly turbulent phase will help us understand how galaxy disks form and evolve (e. g. in cold flows), or how AGN feedback may influence galaxy hosts.

<u>WFIRST:</u> Large galaxy surveys such as SDSS and BOSS have already begun to turn up rare objects in the correct area of phase-space in the local (z < 0.2) universe (see comparison of Fig1a with b). WFIRST's near-IR GRISM capability will allow for a similar exploration at much higher z, where turbulent galaxy disks may be more common. Assuming a GRISM spectrograph with 160 < R < 260 in the wavelength range 1.1 to 2 microns, WFIRST will detect galaxies over a range 1 < z < 2 in the required lines. Because the [OI]/Hα ratio is so extreme in pure shocked systems, it may not even be necessary to detect the [OIII] and Hβ lines to form a sample of extreme shock-dominated systems. Based on simulations and studies using a similar GRISM on HST/WFC3 (WISP), 2000-3000 emission line galaxies/sq. degree would be detectable (> 1 x $10^{-16}$ ergs/s/cm²) over the required range. With large planned deep redshift BAO surveys envisaged in the WFIRST Interim Report covering ~2000-3500 sq. degree of sky/yr, even if only 0.5% of all emission-line galaxies are shocked galaxies (probably conservative given that 7% of SDSS galaxies are LINERs), then we expect to find tens of thousands of shocked galaxies over a few years of operation. These BAO samples will allow us to investigate how these galaxies change with redshift and local environment—for example are they correlated with galaxy over-density or filaments? Significant new ancillary data sets (e. g. deep SKA-precursor radio galaxy surveys) will also be available then, allowing for smaller, deeper, targeted GO searches of specific dense proto-clusters, or quasar environs.

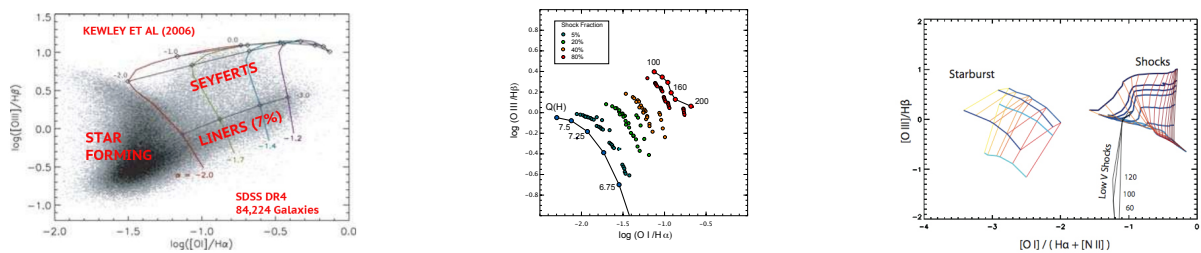

*Fig.1a Classical BPT diagram showing separation of LINERs (7%) from Seyferts and SF galaxies (Kewley et al. 2006). Fig1b. Some LINERS may be created in high-speed shocks (Rich et al. 2012). Models show range of shock velocities and mixing fractions with SF. Fig1c. Even when Hα+[NII] is blended, the shock models separate cleanly from the starburst sequence on the x-axis.*

Key Requirements  for GRISM Survey

Depth :– 1-2 x$10^{-16}$ ergs s⁻¹cm⁻² detection threshold;  Areal coverage:- 2000-3000 deg² (BAO type planned coverage)
Spectral Resolution (R > 150) ;  Wavelength Coverage – 1.1-2 μm

Julian Merten, Jason Rhodes (JPL / Caltech), jmerten@caltech.edu

# Mapping the Distribution of Matter in Galaxy Clusters

## Background

Clusters of galaxies are important tracers of cosmic structure formation. All of their mass components including dark matter, ionized gas and stars are directly or indirectly observable. Furthermore, the complicated effects of baryons are less dominant, although not negligible (Duffy+10). This allows us to directly compare numerically simulated galaxy clusters with real observations. This comparison is usually done via parametrized 1D density profiles, like the NFW profile (Navarro+96); where the distribution of parameters in NFW fits to simulated halos (Bhattacharya+11) and observed halos (Fedeli+12) are compared. However, these comparisons do not make use of the full 2D density distribution as can be inferred from gravitational lensing, X-ray or SZ observations of clusters and which show that individual clusters are usually non-spherical and highly sub-structured. We therefore propose a full 2D, morphological characterization of galaxy clusters by means of e.g. mathematical morphology (Serra+65) or Minkowski functionals (Kratochvil +12). These techniques will be applied to simulated clusters and a large sample of clusters observed with WFIRST. A comparison of the two will give unprecedented insight into the main mechanisms of structure formation.

## WFIRST

Key to such analysis is a map of the mass distribution in galaxy clusters created with the technique of gravitational lensing (Bartelmann11). Methods which combine weak and strong gravitational lensing are particular successful (Bradac+06, Merten+10, Merten+11, Meneghetti+11) and only space-based observations deliver the depth and resolution needed for a detailed reconstruction of a galaxy cluster. In the regime of weak gravitational lensing, space-based observations result in a four or more times higher density of background galaxies, which can be used to map the distribution of matter in the full cluster field. This high density of background objects is key for a reconstruction on high spatial resolution (Massey+12). In the regime of strong gravitational lensing, only the crisp images from space are able to resolve fine structures in strongly lensed galaxies within the full cluster core. These structures are key to trace substructure in gravitational lens systems (Postman+11). In the figure above, we highlight all these requirements with the example of the complex cluster Abell 2744 (Merten+11). The reconstruction in panel (1) is based on ground-based weak lensing and shows only low S/N encoded by the color scale. When adding weak lensing from space in panel (2), the S/N increases significantly in the areas where these observations are available (indicated by the white frames). Adding space-based strong lensing in the panels (3) and (4) provides the resolution needed to compare to a simulation on the the basis of 2D morphology. WFIRST matches all the requirements above and provides the tremendous advantage of a wide FOV, which is similar to the best ground-based telescopes, but combines it with the advantages of a space-based observatory.

## Key Requirements

FOV – to cover the full area of the merger scenario
Depth – to measure shapes of a large number of background galaxies (60-100 arcmin^-2)
Multi-band – to reliably distinguish foreground from background galaxies through photo-z

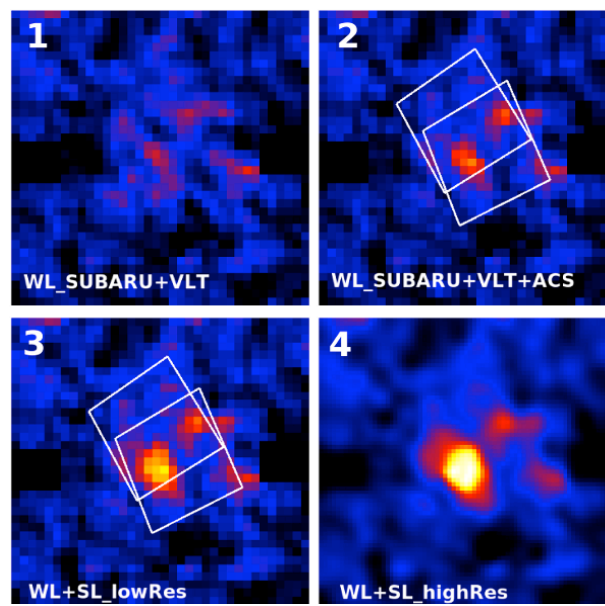

Julian Merten, Jason Rhodes (JPL / Caltech), jmerten@caltech.edu
**Merging Clusters of Galaxies**

Background

Clusters of galaxies are the ideal tracers of cosmic structure formation. They are made of dark matter (~75%), hot ionized gas (~23%) and stars (~2%), and all these fundamental mass components are directly or indirectly observable in the optical, X-ray or radio. Particularly interesting are merging clusters of galaxies, where all mass components are interacting directly during the creation of cosmic structure. Multiple such mergers have been observed, with the Bullet Cluster being the most prominent example (Clowe+06). Paired with follow-up numerical simulations, such systems gave important insight into the behavior of the baryonic component (Springel+07) and set upper limits on the dark-matter self-interacting cross-section (Randall+08), which is of great importance in the search for the nature of dark matter. Recently, more complicated merging systems have been identified (Merten+11, Clowe+12, Dawson+12) offering a great opportunity to boost our understanding of dark matter.

WFIRST

Key in the analysis of a cluster merger is the precise mapping of the matter distribution and finding the exact positions of peaks in the distribution of the different mass components. In the case of dark matter, weak and strong gravitational lensing are the main tools of interest. As has been shown for many merger cases (Bradac+09, Merten+11, Jauzac+12, Clowe+12), space-based observations are required to achieve the necessary spatial resolution. This is true for strong gravitational lensing in order to resolve fine features in the shape of e.g. gravitational arcs (Postman+11) and for weak gravitational lensing where a high effective density of weakly lensed background galaxies is required (Massey+12). Currently, only the HST allows for such studies, but they are limited by its small FOV. The figure below shows an example. The complex merger Abell 2744 is shown on a Subaru/Suprimecam FOV, together with the WFC3/UVIS and ACS footprints (pink). As indicated by the white mass contours, the full merger scenario exceeds the area which was mapped to high resolution shown by the colorful overlay (Merten+11). Mapping the full field of interest with HST would need many separate pointings. But as is indicated by the red frame in the figure above, the full merger field can easily be mapped with a single WFIRST pointing with all advantages of a space-based observation mentioned above.

Key Requirements

FOV – to cover the full area of the merger scenario
Depth – to measure shapes of a large number of background galaxies (60-100 arcmin^-2)
Multi-band – to reliably distinguish foreground from background galaxies through photo-z

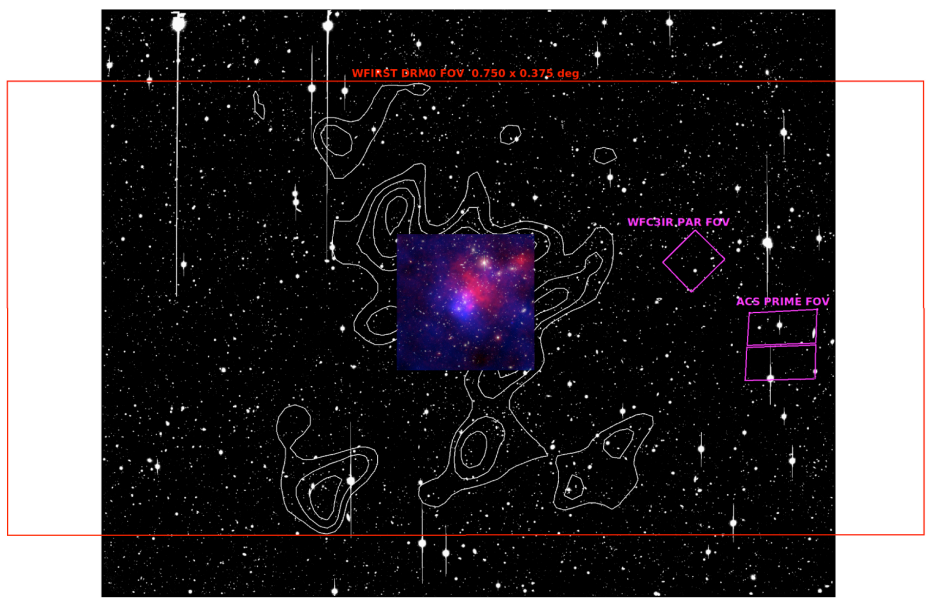

Priyamvada Natarajan (Yale University); priyamvada.natarajan@yale.edu
## Group-Scale Lenses: Unexplored Territory

**Background**

With the precision cosmology enabled by instruments like WMAP, we now have extremely robust constraints on the cold dark matter model and the structure formation paradigm. These results suggest that the bulk of the mass in the universe is locked on group scales, with typical values of ~ $10^{13}$ solar masses. The full census of properties of groups is currently incomplete. Some key questions that remain to be settled are whether groups have a large-scale dark matter halo or whether the individual galaxies that comprise them have independent dark matter halos.

The group environment also enables probing the nature of interaction, if any, between dark matter and the baryonic component – whether and how significant the adiabatic contraction in the center of the halo is can be addressed with a large sample of group scale lenses. Finding groups using observed strong lensing features in large surveys has been demonstrated to be a powerful technique as recent results from the CFHTLS SARCS survey suggest (More et al. 2012 and references therein). Multiple image separations ranging from 10 – 20 arcsec corresponds to group scale masses and can be unambiguously detected.

**WFIRST**

A proposed multi-band survey by WFIRST will enable the discovery of several hundred hitherto undetected groups. CFHTLS found 30-45 arcs from group scale lenses in a survey of 170 sq. degrees. A 1000 – 2000 sq. degree WFIRST survey will provide between 200 – 400 groups. Expected image separation distributions are expected to range from 10 – 20 arcsec.

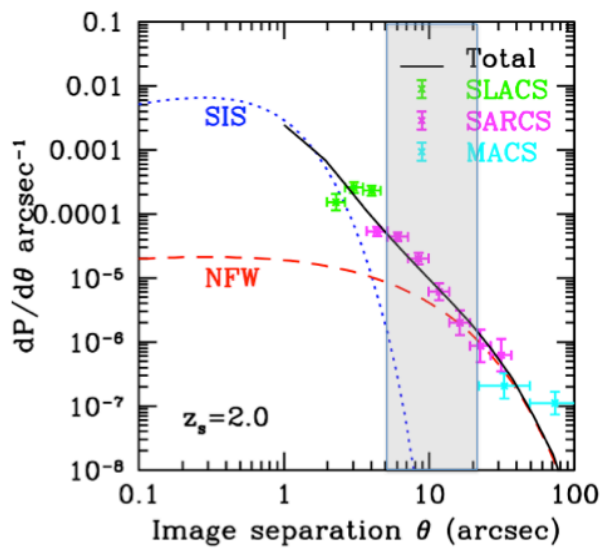

Caption: The targeted image separation range for group scale lenses is marked in gray.

**Key Requirements**

Wavelength coverage: I, Y, J, H, (K) – I band for identifying the arcs and as many bands as possible for optimal photo-z determination.

Megan Donahue and Mark Voit (Michigan State University), donahue@pa.msu.edu

**Finding and Weighing Distant, High Mass Clusters of Galaxies**

Background

The mean density of matter and dark energy and the initial fluctuation spectrum determine the abundance and growth of clusters of galaxies (e.g Voit 2005). The most dramatic effects are experienced by the abundance of the most massive clusters. Furthermore, the abundance of the most extremely massive clusters tests the assumption of Gaussianity of the power spectrum at Mpc scales. The key to cluster cosmology lies in accurate estimates of both the number counts (as a function of redshift) and the gravitating masses of clusters of galaxies. Finding clusters by their projected lensing mass, by the overdensity of red sequence galaxies, and by their Sunyaev-Zeldovich decrement on the CMB would provide mutual verification of cluster existence, hot gas (baryonic) content, presence of a prominent red sequence. Spitzer IRAC surveys have discovered over one hundred z>1 candidates in 7.25 sq degrees (Eisenhardt et al. 2008), yielding at least one massive cluster at z=1.75 (Brodwin et al. 2012; Stanford et al. 2012).

WFIRST

Survey fields obtained for WFIRST weak lensing shear and/or baryon acoustic oscillation studies will also allow a census of massive clusters. Photometric redshift observations, either ground-based or from WFIRST, would be required to estimate the redshift of newly discovered clusters.

Key Requirements

Depth – Well-dithered exposures sufficient to obtain shape measurements for galaxies at z>z_cluster; shape measurements benefit from 3-4 repeats

Field of View – A wide-field survey of 8300 sq degrees include 100 $10^{14}$ $h^{-1}$ and 2800 $10^{13.7}$ $h^{-1}$ z=2-2.5 clusters; for z=1.5-2.0, the numbers rise to 1660 and 20,000 clusters for the same masses, respectively; the most massive clusters are the most rare, the larger the survey, the better the probe of the high-sigma tail of the mass distribution

Wavelength Coverage – NIR filters

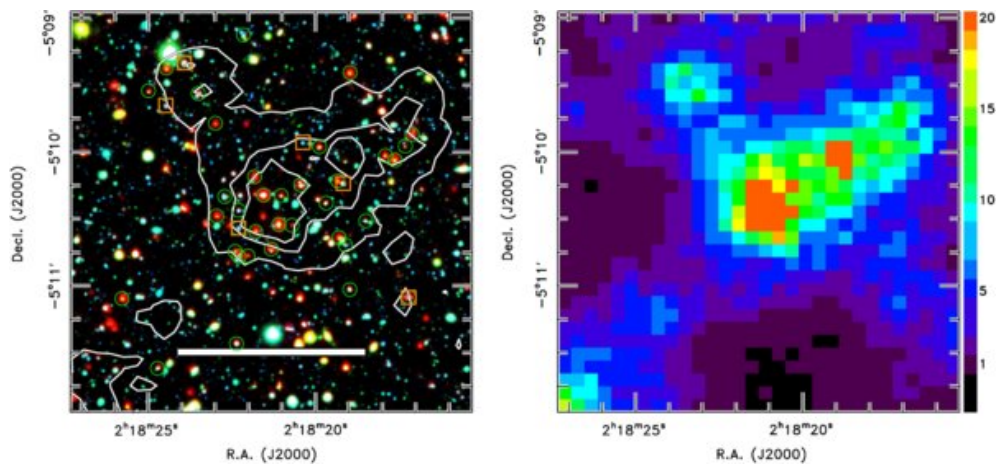

Caption: Left panel shows a false-color image of a Spitzer-selected cluster at z=1.6 (Papovich et al. 2010). Blue corresponds to the Suprime-Cam B band, green to the Suprime-Cam i band, and red to the Spitzer 4.5 μm band. The right panel shows the surface density of galaxies, color coded in units of standard deviations above the mean.

Megan Donahue (Michigan State University), donahue@pa.msu.edu

# The Evolution of Massive Galaxies: the Formation and Morphologies of Red Sequence Galaxies

### Background

The most massive galaxies in the present day universe are red, usually dormant, elliptical galaxies. They appear in the color-magnitude diagrams of galaxies in clusters of galaxies as the red sequence (RS). The red sequence represents a distinct population of galaxies whose stars formed at high redshift (z>3) and have passively evolved since then. These galaxies also experience significant interactions and mergers over their lifetime. The red sequence feature is so prominent that it can be used to discover distant clusters of galaxies and to estimate their redshifts. Current studies suggest that while the stars in these galaxies formed early, the assembly of the galaxies we see locally occurred relatively late. Studies of the RS galaxies as a function of redshift, cluster mass, and local environment are critical tests of our understanding of galaxy evolution.

### WFIRST

WFIRST will enable sensitive IR color-magnitude diagrams of RS galaxies in clusters and in the field at redshifts representing the epoch at which the red sequence is beginning to assemble, z~1.5-2.5. The spatial resolution will allow galaxy sizes to be estimated; as mergers progress, the stellar orbits get progressively more puffed out. The best photometric redshifts will be obtained from filters which span the 4000 Angstrom break and which are medium-width (not too broad, see FourStar results from Spitler et al. 2012.)

### Key Requirements

Depth – Well dithered exposures sensitive to K~25-26 (AB)

Field of View – Targeted follow-up of high-z candidates (from Sunyaev-Zeldovich surveys or other techniques). Cluster size is ~ 1-2 arcmin, ~ independent of z.

Field of View – Survey for z=1.5-2.5 clusters: ~ 100 sq degrees would include ~20 clusters at M~10^14 h^-1 solar masses if you had to find the clusters first.

Wavelength Coverage – Medium-band NIR filters.

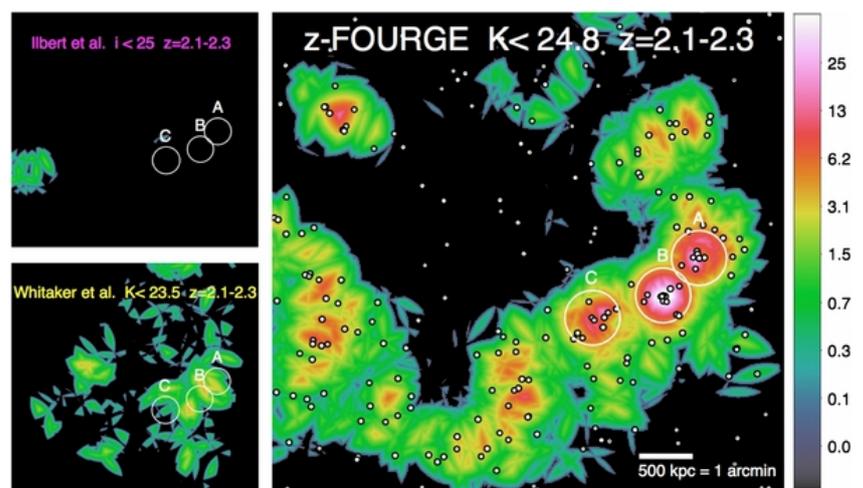

Caption: Nearest-neighbor surface density maps for z = 2.1-2.3 in a 9' × 9' region in the COSMOS field (Spitzer et al. 2012).

Daniel Stern (JPL), daniel.k.stern@jpl.nasa.gov

## Obscured Quasars

### Background

Obscured, or type-2 quasars are expected to outnumber unobscured, type-1 quasars by factors of 2-3. They are predicted by models of active galactic nuclei (AGN), and are required to explain the hard spectrum of the cosmic X-ray background. Until recently, however, our census of this dominant population of AGN has been lacking since such systems are difficult to identify in optical and low-energy X-ray surveys. High-energy missions to date have had limited sensitivity, while the recently launched NuSTAR mission has a very limited field-of-view. Mid-infrared surveys, initially with Spitzer (e.g., Lacy et al. 2004; Stern et al. 2005) and more recently with WISE (Stern et al. 2012; Assef et al. 2013) have dramatically changed the situation. In its shallowest fields, WISE identifies 60 AGN candidates per deg2 with a reliability in excess of 95%. In higher latitude, deeper parts of the WISE survey, this surface density rises to >100 AGN per deg2. Comparably powerful AGN identified in optical surveys have surface densities of only ~20 per deg2 (e.g., Richards et al. 2006), implying that WISE has finally realized the efficient identification of the dominant luminous AGN population across the full sky.

### WFIRST

While mid-infrared observations with WISE have identified this dominant population of obscured, luminous quasars, WFIRST will be required to characterize its properties. Working with deep ground-based optical data, WFIRST will provide photometric redshifts for this population, something not possible from the mid-infrared data alone since the identification relies on the power-law mid-infrared spectra of luminous AGN. Photometric redshifts will allow us to probe the cosmic history of obscured black hole growth, and relate it to galaxy formation and evolution. AGN feedback is expected to play an important role shaping the present-day appearances of galaxies. Measuring the clustering amplitude of obscured and unobscured quasars will probe AGN unification scenarios. For the classic orientation-driven torus model, clustering should be the same for both populations. On the other hand, if obscured AGN are more common in merging

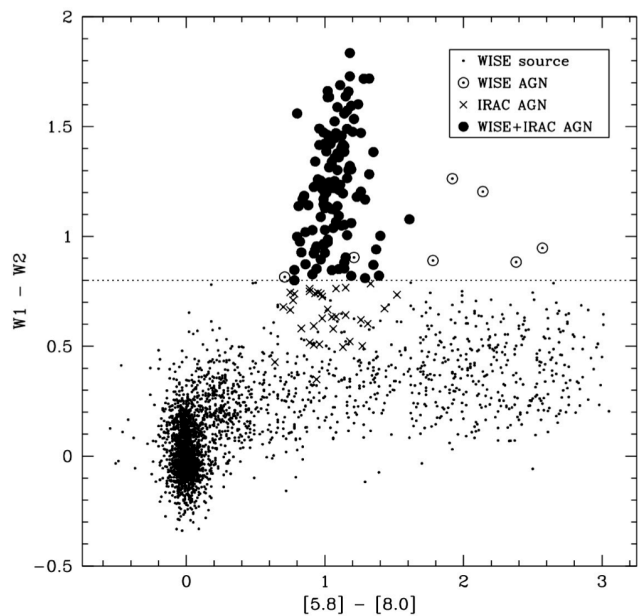

Caption: Mid-infrared color-color diagram, illustrating that a simple WISE color cut of W1-W2 ≥0.8 identifies 62 AGN per deg2 with 95% reliability assuming the complete reliability of the Stern et al. (2005) Spitzer AGN selection criteria. From Stern et al. (2012).

systems, with the obscuration caused by galactic-scale material, then the clustering amplitudes are expected to differ.

### Key Requirements

Depth – Provide robust photometric redshifts for sub-L* populations to z~2
Field of View – Wide-area surveys required for clustering analysis
Wavelength Coverage – >2 NIR filters for robust photometric redshifts

Daniel Stern (JPL), daniel.k.stern@jpl.nasa.gov

# The Faint End of the Quasar Luminosity Function

## Background

The primary observable that traces the evolution of quasar (QSO) populations is the QSO luminosity function (QLF) as a function of redshift. The QLF can be well represented by a broken power-law: $\Phi(L) = \phi_*/[(L/L_*)\alpha + (L/L_*)\beta]$, where $L_*$ is the break luminosity. Current measurements poorly constrain the corresponding break absolute magnitude to be $M_* \simeq -25$ to $-26$ at $\lambda=1450$ Å. The bright-end slope, $\alpha$, appears to evolve and flatten toward high redshift, beyond $z \sim 2.5$ (Richards et al. 2006). The faint-end slope, $\beta$, is typically measured to be around $-1.7$ at $z\sim2.1$; it is poorly constrained at higher redshift, but appears to flatten at $z \sim 3$ (Siana et al. 2008). At yet higher redshift, however, the situation is much less clear due to the relatively shallow flux limits of most surveys to date. The true shape of the QLF at $z > 4$ is still not well measured, and the evolution of $L_*$ and the faint-end slope remain poorly constrained (Glikman et al. 2010, 2011; Ikeda et al. 2010). Studying the faint end of the high-redshift QLF is important for understanding the sources responsible for re-ionizing the Universe. While Vanzella et al. (2010), studying faint $z\sim4$ galaxies in the GOODS fields, finds that Lyman-break galaxies (LBGs) account for <20% of the photons necessary to ionize the intergalactic medium (IGM) at that redshift, Glikman et al. (2011) find that QSOs can account for 60±40% of the ionizing photons. Furthermore, new quasar populations appear when one studies faint, high-redshift quasars. Glikman et al. (2007), in their initial work on the faint end of the high-redshift QLF, found strong NIV] 1486 emission in ~10% of the QSOs surveyed, likely associated with early starbursts.

## WFIRST

The deep, wide-field imaging data from WFIRST DE surveys will provide critical information for constraining the faint end of the high-redshift QLF. While at $z\sim4$, deep ground-based data from surveys such as LSST will be essential, WFIRST grism spectroscopy will play an essential role in confirming the high-redshift AGN nature of the photometric candidates. At yet higher redshifts, $z>7$, deep surveys from WFIRST alone will probe the earliest epochs of nuclear activity.

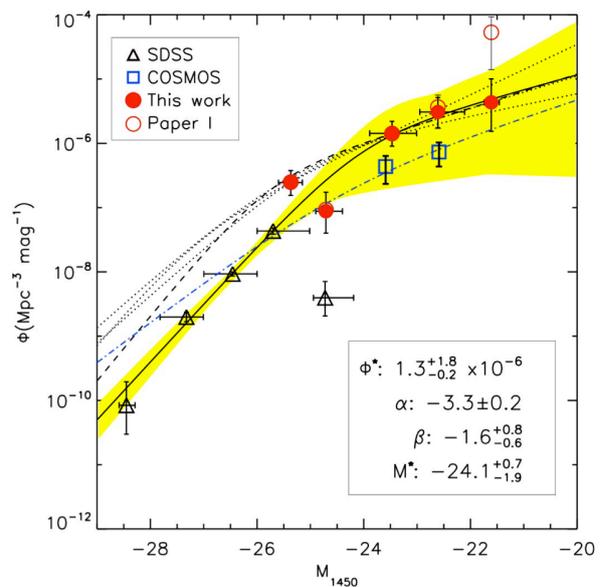

Caption: QLF at $z\sim4$ from Glikman et al. (2011). Note the substantial uncertainties below the knee in the QLF.

## Key Requirements

Depth – To study sub-L* QSO populations at high redshift

Morphology – To separate stars from galaxies (though note that faint QSOs at high redshift are likely to show significant host galaxy light, and thus morphology should be used to characterize the populations, but not as a selection criterion)

Grism – For spectroscopic confirmation and characterization

Field of View – Wide-area surveys required to identify rare populations

Wavelength Coverage – >2 NIR filters to robustly identify QSOs at the highest redshifts

Xiaohui Fan (University of Arizona), fan@as.arizona.edu

**Strongly Lensed Quasars**

Background

Strong gravitational lensing, when multiple images of distant objects are produced by massive objects in the foreground, is a powerful and unique tool for both cosmology and galaxy/AGN physics (e.g., Oguri et al. 2012). Applications of strongly lensed quasars include: using time delay among different lens components to measure $H_0$ and to constrain the expansion history of the Universe (Fassnacht et al. 2002); using lens models to measure galaxy mass and structure (Bolton et al. 2006); using flux ratio anomalies to probe dark matter halo properties and the existence of halo subtructure (Keeton et al. 2006); using AGN microlensing to probe accretion disk structure (Kochanek et al. 2006); using the quasar lensing fraction to study magnification bias and the quasar luminosity function (Richards et al. 2006); using high spatial resolution imaging of lensed systems to study properties of quasar host galaxies and environments (Peng et al. 2006); using rare examples of quasars acting as gravitational lenses to study the cosmic history of the black hole-bulge relation (Courbin et al. 2012). However, strong lensing is a rare event, requiring wide-field, high resolution imaging to establish a large sample for statistical studies, and to uncover those with ideal lensing configurations for cosmological tests (so-called "Golden Lenses").

WFIRST

The deep, wide-field imaging data from WFIRST surveys will provide a treasure trove for lensing studies. With a pixel size of ~0.1", WFIRST will resolve any multiply imaged system with separation >0.2". This is expected to account for the vast majority of all strong quasar lenses, and will allow accurate photometry of lensed image components as well as lensing galaxies. Multicolor photometry will provide accurate photometric redshifts for sources and lenses. Figure 1 presents the expected number/redshift distributions of lensing galaxies (lenses) and lensed quasars (sources) in a total survey area of 20,000 deg, with a limiting magnitude of ~24 AB and image separation of >0.5". This represents a nearly two orders of magnitude increase in size from current samples (Inada et al. 2012). Strong lensing is also one of the key areas for LSST science. WFIRST will have strong synergy with LSST, providing deeper imaging with higher spatial resolution and more accurate photometry and photometric redshifts.

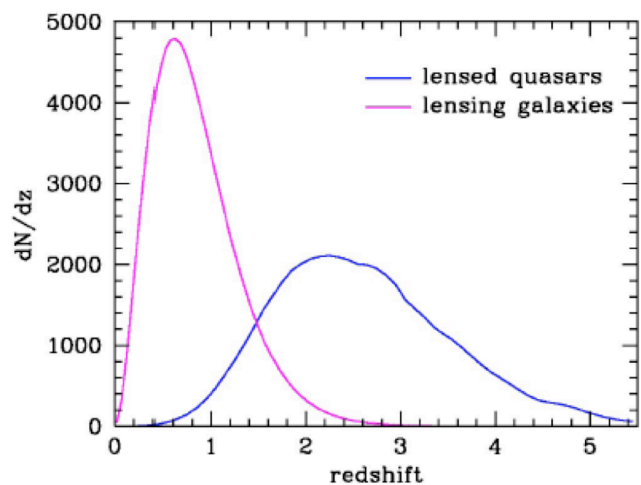

Key Requirements

Depth – To sample a wide range of quasar lum. and to detect faint lensed image components
Spatial Resolution – This is the key: a pixel size and resolution <0.3" is needed to uncover the majority of lenses and to allow accurate photometry of lensed components
Field of View – Wide-area surveys required to establish large sample and to find the rare unique systems
Wavelength Coverage – >2 NIR filters for photo-z of both lenses and sources

Xiaohui Fan (University of Arizona), fan@as.arizona.edu

## High-Redshift Quasars and Reionization

### Background
Luminous quasars at high redshift provide direct probes of the evolution of supermassive black holes (BHs) and the intergalactic medium (IGM) at early cosmic time. The detection of z>7 quasars (e.g., Mortlock et al. 2011) indicates the existence of billion solar mass BHs merely a few hundred million years after the Big Bang, and provides the strongest constraints on the early growth of supermassive BHs and their environments. Spectroscopy of the highest redshift quasars reveals complete Gunn-Peterson (1965) absorption, indicating a rapid increase in the IGM neutral fraction and an end of the reionization epoch at z=6-7 (Fan et al. 2006). Current observations suggest a peak of reionization activity and emergence of the earliest galaxies and AGNs at 7<z<15, highlighting the need to expand quasar research to higher redshift.

### WFIRST
While ground-based surveys such as LSST and VISTA will make progress in the z=7-8 regime in the coming decade, strong near-IR background from the ground will limit observations to the most luminous objects and to z<8. Wide-field, deep near-IR survey data offered by WFIRST will fundamentally change the landscape of early Universe investigations. Figure 1 shows the predicted number of high-redshift quasars in a WFIRST survey based on current measurements at z~6 (Jiang et al. 2008; Willott et al. 2010) and extrapolation to higher redshift with a declining number density following the trend seen at z=3-6. WFIRST should allow robust identifications of a large sample of reionization-epoch quasars up to z>10, if they exist at those epochs. WFIRST grism will provide direct spectroscopic confirmation and characterization of these high-redshift quasars, while JWST and next-generation extremely large telescope high-resolution spectroscopic observations will measure IGM and BH properties. Key questions to be addressed by these observations are: (1) when did the first generation of supermassive BHs emerge in the Universe; (2) how and when did the IGM become mostly neutral; (3) did quasars and AGNs play a significant role in the reionization process?

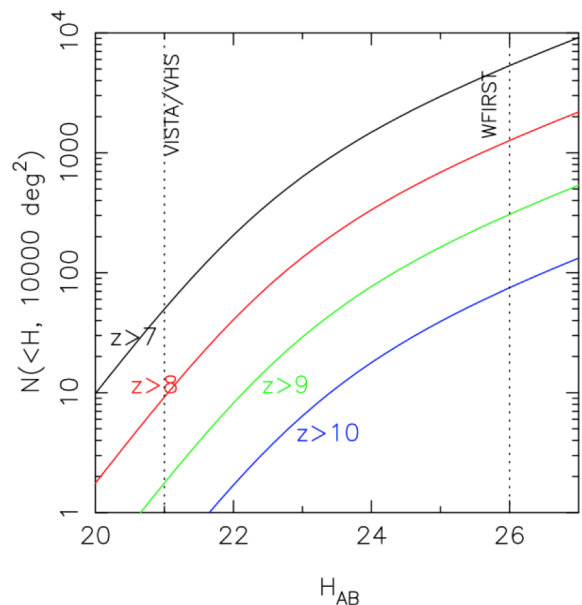

### Key Requirements
Depth – To study sub-L* QSO populations at high redshift
Morphology – To separate stars from galaxies; one of the main contaminations is expected to be low-redshift red galaxies
Grism – For spectroscopic confirmation and characterization
Field of View – Wide-area surveys required to identify rare populations
Wavelength Coverage – >2 NIR filters to robustly identify QSOs at the highest redshifts

Harry Teplitz (IPAC), hit@ipac.caltech.edu

# Characterizing the sources responsible for Reionization

Star–forming galaxies are likely responsible for reionizing the Universe by z~6, implying that a high fraction of HI–ionizing (Lyman continuum; LyC) photons escape into the IGM. Measurements of galaxies at z<3.5 show the average escape fraction, $f_{esc}$, to be very low or undetected at all redshifts. At redshift z~3, high $f_{esc}$ ($\geq$ 50%) have been reported in only about 10% of LBGs and Lyα emitters though significant mysteries remain for even these few detections ($f_{esc}$ > unity, and unknown numbers of interlopers).

Current studies suggest that $f_{esc}$ may evolve with redshift (Figure 1) and/or be higher in low mass galaxies. A recent study of the radiation transport of LyC at z= 3-6 in galaxies drawn from cosmological SPH simulations predicted substantial LyC ($f_{esc}$ =8–20%) emission from galaxies with halo masses $M_{halo}$ <~ $10^{10}$ M*, but little or nothing from more massive systems. In addition, $f_{esc}$ is found to increase with decreasing metallicity and SFR – both of which are thought to positively correlate with $M_{halo}$.

The LyC is best studied at z<3 (in the UV) due to the increasing opacity of the IGM to LyC photons higher z. However, current studies are severely limited by the need to find appropriate analogs to the low mass/metallicity objects that are the likely sources of LyC photons during Reionization.

Figure 1: $L_{LyC}/f_{1500}$ measurements (versus UV luminosity) at z ~ 1 and z ~ 3 corrected for the average IGM attenuation at the relevant wavelengths (Siana et al. 2010).

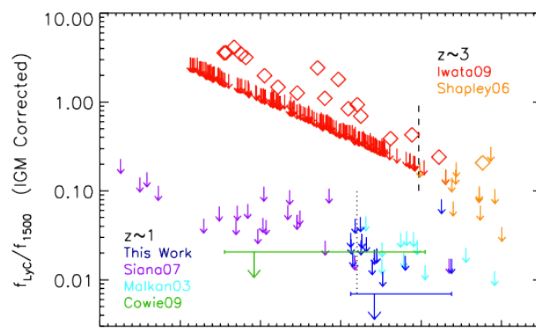

## WFIRST
A wide-area NIR grism survey with WFIRST presents an important opportunity to find LyC-emitting sources through their rest-frame optical emission lines.  HST grism surveys discovered a population of young, low-mass starbursts, selected based on their very strong emission-lines (EW[Hα]$_{Rest}$ > 200Å). These galaxies are ideal for testing the hypothesis of a mass dependent $f_{esc}$: they are among the most metal poor objects known at these redshifts, and their average emission–line corrected stellar mass is $10^7$–$10^8$M*, i.e., ~30 times less massive that typical M* galaxies at the same redshifts.  HST surveys are severely limited by lack of spatial coverage in finding such rare sources.  A WFIRST survey would enable UV follow-up -- either archival (GALEX, HST, SWIFT-UVOT) or with new observations from ground-based U-band (at z~3) and possibly future UV telescopes. Ground-based observations would be particularly helped by space-based source identification that would spatially resolve possible foreground interlopers.

In addition, WFIRST would allow us to identify obscured Type-2 AGNs. The possibility that faint AGNs play a substantial role in Reionization has not yet been ruled out observationally.

## Key Requirements
Grism – To find low mass/metallicity line-emitters to enable (possibly archival) UV follow-up
Depth – These sources are faint
Field of View – Wide-area surveys required to identify rare populations

Claudia Scarlata (UMN, scarlata@astro.umn.edu) and Harry Teplitz (IPAC, hit@ipac.caltech.edu)

# Resolved stellar population studies in z~2 SF galaxies

Observations show that "normal" SF galaxies were in place at z~0.5, with stellar population and scaling relations consistent with passive evolution into the stable population observed locally. Looking back to z > 2, dramatic changes appear. Massive star-forming galaxies along the so-called main sequence at these epochs tend to be thick, clumpy disks, forming stars at rates (100 M☉/yr) much higher than is observed in the thin, quiescent disks observed at z < 0.5.

Gravitational instabilities in gas–rich turbulent disks can produce dense "clumps" supported by cold streams of gas. These structures could eventually migrate into a bulge component over time scales < 10 dynamical times, or < 0.5 Gyrs. As the clumps merge, the growing spheroid/bulge will have a stabilizing effect on the disk, and eventually it will be massive enough (~20% of the galaxy stellar mass) to prevent further fragmentation. Constraining this self-regulating process requires accurate measurements of the ages of both the clumps and the diffuse stellar component.

*Figure 1: z=2.5 galaxy from Elmegreen et al. (2005). WFIRST+IFU will provide low resolution spectra over scales < 2Kpc, accurately allowing age-dating of the stellar population.*

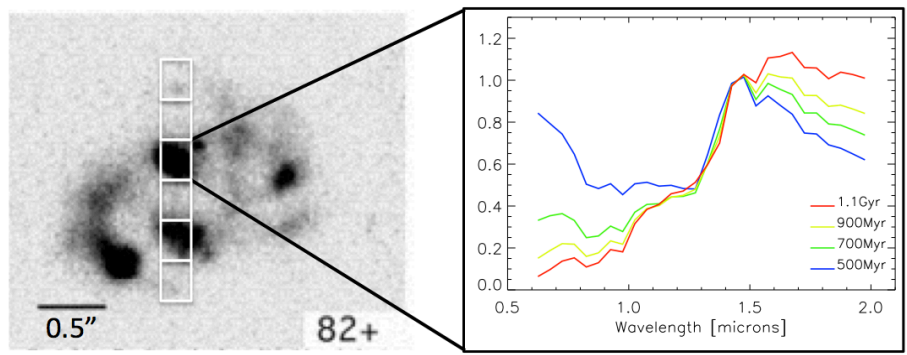

## WFIRST

A wide-area NIR grism slitless survey with WFIRST will discover a large number of z~2 star-forming galaxies via the detection of their Hα and [OIII] emission lines. These objects are ideal targets for followup with the WFIRST IFU spectrometeter, which will provide rest-frame optical spectra on spatial scales <2 Kpc. The low-resolution IFU spectra will allow us to constrain the shape of the age sensitive part of the spectral energy distribution of the z~2 galaxies in a much greater detail than with broad band imaging. This, in turn, will allow for an accurate measurement of the spatially resolved stellar population properties, crucial to constrain galaxy formation models. Moreover, the stellar population maps resulting from the IFU data will be used as input to interpret the higher spectral resolution slitless grism data in an iterative process to obtain maps of emission-line intensities, which will be used to derive SFR and dust extinction surface density.

## Key Requirements

Grism – To find z~2 emission line galaxies
IFU – depth & resolution needed to derive spatially-resolved stellar population properties

# Near Infrared Counterparts of Elusive Binary Neutron Star Mergers

Mansi Kasliwal (Carnegie), Samaya Nissanke (Caltech), Christopher Hirata (Caltech)

Background

Within the next decade, ground-based gravitational wave (GW) detectors are expected to routinely detect the mergers of binary compact objects (neutron stars and black holes). The NS-NS and NS-BH mergers contain ultra-dense (>$10^{14}$ g/cm$^3$) matter and hence provide a laboratory for some of the most extreme astrophysical processes in the Universe. Electromagnetic observations, sensitive to the composition and thermodynamic state of the matter, will be required to complement the GW signal, which constrain the NS and BH masses. Except in the lucky case of a gamma ray burst that happens to be beamed toward us, we will have to rely on isotropic electromagnetic counterparts. Tidal debris or accretion disk winds are expected to expel neutron-rich material from the merger resulting in an isotropic radioactivity-powered remnant. Dubbed "kilonova", the counterparts should be detectable in the optical/NIR with durations of order days and luminosities of order $10^{40}$—$10^{42}$ erg/s. Follow-up of binary mergers will be challenging: GW detectors lack intrinsic angular resolution, and simultaneous fitting of an event in several detectors can only localize a source to an error ellipse of a few to hundreds of deg$^2$. This ellipse will contain not just the binary merger counterpart, but also many unrelated "false positive" transients. The eventual goal is to obtain not just a detection, but also a light curve and spectroscopy to constrain the velocity, opacity, and composition of the debris.

WFIRST

Theoretical predictions for the counterpart suggest that the emission peaks in the NIR due to opacity considerations (Barnes and Kasen 2013). Hence, the exquisite wide-field mapping speed, red sensitivity and rapid response capability of WFIRST in geosynchronous orbit are well-suited to this search. Specifically, we suggest following up binaries localized within 6 deg$^2$ and detection S/N>12 by a network of 5 GW detectors. During a 5-year mission, we expect ~30 such binaries within the WFIRST field of regard at a median distance of 230 Mpc (Nissanke, Kasliwal and Georgieva 2012). At this distance, a depth of $H_{AB}$=24.6 mag corresponds to $M_{H,AB}$=−12.2 or $\lambda L_\lambda$ = 6×$10^{39}$ erg/s. The rate of NS-NS mergers is currently uncertain by 1—2 orders of magnitude: the follow-up strategy and criteria for a ToO trigger will be adjusted based on the actual event rate. One possible follow-up plan, requiring a total of 27 hours per binary (including overheads), would involve 2 filter (J+H to mag 24.6 AB at S/N=10) and grism (S/N=5 per synthetic R=30 element at $H_{AB}$=22) observations of the entire error ellipse as quickly as possible. After this, we envision imaging mode (J+H) revisits at 4 epochs from t ~ 1—30 days, and a later reference epoch. Once the counterpart candidate list is reduced to ≤10 candidates based on the grism spectrum and the photometric evolution, we would use the IFU to acquire a high S/N spectrum (S/N=5 per synthetic R=30 element at mag 24.9 AB).

Key Requirements

Wide-field, high sensitivity NIR camera for counterpart search and detection

Field of regard – only events in an accessible region of sky (59% for WFIRST) can be followed up

Target of opportunity capability, rapid response time (several hours)

Spectroscopy – grism (for prompt, blind spectra) + IFU (deep spectra of candidate counterparts)

Mike Gladders (The University of Chicago), gladders@oddjob.uchicago.edu

## Strong Lensing

Background: Strong lensing of distant sources (galaxies, AGN, SNe) provides deep insight into the distant universe. The increased S/N from lensing magnification allows for precise photometry over a much broader range of wavelengths than would otherwise be accessible, for luminous and highly magnified sources. At the same time, strong lensing allows us to probe more deeply into the galaxy LF than any other technique; the current Hubble Frontier Fields are designed to exploit this effect in the reionization epoch. Strong lensing of distant galaxies also provides access to unique spatial scales; at Hubble-like resolutions structures smaller than 100 parsecs can be resolved. Strong lensing is also a powerful probe of the mass structure of the foreground lenses, allowing tests of ΛCDM predictions of halo and sub-halo properties and/or the detailed physics operating in halo cores over several decades of halo mass. Strong lensing of variable sources also allows for a geometric cosmological test through time delays.

WFIRST: Currently, the total number of known strong lenses is hundreds. Efforts to exploit ground-based surveys (e.g., DES, LSST) will increase this by hundreds more to perhaps a thousand. The strong lens sample in the COSMOS HST data – 67 candidates (Faure et al. 2008) in 1.6 deg$^2$ of imaging comparable in depth to the WFIRST HLS – suggests that the WFIRST strong lens sample will number in the tens of thousands, even if one restricts the sample to the 10-20% of such candidate systems which are unambiguous and `interesting' (i.e. clearly multiply-imaged, and potentially model-able). Defining this sample from the WFIRST HLS represents a fantastic opportunity for citizen science. Strong lensing is visually compelling and unique, and offers a direct connection to the physics of General Relatively that frames modern cosmology.

The bright end of this sample will have redshifts from the HLS grism data – on order of a thousand systems; that will represent an order of magnitude increase in arc redshift statistics, and will likely not be challenged by ground-based efforts since the sky density of systems will still be such that there is little multiplex advantage for all but the widest-field spectrographs. WFIRST-quality multi-band imaging, and grism spectroscopic redshifts will yield an abundance of systems that can be modeled in detail. Likely the limiting factor will be people and not data; the current approach of a hand-tooled model for each system will have to be automated and refined if we are to fully exploit the strong lens data that WFIRST will yield.

The bulk of the sample will not have redshifts; statistical analyses will dominate. Detailed image-level simulations will play a critical role in defining selection functions. One unique problem will be the lack of source-plane images with sufficient resolution. The best path forward will be some form of `complexification' applied to WFIRST imaging (e.g., Groeneboom & Dahle 2014).

Lensing statistics aside, WFIRST offers a truly unique opportunity to coherently probe the mass structure of the largest halos over a prodigious radial and density range. In pointed GO mode, we will be able to study massive clusters from the strong lensing core, through the intermediate flexion regime and out to the weak lensing regime, anchored in the true `field', in one image.

Key Requirements

*Image Quality* – the information density in best possible image quality enables lens modeling; coupled flexion and weak lensing measurements at larges scales also demand good IQ

*Grism* – for redshifts; ground-based spectroscopy will complement in wavelength coverage

*Field of View* – A wide-area survey – i.e. the HLS - is required to identify these rare populations. Single-shot wide FOV also uniquely enables mass probe of massive haloes over full radial range

*Wavelength Coverage* – multi-band imaging, with broadest possible coverage, maximizes lens-modeling return, and enables best photo-zs for weak lensing reconstructions.

http://adsabs.harvard.edu/abs/2008ApJS..176...19F
http://adsabs.harvard.edu/abs/2014ApJ...783..138G

**Appendix E    Pasadena 2014 Conference Summary**

The *Wide-Field InfraRed Surveys: Science and Techniques* conference was held in Pasadena, CA on November 17-20, 2014. At least 220 scientists from around the world attended, engaging in lively discussions both in session and during the breaks. The success of the conference reflected an increasingly engaged and supportive community-at-large looking forward to WFIRST-AFTA and its science.

The conference was intended as a forum for scientists to consider recent progress from missions such as Spitzer, Kepler, Hubble, Planck, WISE and Herschel, as it affects the design and utilization of the next generation of space-based infrared surveys, including WFIRST. Invited talks exposed the participants to the capabilities currently envisaged for WFIRST-AFTA, and assessed the mission's promise in the context of the anticipated performance of Gaia, JWST and Euclid, as well as ground-based surveys, especially LSST and large-scale radio surveys such as the next generation VLASS. Because of the wide range of topics addressed by future infrared survey missions, from dark energy to exoplanets, and because of the increasing importance of additional community-driven science, the emphasis was on bringing together these many scientific constituencies to present their perspectives and inform the trade studies currently being pursued by the WFIRST-AFTA Science Definition Team and Study Office. As the conference unfolded, WFIRST-AFTA was the clear focus for the 220 participants, of which 15% were based at non-US institutions.

The conference was mostly convened in plenary sessions, with about 30% of the time spent in parallel sessions that addressed Dark Energy, Exoplanets, Milky Way and Local Group, and Beyond the Local Group topics. Two panel discussions were held in plenary session; one focused on the placement of the high-latitude survey fields in view of existing and planned surveys and optimizing the science yield in areas besides dark energy; the other explored anticipated Community Research Support Needs for maximizing science return from the WFIRST-AFTA mission. A total of 120 talks were delivered, including a NASA HQ perspective delivered by Paul Hertz, Director of the Astrophysics Division at NASA HQ, via video link, and a set of closing remarks by Alan Dressler. The conference website contains more detail about the program, abstracts, participants and organizers. PDF versions of the talks are also available from the same website: http://conference.ipac.caltech.edu/wfirs2014/.

The richness of the conference and of its multiple science threads makes it nearly impossible to provide a complete and balanced summary. The major themes of dark energy, exoplanet discovery and characterization, and Galactic and extragalactic astrophysics were dominant. The following paragraphs however sample some of the themes that were noteworthy for eliciting particular interest among participants, or which were novel or unanticipated.

<u>Additional Bulge Science</u>: There was much interest in the potential for additional science possible using the data from the baseline microlensing survey of the Galactic Bulge. Talks by Andy Gould and Mike Rich highlighted a number of science pursuits made possible by the exquisite astrometric and photometric accuracies possible with WFIRST-AFTA: under stellar astrophysics, parallax and proper motion studies, ages and [Fe/H] estimates, and astroseismology for millions to tens of millions of stars with WFIRST-AFTA. In addition to studies of the dynamics and subpopulations in the bulge leading to better understanding of its assembly history, the survey will provide a sensitive search for isolated or paired low-mass Black Holes, neutron stars or other exotic objects, and within the Solar System, precision orbits and binary fraction for thousands of Kuiper Belt Objects. It should be stressed that this was only a first look at the potential science yield from the microlensing survey.

<u>Exoplanet Science with WFIRST-AFTA</u>: A wide variety of exoplanet science topics and techniques were discussed in the plenary and parallel sessions. A plenary talk highlighted how WFIRST-AFTA's microlensing census of planets around the snowline will be critical to testing theories of planet formation. The parallel session on science addressed the current state of the art in microlensing; coronagraphic imaging of planets and disks; and the need for a resolving power of R~70 for retrieval of atmospheric properties of giant planets. There was discussion of the potential of a starshade for WFIRST-AFTA in an L2 orbit to test technology and to increase the number of small planets accessible to the mission. Parallel sessions on science requirements and precursor observations addressed the importance of: HST observing in the Bulge fields for initial characterization; ground-based microlensing to build the US community; radial velocity programs and Gaia as providers of targets for the WFIRST-AFTA coronagraph.

General Investigator and General Observer Science: Discussions during the conference explored how science beyond the areas of dark energy and microlensing will be possible with WFIRST-AFTA. Guest Investigators will primarily exploit the data acquired for the microlensing survey or the main dark energy high-latitude survey for other topics such as stellar astrophysics (see above) or galaxy evolution. Guest Investigators would propose to do unique or novel science with the existing survey data, potentially making available special tools or methods that could be distributed to the community. In the parallel and plenary sessions, participants were eager to discuss the potential for the high-latitude survey to find extremely young galaxies at high-redshift (z > 5), track the assembly of galaxies at later times, understand how galaxy properties are influenced by environment, and how best to use the grism data and targeted GO programs to diagnose star-formation and galactic feedback over a wide range of cosmic time. There was also an interesting discussion on how to recognize and maximally exploit the large number of strong gravitational lenses expected in the WFIRST-AFTA database. Guest Observers would follow the traditional proposal process to conduct observations with WFIRST-AFTA separate from the main surveys, targeting for instance galaxies in the Local Group, or specific high-redshift structures. Because both research approaches will be supported by WFIRST-AFTA, the design of the main surveys should be cognizant of the archival value of the data.

Filter Set Options: One manifestation of the robust community engagement in the WFIRST-AFTA mission is that many of the speakers addressing GI/GO science topics suggested that their research would benefit from tuning of the filter bandpasses. In many cases, the suggestion was to extend the wavelength coverage of the mission, either redward to mitigate extinction or reach higher redshifts, or blueward for improved color discrimination for both stars and galaxies.

The Data Volume Challenge: Conference participants anticipated the shear volume and richness of the data expected from WFIRST-AFTA as a new qualitative challenge, and discussed how best to prepare for it. Early availability of simulated WFIRST-AFTA data was generally acknowledged as critical to preparing the community, and LSST data access and exploitation were also seen as contributing significantly to this preparation. Eventually, the archives built to serve WFIRST-AFTA data will require massive power, as well as agility and flexibility to evolve novel tools in response to the needs of users extracting new science from huge data sets.

## Appendix F Exoplanet Detectability with the WFIRST-AFTA Coronagraph

The Coronagraph Instrument (CGI) on WFIRST-AFTA will ultimately be judged on its scientific yield, that is, its ability to directly detect and spectrally characterize planets and disks around nearby stars. In this Appendix we outline the basis for estimating the expected CGI science yield for three groups of targets, known radial-velocity (RV) planets, "new" as-yet unknown blind-search planets, and debris disks. We find that all three groups can be detected in sufficient numbers to produce substantial advances in the science of nearby exoplanets and disks.

### F.1. RV Catalog

The list of known RV planets used here contains 528 targets. Most of these range in distance from about 1 - 600 pc, in mass from 3 - 6000 $M_{Earth}$, and in semi-major axis from 0.04 - 10 AU. The radius of each planet is estimated by using an empirical relationship between mass and radius based on those planets with known values of each. The planet brightness is estimated for a visible geometric albedo of 40%, a Lambert scattering function, a circular orbit, an orbital inclination of 60 degrees, and a mean anomaly of 70 degrees (i.e., slightly brighter than at maximum elongation, and slightly closer to the star). The star brightness is estimated on the basis of its visual magnitude and color. The ratio of planet to star brightness is called the contrast. We assume that the orbital period and phase are so well known that we can estimate the optimum calendar date on which to search for the planet. Even over a broad range of inclinations, with a current RV solution it is possible to determine a range of dates over which the planet will have best observability (typically between 60 and 90 degrees phase). The main remaining uncertainty is the position angle on the sky, which requires that the search range be able to cover 360-degrees of azimuth, either in one observation or in several sequential ones.

### F.2. Science Yield Model

The purpose of our science yield model is to predict the fraction of planet light incident on the telescope that is delivered to the focal plane detector, the shape of that spot (the point spread function, PSF), the instrumental polarization distortion, the brightness of the background diffracted "speckles", and the degree to which the speckles vary with time. Our science yield model comprises four parts: an optical model, a coronagraph model, a detector model, and a telescope model.

### F.2.1 Optical Model

Our optical model is the standard "PROPER" program written by John Krist and employed by him for all of the instrument performance calculations in this section. PROPER propagates a wavefront through all of the telescope and CGI optical surfaces, in the near and far fields, using Fourier-based Fresnel and angular spectrum methods. For the purpose of calculating science yield, the key results from PROPER are the azimuth-averaged contrast values of the background speckle field, the fraction of planet light in the core of the PSF (here taken to be the area enclosed by the full-width at half-maximum, FWHM), all for each of two orthogonal linear polarizations, as a function of radial angle on the sky, for selected values of the RMS angular jitter of the telescope pointing direction, for a given power spectral density (PSD) of polishing errors and amplitude errors on the telescope mirrors, and integrated over the full wavelength band. Polarization is accounted for by noting that the portion of the incident wavefront that is reflected from the angle mirrors (including the primary mirror) will suffer a phase delay between the radial and tangential components, which in turn negatively affects the coronagraph performance. Therefore, we currently plan to separate the x and y polarizations with a Wollaston prism placed just before the direct-imaging detector; the coronagraph deformable mirror settings can then be tuned to optimize contrast in either polarization. The spectrograph may use either a single fixed polarizer or none; simulations show the shaped pupil coronagraph is less sensitive to these low-order differential aberrations. Thus the dark hole will be achieved, and the science measurements taken, in one of these polarizations, even though both polarizations will be separately recorded.

### F.2.2 Coronagraph Model

Our coronagraph model is a mostly shared optical system hosting two coronagraph architectures that were chosen as the prime instrument after a community-wide down-selection process in 2013, with a third architecture as a study-phase backup. The prime instruments are the Hybrid-Lyot Coronagraph (HLC) and the Shaped-Pupil Coronagraph (SPC); the backup instrument is the Phase-Induced Amplitude Apodization (PIAA) coronagraph. All three architectures are being developed in parallel in laboratories for the WFIRST mission. Each uses a combination of phase and ampli-tude control of the wavefront in the pupil, near-pupil, and image planes to achieve contrast values on the order of $10^{-9}$ over a range of azimuthal and radial angles on the sky. The coronagraph systems feed both a direct imager and an Integral Field Spectrograph (IFS).

### F.2.3    Detector Model

Our detector model is centered on the well-known Electron-Multiplying CCD (EMCCD) from e2v. Model parameters include pixel size, wavelength dependence of quantum efficiency, and parameters for each of three read modes: ordinary CCD-type operation (zero gain), electron-multiplying analog operation, and electron-multiplying photon-counting operation. All three modes are accessible in flight, but for the present estimate we use only the latter, most sensitive, photon-counting mode. Key parameters of interest, as used in the current simulations, include dark current (0.0005 electrons/sec per pixel), clock-induced charge (0.001 electrons per pixel per read), gain factor (500), read noise (16 electrons RMS per read), integration time per snapshot (100 sec), and cosmic-ray effects (not included in this estimate).

### F.2.4    Telescope Model

Our telescope model includes the shape of the pupil (including the central obscuration and six spider arms), the off-axis angle of the CGI, the mirror prescriptions, the model Power Spectral Density (PSD) of phase and amplitude errors of the primary mirror (PM) and secondary mirror (SM), the model RMS pointing jitter, the coating reflectivity as a function of wavelength, the polarization transformation as a function of wavelength and position within the pupil, the powered optics between the PM/SM and the CGI, and the model wavefront deformation owing to thermal changes in the shape and location of telescope mirrors as a function of pointing angle, recent slew angle history, Earth shadowing, and orbit phase with respect to Earth.

### F.2.5    Exoplanet Detection Model

The expected science yield of the CGI is estimated separately for each coronagraph architecture as follows. For each wavelength band we estimate the count rate of electrons, in the FWHM of the planet image, for the given radial (sky) location of the planet, using the factors of transmission and image size as supplied by PROPER. We make this estimate for each of three values of RMS pointing jitter (0.4, 0.8, and 1.6 mas) and for two values of post-processing factor (1/10 and 1/30), where the latter parameter gives the factor by which we expect to reduce the RMS spatial variation of the speckle contrast (equal to the average contrast, owing to speckle statistics) by mathematical processing of the image data on the ground. We also estimate the noise associated with the detection of the planet image as the square root of the sum of the total electron count in the FWHM of the image (planet counts, speckle counts, detector counts, and zodiacal background and foreground counts, all increasing with time) plus the fixed noise floor from the speckles reduced by the post-production factor (not increasing with time).

### F.3.    RV Planet Yield Estimates

We solve for the time required to achieve a signal-to-noise ratio (SNR) value of 5.0. Note that the fixed noise floor makes the faintest planets undetectable regardless of integration time. This (assumed) fixed noise floor is therefore an important limitation on our ability to detect planets. A key focus of future work will be to substantiate and reduce this factor given more realistic simulations of the telescope's thermal response. The post-production factors are chosen based on our experience with subtracting PSFs from images from ground-based coronagraphs, as well as from images made by HST. A key focus of future work will be to substantiate and reduce this factor given more realistic simulations of the telescope's thermal response.

Once this procedure is completed for a given spectral band, we repeat it for all bands, 5 bands for the direct imaging channel and 3 bands for the Integral Field Spectrograph (IFS). See Table F-2 for the parameters of all 7 bands. Typically the short-wavelength bands produce the most detections. This is due to diffraction, which places the response window at smaller angles for shorter wavelengths. Since planets tend to cluster at small angular separations, more are captured in short bands.

For planets that are detectable (SNR > 5), we solve for the integration time given the different assumptions regarding telescope jitter and post-production factor. Planets that are detectable within a maximum integration (typically 1 day) are retained. Finally, we generate a graphical display of the results in the form of a plot of contrast as a function of separation angle for detectable planets, for each coronagraph architecture and spectral band. A brief summary of the overall performance of the three architectures is shown in Table F-1. A popular way to indicate the angular range of a coronagraph is to give the inner and outer working angle (IWA, OWA); however, these numbers can be misleading because detection is not a simple binary function of angle. For some architectures the contrast smoothly drops through the inner parts of the dark region. Instead, we show a practical example of the range, using the known RV planets, in columns 2 and 3, where the angular separation of the closest and farthest detectable planet from the parent star is listed. Likewise, the best achievable contrast can be a mis-leading value, as the contrast floors of the architectures here typically vary by a factor of about 100. Thus, we instead use a common example target, 47 UMa c, which has a planet/star contrast of 43 $\times 10^{-10}$, and can be detected by all three architectures in Table F-1. The table gives the 5-sigma sensitivity limit (contrast floor) for each example and the integration time. It is clear from this table that these parameters can vary over a large range in a practical case, so the nominal criterion is not these performance parameters per se. Rather it is the total number of planets that can be detected, as shown in Figure F-1, Figure F-2, and Figure F-3.

| Name | Closest Planet (mas) | Farthest Planet (mas) | 47 U Ma c: a typical planet | | |
|------|------|------|------|------|------|
| | | | Contrast of Planet (e-10) | Contrast Floor (e-10) | Integration Time (hours) |
| HLC | 110 | 450 | 43 | 0.5 | 3.4 |
| SPC | 140 | 320 | 43 | 3 | 4.3 |
| PIAA | 40 | 370 | 43 | 18 | 1.4 |

**Table F-1:** Representative values of the angular range of each coronagraph architecture, and expected values of planet contrast, 5-sigma floor contrast, and integration time for SNR = 5 for a typical planet.

| Band no. | $\lambda_0$ (nm) | $\Delta\lambda_{FWHM}/\lambda_0$ (%) | Science Purpose | Polarization | Channel | Coron. |
|------|------|------|------|------|------|------|
| 1 | 465 | 10.0 | continuum, Rayleigh | Pol. | Imager | HLC |
| 2 | 565 | 10.0 | continuum, Rayleigh | Pol. | Imager | HLC |
| 3 | 660 | 18.0 | $CH_4$ average; disks | Pol. | Imager | HLC |
| | | | $CH_4$ spectrum | Unpol. | IFS | SPC |
| 4 | 835 | 6.0 | $CH_4$ continuum | Pol. | Imager | SPC |
| 5 | 885 | 5.6 | $CH_4$ absorption | Pol. | Imager | SPC |
| 6 | 770 | 18.0 | $CH_4$ spectrum | Unpol. | IFS | SPC |
| 7 | 890 | 18.0 | $CH_4$ spectrum | Unpol. | IFS | SPC |

**Table F-2:** Science filter bands. For each filter, the columns give the central wavelength, ratio of full-width at half-maximum (FWHM) to central wavelength, nominal science purpose, polarization state, destination focal plane, and coronagraph architecture to be used for that filter and destination.

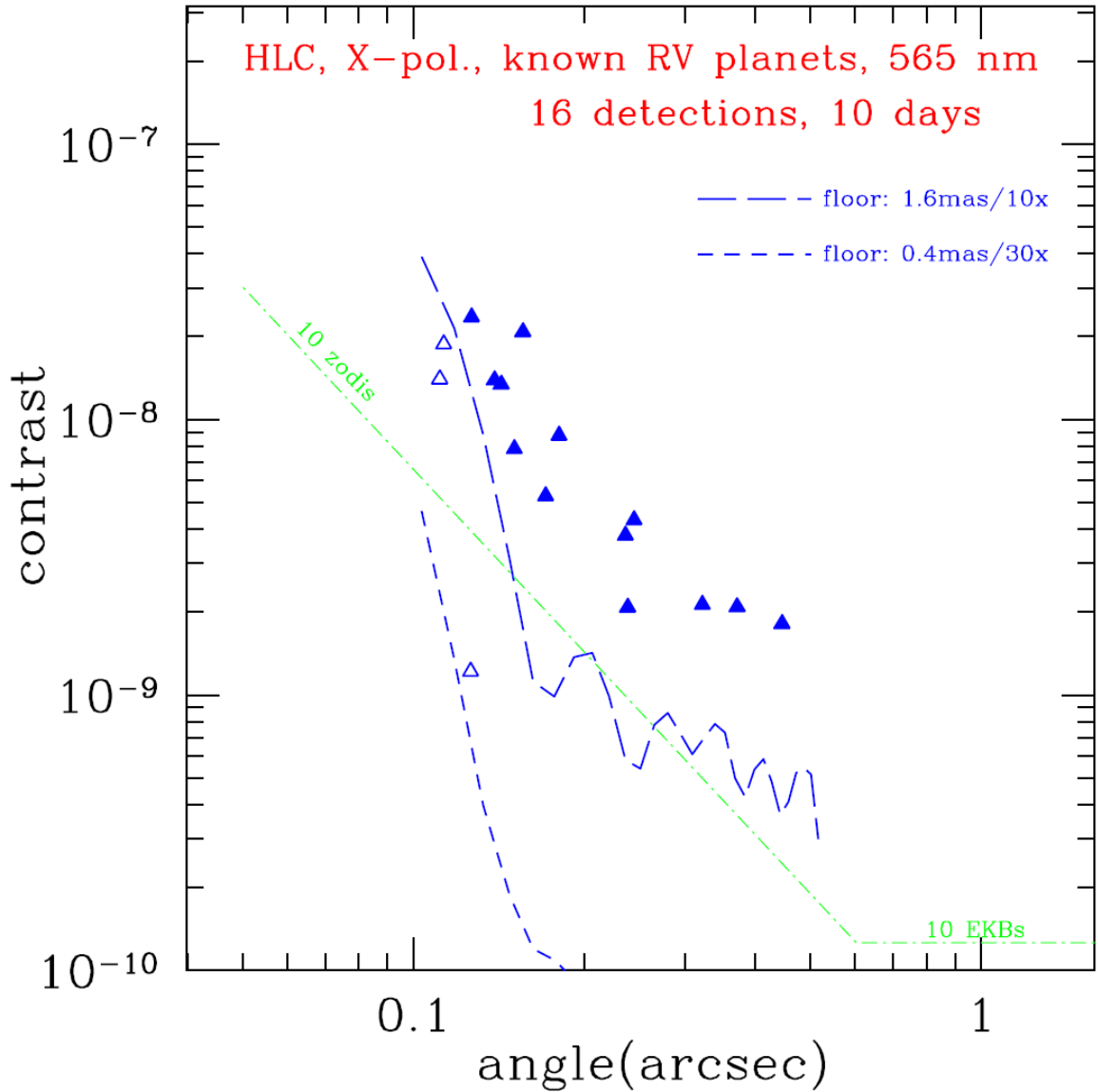

**Figure F-1:** Science imaging yield of known RV planets that are detectable with the HLC, in the 565-nm band (10%), for a single polarization (triangles). The short-dash line (lower curve) is the 5-sigma speckle noise detection floor for the case of a telescope pointing jitter of 0.4 mas RMS residual uncorrected angle and after a post-processing factor of 30 times reduction in the spatial rms speckle noise is applied. The upper long-dash line is similar for a pointing jitter of 1.6 mas and a post-processing factor of only 10. The HLC has a 360-degree azimuth field in a single snapshot. The actual detection sensitivity for a given exposure will also depend on Poisson photon noise as a function of stellar magnitude and integration time; the speckle level sets a floor. The inner and outer working angles are effectively set by the angle limits of these floors as plotted. The solid triangle symbols are detections that are expected to take less than one day each, for the worst-case floor (upper); the open triangles are the extra planets that are detectable for the best-case floor (lower). The dashed green line indicates the estimated contrast of a target zodi cloud that has 10 times more dust than the Solar System, with an estimated continuation into the Edgeworth-Kuiper belt at larger angles, for a target system at 10 pc. For the HLC case shown, we expect to detect about 16 RV planets (S/N=5) in a total integration time of 10 days in a single polarization in this band. Most planets will have an integration time of less than 1 day.

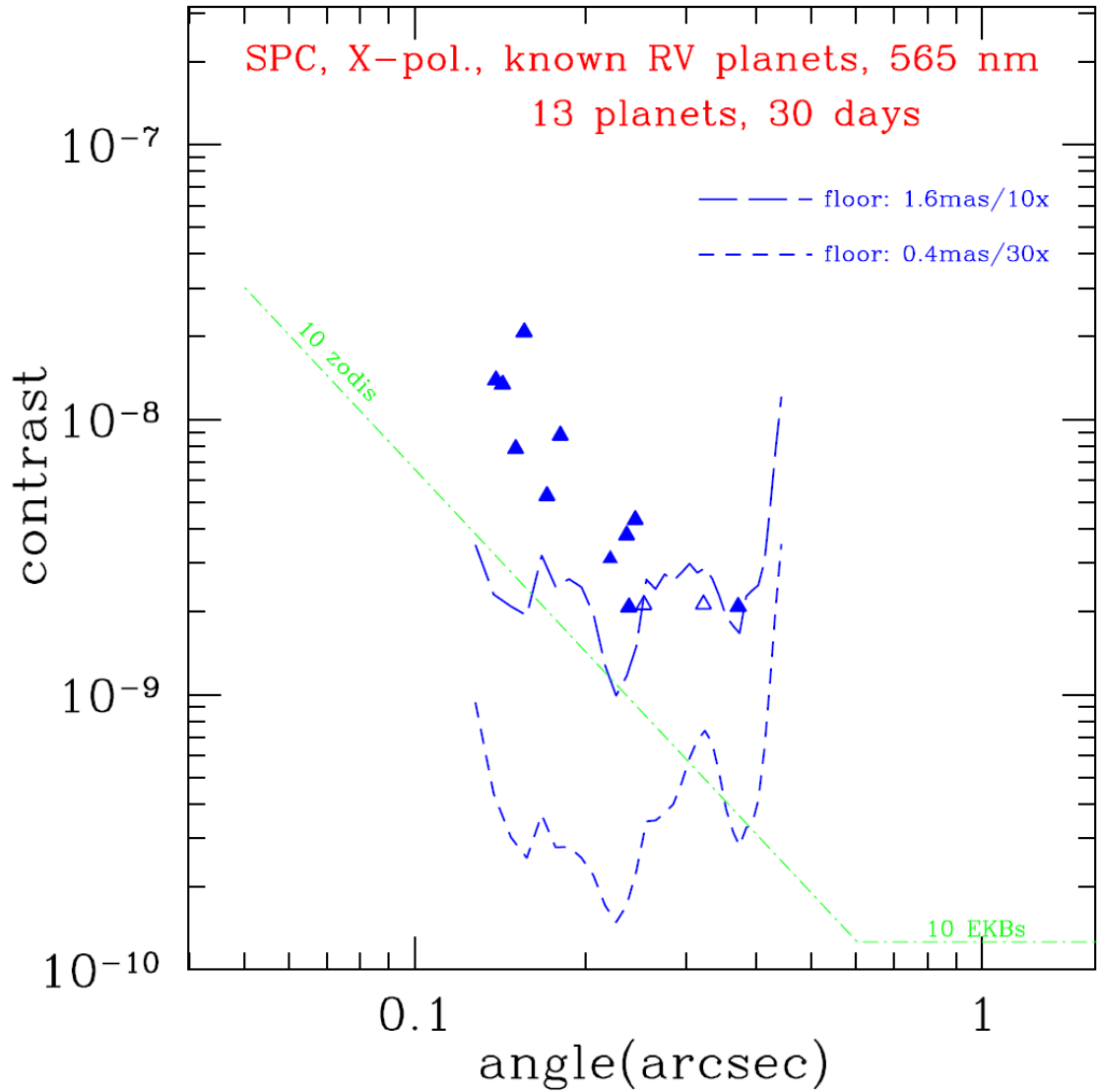

**Figure F-2:** Science imaging yield of RV planets with the SPC, otherwise the same as Figure F-1. Here the science yield is 13 RV-planet detections. Each of which could be done in less than a day if the position angle of the planet were known. However, because it is not known, and because the SPC can only observe an azimuth range of about 120 degrees per snapshot, it will take three such snapshots to cover the full azimuth range, and therefore the total observing time will be about 30 days. The azimuth range of each snapshot with the SPC is only 2×65 degrees, so 3 snapshots are required to cover the entire 360-degree azimuth range, when searching for a planet; however once a planet is discovered, it can then choose the appropriate azimuthal orientation for subsequent snapshots, either to improve the SNR or to observe at a different wavelength. Another difference with respect to the HLC is that the SPC has a higher contrast floor, which reduces the number of detectable planets. A compensating factor is that the SPC is less sensitive to telescope jitter, which gives it a better performance margin in flight and potentially higher spectral bandwidth. The baseline for WFIRST-AFTA is to use the HLC for initial discovery at short wavelength and the SPC for long-wavelength characterization.

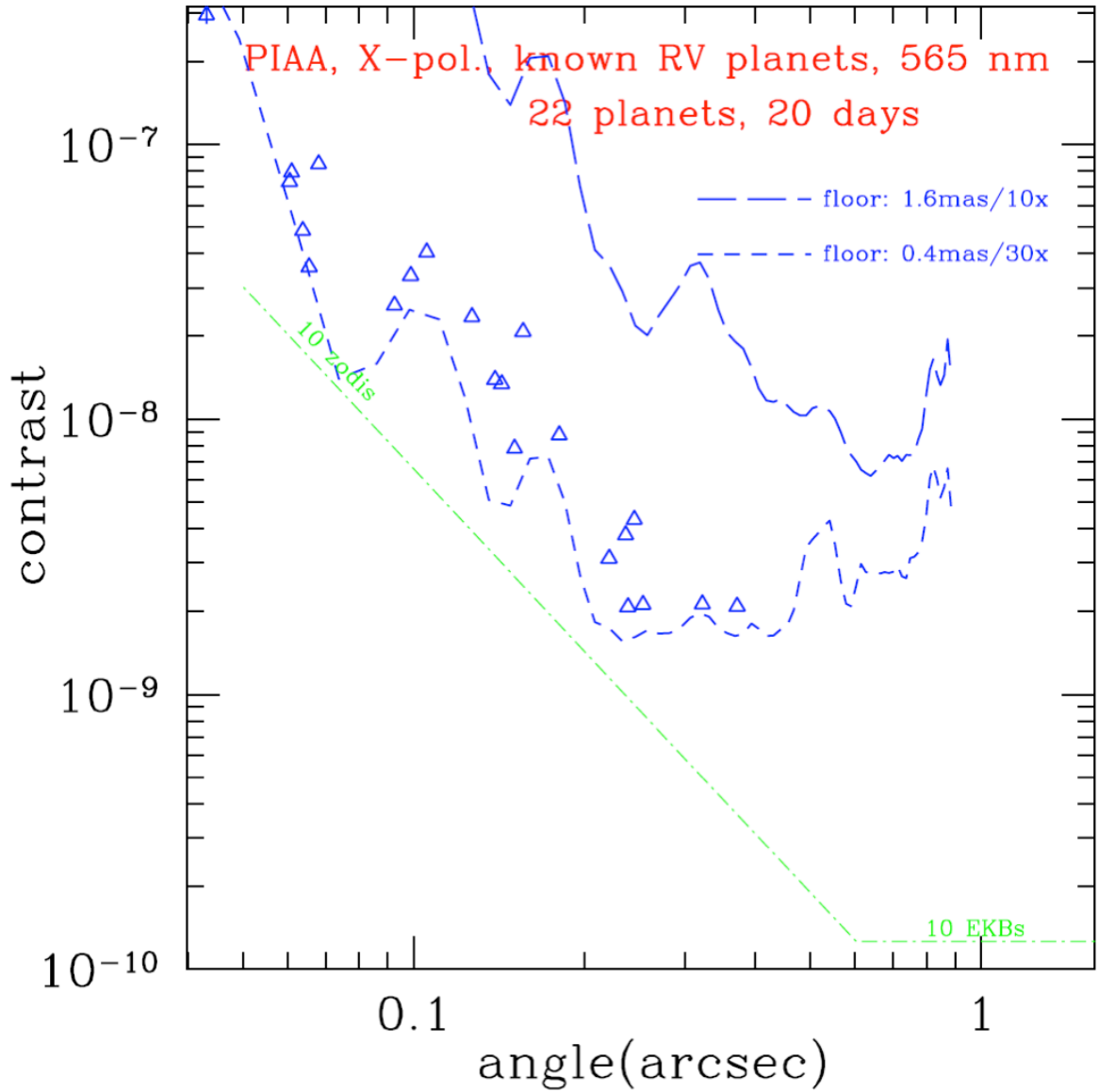

**Figure F-3:** Science imaging yield of the backup architecture, the PIAA coronagraph. This architecture has the twin advantages of having a greater overall throughput as well as having a theoretically smaller inner working angle, allowing it to access considerably more planets close to their star. However an accompanying disadvantage is that it is more sensitive to telescope jitter, which with the adopted values of jitter produces higher 5-sigma floor contrasts, as indicated in the figure. Therefore many planets that were detected by a large contrast margin with the HLC or SPC, are now closer to the PIAA contrast limit, and will therefore require the best possible operating conditions (lowest jitter and most aggressive post-production factor). This version of PIAA also has only one deformable mirror, hence the azimuthal field is 180 degrees, so two snapshots are required in order to fully examine a new system. The higher throughput and smaller inner working angle result in a greater number of detectable RV planets, and at a faster rate per planet than for the HLC or SPC, but operating with a relatively smaller margin.

## F.4. Detectability of Currently-Undiscovered Planets

Analysis of current Doppler data (Howard & Fulton 2014, see also Figure 2-49 in main text) shows that current Doppler surveys have limited sensitivity at the angular separations for which WFIRST-AFTA is sensitive; for example, the 50% Doppler completeness level for the median WFIRST-AFTA target that also has Doppler RV data is ~100 Earth masses at 1 AU, in comparison to WFIRST-AFTA which could detect a 1-Earth mass (or radius) planet at 1 AU, for a typical orbital inclination. Other surveys, and in particular data from the Kepler spacecraft, show that there are a large number of planets below this size. For example, planets from 2-4 Earth radius may be the most common type of planet in the universe (Fressin et al. 2013). Therefore, we also estimated the ability of the CGI to search for as-yet unknown exoplanets, including exoplanets discovered by a future precision RV program.

### F.4.1 25-Star Survey

The first evaluation is for a short program (25 stars, 50 days). We describe here a preliminary estimate of the number of planets that could be seen for one particular extrapolation of the Kepler statistics. Fully evaluating the detection yield - e.g., determining the optimal strategy for revisiting a given star, the best use of partial RV coverage, etc. - will require the development of a complete coronagraph DRM in the next phase. Section F.4.2 below presents an analysis of an extended mission with target optimization.

For our analysis, the frequency of planets, in terms of planet radius and period, is estimated from an analysis of, and extrapolation from, the Kepler planet discoveries (Traub 2015 in prep). To do this, we estimated Kepler's bias against small planets at long periods, owing to instrument noise and stellar faintness, and simulated the mission using this bias function. We hypothesized that the period-radius distribution in the population can be approximated by a separable product of a broken power law in period times a broken power law in radius. We find that this distribution function can fit the observed data with a chi-square of about 2, over a factor of about 100 in period and 50 in radius. This analytic fit produces ~4 planets per star.

The Turnbull catalog of stars out to 30 pc is used as a base sample, but trimmed to have a population of stars that is close to the Kepler population (although this last step seems to make essentially no difference to the result). We randomly assign planets to these nearby stars using the period-radius probability distribution from the Kepler catalog. We then use these results as an input catalog to the same program that is used to calculate the detectability of RV planets. We position the planets at a near-optimal phase (as was done with the RV planets) for convenience in calculation. If these new planets are targeted based on future RV work this is a reasonable assumption; otherwise, multiple visits will be needed for WFIRST to discover planets that are close to the inner working angle and hence only visible for a fraction of their orbit. For brighter planets, most can be detected in as few as 2 or 3 snapshots spaced by several months, though for the lowest-radius planets (which are typically only visible at small separations) detection completeness will be lower.

Using this distribution we simulated a random search for currently-unknown planets around nearby stars. The baseline search targeted 25 stars manually selected from the Turnbull catalog. Each star was visited 4 times and observed for 0.5 days. We split the planets into three groups: giant, sub-Neptunes, and super-Earths. These groups are defined as follows. "Super-Earths" are those with a radius between 0.5 and 2.0 times the Earth's radius and are assigned a geometric albedo of 20%. The sub-Neptune planets, very common in the Kepler population, are between 2 and 4 $R_E$ and are assigned a geometric albedo of 40%. The giant planets, from 4 to 15 $R_E$, are also given the same geometric albedo of 40%.

We simulated both an optimistic performance case (0.4 mas jitter, x30 speckle suppression) and a pessimistic case (1.6 mas jitter, x10 speckle suppression). The results are summarized in Table F-3. In this population, giant planets are relatively rare, and hence few are discovered. Note that the strategy has not been fully optimized. Although the expected number of Super-Earths found is small, any such discovery would be of extraordinary value.

A longer survey would yield more planets, as would a survey that is optimized using techniques such as Savransky et al. (2010) and Stark et al. (2014), or as one that selects targets based on Doppler information, presence of other planets, etc. In these cases, WFIRST-AFTA could expect a yield of ~3 planets in the 1-2 Earth-radius class, as well as a large population of Neptune or sub-Neptune planets.

### F.4.2 Long-Term RV Monitoring

We know that the current population of known RV planets in the WFIRST range is incomplete; new RV discoveries made between now and the launch of WFIRST will increase the potential population of characterizable planets. We expect a NASA-NSF-funded RV spectrometer to come online at the WIYN telescope on Kitt Peak in 2016 and for it to be dedicated to searching for RV planets in connection with new missions, including WFIRST-AFTA. We model the expected yield of this work using the same Kepler-derived distribution function of planets described above, allowing for the expected capabilities of such an instrument, and submitting the RV-discovered planets to our model of the HLC coronagraph on WFIRST-AFTA. We find that 12 of these new planets are detectable with WFIRST-AFTA with exposure times of <1 day (see Figure F-6).

| Planet Type | Radius (Earths) | Number Discovered in 50-day Survey | |
| --- | --- | --- | --- |
| | | Optimistic | Pessimistic |
| Giant | 4 to 15 | 1.4 | 1.1 |
| Sub-Neptune | 2 to 4 | 1.9 | 1.1 |
| Super-Earth | 1 to 2 | 1.3 | 0.5 |
| Total | 1 to 15 | 4.6 | 2.7 |

**Table F-3: Predicted number of planets discovered in a 50-day blind survey, evaluated for an optimistic case (pointing jitter of 0.4 mas RMS after compensation by the coronagraph fast steering mirror and post-processing noise reduction factor of 30 times) and a pessimistic case (1.6 mas jitter and 10 times factor). This prediction is for a totally blind survey, with no foreknowledge of the target stars, however if RV results are available to guide the target selection and timing of observation, these rates could be dramatically increased.**

### F.4.3   Deep Optimized New-Planet Search

The ability to discover new planets in the case above is limited by the time available for the search. We simulated an extended new-planet search using techniques based on Savransky et al. (2010). This survey targets 46 stars, selected from the Turnbull catalog through an algorithm that calculates fractional completeness for each target as a function of mass and semi-major axis and prioritizes targets based on the expected number of planet discoveries per target for a Kepler-based exoplanet distribution. (This analysis used a radius distribution based on Fressin et al. (2013) rather than Traub 2015, but the two distributions are very similar over the range covered here.) This extended survey was allocated 180 days.

Although carrying out this survey completely blind would require 180 days, if targets can be prioritized based on additional information (presence of other planets, weak Doppler signals, stellar properties) it could be executed in significantly shorter time. A preliminary simulation indicates that this survey could obtain full R=70 spectra of ~9 of these newly-discovered smaller planets. Final evaluation of likely planet yield will require development of a full DRM in simulations

Thus, the prospect for detecting new planets around the nearest stars is very promising. These detections by themselves are likely to lead to new science results. Although WFIRST-AFTA's ability to characterize the smallest planets will be limited – most will be faint, and also at such small separations they will be undetectable at longer wavelengths - these detections are also likely to provide known targets for a future mission dedicated to terrestrial habitable-zone planets.

| Planet Type | Radius (Earths) | Number Discovered in 180-day Survey | |
| --- | --- | --- | --- |
| | | Optimistic | Pessimistic |
| Giant | 4 to 15 | 1.9 | 1.2 |
| Sub-Neptune | 2 to 4 | 6.3 | 1.3 |
| Super-Earth | 1 to 2 | 3.8 | 0.14 |
| Total | 1 to 15 | 12.0 | 2.6 |

**Table F-4: Predicted number of planets discovered in a 180-day blind survey, otherwise as in Table F-3. The values in this table were estimated completely independently from those in Table F-3, so it is remarkable that the values for at least the Optimistic case are in reasonable accord.  In future work, these methods of estimation will be revised and reconciled to eliminate any remaining differences.**

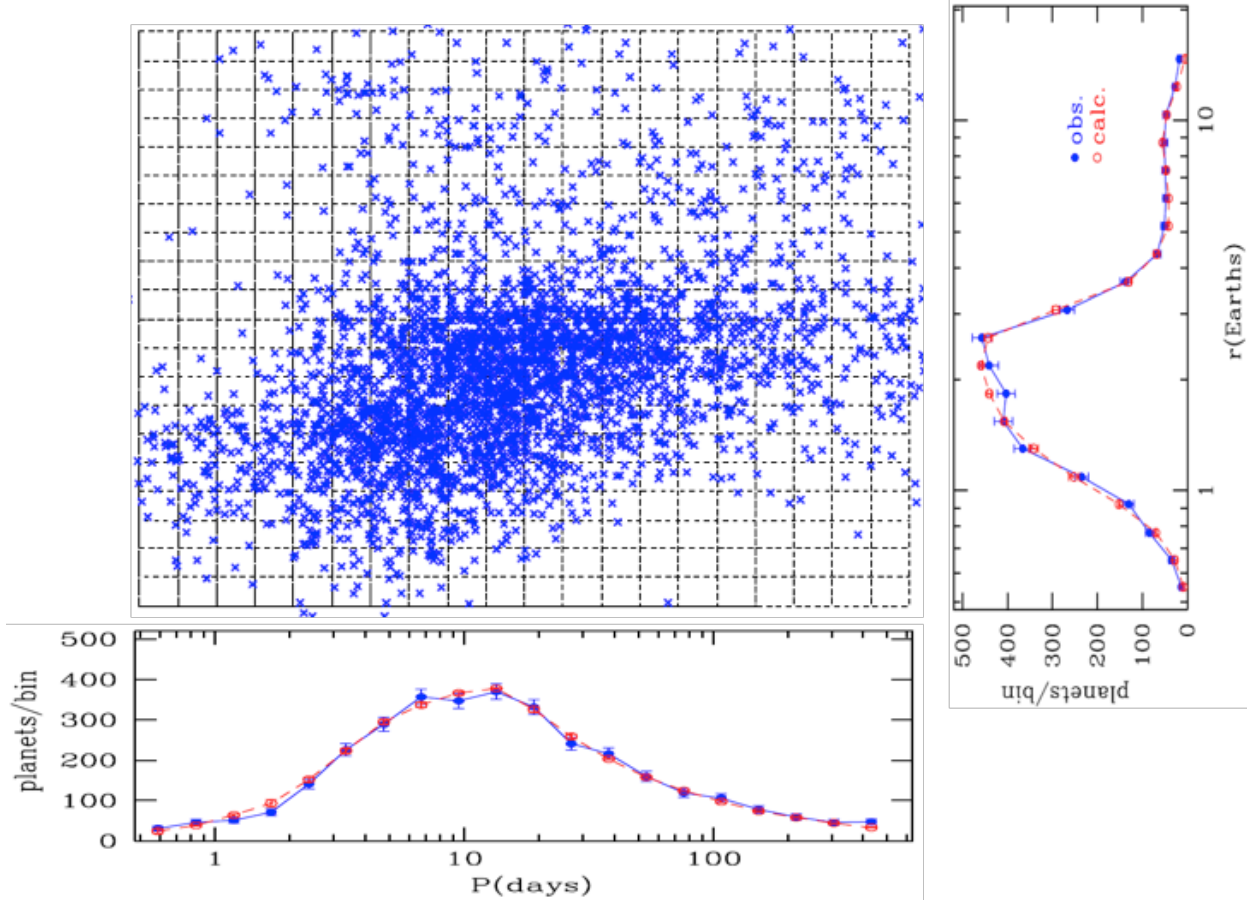

**Figure F-4: (Upper left)** Kepler planet detections are shown as a function of period in units of days (horizontal axis) and planet radius in units of Earth radii (vertical axis). **(Bottom)** The number of planets in each period bin is added up and displayed as blue points, with statistical error bars indicated. The red points are a model fit to the data, allowing for planets that are not seen because they do not transit their star, and also allowing for the reduced sensitivity of the Kepler mission at long periods and for small planets. **(Right)** The number of planets in each radius bin is projected (blue points), and the functional fit is shown by red points. The excellent functional fit to the observed number of planets is our basis for asserting that the model used here for the number of planets per star, as a function of period and radius, is well determined, and can be used to project the expected yield of new planets from the WFIRST-AFTA coronagraph.

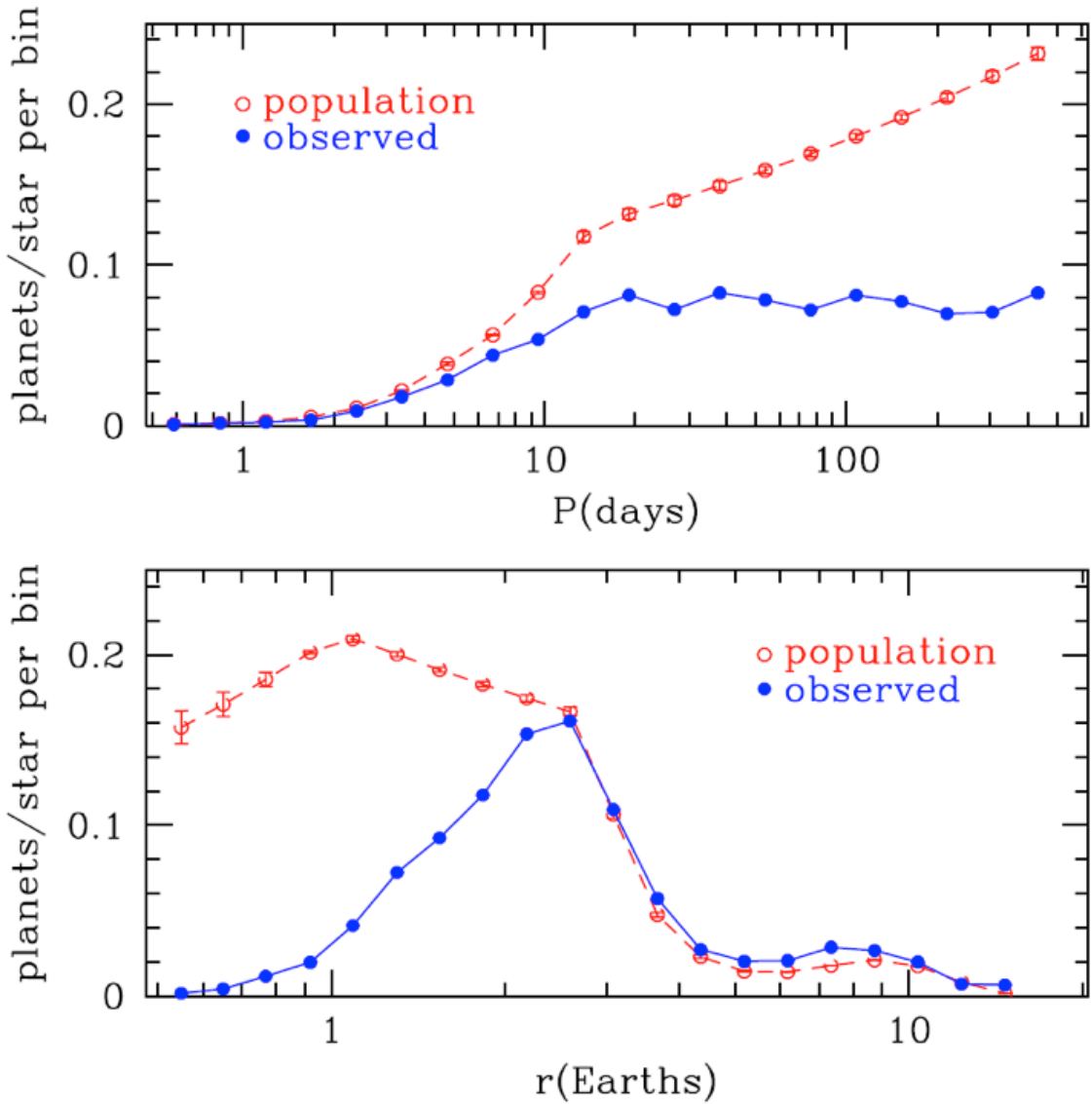

**Figure F-5:** Distribution of the number of planets per star, per bin, as a function of period and radius, in the Kepler population, as seen through the filter of the Kepler mission (blue "observed" points) and without any filtering (red "population" points), as estimated from the observed Kepler planets in combination with a model of the transit probability and the instrument plus star noise. Kepler's total noise (star plus instrument) masks planets at long periods and small sizes, which leads to the gap between the two curves in each panel. The governing law assumed here is that the distribution function is the product of a period-dependent factor times a radius-dependent factor, an assumption that is borne out by the quality of the fits in Figure F-4.

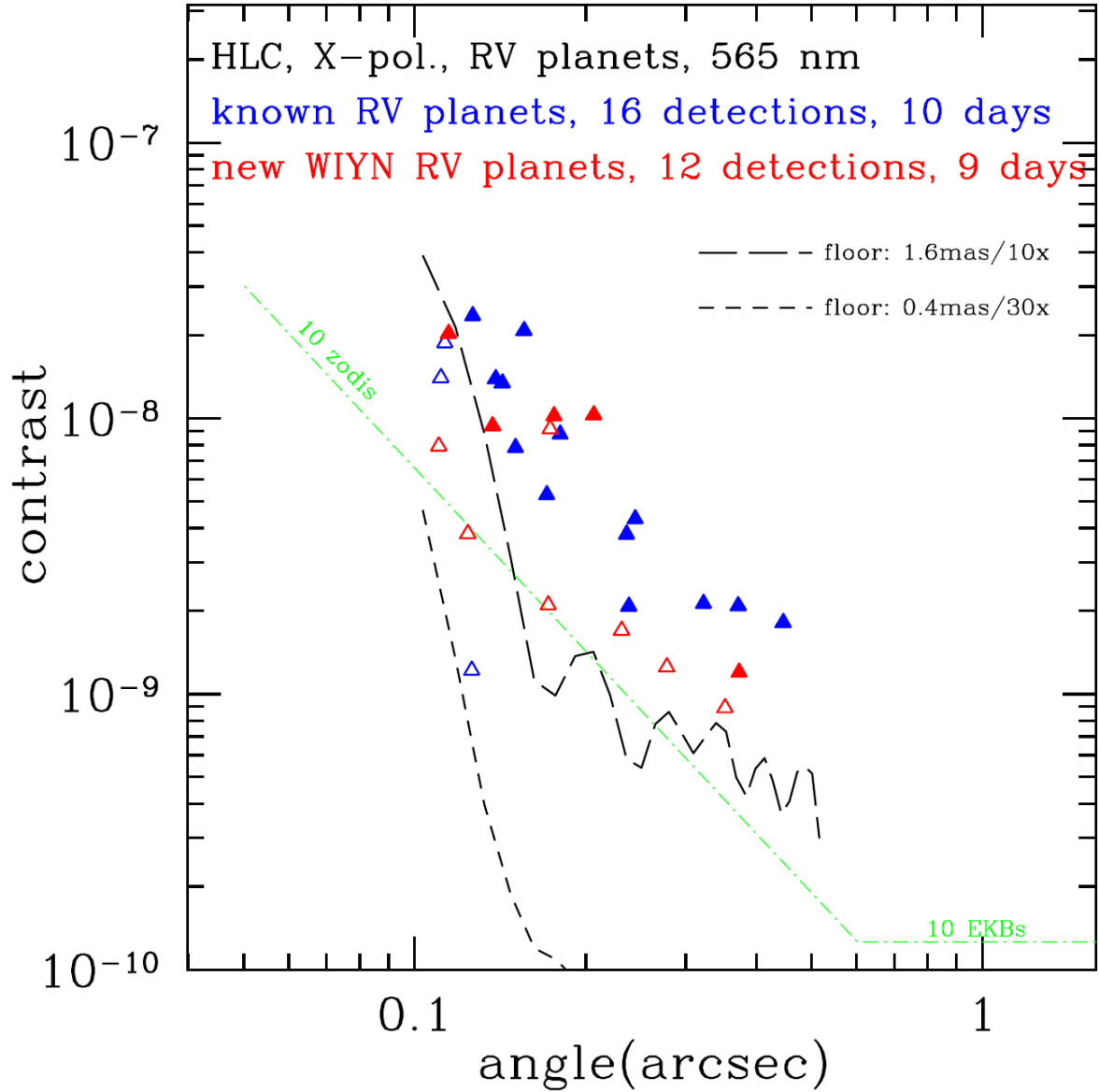

**Figure F-6: Same as Figure F-1 but with a simulated population of planets discovered through an extended WIYN RV survey plotted in red.**

## F.5. Mission Time Estimate

Starting with the integration times for RV planets and new planets given above, we can make a zeroth-order estimate of how the CGI could spend its one-year allocation of time on the WFIRST-AFTA mission. There are six main categories to consider: (1) RV planet images; (2) RV planet spectra; (3) disk images; (4) new planet searches; (5) GO programs; (6) overhead. A nominal time estimate for each of these categories is sketched below. Note that this is not a full DRM, merely a strawman allocation of coronagraph mission time.

### F.5.1 Time for Photometric Characterization of Known RV Planets

The total time required for a first-time detection of an RV planet is mainly determined by just the integration time assuming that the planet is in an approximately optimum orbital position, plus whatever overhead is needed to tune the dark hole on a brighter star. We can be relatively sure that we can predict the right time to observe an RV planet because we already have in hand the ephemeris from RV observations, with only the inclination of the orbit and the position angle missing. For the current work, we assume a statistically average inclination of 60 degrees. For the HLC, with a full 360 degrees of azimuth in a single snapshot, only one snapshot of a target needs to be taken. Numerical simulations show that all of the RV planets can be detected on the first try if they are observed when they are in the orbital range from about 45 to 90 degrees, where 90 is maximum elongation. This range occurs 25% of the time (45/180 = 0.25), allowing for orbital symmetry. Most RV periods are in the range of 2 to 10 years, so the time when a planet is in an optimum orbital window is easily predicted beforehand, particularly as ongoing Doppler measurements refine the orbital solution. So we start by assuming that we can detect 16 planets in 10 days of observing time, at 565 nm, in x-polarization, with no additional searching time. We add 10 days more for measuring the y-polarization. Add another 10 days for observing the same planets in the 465 nm band, single polarization. Add another 10 days for the 660 band. All of these are in imaging mode. Simulations stand behind these time estimates. The total is 40 days for photometric characterization RV planets.

### F.5.2 Time for Spectra of Known RV Planets

We use the SPC and the IFS for spectra in the 660, 770, and 890 nm bands, at a spectral resolution of about 70. Two factors govern the yield of spectra: first, the inner working angle moves outward as the wave-length increases, cutting off planets that are too close to their star; second, the quantum efficiency of the detector decreases at longer wavelengths. The net result is that we expect to obtain spectra of RV planets at the rates of about 7 spectra in 38 days in the 660 nm band, 6 spectra in 40 days in the 770 nm band, and only 3 spectra in 20 days in the 890 nm band. All spectra will be in the unpolarized mode.

### F.5.3 Time for Disk Images

Disks will already be included in the search for RV planets, as we cannot avoid seeing whatever disks are already there, within the limits of our integrations time per target. In Figure F-1, Figure F-2, and Figure F-3, we indicate by a dashed green line the approximate brightness of a zodi disk like the Solar System's, but with a 10-times greater particle density, for a Sun-like star at 10 pc. For the fixed solid-angle of a pixel in the focal plane, the surface brightness of a disk is independent of target distance, but the angular size will scale as the inverse of distance, and the contrast will increase linearly with distance. The green line is for a single PSF, so the signal from a distributed disk will be determined by the ensemble of PSFs across the disk, improving the detection threshold by a factor of several. Thus we expect that the CGI will be able to detect a disk at roughly the 1-zodi level, given all these caveats. In addition to disks around RV targets, we may wish to look at a score or more of other targets, e.g., selected on the basis of infrared observations, or as a control population of stars without Doppler planets, for which we allocate an extra 10 days. The total is thus 10 dedicated days for disks.

### F.5.4 Time for New Planet Searches

For as-yet undiscovered ("new") planets, a precise estimate of search time requires a full DRM, which will be carried out in future phases. We initially allocate 50 days for this search, plus 20 days for acquisition/overhead/follow-up of candidate planets. As discussed above, an extended 180-day search could significantly increase yield, including time for spectrophotometric follow-up.

### F.5.5 GO Program

The WFIRST mission allocates about 25% of the time to a General Observer (GO) program, so for the CGI this is 91 days.

## F.5.6 Overhead

We assume that we will need 1 day to set up the dark hole for each RV planet that we observe for the first time. Then, for a given RV target, the subsequent observations in a different polarization and different filters should take less time to tune the dark hole, for which we assume a quarter-day each for roughly 160 targetings. (We do not count here an estimated 5 days of coronagraph time for a dedicated checkout period, which occurs before the official 6-year mission starts). This leaves us with a grand total of 56 days of overhead, conservatively estimated.

## F.5.7 Total Time

Adding the above time allocations for the CGI, we find a total of 365 days required for the proposed program, as shown in Table F-5.

| L1 Product | Rqmt. | Performance Assessment | Inst. | Comment | Clock Time (days) |
|---|---|---|---|---|---|
| Exoplanet Images | >12 planets | 16 RV planets in 10 days at 565 nm, x-pol. | HLC | HLC meets rqmts, with margin | 40 |
| | | 16 RV planets in 10 days at 465 nm, x-pol. | | | |
| | | 16 RV planets in 10 days at 565 nm, y-pol. | | | |
| | | 16 RV planets in 10 days at 660 nm, x-pol. | | | |
| Exoplanet Spectra | >6 planets | 7 RV spectra in 38 days, 660 nm, unpol. | SPC | SPC meets rqmts, with margin | 98 |
| | | 6 RV spectra in 40 days, 770 nm, unpol. | | | |
| | | 3 RV spectra in 20 days, 890 nm, unpol. | | | |
| Disk Images | Several disks | During all imaging observations | HLC | Probably OK, need detailed simulations | 10 |
| | | Additional 10 days dedicated to disks | | | |
| New Planet Searches | None | Expect ~2.5 planets per star accessible range of radii and periods | HLC | Need detailed simulations | 70 |
| GO Programs | 25% of Time | General Observer projects | HLC, SPC | Community proposals | 91 |
| Overhead | None | 16 days for each initial RV obs. | HLC | Set-up times and stability need detailed simulations | 56 |
| | | ~0.25 days per target, for ~160 additional observations | HLC | | |
| Total Time | | | | | 365 |

**Table F-5: Time allocation and performance against requirements for the CGI, for each major category of observation. To be revised when detailed engineering overhead times are available, and a realistic DRM can be calculated.**

## Appendix G    Precursor Observations

The NASA Exoplanet Exploration Program Analysis Group (ExoPAG) recently completed a study entitled "Preparing for the WFIRST Microlensing Survey" that identified several precursor observation programs that will enable and enhance WFIRST microlensing science, and reduce the mission's scientific risk (Yee et al. 2014). The most important of these are:

1. **A Ground-Based, Near-IR, Microlensing Survey.** The near-IR microlensing rate as a function of Galactic position is a critical input for the selection of the WFIRST-AFTA exoplanet fields and design of the survey strategy, but the microlensing rate has never been measured in the IR. Galactic models are currently insufficient for accurate predictions of the microlensing lensing rate and optical microlensing surveys are insensitive to microlensing in the higher extinction regions closer the Galactic plane where the microlensing rate is expected to be higher. Thus, a near-IR microlensing survey of the central Galactic bulge is needed.

2. **Hubble Space Telescope Precursor Observations.** The infrared channel of HST's Wide Field Camera 3 camera provides images with similar angular resolution, pixel sampling and passbands to WFIRST-AFTA. Thus, it is the ideal instrument to provide test observations for development of the WFIRST-AFTA exoplanet microlensing survey photometry-astrometry pipeline. In order to simulate both the photometric and astrometric signals, we will need well-dithered, multi-epoch data. Optical and IR HST observations spanning the candidate WFIRST-AFTA exoplanet microlensing fields will provide deep color-magnitude diagrams that can be combined with the IR microlensing rate measurements to build Galactic bulge models that can be used to make accurate predictions of the microlensing rates. Optical HST observations of previous planetary microlensing events can be used to infer the distance to the host stars, and thus determine the Galactic distribution of planets, which will in turn inform the choice of the WFIRST-AFTA target fields. Finally, WFIRST-AFTA has the potential of measuring astrometric microlensing for many of the faint planetary host stars and stellar remnants. This capability could be enhanced with precursor HST observations of a large fraction of the WFIRST-AFTA exoplanet microlensing fields. Observations of the exoplanet microlensing fields will provide long timeline baseline data for testing astrometric measurements from WFIRST-AFTA observations alone.

**Appendix H    LSST/Euclid/WFIRST U. Penn Report**

# The Whole is Greater than the Sum of the Parts: Optimizing the Joint Science Return from LSST, Euclid and WFIRST

January 30, 2015

B. Jain,[1] D. Spergel,[2] R. Bean, A. Connolly, I. Dell'antonio, E. Gawiser, N. Gehrels, L. Gladney, K. Heitmann, G. Helou, C. Hirata, S. Ho, Ž. Ivezić, M. Jarvis, S. Kahn, J. Kalirai, A. Kim, R. Lupton, R. Mandelbaum, P. Marshall, J. A. Newman, M. Postman, J. Rhodes, M. A. Strauss, J. A. Tyson, W. M. Wood-Vasey

## Contents

---

[1]bjain@physics.upenn.edi
[2]dns@astro.princeton.edu

# 1   Where will we be in 2024?

Astronomy in 2024 should be very exciting! LSST and Euclid, which should each be in the midst of their deep surveys of the sky, will be joined by WFIRST. With higher resolution and sensitivities than previous astronomical survey instruments, they will reveal new insights into areas ranging from exoplanets to the nature of dark energy. At the same time, JWST will be staring deeper into the early universe than ever before. Advanced LIGO should be detecting frequent collisions between neutron stars. ALMA will be operating at all of its planned frequencies, and the new generation of very large optical ground based telescopes should be revolutionizing ground-based optical astronomy. In parallel, advances in computational capabilities should enable observers to better exploit these complex data sets and theorists to make detailed time-dependent three-dimensional models that can capture much of the physics needed to explain the new observations.

The focus of this report is an exploration of some of the opportunities enabled by the combination of LSST, Euclid and WFIRST, the surveys that will be an essential part of the next decade's astronomy. The sum of these surveys has the potential to be significantly greater than the contributions of the individual parts. As is detailed in this report, the combination of these surveys should give us multi-wavelength high-resolution images of galaxies and broadband data covering much of the stellar energy spectrum. These stellar and galactic data have the potential of yielding new insights into topics ranging from the formation history of the Milky Way to the mass of the neutrino. However, enabling the astronomy community to fully exploit this multi-instrument data set is a challenging technical task: for much of the science, we will need to combine the photometry across multiple wavelengths with varying spectral and spatial resolution. Coordination will be needed between the LSST, Euclid, and WFIRST projects in order to understand the trades between overlapping areal coverage, filter design, depth and cadence of the observations, and performance of the image analysis algorithms. We will need to provide these data to the community in a highly usable format. If we do not prepare the missions for this task in advance, we will limit their scientific return and increase the cost of the eventual effort of fully exploiting these data sets. The goal of this report is to identify some of the science enabled by the combined surveys and the key technical challenges in achieving the synergies.

## 2 Background on missions

This section covers the science goals and technical overview of the three missions.

### 2.1 WFIRST

WFIRST is the top-ranked large space mission of the 2010 New Worlds New Horizon (NWNH) Decadal Survey. With the addition of a coronagraph, WFIRST would also satisfy the top ranked medium space priority of NWNH. The mission is designed to settle essential questions in dark energy, exoplanets, and infrared astrophysics. The mission will feature strategic key science programs plus a vigorous program of guest observations. A Science Definition Team and a Study Office at the Goddard Space Flight Center and Jet Propulsion Laboratory are studying the mission with reports[3] in 2013, 2014, and 2015.

WFIRST is now baselined with an existing 2.4 m telescope NASA acquired from the National Reconnaissance Office (NRO), a telescope that became available after the completion of NWNH. This configuration is referred to as WFIRST-AFTA (Astrophysics Focused Telescope Asset). The mission consists of telescope, spacecraft, a Wide-Field Instrument (WFI), an IFU spectrometer and a coronagraph instrument, which includes an IFS spectrometer. The WFI operates with multiple bands covering 0.7 to 2.0 micron band with extension to 2.4 microns under study. It has a filter wheel for multiband imaging and a grism for wide-field spectroscopy (R=550-800). The pixel scale is 0.11 arcsec, which fully samples the PSF at the H band. The IFU has a 3 arcsec FoV and R=75 resolution. The coronagraph operates in the visible 400 - 1000 nm band. It has a 2.5 arcsec FoV, $10^{-9}$ effective contrast and 100-200 mas inner working angle. The IFS has R=70 resolution.

WFIRST will measure the expansion history of the Universe as a function of cosmic time and the growth rate of the large-scale structure of the Universe as a function of time to test the theory of general relativity on cosmological scales and to probe the nature of dark energy. It will employ five different techniques: supernovae, weak lensing, baryon acoustic oscillations (BAO), redshift space distortions (RSD), and clusters. With its 2.4-meter primary mirror, the mission will measure more than double the surface density of galaxies detected by the Euclid mission and push these measurements further in the NIR. This higher density will enable more detailed maps of the dark matter distribution, measurements not only of two-point statistics but also higher order statistics, and multiple tests of systematic dependance of cosmological parameters on galaxy properties. The NRC noted in its recent review[4]: "For each of the cosmological probes described in NWNH, WFIRST/AFTA exceeds the goals set out in NWNH."

For exoplanets, WFIRST will complete the census of exoplanets begun by Kepler. It will make microlensing observations of the galactic bulge using the WFI to discover several thousand planets at the orbit of Earth and beyond and will be sensitive to planets as small as Mars. WFIRST's microlensing survey will have the unique capability to detect free-floating planets, thus, enabling astronomers to determine the efficiency of planet formation. The coronagraph will directly image and characterize tens of Uranus to Jupiter mass planets around nearby stars and study debris disks.

WFIRST will also be a worthy successor to the Hubble Space Telescope. With 200 times the field-of-view of HST, and the same size primary mirror, it will conduct a rich program of general astrophysics and reveal new insights and discoveries on scales ranging from the nearest stars to the most distant galaxies.

Table 1 gives the WFIRST capabilities for surveys and SN monitoring observations.

*High Latitude Survey (HLS):* The nominal HLS will be for 2 years with 1.3 years for imaging and 0.6 years for grism spectroscopy. The coverage will be 2200 deg$^2$ of high Galactic latitude sky within the LSST footprint. The imaging will have 2 passes over the survey footprint in each of the 4 imaging filters (J, H, F184 [for shapes] and Y [for photo-z's]). Data from LSST will be required to provide optical filters for

---

| Attributes | WFIRST Capability |
|---|---|
| Imaging survey | J ∼ 27 AB over 2400 sq deg |
| | J ∼ 29 AB over 3 sq deg deep fields |
| Multi-filter photometry | Filters: z, Y, J, H, F184 (Ks), W (wide) |
| Slitless wide-field spectroscopy | 0.28 sq deg, R∼600 |
| Slit multi-field spectroscopy | IFU, R∼70 |
| Number of SN Ia | $2 \times 10^3$ to z∼1.7 |
| Number galaxies with spectra | $2 \times 10^7$ |
| Number galaxies with shapes | $4 \times 10^8$ |
| Number of galaxies detected | few $\times 10^9$ |
| Number of massive clusters | $4 \times 10^4$ |

Table 1: **WFIRST capabilities in a nominal ∼ 2.5 year dark energy survey**

photo-z determination. Each pass will include four 184 sec exposures in each filter (with five exposures in J band) with each exposure offset diagonally by slightly more than a detector chip gap. This pattern is repeated across the sky. The spectroscopic survey will have four passes total over the survey footprint with two "leading" passes and two "trailing" passes to enable the single grism to rotate relative to the sky. Each pass includes two 362 sec exposures with offset to cover chip gaps. The two "leading" passes (and two "trailing") are rotated from each other by ∼ 5°. It is anticipated that the sky coverage of the HLS will be significantly expanded in the extended phase of the mission after the baseline first 6 years. WFIRST has no consumables that would prohibit a mission of 10 years or longer, and is being designed with serviceability in mind to enable the possibility of an even longer mission.

*Supernova Survey:* A six month SN Ia survey will employ a three-tier strategy so to track supernova over a wide range of redshifts:

- Tier 1 for z<0.4: 27.44 deg² Y=27.1, J= 27.5

- Tier 2 for z<0.8: 8.96 deg² J=27.6, H= 28.1

- Tier 3 for z<1.7: 5.04 deg² J=29.3, H=29.4

Tier 3 is contained in Tier 2 and Tier 2 is contained in Tier 1. Each of these fields will be visited every 5 days over the 6 months of the SN survey. The imager is used for SN discovery and the IFU spectrometer is used to determine SN type, measure redshifts, and obtain lightcurves. Each set of observations will take a total of 30 hours of combined imaging and spectroscopy. The fields are located in low dust regions ≤ 20° off an ecliptic pole. A final revisit for each target for spectroscopy will occur after the SN fades for galaxy subtraction.

*Guest Observer Program:* A significant amount of observing time will be awarded to the community through a peer-selected GO program. An example observing program prepared by the SDT has 25% of the baseline 6 years of the mission, or 1.5 years, for guest observations. Significantly more time would be awarded in an extended phase after the 6th year. The GO program is expected to cover broad areas of science from the solar system to Galactic studies to galaxies to cosmology. The April 2013 SDT study contains a rich set of ∼50 potential GO science programs that are uniquely enabled by WFIRST. Astronomical community members provided a wide-ranging set of one-page descriptions of different GO programs that highlight the tremendous potential of WFIRST to advance many of the key science questions formulated by the Decadal survey. An important contribution that would likely come from key projects in the GO program would be to have deep IR observations of selected few-square-degree fields. These ultra deep drilling fields could reach limits of J=30 AB.

## 2.2 LSST

The Large Synoptic Survey Telescope (LSST) is a large-aperture, wide-field, ground-based facility designed to perform many repeated imaging observations of the entire southern hemisphere in six optical bands (u, g, r, i, z, y). The majority of the southern sky will be visited roughly 800 times over the ten-year duration of the mission. The resulting database will enable a wide array of diverse scientific investigations ranging from studies of moving objects in the solar system to the structure and evolution of the Universe as a whole (Ivezic et al 2014).

The Observatory will be sited atop Cerro Pachon in Northern Chile, near the Gemini South and SOAR telescopes. The telescope incorporates a 3-mirror astigmatic optical design. Incident light is collected by the primary, which is an annulus with an outer diameter of 8.4 m, then reflected to a 3.4-m convex secondary, onto a 5-m concave tertiary, and finally into three refractive lenses in the camera. The total field of view is 9.6 deg$^2$ and the effective collecting aperture is 6.6 m in diameter. The design maintains a 0.2 arc-second system point spread function (PSF) across the entire spectral range of 320 nm to 1050 nm. The etendue (the product of collecting area and field of view) of the system is several times higher than that of any other previous facility.

The telescope mount assembly is a compact, stiff structure with a high fundamental frequency that enables fast slew and settle. The camera contains a 3.2 Gigapixel focal plane array, comprised of roughly 200 4K × 4K CCD sensors with 10 $\mu$m pixels. The sensors are deep depleted, back-illuminated devices with a highly segmented architecture that enables the entire array to be read out in 2 s or less.

Four major science themes have motivated the definition of science requirements for LSST:

- Taking a census of moving objects in the solar system.

- Mapping the structure and evolution of the Milky Way.

- Exploring the transient optical sky.

- Determining the nature of dark energy and dark matter.

These four themes stress the system in different ways. For the science of dark energy and dark matter, LSST data will enable a variety of complementary analyses, including measurements of cosmic shear power spectra, baryon acoustic oscillations, precision photometry of Type Ia supernovae, measurements of time-delays between the multiple images in strong lensing systems, and the statistics of clusters of galaxies. Collectively, these will result in substantial improvements in our constraints on the dark energy equation of state and the growth of structure in the Universe, among other parameters.

The main fast-wide-deep survey will require 90% of the observing time and is designed to optimize the homogeneity of depth and number of visits. Each visit will comprise two back-to-back 15 second exposures and, as often as possible, each field will be observed twice per night, with visits separated by 15-60 minutes. Additional survey areas, including a region within 10 degrees of the northern ecliptic, the South Celestial Pole, and the Galactic Center will be surveyed with either a subset of the LSST filter complement or with fewer observations. The remaining 10% of the observing time will be used on mini-surveys that improve the scientific reach of the LSST. Examples of this include the "Deep Drilling Fields" where each field receives approximately 40 hour-long sequences of 200 exposures. When all 40 sequences and the main survey visits are coadded, this would extend the depth of these fields to $r \sim 28$. The LSST has identified four distant extragalactic survey fields to observe as Deep Drilling Fields: Elias S1, XMM-LSS, Extended Chandra Deep Field-South, and COSMOS.

Table 2 gives the LSST baseline design and survey parameters

The rapid cadence of the LSST will produce an enormous volume of data, roughly 15 Terabytes per night, leading to a total dataset over the ten years of operations of over a hundred Petabytes. Processing such a large dataset and archiving it in useful form for access by the community has been a major design consideration for the project. The data management system is configured in three layers: an infrastructure layer consisting of the computing, storage, and networking hardware and system software; a middleware layer, which handles distributed processing, data access, user interface, and system operations services; and an applications layer, which includes the data pipelines and products and the science data archives. There will be both a mountain summit and a base computing facility (located in La Serena, the city nearest the telescope site), as well as a central archive facility in the United States. The data will be transported over existing high-speed optical fiber links from South America to the U.S.

| Attributes | LSST Capability |
|---|---|
| Final f-ratio, aperture | f/1.234, 8.4 m |
| Field of view, étendue | 9.6 deg$^2$, 319 m$^2$deg$^2$ |
| Exposure time | 15 seconds (two exposures per visit) |
| Main survey area | 18,000 deg$^2$ |
| Pixel count | 3.2 Gigapix |
| Plate scale | 50.9 $\mu$m/arcsec (0.2" pix) |
| Wavelength coverage | $320 - 1050$ nm, $ugrizy$ |
| Single visit depths, design | 23.7, 24.9, 24.4, 24.0, 23.5, 22.6 |
| Mean number of visits | 56, 80, 184, 184, 160, 160 |
| Final (coadded) depths | 25.9, 27.3, 27.2, 26.8, 26.3, 25.4 |

Table 2: **LSST Capabilities for the fast-wide-deep main survey**

The observing strategy for the LSST will be optimized to maximize observing efficiency by minimizing slew and other down time and by making appropriate choices of filter bands given the real-time weather conditions. The final cadence selection will be undertaken in consultation with the community. A prototype simulator has been developed to help evaluate this process, which will be transformed into a sophisticated observation scheduler. A prototype fast Monte Carlo optical ray trace code has also been developed to simulate real LSST images. This will be further developed to aid in testing science analysis codes. The LSST cadence will take into account the various science goals and can also be refined to improve the synergy with other datasets.

LSST anticipates first light in 2020 and the start of the 10-year survey in 2022. It is fully funded for construction as of the summer 2014.

## 2.3 Euclid

Euclid is a medium class mission within the European Space Agency's (ESA) 'Cosmic Vision' program[5]. Euclid was formally selected in 2011 and is currently on schedule for launch in 2020. After traveling to the L2 Lagrange point and a brief shakeout and calibration period, Euclid will undertake an approximately 6-year survey aimed at "mapping the geometry of the dark Universe." Euclid will pursue four primary science objectives (Laureijs et al 2011):

- Reach a dark energy FoM $> 400$ using only weak lensing and galaxy clustering; this roughly corresponds to 1% errors on $w_p$ and $w_a$ of 0.02 and 0.1, respectively.

- Measure $\gamma$, the exponent of the growth factor, with a $1\sigma$ precision of $< 0.02$, sufficient to distinguish General Relativity and a wide range of modified-gravity theories.

---

[5]*http://sci.esa.int/cosmic-vision/*

- Test the Cold Dark Matter paradigm for hierarchical structure formation, and measure the sum of the neutrino masses with a $1\sigma$ precision better than 0.03 eV.

- Constrain $n_s$, the spectral index of the primordial power spectrum, to percent accuracy when combined with Planck, and to probe inflation models by measuring the non-Gaussianity of initial conditions parameterized by $f_{NL}$ to a $1\sigma$ precision of $\sim$2.

Euclid is optimized for two primary cosmological probes: weak gravitational lensing and galaxy clustering (including BAO and RSD). Therefore, Euclid will measure both the expansion history of the Universe and the growth of structure. Euclid comprises a 1.2m Korsch 3 mirror anastigmatic telescope and two primary science instruments. The visible instrument (VIS) consists of 36 4k $\times$ 4k CCDs and will be used to take images in a single, wide (riz) filter for high precision weak lensing galaxy shape measurements. The light entering the telescope will be split via a dichroic to allow for simultaneous observations with the Near Infrared Spectrometer and Photometer (NISP), which will take 3 band imaging (Y, J, H) and grism spectroscopy in the 1-2 $\mu$m wavelength range. The NISP imaging is aimed at producing high quality photometric redshifts when combined with ground-based optical data from a combination of telescopes in the southern and norther hemispheres and the spectroscopy is aimed at producing accurate maps of galaxy clustering over 2/3 of the age of the Universe. NISP will contain 16 H2RG 2k $\times$ 2k NIR detectors procured and characterized by NASA. Together, VIS and NISP will survey the darkest (least obscured by dust) 15,000 square degrees of the extragalactic sky, providing weak lensing shapes for over 1.5 billion galaxies and emission line spectra for several tens of millions of galaxies, while taking full advantage of the low systematics afforded by a space mission.

While the Euclid hardware and survey design are specifically optimized for dark energy studies, the images and catalogs will enable a wide range of ancillary science in cosmology, galaxy evolution, and other areas of astronomy and astrophysics. These data will be made publicly available both in Europe via ESA and The Euclid Consortium and in the US via the Euclid NASA Science Center at IPAC (ENSCI) after a brief proprietary period. Small areas of the Euclid survey (suitable for testing algorithms and pipelines but not large enough for cosmology) will be released at 14, 38, 62, and 74 months after survey operations start (these are referred to as "quick releases"). The full survey data will be released in three stages: circa 2022, 2500 square degrees; circa 2025, 7500 square degrees; circa 2028, 15,000 square degrees.

# 3 Enhanced science

The main dark energy probes established over the last decade are described next in the context of the three missions. They are: Type Ia supernova, galaxy clustering (baryon acoustic oscillations and redshift space distortions), weak lensing, strong lensing and galaxy clusters. The biggest advantage from combining information from the three missions is in the mitigation of systematic errors, especially via redshift information. For the observational and theoretical issues underlying the discussion here we refer the reader to recent review articles in the literature (Weinberg et al 2013; Joyce et al 2014).

LSST, WFIRST and Euclid each have their own systematic errors, most of which are significantly reduced via the combination of the survey datasets. The systematic errors affecting each of the surveys individually arise from their incomplete wavelength coverage (creating photo-z systematics), their differences in imaging resolution and blending (creating shear systematics), or from different biases in galaxy sample selections. In most respects the surveys are complementary in the sense that one survey reduces or nearly eliminates a systematic in the other. To achieve this synergy requires some level of data sharing; we discuss below the cases where catalog level sharing of data is sufficient and others where it is essential to jointly process the pixel data from the surveys.

This section describes each dark energy probe, with a focus on mitigation of systematics, and returns to photometric redshifts (photo-z's). The last three subsections are devoted to stellar science, galaxy and quasar science, and the variable universe.

## 3.1 Weak lensing

A cosmological weak lensing analysis in any of these surveys will use tomography, which involves dividing the galaxy sample into redshift slices and measuring both the auto-correlations of galaxies within slices and cross-correlations of galaxies in different slices. Tomography gives additional information about structure growth beyond a strictly 2D shear analysis. However, there are a number of requirements on the data for a tomographic weak lensing analysis to be successful. First, we require a good understanding of both additive and multiplicative bias in the shear estimates, including how they scale with redshift. Second, we require a strong understanding of photometric redshift bias and scatter. This is true in general, but it becomes even more important when trying to model theoretical uncertainties such as the intrinsic alignments of galaxy shapes with large-scale structure.

There are a number of ways in which the weak lensing science from each of LSST, WFIRST and Euclid would benefit by sharing data with the other surveys and thus improving our understanding of shear estimates and/or the photometric redshift estimates. This data sharing may include catalog-level data or possibly image data. Although the latter would pose additional complications in terms of the data reprocessing required to appropriately incorporate both ground and space images, the gains may be worth the effort where confusion or image blending are severe.

The main benefit to LSST from either WFIRST or Euclid will be in terms of the calibration of shear estimates. The space-based imaging will have much better resolution due to the much smaller PSF size. The shapes of galaxies measured from space images will suffer less from model bias and noise bias (at a fixed depth) than the shapes of the same galaxies measured from ground images. A comparison of the shear signal of the population of galaxies observed by both LSST and either Euclid or WFIRST can help quantify these possible systematic errors in the LSST estimates. While each survey will have its own independent scheme to correct for calibration biases, a cross-comparison of this sort can be used to validate those schemes. Any additional correction that is derived could then be extended to the rest of the LSST area where there is no such overlap. Note that such a comparison would use a matched sample of galaxies in LSST and one of the space-based surveys, but would not involve comparing per-galaxy shape estimates. With different effective resolution, wavelength, and possibly shear estimation methods, there is no reason to expect agreement in per galaxy shears (even allowing for variation due to noise). Instead, such a comparison would utilize the reconstructed lensing shear fields from the different surveys. WFIRST may be more effective than Euclid for this purpose because its narrower passbands mean that it does not suffer from the wide-band chromatic PSF issue mentioned below.

Another benefit to LSST relates to blended galaxies. The higher resolution space-based images make it easier to reliably identify blended galaxies than in ground-based images. The performance of deblending algorithm in LSST image processing pipeline can be significantly improved by providing higher-resolution space-based data because of order half of all observed galaxies will be significantly blended with another galaxy. With space-based catalogs, one can at least identify which objects are really multiple galaxies or galaxy-star blends and do forced fitting using the space-estimated positions. Euclid will be more effective than WFIRST for this purpose because its footprint has a greater overlap with that of LSST.

One benefit to Euclid from combination with LSST relates to chromatic effects. The diffraction-limited PSF is wavelength dependent and therefore differs for stars and galaxies due to their different SEDs. Euclid's very wide band imaging means that with its optical imaging alone, it will not be able to correctly estimate the appropriate PSF for each galaxy. Multi-band photometry is required to correct for this effect. Without correctly accounting for the color-dependent PSF, the systematic will dominate Euclid's WL error budget, so LSST photometry will be very useful in reducing this systematic error. Note that LSST cannot help with chromatic effects involving color gradients within galaxies, for which higher resolution imaging data will be necessary to estimate corrections; but LSST can provide information about chromatic effects involving the average galaxy colors.

Finally, there are some shear systematics tests that can be done in a joint analysis but not in any one survey individually. For example, one route that has been proposed within surveys to mitigate additive shear systematics is to cross-correlate shear estimates in different bands or in different exposures rather than auto-correlate shear estimates. However, some sources of additive systematics may persist across bands or exposures, which would make them quite difficult to identify within the data from the survey alone. Indeed, in essentially every survey that has been used for weak lensing analysis to date, the preliminary results indicated some unforeseen systematics. Despite care taken in the design and planned operations of new surveys, we cannot reliably exclude the possibility of unforeseen systematics.

In the case of these three surveys, the systematic errors from each survey are expected to be very different. Thus, cross-correlating the shear estimates from one survey with those of another should remove nearly all shear systematic error contributions, leaving only the true weak lensing signature. If this cross-correlation analysis reveals differences from the analysis within individual surveys, it could signal a problem that needs to be investigated and mitigated. Furthermore, the surveys have different redshift distributions for the lensed galaxies, so a joint cosmological analysis using data from two or more of them will be complementary. The joint analysis would also be better able to constrain nuisance parameters like intrinsic alignments and photometric redshift errors.

As described below, the principal way in which Euclid and WFIRST would benefit from sharing information with LSST is in the photometric redshift catalog. Without this additional data, Euclid and WFIRST will not be able to complete their weak lensing science goals. LSST will also gain from the IR band photometry of WFIRST and Euclid in terms of photometric redshift determination.

The easiest form of data sharing between surveys is if only catalog information is shared. For some of the gains mentioned above, this level of sharing is completely sufficient: cross-correlations and shear calibration, for example. The deblending improvements would require an iterative reprocessing of the data, using a space catalog as a prior and then reprocessing the LSST data with that extra information. A similar iterative reprocessing would be necessary for Euclid to properly account for its color-dependent PSF using the LSST SED estimates as a prior. For the photometric redshift improvement, one could just use the multiple catalogs, but to do it properly, it is important to have access to the pixel data from both experiments at once to properly estimate robust colors for each galaxy. It will be a significant technical challenge to develop a framework that will work with the different kinds of image data in a coherent analysis.

## 3.2 Large-scale structure

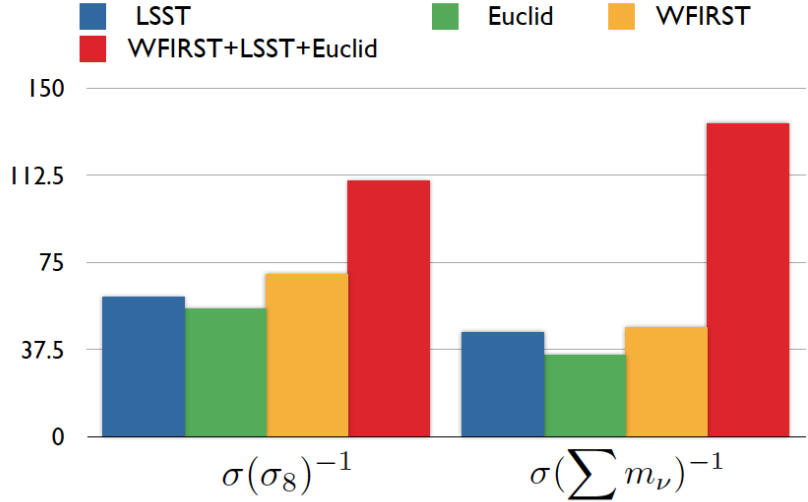

Figure 1: The chart shows how the complementarity of LSST, Euclid and WFIRST contributes to significant improvement in constraints on cosmological parameters. As described in the text, the improved constraints on $\sigma_8$ come from the mitigation of intrinsic alignment and other systematics in weak lensing; the improved constraints on the sum of neutrino masses $\sum m_\nu$ (in eV) comes from the combination of the weak lensing, CMB convergence maps, and galaxy clustering, in particular by reducing the multiplicative bias in shear measurement. Note that the space based surveys are assumed to have used ground based photometry to obtain photo-z's.

Measurements of large-scale structure probe dark energy via baryonic acoustic oscillation features in the galaxy power spectrum. In addition, the full shape and amplitude of the power spectrum measure the clustering of matter, if one can also measure the biasing of galaxies relative to the matter. This provides new avenues to measure the properties of dark energy and probe gravity.

The upcoming generation of spectroscopic surveys will make detailed maps of the large-scale structure of the universe. DESI, PFS, Euclid and WFIRST will focus on measuring galaxy clustering by obtaining spectra of tens of millions of galaxies over redshift ranges from $z \sim 0.4 - 3.5$. The surveys complement each other using a variety of spectral lines (H$\alpha$, OII, CII) and galaxy types (ELG and LRGs) as tracers and through having distinct, but overlapping, redshift ranges. DESI and Euclid will be wider surveys, while PFS and WFIRST will go deeper on smaller areas of the sky.

The combination of spectroscopic and lensing data enables a test of general relativity on cosmological scales. The spectroscopic data provides information about galaxy positions and motions determined inertial masses moving in the local gravitational potential. The imaging data, by contrast, shows the effect of space-time curvature on the trajectories of photons. If general relativity is the correct description of gravity then these will agree. Spectroscopic and photometric galaxy cluster surveys from WFIRST, LSST, Euclid and DESI, and others, and their cross-correlation with contemporary CMB polarization (CMB lensing) and temperature (CMB lensing and kinetic Sunyaev-Zel'dovich) data will provide coincident, complementary dynamical and weak lensing tests of gravity in halos two to three orders of magnitude more massive than the galaxies within them. Thus we expect powerful tests of gravity by a joint analysis of the three surveys.

We study the complementarity of WFIRST, LSST and Euclid by forecasting constraints on matter clustering (we pick $\sigma_8$ for convenience, which is sensitive to dark energy and is partially degenerate with the parameters, $n_s$, $\Omega_m$), and the sum of neutrino masses in the Universe. While we concentrate on these particular parameters, there are many other possibilities considered by the community.

Figure 1 shows constraints on $\sigma_8$ and the sum of neutrino masses $\Sigma m_\nu$ from the three surveys individually and in combination. We include a CMB prior from Planck, and marginalize over cosmological (and galaxy bias) parameters when appropriate. For the space surveys we assume that photo-z's will be obtained with the aid of ground based data. WFIRST's densely sampled spectroscopic galaxies will provide a detailed characterization of filamentary structure that is complementary to the lensing measurements provided by imaging surveys. This allows us to understand one of the most elusive systematics in weak lensing surveys – the intrinsic alignment (IA) of galaxies. One can test and validate IA models and marginalize over the free parameters in specific models when we can have a 3D map of the universe via galaxy spectra. In LSST, the lensing sources spans a large redshift range between 0 and 3 and a large luminosity range. With WFIRST and Euclid, we will have spectroscopic sources spanning a large part of this redshift range, which enables a very good construction of filament map of the surveyed volume. WFIRST offers a much higher density of sources while LSST and Euclid will cover a much large area of sky. The joint analysis leads to significant improvement in the constraint on $\sigma_8$, as shown on the left in Figure 1.

We also consider the constraints on massive neutrinos which suppress the growth of structure below the free-streaming scale. The combination of WFIRST and Euclid's spectroscopic sample, along with CMB lensing convergence, can be used to calibrate the shear multiplicative bias in LSST sources. We follow Das, Evrard & Spergel (2013) in methodology and calculate the improvement in constraints first in the multiplicative bias in shear, and propagate this to the constraints on the sum of neutrino masses that relies on the large sky coverage of LSST. We assume that a Stage III level CMB convergence field will be available. We plot the improvement in $\Sigma m_\nu$ on the right in Fig 1.

### 3.3 Galaxy clusters

LSST, Euclid and WFIRST will use galaxy clusters (primarily via gravitational lensing mass measurements) as part of their cosmological measurements. However, both for cosmology and the study of galaxy evolution in clusters, a joint analysis of the data can greatly improve the reach of each mission. Fundamentally, this is because each of the missions observes different aspects of the emission from clusters. LSST provides optical colors, useful for cluster identification via photometric redshifts, as well as (for group-scale structures) time delays for lensed objects. Euclid provides shallow NIR photometry and somewhat deeper NIR spectroscopy, as well as higher resolution optical imaging. WFIRST will provide much better resolution images and deep NIR photometry and spectroscopy over a smaller area than Euclid. Because of the complementarity of these measurements, combining the data has implications for cluster finding, redshift determination, mass measurement and calibration, and the study of galaxy evolution in clusters.

*Galaxy cluster finding:* Over the past decade, techniques have been developed that select galaxy groups and clusters very effectively by isolating overdensities simultaneously in projected position and color space (for example redMaPPer). These techniques have been optimized for optical wavelengths. LSST will be extremely efficient at finding groups and clusters of galaxies and estimating their redshifts (although eROSITA will also be an efficient cluster finder, the redshift information available from the LSST imaging will allow selection of samples at fixed redshift for study). Although redMaPPer is effective to $z \sim 1$, extending its redshift range will require infrared imaging, which would be provided by Euclid or WFIRST. Given that the number density of detected (as opposed to resolved) sources in LSST and WFIRST are going to be similar, and given the importance of consistent photometry measurements across bands, simultaneous measurements of the pixel data would yield more accurate selection of clusters.

*Redshift Determination:* As discussed below, the photo-z's of galaxies will be greatly improved by combining the optical and infrared imaging of the three missions. Thus, joint analysis of the LSST+WFIRST/Euclid data gives more accurate cluster sample selection, and allows accurate determination of the background galaxy redshift distribution, which is the largest single source of systematic uncertainty in determination of the mass function from weak lensing mass determinations.

*Strong and Weak Lensing:* Measuring the weak lensing shear requires high resolution imaging. WFIRST will be able to resolve many more galaxies than LSST, and as discussed above will aid the shear calibration of LSST. In addition, the conversion of shear to mass depends on redshift information. In this way, the combination of LSST, Euclid, and WFIRST data will provide much more accurate mass measurements than either will alone.

Cluster arc tomography is greatly improved by the joint pixel-level analysis of LSST, Euclid and WFIRST data. For instance, WFIRST will have the angular resolution to discover $\sim 2000$ strong lensing clusters in its survey footprint. However, most of the arcs will be too faint (and too low surface brightness) to have spectroscopic redshifts. By contrast, LSST will reach surface brightness limits ($> 28.7$ Mag per square arcsecond) sufficient to detect the arcs and obtain photometric redshift estimates, but will not have the resolution to separate the arcs cleanly from the cluster galaxies in many cases. Thus, the measurement of the tomographic signal from multiple arc systems, which is a strong test of cosmology, will be much stronger from the joint analysis than from Euclid, WFIRST or LSST alone. Once again, the added value of the pixel data is great, because cluster cores are very dense environments and obtaining valid photometric redshifts will require use of the space-based imaging to disentangle the light contribution form the arcs and cluster galaxies.

*Systematics:* One of the sources of systematic uncertainty in weak lensing shear measurements from galaxy clusters comes from the blending of sources both with other sources and with cluster galaxies. The source-source blending is in many ways similar to that encountered in large-scale weak lensing measurements, but in clusters, there is an additional source of bias from blending with cluster galaxies. This is both radially dependent and (because cluster concentration evolves with redshift) redshift dependent, and directly affects the normalization of the cluster mass function. Using the WFIRST data to directly measure the deblending bias in the clusters in common between the two surveys would make it possible to correct for the bias.

## 3.4 Supernovae

Each individual Type Ia supernova (SN Ia) is potentially a powerful probe of relative distance in the Universe. Thousands of them provide detailed information about the expansion history of the Universe, in particular allowing us to probe the accelerated expansion and the nature of dark energy. Hundreds of thousands of them will allow us to probe the nature of dark energy in different regions and environments. A set spanning the past 10 billion years of cosmic history will allow for a rich and powerful exploration of the nature of the expansion and more recent acceleration of the Universe and thus reveal key insights into the key constituents.

Ground-based SN Ia searches have provided almost one thousand well-observed SNe Ia in 20 years of work. Increasingly powerful ground facilities will rapidly increase this number to thousands per year (e.g., Pan-STARRS, PTF and DES) and then hundreds of thousands per year in the era of LSST. For a smaller subset (hundreds) of these supernovae, even calibrated spectrophotometric time-series have been obtained (at the nearby SN Factory). However, the rest-frame emission of SNe Ia is line-blanketed at UV wavelengths. Thus, by $z \sim 1$ their light has already redshifted out of the optical atmospheric window easily accessible from the ground. In addition, because the relative distances across redshifts are estimated based on relative flux differences across wavelength, accurate cross-wavelength calibration is critical to use supernovae to determine the properties of dark energy.

The stability of WFIRST in space yields significant benefits in both wavelength coverage and flux calibration for thousands of SNe Ia/year from $0.1 < z < 1.7$. The accessibility of the ground allows for the large aperture and high-etendue of the LSST system that can observe 10,000s of SNe Ia per year from

$0.03 < z < 1$.

Measuring distances from $z \sim 0.03$ to $z \sim 1.7$ will connect the era when dark energy first made its presence known, through the transition to an accelerating expansion, and up to the present day. These are the key redshifts over which we will discover hints to the underlying nature of dark energy, and the combination of the high-redshift reach and calibration of WFIRST and the huge numbers and volume coverage of LSST will yield a strong and powerful pairing for SN Cosmology.

The key question for this work will be the control of systematic uncertainties as we explore a large span of time and changing populations of host-galaxy environments in which the supernovae are born. The combination of WFIRST, LSST, and other ground-based follow-up instruments will make possible unprecedented rich characterization of each supernova so that low- and high-redshift supernovae can be matched at a more detailed level than with the broad "Type Ia" definition. For example, a time series of detailed spectrophotometry could be obtained for the same rest-wavelengths for every supernova.

Modern SN distance analyses are already joint analyses. The current Hubble Diagrams are constructed from supernovae over different surveys (e.g., Betoule et al. 2014 combined the Supernova Legacy Survey (SNLS), SDSS Supernova Survey, and the current heterogeneous sample of nearby $z < 0.05$ SNeIa). But a true joint analysis should begin at the pixel level and take advantage of all the internal details about calibration and photometry within each survey. Specifically, when combining SNeIa from different data sources, a joint analysis is beneficial because of: 1. Consistent pixel-level analysis, in particular photometric extraction; and, 2. Deep cross-checks of photometric calibration not possible with just extracted SN light curves.

Supernova surveys have distinct science requirements both on supernova discovery and early classification, and on the quality of distance modulus and redshift determinations of discovered supernovae. Each science requirement propagates into hardware and survey requirements. Technical requirements for supernova discovery (e.g., large field of view, detection of new point sources) can have a more cost-effective ground-based solution, whereas requirements for photometry and spectroscopy (e.g., high signal-to-noise SN spectral features, infrared wavelength coverage) are only fulfilled with a space observatory. A case in point is the baseline WFIRST plan where an imaging survey is conducted to identify supernovae for targeted spectroscopic observations. The large field of view of LSST and the availability of SN photons detectable by LSST out to $z = 1.2$ should make the LSST the better instrument for SN discovery, especially at low redshifts where the angular density of potential host galaxies is low. Assuming the triggering logistics work, this might free up WFIRST to do what it is better at: obtaining supernova spectrophotometric time series.

In addition, there is a benefit of observing the same SN with multiple overlapping surveys. Such observations provide:

- Direct cross-instrumental calibration on the source of interest. (It is important to note, however, that this calibration is not the same as the cross-wavelength calibration that is required when comparing supernovae at very different wavelength.)

- Expanding observations of the wavelength/temporal range of the supernova.

- A quantitative assessment of systematic errors.

Finally, supernova cosmology analyses require a suite of external additional resources beyond the direct observations of high-redshift supernovae. Thus, planning for a joint LSST-WFIRST SN cosmology analysis will yield significant gains in combining resources to obtain external data and theory needs. Some key examples include:

- Low-z SN set, including possible spectrophotometric time-series follow up.

- Fundamental flux calibrators (which *can* provide the cross-wavelength calibration that is required when comparing supernovae at very different wavelength).

- High-resolution spectroscopy of host galaxies.

- Improved SN Ia theory and empirical modeling.

## 3.5 Strong lensing

The LSST, Euclid and WFIRST datasets will be particularly complementary in the field of strong gravitational lensing. Galaxy-scale strong lenses can be used as probes of dark energy (via time delay distances or multiple source plane cosmography). The time delays between multiple images in a strong gravitational lens system depend, as a result of the lens geometry, on the underlying cosmology – primarily the Hubble constant but also, in large samples where internal degeneracy breaking is possible, the dark energy parameters. (Refsdal 1964, Suyu et al 2014). Lens systems with sources at multiple redshifts may also provide interesting and competitive constraints on cosmological parameters, via the ratio of distance ratios in each system.

Strong lenses also provide unique information about dark matter, via the central density profiles of halos on a range of scales, and the mass function of sub-galactic mass structures that cause measurable perturbations of well-resolved arcs and Einstein Rings. Both this science case and strong lensing cosmology will benefit greatly from the large samples of lenses detectable in these wide field imaging surveys; we sketch these cases out below in Section 4, and identify how high fidelity space-based imaging and spectroscopy will enable new measurements.

## 3.6 Photometric redshifts

Photometry provides an efficient way to estimate the physical properties of galaxies (i.e. their redshifts, spectral types, and luminosities) from a small set of observable parameters (e.g. magnitudes, colors, sizes, and clustering). LSST will depend on these photometric redshifts for all of its major probes of dark energy as it is infeasible to obtain spectroscopic redshifts for the bulk of objects in the LSST sample with any reasonable amount of telescope time. For the WFIRST and Euclid missions, three of the key applications will rely on photometric redshifts: measures of the mass power spectrum from analyses of the weak lensing of faint galaxies, breaking of the degeneracies between possible redshift solutions for galaxies exhibiting only a single emission line in grism spectroscopy, and the identification of high redshift Type Ia supernovae based on the photometric redshifts of their host galaxies. In addition, photometric redshifts will play a key role in much of the LSST, WFIRST, and Euclid extragalactic science where we wish to constrain how the demographics of galaxies change over time.

The high efficiency of photometric redshifts comes at a cost. The primary features that enable photometric redshift estimation are the transitions of the Balmer (3650 Å) and Lyman (912 Å) breaks through a series of photometric filters. Uncertainties in isolating the position of the breaks due to the low spectral resolution of the photometry give rise to a scatter between photometric redshift estimates and the true redshifts of galaxies. Confusion between breaks (due to sparse spectral coverage and incomplete knowledge of the underlying spectral energy distributions) leads to catastrophic photo-z errors, as seen in Figure 2.

Existing studies have demonstrated that while photometric redshift accuracies of $\sigma_z \sim 0.007(1 + z)$ are possible for bright objects with many filters, uncertainties of $\sigma_z \sim 0.05(1 + z)$ are more typical when restricted to 5-6 bands of deep imaging. Catastrophic photometric redshift errors (where the difference between the photometric and spectroscopic redshifts, $\Delta z$, exceeds the $3 - \sigma$ statistical error) typically occur in well over 1% of cases. As the photo-z scatter and catastrophic failure rate increase, information is degraded and dark energy constraints will weaken. Furthermore, if there are nonnegligible ($> \sim 0.2\%$) systematic offsets in photo-z's or if the $\Delta z$ distribution is mischaracterized, dark energy inference will be biased at levels comparable to or greater than expected random errors. As a result, careful calibration and validation of the photometric redshifts will be necessary for the LSST, WFIRST, and Euclid surveys.

### 3.6.1 The impact of Euclid and WFIRST near-infrared data on LSST photometric redshifts

The LSST filter system covers the $u, g, r, i, z$ and $y$ passbands, providing substantial leverage for redshift estimation from $z = 0$ to $z > 6$. For the LSST "gold" sample of galaxies, $i < 25.3$, Figure 2 shows how the HLS for WFIRST, with four bands comparable in depth to the 10 year LSST survey, (i.e. $5\sigma$ extended source depths of $Y_{AB}$=25.6, $J_{AB}$=25.7, $H_{AB}$=25.7, and $F184_{AB}$=25.2), will significantly improve on the photometric redshift performance from LSST alone. For example, at redshifts $z > 1.5$, where the Balmer break transitions out of the LSST $y$ band and into the WFIRST and Euclid infrared bands, the inclusion of the WFIRST data results in a reduction in $\sigma_z$ by a factor of more than two ($1.5 < z < 3$), and a reduction in the fraction of catastrophic outliers to <2% across the full redshift range. Euclid's three-band NIR photometry, while shallower, will have a much greater overlap with LSST and will also provide a quantitative improvement in LSST photo-z's. Combining the LSST, WFIRST and Euclid photometric data effectively will depend, however, on the details of the respective filter systems, their signal-to-noise, our ability to extract unbiased photometric measurements from extended sources (e.g. the deblending of sources using the higher spatial resolution of the WFIRST data), and the accuracy of the photometric calibration of the data both across the sky and between the near-infrared and optical passbands.

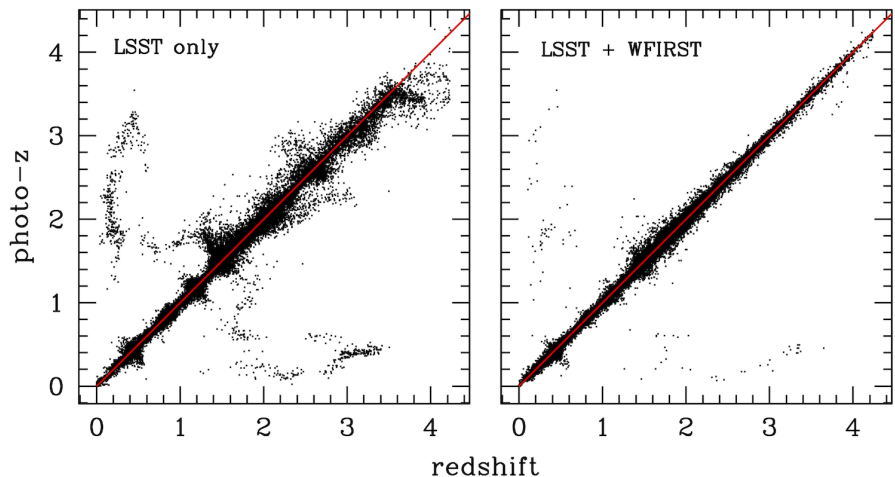

**Figure 2:** A comparison of the relative photometric redshift performance of the LSST optical filters (left panel) with a combination of LSST and WFIRST filters (right panel). The simulated data assumes a 10-year LSST survey and a "gold sample" with $i < 25.3$. The addition of high signal-to-noise infrared data from WFIRST reduces the scatter in the photometric redshifts by roughly a factor of two (at redshifts $z > 1.5$) and the number of catastrophic outliers by a factor of three. These simulations do not account for deblending errors or photometric calibration uncertainties, and assume complete knowledge of the underlying spectral energy distributions of galaxies as an ensemble.

### 3.6.2 Mitigating systematics with WFIRST and Euclid spectroscopy

The optimization of photometric redshift algorithms and the calibration of photometric redshift uncertainties both require spectroscopic samples of galaxies. If simple algorithms are used, more than 100 spectroscopic survey regions (of $\sim$0.25 deg$^2$) with at least 300-400 spectroscopic redshifts per region may be required to optimize a photometric redshift algorithm (whether by refining templates and photometric zero points or as input for machine learning algorithms) to ensure that their accuracy is not limited by sample variance in the spectroscopic training set (Cunha et al. 2012); with techniques that take this variance into account, 15-30 fields may be sufficient (Newman et al. 2014). An ideal training set would span the full range of properties (including redshifts) of the galaxies to which photometric redshifts will be applied. To the degree to which we do not meet this goal, we can expect that photometric redshift errors will be degraded, weakening constraints on dark energy as well as other extragalactic science.

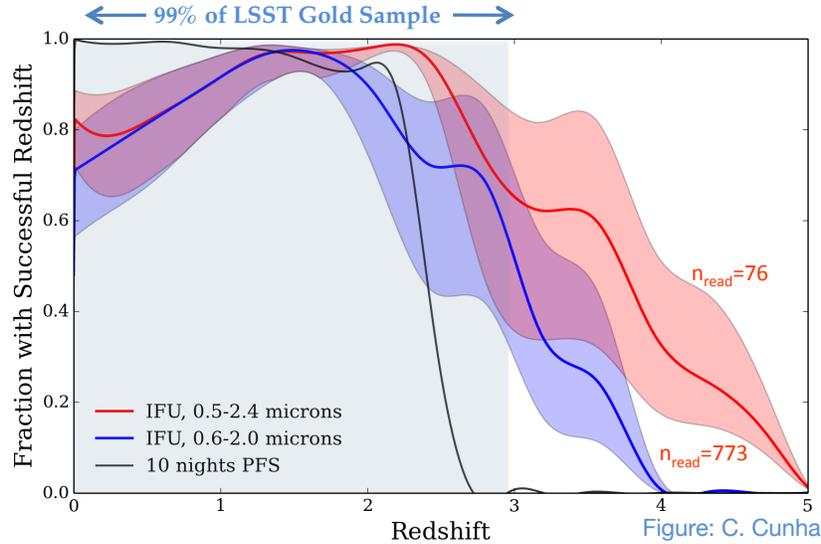

**Figure 3:** Predictions of the fraction of LSST weak lensing sample objects that would yield a secure (multiple-confirmed-feature) spectroscopic redshift, based either on 1440-second exposure time with *WFIRST* (colored regions) or 10 nights' open-shutter-time spectroscopy with the Subaru/PFS spectrograph (black curve) *WFIRST* IFU spectroscopy would provide training redshifts for objects at higher $z$ than are easily accessible from the ground, particularly if read noise per pixel is small (the colored regions indicate a range of feasible scenarios). Longer exposure times (e.g., in supernova fields or by optimized dithering strategies) could enhance the success rate further.

Current spectroscopic samples fare well; surveys to $R = 24.1$ or $i = 22.5$ (more than two magnitudes shallower than the LSST "gold sample") with 8-10m telescopes have obtained >99% secure redshifts for 21-60% of all targeted galaxies, and >95% secure redshifts for 42-75% of the galaxies (incorrect redshift rates above 1% would lead to photo-z systematics that exceed LSST requirements). WFIRST spectroscopy can address these limitations in a number of ways. Depending on the final configuration and dithering strategy for *WFIRST*, an object in the LSST weak lensing sample will fall on the IFU field-of-view at least 10% (and up to $\sim 100\%$) of the time. Most objects in this sample would yield a successful redshift with a $\sim 1440$ sec exposure (see Figure 3). For an IFU with a $3^{\prime\prime} \times 3^{\prime\prime}$ field-of-view, at least 10,000 spectra down to the LSST weak lensing depth, corresponding to roughly 20,000 down to the *WFIRST* limits, would be measured concurrently with the WFIRST HLS (with minimal impact from sample/cosmic variance). This spectroscopy would have very different incompleteness from ground-based samples, allowing a broader range of galaxies to have well-trained photometric redshifts, with accuracy limited by the imaging depth rather than our knowledge of galaxy SEDs.

If our spectroscopic samples do not have a success rate approaching 100%, accurate characterization of photo-z errors (required for dark energy inference) is likely to instead be based on the cross-correlation between spectroscopic and photometric samples (cf. Newman 2008). To meet LSST calibration requirements, such an analysis will require secure ($< 1\%$ incorrect) spectroscopic redshifts for $\sim 100,000$ galaxies spanning the full redshift range of the photometric samples of interest and an area of at least a few hundred square degrees. Euclid and WFIRST grism spectroscopy will provide spectra for many millions of galaxies over wide areas (e.g., $2 \times 10^7$ galaxies in the WFIRST HLS). However, this spectroscopy will provide the highly secure, multiple-feature redshifts required for training and calibrating photo-z's in only very limited redshift ranges. As a result, Euclid and WFIRST grism spectroscopy may contribute to photometric redshift calibration in combination with other datasets spanning the remaining redshift range (e.g., from DESI), but cannot solve this problem on their own.

## 3.7 Stellar science

New insights on stellar populations in the Milky Way and nearby galaxies depend critically on gains in photometric sensitivity and resolution. Thus far, the state of the art imaging surveys of our Galaxy have involved relatively modest tools with respect to today's capabilities. For example, the SDSS survey is both shallow and operates at low resolution, and IR surveys such as 2MASS and WISE are based on even smaller telescopes operating at even lower resolution.

The combination of LSST, Euclid, WFIRST and Gaia will transform our view of the Milky Way and nearby galaxies. LSST will map 18,000 square degrees of the night sky every few days, across 6 visible light filters. The imaging depth of LSST will be over 100 times (5 magnitudes) deeper than SDSS, and at higher resolution. WFIRST will overlap some of the same footprint, but will also extend the spectral range of Milky Way surveys to near-infrared wavelengths at even higher spatial resolution (0.1 arcsecond pixels). Euclid provides much higher spatial resolution than LSST in the visible (aids deblending sources and crowded field photometry) and Gaia addresses the single biggest contribution to the error budget of most stellar population studies aimed at characterizing physical properties; knowledge of the distance of sources.

Much of the stellar population science case for LSST (e.g., from the Science Book) overlaps that from WFIRST (e.g., the community science white papers in the 2013 WFIRST SDT report). These include studies of the initial mass function, star formation histories for Milky Way components, near field cosmology through deep imaging and proper motions of dwarf satellites, discovery and characterization of halo substructure in nearby galaxies, and more. Given the wavelength complementarity alone, most stellar population investigations will be aided by the panchromatic baseline of LSST (and Euclid) + WFIRST (e.g., interpreting SEDs into fundamental stellar properties, increasing the baseline for star formation history fits from multi-band color-magnitude diagrams, confirming memberships of very blue or red stars in either data set, etc.). However, the overall quality of all three data sets for general investigations can be greatly enhanced (and, in a uniform way) by jointly re-processing the pixel data sets for LSST, Euclid, and WFIRST.

To highlight the technical requirements for this joint processing, we point to three LSST and WFIRST "killer apps" in stellar population science,

- Establishing and characterizing the complete spectrum of Milky Way satellites

- Testing halo formation models by measuring the shapes, substructure content, masses, and ages of a large set of stellar halos within 50 Mpc

- Measuring the star formation history, Galactic mass budget, and spatially and chemically-dependent IMF

In each of these cases, LSST and Euclid have the advantage with respect to field of view (and cadence). WFIRST has the advantage of higher photometric sensitivity to red giant branch and low mass star tracers, and much higher resolution for star-galaxy separation in the IR. Joint processing of the three data sets offers the following rewards,

1. By using the WFIRST astrometry to deblend sources and feed new positions of stars to LSST (and to a lesser extent, Euclid), the LSST and Euclid photometry can be made to go deeper. In the actual processing, the pixels from LSST, Euclid and WFIRST would be analyzed simultaneously to detect and perform photometry on all detections in any of the more than dozen LSST+Euclid+WFIRST bandpasses.

2. WFIRST photometry alone will be critically insensitive to hot stars (e.g., white dwarfs in the Milky Way, horizontal branch stars in the Local Group). In most halo and sparse populations, these objects will be easily detected by LSST, and to a shallower depth, Euclid. Similar to 1.), the LSST photometry can be used to obtain new measurements in the WFIRST data set and to get an infrared color for these sources.

3. As a result of the above, many more sources in the overlapping footprint of the three telescopes will have high-precision panchromatic photometry. This increased wavelength baseline will yield a much more accurate characterization of the underlying stellar populations (e.g., the sensitivity to age, reddening, and distance variations increases on this baseline). Note, this is different from matching LSST, Euclid, and WFIRST catalogs since the goal here is to detect sources that otherwise would not have been measured in one of the three data sets. The biggest gain in our interpretation of the stellar populations to derive physical properties will come from accurate distances to the stellar populations from Gaia. The different photometric sensitivities is not a concern here, since there are bright tracers in most populations that Gaia can target.

4. The WFIRST morphological criteria of faint sources provides a crucial assessment of LSST's star-galaxy separation.

Many of these joint analyses can be done at the catalog level while others likely require working with the pixels from each of the surveys.

## 3.8 Galaxy and quasar science

The WFIRST photometry will go to a similar depth (on an AB scale) as the full coadded LSST images. Euclid NIR data, while shallower, will overlap a much larger area of the LSST survey. Thus the combination of LSST, Euclid and WFIRST will give high S/N 9-band photometry stretching from 3000 Å to 2 microns (a factor of six in wavelength, twice that of either WFIRST/Euclid or LSST alone) for the Gold sample ($i < 25.3$) of hundreds of millions of galaxies. This broad wavelength coverage probes the SEDs of galaxies beyond 4000A to redshifts 3 and beyond, allowing detailed determination of star formation rates and stellar masses during the cosmic epoch when galaxies assembled most of their stars. (Unobscured) AGN will be identified both from their SEDs and from their variability in the repeat LSST imaging.

The Hubble Space Telescope has been able to obtain comparably deep photometry over a similar wavelength range over tiny areas of sky; for example, the Cosmic Assembly Near-infrared Deep Extragalactic Legacy Survey (CANDELS) covers only 0.2 deg$^2$, while WFIRST will cover an area ten thousand times larger. With this much larger sky coverage, the combination of LSST and WFIRST can explore the evolution of galaxy properties over this broad range of cosmic history.

The weak lensing signature from individual galaxies is too small to be measured, but stacking analyses can be done in fine bins of stellar mass and redshift, exploring the relationship between dark matter halo mass, stellar populations, galaxy morphology, AGN activity and environments during the peak of galaxy assembly. The exquisite resolution of the WFIRST images will also allow the role of mergers in galaxy evolution to be quantified, with galaxy pairs separated by as little as 3 kpc discernible over essentially the entire redshift range.

The multi-band photometry will allow both quasars and galaxies at much higher redshifts to be identified as well. Current studies with HST have allowed identification of a handful of galaxy candidates to redshift 8. WFIRST + LSST will expand these studies over 10,000 times the area, looking for objects that are dropouts in the u through y filters. The wide area is particularly important for discovering the rare most luminous objects at these redshifts, that can be followed up in detail with JWST and the next generation of ∼30-meter telescopes. The study of such objects probes the exit from the Dark Ages, as the intergalactic medium became reionized. Understanding the formation of the first galaxies and the reionization process in detail requires putting these objects in their large-scale structure context, which is only possible with samples selected over large solid angles. In particular, the expanding bubbles by which reionization is thought to progress are thought to subtend tens of arcminutes on the sky, requiring wide-field samples to explore them.

## 3.9   The variable universe

Time domain astronomy encompasses an extremely varied suite of astrophysical topics, from the repeated pulsations of regular variable stars that serve as tracers of Galactic structure and cosmological distances (e.g. RR Lyrae and Cepheids), to transient events that by their nature occur either unpredictably or only once (e.g. stellar flares, supernovae, tidal disruption events). As suggested by these examples, transients and variables occur both locally and at the most distant reaches of our universe, and so comprise a population with both a very high dynamic range in luminosity and a wide span of colors. Time domain astronomy is a rapidly growing field: the rich public dataset provided by LSST will enable a large part of the astronomy community to analyze observations of transient and/or variable objects. The potential synergies between LSST, Euclid and WFIRST are as compelling and varied as time domain astronomy itself, but here we highlight some compelling examples.

The combined WFIRST, Euclid and LSST datasets would greatly facilitate the characterization and classification of transient and variable objects. One of the challenges imminent in the coming LSST era is the identification of samples of variable and transient events to be targeted for additional observations. From the transient side, the anticipated high volume of the LSST alert stream requires rapid triage of large numbers of events to identify the most important transients: those events which, if not pursued with followup observations immediately, will be opportunities lost forever. Furthermore, the high volume of events also necessitates the prioritization of the most compelling sources to followup, which in turn requires rapid characterization of an event (where characterization here is distinguished from actual classification into a known category of object). The WFIRST and Euclid datasets will assist with these challenges in a number of ways. First, the deep infrared observations will greatly assist in the identification of the "foreground fog" of very red Galactic sources: low mass stars and brown dwarfs. While LSST's depth will identify many of these objects in quiescence, some percentage will escape even LSST on account of their very red colors and low luminosities. Therefore, flares on these objects may still create what appear to be cosmological transient events in spite of their local origin. The deep, panchromatic characterization of the static sky will enable the association of transient events with co-located sources, whether those sources are quiescent detections of the same source, or possible host galaxies of the event whose properties may give clues for a deeper understanding of those transients.

As pointed out in the previous subsection on stellar science, the strength of the combined datasets goes far beyond matching the WFIRST, Euclid and LSST catalogues. Rather it increases the depth of the photometry by making it possible to detect sources that would otherwise not have been detected. The ability to identify (or rule out) infrared counterparts may also be helpful in confirming some of the most elusive transients. For example, potential electromagnetic counterparts to gravitational wave sources are expected by current models to emit their peak emission in near IR wavelengths. The association of any gravitational wave event detection with a known galaxy greatly reduces the search volume for EM counterparts to gravitational wave events, and the elimination of faint red background galaxies as putative associated transient kilonova emission will help confirm potential multi-messenger events.

# 4 Coordinated observations

A significant fraction of WFIRST time will be devoted to a Guest Observer program. In this section of the report, we consider a number of potential GO programs that will likely follow-up on or extend LSST observations. While the GO programs will likely be selected by a future Time Allocation Committee (TAC), we will benefit by preparing for likely observing scenarios. Here, we consider three examples of joint LSST/WFIRST studies: WFIRST follow-up of strong lenses and lensed AGNs found in LSST, joint observations of deep drilling fields, and joint WFIRST/LSST searches for the afterglows of gravitational wave bursts.

## 4.1 Measuring dark energy with strong lens time delays

As described in the previous section, time delays between multiply imaged sources are a powerful cosmological probe. LSST will enable time delay distance measurements towards hundreds of lensed AGN and supernovae (Oguri & Marshall 2010, LSST Science Book). The detection of candidate systems, which have image separations of just 1-3 arcseconds, and confirmation of them by lens modeling, depends critically on the quality and depth of the imaging data, suggesting that there is much to be gained by searching a combined LSST/Euclid/WFIRST imaging dataset. Measuring cosmological parameters with this sample will require high signal to noise ratio and high resolution imaging and spectroscopy of any Einstein Rings discovered in order to constrain the lens mass distributions. By integrating longer on the most interesting systems, WFIRST can achieve background source densities of over 200 sources/arcmin$^2$, which will enable detailed weak lensing mass maps of these systems.

Joint analysis of all LSST and wide field infrared survey data will improve the completeness and purity of the lens searches in the following ways: 1) a combined catalog of LSST and space-detected objects will produce a more accurate initial list of target lenses, and 2) joint analysis of all the available cutout images will enable better lens modeling, leading to a much cleaner sample of candidates to be confirmed in the followup observations.

Being able to model large proportions of the lens sample precisely with high resolution space-based imaging will enable the surveys themselves to constrain the structure and evolution of massive galaxies across a homogeneously-selected sample, providing important prior information needed by other cosmology probes.

The joint optical and infrared photometry provided by LSST, Euclid and WFIRST will constrain the mass environments of the strong lenses (including the line of sight environment) by providing accurate photometric redshifts and stellar masses for all the relevant galaxies in the lightcone (e.g., Greene at al 2013, Collett et al 2013).

## 4.2 Probing dark matter with grism spectroscopy of lensed AGN

Measuring the mass function of dark matter subhalos down to small masses (e.g., $10^6$ solar masses) is a unique test of the CDM paradigm, and of the nature of dark matter in general. The CDM mass function is expected to be a power law rising as $M^{-1.9}$ down to Earth-like masses, while a generic feature of self-interacting dark matter and warm dark matter models is to introduce a lower mass cutoff. Current limits set that cutoff at somewhere below $10^9 M_\odot$. We know from observations of the environs of the Milky Way that there is a shortage of luminous satellites at those masses, but this may be entirely due to the physics of star formation at the formation epochs of these satellites. Dark Milky Way satellites at lower masses may be detectable through their dynamical effects on tidal streams, but statistical studies of the one galaxy we live in will always carry high uncertainty.

A powerful way to detect subhalos independent of their stellar content and in external galaxies is via the study of large samples of strong gravitational lenses, in particular the so-called flux ratio anomalies (see Treu 2010, and references therein). Small masses located in projection near the four images of quadruply-lensed quasars cause a strong distortion of the magnification and therefore alter the flux ratios. However, anomalous flux ratios can also be caused by stellar microlensing if the source is small enough. One solution is to observe the narrow line region emission lines, where the lensed quasar emission is sufficiently extended to be unaffected by microlensing. Large samples of thousands of lenses could achieve sensitivity down to $10^6 - 10^7$ solar masses required to probe the nature of Dark Matter. Unfortunately, at the moment only 3 dozen quads are known, and only a handful of those are bright enough to be observable from the ground (e.g. Chiba et al 2005).

Discovering thousands of quads will require surveys covering thousands of square degrees at sub-arcsecond resolution (Oguri & Marshall 2010). Between them, Euclid, WFIRST and LSST should discover over 10,000 lensed AGN, one sixth of which will be quad systems. Euclid and WFIRST have the potential to observe of order a thousand of these quads spectroscopically themselves, using narrow near infrared emission lines to avoid the microlensing and enabling the wholesale application of the lens flux ratio experiment (Nierenberg et al. 2014).

## 4.3 Deep drilling fields

The LSST collaboration plans several "deep drilling" fields. While this aspect of the LSST survey is still under design, the broad plan is to reach $AB \approx 28$ in $ugriz$ bands (and shallower in $y$), which is comparable to the medium depth planned for the WFIRST SN survey (J=27.6 and H=28.1 over 8.96 $\deg^2$). The WFIRST deep SN fields will be deeper with depths of J=29.3 and H=29.4 over 5.04 $\deg^2$. LSST has currently identified four selected fields, each with areas of 9.6 $\deg^2$: ELAIS S1, XMM-LSS, COSMOS, and Extended Chandra Field South. With three days of observation per band, WFIRST can cover each of these deep fields to AB = 28.

The goals of these fields (Gawiser et al., white paper; Ferguson et al., white paper) are to (1) Test and improve photometric redshifts critical for LSST Main Survey science. (2) Determine the flux distribution of galaxy populations dimmer than the Main Survey limit that contribute to clustering signals in the Main Survey due to lensing magnification. (3) Measure clustering for samples of galaxies and Active Galactic Nuclei (AGN) too faint to be detected in the Main Survey (4) Characterize ultra-faint supernova host galaxies. (5) Characterize variability-selected AGN host galaxies. (6) Identify of nearby isolated low-redshift dwarf galaxies via surface-brightness fluctuations. (7) Characterize of low-surface-brightness extended features around both nearby and distant galaxies. (8)Provide deep "training sets", for characterizing completeness and bias of various types of galaxy measurements in the wide survey (e.g. photometry, morphology, stellar populations, photometric redshifts). All of these goals will be clearly enhanced by complementary data in the infrared. As already outlined in the report, infrared data will significantly improve the certainty of photo-z's. WFIRST high spatial resolution will be particularly powerful in these deeper fields where confusion is likely to be significant. The deblending, color correction and morphology arguments in sec 3 apply even more strongly at the greater depth reached in these fields.

Indeed, WFIRST provides great synergy by providing a high-resolution detection image that will allow LSST to go past the ground-based confusion limit. This in turn will motivate LSST to go deeper in $gri$ than currently planned for the Deep Drilling Fields; by matching the LSST Main Survey filter time distribution, LSST will reach deeper than AB$\sim$29 in these three filters rather than the current baseline of AB$\sim$28.5, providing a better match to the deepest WFIRST imaging.

It is worth noting that the VISTA VIDEO deep NIR survey regions are located at the centers of the already-approved LSST DDFs, and these will generate initial spectroscopy and multi-wavelength coverage even before WFIRST launches, making them optimal regions for covering deeper in NIR. However, if WFIRST is able to cover all (or most) of the 9.6 deg$^2$ LSST field-of-view, LSST can respond by skipping the significant fraction of time spent in $y$ band in the current LSST "near-uniform DDF" approach. Using WFIRST imaging in lieu of LSST imaging in $y$ band (and possibly $z$ band) also offers a big improvement in both resolution and depth, as LSST only reaches AB=28.0,27.0 in $z$ and $y$ in the "near-uniform DDF" approach spending 60% of DDF observing time in those filters; significantly deeper $ugri$ images could be obtained if this time is saved. Full areal coverage by WFIRST of as many LSST DDFs as possible, in $zyJH$, will motivate LSST to devote its observing time in those fields to the deepest possible $ugri$ imaging, enabling a wide range of science with these "ultra-deep drilling" fields.

## 4.4 Coordinated supernovae observations

If the WFIRST SN medium and deep surveys target the same fields as one of the LSST deep drilling fields, then coordinated observations would yield 8 band supernova light curves. The WFIRST IFU measurements of these supernovae would yield a powerful training set that could greatly enhance the value of the large LSST SN sample. These coordinated observations would not require any increase in observing time by either project but would enhance the return of both sets of observations.

## 4.5 Variable universe: follow-up on novel targets

LSST's cadence and depth will allow it to discover new classes of variable objects and "golden" transients, examples of known classes of objects that have nearly ideal properties for particular aspects of astronomical study. WFIRST's resolution, spectral coverage, and wavelength coverage will make it a premier facility for following up on some of these objects in the near infrared. Since studies of individual targets do not take advantage of WFIRST's large field of view, large ground-based telescopes with high quality adaptive optics will also be important tools for follow-up observation.

There are a host of potential targets for follow-ups. Examples include:

- "Macronovae" produced by the coalescence of binary neutron stars (Li & Paczynski 1998, Kulkarni 2005, Bildsten et al. 2007). These explosive events likely peak in the near-infrared and would be potential targets for both WFIRST imaging and IFU follow-up.

- LSST and WFIRST will be our most powerful combination of telescopes for detecting afterglows of gravitational mergers (Kasliwal and Nissanke 2014; Gehrels et al. 2014). Given the LIGO's large error box for each event, there will be many variable objects found at the LSST/WFIRST depths, a program of joint monitoring (perhaps, combined with the deep field program described above) will enable a better characterization of the rich zoo of transients.

- Gamma Ray Bursts (Rossi et al. 2007).

# 5 Data management and analysis challenges

The sections above make clear that in essentially every area of astronomy and cosmology, joint analysis of the LSST, WFIRST and Euclid data will provide significant benefits. These will be made possible by linking the data from the different surveys and providing a common access point for interrogating the data in a user-friendly way with the appropriate tools. This would enable scientists to explore connections between the data sets and stimulate novel community research leveraging the combined data. The publication of detailed, accurate documentation along with the cross-referencing and associated data releases would be the most effective way to encourage and support the community in pursuing such research. This section describes the issues involved in carrying out such a program for the photometry and simulations required to analyze and interpret the data.

## 5.1 Multi-resolution photometry

There are quite a few challenges to producing accurate, consistent photometry across different filters with widely different resolutions, as would be the situation for combining LSST data with either WFIRST or Euclid.

We can assume that we will have a catalogue of the positions of objects detected in at least one band (including single-band detections in u or H), and that we know the PSF in each band. There are issues with respect to the chromatic effects within a single band as well as spatial variation of the PSF across the field of view, which are important, but we can neglect them for this discussion. We also assume that blending issues have been taken into account, so we have a set of pixels in each band that correspond to a single object.

For point sources, photometry is relatively straightforward. Because all point sources have an identical profile (again, ignoring for now the spatial variation of the PSF) we may use any aperture to measure the relative flux of all point sources in an image (a PSF aperture is optimal). We may then estimate the image's zero-point which gives us total fluxes for all sources. We may repeat this in all the bands, resulting in a catalogue of fluxes and colors.

Measuring the fluxes of galaxies is harder. It is not possible to simply measure the total flux in an aperture large enough to include the entire galaxy as the measurements are too noisy, and this is not even a well-defined concept. On the other hand, a small aperture measures a different fraction of the flux of different galaxies, so we cannot simply apply an aperture correction as in the stellar case. Even if a galaxy has the same profile in each band these fractions are different if the PSF is band-dependent; if the profiles vary the difficulties are multiplied.

If we can model galaxies adequately, this alleviates the difficulties of flux measurement. The PSF-convolved model flux is a reasonable definition of the galaxy's total flux in each band. However using model-based photometry is not a panacea, as if the model must be correct. For example, choosing a model in one band and applying it in another will not yield the correct color for a galaxy with a red bulge and a blue disk, a scale length that's a function of band, spiral arms, HII regions, or any other kind of substructure with a different color from the overall galaxy light.

If the seeing is (or can be made) the same in all bands we are allowed to choose a single model. The resulting fluxes are not the galaxy's total fluxes, as different components are not necessarily weighted correctly, but these "consistent fluxes" do represent the fluxes of a certain sample of the galaxy's stellar population. However, degrading all bands to a lowest common denominator would destroy much of the extra information present in the higher-resolution space data. Determining the optimal way to estimate consistent fluxes while keeping as much of this high-resolution information as possible is still an open research question.

Given the complexity of the problem of measuring galaxy photometry across a decade in wavelength with varying PSFs, there is no optimal approach. Different science questions will require different ways of combining the multi-color data.

- *Photometric Redshifts:* We may use an object's total or consistent fluxes to find a location in multi-color space and hence its redshift probability function, $p(z)$. If $p(z)$ is unimodal and sharp, the two estimators predict nearly the same photo-z. However, in general this will not be so and $p(z)$ will be different for the two choices of flux.

- *Choice of Galaxy Model:* In the rest-frame optical, a simple Sersic or constrained bulge-disk model is probably sufficient to return reliable fluxes (such models down-weight the contribution from localized star-forming regions, changing p(z) at least in principle). However, modeling galaxies across the broad redshift range covered by WFIRST, Euclid, and LSST will likely require more complex models (including, for example, the uv-bright knots that contribute the bulk of the flux in the far-uv). Constraining these models is going to be hard; one approach would be to demand that all components have a consistent $p(z)$.

- *Photometry of Blended Objects* If our models indeed accurately describe our galaxies then the de-blending problem is no different from the galaxy photometry problem; instead of fitting multiple components to a single galaxy we fit multiple components to multiple galaxies, relaxing but not abandoning the constraints on components' $p(z)$. In practice it is not clear how well this will work, and there are intermediate schemes (similar to the deblender used in SDSS) that fit simplified models to the ensemble of objects to assign the flux to the child objects and then proceed with photometry. This is an unsolved problem, but given the upcoming data, we will need to must significant progress in the coming years.

## 5.2 Joint simulations

WFIRST, Euclid and LSST will generate a remarkable set of observations in the optical and near-infrared wavebands. Their cosmological interpretation, especially in combination, will be very powerful but also very challenging. Many systematic effects will have to be disentangled from fundamental physics to fully exploit the power of the measurements. Simulations of all aspects of the experiment – at the cosmological scale, of the instruments, and of the data processing and analysis software – are critical elements of the systematics mitigation program. Joint analyses of multiple surveys require these simulations to have consistent interfaces, to enable the same realizations of cosmological volumes or multi-wavelength simulated galaxies to be passed through instrument-specific simulators, and ultimately to ensure that the correct correlations and joint statistical properties are represented when mock analyses are performed on the simulated output catalogs.

We focus here on the role of coordinated simulations in two areas – *cosmological simulations* and *instrument/pipeline simulations*.

### 5.2.1 Cosmological simulations

Cosmological simulations play key roles within the interpretation of large scale structure surveys: (i) they provide controlled testbeds for analysis pipelines and tools for survey design and optimization via sophisticated synthetic skies populated with galaxies targeted by the specific survey, (ii) they provide predictions for different fundamental physics effects, including dynamical dark energy, modified gravity, neutrinos, and non-trivial dark matter models, (iii) they provide modeling approaches for astrophysical systematics, including gas dynamics, star formation, and feedback effects, and (iv) they provide important information about the error estimates via covariance studies. We will briefly elaborate on the synergies between the simulation programs required for WFIRST and LSST in these four areas in the following.

*Synthetic sky maps* — In order to build synthetic sky maps for Euclid, WFIRST and LSST, large-volume $N$-body simulation with very high force and mass resolution are essential. Next, the simulations have to be populated with galaxies using sub-halo abundance matching or semi-analytic modeling techniques, tuned to match the observed population of galaxies. While the wavelength ranges of Euclid, WFIRST and LSST will be different, a common code infrastructure will be beneficial in some areas (e.g., in exploring model parameter space) and critical in others (e.g., simulating realistic merged multi wavelength catalogs).

*Precision predictions for fundamental cosmological statistics* — The precision requirements for cosmological observables and the fundamental physics to be explored is very similar for Euclid, WFIRST and LSST and a joint program in this area would be very fruitful. Both surveys are promising observations of large scale structure measurements at the sub-percent level, a tremendous challenge for cosmological simulations. Besides the precision challenge, different fundamental physics effects have to be explored systematically for both surveys.

*Astrophysical systematics* — Some of the systematics that have to be accounted for to interpret the observations from Euclid, WFIRST and LSST are common, such as the effects of baryonic physics on the weak lensing shear power spectrum. Since the accuracy requirements are similar, a joint program would be beneficial to both surveys. However, there are significant differences because of the differing ground and space-based nature of the two programs, different instrumentation and associated wavebands, and different galaxy populations and densities. Nevertheless, the underlying sky catalogs can be generated from the same set of simulations.

*Covariance matrices* — Since the Euclid, WFIRST and LSST surveys will have overlapping volumes, combining cosmological constraints from their data sets will require understanding the full covariance matrix of all observables extracted from the data sets, using simulations of the entire relevant volume. While the accuracy requirements for these simulations are much less stringent than for other applications, the large number of simulations required will represent a computational challenge.

## 5.2.2 Instrument and pipeline simulations

Instrument and pipeline simulations are also a critical element of any precision large scale structure program: they inform the relationship between the astronomical "scene" in the Universe and the catalogs, parameters, and other information that is recorded in the processed data. They also provide an opportunity to exercise the data reduction and analysis tools prior to first light. The instrument and pipeline tools must be considered together, since in a precision experiment the pipeline is an integral part of the measurement process. Unlike cosmological simulations, these tools are highly instrument-specific: for example, the atmosphere is only included for ground-based observations, the patterns of ghosts and diffraction features are completely different for the LSST, Euclid and WFIRST configurations, and internal optical effects, cross-talk, and readout artifacts are fundamentally different in silicon CCDs versus NIR arrays.

Despite the differences in instrument simulators, it is critical that they be able to take compatible input data. Precision measurements of galaxy clustering and weak lensing depend on not just counting objects and measuring shapes, but understanding selection biases, blending effects, and error distributions of measured parameters. The combination of WFIRST, Euclid and LSST data will be used to measure photometric redshifts, and for both object-by-object and statistical comparisons of number density and shear. Simulation of these applications requires that the same objects be fed through both pipelines, and hence that the input scene (simulated stars and galaxies) be consistent and have the proper correlations of object properties across wavebands. While proprietary pipelines do not need to be shared outside of the individual constortia, the inputs of these pipelines must remain compatible.

# 6 Conclusion

The scientific opportunity offered by the combination of data from LSST, WFIRST and Euclid goes well beyond the science enabled by any one of the data sets alone. The range in wavelength, angular resolution and redshift coverage that these missions jointly span is remarkable. With major US investments in LSST and WFIRST, and partnership with ESA in Euclid, we have an outstanding scientific opportunity to carry out a combined analysis of these data sets. It is imperative for us to seize it, and together with our European colleagues, prepare for the defining cosmological pursuit of the 21st century.

The main argument for conducting a single, high-quality reference co-analysis exercise and carefully documenting the results is the complexity and subtlety of systematics that define this co-analysis. Falling back on many small efforts by different teams in selected fields and for narrow goals will be inefficient, leading to significant duplication of effort.

For much of the science, we will need to combine the photometry across multiple wavelengths with varying spectral and spatial resolution – a technical challenge. As described in Section 3, the joint analysis can be carried out in ways that have different computational demands. The most technically demanding joint analysis is to work with pixel level data of the entire area of overlap between the surveys. Many of the goals of a joint analysis require such a pixel-level analysis. If pixel-level joint analysis is not feasible, catalog-level analysis can still be beneficial, say to obtain calibrations of the lensing shear or the redshift distribution of galaxies. Hybrid efforts are also potentially useful, for example using catalog level information from space for deblending LSST galaxies, or using only a mutually agreed subset of the data for calibration purposes. However the full benefits of jointly analyzing any two of the surveys can be reaped only through pixel-level analysis.

The resources required to achieve this additional science are outside of what is currently budgeted for LSST by NSF and DOE, and for WFIRST (or Euclid) by NASA. Funding for this science would most naturally emerge from coordination among all agencies involved, and would be closely orchestrated scientifically and programmatically to optimize science returns. A possible model would be to identify members of the science teams of each project who would work together on the joint analysis. The analysis team would ideally be coupled with an experienced science center acting as a focal point for the implementation, and simultaneously preparing the public release and documentation for broadest access by the community.

**Appendix I    Foreign Statements of Interest**

## I.1.    Introduction

Outside the US, there is strong interest in participating in the WFIRST-AFTA mission, as part of both the Coronagraph Instrument (CGI) and the Wide-Field Instrument (WFI). For the past year, scientists in Japan, Canada, and the United Kingdom and Europe have been having scientist-to-scientist discussions with colleagues in the US about possible modes of participation. In each of these national areas, scientists have come together to make informal suggestions for future participation as part of the CGI and WFI. These are their statements of interest.

## I.2.    Japanese interest in CGI and WFI

WFIRST-AFTA provides unprecedented opportunities in the many fields of astronomy and astrophysics, especially for precision cosmology, microlensing and high-contrast studies of exoplanets. Deep and wide-field surveys produce magnificent data to study of extragalactic and Galactic sciences. We appreciate the opportunity to have a representative in WFIRST-AFTA SDT from its start in 2013 under the agreement of NASA/JAXA. The activity is well supported by the Space Science Steering Committee of JAXA/ISAS and it is highly visible in Japanese astronomical community. WFIRST-AFTA indeed rouses great interests of Japanese astronomers in many fields from cosmology, galaxy formation and evolution, Galactic bulge science, astrometry, exoplanets, with expertise of deep/wide surveys and high-contrast imaging so far conducted with Subaru Telescope as well as the microlensing observations with MOA telescopes. The Japanese activity of WFIRST-AFTA Coronagraph (WACO) Working Group including the funds for preliminary Research and Developments is officially supported by JAXA/ISAS Space Science Steering Committee. This activity will be enlarged to the WFIRST-AFTA Working Group so that future contributions in developing WFIRST-AFTA hardware/software as well as science cases in various fields from exoplanets to cosmology will be developed. Following are the important items of WFIRST-AFTA project and its development to which Japanese astronomers can significantly contribute.

### CGI (Japan)

As hardware contributions, Japan has been developing and can develop instrumentation that could improve the science yield of the WFIRST-AFTA coronagraph. One of the possible contributions is the polarization module in the coronagraph; Japan could furnish a polarization module and insertion mechanism for the OMC, to enhance debris/protoplanetary disk science where polarization measurements are a valuable observing tool and allow better inner working angle than the conventional PSF intensity subtraction. Other possible contributions include innovative masks and analysis techniques for coronagraph. Japan has developed innovative mask fabrication techniques, specifically photonic crystals and super ink-jet printing. These are potential risk-reduction technologies for the mission, and the masks could be evaluated on the vacuum testbeds for such optics at JAXA/ISAS. Japan has also developed new analysis techniques for coronagraphs; these include calibration methods PSF subtraction, Speckle Area Nulling (SAN) method as a dark-hole control algorithm, the use of low order wavefront sensor telemetry for PSF reconstruction, and other related analysis tools. Japan could also provide a module and insertion mechanism, using transmissive elements, to be moved in or out of the OMC optical train with minimal impact on the overall design. Japan could provide a PIAA module using reflective elements, to be mounted on the back side of the OMC optical bench, sharing much of the OMC's optical system. This module would be optimized for operation in the red (0.6-1.0 micron) and possibly the near-infrared (0.6-1.6 micron). This takes advantage of the better Inner Working Angle of a PIAA design (IWA~1.5 micron/D), and would have comparable angular resolution to the OMC in the blue (IWA ~2.5 $\lambda$/D) analysis tools.

### WFI (Japan)

**Coordinated ground-based observations:** Coordinating ground-based wide-field and deep spectroscopy as well as deep optical imaging are important in maximizing the WFIRST-AFTA output. Deep spectroscopic data provides unique calibration for photometric redshift which is essential in precision cosmology and study of structure formation. The Hyper Suprime-Cam (HSC) and Prime Focus Spectrograph (PFS), which are the new wide-field instruments of 8-m Subaru Telescope, would complement the science of WFIRST-AFTA. The large-scale observation program with HSC is now ongoing and will be completed until around 2019. The HSC program will deliver the datasets with 5 broad-bands (grizy) and several narrow-band filters over the sky area of 28 deg$^2$ (deep layer) and 3.5 deg$^2$ (ultradeep, includ-ing a narrow-band filter centered at 1μm). These multi-color data sets that will be available well before WFIRST-AFTA launch would be very complementary to the WFIRST-AFTA NIR data. The similar-size large-scale PFS program is planned to start from around 2019 for 5 years duration. The combination of the Subaru aperture and the PFS's wide FoV and multiplex can deliver a unique spectroscopic data set of faint galaxies for calibrating photometric redshifts of the WFIRST/AFTA weak lensing program as well as for better characterizing spectral energy distribution of galaxies when combined with WFIRST-AFTA NIR data. With support of the Subaru community, we are also interested in exploring a coordination to carry out additional HSC and/or PFS surveys for the regions in the northern hemisphere that are needed for the WFIRST-AFTA, if necessary.

**Microlensing / Galactic Bulge Science:** The MOA collaboration has been one of the major microlensing survey teams for more than 10 years, and their experiences can contribute to the analysis of the microlensing exoplanet search by WFIRST-AFTA. The state-of-the-art difference imaging pipeline and fast light-curve modeling algorithm can be adopted to maximize the exoplanet yield from the massive datasets by WFIRST-AFTA. We are planning to conduct a precursor ground-based IR microlensing exoplanet survey by a dedicated 1.8m wide-field telescope in Namibia. This survey will directly search for the best observational fields for the WFIRST-AFTA, i.e. the highest event rate, in the galactic bulge before the launch. After the launch, the simultaneous observations from this ground-based telescope and WFIRST enable us to measure the mass of the planets by the parallax measurements. As for the galactic bulge science, the microlensing event rate and the timescale distributions can be used to constrain the bulge model and we have many experiences on such studies. The high precision astrometry measurements of the stars in the galactic bulge fields by WFIRST-AFTA are powerful tools to study the structure of the bulge and their formation and evolutional history. However, a high precision astrometry is mostly systematic limited. Significant amount of efforts for studying various systematics on the micro-arcsec astrometry has been invested by the Japanese planned IR astrometric satellite, JASMINE. The precursor mission, nano-JASMINE will be launched very soon and their experiences should be useful for the planning and the analysis of the astrometric measurements by WFIRST-AFTA.

Motohide Tamura (University of Tokyo/NAOJ), coordinator for CGI
Takahiro Sumi (Osaka University)
Masahito Takada (Kavli IPMU, University of Tokyo)
Toru Yamada (Tohoku University), JAXA/ISAS representative for WFIRST-AFTA
JAXA/ISAS WACO Working Group
Japanese WFIRST-AFTA Research Group

### I.3. Canadian interest in CGI and WFI

**Introduction - WFIRST in the context of strategic planning in Canada:** Within the Canadian astronomical community there is very strong interest in and support for WFIRST. The Canadian Long-Range Plan (LRP) for Astronomy of 2010 ranked both ground-based and space-based projects. The top priority for space astronomy was the recommendation that "Canadian astronomers participate in a major wide-field Dark Energy satellite mission. Joining Euclid or WFIRST as a significant partner would fulfill this recommendation, provided that we can (i) negotiate a partnership in the leading mission, and (ii) identify a contribution to the satellite instrumentation." Note that Euclid has proceeded without a Canadian contribution to instrumentation. Furthermore, WFIRST in its current 2.4m AFTA format, offers additional science opportunities beyond the dark energy projects. In particular, the addition of a coronagraph module also allows for Canadian strengths in this area to be applied to WFIRST, as described below.

**Canadian science interests related to WFIRST - Wide-field imaging:** Canada has traditional strength in wide-field imaging by virtue of the Canada-France-Hawaii Telescope (CFHT), and in particular Megacam, the CFHT Legacy Survey (CFTHLS) and the CFHT Large Programs. The Supernova Legacy Survey, based on the CFHTLS-Deep, is the most advanced SN survey in the world at present. It is a joint project between Canadian and French astronomers. Similarly, the CFHTLenS analysis of the Wide component of the CFHTLS is the current state-of-the-art weak lensing survey. Both of these programs have had a very significant scientific impact. Participation in WFIRST would allow existing team members and their students and postdocs to apply their expertise to SN and weak lensing surveys of the 2020s. More broadly, the 25% GO time will also be of great interest to Canadian astronomers. Finally, as described below, the Canadian astronomical community has ample expertise in instru-mentation related to the wide-field imaging module of WFIRST.

**Canadian science interests related to WFIRST - Exoplanets:** WFIRST is of great interest to the exoplanet community in Canada, which views the coronagraph instrument as a key stepping stone in the quest to directly image and characterize planetary systems similar to our own, a line of research in which Canada has been very interested, active, and has acquired internationally recognized expertise. Participation in WFIRST would build upon the space science and space instrumentation expertise in the community developed in the context of the FGS-NIRISS instrument for the Webb telescope. Indeed, the two main science drivers of the NIRISS instrument, "the detection of first-light galaxies and characterization of exoplanets" are intricately connected to the main science goals of WFIRST. In addition, the Canadian community has invested in several ground-based exoplanet-focused enterprises, most notably GPI (high-contrast exoplanet imager for Gemini South), NFIRAOS/IRIS/PSI (near-infrared adaptive optics imaging instrument for the Thirty Meter Telescope), and SPIRou (radial-velocity exoplanet finder spectrograph for CFHT), which, together with the NIRISS efforts, will help set the stage for the work to be undertaken with the WFIRST coronagraph. The timeline of WFIRST is perfect, as it would come right after these projects.

**Status of WFIRST instrumentation studies in Canada:** Since late 2013, a Canadian astronomer (Hudson) has participated in the WFIRST-AFTA SDT, supported by the Canadian Space Agency (CSA). In 2014, CSA issued an RFI on potential contributions to WFIRST. Based on the response to this RFI and the importance of WFIRST to the Canadian astronomical community as expressed in the Long-Range Plan, CSA issued an RFP for two WFIRST studies, one for contributions to the WFI component, and the other for contributions to the coronagraph. Based on the RFI submissions and other considerations, CSA selected the following instrumentation options for further study.

**Summary:** At the time of writing, these instrumentation studies are ongoing. The WFI study is being led by COM DEV and the coronagraph study by ABB Canada. In the second phase of the studies, a subset of these options will be studied in greater detail. CSA also issued a call for a "Pool of Experts" to assist with the studies which received a strong response (over 20 Ca-nadian researchers). The studies are expected to be completed by mid-2015. The mid-term review of the 2010 LRP is also underway and will update the recommendations of the Canadian LRP.

The findings and recommendations of these studies will be used as a basis to solicit the necessary funding for a contribution to WFIRST (through the CSA). It is hoped that, through these efforts, Canadian astronomers will have the opportunity not only to collaborate on WFIRST instrumentation but also that Canadian astronomical community as a whole will be able to fully access WFIRST data and to participate in WFIRST science team membership.

## WFI (Canada)

**Integral Field Unit (IFU) or subsystem(s):** Canada has accumulated ample expertise in developing near-infrared spectrographs, including an IFU-based one, which will be essential to making a contribution to the WFI IFU module. Specific examples include: (1) the Wide Integral Field Infrared Spectrograph, which is an image slicer-based integral field spectrograph in the near-infrared waveband, similar to what has been envisioned for the IFU module of WFIRST; (2) in partnership with the NASA JWST team, leading the development of a near-infrared multi-object spectrograph using the microshutter array for ground-based wide-field adaptive optics observations; (3) the Near Infrared Echelle Spectrograph for the Keck II telescope with the California Institute of Technology; and (4) the calibration system for the near-infrared instruments and adaptive optics system of the Thirty Metre Telescope. Through these projects, by developing cutting-edge near-infrared spectrographs, most of the expertise required for developing the WFI IFU module is present in Canada.

**Photometric calibration (pre-flight ground or flight):** Several of the primary science goals of WFIRST, including measurements of dark energy using type Ia supernovae, and minimization of bias (and of bias uncertainties) in photometric redshifts for weak lensing measurements, will be limited in precision by calibration uncertainties, particularly those stemming from photometric calibration of the imager. Canadian expertise in multiple areas of calibration, including photometric calibration via traditional techniques of determining and utilizing stellar standards, as well as novel means such as calibrating stellar standards via calibrated artificial sources above the atmosphere, will be critical in mini-mizing these limiting uncertainties and thus maximizing the scientific output of WFIRST.

**Fine Guidance System:** Canada has provided the full fine-guide capability (hardware and software) for the FUSE and JWST missions, via CSA's aerospace contractors, working with science teams. The complex tradeoffs between sensitivity, noise, image quality, wavelength, and guide star availability are well understood, as well as field recognition and target acquisition. Canada is in an excellent and experienced position to provide such capabilities for the WFIRST WFI.

**Wide-field imager:** In addition to the possible contributions to WFIRST instruments described above, the Canadian community has developed a concept for a UV/blue wide-field imager (CASTOR) that could significantly enhance the scientific reach of the proposed WFIRST WFI survey.

### CGI (Canada)

**Integral Field Spectrograph (IFS):** The proposed IFS unit shares many similarities with the IFS successfully developed for the Gemini Planet Imager by Université de Montréal and Université Laval. Canadian space industry has both the technical expertise and the collaborative experience with CSA and NASA for delivering such a science instrument.

**EMCCD detector for the IFS and the imaging camera :** An electron-multiplying charge-coupled device (EMCCD), effectively operating with zero noise, would be an excellent choice for the needed detectors. NuVu Cameras, based in Montreal, is one of only three groups in the world with the required expertise on the control electronics for EMCCDs. With funding from CSA, work is proceeding on qualifying these cameras for space applications.

**Subsystems:** Coronagraph subsystems, more specifically, the coronagraph low-order wave front sensor and the coronagraph filters/masks wheels. There is excellent Canadian expertise in adaptive optics (Canada has led or is an important contributor to the facility-class Gemini/ALTAIR system, the Gemini/GPI and the TMT/NFIRAOS) as well as in building wheel mechanisms for space applications.

**Data reduction pipeline, data processing, and archiving:** Canadian astronomers developed the image processing algorithms that allowed the first direct imaging detection of exoplanets in 2008. This pioneering work has then continued with the development of the data reduction pipeline for the Gemini Planet Imager. Both the WFI and coronagraph studies will also analyze options for data processing and archiving contributions. The Canadian Astronomy Data Centre has a long history of productive collaboration with NASA on HST data management. CADC also archives wide-field images from CFHT as well as data from many other observatories, so there are clear opportunities to enhance WFIRST science potential for the astronomical community.

Michael Hudson, University of Waterloo (CSA representative on SDT, coordinator for WFI and CGI)
Roberto Abraham, University of Toronto
Justin Albert, University of Victoria
Loic Albert, Université de Montréal
Étienne Artigau, Université de Montréal
Pauline Barmby, Western University
Pierre Bastien, Université de Montréal
John Blakeslee, NRC-Herzberg
J. Richard Bond, CITA/University of Toronto
Colin Bradley, University of Victoria
Raymond Carlberg, University of Toronto
Patrick Coté, NRC-Herzberg
Andrew Cumming, McGill University
Rene Doyon, Université de Montréal
Ray Jayawardhana, York University
Gary Hinshaw, University of British Columbia
John Hutchings, NRC-Herzberg
David Lafrenière, Université de Montréal
Jérome Maire, University of Toronto/Dunlap Institute
Christian Marois, NRC-Herzberg
Brenda Matthews, NRC-Herzberg
Jaymie Matthews, University of British Columbia
Alan McConnachie, NRC-Herzberg
Kristen Menou, University of Toronto
Stanimir Metchev, University of Western Ontario
Dae-Sik Moon, University of Toronto
Ralph Pudritz, McMaster University
David Schade, NRC-Herzberg
Simon Thibault, Université Laval
Diana Valencia, University of Toronto
Jean-Pierre Véran, NRC-Herzberg
Yanqin Wu, University of Toronto

## I.4. UK and European interest in CGI and WFI

### CGI (UK and Europe)

In France, Paris Observatory (LESIA), Marseille laboratory (LAM) and Nice Observatory (LAGRANGE) have been involved in coronagraphy and high contrast imaging for exoplanets and disks science. These activities were started many years ago for both ground-based and space projects.

**LESIA/Paris Observatory (FR):** LESIA has been working on the concept of the phase mask (Rouan et al. 2000). We have manufactured coronagraphs for NACO/VLT (Boccaletti et al. 2004), SPHERE/VLT (Boccaletti et al. 2008) and MIRI/JWST (Boccaletti et al. 2005, Baudoz et al. 2006, Boccaletti et al. 2014.). We have led the Cosmic Vision proposal to ESA in 2010 « SPICES », a small-scale version of WFIRST for spectro-polarimetric analysis of giant planet (ultimately Super-Earth) atmospheres (Boccaletti et al. 2012, Maire et al. 2012). Several observations of disks and exoplanets using current facilities were published by our team (Boccaletti et al. 2009, 2012, 2013a,b, Mazoyer et al. 2014, Galicher et al. 2013, 2014) or in collaboration (Lagrange et al. 2009, 2010, 2012, Bonnefoy et al. 2011, 2013, Chauvin et al. 2012, Rameau et al. 2013). Recently, we have been developing a dedicated high contrast test bench (THD bench) to study under the same conditions several coronagraphs and post-coronagraphic techniques, individually and in association. The main post-coronagraphic solution we have been working on is the Self Coherent Camera (SCC), which uses the coherence of the stellar light to generate Fizeau fringes in the science focal plane and spatially encode the speckles (Baudoz et al. 2006). The SCC can be used as a focal plane wavefront sensor for active correction on space-based and ground-based telescopes (Galicher et al. 2008, 2010, Baudoz et al. 2010, 2012, Mazoyer et al. 2013) and as a differential imaging technique to discriminate the planet image from the stellar residuals (Baudoz et al. 2006, 2013, Galicher et al. 2007). The SCC has been tested in the laboratory with several types of coronagraphs: Four Quadrant phase Mask (Rouan et al. 2000), Multi-stage Four quadrant phase mask (Baudoz et al 2008, 2010, Galicher et al. 2011) and Dual Zone Phase mask (Soummer et al. 2003, Delorme et al. 2014). Laboratory results on the THD bench proved that the SCC could reach a contrast of a few $10^{-8}$ for all of these coronagraphs at short spectral bandwidth (Mazoyer et al. 2014) and that the contrast is slightly degraded ($4 \times 10^{-8}$) for larger bandwidth (up to 5%, Delorme in preparation) or when measuring the wavefront errors on medium spectral bandwidth (10%, Delorme et al. 2014).

Apart from these developments on high contrast imaging, LESIA and the Observatoire de Paris have strong expertise in building instruments for space missions with for instance participations in Corot, ROSETTA, CASSINI, Gaia)

**LAM – Marseille (FR):** LAM is a leading European space laboratory combining astrophysical research and instrument development for large ground-based and space telescopes, with participation in large space missions (SOHO, Herschel, Corot, Galex, Rosetta, Euclid). The laboratory has been deeply involved in the high contrast imaging instrument VLT-SPHERE (IRDIS camera, high quality aspherical mirrors and high speed WFS camera), developing a deep knowledge on high-contrast imaging and spectrograph systems (Vigan et al. 2013, Antichi et al. 2009). LAM has also a long expertise in phase mask coronagraphy (DZPM, Soummer et al. 2003, N'Diaye et al. 2010, 2012) and associated post coronagraphic wavefront sensing (Vigan et al 2011, Dohlen et al. 2013, Paul et al. 2013, 2014). LAM is specialized in manufacturing very-high quality optics (Hugot et al. 2012) and contributed to the STScI high contrast imaging bench (HiCAT – N'Diaye et al. 2014). LAM has an excellent knowledge in producing active mirrors and wavefront sensing for space (MADRAS and RASCASSE projects, Laslandes et al. 2012, 2013), in close collaboration with space industry (Thales Alenia Space). On all these topics, LAM has a very strong collaboration with the French aerospace laboratory (ONERA). With the support of the French space agency (CNES), numerous PhD thesis are pursued or planned in collaboration with ONERA, LESIA and STScI.

LAM is equipped with several technology facilities, specially conceived for the assembly, integration and tests of future space instrumentation. These facilities also include MOEMS cryogenic test bench and high-contrast imaging bench. LAM has also a deep expertise in detector technology and was at the origin of the spin-off company First Light Imaging (FLI). This company (already under contract with JPL) manufactures cameras with high cadence and low noise designed for wavefront sensing. They are built around E2V CCDs and the company would be able to provide support on the test/design of the detector and the electronics in collaboration with UK (see below).

**LAGRANGE/Nice Observatory (FR):** LAGRANGE is associated to the development of SCExAO at Subaru (Martinache et al. 2014) and in particular of the Low order wavefront sensor (Singh et al. 2014), as well as new techniques for wavefront sensing (Martinache 2013a, Martinache et al. 2012) and postprocessing (Martinache 2013b). LAGRANGE has expertise in designing coronagraphs and high contrast test benches (Martinez et al. 2007, 2008, 2009, 2010, 2011, 2012).

French institutes could participate at scientific and instrumental levels. Importantly, CNES would be willing to support a small contribution if WFIRST got select and is already funding our laboratories for technological developments (THD bench for the SCC, space active optic and WFS benches). First of all, the Maire et al. analysis can be re-applied and extended further to the case of WFIRST-AFTA with the objective to evaluate the science cases (not just detection but rather performance in characterization). We can also participate to the design/optimization of the instrument with emphasizes on the wavefront sensing and correction (low and high orders) and provide very specific components (optics, etc..). The various benches and space facilities can come in support to other US facilities for testing some components. In addition, with our observational expertise (VLT, Subaru) we can contribute to the data reduction strategy/pipeline.

Other countries involved in SPICES have also expressed interests: Obs Padova Italy for the design of the IFS with expertise in SPHERE, Netherland for polarimetry.

Finally, Airbus Defence & Space (Toulouse) has participated in the SPICES study and might be able to contribute on the platform or telescope.

**CEI-OU/MSSL-UCL:** In the UK interests centre on the CCDs and the associated detection chain camera system. These are based around a close alliance between CCD manufacturer e2v, the Centre for Electronic Imaging at the Open University (CEI-OU) and the Mullard Space Science Laboratory, University College London (MSSL-UCL).

e2v and CEI-OU have established a formal cooperative relationship, with e2v funding PhD studentships and postdoctoral fellowships at CEI-OU, and providing devices for analysis, characterisation and testing, particularly with respect to radiation damage for the space environment. Senior engineers interact with the CEI-OU team. e2v in turn benefit from this detailed evaluation of their devices, and are able to implement improvements in both manufacture and operation as a result. MSSL-UCL have worked with e2v and CEI-OU in a number of projects, providing the detection chains to operate the CCDs and provide digital outputs, including those in *Gaia*, *Euclid* and *PLATO*, all of which have massive CCD focal planes, the biggest flown in any space-science instrumentation. MSSL-UCL lead the VISible instrument, one of the two instruments on *Euclid*. VIS has 36 4k x 4k CCDs, and 144 detection chains supported by 12 camera units with low noise and ultra-high stability, and with sophisticated readout schemes developed in consultation with CEI-OU to maximise resilience to radiation damage. In Gaia, CEI-OU worked with MSSL-UCL to space-qualify EMCCDs for the spectrometer instrument, in a large testing programme using 100 CCDs. In *PLATO*, MSSL-UCL is building the camera units for 128 large format CCDs, this time with 256 detection chains in 32 units. Moreover, CEI-OU is already under contract with the WFIRST team for EM CCD characterisation.

**IoA, Cambridge:** The IoA has a long-standing involvement in science teams for space-based coronagraphic imaging, primarily through its expertise in debris disks (e.g., Wyatt 2008, ARAA, 46), the circumstellar disks of asteroids and dust which will be characterized by such techniques and provide noise that may prevent detection of planets. Science team membership includes both European and US projects (SPICES, DDX, EXCEDE which is currently receiving NASA Category III Explorer funding for technology development and maturation, and HOSTS – the NASA-funded LBTI project to characterise the exozodis of nearby stars in preparation for future imaging of Earth-like planets). The group at the IoA also has a strong heritage in the production of science data products from key high value missions. It has recent relative heritage in delivering advanced high throughput processing chains for Gaia, and is leading the exoplanet analysis system for PLATO. It provides data systems for a number of ground-based surveys, imaging (VISTA, VST) and spectroscopic (VLT/FLAMES, Gaia-ESO Survey, 4MOST, WEAVE [includes significant IFS components]). There is a strong focus on exoplanet synergies, with Gaia and PLATO being closely coupled, interfaces to both will be of significant value in the analysis of WFIRST coronagraphic data.

This tight collaboration between manufacturers and scientific institutes provides an unparalleled understanding and capability of optimisation of the detector chain and of its performance in space, which would be of undoubted benefit to the two Coronagraph channels on WFIRST. While science groups such as CEI and MSSL are both enthusiastic and competent to contribute to the Coronagraph, UK funding will require the usual approvals from the UK Space Agency, involving, in the first instance, a Statement of Interest to the UKSA Space Projects Review Panel, and would be subject to strategic and financial considerations currently being explored by UK groups with UKSA.

### WFI (UK and Europe)

WFI will be the pre-eminent astronomical wide field infra-red imager in the short and medium term. It's no surprise therefore that several European groups are eager to participate in the instrument programme, in the ground system for the processing of the data from it, and in scientific programmes of interest to them which would be most effectively pursued with the WFI. An involvement would be informed by current contributions to (among others) *JWST* NIRSpec/MIRI and *Euclid*. Consequently, groups in France, Germany, Italy, Spain and the UK were consulted, coordinated by UK interests, with the request to non-UK groups to consult within their own communities. The list included in France: le Fèvre (LAM), Mellier (IAP); in Germany: Bender (MPE), Rix (MPIA); in Italy: Cimatti (Bologna), Scaramella (Rome); in Spain: Castander (Barcelona); and in the UK: Sharples (CfAI, Durham), Wright (UKATC), Davies (Oxford), Nichol (Portsmouth), Cropper (coordinator, MSSL-UCL). These interests are summarised below for those who have responded.

**LAM:** Instrumental interests in the WFI design, manufacturing, testing; the IFU spectrograph including slicer disperser and mechanical structure, and, more generally, opto-mechanical systems. For the ground system, they have expertise in deep spectroscopy surveys for cosmology and galaxy evolution; IFU spectroscopy processing; multi-wavelength analysis – photometric redshifts, SED fitting and spectral analysis. They are lead institute for the *Euclid*-NISP instrument, and contributed to the *Herschel* SPIRE FTS. Facilities include a 90m$^3$ vacuum chamber which they use for NISP. They also developed to TRL6 an IFU spectrograph for the *SNAP* study with LBNL, with a 4" x 4" field of view and spectral resolving power of 100 – 300.

**IAP:** Interest is in image processing and automated quality control software development and in pipelining. IAP hosts the Terapix data centre for wide field visible (MegaCam at CFHT) and near infrared imagers

(VIRCAM on ESO/VISTA). IAP is responsible for the VIS Organisation Unit in the Science Ground Segment of *Euclid*, of the VIS component of the Simulation Organisation Unit and of the wide field ground based survey of the *Euclid* northern sky with MegaCam at CFHT. IAP currently hosts the *Planck* HFI data center, producing the official releases for ESA. IAP is also strongly involved in the data processing and analysis pipelines of the *JWST* NIRSPEC instrument.

**MPE:** Depending on time scales and support by the German funding agency DLR, MPE could envisage contributing similar components as for *Euclid*, i.e. lenses, mounts including tests etc.

**ICE, IFAE Barcelona, & CIEMAT, IFT Madrid:** Interest is in possible participation in the WFI mechanics, electronics and/or control software. They currently are responsible for the Filter Wheel Assembly of *Euclid*-NISP and the Simulations Organization Unit in the Science Ground Segment; they also host the *Euclid* Spanish Data Centre. Besides *Euclid*, they are currently involved in several ground-based Dark Energy experiments – DES, PAU, DESI – in which they participate in management, planning, instrumental development, software development and scientific coordination and exploitation. Their scientific interests are focused on Dark Energy related research.

**CfAI Durham:** The primary interest is in the Integral Field Unit. Their relevant heritage is participation in the optical & opto-mechanical design, complete optics manufacture and test and cryogenic testing and qualification of the *JWST* NIRSpec IFU under contract to Astrium GmbH, and in collaboration with SSTL. Engineering, flight and flight spare models were successfully delivered to Astrium for integration into NIRSpec. They also performed the complete optical & opto-mechanical design, optics manufacture and metrology and subsystem I&T for the 24 IFU units for the VLT-KMOS spectrograph. Their capabilities include optical and opto-mechanical systems design, a fully equipped diamond machining and optical metrology facility as well as sub-system test and integration.

**MSSL-UCL:** The primary interest is in the potential WFI Calibration Unit. The applicable heritage is the *JWST* NIRSpec Calibration Unit engineering, flight and flight spare models provided under contract to Astrium GmbH. This unit provided 11 bands of highly uniform illumination in the 1–5 µm range, and included comb filters for the wavelength calibration. MSSL has instruments in over 50 missions with NASA (*Cassini, Swift*), ESA, JAXA, USSR/Russia, China and India. Cryogenic missions include *ISO* LWS, *Herschel* SPIRE, *JWST* NIRSpec, *Gaia* and *Euclid* (VIS instrument lead, VIS scientist on *Euclid* Science Team; weak lensing science working group lead). Their scientific interests include galaxy evolution astrophysics and cosmology, particularly weak lensing, via their *Euclid* roles.

**ICG Portsmouth:** Interest is in IFU spectroscopy processing, and then in scientific aspects concentrated in supernova classification and analysis; redshift determination from slitless spectroscopy; and precision large scale structure measurements. They developed the IFU data processing pipeline for SDSS MaNGA. They have significant science leadership roles in DES, SDSS, DESI and *Euclid* (lead of galaxy clustering science working group and BAO scientist on the *Euclid* Science Team).

**UKATC:** Interests are in opto-mechanical aspects of the WFI camera, IFU, mechanisms and potential calibration sources. They have extensive expertise in cryogenic instrumentation for space (*JWST*-MIRI lead institute) and ground (VLT-KMOS; SCUBA2).

**Oxford:** Interest is in designing, building, testing and using IFUs, as well as red/infrared spectrograph design and construction (through FMOS, SWIFT & KMOS). They have delivered extensive scientific surveys based on IFU data (SAURON + ATLAS-3D + SWIFT) and have expertise in terms of handling IFU data and bringing it to a stage ready for scientific exploitation. They are involved in SDSS MaNGA having fabricated the V-groove blocks that hold the fibres in place at the spectrograph slit. Oxford leads the *Euclid* science working group for supernovae.

European participation could be through an ESA Mission of Opportunity, or via bilaterals between national agencies and NASA. Initial discussions have taken place with ESA and some national agencies, and these are being pursued subsequent to the WFIRST SDT Meeting on 20/21 Nov. An MoO with ESA would require national agency sponsorship, and might follow the precedent established with *Euclid* with a reciprocal involvement by European scientists, together with instrument and ground system roles. Pending further consultations, expectations should be maintained at a realistic level, and will be dependent on the outcomes from the recent ESA Ministerial.

**CGI Contributors:**
Anthony Boccaletti, LESIA - Paris Observatory, FR (co-ordinator for CGI)
Pierre Baudoz, LESIA - Paris Observatory, FR
Riccardo Caudi, Padova Observatory, IT
Kjetil Dohlen, LAM - Marseille Observatory, FR
Marc Ferrari, LAM - Marseille Observatory, FR
Raphaël Galicher, LESIA - Paris Observatory, FR
Jones Geraint, UCL, UK
Raffaele Gratton, Padova Observatory, IT
Andrew Holland, CEI Open University, UK
Emmanuel Hugot, LAM - Marseille Observatory, FR
Frantz Martinache, LAGRANGE - Nice Observatory, FR
Patrice Martinez, LAGRANGE - Nice Observatory, FR
Jean Schneider, LUTH - Paris Observatory, FR
Arthur Vigan, LAM - Marseille Observatory, FR
Mark Wyatt, Cambridge University, UK

**WFI Contributors:**
Mark Cropper, Mullard Space Science Lab (MSSL), University College London, UK (coordinator for WFI)
Ralf Bender, Max-Planck-Institut für extraterrestrische Physik (MPE), Germany
Francisco Castander, Institut de Ciències de l'Espai (ICE), University of Barcelona, Spain
Roger Davies, Oxford Astrophysics, Oxford University, UK
Ignacio Ferreras, Mullard Space Science Lab (MSSL), University College London, UK
Olivier le Fèvre, Laboratoire d'Astrophysique de Marseille (LAM), France
Yannick Mellier, Institut Astrophysique de Paris (IAP), France
Bob Nichol, Institute of Cosmology and Gravitation (ICG), University of Portsmouth, UK
Ray Sharples, Centre for Advanced Instrumentation (CfAI), Durham University, UK
Gillian Wright, United Kingdom Astronomy Technology Centre (ATC), UK

**Appendix J    Science Synergies with Future Missions**

# Wide Field Infrared Survey Telescope – Astrophysics Focused Telescope Assets (WFIRST-AFTA)

## *Science Opportunities Enabled by Synergistic Observations with Future Missions*

Jan 2015

http://wfirst.gsfc.nasa.gov/

WFIRST observations will be synergistic with other major astronomical facilities planned for the next decade. The WFIRST science program has many unique aspects that will enhance the science return from upcoming space missions including JWST and Euclid, wide imaging and spectroscopic surveys being undertaken by Subaru, and ambitious upcoming ground based facilities including LSST and multiple 30 meter telescopes. In this appendix, we include a few of the science opportunities enabled by such synergistic observations. These ideas were submitted to the WFIRST SDT by members of the astronomical community.

Jason Rhodes (JPL), Jason Kalirai (STScI), Eric Gawiser (Rutgers)
jason.d.rhodes@jpl.nasa.gov

**LSST Cadence**

The 2200 square degree HLS will need to be complemented by near UV and ground based photometry in order to provide photometric redshifts for the WFIRST-AFTA weak lensing experiment. These photometric redshifts will of course be an important WFIRST-AFTA data product for a wide range of other science areas, especially in galaxy evolution and the study of high redshift and rare objects. The unprecedented combination of area and depth of the WFIRST-AFTA HLS creates a challenge in getting optical and near UV imaging to the depth required to produce quality photo-z's for the 50-60 lensed galaxies per square arcminute that WFIRST-AFTA will observe. As discussed in §2.2.3.4, the WFIRST-AFTA weak lensing analysis will likely require the full ten year LSST depth over the 2200 square degree HLS. It would be beneficial to both LSST and WFIRST-AFTA to reach this full depth over the area of the WFIRST-AFTA HLS on a time scale contemporary with the WFIRST-AFTA HLS, rather than waiting until the full LSST survey is completed in 2032.

**Benefits to LSST.** The combination of full-depth LSST data and WFIRST-AFTA HLS NIR data will provide the gold standard in photo-z's. Furthermore, WFIRST-AFTA grism observations over the same area will provide many millions of high quality slitless spectra and WFIRST-AFTA's IFU can be run in parallel with WFI observations to provide even higher precision spectra of faint galaxies (~20k to LSST depth and ~30k to WFIRST depth). Thus, the WFIRST-AFTA photometric data will help to provide better LSST photo-z's for their analysis but WFIRST-AFTA will also provide many of the spectra needed for a training set to calibrate the photo-z's for both missions. A further benefit to LSST might be the reduced need for LSST observations at the reddest end of the LSST wavelength range (the *z* and *y* filters), where both the atmosphere and the physics of CCDs make ground-based observations less efficient than what WFIRST-AFTA can achieve. Finally, the joint processing of LSST and WFIRST-AFTA data will provide better object deblending parameters than LSST can achieve alone; WFIRST-AFTA will be able to provide a morphological prior for the deblending of LSST images (see § 2.7.3).

**Cadence Suggestion.** Given the aforementioned benefits to LSST and the requirement for deep optical photometry for WFIRST-AFTA, it would be beneficial for the entire HLS to lie within the LSST observing area and to complete LSST and WFIRST-AFTA surveys of that area on comparable time scales. While the development of an observation schedule for WFIRST-AFTA remains a task for the Project and selected Science Teams, the likely timelines of the two projects indicate that reaching the full WFIRST-AFTA depth over this area will require LSST to accelerate observations on that part of the sky. Since the HLS area is roughly 1/8 as large as the LSST "Main Survey" region, this could be achieved by devoting 1.25 years of LSST observations to the HLS area, assuming that it covers a wide enough range of Right Ascension. More practically, it could be achieved by devoting 25% of LSST observing time to this area during each of the first 5 years of the LSST survey, which doubles the time it would naturally be observed during those years at a modest reduction in coverage of the rest of the Main Survey area during that time period. As LSST and WFIRST progress, there is a mutual benefit in continuing discussions about the optimal joint observation schedule.

**Other Cadence Issues**. There are other observation cadence symmetries between WFIRST-AFTA and LSST. These lie in areas where time domain is a key issue such as supernova observations and exoplanet microlensing. It is also imperative to ensure that the deep fields for both observatories are coordinated. Synergies for panchromatic data sets in GO programs and high resolution imaging of stellar populations should also be explored.

Ian Dell'Antonio, Gillian Wilson, Tommaso Treu for the Cosmology ISDT (ian_dell'antonio@brown.edu)

**WFIRST/TMT Synergies: Cosmology and Fundamental Physics**

In the past 10 years, progress in cosmology has been greatly aided by the synergy between ground-based and space OIR facilities (HST and Spitzer). One need only think of the measurement of the acceleration (and early-time deceleration) of the Universe that has been possible due to the measurement of type Ia supernovae (Perlmutter & Riess 1999; Riess et al. 2004), or of the combined use of ground-based and space-based weak lensing to discover and characterize "bullet-like" clusters e.g, Clowe et al 2006; Dawson et al. 2012. In many cases, the ground-based telescopes, with their larger collecting area and field of view have served as discovery telescopes and spectroscopic engines, and space-based observatories with their superior resolution and lower sky background have served to extend the redshift range to higher redshift and provide superior resolution.

In the TMT/WFIRST era, this synergy will continue, this time relying on the superior area coverage and photometric stability of the space-based WFIRST, combined with the greater resolution and broader wavelength coverage of TMT. Although other topics in cosmology will of course benefit from the synergistic capabilities of TMT and WFIRST, we highlight four areas that will greatly benefit. They are: Cluster scale cosmology (tomography, mass function evolution and high redshift cluster formation); galaxy-scale cosmography (gravitational time delays); supernova cosmology; and "near-field" cosmology, in particular surveys of the MW satellites to understand the properties of dark matter.

Cluster Scale Cosmology

WFIRST is expected to discover ~40000 M$> 10^{14}$ M$_{sun}$ clusters out to z~3 in the 2000 deg$^2$ area of the high-latitude survey (WFIRST-AFTA Final Report), offering new possibilities for constraining cosmology using the positions of multiply-imaged background sources using the so-called "strong gravitational lensing angular diameter cosmography" technique. The positions of the multiple images depend on the lens mass distribution. They also depend on the angular diameter distance ratios between the lens, source, and observer, thereby encapsulate information about the underlying cosmology (Paczynski & Gorski 1981; Blandford & Narayan 1992; Link & Pierce 1998; Golse et al 2002) Recent feasibility studies have demonstrated the power of this method (Jullo et al 2011; D'Aloisio & Natarajan 2011), which is independent of the value of the Hubble parameter, and complementary to CMB and SNIa techniques. WFIRST will discover samples of clusters at higher redshift than e.g., DES, LSST, Euclid enabling a longer redshift lever arm for measuring the time evolution of the dark energy equation of state. However, TMT will be needed to follow up the multiple images and obtain spectroscopic redshifts for them. In particular, the much higher resolution of TMT's IRIS imager (roughly 5 times the resolution of the Frontier Field images, and more than 10 times the resolution of WFIRST) will allow many more lensed pairs to be discovered and a much more accurate mass model to be constructed.

While mass estimates for the new sample of WFIRST clusters will be able to be determined from the weak gravitational lensing signal, there will be a bias inherent in these mass estimates. This is because to relate the lensing signal to the mass, one needs to determine the value for the critical surface density, which requires knowledge of the redshifts of the faint source galaxies. While the background redshift distribution can be estimated photometrically, it will be preferable to reduce the uncertainties in cluster masses by determining the redshift distribution spectroscopically. This will be particularly important for the higher redshift WFIRST clusters for which the background galaxies will be faint, their source density low, and the cluster-to-cluster variation high. While the redshifts of some background galaxies will be directly determinable by WFIRST, TMT is capable of reaching much deeper flux limits than WFIRST, and will be able to determine spectroscopic redshifts for a much higher fraction of the background galaxy population.

Galaxy Scale Cosmography

It is not just cluster scale lenses that benefit greatly from the synergy between WFIRST and TMT. Time delay measurements of galaxy scale strong gravitational lenses have become an important tool for cosmology (e.g. Suyu et al. 2013). In particular, the combined measurement of time delays of multiply-imaged quasars, the redshifts of the quasars and lens galaxy as well as the internal lens galaxy dynamics and detailed images of the lensed quasar host galaxy are required for the tightest constraints on the evolution of the scale factor. The combination of WFIRST and TMT is ideally suited for this task. While large ground-based surveys such as LSST will cover more sky, the majority of lensed systems are expected to have small Einstein radii (<0.5"), and these will be underrepresented in the LSST sample. The WFIRST high latitude survey will discover hundreds of galaxy-scale lensed AGN; although the cadence of the survey is unlikely to lead to time delay measurements, these will be obtained from the LSST time series (using high resolution images to deblend the ground based images; Tewes et al. 2013; Liao et al. 2014) or for the fainter systems with WFIRST GO mode or more likely through TMT imaging with IRIS. Crucially, the superior resolution of IRIS would allow precise mapping of the QSO centroids (with <50 microarcsec positional uncertainty), as well as precise maps of the arcs from the background quasar host galaxy. These are needed to tie down the mass model. In addition, IRIS's IFU would allow the lens galaxy kinematics to be determined. In a sample of >100 galaxy lenses, we expect to also find multiply lensed Supernovae. Lensed Supernovae provide independent tests of the mass model via the measure of the absolute magnification at the position of each supernova image. TMT spectroscopy will be critical to establishing the supernova type and redshift.

Supernova Cosmology

Lensed supernovae are not the only type of supernova for which the combination of WFIRST and TMT can provide the critical cosmological information. The WFIRST Supernova Survey will potentially detect supernovae out to z~4. However, obtaining spectra for the faintest ones will require better NIR sensitivity than is possible with the current generation of 10m-class telescopes. AO-assisted spectroscopy with IRMS on TMT would be the ideal instrument to characterize the supernova spectrum and test for any evolutionary effects at these highest redshifts. In this way, TMT spectroscopic observations can provide support for WFIRST lightcurve measurements of distant supernovae.

Near Field Cosmology

Another cosmological puzzle that can be addressed best with the combination of TMT and WFIRST is the puzzle of the dark matter content of the Milky Way satellite galaxies. The dark matter profile of these galaxies is important both as a test of the LCDM model which predicts the existence of density cusps in these dark matter dominated systems (see for example the review by Pontzen & Governato 2014), and because it affects the predicted rate of dark matter annihilation. Determining the annihilation rate (the so called "J" factor) is critical in constraining the possible masses of thermal relic WIMPS (Geringer-Sameth & Koushiappas 2013). Assuming that part of the high galactic latitude survey of WFIRST is carried out in the northern galactic cap, then these dwarfs will most likely be observed by WFIRST as a matter of course. If not, it will be necessary to secure pointed observations with WFIRST as part of the GO program.

Measuring the density profile of the faint Milky Way halos is complicated by two factors. First, we typically only have radial velocities to work with to reconstruct the 3-d dynamical profile. Strong constraints on the profile in this case can require thousands of radial velocities. For many of these systems, however, only tens or hundreds of stars have been identified. These closest (and thus most interesting from the point of view of dark matter detection) dwarfs are much too large on the sky for efficient mapping with TMT's imager (or with HST). At the same time, the photometric precision and high resolution of a space based imager can allow a much cleaner look at the color-magnitude diagram for stars in the dwarfs, allowing many more stars to be identified over the entire tidal radius of the dwarfs. Once the candidate stars are identified, radial velocities must be obtained. For the brighter stars currently targeted, this requires 8-10m telescopes (e.g. Walker 2009). For the fainter stars, TMT will be required.

The superb astrometric precision that will be afforded by TMT using IRIS opens another possibility, that of measuring stellar proper motions and recovering star-by-star 3-d proper motions. Given the accuracy of astrometric precision already achieved by Keck's AO system, and the astrometric improvement afforded by the smaller diffraction limit of TMT, it will be possible to make 30 microarcsec accuracy measurements of the stellar positions. By monitoring the dwarf galaxy stars over the period of three years, proper motions with uncertainties of ~7km/s can be achieved for a dwarf at 100 kpc distance. With the 3-d motions, the largest source of degeneracy in the dynamical model is removed, and the uncertainty on the slope of the density profile will be reduced by at least a factor of 5. Combining the power of WFIRST and TMT on the profiles of the dwarfs will reduce the uncertainty on the profile parameters for a fixed parametric model by more than an order of magnitude and help resolve the core/cusp problem as well as severely constraining the most popular models of dark matter (or aiding in its discovery!)

## Conclusive Identification of Gravitational Wave Sources

(Masaomi Tanaka, NAOJ, masaomi.tanaka@nao.ac.jp,

and TMT ISDT for Time-domain science)

Background

The next gravitational wave (GW) detectors, such as advanced LIGO, advanced Virgo, INDIGO and KAGRA, are planned to start operation from around 2017. They are expected to directly detect GWs from neutron star (NS) mergers at a distance within about 200 Mpc. To fully understand the nature of GW sources, follow-up observations of electromagnetic counterparts are extremely important since the position of the sources can only be moderately determined with a localization of about 10-100 deg2 with the GW detection alone (e.g., Nissanke et al. 2013, ApJ, 767, 124). Such multi-messenger observations will provide a new opportunity to study the origin of r-process elements and high-density equation of state.

WFIRST & TMT

The expected emission powered by radioactive decay of r-process nuclei peaks at the red edge of optical or near infrared wavelengths (Barnes & Kasen 2013, ApJ, 775, 18; Tanaka & Hotokezaka 2013, ApJ, 775, 113). At 200 Mpc, the expected brightness is 22 - 25 mag in i or z bands and 21-24 mag in NIR JHK bands (in AB magnitude). WFIRST has ideal characteristics (wide field of view and deep sensitivity) to detect electromagnetic counterpart of GW sources. Such wide field survey may detect more supernovae than GW sources within the localization area. To conclusively identify a detected transient as a compact star merger, we have to take spectra to see if the object shows a predicted extremely broad-line, red spectrum in the optical and NIR wavelengths. Therefore, synergetic observations between WFIRST and TMT are crucial, i.e., the detected source can be conclusively identified as a true GW source by low/medium resolution (R~500) spectroscopic follow-up observations with TMT MOBIE/IRIS.

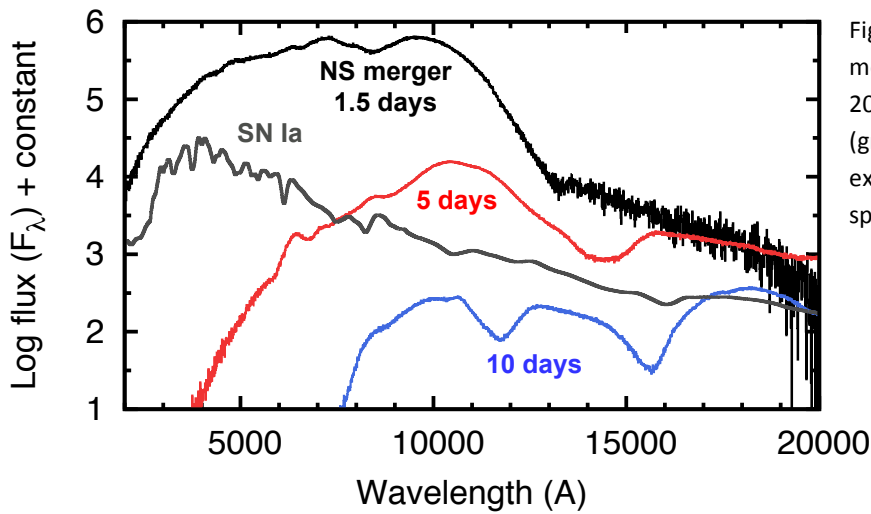

Figure: Expected spectrum of NS merger (Tanaka & Hotokezaka 2013). Compared with supernova (gray line), it is expected to show extremely broad-line, red-color spectrum.

# Identification of First Supernovae

(Masaomi Tanaka, NAOJ, masaomi.tanaka@nao.ac.jp,

and TMT ISDT for Time-domain science)

Background

Identification of first stars in the Universe is one of the great challenges in the modern astronomy. Although identification of single first stars is not possible even with deep observations with JWST in the 2020s, detection of supernova explosion of first stars can be feasible. Especially, bright subclasses of supernovae (superluminous supernovae and pair-instability supernovae) are one of the most promising targets (e.g., Scannapieco et al. 2005, ApJ, 633, 1031; Quimby et al. 2011, Nature, 474, 487, Whalen et al. 2013, ApJ, 777, 110). Identification of high-z supernovae enables to probe initial mass function in the early Universe. Since superluminous supernovae and pair-instability supernovae are thought to arise from very massive progenitors, a number ratio of these types of supernovae to normal supernovae (or ratio to the total star formation rate) will provide unique information of the initial mass function.

WFIRST & TMT

For example, if WFIRST perform deep NIR survey (26 AB mag/visit for > 10 deg^2), more than 10 superluminous supernovae can be detected at z>10 (Tanaka et al. 2013, MNRAS, 435, 2483). Color selection will give reasonable estimates for the redshift and explosion types. However, in order to study if the rate of such special types of supernovae increase/decrease relative to the total star formation rate, which gives the information of the initial mass function, spectroscopic identification for at least some subsamples is necessary. Synergetic spectroscopic observations with TMT MOBIE/IRIS will give firm identification of such high-z supernovae. Especially superluminous supernovae and pair instability supernovae have characteristic features in the rest-UV wavelengths (bright continuum and metal absorption, respectively), deep optical spectroscopy, in addition to low-resolution NIR spectroscopy by WFIRST, is essential to classify these types of high-z supernovae.

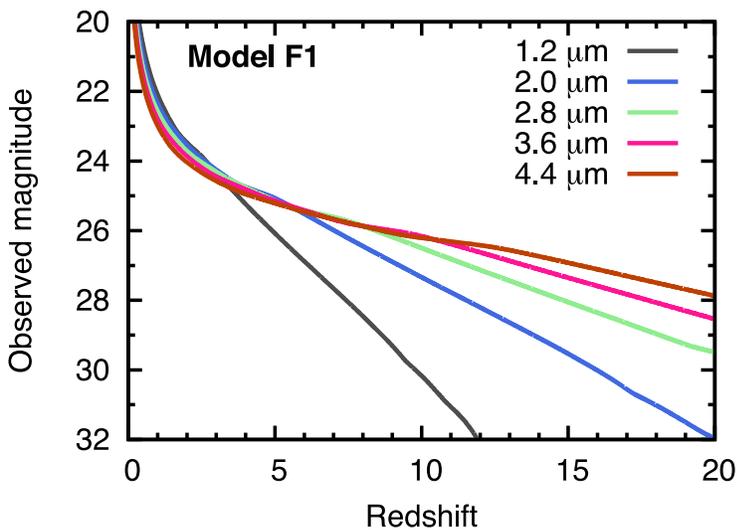

Figure: Brightness of superluminous supernovae as a function of redshift, based on numerical simulations by Moriya et al. (2013, MNRAS, 428, 1020). Figure from Tanaka et al. 2013.

Linhua Jiang (China)

**High-redshift (z>6) Quasars and Cosmic Reionization**

High-redshift (z>6) quasars provide a powerful tool to study the early universe. They are essential in understanding black hole accretion, galaxy evolution, and the IGM state in the first billion years of cosmic time. In recent 15 years, more than 70 z>6 quasars has been discovered (See a series of papers by X. Fan, by L. Jiang, and by C. Willott). However, these quasars only represent the most luminous quasars at z>6 (brighter than the characteristic luminosity in the quasar luminosity function, QLF). The combination of wfirst and tmt will allow us to perform a systematic search/study of much fainter quasars at z>6.

WFIRST, especially WFIRST+LSST, will help us very efficiently select z>6 quasar candidates down to 25-26 AB mag (Y band). Relatively bright candidates (Y<22 AB mag) can be easily identified by 5m class telescopes. But identification of fainter candidates can be very difficult, or more precisely, expensive (the spatial density of these objects is very low so that it's impossible to observe >1 object in the same field with masks). Currently it takes the MMT (6.5m) about 15+ min to identify ~22 mag candidates (almost the faintest known so far) in the SDSS stripe 82. With the same int. time, Keck can reach ~23 mag, and TMT can reach ~24.5 mag. So for faint z>6 quasars, TMT is very efficient.

Two main science cases from faint quasars: 1) allow us to study regular quasars (not super luminous ones) and black holes at z>6, including emission line properties, black hole masses, host galaxies, dust properties, etc. 2) measure the QLF to the faint end, and determine total ionizing photons from quasars. We know that one key question is, what sources are responsible for cosmic (H) reionization. People have tentatively ruled out quasars due to their very low spatial density. However, the QLF in the bright end is very steep, meaning that if the turn over point in the QLF is very faint, the ionizing photons from the whole quasar population will increase dramatically. This can only be determined by deep spectroscopic survey of very faint quasars at z>6.

Yue Shen (China)

**High-z Quasar Demography in the Mass-Luminosity Plane**

The growth of the earliest SMBHs (at z>6) is one of the least understood topics in high-z galaxy formation sciences. Demographic studies are traditionally based on the quasar luminosity function (LF). With the development of spectroscopic methods to estimate the mass of these high-z quasars (e.g., virial BH mass estimators), it is now possible to study the abundance of these quasars in the mass-luminosity plane (Shen & Kelly 2012, Kelly & Shen 2013). Unlike the 1D LF, the quasar mass-luminosity plane constrains both the current activity (instantaneous luminosity) and integrated accretion history (BH mass) for the population of high-z quasars. Thus the 2D M-L plane can be used to understand the emergence/growth of high-z quasars and the average evolution of their accretion rate, in a way that cannot be achieved by studying the LF alone (which loses information by projecting the 2D distribution function onto the luminosity axis). Current instrumentation is able to constrain the M-L plane of z~6 quasars down to H_ab~20 with dedicated NIR spectroscopic programs on 6-10m telescopes.

A wide-area survey of WFIRST will provide a large sample of z>6 quasars down to sufficiently faint luminosities, and start to unveil the progenitor population of the most massive z>6 quasars (with BH mass > a few 1d9 $M_{Sun}$) discovered today. By mapping the M-L plane from high-z till z~6 to these faint luminosities, we will be able to monitor the growth of these high-z quasars in "real time" (as the cosmic time span at these high redshifts is short), and truly understand how these SMBHs grow so rapidly at early times.

Requirements: A wide-area WFIRST/AFTA survey with grism spectroscopy for quasar confirmation. Expected quasar number density is ~100 per 1000 deg^2 down to H_ab<24 at z>7. This flux limit is ~ 3 mag fainter than the current limit with VISTA. TMT/IRMS follow-up can provide K-band spectra (covering restframe MgII at z~7 to use as the BH mass estimator) with sufficient quality to measure spectral properties (S/N>10 at K_ab~24 in 5-hr integration). A TMT program with tens of hrs will assemble a decent sample of more than tens of z~7 quasars with BH mass estimates down to K_ab~23, and map the distribution in the quasar M-L plane at z~7.

<u>Key References</u>
Shen & Kelly, 2012, ApJ, 2012ApJ...746..169S
Kelly & Shen, 2013, ApJ, 2013ApJ...764...45K
WFIRST-AFTA Final Report
TMT instrument documentation

Yue Shen (China)

**Evolution of the BH Mass - Galaxy Scaling Relations**

A key science case with TMT in the next decade is to measure the evolution of the BH mass - galaxy scaling rations up to high redshift. This measurement will help us understand the nature of the co-evolution of galaxies and SMBHs, and whether or not AGN feedback plays an important role in galaxy formation.

Dynamical BH masses with spatially resolved spectroscopy are only possible at low-redshift (except maybe for 1d10 $M_{Sun}$ BHs). Beyond z>0.3, the primary method to estimate SMBH masses is reverberation mapping (RM) and its derivative, single-epoch virial mass estimators. However, these techniques only apply to AGN/quasars, and measuring host galaxy properties in these active galaxies are notoriously difficult. For host luminosity and stellar mass, WFIRST/AFTA, when combined with deep optical imaging from the ground or space, will provide superb image quality and depth that will yield excellent measurements for numerous quasar hosts up to z~3 and to very faint quasar luminosities. For (integrated) host stellar velocity dispersion, TMT/MOBIE/IRMS/IRIS can measure sigma_* to z~4 for large samples of quasars. Therefore with WFIRST+TMT, we will revolutionize this field by compiling host measurements for large quasar samples to much fainter luminosities and high redshifts.

TMT will be absolutely necessary to push on stellar velocity dispersion measurements in faint quasar hosts. Currently, the best quasar sample with host sigma_* measurements is achieved with the SDSS-RM program (Shen et al. 2014, 2015), which has a sample size of ~90 quasars, probes one mag fainter than the earlier Keck samples (Woo et al. 2008), and covers to z~1 with ~60 hrs of optical spectroscopy with the 2.5m SDSS telescope. With TMT, we will be able to probe yet another 1-2 magnitudes fainter with only 1 hr integration, and expand to higher redshift with NIR spectroscopy. Hence a program of tens of hrs on TMT will compile an unprecedented sample to study the evolution of the M-sigma relation. By the time TMT becomes a reality, we should have much improved BH mass estimators from dedicated RM programs (such as the SDSS-RM program).

Requirements: Targeting the SDSS-RM field (14:14:49.00 +53:05:00.0), which is a 7 deg^2 field within CFHT-LS W3 and fully covers the AGEIS field and the CFHT-LS D3 field. This field also coincides with the PanSTARRS 1 Medium Deep Survey field MD07, with hundreds of multi-band photometric epochs during 2010-2013. This SDSS-RM field is being regularly monitored with SDSS BOSS spectroscopy (30-60 total epochs, 2014-2017), ground-based optical imaging (CFHT, Bok, etc., 2014-), NIR imaging (UKIRT), Spitzer (Cyc 11-12, 20-month baseline), and XMM. The SDSS-RM program monitors 849 quasars (0.1<z<4.5; to i<21.7) photometrically and spectroscopically in order to measure broad-line lags and RM BH masses for more than tens of quasars (potentially hundreds of RM measurements combined with the early PanSTARRS light curves), and deep spectroscopy for excellent single-epoch virial BH mass estimates for the full sample. The multi-wavelength synergy in this field will maximize quasar sciences (quasar variability, host galaxies, quasar absorption lines, etc.). WFIRST+TMT coverage of this field will further enhance the science yields on quasars and SMBHs.

<u>Key References</u>
Shen, et al., 2014, ApJS, in press, 2014, arXiv1408.5970S
Shen, et al., 2015, submitted
Woo, et al., 2008, 2008ApJ...681..925W
WFIRST-AFTA Final Report
TMT instrument documentation

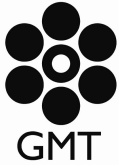

## Giant Magellan Telescope Project

### Galaxy Evolutionary Studies with WFIRST and The Giant Magellan Telescope

#### P. McCarthy (GMTO)

**Introduction**

The WFIRST high-latitude survey will provide a large and homogenous sample of faint galaxies selected at wavelengths that most accurately trace stellar mass. This data set will enable a wide range of studies of galaxy formation, mass assembly and star formation over the full galaxy-building epoch. The low-resolution spectroscopic capability offered by slitless grisms has been shown to be highly effective in selecting galaxies with well-determined redshifts and a range of star forming histories – from extreme starburst, to quenched galaxies and old passively evolving systems. The Giant Magellan Telescope and other extremely large telescopes on the ground can leverage this dataset by providing deep spectroscopy, diffraction-limited imaging at the 20 mas scale and spatially resolved 2-D spectroscopy with 20-100 mas sampling.

**Studies of Extreme Star Forming Galaxies**

WFC3 on HST and powerful ground-based near-IR spectrographs have extended classical nebular spectroscopy to the epoch of galaxy building, 1.5 < z < 4. The WFC3 grisms have shown the power of slitless spectroscopy to reveal galaxies with high equivalent width emission lines and, hence, high specific star formation rates. The R ~ 200 spectra provided by WFC3 provide samples for further study with large ground based spectrographs at resolutions of a few thousand.

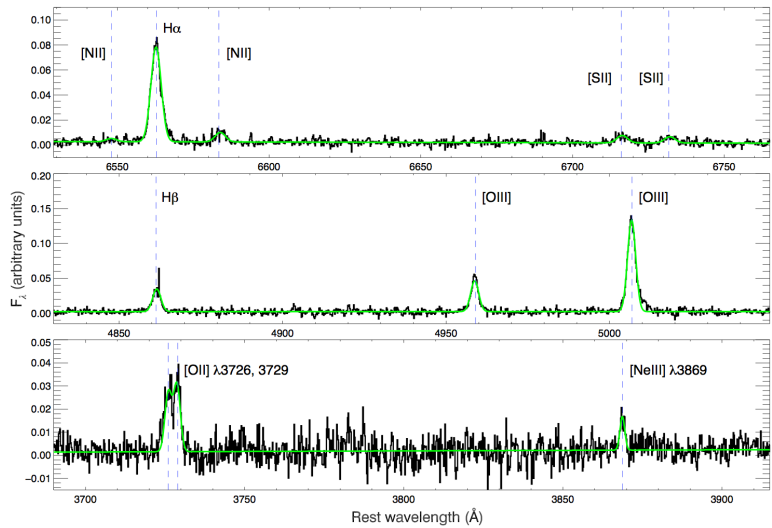

Figure 1. Composite spectrum of star forming galaxies selected with the WFC3 grism on HST and observed with the FIRE near-IR echelle on Magellan (Masters et al. 2014, ApJ, 785, 153).

Moderate resolution spectra obtained with sensitive spectrographs on the ground provide access to weak emission lines that are sensitive to abundances and physical conditions in the HII regions. Spectroscopy of moderate sized samples has revealed evolution in line ratios and the topology of the diagnostic

GMTO Corporation is a not-for-profit organization

diagrams (e.g. [NII]/Hα vs. [OIII]/Hβ). The origin of the anomalous line ratios is not well understood at this time.

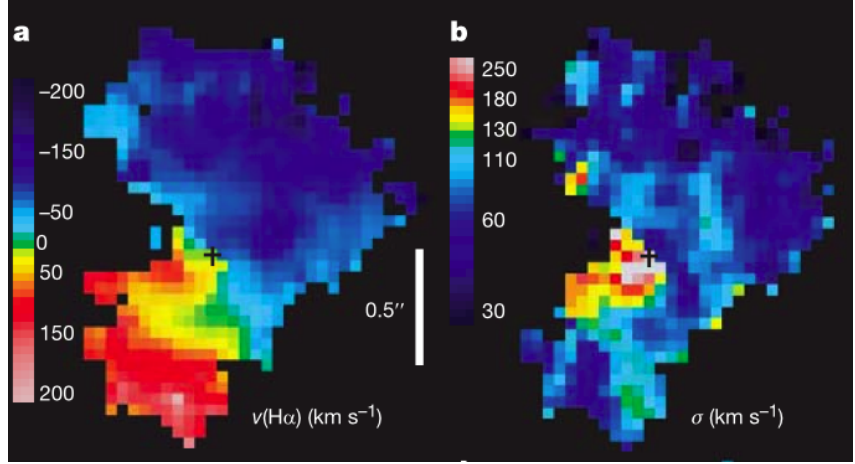

**Figure 2. Velocity field and dispersion map for a massive star-forming galaxy observed with an AO fed integral field spectrograph on the VLT (Genzel et al. 2006, Nature, 422 786).**

Spatially resolved spectroscopy with integral field spectrographs is providing us with dynamical masses and resolved ionization maps of star forming galaxies. The combination of adaptive optics and IFU spectrographs on 6-10m telescopes allows us to probe galaxies on kpc scales. These observations, however, are photon-starved. The ELTs will have enough light grasp to enable 2-D spectroscopy of large samples of galaxies. At present we lack large and robust data sets of star forming galaxies with suitable guide stars for AO spectroscopy with high Strehl ratios. The WFIRST wide-field survey will provide a large enough sample that we will be able to select desirable combinations of galaxy properties *and wavefront control* stars for resolved spectroscopy on sub-kpc scales with adaptive optics.

The James Webb Space Telescope will have a large impact in the study of star forming galaxies over a wide range of redshifts, but it will not provide samples as large as those resulting from the WFIRST survey. Rare objects, systems with extreme environments and, perhaps, pristine abundances, will be revealed in the large WFIRST survey date. Deep moderate-resolution spectroscopy with near-IR spectrographs on the GMT and other ELTs will provide a powerful combination for advancing our understanding of the star formation in galaxies at the peak of the galaxy building epoch.

**Quenched and Passive Galaxies**

One of the outstanding challenges in contemporary astrophysics is understanding the origin of the Hubble sequence and the bimodal galaxy distribution. Quenched systems, those galaxies in which star formation has abruptly ended, putting them on the path to the red sequence, are easily recognizable from their strong Balmer sequence absorption features. Adequate spectra, however, are only available from massive systems. WFIRST will sample large numbers of passive and quenched systems over a range of stellar masses. Efficient near-IR spectrographs on the ground can provide high quality spectra needed to derive precise ages and star formation histories.

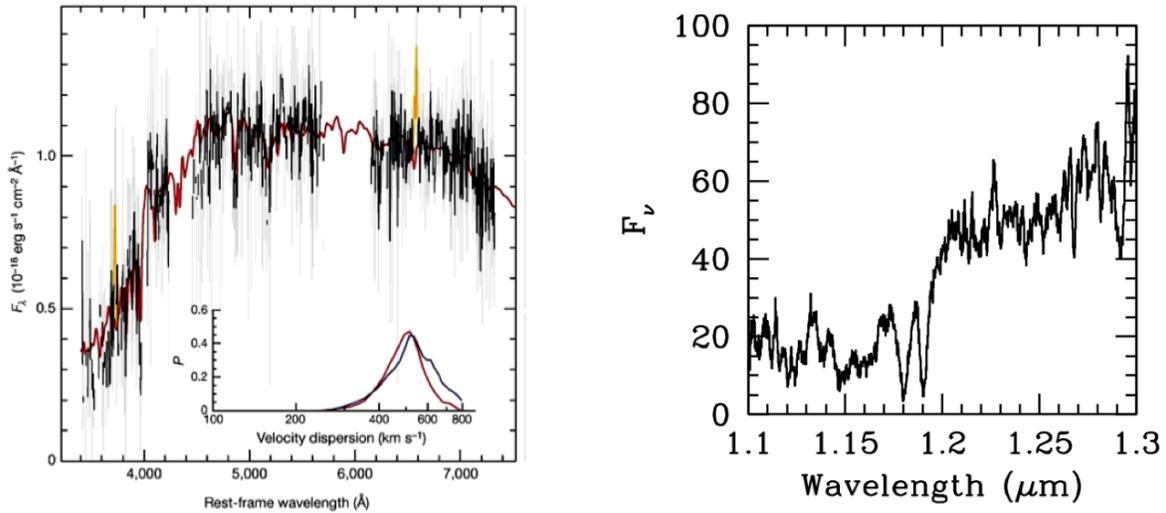

Figure 3. (left) Gemini GNIRS Spectrum of a compact passive galaxy and the derived likelihood function for its velocity dispersion from van Dokkum et al. 2009, Nature, **406**, 717. On the right we show a simulated spectrum with the GMT centered around the Ca II H&K features that provide most of the weight for the velocity dispersion. The GMT and other ELTs can be used to determine dynamical masses for passive galaxies and cluster cores revealed by the WFIRST wide-area survey.

The key to connecting theory and observation and improving our understanding of the growth of galaxies is accurate stellar masses. Velocity dispersions have been measured for a few compact passive galaxies, but only for those with stellar masses well beyond the knee of the mass function. The grism on WFIRST will reveal large numbers of systems suitable for high precision determinations of velocity dispersions with near-IR spectrometers on ELTs. GMT's red and near-IR spectrographs can play a key role in this area by tying down the dynamical masses of passive and quenched galaxies for with masses equal to, or less, than $M^*$.

The WFIRST wide-area survey will also produce large number of distant galaxy clusters and bound groups that are the sites of galaxy transformations. Red and near-IR spectrographs on the GMT can provide both velocity dispersions for individual galaxies and for groups of galaxies undergoing rapid evolution.

**Summary**

In addition to its core missions of BAO, weak lensing, and exoplanet science, WFIRST will provide a rich legacy of unparalleled data for studies of galaxy evolution. The GMT and other extremely large telescopes on the ground can leverage the powerful imaging and low-resolutions spectra produced by WFIRST by providing high-sensitivity intermediate resolution spectroscopy of faint galaxies, galaxy clusters and rapidly evolving groups.

### Synergy Between WFIRST-AFTA and the Giant Magellan Telescope: Spectroscopic Probes of Reionization

*Steven L. Finkelstein (UT Austin)*

The reionization of the intergalactic medium (IGM) marks the last major phase transition in the Universe, and provides a key glimpse into the formation and evolution of the first luminous sources after the Big Bang. By the middle of the next decade, a number of new facilities will be coming online, including the Giant Magellan Telescope (GMT), the Large Synoptic Survey Telescope (LSST), the Square Kilometer Array (SKA), and WFIRST-AFTA, which, when combined, will allow a detailed probe of the evolution of the neutral fraction in the IGM during reionization. Two highly complementary techniques available for probing reionization are faint Ly$\alpha$ spectroscopy and 21cm tomography. As Ly$\alpha$ photons are resonantly scattered by neutral hydrogen, Ly$\alpha$ emission effectively maps ionized regions in the universe, escaping when the size of the ionized bubble allows the Hubble flow to redshift the photons away from resonance. The 21cm line observations, specifically the highly sensitive observations to come from the Square Kilometer Array in the next decade, will provide high sensitivity, high resolution maps of the neutral gas in the IGM. This complementary view will provide some of the first maps of the spatial distribution of reionization, further constraining the nature and distribution of the key ionizing sources.

To understand the requirements, we need to make some assumptions on the scale. If we assume that the SKA has a resolution of 1′, a conservative assumption is that SKA observations will resolve neutral regions on linear scales of $\gtrsim$10′. This corresponds to a physical size of ~3 Mpc at $z = 8$, comparable to the expected size of ionized bubbles in the early universe (e.g., Furlanetto 2006). This is also the size of the deepest *HST* survey fields (e.g, GOODS, CANDELS, etc.). Thus, to obtain a statistical sample of ionized and neutral regions, a much wider survey is needed. While *JWST* will have the capability of detecting galaxies over the key redshift range of $6 < z < 10$, its field-of-view is small, and thus multiple wide field surveys ($\gg 1$ degree) are unlikely. The High Latitude Survey (HLS) component of WFIRST will image 2000 deg$^2$ to $m_{AB} \approx 27$ from $0.8 - 2.0$ $\mu$m. When combined with LSST's deep optical data, the WFIRST survey is sufficient to photometrically select galaxies at $6 < z < 12$. In Table 1 I show the estimated numbers of $6 < z < 10$ galaxies to be discovered in the WFIRST HLS survey at two different magnitude limits.

Spectroscopic followup of these WFIRST sources is necessary both to prove their validity, but also to obtain measurements of the Ly$\alpha$ emission strength to provide the counter-image to SKA's neutral hydrogen maps. Although the GMT has the smallest mirror of the three giant segmented mirror telescopes, it will have instruments with the largest field-of-view. The GMT collaboration is currently exploring multiple options for near-infrared spectroscopy at first light, including a single object cross-dispersed $JHK$ spectrograph, and a multi-object 1-1.3$\mu$m $J$-band spectrograph. The single-object spectrograph will be highly useful for observing the brightest ($m_{AB} < 25$) galaxies, which, as shown in Table 1, have very small space densities, unsuitable for a multi-object spectrograph (MOS). These galaxies will also be easily observable by *JWST*, should these missions fly simultaneously. However, the collecting area of the GMT is 11.5$\times$ larger than *JWST*, allowing fainter spectroscopy. Though night sky lines will be present, the high-resolution (R~6000) of the instrument, and the faint inter-line sky background (Sullivan & Simcoe 2012), will result in such observations being at least as efficient, if not more so, than similar observations with *JWST*.

The true synergy between the GMT and WFIRST is in multi-object spectroscopy. Observing several sources simultaneously with a 22m (effective aperture) telescope will reach extremely faint flux levels for Lyα emission, even for the faintest galaxies in the WFIRST survey. The current MOS design has an 8′ field-of-view, perfectly suited to the expected angular size of ionized bubbles during reionization, and >5× larger in area than the equivalent spectrograph on *JWST*. As deep GMT observations will reach $5\sigma$ line flux limits of a few $\times 10^{-19}$ erg s$^{-1}$ cm$^{-2}$, given the numbers in Table 1 a single GMT observation will observe $\sim$80 galaxies at $6 < z < 8$ *simultaneously* to a rest-frame Lyα equivalent width of 5 Å at the WFIRST magnitude limit.

*Summary*: The Giant Magellan telescope occupies a unique region of the aperture size – field-of-view parameter space amongst the next generation of telescopes. A deep GMT multi-object spectroscopic survey of WFIRST-identified high-redshift galaxies will allow the deepest, most detailed spatial maps of Lyα emission in the distant universe. Combining these luminous observations with observations of the "dark side", the 21cm H I maps from the SKA, will provide a complete view of the reionization history of the universe.

References: Bouwens et al. 2014, ApJ Submitted ⋄ Finkelstein et al. 2014, ApJ Submitted ⋄ Furlanetto 2006, New Astronomy Reviews, 50, 157 ⋄ McLeod et al. 2014, MNRAS Submitted ⋄ Sullivan & Simcoe 2012, PASP, 124, 1336

**Table 1: $z = 6$–8 Galaxies in the WFIRST-AFTA HLS Survey**

| Redshift | N($m < 25$) | $n(m < 25)$ (arcmin$^{-2}$) | N($m < 27$) | $n(m < 27)$ (arcmin$^{-2}$) |
|---|---|---|---|---|
| 6 | $1.1 \times 10^5$ | $1.5 \times 10^{-2}$ | $6.1 \times 10^6$ | 0.8 |
| 7 | $0.3 \times 10^5$ | $3.6 \times 10^{-3}$ | $2.7 \times 10^6$ | 0.4 |
| 8 | 1800 | $2.4 \times 10^{-4}$ | $7.1 \times 10^5$ | 0.1 |
| 9 | 18 | $2.4 \times 10^{-6}$ | $1.5 \times 10^5$ | $2.1 \times 10^{-2}$ |
| 10 | 2 | $3.2 \times 10^{-7}$ | $2.7 \times 10^4$ | $3.9 \times 10^{-3}$ |

These numbers assume the luminosity function of Finkelstein et al. (2014) at $z = 6$–8, McLeod et al. (2014) at $z = 9$, and Bouwens et al. (2014) at $z = 10$. We assume a volume probed by a 2000 deg$^2$ survey, with $\Delta z = 1$, with a selection completeness of 90%.

# WFIRST and the Prime Focus Spectrograph

This white paper discusses the complementary nature of WFIRST with the Prime Focus Spectrograph (PFS), an upcoming multi-object spectrograph that will go on the Subaru telescope. With 2400 fibers, PFS will be the most highly multiplexed spectrograph on an 8m class telescope. Combined with the very wide wavelength coverage from $0.38 - 1.3\mu m$, PFS is in a unique position as a spectroscopic survey machine that can strongly complement upcoming space missions such as WFIRST. This document explores some of the obvious possibilities for complementarity, but is by no means exhaustive.

PFS will be a facility instrument on Subaru. We are planning a Strategic Survey Proposal (SSP; Takada et al. 2014) with three pillars. A 1400 deg$^2$ survey of emission line galaxies has the primary goal of measuring the baryonic acoustic oscillations out to $z \sim 1.6$ (check), a galactic archeology component will measure the kinematics and chemical abundances of stars in the halos of the Milky Way and Andromeda, and a 25 deg$^2$ galaxy evolution survey will study the evolution of continuum and emission-line selected galaxies from the first galaxies at $z \sim 7$ to the present. With three-five hour integrations we expect to reach limiting continuum magnitudes of $J = 23.5$ mag and line fluxes of $\sim 10^{-17}$ erg/s/cm$^2$ in the NIR arm. The galaxy evolution survey will exploit *grizy* imaging from the ongoing HyperSuprime Camera survey (cite?) in four deep fields: eCOSMOS, XMM-LSS, DEEP2-3, and ELAIS-N1. The combination of the galaxy evolution survey with WFIRST data would be very powerful:

- WFIRST will provide rest-frame optical galaxy morphologies for the continuum-selected galaxies at $0.7 < z < 2$ galaxies studied by PFS, allowing an unprecedented mapping between bulge growth, star formation rate, and galactic outflows that may finally address the mechanisms that shut off star formation in galaxies (Barro et al. 2013).

- Galaxy-galaxy lensing measurements enabled by WFIRST will allow mapping of the average dark matter halo properties of $z \sim 1.5 - 2$ galaxies as a function of physical properties derived from the spectra (e.g., Leauthaud et al. 2012).

- The integral-field unit will measure stellar population gradients in some of our galaxies; these can be examined both as a function of the nuclear spectra (e.g., emission line ratios, absorption line outflows) and as a function of galaxy structure from the imaging.

In the longer term, the wide wavelength coverage in the blue and high spectral resolution of PFS will provide an important counterpoint to the grism available on WFIRST.

- As emphasized by Newman et al. (2014), upcoming dark energy surveys place extremely stringent requirements on photometric redshifts. PFS is apparently the *only* instrument (prior to the 20m GMT) that has the combination of sensitivity and multiplexing needed to build up adequate photometric redshift training samples.

- One might imagine performing a PFS survey over the HLS area to $Y \sim 22$ AB mag over a couple of years to provide a high fidelity grid of spectroscopic redshifts for continuum-selected galaxies down to , to try and understand the bias of the emission-line galaxies observable by the WFIRST grism. From such a survey, one could improve photometric redshifts through clustering (e.g., Newman et al.; Menard et al.) and do any number of exciting astrophysical experiments as well.

- PFS will also have a moderate resolution ($R \sim 5000$) red arm ideal for measuring individual stellar abundances (Kirby et al.). The combination of PFS spectroscopy and WFIRST imaging would make a very powerful combination for Local Group studies.

We should plan now to maximize the joint science returns between planned PFS and WFIRST surveys.

Michael Strauss (Princeton University), strauss@astro.princeton.edu

## Synergy Between LSST and WFIRST

Background

WFIRST and LSST were the two highest priority space and ground-based missions in the 2010 Decadal Survey. The primary mission of both facilities is to carry out wide-field imaging surveys of the sky in multiple bands; these surveys will enable a wide range of science, including weak lensing, large-scale structure in the distribution of galaxies, supernovae, the evolution of galaxies, and stellar populations. LSST is an optical telescope, and will do deep repeated broad-band photometry; the near-infrared and spectroscopic capabilities of WFIRST will allow much more accurate photometric and spectroscopic galaxy redshifts in regions of survey overlap.

WFIRST

The spectroscopic survey of WFIRST will allow redshifts to be measured for the brighter galaxies in the survey area. To go deeper will require photometric redshifts, which photometry in the near-infrared alone will not allow. LSST will survey 18,000 square degrees in the Southern Hemisphere in six filters ($u,g,r,i,z,y$) to a point-source depth of $r\sim$27.5. The overlap with WFIRST should be several thousand square degrees, in which we will have photometry from the $u$-band to H or K, allowing for excellent photometric redshifts. WFIRST image quality will allow galaxy shapes to be measured to fainter magnitudes than LSST can go; combining with the photometric and spectroscopic redshifts will allow much enhanced cosmic shear measurements as a function of cosmic time. The broad wavelength coverage will also allow accurate estimates of star formation rates, stellar masses, and dust content for galaxies at a wide range of redshifts, as well as for stellar populations in the Milky Way through SED fitting. The multi-band photometry will reveal and characterize rare populations of stars and galaxies, including "dropouts" at high redshift. LSST will visit any given point on the sky roughly 1000 times over ten years, giving detailed information on variability and proper motions. If the operations of WFIRST and LSST overlap in time, there will be an opportunity for combining the data streams to look for variability in both. LSST discoveries can also be followed up by WFIRST in its GO program.

Key Requirements

Depth – To match the photometric S/N in J/H/K for galaxies seen in LSST
Sky Coverage – To overlap the LSST footprint over at least 1000 square degrees
Resolution – Approach the diffraction limit with well-sampled images to measure galaxy shapes
Grism – For spectroscopic redshifts of large populations of galaxies.
Wavelength Coverage – >2 NIR filters to extend the galaxy SEDs to beyond the observed optical
Data Processing – Allow joint processing with LSST data to put photometric measurements on a common scale to optimize matching and cross-comparison of the two datasets.

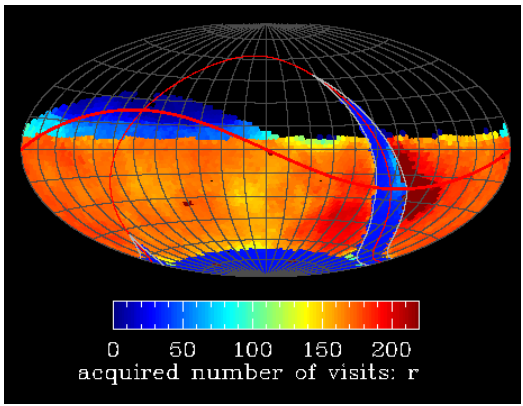

A realization of the sky coverage of LSST in the r-band, represented in Equatorial coordinates. The color represents the number of visits across the sky in a simulation of LSST operations: low Galactic latitudes, the Southern equatorial cap, and regions in the North along the ecliptic receive fewer visits. There is similar coverage in the u, g, i, z, and y filters. From the LSST Science Book (2009).

P. Capak (Caltech), capak@astro.caltech.edu

## Synergies between Euclid and WFIRST

The Euclid and WFIRST missions are complementary and have a number of synergies for both cosmological measurements and for extragalactic science. These are summarized below:

1) A large uncertainty for both Euclid and WFIRST are systematics and assumptions in the Weak Lensing (WL) and Baryon Acoustic Oscillation (BAO) measurements (e.g. Refregier et al. 2004 ApJ, 127, 310, Hirata & Eisenstein 2009, astro2010,127). Euclid and WFIRST probe different optimizations for these measurements, mitigating many of these systematic effects. In particular, the weak lensing measurements will use different detector technology and probe different wavelengths providing a truly independent measurement of the shear signal. In addition both the Lensing and BAO surveys will probe different galaxy populations with different biases, yielding different systematic effects in the measurements.

2) WFIRST will add 1-2um images of similar depth and resolution as those provided by Euclid in the optical (0.6-1um) enabling a range of galaxy evolution science. Star formation is known to drop precipitously at z<2 (Hopkins et al. 2006, ApJ, 651, 142), and this is believed to be a largely secular (internal) process best probed with large statistics (e.g. Peng et al. 2010, ApJ, 721, 193, Wuyts et al. 2011, ApJ, 742, 96). Studying what is causing this global decline in star formation requires resolved high-resolution rest frame UV measurements that probe the instantaneous star formation rate and resolved rest frame optical measurements to probe the accumulated stellar mass. WFIRST will not independently probe the rest frame Ultra-Violet (UV) at z<~1.8 and Euclid has poor sensitivity and resolution to the rest-frame optical light at z>0.7, so both instruments will be required to conduct these studies.

At z>4 the global star formation rate density is rapidly increasing, peaking at z~2 (Hopkins et al. 2006, ApJ, 651, 142). However, WFIRST alone will only be able to select and study these first galaxies at z>8 due to its lack of an optical channel, while Euclid will not have the near-IR sensitivity to probe "typical" systems at z>5. So, again both instruments are required to study these first epochs of galaxy formation.

3) If WFIRST includes longer wavelength (1.5-2.5um) coverage the spectra will complement the 0.9-2um spectra from Euclid. This increased wavelength coverage will expand the redshift range over which standard spectral line diagnostics (e.g. Kewley & Ellison 2008, ApJ, 681, 1183) can be used from 1<z<2 with Euclid and 2<z<2.8 with WFIRST to 1<z<2.8 with the combined instruments. This redshift range is crucial because star formation is known to peak and start declining in the 1<z<3 redshift range and the combination of Euclid and WFIRST will enable studies of ionization parameters, AGN content, and metalicity during this key epoch.

**Appendix K    Microlensing Path Forward**

The basic input ingredients that are required for estimating the yield of a microlensing exoplanet survey are (1) a model for the spatial, kinematic, and luminosity distribution of source stars, (2) a model for the spatial, mass, and kinematic distributions of the host lenses, which are then used to estimate microlensing event rates and event parameter distributions, and (3) a model for the probability distribution of planets as a function of planet mass and semi-major axis, and host mass and distance.

Unfortunately, for the regimes of interest for WFIRST-AFTA, the properties of the populations of sources and host lenses are poorly constrained by empirical data, leading directly to relatively large uncertainties in the final yields. (We note, however, that these uncertainties are shared amongst all of the different WFIRST designs. We therefore expect the relative yields between the designs to be substantially more robust.) In particular, the magnitude distribution of source stars in the fields of interest has not been measured in the pass bands and to the faint magnitudes that will be probed by WFIRST-AFTA. Similarly, the microlensing event rates for some of the fields of interest have not been measured, as these fields typically suffer high optical extinction, and to date all microlensing surveys have been performed in the far optical.

Another, somewhat subtler, source of uncertainty in the planet yields that may significantly impact the choice of target fields is the relative frequency of planets in the Galactic bulge versus disk. The relative contribution of the bulge and disk lenses to the event rate varies as a function of Galactic latitude b, with bulge lenses expected to dominate at low |b|. If the Galactic bulge happens to be devoid of planets, e.g., because of an extreme radiation environment during a starburst-like bulge formation event (Thompson 2013), then it would likely be desirable to avoid such low latitude fields.

These uncertainties motivate the precursor observations proposed in Appendix G, specifically a ground-based, near-IR survey, and a variety of optical and near-IR HST observations of (a subset of) the WFIRST-AFTA target fields.

In addition, there are several additional activities that are needed to properly prepare for the WFIRST-AFTA microlensing survey. These include refined predictions of the science yield of WFIRST-AFTA, which may in turn affect the survey strategies and mission requirements, and focused development of the US microlensing community, which will ensure that the maximum amount of science is extracted from the microlensing survey.

- **Detailed properties of detected systems.** Detailed and quantitative estimates of the ability to measure detailed properties of the detected planetary systems are needed, including expectations for measurement uncertainties for the masses, distances, and orbital parameters of the planets from the WFIRST-AFTA data alone, and by combining WFIRST-AFTA data with contemporaneous or follow-up observations from the ground and other satellites.

- **Refined yields.** The yields for planets near the edge of the sensitivity of WFIRST-AFTA suffer from additional uncertainties beyond those arising from the sources discussed above. These uncertainties arise from the fact that there is a relatively strong dependence of number of detections on the minimum detection threshold in these regimes, which include planets with very small or very large separations, and very low mass planets. As a result of the strong scaling with detection threshold, small differences in the assumptions and approximations needed to make these predictions result in large changes in the estimates of the number of detected planets.

- **Yields of Habitable Planets.** Predictions for the yield of habitable planets suffer from all of the uncertainties above, but are also sensitive to additional assumptions, such as the mass-bolometric luminosity relationship for stars in the bulge and disk, the age and metallicity of the stars in the bulge and disk, and the precise definitions for the mass and semi-major axes boundaries of the habitable zone. Therefore, the yields of habitable planets presented in 2.4.2.7.2 are quite uncertain, and substantially more work needs to be done to provide robust estimates of the habitable planet yield.

- **Development of Exoplanet Microlensing Expertise.** The US microlensing community is quite small with only 2 PhD granting institutions active in microlensing research, and only 4 US researchers beyond the postdoctoral level. NASA would be better prepared for the WFIRST-AFTA exoplanet microlensing survey if it would support some of the current and near future exoplanet microlensing programs at a higher level. These programs include ground-based exoplanet microlensing surveys and follow-up observations, space-based microlensing parallax observations with Spitzer and Kepler, and high angular resolution follow-up observations with the Hubble Space Telescope and Adaptive Optics systems to develop the WFIRST-AFTA method for measuring planet and host star masses. Research to develop tools that reduce barriers for other scientists to enter the microlensing field would be especially welcome.

## Appendix L    Acronym List

| | |
|---|---|
| 2MASS | 2 Micron All Sky Survey |
| 4MOST | 4 Metre Multi-Object Spectroscopic Telescope |
| ΛCDM | Lambda Cold Dark Matter |
| A&A | Astronomy and Astrophysics |
| ACS | Advanced Camera for Surveys |
| ACS | Attitude Control System |
| AFTA | Astrophysics Focused Telescope Assets |
| AGB | Asymptotic Giant Branch |
| AGN | Active Galactic Nuclei |
| AIT | Assembly, Integration and Test |
| AJ | Astronomical Journal |
| aka | Also Known As |
| ALMA | Atacama Large Millimeter/submillimeter Array |
| ALTAIR | ALTitude conjugate Adaptive optics for the InfraRed (Gemini instrument) |
| AMS | Aft Metering Structure |
| AO | Adaptive Optics |
| AO | Announcement of Opportunity |
| AOX | AOA Xinetics |
| APh | Astroparticle Physics |
| ApJ | Astrophysical Journal |
| ApJS | Astrophysical Journal Supplement Series |
| APLC | Apodized Pupil Lyot Coronagraph |
| AU | Astronomical Unit |
| BAO | Baryon Acoustic Oscillations |
| BDBS | Blanco DECam Bulge Survey |
| BH | Black Hole |
| BOSS | Baryon Oscillation Spectroscopic Survey |
| C&DH | Command and Data Handling |
| CADC | Canadian Astronomy Data Centre |
| CANDELS | Cosmic Assembly Near-infrared Deep Extragalactic Legacy Survey |
| CASTOR | Cosmological Advanced Survey Telescopes for Optical and uv Research |
| CCD | Charged Coupled Device |
| CCE | Cryocooler Electronics |
| CDM | Cold Dark Matter |
| CDR | Critical Design Review |
| CDS | Correlated Double Sample |
| CEI-OU | Centre for Electronic Imaging at the Open University |
| CfAI | Center for Advanced Instrumentation |
| CFHT | Canada-France-Hawaii Telescope |
| CFP | Call for Proposal |
| CFTHLS | CFHT Legacy Survey |
| CGI | Coronagraph Instrument |
| CIB | Cosmic Infrared Background |
| CIC | Clock Induced Charge |
| CIEMAT | Centro de Investigaciones Energéticas, Medioambientales y Tecnológicas |
| CITA | Canadian Institute for Theoretical Astrophysics |
| CLASH | Cluster Lensing And Supernova survey with Hubble |
| CMB | Cosmic Microwave Background |
| CMD | Color-Magnitude Diagram |
| CNES | Centre National d'Études Spatiales |
| CoC | Center of Curvature |
| Co-I | Co-Investigator |
| COL | Collimator |
| COR | Coronagraph |
| COSMOS | Cosmic Evolution Survey |
| CPT | Comprehensive Performance Test |
| CSA | Canadian Space Agency |
| CSB | Collimator Sub-Bench |
| CTE | Coefficient of Thermal Expansion |
| DC | Direct Current |
| DES | Dark Energy Survey |
| DESI | Dark Energy Spectroscopic Instrument |
| DETF | Dark Energy Task Force |
| DDX | Debris Disk Explorer |
| DECam | Dark Energy Camera |
| DLR | Deutsches Zentrum für Luft-und Raumfahrt |
| DM | Deformable Mirror |
| DOE | Department of Energy |
| DOF | Degree of Freedom |
| DRM | Design Reference Mission |
| DRM1 | Design Reference Mission #1 |
| DRM2 | Design Reference Mission #2 |
| DSN | Deep Space Network |
| DTAC | Detector Technology Assessment Committee |
| eBOSS | Extended Baryon Oscillation Spectroscopic Survey |
| EDU | Engineering Development Unit |
| EE | Encircled Energy |
| ELG | Emission Line Galaxy |
| ELT | Extremely Large Telescope |
| EM | Electron Multiplying |
| EMCCD | Electron Multiplying Charge Coupled Device |
| EMI/EMC | Electromagnetic Interference/Electromagnetic Compatibility |
| ENSCI | Euclid NASA Science Center at IPAC |
| EPO | Education and Public Outreach |
| eROSITA | extended ROentgen Survey with an Imaging Telescope Array |
| EROS | Expérience pour la Recherche d'Objets Sombres |
| ESA | European Space Agency |
| ESO | European Southern Observatory |
| ETC | Exposure Time Calculator |
| ETE | End to End |
| EW | Element Wheel |
| EXCEDE | The EXoplanetary Circumstellar Environments and Disk Explorer |
| ExoPAG | Exoplanet Exploration Program Analysis Group |
| F2 | Fold Flat #2 |
| FEM | Finite Element Model |
| FDF | Flight Dynamics Facility |
| FGS | Fine Guidance Sensor |
| FLAMES | Fibre Large Array Multi Element Spectrograph |
| FLI | First Light Imaging |
| FMA | Fold Mirror Assembly |
| FMOS | Fiber Multi Object Spectrograph |
| FOA | Forward Optics Assembly |
| FoM | Figure of Merit |
| FoV | Field-of-View |
| FP | Focal Plane |
| FPA | Focal Plane Assembly |
| FPE | Focal Plane Electronics |
| FPGA | Field-Programmable Gate Array |
| FSA | Forward Structure Assembly |
| FSM | Fast Steering Mirror |
| FSW | Flight Software |
| FT | Functional Test |
| FUSE | Far Ultraviolet Spectroscopic Explorer |
| FWHM | Full-Width Half-Maximum |
| FY | Fiscal Year |
| GALEX | Galaxy Evolution Explorer |
| Gbps | Gigabits per Second |
| GC | Galaxy Clustering |
| GEO | Geosynchronous Orbit |
| GGL | Galaxy-Galaxy Lensing |
| GI | Guest Investigator |
| GMT | Giant Magellan Telescope |
| GO | Guest Observer |
| GOODS | Great Observatories Origins Deep Survey |
| GOODS-S | Great Observatories Origins Deep Survey-South |
| GPI | Gemini Planet Imager (Gemini instrument) |
| GPM | Global Precipitation Measurement |
| GR | General Relativity |
| GRS | Galaxy Redshift Survey (includes BAO & RSD) |
| GSE | Ground Support Equipment |
| GSFC | Goddard Space Flight Center |
| GSMT | Giant Segmented Mirror Telescope |
| GW | Gravitational Wave |
| GW | Guide Window |
| HGA | High Gain Antenna |
| HCIT | High Contrast Imaging Testbed |
| HgCdTe | Mercury Cadmium Telluride |
| HK | Housekeeping |
| HLC | Hybrid Lyot Coronagraph |
| HLS | High-Latitude Survey |
| HOSTS | Hunt for Observable Signatures of Terrestrial planetary Systems |
| HQ | Headquarters |
| HSC | Hyper Suprime-Cam |
| HST | Hubble Space Telescope |
| H/V | Horizontal and Vertical |
| HV | High Voltage |
| HZ | Habitable Zone |
| I&T | Integration and Test |
| IA | Intrinsic Alignment |
| IAP | Institut Astrophysique de Paris |

| IAU | International Astronomical Union |
|---|---|
| IC | Instrument Carrier |
| ICDH | Instrument Command and Data Handling |
| ICE | Institut de Ciències de l'Espai |
| ICG | Institute for Cosmology and Gravitation |
| ICL | Intracluster Light |
| IDL | Interactive Data Language |
| IDRM | Interim Design Reference Mission |
| IFAE | Institut de Física d'Altes Energies |
| IFS | Integral Field Spectrograph |
| IFT | Instituto de Física Teórica |
| IFU | Integral Field Unit |
| IGM | Intergalactic Medium |
| IM | Integrated Modeling |
| IMF | Initial Mass Function |
| IMO | Inverted Mode Operation |
| IndIGO | Indian Initiative in Gravitational-wave Observations |
| IOA | Institute of Astronomy |
| IOC | Instrument Operations Center |
| iPTF | Intermediate Palomar Transient Factory |
| IR | Infrared |
| IRAC | Infrared Array Camera |
| IRIS | Infrared Imaging Spectrograph (TMT instrument) |
| IRMS | Infrared Multi-object Spectrometer (TMT instrument) |
| IRSA | Infrared Science Archive |
| ISAS | Institute of Space and Astronautical Science |
| IUBMB | International Union of Biochemistry and Molecular Biology |
| IWA | Inner Working Angle |
| JASMINE | Japan Astrometry Satellite Mission for Infrared Exploration |
| JAXA | Japanese Aerospace Exploration Agency |
| JDEM | Joint Dark Energy Mission |
| JM | Jitter Mirror |
| JMAPS | Joint Milli-Arcsecond Pathfinder Survey |
| JPL | Jet Propulsion Laboratory |
| JRASC | Journal of the Royal Astronomical Society of Canada |
| JWST | James Webb Space Telescope |
| KAGRA | Kamioka Gravitational Wave |
| KBO | Kuiper Belt Object |
| KDP | Key Decision Point |
| KMOS | K-band Multi Object Spectrograph |
| KSC | Kennedy Space Center |
| L2 | Sun-Earth 2$^{nd}$ Lagrangian Point |
| LAM | Laboratoire d'Astrophysique de Marseille |
| LAT | Large Area Telescope |
| LBT-I | Large Binocular Telescope Interferometer |
| LCDM | Lambda Cold Dark Matter |
| LDCM | Landsat Data Continuity Mission |
| LESIA | Laboratoire d'Études Spatiales et d'Instrumentation en Astrophysique |
| LF | Luminosity Function |
| LIGO | Laser Interferometer Gravitational-Wave Observatory |
| LMC | Large Magellanic Cloud |
| LOM | Linear Optical Model |
| LOS | Line of Sight |
| LOWFS | Low-Order Wavefront Sensor |
| LOWFS/C | Low-Order Wavefront Sensor and Control |
| LRD | Launch Readiness Date |
| LRG | Luminous Red Galaxy |
| LRO | Lunar Reconnaissance Orbiter |
| LRP | Long-Range Plan |
| LSST | Large Synoptic Survey Telescope |
| M3 | Tertiary Mirror |
| M4 | Collimator Mirror (in the coronagraph instrument) |
| MACE | Mechanism and Attitude Control Electronics |
| MACHO | MAssive Compact Halo Objects |
| MACS | MAssive Cluster Survey |
| MADRAS | Mirror Active, Deformable and Regulated for Applications in Space |
| MANGA | Mapping Nearby Galaxies at APO |
| MAST | Mikulski Archive for Space Telescopes |
| Mbps | Megabits per Second |
| MCE | Mechanism Control Electronics |
| MCR | Mission Concept Review |
| MDR | Mission Definition Review |
| MEL | Master Equipment List |
| MGSE | Mechanical Ground Support Equipment |
| MIRI | Mid Infrared Instrument (JWST instrument) |
| ML | Microlensing |
| M-L | Mass-Luminosity |
| MLI | Multi-Layer Insulation |
| MMS | Multimission Modular Spacecraft |
| MMT | Multiple Mirror Telescope |
| MNRAS | Monthly Notices of the Royal Astronomical Society |
| MOA | Microlensing Observations in Astrophysics |
| MOBIE | Multi-Object Broadband Imaging Echellette (TMT instrument) |
| MOC | Mission Operations Center |
| MoO | Mission of Opportunity |
| MOS | Multi-Object Spectrograph |
| MPE | Max-Planck-Institut für extraterrestrische Physik |
| MPIA | Max-Planck-Institut für Astronomie |
| MPix | Megapixel |
| MRS | Module Restraint System |
| MS-DESI | Mid-Scale Dark Energy Spectroscopic Instrument |
| MSSL-UCL | Mullard Space Science Laboratory, University College London |
| MUF | Model Uncertainty Factor |
| MW | Milky Way |
| NACO | Nasmyth Adaptive Optics System (NAOS) Near-Infrared Imager and Spectrograph (CONICA) |
| NAOJ | National Astronomical Observatory of Japan |
| NAS | National Academy of Science |
| NASA | National Aeronautics and Space Administration |
| NFIRAOS | Narrow Field Infrared Adaptive Optics System (TMT instrument) |
| NGC | New General Catalog |
| NICMOS | Near Infrared Camera and Multi-Object Spectrometer |
| NIR | Near Infrared |
| NIRCam | Near Infrared Camera |
| NIRISS | Near-Infrared Imager and Slitless Spectrograph (JWST instrument) |
| NISP | Near Infrared Spectrometer and Photometer (Euclid Instrument) |
| NRA | NASA Research Announcement |
| NRC | National Research Council |
| NRO | National Reconnaissance Office |
| NS | Neutron Star |
| NSF | National Science Foundation |
| NWNH | New Worlds, New Horizons in Astronomy and Astrophysics |
| OA | Optical Alignment |
| OAP | Off-Axis Parabola |
| OB | Optical Bench |
| OBA | Outer Barrel Assembly |
| OBE | Outer Barrel Extension |
| OMC | Occulting Mask Coronagraph |
| ONERA | Office National d'Études et de Recherches Aérospatiales |
| OPD | Optical Path Difference |
| OTA | Optical Telescope Assembly |
| OVU | Optical Verification Unit |
| OWA | Outer Working Angle |
| PAF | Payload Attach Fitting |
| PanSTARRS | Panoramic Survey Telescope & Rapid Response System |
| PASP | Publication of the Astronomical Society of the Pacific |
| PAU | Physics of the Accelerating Universe |
| PBS | Polarizing Beam Splitter |
| PDR | Preliminary Design Review |
| PFS | Prime Focus Spectrograph |
| PHAT | Panchromatic Hubble Andromeda Treasury |
| Photo-z | Photometric Redshift |
| PI | Principal Investigator |
| PIAA-CMC | Phase-Induced Amplitude Apodization Complex Mask Coronagraph |
| PID | Proportional-Integral-Derivative |
| PISN | Pair Instability SuperNovae |
| PLATO | PLanetary Transits and Oscillations of stars |
| PM | Primary Mirror |
| PMA | Primary Mirror Assembly |
| PMN | Lead Magnesium Niobate |
| PNAS | Proceedings of the National Academy of Sciences of the United States of America |
| PSD | Power Spectral Density |
| PSDU | Power Switching & Distribution Unit |
| PSF | Point Spread Function |
| PSI | Planetary System Imager |
| PSR | Pre-Ship Review |
| PZCS | Photo-Z Calibration Survey |
| PZT | Lead Zirconate Titanate |
| QE | Quantum Efficiency |
| QLF | Quasar Luminosity Function |
| QSO | Quasi-Stellar Object (Quasar) |
| QUEST | Quasar Equatorial Survey Team |
| R1 | Relay #1 |
| R2 | Relay #2 |
| R3 | Relay #3 |
| RAAN | Right Ascension of the Ascending Node |
| RASCASSE | Réalisation d'un Analyseur de Surface d'onde pour le Contrôle de miroirs Actifs Spatiaux sur Sources Etendues |
| RM | Reverberation Mapping |
| RMS | Root Mean Square |
| ROIC | Readout Integrated Circuit |
| RSD | Redshift Space Distortion |
| RV | Radial Velocity |
| RWA | Reaction Wheel Assembly |
| S/C | Spacecraft |
| S/N | Signal/Noise |
| SAN | Speckle Area Nulling |
| SBCC | Single Board Computer Card |
| SCA | Sensor Chip Assembly |
| SCC | Self Coherent Camera |
| SCE | Sensor Cold Electronics |
| SCExAO | Subaru Coronagraphic Extreme Adaptive Optics |
| sCMOS | Scientific Complementary Metal-Oxide Semiconductor |
| SCU | SCE Control Unit |
| SCUBA | Submillimetre Common-User Bolometer Array second generation) |
| SDO | Solar Dynamics Observatory |
| SDT | Science Definition Team |
| SSDIF | Spacecraft Systems Development and Integration Facility |
| SDSS | Sloan Digital Sky Survey |
| SED | Spectral Energy Distribution |
| SES | Space Environment Simulator |
| SF | Star-Forming |
| SFR | Star Formation Rates |
| SiC | Silicon Carbide |
| SIR | Systems Integration Review |
| SIT | Science Investigation Team |
| SKA | Square Kilometer Array |
| SM | Secondary Mirror |
| SMA | Secondary Mirror Assembly |
| SMBH | Supermassive Black Hole |
| SMC | Small Magellanic Cloud |
| SN | Supernova |
| SNe | Supernovae |
| SNLS | Supernova Legacy Survey |
| SNR | Signal to Noise Ratio |
| SOC | Science Operations Center |
| SOHO | Solar and Heliospheric Observatory |
| SP | Shaped Pupil |
| SPC | Shaped Pupil Coronagraph |
| SPHERE | Spectro-Polarimetric High-contrast Exoplanet Research (VLT instrument) |
| SPICES | Spectro-Polarimetric Imaging and Characterization of Exoplanetary Systems |
| SPIRE FTS | Spectral and Photometric Imaging Receiver Fourier-Transform Spectrometer |
| SPIRou | SpectroPolarimetre Infra-Rouge |
| SPLASH | Spitzer Large Area Survey with Hyper-Suprime-Cam |
| SPLC | Shaped Pupil Lyot Coronagraph |
| SRR | System Requirements Review |
| STE | Special Test Equipment |
| STM | Safe to Mate |
| STOP | Structural/Thermal/Optical |
| STScI | Space Telescope Science Institute |
| SuMIRe | Subaru Measurement of Images and Redshifts |
| SWEEPS | Sagittarius Window Eclipsing Extrasolar Planet Search |
| SWG | Science Working Group |
| SZ | Sunyaev-Zel'dovich |
| TBD | To Be Determined |
| TAC | Technology Assessment Committee |
| TAC | Time Allocation Committee |
| TBR | To Be Resolved |
| TDEM | Technology Development for Exoplanet Missions |
| TDP | Technology Development Plan |
| TESS | Transiting Exoplanet Survey Satellite |
| TMM | Thermal Math Model |
| TMT | Thirty Meter Telescope |
| TNO | Trans-Neptunian Objects |
| TRL | Technology Readiness Level |
| TVAC | Thermal Vacuum |
| UDF | Ultra Deep Field |
| UK | United Kingdom |
| UKATC | United Kingdom Astronomy Technology Centre |
| UKIDSS | UKIRT Infrared Deep Sky Survey |
| UKSA | United Kingdom Space Agency |
| ULE | Ultra Low Expansion Fused Silica |
| US | United States |
| UV | Ultraviolet |
| Vdc | Volts Direct Current |
| VIS | Visible Instrument (Euclid instrument) |
| VISTA | Visible and Infrared Survey Telescope for Astronomy |
| VLASS | Very Large Area Sky Survey |
| VLT | Very Large Telescope |
| VNC | Visible Nulling Coronagraph |
| VST | VLT Survey Telescope |
| VVC | Vector Vortex Coronagraph |
| VVV | Vista Variables in the Via Lactea |
| WACO | WFIRST-AFTA Coronagraph |
| WF | Wide-Field |
| WFC3 | Wide Field Camera 3 |
| WFC3/IR | Wide Field Camera 3/Infrared channel |
| WFCS | Wavefront Control System |
| WFE | Wavefront Error |
| WFI | Wide-Field Instrument |
| WFIRST | Wide-Field Infrared Survey Telescope |
| WIM | Wide-field Imaging Mode |
| WIMPS | Weakly Interacting Massive Particles |
| WISE | Wide-field Infrared Survey Explorer |
| WL | Weak Lensing |
| WSM | Wide-field Spectroscopy Mode |
| XMM | X-ray Multi-Mirror Mission |
| Z | Redshift |
| ZTF | Zwicky Transient Facility |
| ZWFS | Zernike Wavefront Sensor |

## Appendix M    References